\documentclass[12pt,twoside,singlespace]{mitthesis}
\usepackage{lgrind}
\usepackage{graphicx}
\usepackage{amsmath}
\usepackage{subfigure}
\begin{document}

%
%
%
%
%
%
%
\title{Recoil Polarization Measurements of the Proton Electromagnetic Form Factor Ratio to High Momentum Transfer}

\author{Andrew James Ruehe Puckett}
\department{Department of Physics}
\degree{Doctor of Philosophy in Physics}
\degreemonth{February}
\degreeyear{2010}
\thesisdate{October 7, 2009}


\supervisor{William Bertozzi}{Professor of Physics}
\supervisor{Charles Perdrisat}{Professor of Physics}
\supervisor{Shalev Gilad}{Senior Research Scientist}
\chairman{Professor Krishna Rajagopal}{Associate Department Head for Education}

\maketitle



\cleardoublepage
\setcounter{savepage}{\thepage}
\begin{abstractpage}
%
%
%
\paragraph{}
The electromagnetic form factors of the nucleon characterize the effect of its internal structure on its response to an electromagnetic probe as studied in elastic electron-nucleon scattering. These form factors are functions of the squared four-momentum transfer $Q^2$ between the electron and the proton. The two main classes of observables of this reaction are the scattering cross section and polarization asymmetries, both of which are sensitive to the form factors in different ways. When considering large momentum transfers, double-polarization observables offer superior sensitivity to the electric form factor. This thesis reports the results of a new measurement of the ratio of  the electric and magnetic form factors of the proton at high momentum transfer using the recoil polarization technique. A polarized electron beam was scattered from a liquid hydrogen target, transferring polarization to the recoiling protons. These protons were detected in a magnetic spectrometer which was used to reconstruct their kinematics, including their scattering angles and momenta, and the position of the interaction vertex. A proton polarimeter measured the polarization of the recoiling protons by measuring the azimuthal asymmetry in the angular distribution of protons scattered in CH$_2$ analyzers. The scattered electron was detected in a large-acceptance electromagnetic calorimeter in order to suppress inelastic backgrounds. The measured ratio of the transverse and longitudinal polarization components of the scattered proton is directly proportional to the ratio of form factors $G_E^p/G_M^p$. The measurements reported in this thesis took place at $Q^2=$5.2, 6.7, and 8.5 GeV$^2$, and represent the most accurate measurements of $G_E^p$ in this $Q^2$ region to date.

\end{abstractpage}


\cleardoublepage

\section*{Acknowledgments}
\paragraph{}
This thesis would not have been possible without the support and efforts of many, many people. It would be impossible to give all due credit to all of the individual efforts that went into designing, planning, staging, and executing this experiment, and the subsequent analysis of the data. Therefore, any omissions from the following acknowledgments are in no way meant to diminish those contributions.

I would like to thank the members of the GEp-III collaboration whose tireless efforts made experiments E04-108 and E04-019 a success. This includes the spokespeople; Ed Brash, Mark Jones, Charles Perdrisat and Vina Punjabi, and the other outstanding physicists involved in the core collaboration. I thank Frank Wesselmann for his successful leadership of the testing, installation, commissioning and operation of the Focal Plane Polarimeter, and his significant contribution to the data analysis in the form of the event decoding and tracking software for the FPP drift chambers. I thank Mark Jones for his encyclopedic knowledge of experimental Hall C and for his outsized contribution to making the various parts of the experiment work together as a whole, and for being called upon more often than not to troubleshoot problems with the experiment as they occurred. Mark's many important contributions to the data analysis included the optimization of the HMS reconstruction coefficients. Much of what I have learned about the actual hands-on requirements of conducting a medium-energy accelerator-based nuclear physics experiment, I have learned from Mark. I thank Lubomir Pentchev for his successful stewardship of the testing, installation, commissioning and operation of the BigCal detector, for many essential discussions during the analysis of the data, and for checking my work in many important areas. Lubomir also deserves credit as the preeminent expert on the calculation of spin precession in the HMS, for performing the COSY calculations of the precession matrix elements and designing the non-dispersive optical studies which reduced the systematic uncertainty in the calculation. Lubomir's understanding of the data analysis requirements of recoil polarization experiments is unsurpassed. 

I thank Ed Brash, Charles Perdrisat and Vina Punjabi for their excellent leadership of the collaboration and management of the planning and execution of the experiment, and also for their many important contributions to the analysis of the data. I wish to thank Charles in particular for giving me the opportunity to participate in this important research, and for his mentorship and support of my growth as a scientist while at JLab, and for reading and proofreading this thesis on a rather compressed timetable. I also thank Charles for serving on my thesis committee and for attending my thesis defense at MIT. I thank the collaborators from Dubna and Protvino for their construction of and operational support for, respectively, the polarimeter and the calorimeter. I thank my fellow Ph.D. students, Mehdi Meziane of William and Mary and Wei Luo of Lanzhou for their contributions to the experiment and the data analysis, for their camaraderie during this shared effort spanning many years, and for the many memorable hours spent in the Hall C counting house monitoring the experiment. Mehdi's efforts were essential to the successful operation of the FPP and his exhaustive analysis of the FPP data has been indispensable in understanding its characteristics and performance, and he will complete the analysis of the data at $Q^2=2.5$ GeV$^2$ to finalize these important results. Wei's operational expertise with BigCal, particularly with respect to the maintenance and operation of the high voltage system and the calibration of the PMT high voltages for gain matching purposes was crucial to the success of the experiment. Wei also deserves much of the credit for the adaptation of SIMC to the conditions of this experiment, including the addition of $\pi^0$ photoproduction, and his in-depth analysis of the inelastic data promises exciting results. 

I wish to thank the entire Hall C scientific and technical staff for their operational support of these experiments. I thank Steve Wood for, in particular, his expertise with the Hall C data acquisition system and analysis software. I wish to recognize the efforts of the target group, and particularly Greg Smith and Dave Meekins, for ensuring the successful operation of the scattering chamber and the 20 cm LH$_2$ cell. I thank Steve Lassiter and Mike Fowler for their operational support for the HMS magnets, Howard Fenker, Joe Beaufait and others for their support of the HMS detectors and other general Hall C systems. I wish to thank Tanja Horn and Dave Gaskell for their outstanding contributions to running the experiment, including performing the M\"{o}ller and arc energy measurements. I thank Walter Kellner and his staff for the successful installation of the experiment, including but not limited to the installation of the FPP in the HMS hut, the construction of the electronics bunker, and the installation and movement of BigCal. I thank Hamlet Mrktchyan, Albert Shahinyan and others for their efforts, including but not limited to the commissioning of the HMS hodoscopes and the enormous cabling effort required to instrument BigCal. I thank all the students who aided these and other efforts. 

I also wish to thank the accelerator staff and operators for their tireless and outstanding work delivering the high-energy, high-intensity, high-polarization, 100\% duty-factor, stable CW electron beam to Hall C. I thank all the scientists and students who took shifts on the experiment so that it could run continuously. I wish to thank Earle Lomon for helpful discussions of the Vector Meson Dominance model which has successfully anticipated the new results.

I wish to thank Shalev Gilad for his mentorship, support, and supervision of my graduate work while at MIT and at Jefferson Lab. His regular visits to Jefferson Lab went a long way in helping me to stay on track. I wish to thank Bill Bertozzi for accepting me into the Nuclear Interactions Group and for his unwavering and always appreciated support and advocacy of my Ph.D. candidacy, and for teaching me a great deal of what I have learned about nuclear physics during my time at MIT. I thank Bill and Shalev for their guidance of my research direction and encouragement of my efforts, and the wealth of valuable advice they have given me throughout this effort.

I am grateful to all of the friends and colleagues I have come to know at MIT and at JLab. I am grateful to family members who have supported my efforts and been there for me when I needed them. Finally, and most importantly, I wish to thank my beloved wife of nearly two years, the love of my life, Emily. The best thing that has ever happened or will ever happen to me is the fact of your presence in my life. The completion of this challenging endeavor would not have been possible without your love, support, and understanding, and I am eternally grateful to you for staying with me and enduring the various challenges to reach the long-awaited conclusion of this graduate school adventure. I recognize that you have put your life on hold in a significant way to see us through to this Ph.D., and for that I am forever indebted; I have a lot of ground to make up in supporting your goals and aspirations, and I know that together we can find a way for you to do what makes you happy. I know that our bond can never be broken, and I could not be a more happily married man. I dedicate this thesis to you.


\pagestyle{plain}
\tableofcontents
\newpage
\listoffigures
\newpage
\listoftables

\chapter{Introduction}
\label{chapter1}
\paragraph{}
The nucleon occupies a position of fundamental importance in the physics of strongly interacting matter. It is the building block of the nucleus, and it is the only stable baryon. The challenge of understanding its structure and dynamics has engaged experimental and theoretical physicists in an effort that has spanned generations and continues to this day. The first clue that nucleons are not pointlike, elementary particles came from Otto Stern's measurements of the magnetic moment of the proton and the deuteron in 1933\cite{Stern1933}\footnote{The first measurement of the free neutron magnetic moment using magnetic resonance methods was performed by Bloch and Alvarez in 1940\cite{BlochAlvarez1940}.}. These measurements showed drastic deviation from the expected value in the Dirac equation for a ``point'' spin-$\frac{1}{2}$ particle, implying that protons and neutrons are composite and have internal structure. Since that groundbreaking discovery, a modern understanding of the nucleon has taken shape, answering many questions and raising many more.  

In the Standard Model of elementary particles and interactions, protons and neutrons are composed of elementary fermions called quarks, which are bound together by strong ``color'' interactions. Quarks come in six different ``flavors'': u(``up''), d(``down''), c(``charm''), s(``strange''), t(``top''), and b(``bottom''). They are grouped into families by increasing mass, as shown in table \ref{quarktable}. Each family has an ``up'' member with charge $+2/3e$ and a ``down'' member with charge $-1/3e$. 
\begin{table}[h]
  \label{quarktable}
  \begin{center}
    \begin{tabular}{|c|c|c|c|}
      \hline $Q=+\frac{2}{3} e$ & u & c & t \\
      \hline $Q=-\frac{1}{3} e$ & d & s & b \\ 
      \hline
    \end{tabular}
  \end{center}
  \caption{Quark families in the Standard Model}
\end{table}
The distinguishing characteristic of quarks is that they have color and therefore feel color forces, as opposed to other elementary fermions, which are collectively referred to as leptons. Both quarks and leptons experience electromagnetic and weak interactions. While leptons all exist as free particles in nature or can be readily produced in the laboratory, quarks only occur in ``colorless'' combinations that fall into one of two categories: 
\begin{enumerate}
  \item \emph{Mesons}, which include pions and kaons, are understood to be quark-antiquark bound states
  \item \emph{Baryons}, including nucleons, are understood to be bound states of three quarks.
\end{enumerate}
This dichotomy is a consequence of the fact that quarks come in three different colors, called red, green, and blue, which form an SU(3) symmetry group that is thought to be an exact symmetry of nature. To form an SU(3) singlet state requires, at a minimum, either three quarks to form the singlet $3 \otimes 3 \otimes 3$ state, or a quark and an antiquark with color and anticolor to form a singlet $3 \otimes \bar{3}$ state. 

The history of quarks and color begins with the advent of modern particle physics, in which rapid advances in particle accelerator technology provided increasingly energetic collisions for physicists to study, and the number of known strongly interacting particles proliferated rapidly. The growing ``zoo'' of baryons and mesons inspired efforts to classify them, i.e., to write down the ``periodic table'' of subnuclear constituents from which all the newly discovered particles could be built; in other words, the quark families of the Standard Model.   

It was the nearly identical masses of the proton and neutron that originally led physicists to the conclusion that they are different quantum states of a single entity, the nucleon. In the language of spin, the proton and the neutron, respectively, are the ``isospin-up'' and ``isospin-down'' states of an isospin-$1/2$ system. In the basic quark model, they are composed of u(``up'') and d(``down'') quarks, so named because they have isospin up and down, respectively. The proton contains two up quarks and one down quark, with net charge $2\left(\frac{2}{3}e\right) - \frac{1}{3}e = +e$ and isospin $+1/2$, while the neutron has two down quarks and one up quark, for a net charge of $\frac{2}{3}e - 2\left(\frac{1}{3}e\right) = 0$ and isospin $-1/2$. Carrying this idea further, as Gell-Mann did with his ``Eightfold-Way'' symmetry, the nucleons can be grouped together with all baryons that can be built from the three lightest quark flavors (u, d, and s) according to an approximate flavor SU(3) symmetry, reflecting the approximate mass equality of the lightest three quark flavors. Building a three-quark state from the basic triplet of flavors gives 
\begin{equation}
  3 \otimes 3 \otimes 3 = 10 \oplus 8 \oplus 8 \oplus 1 \nonumber
\end{equation}
unique flavor states grouped according to their exchange symmetry properties. Since baryons are fermions, their overall flavor-spin-orbital-color wavefunction must be antisymmetric under the exchange of any two quarks. Since only colorless quark combinations occur in nature, all baryons are in the color-singlet state, so that the combined flavor-spin-orbital state must be symmetric to guarantee overall antisymmetry. For the lightest, ground state baryons, it is assumed that the orbital wavefunction of the three quarks is a symmetric S-state, so that the combined spin-flavor state must also be symmetric. It is found that the lightest observed baryons can be grouped into a spin-$3/2$ decuplet\footnote{In fact, this was one of the original motivations for introducing the color quantum number, since it would be impossible to preserve the Pauli principle for the decuplet states without it.} and a spin-$1/2$ octet. The decuplet results from the combination of the ten symmetric flavor states with the symmetric spin states resulting when three spin-$1/2$ quarks are combined to give spin-$3/2$. The octet results from the symmetric product of the eight mixed-symmetry flavor states with the appropriate mixed-symmetry spin states resulting when three quark spins are combined to give total spin-$1/2$. The nucleons fall within the octet, and their spin-flavor wavefunction can be written down simply by respecting the required symmetry. For example, a proton with spin up has the state
\begin{equation}
  \left|p, \uparrow \right\rangle = \sqrt{\frac{1}{2}}\left[\left|p_S\right\rangle \left|\chi_S, \uparrow \right\rangle + \left|p_A\right\rangle \left|\chi_A, \uparrow \right\rangle \right] \nonumber 
\end{equation}
where $p$ and $\chi$ denote the flavor and spin states, respectively, while the subscripts S and A refer to symmetry and antisymmetry under exchange of the first two quarks. The flavor combination for the proton is two up quarks and a down quark. For the antisymmetric state, the flavor wave function is $\left|p_A\right\rangle = \sqrt{\frac{1}{2}}\left| udu - duu \right\rangle$ while for the symmetric state, the correct flavor combination is $\left|p_S\right\rangle = \sqrt{\frac{1}{6}}\left| udu + duu - 2uud \right\rangle$ which is obtained by requiring orthogonality to both $\left| p_A \right\rangle$ and the totally symmetric decuplet uud state. Similarly, the antisymmetric and symmetric spin-$1/2$ states are $\left|\chi_A, \uparrow \right\rangle = \sqrt{\frac{1}{2}}\left| \uparrow \downarrow \uparrow - \downarrow \uparrow \uparrow \right\rangle$ and $\left|\chi_S, \uparrow \right\rangle = \sqrt{\frac{1}{6}}\left| \uparrow \downarrow \uparrow + \downarrow \uparrow \uparrow - 2\uparrow \uparrow \downarrow \right\rangle$, respectively. The total spin-flavor state of a proton with spin up, then, is: 
\begin{eqnarray}
  \left|p, \uparrow \right\rangle &=& \sqrt{\frac{1}{18}}\left[2\left(\left|u\uparrow u\uparrow d\downarrow \right\rangle + \left| u\uparrow d\downarrow u\uparrow \right\rangle + \left| d\downarrow u\uparrow u\uparrow \right\rangle \right) - \right. \nonumber \\
    & & \left(\left| u\uparrow u\downarrow d\uparrow \right\rangle + \left| u\downarrow u\uparrow d\uparrow \right\rangle + \left| u\uparrow d\uparrow u\downarrow\right\rangle + \left| u\downarrow d\uparrow u\uparrow \right\rangle \right. + \nonumber \\
    & & \left. \left. \left| d\uparrow u\uparrow u\downarrow \right\rangle + \left| d\uparrow u\downarrow u\uparrow \right\rangle \right) \right] \nonumber \\
  &=& \sqrt{\frac{1}{18}}\left[2\left| u\uparrow u\uparrow d\downarrow \right\rangle - \left| u\uparrow u\downarrow d\uparrow \right\rangle - \left| u\downarrow u\uparrow d\uparrow \right\rangle \right. \nonumber \\ 
    & & \left. + \quad\mbox{flavor permutations}\quad \right] \label{protonquarkwavefunc}
\end{eqnarray}
The proton's magnetic moment is obtained from this quark model spin-flavor wavefunction as follows: 
\begin{equation}
  \label{momentdef}
  \mu_p = \sum_i \left\langle p,\uparrow \right| \frac{Q_i}{2m_i} \sigma_3 \left| p,\uparrow \right\rangle
\end{equation}
Substitution of the proton's spin-flavor wavefunction \eqref{protonquarkwavefunc} into the definition of the magnetic moment \eqref{momentdef} gives:
\begin{eqnarray}
  \mu_p &=& 3 \times \frac{1}{18}\left[4\left(2\left(\frac{2}{3}\frac{e}{2m_u}\right) - \left(-\frac{1}{3}\frac{e}{2m_d}\right)\right)+2\left(-\frac{1}{3}\frac{e}{2m_d}\right)\right] \nonumber \\
  &=& \frac{1}{6} \left[\frac{16}{3}\frac{e}{2m_u} + \frac{2}{3}\frac{e}{2m_d}\right] = \frac{8}{9} \frac{e}{2m_u} + \frac{1}{9}\frac{e}{2m_d} \nonumber
\end{eqnarray}
The neutron wavefunction is obtained by interchanging $u \leftrightarrow d$, so the neutron magnetic moment is given by
\begin{eqnarray}
  \mu_n &=& 3 \times \frac{1}{18}\left[4\left(2\left(-\frac{1}{3}\frac{e}{2m_d}\right) - \frac{2}{3}\frac{e}{2m_u}\right)+ 2\left(\frac{2}{3}\frac{e}{2m_u}\right)\right] \nonumber \\
  &=& \frac{1}{6}\left[-\frac{8}{3}\frac{e}{2m_d} - \frac{4}{3}\frac{e}{2m_u}\right] = -\frac{2}{9}\frac{e}{2m_u} - \frac{4}{9}\frac{e}{2m_d} \nonumber 
\end{eqnarray}
If the up and down quark masses are naively assumed to be $m_u = m_d \approx \frac{1}{3} M_N$, then the quark model predictions for the nucleon magnetic moments are: 
\begin{eqnarray}
  \mu_p &=& \frac{3e}{2M_N} \nonumber \\
  \mu_n &=& \frac{-2e}{2M_N} \nonumber \\
  \frac{\mu_n}{\mu_p} &=& -\frac{2}{3} \nonumber 
\end{eqnarray}
In nuclear magnetons, then, the quark model predicts $\mu_p = 3$, compared with $\mu_p = 2.79$ experimentally, and $\mu_n = -2$, compared with $\mu_n = -1.91$ experimentally. Moreover, the experimental ratio $\frac{\mu_n}{\mu_p} = -.685$ is also very close to the quark-model prediction. The magnetic moments of the other octet baryons calculated in this model are in similar rough agreement with their experimental values, which is quite remarkable given its simplicity. Another success of the basic quark model is that the main features of the baryon mass spectrum can be reproduced using a basic spin-spin interaction between the quarks' color-magnetic moments analogous to the hyperfine splitting of spectral lines in atoms, though this again requires some ad hoc assumptions about the masses of the constituent quarks. 

The considerable success of the early quark model in explaining gross features of the observed spectrum of baryons and mesons suggests an important role for these ``constituent" or ``valence" quarks in going beyond the qualitative description furnished by the quark model to a quantitative understanding of the strong interaction. The theory of color forces is called Quantum Chromodynamics (QCD) and is based on invariance of the Lagrangian under local SU(3) gauge transformations of the quark color fields. While QCD is generally accepted as the correct theory of the strong interaction, a complete understanding of nucleon and nuclear structure in terms of QCD has eluded physicists for decades, owing to several complicating characteristics of the theory: 
\begin{enumerate}
\item{} The coupling strength between quarks and gluons, the gauge bosons of the color interaction, is strong and increases at large distances. This property, called \emph{confinement}, is responsible for the fact that only colorless mesons and baryons are found in nature. In fact, a free quark has never been observed experimentally. Instead, the existence of quarks was established in deep inelastic electron-proton scattering experiments at SLAC in the late 1960s, for which the 1990 Nobel Prize in Physics was awarded to Friedman, Kendall, and Taylor. 
\item{} The flipside of the confinement property is that QCD is an ``asymptotically free'' theory. In other words, the coupling strength of the interaction decreases at short distance, enabling a perturbative description of quark interactions at sufficiently high energies. The discovery of this property by Gross, Politzer and Wilczek was awarded the Nobel Prize in Physics in 2004. The agreement of perturbative QCD (pQCD) predictions with experimental data on the evolution of the proton structure function $F_2$ in a remarkably large dynamic range is among the strongest evidence in favor of QCD.   
\item{} Unlike in QED, in which the gauge bosons (photons) are electrically neutral, the gluons possess color charge and are thus self-interacting, further complicating efforts to make empirical predictions from QCD, especially in the strong-coupling regime. This self-interaction of gluons is a consequence of the non-Abelian character of the SU(3) gauge symmetry of QCD.
\end{enumerate}
 
Despite the lack of a complete and self-consistent description of nuclear (and nucleon) structure and dynamics from the first principles of QCD, very successful models exist in which mesons and baryons take the place of quarks and gluons as the degrees of freedom in effective field theories, in which the residual strong force between baryons with no net color is mediated by meson exchange. A great deal of the empirical knowledge of nuclei can be described within this framework. One of the most challenging outstanding problems of nuclear physics is to bridge the gap in theoretical understanding between the meson-baryon models that hold at low energies and the successful pQCD description at very high energies. On the experimental side, it requires, in principle, that key properties of the nucleon be measured from low energies all the way up to the onset of pQCD-consistent behavior, in order that theoretical predictions can be rigorously confronted with data in the region of transition between the two limiting cases. One such measurement is the subject of this thesis.
\section{Elastic $eN$ Scattering: Formalism}
\paragraph{}
Ever since Rutherford, Geiger, and Marsden discovered the atomic nucleus by measuring the distribution of alpha particles scattered from gold foils (\cite{GeigerMarsden1909}, \cite{Rutherford1911}, and \cite{GeigerMarsden1913}), scattering experiments have been the method of choice of nuclear and particle physicists to examine the microscopic structure of matter. In the mid-1950s, a series of experiments led by Robert Hofstadter at Stanford established electron scattering as a powerful technique for exploring nuclear structure(\cite{Hof55}, \cite{Hof56}, \cite{HofRev56}). In particular, this work firmly established that the proton has an extended charge distribution and measured its size. Hofstadter was awarded the Nobel Prize in Physics in 1961 for his pioneering work. To this day, electron scattering remains one of the most powerful techniques to study nuclear and nucleon structure. The interaction of a beam of electrons with a nuclear target is well understood and precisely calculable within Quantum Electrodynamics (QED), making possible a straightforward interpretation of the results in terms of the underlying nuclear physics. One of the main drawbacks of the technique is that it can only reveal the \emph{electromagnetic} structure of a nucleon or nucleus, since electrons interact only with the charged quarks. The gluon structure, then, can only be probed indirectly through the quark structure. Also, the small cross sections associated with electromagnetic processes necessitate a very high-luminosity experiment, with intense electron beams and dense nuclear targets. Despite these limitations, electron scattering or lepton scattering more generally is unparalleled as a precision probe of nuclear structure.

Of all the things that can happen when an energetic electron beam scatters from a nucleon target, arguably the simplest and most basic is elastic scattering, that is, the reaction $e + N \rightarrow e + N$ in which the struck nucleon stays intact and in its ground state and the energy and momentum of the electron-nucleon system is conserved. The cross section for elastic scattering is characterized by form factors, which are fundamental properties of the nucleon representing the effect of its structure on its response to electromagnetic probes such as electrons. The electron and the nucleon are both spin-$1/2$ objects. The electron, at any length/energy scale presently accessible to experiment, and for all practical purposes of this thesis, is a structureless, point particle; its magnetic moment is determined entirely by its spin. The nucleon, on the other hand, has a complex internal structure; its magnetic moment is determined by the combined spin and orbital angular momentum of its three valence quarks and the surrounding sea of transient quark-antiquark pairs and gluons that fluctuate in and out of existence in the strong color field of the valence quarks. Since the magnetic moment of the electron interacts with the magnetic moment distribution of the nucleon in the scattering process, it is useful to ask how the reaction depends on the initial and final spin orientations of both the electron and the nucleon. As it turns out, the spin dependence of the scattering amplitude gives rise to a set of polarization observables directly related to the form factors. These observables can be used as an alternative or a complement to cross section measurements in determining the form factors.
\begin{figure}
  \setlength{\unitlength}{1.0in}
  \begin{picture}(6.0,5.0)			
    \put(1.5,0.25){\includegraphics[width=.50\textwidth]{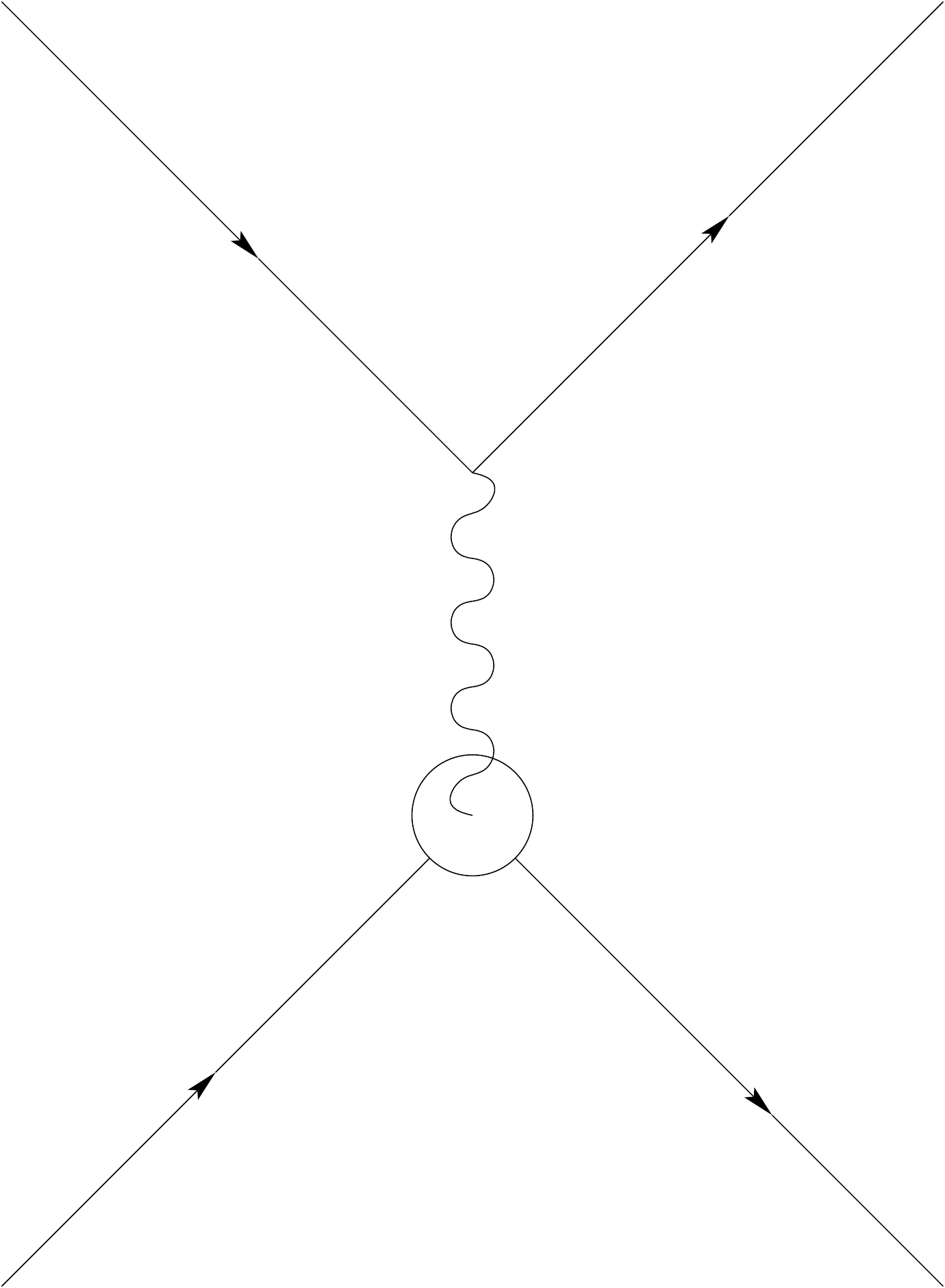}}
    \put(1.25,0.05){$\left| N(p)\right\rangle$}
    \put(4.4,0.05){$\left\langle N(p')\right|$}
    \put(1.25,4.45){$\left| e(k)\right\rangle$}
    \put(4.4,4.45){$\left\langle e(k')\right|$}
    \put(3.1,2.25){$\gamma^*$}
  \end{picture}
  \label{diagram1}
  \caption{Leading-order Feynman diagram for $e + N \rightarrow e + N$}			
\end{figure}
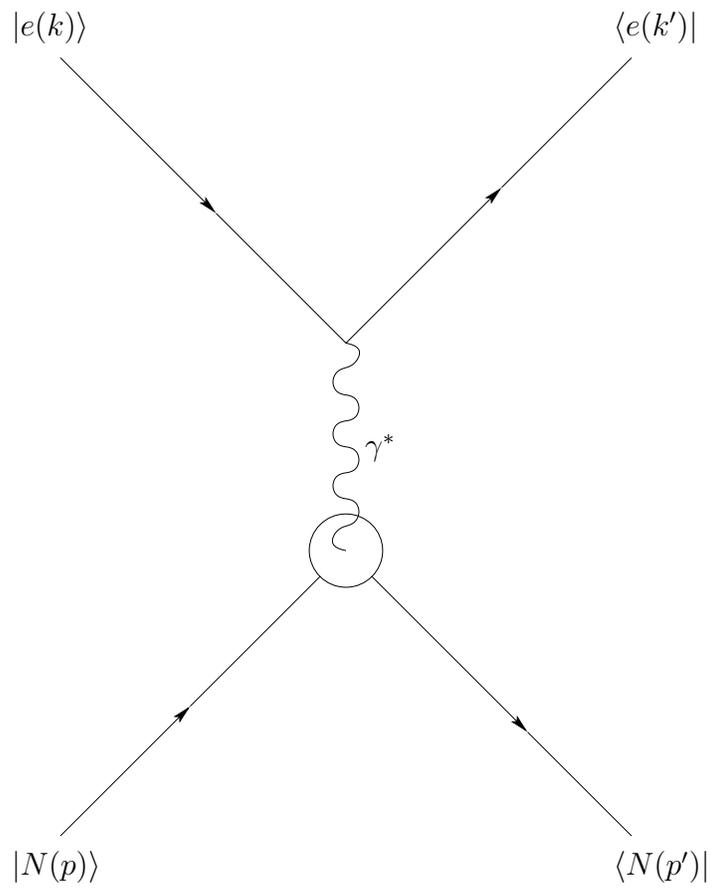

Figure \ref{diagram1} shows the leading-order Feynman diagram for the elastic scattering process. At leading order in $\alpha$, the scattering occurs through the exchange of a single photon. Because the fine-structure constant of electromagnetism is so small ($\alpha \equiv \frac{e^2}{4\pi \hbar c} = \frac{1}{137.03599911}$ experimentally\cite{PDG2008}), first-order perturbation theory in $\alpha$ is a very good approximation to the real physical process. The circle drawn around the nucleon vertex in figure \ref{diagram1} indicates that the QED vertex factor is to be modified to take into account the nucleon's internal electromagnetic structure. According to the Feynman rules for QED\footnote{See e.g. \cite{PeskinSchroeder}, Section 4.8, or \cite{QandL}, Table 6.2}, the invariant amplitude for elastic $eN$ scattering can be read off from the diagram as follows\footnote{From this point on, unless otherwise noted, natural units will be assumed, i.e., $\hbar = c = 1$.}: 
\begin{equation}
  -i\mathcal{M} = \bar{u}(k')\left(ie\gamma^{\mu}\right)u(k)\left(-i\frac{g_{\mu\nu}}{q^2}\right)\bar{u}(p')\left(-ie\Gamma^{\nu}\right)u(p)
  \label{amplitude}
\end{equation}
In equation \ref{amplitude}, $k$ and $k'$ are the initial and final electron four-momenta, $p$ and $p'$ are the initial and final proton four-momenta, and $q^2 = (k-k')^2 = (p'-p)^2$ is the four-momentum transfer squared and also the invariant mass of the virtual photon. $g_{\mu \nu}$ is the Minkowski metric tensor. The de Broglie wavelength of the virtual photon can be thought of as the resolution with which it ``sees'' the nucleon structure. This wavelength is given by $\lambda = \frac{\hbar}{Q}$, where $Q \equiv \sqrt{\left|q^2\right|}$. With $\hbar c \approx 197\ MeV \cdot fm$, it is found that $\lambda \approx .2\ fm / Q$, with $Q$ in GeV/c. The $u$'s and $\bar{u}$'s are free-particle Dirac spinors and their adjoints\footnote{$\bar{u} = u^{\dagger}\gamma^{0}$}, respectively. $-ie\Gamma^{\nu}$ is the modified nucleon vertex factor. In its most general possible form, it is expressed as a linear combination of bilinear covariants of the Dirac equation\footnote{Axial vector terms including $\gamma^5$ are not allowed, as they would violate parity conservation, which is known to hold in electromagnetic interactions.} and the independent four-vectors $p^{\nu}$ and $p'^{\nu}$:
\begin{eqnarray}
  \Gamma^{\nu} &=& K_1 \gamma^{\nu} + iK_2 \sigma^{\nu \alpha}(p' - p)_{\alpha} + iK_3 \sigma^{\nu \alpha}(p' + p)_{\alpha} + \nonumber \\
  & &K_4 (p' - p)^{\nu} + K_5 (p' + p)^{\nu} \label{current1}
\end{eqnarray}
where the tensor $\sigma^{\mu \nu} \equiv \frac{i}{2}\left[\gamma^{\mu},\gamma^{\nu}\right]$. The structure factors $K_i$ are all functions of $q^2$. Demanding Lorentz invariance of the single photon exchange amplitude requires that the form factors are functions of only one variable, $q^2$. All other Lorentz scalars involved in the problem can be expressed in terms of $q^2$ by energy and momentum conservation. The terms involving $\sigma^{\nu \alpha}$ are multiplied by $i$ so that the invariant amplitude is real-valued\footnote{Because no other diagrams interfere with figure \ref{diagram1} at this order in $\alpha$, one is free to choose the phase of the amplitude \eqref{amplitude} arbitrarily, since it will not affect any physical observables. However, once a phase convention is adopted, it must be applied consistently when calculating the contribution of any higher-order diagrams to the process.}. From \eqref{current1} it appears that there are five independent form factors; however, not all of the terms are independent.  Keeping in mind that the vertex factor is sandwiched between spinors that obey the Dirac equation, the number of independent form factors can be reduced to three:
\begin{eqnarray}
  \label{array1begin}
  \sigma^{\mu \nu}(p'_{\nu} + p_{\nu}) &=& \frac{i}{2}\left[\gamma^{\mu}(\not{p}' + \not{p}) - (\not{p}' + \not{p})\gamma^{\mu}\right] \nonumber \\
  &=& \frac{i}{2}\left[-M\gamma^{\mu} + 2p'^{\mu} + M\gamma^{\mu} -M \gamma^{\mu} + M\gamma^{\mu} -2p^{\mu}\right] \nonumber \\
  &=& i(p' - p)^{\mu} \\
  \sigma^{\mu \nu}(p'_{\nu} - p_{\nu}) &=& \frac{i}{2}\left[\gamma^{\mu}(\not{p}' - \not{p}) - (\not{p}' - \not{p})\gamma^{\mu}\right] \nonumber \\
  &=& \frac{i}{2}\left[-M\gamma^{\mu} + 2p'^{\mu} - M\gamma^{\mu} - M \gamma^{\mu} - M\gamma^{\mu} + 2p^{\mu}\right] \nonumber \\ 
  &=& i(p'^{\mu} + p^{\mu} - 2M\gamma^{\mu}) \label{array1end}
\end{eqnarray}
In equations \eqref{array1begin}-\eqref{array1end}, the anticommutation relation for Dirac gamma matrices $\left\{\gamma^{\mu},\gamma^{\nu}\right\} = 2g^{\mu \nu}$, the Dirac equation for free-particle spinors ($(\gamma^{\mu}p_{\mu}-M)u = 0$ and $\bar{u}(\gamma^{\mu}p_{\mu} - M) = 0$), and the standard notation $\not{a} \equiv \gamma^{\mu} a_{\mu}$ have all been applied. Equations \eqref{array1begin}-\eqref{array1end} show that the term in \eqref{current1} proportional to $p'^{\mu} + p^{\mu}$ can be absorbed into a combination of the terms proportional to $\gamma^{\mu}$ and $\sigma^{\mu \nu}(p'_{\nu} - p_{\nu})$, and the term proportional to $\sigma^{\mu \nu}(p'_{\nu} + p_{\nu})$ can be absorbed into the $p'^{\mu} - p^{\mu}$ term, leaving just three independent form factors and the following expression for the vertex factor (with $q = p' - p$ by 4-momentum conservation): 
\begin{equation}
  \Gamma^{\nu} = F_1 \gamma^{\nu} + i\frac{F_2}{2M} \sigma^{\nu \alpha}q_{\alpha} + F_3 \frac{p'^{\nu} - p^{\nu}}{M} \nonumber 
\end{equation}
Finally, current conservation at the nucleon vertex requires that $q_{\mu} J^{\mu} = 0$, which implies: 
\begin {eqnarray}
  q_{\mu} J^{\mu} &=& \bar{u}(p')\left[F_1 \not{q} + i\frac{1}{2M} F_2 q_{\mu} \sigma^{\mu \nu} q_{\nu} + q^2 F_3\right]u(p) = 0 \nonumber \\
  &\Rightarrow & F_3 = 0
\end{eqnarray}
The first term can be shown to vanish by applying the Dirac equation to the spinors sandwiching $\not{q}$. The second term is zero because $\sigma^{\mu \nu}$ is totally antisymmetric, while $q_\mu q_\nu$ is symmetric, leaving $q^2 F_3 = 0$. So in the one photon exchange approximation, the nucleon current is characterized by just two independent form factors, which are functions of $q^2$: 
\begin{equation}
  \label{nvertex}
  \Gamma^{\nu} = F_1(q^2) \gamma^{\nu} +  F_2(q^2)i\sigma^{\nu \alpha} \frac{q_{\alpha}}{2M} 
\end{equation}
Substituting \eqref{nvertex} into \eqref{amplitude} for the scattering amplitude gives:
\begin{equation}
  \label{amplitude2}
  \mathcal{M} = \frac{e^2}{q^2} \bar{u}(k') \gamma^{\mu} u(k) g_{\mu \nu} \bar{u}(p')\left[F_1(q^2) \gamma^{\nu} + F_2(q^2)i\sigma^{\nu \alpha} \frac{q_{\alpha}}{2M}\right]u(p)
\end{equation}
The convention \eqref{nvertex} for the nucleon vertex factor is commonly used in the literature on nucleon form factors; $F_1$ and $F_2$ are known as the Dirac and Pauli form factors, respectively. Another commonly used choice of the form factors uses the linear combinations
\begin{eqnarray}
  \label{SachsDefBegin} G_E(q^2) &\equiv& F_1(q^2) - \tau F_2(q^2) \\
  G_M(q^2) &\equiv& F_1(q^2) + F_2(q^2) \\
  \tau &\equiv& \frac{Q^2}{4M^2} = \frac{-q^2}{4M^2}
  \label{SachsDefEnd}
\end{eqnarray}
which are known as the Sachs electric and magnetic form factors, respectively. These form factors have the advantage that the scattering cross section has only terms proportional to $G_E^2$ and $G_M^2$, and no terms proportional to $G_E G_M$, making an extraction of $G_E$ and $G_M$ separately from a cross section measurement that much simpler. They also have the advantage of an intuitive physical interpretation, at least in a specific reference frame in the low-energy limit. 

The nucleon vertex factor can be reexpressed in terms of Sachs form factors and simplified. Substituting \eqref{array1end} into \eqref{nvertex} gives: 
\begin{eqnarray}
  \Gamma^{\nu} &=& F_1 \gamma^{\nu} + \frac{1}{2M} F_2 \left(2M\gamma^{\nu} - p'^{\nu} - p^{\nu}\right)	\nonumber \\
  \Gamma^{\nu} &=& (F_1 + F_2) \gamma^{\nu} - \frac{1}{2M} F_2 \left(p'^{\nu} + p^{\nu} \right) \nonumber 
\end{eqnarray}
Using the definitions \eqref{SachsDefBegin}-\eqref{SachsDefEnd}, $\Gamma^\nu$ becomes:
\begin{eqnarray}
  \Gamma^\nu &=& G_M \gamma^{\nu} + \frac{G_E - G_M}{2M\left(1+\tau\right)} \left(p'^{\nu} + p^{\nu} \right) \label{SachsNvertex}
\end{eqnarray}
Finally, the scattering amplitude in terms of Sachs form factors is given by
\begin{equation}
  \label{MasterAmplitude}
  \mathcal{M} = \frac{e^2}{q^2} \bar{u}(k') \gamma^{\mu} u(k) \bar{u}(p')\left[G_M \gamma_{\mu} + \frac{G_E - G_M}{2M\left(1+\tau\right)} \left(p'_{\mu} + p_{\mu} \right)\right]u(p)
\end{equation}
This formula, equation \eqref{MasterAmplitude}, will be the point of departure for all the derivations that follow.
\section{Elastic $eN$ Scattering: Rosenbluth Cross Section}
\label{sec:Rosenbluth}
\paragraph{}
The cross section for elastic $eN$ scattering in terms of the invariant amplitude $\mathcal{M}$ is given by Fermi's Golden Rule (see e.g. \cite{QandL}, section 4.3, p. 88): 
\begin{eqnarray}
  \label{crosssect1}
  d\sigma &=& \overline{\left|\mathcal{M}\right|^2} \frac{dQ}{F} \\
  dQ &\equiv& \frac{d^3p'}{(2\pi)^3 2p'^0} \frac{d^3k'}{(2\pi)^3 2k'^0}(2\pi)^4\delta^{(4)}(k + p - k' - p') \nonumber \\
  F &\equiv& 4k \cdot p = 4ME_e \nonumber 
\end{eqnarray}
$dQ$ is the Lorentz invariant phase space factor representing the density of final states available to the outgoing electron and proton, while $F$ is the incident flux in the lab frame, in which the target proton is at rest. The electron mass is neglected throughout; this is a very good approximation in all situations encountered in this thesis, which deals with incident and scattered electron energies roughly in the range of 0.5-6 GeV, while the mass of an electron is about .511 MeV or at most one part in $10^3$ of any of the energies involved. The bar over the square of the scattering amplitude indicates that it is to be averaged over the spin states of the initial particles and summed over the spin states of the outgoing particles; in other words, \eqref{crosssect1} refers to the \emph{unpolarized} cross section. 

The square of the invariant amplitude \eqref{MasterAmplitude} averaged over initial spins and summed over final spins can be written as: 
\begin{eqnarray}
  \overline{\left|\mathcal{M}\right|^2} &=& \frac{e^4}{q^4} L^e_{\mu \nu} W_N^{\mu \nu} \label{tensordef} \\
  L^e_{\mu \nu} &\equiv& \frac{1}{2} \sum_{s} \sum_{s'} \bar{u}^{(s')}(k') \gamma_\mu u^{(s)}(k) \bar{u}^{(s)}(k) \gamma_\nu u^{(s')}(k') \\
  W_N^{\mu \nu} &\equiv& \frac{1}{2} \sum_{s} \sum_{s'} \bar{u}^{(s')}(p')\left[G_M \gamma^\mu + \frac{G_E-G_M}{2M(1+\tau)}(p'^\mu+p^\mu)\right]u^{(s)}(p) \times \nonumber \\
  & & \bar{u}^{(s)}(p)\left[G_M \gamma^\nu + \frac{G_E-G_M}{2M(1+\tau)}(p'^\nu+p^\nu)\right]u^{(s')}(p')
\end{eqnarray}
The completeness relation for the sum over spin states is (see \eqref{unpolarizedspinsum}) 
\begin{equation}
  \sum_{s} u^{(s)}(p)\bar{u}^{(s)}(p) = \not{p} + M
\end{equation}
Again neglecting the electron mass, the electron tensor $L^e_{\mu \nu}$ is easily evaluated using standard trace technology (see \eqref{Casimirtrick}): 
\begin{eqnarray}
  L^e_{\mu \nu} &=& \frac{1}{2}\mbox{Tr} \left[\gamma_\mu \not{k} \gamma_\nu \not{k'} \right] \nonumber \\
  &=& 2\left(k_\mu k'_\nu + k_\nu k'_\mu - (k\cdot k')g_{\mu \nu}\right) \label{leptontensorunpol}
\end{eqnarray}
Note that it is symmetric, $L^e_{\nu \mu} = L^e_{\mu \nu}$. Using \eqref{Casimirtrick}, the nucleon tensor $W_N^{\mu \nu}$ evaluates to:  
\begin{eqnarray}
  W_N^{\mu \nu} &=& \frac{1}{2} \mbox{Tr}\left[\left\{G_M \gamma^\mu + \frac{G_E-G_M}{2M(1+\tau)}(p'^\mu+p^\mu)\right\}(\not{p}+M) \times \right. \nonumber \\ 
    & & \left. \left\{G_M \gamma^\nu + \frac{G_E-G_M}{2M(1+\tau)}(p'^\nu+p^\nu)\right\}(\not{p}'+M)\right] \nonumber \\
  &=& (i) + (ii) + (iii) \nonumber \\
  (i) &\equiv& \frac{1}{2} G_M^2 \mbox{Tr} \left[\gamma^\mu(\not{p} + M)\gamma^{\nu}(\not{p}'+M)\right] \nonumber \\
  (ii) &\equiv& \frac{1}{2} \frac{G_M(G_E-G_M)}{2M(1+\tau)}\left[(p'^\nu+p^\nu)\mbox{Tr}\left\{\gamma^\mu (\not{p}+M)(\not{p}'+M)\right\} + \right. \nonumber \\
    & & \left. (p'^\mu+p^\mu)\mbox{Tr}\left\{(\not{p}+M)\gamma^\nu(\not{p}'+M)\right\}\right] \nonumber \\
  (iii) &\equiv& \frac{1}{2}\left(\frac{G_E-G_M}{2M(1+\tau)}\right)^2(p'^\mu+p^\mu)(p'^\nu+p^\nu)\mbox{Tr}\left[(\not{p}+M)(\not{p}'+M)\right] \nonumber
\end{eqnarray}
The expression for $W^{\mu\nu}$ above is grouped into smaller, more easily digestible terms $(i)$, $(ii)$ and $(iii)$ which are evaluated separately below and then added. Using the trace theorems of \ref{tracetech}, these three terms become
\begin{eqnarray}
  (i) &=& 2G_M^2\left[p^\mu p'^\nu + p^\nu p'^\mu + (M^2 - p\cdot p')g^{\mu \nu}\right] \nonumber \\
  (ii) &=& 2\frac{G_M(G_E-G_M)}{1+\tau}(p'^\mu+p^\mu)(p'^\nu+p^\nu) \nonumber \\
  (iii) &=& 2\left(\frac{G_E-G_M}{2M(1+\tau)}\right)^2(p'^\mu+p^\mu)(p'^\nu+p^\nu)(M^2 + p \cdot p') \nonumber \\
  q^2 = (p'-p)^2 &=& 2M^2 - 2p \cdot p' \Rightarrow M^2 + p \cdot p' = 2M^2(1+\tau) \nonumber \\
  (ii) + (iii) &=& 2\left[\frac{G_M(G_E-G_M)}{1+\tau}(p'^\mu+p^\mu)(p'^\nu+p^\nu) + \right. \nonumber \\
    & & \left. \frac{G_E^2 + G_M^2 - 2G_E G_M}{2(1+\tau)}(p'^\mu+p^\mu)(p'^\nu+p^\nu)\right] \nonumber \\
  &=& 2\left[\frac{G_E^2 - G_M^2}{2(1+\tau)}(p'^\mu+p^\mu)(p'^\nu+p^\nu)\right] \nonumber 
\end{eqnarray}
Note that in the sum $(ii)+(iii)$, the terms proportional to $G_E G_M$ cancel, leaving only separate $G_E^2$ and $G_M^2$ terms, so that, finally, $W_N^{\mu \nu}$ becomes:
\begin{eqnarray}
  W_N^{\mu \nu} &=& 2G_M^2\left[p^\mu p'^\nu + p^\nu p'^\mu + (M^2 - p\cdot p')g^{\mu \nu}\right] + \nonumber \\
  & & \frac{G_E^2 - G_M^2}{1+\tau}(p'^\mu+p^\mu)(p'^\nu+p^\nu)
\end{eqnarray}
The nucleon tensor is also symmetric. Before proceeding any further, it is useful to write down some kinematic identities which will be useful for expressing the cross section in terms of lab frame quantities, and define the invariant Mandelstam variables $s$, $t$, and $u$:
\begin{eqnarray}
  s \equiv (k+p)^2 = M^2 + 2k \cdot p &=& M^2 + 2ME_e = (k' + p')^2 = M^2 + 2 k' \cdot p' \nonumber \\
  t = -Q^2 = (k-k')^2 &=& -2k \cdot k' = (p' - p)^2 = 2(M^2 - p' \cdot p) \nonumber \\
  u = (k-p')^2 = (p-k')^2 &=& M^2 - 2k \cdot p' = M^2 - 2k' \cdot p \nonumber \\
  &=& M^2 - 2ME'_e \nonumber \\
  k' \cdot k &=& \frac{Q^2}{2} \nonumber \\
  p' \cdot p &=& M^2 + \frac{Q^2}{2} \nonumber \\
  k \cdot p = k' \cdot p' &=& ME_e \nonumber \\
  k \cdot p' = k' \cdot p &=& ME'_e = ME_e - \frac{Q^2}{2}
  \label{Mandelstam}
\end{eqnarray}
Here $E_e$ and $E'_e$ refer to the lab-frame energies of the incident and scattered electron, respectively. They are related by energy and momentum conservation. Choosing the incident electron direction along the $z$ axis, and defining the scattering plane as the $xz$ plane, with the electron scattering in the positive $x$ direction, the four-momenta of the particles involved in the reaction are:
\begin{eqnarray}
  k^\mu &=& (k^0,\mathbf{k})=(E_e,0,0,E_e) \label{labfourmomenta} \\
  p^\mu &=& (p^0,\mathbf{0})=(M,0,0,0) \nonumber \\
  k'^\mu &=& (k'^0,\mathbf{k'})=(E'_e,E'_e\sin{\theta_e},0,E'_e\cos{\theta_e}) \nonumber \\
  p'^\mu &=& (p'^0,\mathbf{p'})=(E_p,-p_p \sin{\theta_p},0,p_p\cos{\theta_p}) \nonumber   
\end{eqnarray}
The following important relationships among the various kinematic quantities follow from energy and momentum conservation and the relativistic energy-momentum relation:
\begin{eqnarray}
  Q^2 &=& 2k \cdot k' = 2E_eE'_e(1-\cos{\theta_e}) = 4E_eE'_e\sin^2{\frac{\theta_e}{2}} \nonumber \\
  E_p &=& M + \nu \nonumber \\
  \nu &\equiv& E_e - E'_e \nonumber \\
  Q^2 &=& 2(p' \cdot p - M^2) = 2(M(M+\nu) - M^2) = 2M\nu \nonumber \\
  E'_e \sin{\theta_e} &=& p_p \sin{\theta_p} \nonumber \\
  E_e - E'_e &=& \frac{E_eE'_e}{M}(1 - \cos{\theta_e}) \nonumber \\
  \Rightarrow E'_e &=& \frac{E_e}{1 + \frac{E_e}{M}(1-\cos{\theta_e})} = \frac{E_e}{1 + \frac{2E_e}{M}\sin^2{\frac{\theta_e}{2}}} \nonumber \label{Eescat}
\end{eqnarray}
The contraction of the electron and nucleon current tensors is given in terms of the relevant four-momenta by: 
\begin{eqnarray}
  L^e_{\mu \nu} W_N^{\mu \nu} &=& 2\left(k_\mu k'_\nu + k_\nu k'_\mu - k \cdot k' g_{\mu \nu}\right) \times \nonumber \\
  & & \Bigg \{ 2G_M^2\left[p^\mu p'^\nu + p^\nu p'^\mu + (M^2 - p \cdot p')g^{\mu \nu}\right] + \nonumber \\
  & & \frac{G_E^2 - G_M^2}{1+\tau}(p'^\mu+p^\mu)(p'^\nu+p^\nu) \Bigg\} \nonumber \\
  &=& 4G_M^2\Bigg[2\{(p \cdot k) (p' \cdot k') + (p \cdot k') (p' \cdot k)\}-2(p\cdot p') (k\cdot k') - \nonumber \\  
    & & 4M^2 k\cdot k' + 4(p\cdot p') (k\cdot k') + 2M^2 k\cdot k' - 2(p \cdot p') (k\cdot k')\Bigg] + \nonumber \\
  & & 2\frac{G_E^2-G_M^2}{1+\tau}\left[2\{k \cdot (p' + p) k'\cdot (p'+p)\} - k\cdot k'(p'+p)^2\right] \nonumber \\
  &=& \label{tensorcontraction} 8G_M^2\left[(p \cdot k) (p' \cdot k') + (p \cdot k') (p' \cdot k) -M^2 k\cdot k'\right] + \nonumber \\
  & & 4\frac{G_E^2-G_M^2}{1+\tau}\left[\{k \cdot (p' + p) k'\cdot (p'+p)\} - k\cdot k'(M^2 + p' \cdot p)\right]
\end{eqnarray}
Using the kinematic relations \eqref{Mandelstam}, the contraction (\ref{tensorcontraction}) in terms of lab frame quantities and the invariant $Q^2$ is given by:
\begin{eqnarray}
  L^e_{\mu \nu} W_N^{\mu \nu} &=& 8G_M^2\left[M^2E_e^2 + \left(ME_e - \frac{Q^2}{2}\right)^2 -\frac{Q^2M^2}{2}\right] + \nonumber \\
  & & 4\frac{G_E^2 - G_M^2}{1+\tau}\left[\left(2ME_e - \frac{Q^2}{2}\right)^2 -Q^2M^2(1+\tau)\right] \nonumber \\
  &=& 8G_M^2\left[2M^2E_e^2 - Q^2ME_e + \frac{Q^4}{4} -\frac{Q^2M^2}{2}\right] + \nonumber \\
  & & 4\frac{G_E^2-G_M^2}{1+\tau}\left[4M^2E_e^2 - 2Q^2ME_e + \frac{Q^4}{4} - Q^2M^2(1+\tau)\right] \nonumber \\
  &=& 8\left[\begin{array}{c} G_M^2\left[2M^2E_e^2 - Q^2ME_e + \frac{Q^4}{4} -\frac{Q^2M^2}{2}\right] + \\ \frac{G_E^2-G_M^2}{1+\tau}\left[2M^2E_e^2 - Q^2ME_e - \frac{Q^2M^2}{2}\right]\end{array} \right] \nonumber \\
  &=& 8\left[Q^2M^2\tau G_M^2 + \frac{G_E^2+\tau G_M^2}{1+\tau}\left(2M^2E_e^2 - Q^2ME_e - \frac{Q^2M^2}{2}\right)\right] \nonumber \\
  &=& 8M^2\left[Q^2\tau G_M^2 + \frac{G_E^2+\tau G_M^2}{1+\tau}\left(2E_e^2 - Q^2\frac{E_e}{M} - \frac{Q^2}{2}\right) \right] \label{tensorcontraction2}
\end{eqnarray}
It is possible to rewrite (\ref{tensorcontraction2}) in a simpler form using the kinematic relations above. In particular, the quantity in parentheses in equation (\ref{tensorcontraction2}) becomes: 
\begin{eqnarray}
  2E_e^2 - Q^2\frac{E_e}{M} - \frac{Q^2}{2} &=& 2E_e^2 - \frac{Q^2}{2}\left(\frac{2E_e}{M} + 1\right) \nonumber \\
  \frac{Q^2}{2\sin^2{\frac{\theta_e}{2}}}\frac{E_e}{E'_e} &=& \frac{4E_eE'_e\sin^2{\frac{\theta_e}{2}}}{2\sin^2{\frac{\theta_e}{2}}}\frac{E_e}{E'_e} = 2E_e^2 \nonumber \\
  \frac{E_e}{E'_e} &=& 1 + \frac{2E_e}{M}\sin^2{\frac{\theta_e}{2}} \nonumber \\
  2E_e^2 - Q^2\frac{E_e}{M} - \frac{Q^2}{2} &=& \frac{Q^2}{2\sin^2{\frac{\theta_e}{2}}}\left(1 + \frac{2E_e}{M}\sin^2{\frac{\theta_e}{2}}\right) -\frac{Q^2}{2}\left(\frac{2E_e}{M} + 1\right) \nonumber \\
  &=& \frac{Q^2}{2\sin^2{\frac{\theta_e}{2}}}\left(1 + \frac{2E_e}{M}\sin^2{\frac{\theta_e}{2}} - \left(\frac{2E_e}{M} + 1\right)\sin^2{\frac{\theta_e}{2}}\right) \nonumber \\
  &=& \frac{Q^2}{2}\cot^2{\frac{\theta_e}{2}}
\end{eqnarray}
With this useful result, the squared amplitude becomes: 
\begin{eqnarray}
  \overline{\left|\mathcal{M}\right|^2} &=& \frac{4e^4}{q^4}Q^2M^2\left[\frac{G_E^2+\tau G_M^2}{1+\tau}\cot^2{\frac{\theta_e}{2}} + 2\tau G_M^2\right] \label{finalampsquaredfirst} \\
  q^2 &=& -Q^2 \Rightarrow q^4 = Q^4 \nonumber \\
  \overline{\left|\mathcal{M}\right|^2} &=& 4e^4 \frac{M^2}{Q^2}\left[\frac{G_E^2+\tau G_M^2}{1+\tau}\cot^2{\frac{\theta_e}{2}} + 2\tau G_M^2\right] \label{finalampsquared}
\end{eqnarray}
The next step is to calculate the cross section using this amplitude: 
\begin{equation}
  d\sigma = \frac{\overline{\left|\mathcal{M}\right|^2}}{64\pi^2ME_e}\frac{d^3p'}{p'^0}\frac{d^3k'}{k'^0}\delta^{(4)}(k + p - k' - p') 
\end{equation} 
Integrating over all possible outgoing proton momenta, the delta function sends $\mathbf{p'} \rightarrow \mathbf{k} - \mathbf{k'}$, leaving:
\begin{equation}
  d\sigma = \left. \frac{\overline{\left|\mathcal{M}\right|^2}}{64\pi^2ME_e}\frac{d^3k'}{E'_e E'_N} \delta(E_e + M - E'_e - E'_N)\right|_{\mathbf{p'} = \mathbf{k} - \mathbf{k'} \equiv \mathbf{q}}
\end{equation}
Using the identity $\delta(f(x)) = \frac{\delta(x-x_0)}{\left|f'(x_0)\right|},\ (f(x_0) = 0)$, and temporarily reinstating the electron mass in order to recall the functional relationship between its energy and its momentum, the delta function can be rewritten as:
\begin{eqnarray}
  E'_e + E'_N &=& \sqrt{\mathbf{k'}^2 + m_e^2} + \sqrt{\mathbf{q}^2 + M_N^2} \nonumber \\
  \mathbf{q}^2 &=& (\mathbf{k}-\mathbf{k'})^2 = \mathbf{k}^2 + \mathbf{k'}^2 - 2\left|\mathbf{k}\right|\left|\mathbf{k'}\right|\cos{\theta_e} \nonumber \\
  \frac{d(E'_e + E'_N)}{d\left|\mathbf{k'}\right|} &=& \frac{\left|\mathbf{k'}\right|}{\sqrt{\mathbf{k'}^2 + m_e^2}} + \frac{\left|\mathbf{k'}\right| - \left|\mathbf{k}\right|\cos{\theta_e}}{\sqrt{\mathbf{q}^2 + M_N^2}} = \frac{\left|\mathbf{k'}\right|}{E'_e} + \frac{\left|\mathbf{k'}\right| - \left|\mathbf{k}\right|\cos{\theta_e}}{E'_N} \nonumber 
\end{eqnarray}
Returning now to the approximation of massless electrons, the delta-function becomes:
\begin{eqnarray}
  \frac{d(E'_e + E'_N)}{d\left|\mathbf{k'}\right|} &=& 1 + \frac{E'_e - E_e\cos{\theta_e}}{M + \nu} \nonumber \\
  \delta(E_e+M - E'_e - E'_N) &\rightarrow& \frac{M+\nu}{M+\nu + E'_e - E_e\cos{\theta_e}}\delta\left(\left|\mathbf{k'}\right|+\frac{Q^2}{2M} - E_e\right)
\end{eqnarray} 
Using this form of the delta function, the cross section reduces to:
\begin{eqnarray}
  d^3k' &=& \mathbf{k'}^2 d\left|\mathbf{k'}\right| d\Omega_e \nonumber \\
  \left|\mathbf{k'}\right| &=& E'_e \nonumber \\
  \frac{d\sigma}{d\Omega_e} &=& \frac{\overline{\left|\mathcal{M}\right|^2}}{64\pi^2ME_e} \frac{E'_e d\left|\mathbf{k'}\right|}{M+E_e(1-\cos{\theta_e})}\delta\left(\left|\mathbf{k'}\right|+\frac{Q^2}{2M} - E_e\right)
\end{eqnarray}
Integrating now over all outgoing electron momenta, and recognizing the factor in the denominator $M(1+\frac{E_e}{M}(1-\cos{\theta_e})) = \frac{E_e}{M E'_e}$, the cross section becomes:
\begin{equation}
  \frac{d\sigma}{d\Omega_e} = \frac{\overline{\left|\mathcal{M}\right|^2}}{64\pi^2M^2}\left(\frac{E'_e}{E_e}\right)^2
\end{equation}
Substitution of the result (\ref{finalampsquared}) for the spin-averaged, squared scattering amplitude, and using $\alpha \equiv e^2/4\pi$, the final expression for the cross section is obtained.
\begin{eqnarray}
  \frac{d\sigma}{d\Omega_e} &=& \frac{1}{64\pi^2M^2}\left(\frac{E'_e}{E_e}\right)^2\frac{64\pi^2\alpha^2M^2}{Q^2}\left[\frac{G_E^2+\tau G_M^2}{1+\tau}\cot^2{\frac{\theta_e}{2}} + 2\tau G_M^2\right] \nonumber \\ 
  &=& \frac{\alpha^2}{Q^2} \left(\frac{E'_e}{E_e}\right)^2\left[\frac{G_E^2+\tau G_M^2}{1+\tau}\cot^2{\frac{\theta_e}{2}} + 2\tau G_M^2\right] \nonumber \\
  \frac{d\sigma}{d\Omega_e} &=& \frac{\alpha^2}{4E_e^2\sin^4{\frac{\theta_e}{2}}} \frac{E'_e}{E_e}\cos^2{\frac{\theta_e}{2}} \left[\frac{G_E^2+\tau G_M^2}{1+\tau} + 2\tau G_M^2\tan^2{\frac{\theta_e}{2}}\right] \label{RosenbluthFormula}
\end{eqnarray}
In this form, equation \eqref{RosenbluthFormula}\footnote{Equation \eqref{RosenbluthFormula} is commonly known as the Rosenbluth Formula after M. N. Rosenbluth, who first derived it in 1950 \cite{Rosenbluth1950}.}, it is clear that the cross section factors neatly into the product of the Mott cross section, representing spin-$1/2$ electron scattering from a point charge, and a ``structure'' factor determined by the form factors. It is useful to define a ``reduced'' cross section $\sigma_r$ to isolate the effect of the nucleon's structure: 
\begin{eqnarray}
  \sigma_r &\equiv& \frac{\sigma_{eN}}{\sigma_{Mott}} \label{sigmardef} \\
  \sigma_{Mott} &\equiv& \frac{\alpha^2}{4E_e^2\sin^4{\frac{\theta_e}{2}}} \frac{E'_e}{E_e}\cos^2{\frac{\theta_e}{2}} \\
  (1+\tau)\epsilon \sigma_r &=& \epsilon G_E^2 + \tau G_M^2 \label{sigmar} \\
  \epsilon &\equiv& \left[1 + 2(1 + \tau) \tan^2{\frac{\theta_e}{2}}\right]^{-1}
\end{eqnarray}
It is clear from \eqref{sigmar} that the form factors (or at least the squares of the form factors) can be extracted separately by measuring the electron-nucleon elastic scattering cross section at fixed $\tau$ and varying the scattering angle $\theta_e$. Then, a plot of $\sigma_r$ vs. $\epsilon$ yields a straight line with slope proportional to $G_E^2$ and intercept proportional to $G_M^2$. This procedure is called Rosenbluth or L/T separation\footnote{L/T refers to the separation between longitudinally and transversely polarized (virtual) photons, with the degree of longitudinal polarization of the virtual photon characterized by $\epsilon$}. 
 It is informative to compare the Rosenbluth Formula \eqref{RosenbluthFormula} to the cross section for elastic electron scattering from a structureless, pointlike nucleon. In that case the vertex factor simply becomes $\Gamma^{\nu} \Rightarrow \gamma^{\nu}$, meaning $F_1 = 1$ and $F_2 = 0$ by definition, so that $G_E = G_M = 1$, and the cross section becomes 
\begin{equation}
  \frac{d\sigma}{d\Omega} = \sigma_{Mott} \left[1 + 2\tau \tan^2{\frac{\theta_e}{2}}\right] \label{sigma_point}
\end{equation}
The Mott cross section describes the scattering of spin-$1/2$ electrons from pointlike, spinless charged particles\footnote{See e.g. \cite{QandL}, equation 6.51.}. The $\cos^2{\frac{\theta_e}{2}}$ suppression of large-angle scattering in the Mott cross section is a consequence of helicity conservation in interactions mediated by vector fields at high energies. Because a spinless target lacks a magnetic moment, it cannot flip the spin of the electron, and since the helicity $\vec{\sigma} \cdot \hat{k}$ is conserved, scattering at backward angles is suppressed. On the other hand, when the target is spin-$1/2$, scattering at backward angles is enhanced at high energies by the spin-flip interaction between the magnetic moments, resulting in the $\tan^2 \frac{\theta_e}{2}$ term in \eqref{sigma_point}. 

The physical nucleon is a composite object with a rich substructure. However, in the limit as $Q^2 \rightarrow 0$, the long-wavelength virtual photon has insufficient resolution to be sensitive to this detailed structure. The nucleon should behave like a point particle with charge $ze$ ($z=1$ for the proton or $0$ for the neutron) and magnetic moment $e/2M_N(z+\kappa)$. Demanding point-like limiting behavior fixes the normalization of the Dirac and Pauli form factors at $Q^2 = 0$.

Expressing the current for a Dirac particle ($\Gamma^\mu = \gamma^\mu$) in the form
\begin{equation}
  \bar{u}(p')\gamma^\mu u(p) = \frac{1}{2M}\bar{u}(p')\left[p^\mu + p'^\mu + i\sigma^{\mu \nu}(p'-p)_\nu \right]u(p) \label{Gordon}
\end{equation}
which is simply a rearrangement of \eqref{array1end}, it is clear that the spin-dependent part of the interaction, which arises from the Dirac magnetic moment $e/2M_N$, is contained in the $\sigma^{\mu \nu}$ term, while the spin-independent part of the interaction, which arises from the nucleon charge, is contained in the $(p+p')^\mu$ term. To generalize \eqref{sigma_point} to the case of a nucleon with charge $z$ and anomalous magnetic moment $\kappa$, then, simply requires a factor of $z$ multiplying the charge and (Dirac) magnetic moment terms above and an additional anomalous magnetic moment term $i\kappa \sigma^{\mu \nu}(p'-p)_\nu$:
\begin{equation}
  \left.\bar{u}(p')\Gamma^\mu u(p)\right|_{point} = \frac{1}{2M}\bar{u}(p')\left[z(p^\mu + p'^\mu) + i(z+\kappa)\sigma^{\mu \nu}(p'-p)_\nu \right]u(p) \label{ModifiedGordon}
\end{equation}
Comparing \eqref{ModifiedGordon} to \eqref{nvertex}, the low-$Q^2$ limit of the Dirac and Pauli form factors must be
\begin{eqnarray}
  F_1^N(0) &=& z_N \nonumber \\
  F_2^N(0) &=& \kappa_N \nonumber \\
  F_1^p(0) &=& 1 \nonumber \\
  F_2^p(0) &=& 1.79 \nonumber \\
  F_1^n(0) &=& 0 \nonumber \\
  F_2^n(0) &=& -1.91 \nonumber
\end{eqnarray}
The Sachs form factors simply reduce to the nucleon charges and magnetic moments: 
\begin{eqnarray}
  G_E^p(0) &=& 1 \nonumber \\
  G_M^p(0) &=& \mu_p = 2.79 \nonumber \\
  G_E^n(0) &=& 0 \nonumber \\
  G_M^n(0) &=& \mu_n = -1.91 \nonumber
\end{eqnarray}

In order to give an intuitive meaning to the form factors, it is useful to consider the Rosenbluth Formula in the non-relativistic limit, $Q^2 \ll M^2$. The electric form factor dominates the cross section in this limit, since all $G_M^2$ terms in the cross section are multiplied by $\tau$. Neglecting these terms, the Rosenbluth formula reduces to: 
\begin{equation}
  \frac{d\sigma}{d\Omega_e} = \left(\frac{d\sigma}{d\Omega_e}\right)_{Mott} \left(G_E(q^2)\right)^2 \label{lowQ2Rosenbluth}
\end{equation}
In this limit, the energy transfer $\nu$ is negligible, as is the recoil momentum of the proton, so that the reaction can be viewed as scattering of the electron by the static charge distribution of a stationary proton. Comparing (\ref{lowQ2Rosenbluth}) to the cross section for scattering from a static charge distribution (see \cite{QandL}, equation 8.1), $G_E$ can be identified with the Fourier transform of the proton's charge distribution $F(\mathbf{q}) = \int \rho(\mathbf{x}) e^{i\mathbf{q}\cdot \mathbf{x}}d^3\mathbf{x}$. Expanding the exponential in powers of $\mathbf{q}$ gives 
\begin{equation}
  G_E = \int \rho(\mathbf{x}) \left(1 + i\mathbf{q} \cdot \mathbf{x} - \frac{(\mathbf{q} \cdot \mathbf{x})^2}{2} + \ldots \right) d^3 \mathbf{x}
\end{equation}
For a spherically symmetric charge distribution, $\rho = \rho(r \equiv \left|\mathbf{x}\right|)$, this becomes 
\begin{eqnarray}
  G_E &=& \int_0^\infty \rho(r) r^2 dr \int_0^\pi \sin{\theta} d\theta \left(1 + i\left|\mathbf{q}\right|r \cos{\theta} - \frac{1}{2}\mathbf{q}^2 r^2 \cos^2{\theta}+\ldots \right) \\
  G_E &=& 1 - \frac{1}{6} \mathbf{q}^2 \int \left|\mathbf{x}\right|^2 \rho(\left|\mathbf{x}\right|) d^3 \mathbf{x} + \ldots = 1 - \frac{1}{6} \mathbf{q}^2 \left<r^2\right> + \ldots \label{lowQexpansion}
\end{eqnarray}
meaning that, at leading order in $q^2$, the electric form factor simply measures the r.m.s. charge radius of the nucleon. Similar reasoning leads to the interpretation of the magnetic form factor as the Fourier transform of the nucleon's magnetization distribution in the non-relativistic limit. 
\section{Polarization Transfer in Elastic $eN$ Scattering}
\paragraph{}
Given the availability of high-energy, high-intensity polarized electron beams, it is fruitful to examine how the amplitude for elastic scattering depends on the initial and/or final spin orientations of the electron and the nucleon. The density matrix for a mixed ensemble of spin-$1/2$ particles with spins preferentially aligned along a direction $\hat{n}$ is given by: 
\begin{equation}
  \rho \equiv \sum_{s} w^{(s)} \left| \chi^{(s)} \right> \left< \chi^{(s)} \right|
\end{equation}
where $s$ refers to the spin state of the particle and the sum runs over a complete set of spin states, and the weight $w^{(s)}$ is the fraction of particles in state $s$. Taking the polarization direction $\hat{n}$ as the quantization axis, and denoting the degree of polarization as $h$, the weights and the density matrix become
\begin{eqnarray}
  h &\equiv& \frac{N\uparrow - N\downarrow}{N\uparrow + N\downarrow} \\
  w_\uparrow &=& \frac{1}{2}(1 + h) \\ 
  w_\downarrow &=& \frac{1}{2}(1 - h) \\
  \rho &=&  w_\uparrow \left|\hat{n}, \uparrow \right> \left< \hat{n},\uparrow \right| + w_\downarrow \left| \hat{n},\downarrow \right> \left<\hat{n}, \downarrow \right| \\
  \rho &=& \frac{1}{2}\left(1 + h \vec{\sigma} \cdot \hat{n} \right) \label{spinensemble}
\end{eqnarray}
where $\vec{\sigma}$ is the vector of Pauli spin matrices $\sigma_x$, $\sigma_y$, and $\sigma_z$, and $\left|\hat{n},\uparrow \right>$ and $\left|\hat{n},\downarrow \right>$ are, by definition, eigenstates of the operator $\vec{\sigma} \cdot \hat{n}$ with eigenvalues $+1$ and $-1$, respectively.

In polarized elastic scattering, there are four possible polarization observables that one might attempt to measure: 
\begin{enumerate}
\item The polarization of the incident electron beam
\item The polarization of the nucleon target 
\item The polarization of the scattered nucleon
\item The polarization of the scattered electron
\end{enumerate}
If the electron beam is polarized, and the nucleon target is unpolarized, the scattering can impart some polarization to the recoiling nucleon, which can then be measured with a suitable secondary analyzing reaction. This class of experiment is called polarization transfer or recoil polarization and is the subject of this thesis. If both beam and target are polarized, but the polarization of the reaction products is not measured, the corresponding polarization observable is the asymmetry in the scattering cross section resulting from reversal of either the beam or the target polarization while holding the other fixed (for practical reasons, it is usually the beam polarization that is reversed while holding the target polarization fixed). If both the beam and the target are unpolarized, any measured polarization of the scattered nucleon is said to be induced. In the Born (single photon exchange) approximation, the induced polarization turns out to be identically zero.

In the following section, the electron and nucleon current tensors $L_e^{\mu \nu}$ and $W_N^{\mu \nu}$ are calculated, allowing for polarization of the \emph{incident} electron and the \emph{scattered} nucleon. The target nucleon is unpolarized, and the polarization of the scattered electron is not derived, because it won't be measured. The nucleon current tensor is given by
\begin{eqnarray}
  W_N^{\mu \nu} &=& \mathcal{J}_N^\mu \mathcal{J}_N^{\nu *} \\
  \mathcal{J}_N^\mu &=& \bar{u}(p') \Gamma^\mu u(p) \\
  \Gamma^\mu &\equiv& G_M\gamma^\mu + \frac{G_E-G_M}{2M(1+\tau)}(p+p')^\mu 
\end{eqnarray}

Since the product $L^e_{\mu \nu} W^{\mu \nu}_N$ is Lorentz-invariant, the electron and nucleon current tensors may be calculated in any reference frame, as long as both tensors are calculated in the same reference frame. The result is applicable in the lab frame as well. It turns out that for the problem at hand, it is most convenient to work in the Breit or ``brick-wall'' frame in which there is no energy transfer and the nucleon recoils backward along its incident direction; i.e., $\mathbf{p'} = -\mathbf{p}$. Before evaluating $W^{\mu\nu}_{N}$, the kinematics of the Breit frame are presented.
\subsection{Breit Frame Kinematics}
\label{Breitframesection}
\begin{figure}[h]
  \begin{center}
    \includegraphics[angle=90,width=.9\textwidth]{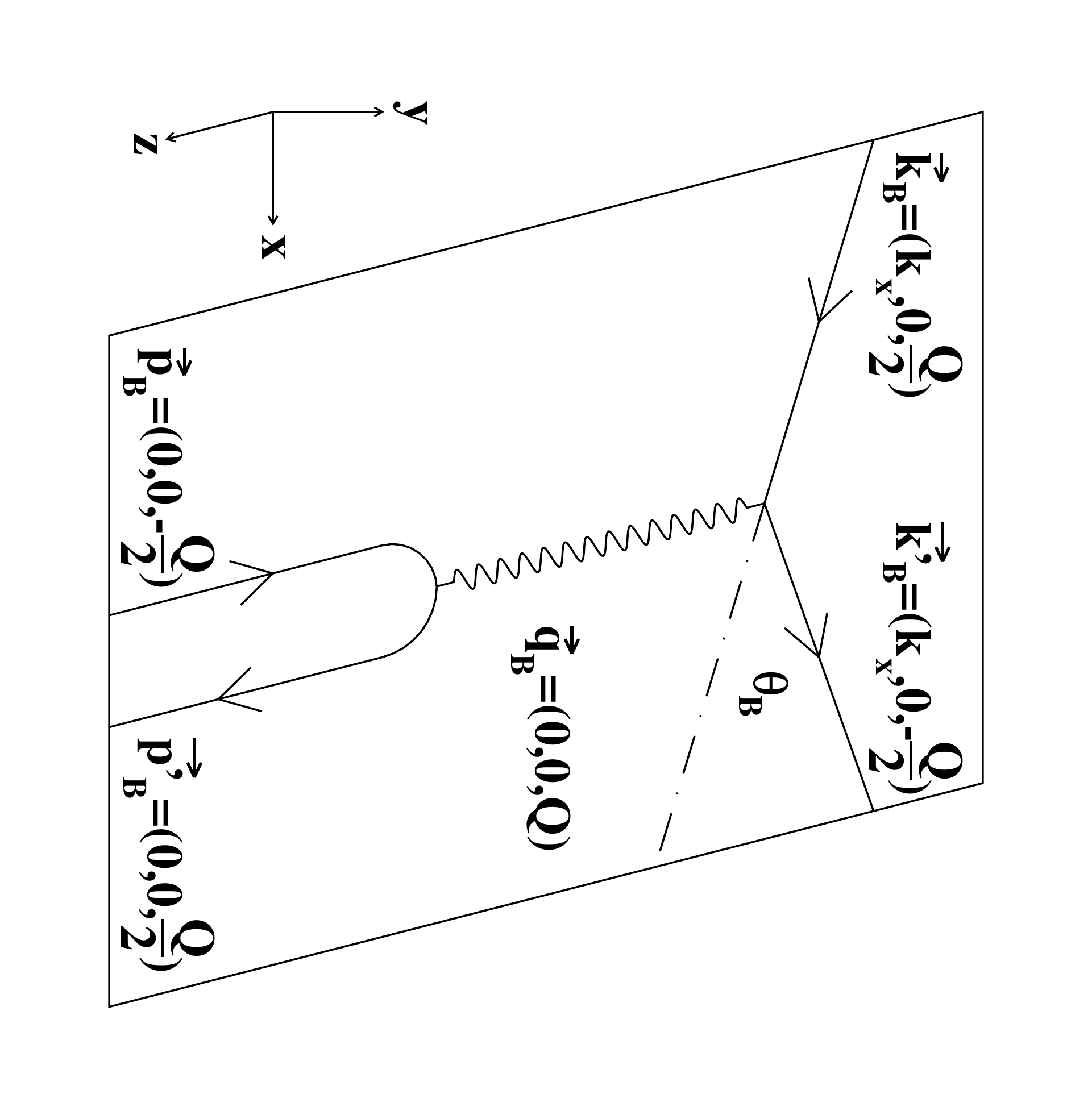}
    \caption{\label{BreitFig} Elastic scattering in the Breit frame}
  \end{center}
\end{figure}

\paragraph{}
The defining characteristic of the Breit frame is $\mathbf{p'} = -\mathbf{p}$. This implies that the energy transfer $\nu$ is zero. 
\begin{eqnarray}
  q^2 &=& (p'-p)^2 = -(\mathbf{p'} - \mathbf{p})^2 = -4\mathbf{p}^2 \nonumber \\
  Q^2 &=& -q^2 = 4 \mathbf{p}^2 \nonumber \\
  \Rightarrow \left|\mathbf{p}\right| &=& \frac{Q}{2} \nonumber \\
  E_N &=& \sqrt{\mathbf{p}^2 + M^2} = \sqrt{M^2(1 + \tau)} = M\sqrt{1+\tau} \nonumber 
\end{eqnarray}
In the following, a coordinate system for the Breit frame is adopted in which the positive $z$ axis coincides with the direction of the transferred momentum. The $x$-axis is taken to be the coordinate transverse to the momentum transfer in the scattering plane, and the $y$-axis is the coordinate transverse to the momentum transfer normal to the scattering plane, and is directed such that $(\hat{x}, \hat{y}, \hat{z})$ forms a right-handed coordinate system. Figure \ref{BreitFig} illustrates the chosen coordinate system. 
With these definitions, the momentum four-vectors of the incident and the scattered nucleon in the Breit frame are:
\begin{eqnarray}
  p^\mu = \left(\begin{array}{c} M\sqrt{1+\tau} \\ 0 \\ 0 \\ -\frac{Q}{2} \end{array}\right) &, & p'^\mu = \left(\begin{array}{c} M\sqrt{1+\tau} \\ 0 \\ 0 \\ \frac{Q}{2} \end{array}\right) \label{proton4momBreit}
\end{eqnarray}

The kinematics of the electron in the Breit frame are also simplified. The y-components of $\mathbf{k}$ and $\mathbf{k'}$ are zero by definition. However, the collision is not required to be ``head-on'' in the Breit frame, so nonzero x-components $k_x$ and $k'_x$ must be allowed. Because the momentum transfer is, by definition, along the z-axis, the x-component of the electron's momentum is conserved, that is, $k'_x = k_x$. Four-momentum conservation fixes the z-component of the incident and outgoing electron momenta $\mathbf{k}$ and $\mathbf{k'}$:
\begin{eqnarray}
  \mathbf{q} &=& Q \hat{z} \nonumber \\
  q_z = Q &=& k_z - k'_z \nonumber \\
  k_z &=& k'_z + Q \nonumber \\
  \nu = 0 &\Rightarrow& \left|\mathbf{k'}\right| = \left|\mathbf{k}\right| \nonumber \\
  \Rightarrow k_z^2 &=& k'^2_z \nonumber \\
  \Rightarrow k'_z &=& -\frac{Q}{2} \nonumber \\
  k_z &=& \frac{Q}{2} \nonumber 
\end{eqnarray} 
 The four-momenta of the incident and scattered electron are thus given by:
\begin{eqnarray}
  k^\mu = \left(\begin{array}{c} \sqrt{k_x^2 + \frac{Q^2}{4}} \\ k_x \\ 0 \\ \frac{Q}{2} \end{array}\right) &,& k'^\mu = \left(\begin{array}{c} \sqrt{k_x^2 + \frac{Q^2}{4}} \\ k_x \\ 0 \\ -\frac{Q}{2} \end{array}\right) \label{Breitelectron1}
\end{eqnarray}
From \eqref{Breitelectron1}, $k$ and $k'$ can be expressed in terms of $Q$ and the scattering angle of the electron in the Breit frame: 
\begin{eqnarray}
  \mathbf{k'} \cdot \mathbf{k} &=& \left|\mathbf{k}\right|\left|\mathbf{k'}\right| \cos{\theta_B} \nonumber \\
  \frac{k_x^2 - \frac{Q^2}{4}}{k_x^2 + \frac{Q^2}{4}} &=& \cos{\theta_B} \nonumber \\
  k_x^2 (1 - \cos{\theta_B}) &=& \frac{Q^2}{4}(1+\cos{\theta_B}) \nonumber \\
  \Rightarrow k_x^2 &=& \frac{Q^2}{4} \cot^2{\frac{\theta_B}{2}} \nonumber \\
  \Rightarrow k_x &=& \frac{Q}{2} \cot{\frac{\theta_B}{2}} \nonumber \\
  k^\mu = \left(\begin{array}{c} \frac{Q}{2\sin{\frac{\theta_B}{2}}} \\ \frac{Q}{2} \cot{\frac{\theta_B}{2}} \\ 0 \\ \frac{Q}{2} \end{array}\right) &,& k'^\mu = \left(\begin{array}{c} \frac{Q}{2\sin{\frac{\theta_B}{2}}} \\ \frac{Q}{2} \cot{\frac{\theta_B}{2}} \\ 0 \\ -\frac{Q}{2} \end{array}\right) \label{Breitelectron2}
\end{eqnarray}
In the lab frame, the nucleon is at rest in the initial state. Temporarily redefining the coordinate system of the lab frame so that the $z$-axis is along the transferred momentum, the $x$ and $y$ axes point in the same direction in the lab frame and the Breit frame. The two frames are related by a boost along the transferred momentum, which coincides with the $z$-axis in this temporary coordinate system. The relativistic boost factor $\gamma$ is given by:
\begin{eqnarray}
  (p_z)_L = 0 &=& \gamma ( (p_z)_B + \beta E_B) \nonumber \\
  \Rightarrow \beta &=& \frac{Q}{2M\sqrt{1+\tau}} = \sqrt{\frac{\tau}{1+\tau}} \nonumber \\
  \gamma = (1-\beta^2)^{-\frac{1}{2}} &=& \sqrt{1+\tau} \nonumber 
\end{eqnarray}
The next step is to derive the boosted four-momenta of the incident and scattered particles in the lab frame, and determine the relation between the lab-frame and Breit frame electron scattering angles. In the following, the subscript $L$ denotes lab-frame quantities, while the subscript $B$ denotes Breit frame quantities:
\begin{eqnarray}
  (k_x)_L = (k'_x)_L &=& (k_x)_B = \frac{Q}{2} \cot{\frac{\theta_B}{2}} \label{electronboost} \\
  (k_y)_L = (k'_y)_L &=& (k_y)_B = 0 \nonumber 
\end{eqnarray}
In the lab frame, a different coordinate system (see \eqref{labfourmomenta}) was defined above in which the $z$-axis is along the direction of the beam momentum $\mathbf{k}$. In the \emph{Breit}-frame coordinate system, the $z$-component of the incident electron momentum in either the lab frame or the Breit frame is given by
\begin{eqnarray}
  k_z &=& \frac{\mathbf{k} \cdot \mathbf{q}}{\left|\mathbf{q}\right|} \nonumber \\
  \mathbf{q} &\equiv& \mathbf{k} - \mathbf{k'} \nonumber \\
  k_z^2 &=& \frac{\left(\mathbf{k}\cdot \mathbf{q}\right)^2}{\mathbf{q}^2} \label{kz2lab}
\end{eqnarray}
Similarly, the x-component of $\mathbf{k}$ is given by:
\begin{eqnarray}
  k_x^2 + k_z^2 &=& \mathbf{k}^2 \nonumber \\
  \Rightarrow k_x^2 &=& \mathbf{k}^2 - k_z^2 \nonumber \\
  k_x^2 &=& \frac{\mathbf{k}^2\mathbf{q}^2 - \left(\mathbf{k} \cdot \mathbf{q}\right)^2}{\mathbf{q}^2} \label{kx2lab}
\end{eqnarray} 
In the lab frame, the expression for $k_x^2$ can be derived as follows:
\begin{eqnarray}
  \mathbf{k}^2 &=& E_e^2 \nonumber \\
  \mathbf{q}^2 &=& E_e^2 + E_e'^{2} - 2E_e E'_e \cos{\theta_e} \nonumber \\
  \mathbf{k}\cdot \mathbf{q} &=& E_e^2 - E_e E_e' \cos{\theta_e} \nonumber \\
  q^2 = \nu^2 - \mathbf{q}^2 &=& -Q^2 \nonumber \\
  \Rightarrow \mathbf{q}^2 &=& Q^2 + \nu^2 = Q^2(1+\tau) \nonumber \\
  \mathbf{k}^2\mathbf{q}^2 - \left(\mathbf{k} \cdot \mathbf{q}\right)^2 &=& E_e^2 E_e'^2 \sin^2{\theta_e} = 4E_e^2E_e'^2\sin^2{\frac{\theta_e}{2}}\cos^2{\frac{\theta_e}{2}} \nonumber \\
  &=& \frac{Q^4}{4} \cot^2{\frac{\theta_e}{2}} \nonumber \\
  \Rightarrow k_{x}^2 &=& \frac{Q^2}{4(1+\tau)} \cot^2 \frac{\theta_e}{2} \label{kx2lab2}
\end{eqnarray}
Since $k_{x,B}=k_{x,L}$, equation \eqref{kx2lab2} implies the following relationship between the lab-frame and Breit-frame electron scattering angles:
\begin{eqnarray}
  k_x^2 = \frac{Q^2}{4} \cot^2{\frac{\theta_B}{2}} &=& \frac{Q^2}{4(1+\tau)} \cot^2{\frac{\theta_e}{2}} \nonumber \\
  \cot^2{\frac{\theta_e}{2}} &=& (1+\tau)\cot^2{\frac{\theta_B}{2}} \label{BreitLabAngles}
\end{eqnarray}

\subsection{Transferred Polarization Components in the Breit Frame}
\paragraph{}
Rather than use the trace theorems of appendix \ref{tracetech} that were used to derive the Rosenbluth formula, one can take advantage of the fact that the nucleon current $\mathcal{J}$ reduces to a particularly simple form in the Breit frame (see appendix \ref{BreitCurrentAppendix} for a full derivation) to derive the polarized component of the nucleon current tensor. The Breit frame expression for $\mathcal{J}_N^\mu$ is given by:
\begin{equation}
  \mathcal{J}_N^\mu = \left(\begin{array}{c} \mathcal{J}_N^0 \\ \vec{\mathcal{J}}_N \end{array} \right) = \left(\begin{array}{c} 2MG_E \chi'^\dag \chi \\ 2iG_M \chi'^\dag (\mathbf{p}\times \mathbf{\sigma}) \chi \end{array}\right)
\end{equation}
In the Breit-frame coordinate system, the incident proton momentum is $\mathbf{p} = -\frac{Q}{2} \hat{z}$, leading to the following expression for the current:
\begin{eqnarray}
  \mathcal{J}_N^\mu &=& \left(\begin{array}{c}2MG_E \chi'^\dag \chi \\ iQG_M \chi'^\dag \sigma_y \chi \\ -iQG_M \chi'^\dag \sigma_x \chi \\ 0 \end{array}\right) \label{breitcurrent}
\end{eqnarray}
The adjoint of the current is given by the complex-conjugate transpose of \eqref{breitcurrent}:
\begin{equation}
  \mathcal{J}_N^{\nu \dag} = \left(2MG_E \chi^\dag \chi',-iQG_M \chi^\dag \sigma_y \chi',iQG_M \chi^\dag \sigma_x \chi',0\right)
\end{equation}
Now it is straightforward to form the tensor product $W_N^{\mu \nu}$. To save space, the outer product $\chi \chi^\dag$ is replaced wherever it appears with the completeness relation $\sum_s \chi^{(s)} \chi^{\dag(s)} = 1$ for the unpolarized target proton with a factor of $1/2$ for the average of initial spin states. The normalization condition $\chi'^\dag \chi' = 1$ for the two-component spinor $\chi'$ of the scattered proton has also been applied.
\begin{eqnarray}
  W_N^{\mu \nu} &=& \left(\begin{array}{cccc} 2M^2 G_E^2  &-iMQG_EG_M\chi'^\dag \sigma_y \chi' & iMQG_EG_M\chi'^\dag \sigma_x \chi' & 0 \\ iMQG_EG_M\chi'^\dag \sigma_y \chi' & \frac{Q^2G_M^2}{2}  & i\frac{Q^2G_M^2}{2} \chi'^\dag \sigma_z \chi' & 0 \\ -iMQG_EG_M \chi'^\dag \sigma_x \chi' & -i\frac{Q^2G_M^2}{2} \chi'^\dag \sigma_z \chi' & \frac{Q^2G_M^2}{2}  & 0 \\ 0&0&0&0 \end{array}\right) \nonumber \\
  & & \label{TotalBreitCurrentTensor}
\end{eqnarray}

In equation \eqref{TotalBreitCurrentTensor}, it is clear that the Breit-frame nucleon current tensor has real, spin-independent diagonal elements with terms proportional to $G_E^2$ and $G_M^2$ which survive the sum over final-state polarizations in calculating the unpolarized amplitude. The polarization effects are contained in the imaginary, off-diagonal, spin-dependent elements. These elements, which vanish when summed over final spin states in the calculation of the unpolarized amplitude, determine the polarization of the scattered nucleon.

To calculate the transferred polarization, the current tensor of the incident polarized electron beam is required. For this calculation, the trace techniques of Appendix \ref{tracetech} are used. After summing over spin states of the scattered electron, $L^e_{\mu\nu}$ becomes:
\begin{eqnarray}
  L^e_{\mu \nu} &=& \bar{u}(k') \gamma_\mu u(k) \bar{u}(k) \gamma_\nu u(k') \nonumber \\
  &=& Tr\left[\gamma_\mu \sum_s u^{(s)}(k)\bar{u}^{(s)}(k) \gamma_\nu \not{k'}\right] \nonumber 
\end{eqnarray}
For the polarized part of $L^e_{\mu \nu}$, the average over initial spin states is to be replaced by the polarized spin sum \eqref{polarizedspinsum}. For the (assumed to be) massless electron, \eqref{polarizedspinsum} reduces to:
\begin{eqnarray}
  2\sum_s u^{(s)}(k) \bar{u}^{(s)}(k) &=& k^0 \gamma^5 \gamma^0 (\vec{\gamma} \cdot \mathbf{h}_e ) + (\mathbf{k} \cdot \mathbf{h}_e)\gamma^5 \gamma^0 + \nonumber \\ 
  & & i\gamma^0 \vec{\gamma} \cdot (\mathbf{k}\times \mathbf{h}_e)-\frac{\mathbf{k}\cdot\mathbf{h}_e}{k^0}\gamma^5(1+\gamma^0)(\vec{\gamma}\cdot \mathbf{k}) \label{leptonpolspinsum}
\end{eqnarray}
The trace above is actually relatively simple to evaluate in a general reference frame, since the electron mass is neglected and the structureless electron vertex factor contains $\gamma_\mu$ rather than $\Gamma_\mu$. Considering each of the four terms of the spin sum \eqref{leptonpolspinsum} separately, one finds:
\begin{enumerate}
\item
  \begin{eqnarray}
    k^0 Tr\left(\gamma^\mu \gamma^5\gamma^0 \vec{\gamma} \cdot \mathbf{h}_e \gamma^\nu \not{k'}\right) &=& 0 \nonumber 
  \end{eqnarray}
\item
  \begin{eqnarray}
    (\mathbf{k} \cdot \mathbf{h}_e) Tr\left(\gamma^\mu \gamma^5 \gamma^0 \gamma^\nu \not{k'}\right) &=& 4i (\mathbf{k} \cdot \mathbf{h}_e)k'_\alpha \epsilon^{\mu \nu 0 \alpha} \nonumber 
  \end{eqnarray}
\item
  \begin{eqnarray}
    i (\mathbf{k} \times \mathbf{h}_e) \cdot Tr \left(\gamma^\mu \gamma^0 \vec{\gamma} \gamma^\nu \not{k'}\right) &=& 0 \nonumber 
  \end{eqnarray}
\item
  \begin{eqnarray}
    -\frac{(\mathbf{k} \cdot \mathbf{h}_e)}{k^0}Tr\left(\gamma^\mu \gamma^5(1+\gamma^0)(\vec{\gamma}\cdot \mathbf{k})\gamma^\nu \not{k'}\right) &=& -\frac{(\mathbf{k} \cdot \mathbf{h}_e)}{k^0} Tr\left(\gamma^\mu \gamma^5(\vec{\gamma}\cdot \mathbf{k})\gamma^\nu \not{k'}\right) \nonumber \\
    &=& -4i \frac{(\mathbf{k} \cdot \mathbf{h}_e)}{k^0} k^jk'_\alpha \epsilon^{\mu \nu j \alpha} \nonumber 
  \end{eqnarray}
\end{enumerate}
Adding the two nonzero terms together and multiplying by the factor of $1/2$ from the completeness relation \eqref{spinensemble}, the trace can be written in the covariant notation:
\begin{eqnarray}
  L_e^{\mu \nu,A} &=& 2i\frac{(\mathbf{k} \cdot \mathbf{h}_e)}{k^0}k'_\alpha\left(k^0 \epsilon^{\mu \nu 0 \alpha} - k^j \epsilon^{\mu \nu j \alpha}\right) = 2i (\hat{k} \cdot \mathbf{h}_e)k_\alpha k'_\beta \epsilon^{\mu \nu \alpha \beta} \nonumber \\
  L^e_{\mu \nu,A} &=& 2i h k^\alpha k'^\beta \epsilon_{\mu \nu \alpha \beta} \label{leptontensorpol}
\end{eqnarray}
where on the second line the longitudinal polarization $h$ of the electron beam has been introduced. The total leptonic current tensor $L^e_{\mu\nu}$ is obtained by adding \eqref{leptontensorunpol} to \eqref{leptontensorpol}\footnote{Note that since the electron beam is now polarized, the factor of $1/2$ for the average over initial spin states is no longer needed. On the other hand, the factor of $1/2$ from the polarized completeness relation \eqref{spinensemble} is now required, so these factors of two cancel and the proper normalization of $L_{\mu\nu}$ is unchanged.}:
\begin{equation}
  L^{e,tot}_{\mu\nu} = 2\left[k_\mu k'_\nu + k_\nu k'_\mu - (k\cdot k')g_{\mu \nu} + i h k^\alpha k'^\beta \epsilon_{\mu\nu\alpha\beta}\right] \label{totalleptontensor}
\end{equation}

An interesting property of the result \eqref{leptontensorpol} is that only the longitudinal component of the incident electron's polarization contributes. This is a consequence of treating the electron as massless, and it is a result that could have been anticipated from the results for the transverse polarization-dependent elements of the nucleon tensor, both of which are proportional to the nucleon mass. Since the effect of any transverse polarization of the electron beam is suppressed by at least a factor of $m_e/E_e$ relative to the effect of longitudinal polarization, polarization transfer experiments at relativistic energies require longitudinally polarized electron beams.

The next step toward calculating the physical polarization transfer observables is to calculate the contraction $L_{\mu \nu} W^{\mu \nu}$ in the Breit frame and then express the result, which is invariant, in terms of lab-frame kinematic quantities and the nucleon form factors. In terms of the Breit frame scattering angle $\theta_B$ and $Q$, the polarized term of $L_{\mu \nu}$ in the Breit frame is given by:
\begin{eqnarray}
  L^e_{\mu\nu,A} &=& 2ih k^\alpha k'^\beta \epsilon_{\mu \nu \alpha \beta} \nonumber \\
  k^\alpha &=& \left(\begin{array}{c} \frac{Q}{2\sin{\frac{\theta_B}{2}}} \\ \frac{Q}{2}\cot{\frac{\theta_B}{2}} \\ 0 \\ \frac{Q}{2} \end{array}\right) ,  k'^\beta = \left(\begin{array}{c} \frac{Q}{2\sin{\frac{\theta_B}{2}}} \\ \frac{Q}{2}\cot{\frac{\theta_B}{2}} \\ 0 \\ -\frac{Q}{2} \end{array}\right) \nonumber \\
  L^e_{\mu \nu,A} &=& 2ih \left(\begin{array}{cccc} 0 & 0 & k^3 k'^1 - k^1 k'^3 & 0 \\ 0 & 0 & k^0 k'^3 - k^3 k'^0 & 0 \\ k^1 k'^3 - k^3 k'^1 & k^3 k'^0 - k^0 k'^3 & 0 & k^0 k'^1 - k^1 k'^0 \\ 0 & 0 & k^1 k'^0 - k^0 k'^1 & 0 \end{array}\right) \nonumber \\
  &=& 2ih \frac{Q^2}{4} \left(\begin{array}{cccc} 0 & 0 & 2\cot{\frac{\theta_B}{2}} & 0 \\ 0 & 0 & -2 \csc{\frac{\theta_B}{2}} & 0 \\ -2\cot{\frac{\theta_B}{2}} & 2 \csc{\frac{\theta_B}{2}} & 0 & 0 \\ 0 & 0 & 0 & 0 \end{array}\right) \nonumber \\
  &=& ihQ^2 \left(\begin{array}{cccc} 0 & 0 & \cot{\frac{\theta_B}{2}} & 0 \\ 0 & 0 & - \csc{\frac{\theta_B}{2}} & 0 \\ -\cot{\frac{\theta_B}{2}} & \csc{\frac{\theta_B}{2}} & 0 & 0 \\ 0 & 0 & 0 & 0 \end{array}\right)
\end{eqnarray}
For completeness, the unpolarized term is:
\begin{eqnarray}
  L^e_{\mu \nu,S} &=& 2k_\mu k'_\nu + 2k_\nu k'_\mu - 2(k\cdot k')g_{\mu \nu} \nonumber \\
  &=& 2k_\mu k'_\nu + 2k_\nu k'_\mu - Q^2 g_{\mu\nu} \nonumber \\
  &=& \left(\begin{array}{cccc} Q^2 \cot^2\frac{\theta_B}{2} & -Q^2 \cot \frac{\theta_B}{2} \csc \frac{\theta_B}{2} & 0 & 0 \\  -Q^2 \cot \frac{\theta_B}{2} \csc \frac{\theta_B}{2} &Q^2 \csc^2 \frac{\theta_B}{2} & 0 & 0 \\ 0 & 0 &Q^2& 0 \\ 0 & 0 &0 & 0 \end{array}\right)
\end{eqnarray}
Like the nucleon tensor $W^{\mu \nu}$, the lepton tensor is the sum of a real-valued, totally symmetric term and a purely imaginary, totally antisymmetric term. Since the contraction of symmetric and antisymmetric tensors always vanishes, the total amplitude is real-valued. Decomposing the contraction into symmetric and antisymmetric terms gives:
\begin{equation}
  L^e_{\mu\nu}W_N^{\mu\nu} = L^e_{\mu\nu,S}W_N^{\mu\nu,S} + L^e_{\mu\nu,A}W_N^{\mu\nu,A}
\end{equation}
The first term reproduces the unpolarized amplitude:
\begin{eqnarray}
  L^e_{\mu\nu,S}W_N^{\mu\nu,S} &=& 2M^2Q^2G_E^2\cot^2\frac{\theta_B}{2} + \frac{Q^4G_M^2}{2}\left(\csc^2\frac{\theta_B}{2} + 1\right) \nonumber \\
  &=& 2M^2Q^2\left[\cot^2\frac{\theta_B}{2}\left(G_E^2 + \frac{\tau G_M^2}{\cos^2 \frac{\theta_B}{2}} \right) + \tau G_M^2 \right]
\end{eqnarray}
Substituting the relation between the Breit-frame and lab-frame electron scattering angles \eqref{BreitLabAngles} in the above result gives 
\begin{eqnarray}
  L^e_{\mu\nu,S}W_N^{\mu\nu,S} &=& 2M^2Q^2\left[\frac{\cot^2\frac{\theta_e}{2}}{1+\tau}\left(G_E^2+\tau G_M^2\left(1+(1+\tau)\tan^2\frac{\theta_e}{2}\right)\right)+\tau G_M^2\right] \nonumber \\
  &=& 2M^2Q^2\left[\frac{G_E^2+\tau G_M^2}{1+\tau}\cot^2\frac{\theta_e}{2} + 2\tau G_M^2 \right] \label{SymmTerm}
\end{eqnarray}
After multiplying by the factor $e^4/q^4$ appearing in the definition of the squared amplitude $\overline{\left|\mathcal{M}\right|^2} = e^4/q^4 L^e_{\mu\nu}W_N^{\mu\nu}$, the above result is equal to the unpolarized result \eqref{finalampsquaredfirst} up to a factor of two. The missing factor of two comes from the fact that the sum over final proton spin states has not yet been applied. For the contraction of the antisymmetric terms, the result is:
\begin{eqnarray}
  L^e_{\mu\nu,A}W_N^{\mu\nu,A} &=& -2hMQ^3G_EG_M\cot \frac{\theta_B}{2}\chi'^\dag \sigma_x\chi' + hQ^4G_M^2 \csc \frac{\theta_B}{2} \chi'^\dag \sigma_z\chi' \label{LmunuWmunu_pol}
\end{eqnarray}
The presence of the expectation values of $\sigma_x$ and $\sigma_z$, with $\chi'^\dag \sigma_x \chi$ in the first term and $\chi'^\dag \sigma_z \chi$ in the second term, is responsible for the polarization of the scattered proton. The absence of a $\chi'^\dag \sigma_y \chi$ term means that there is no net polarization normal to the scattering plane. For this exercise it is useful to write down the Pauli matrices and the eigenvectors of $\sigma_x$, $\sigma_y$, and $\sigma_z$ explicitly: 
\begin{eqnarray}
  \sigma_x = \left(\begin{array}{cc} 0 & 1 \\ 1 & 0 \end{array}\right) , \sigma_y &=&\left(\begin{array}{cc} 0 & -i \\ i & 0 \end{array}\right) , \sigma_z = \left(\begin{array}{cc} 1 & 0 \\ 0 & -1 \end{array}\right) \\
  \chi'(\hat{z},+) = \left(\begin{array}{c}1 \\ 0 \end{array}\right) &,& 
  \chi'(\hat{z},-) = \left(\begin{array}{c} 0 \\ 1 \end{array} \right) \\
  \chi'(\hat{x},+) = \left(\begin{array}{c} \frac{1}{\sqrt{2}} \\ \frac{1}{\sqrt{2}} \end{array} \right) &,&
  \chi'(\hat{x},-) = \left(\begin{array}{c} \frac{1}{\sqrt{2}} \\ -\frac{1}{\sqrt{2}} \end{array} \right) \\
  \chi'(\hat{y},+) = \left(\begin{array}{c} \frac{1}{\sqrt{2}} \\ \frac{i}{\sqrt{2}} \end{array} \right) &,& 
  \chi'(\hat{y},-) = \left(\begin{array}{c} \frac{1}{\sqrt{2}} \\ -\frac{i}{\sqrt{2}} \end{array} \right)
\end{eqnarray}
The amplitudes resulting from \eqref{LmunuWmunu_pol} for purely longitudinal, purely transverse, and purely normal polarization of the scattered proton are considered separately. Longitudinal is understood to mean parallel to the scattered proton's momentum (i.e.; the momentum transfer). Transverse is understood to mean parallel to the reaction plane, but perpendicular to the momentum transfer. Normal is understood to mean perpendicular to the reaction plane, along the direction defined by $\mathbf{q} \times \mathbf{k}$.
\begin{description}
\item[Longitudinal Polarization] For the case in which the scattered proton has spin-up along the $z$ axis, the expectation values of $\sigma_x$ and $\sigma_z$ are given by: 
  \begin{eqnarray}
    \left<\sigma_x\right> = \chi'^\dag(\hat{z},+) \sigma_x \chi'(\hat{z},+) &=& \left(\begin{array}{cc} 1 & 0\end{array}\right)\left(\begin{array}{cc} 0 & 1 \\ 1 & 0\end{array}\right)\left(\begin{array}{c} 1 \\ 0\end{array}\right) = 0 \nonumber \\
	  \left<\sigma_z\right> = \chi'^\dag(\hat{z},+) \sigma_z \chi'(\hat{z},+) &=& \left(\begin{array}{cc} 1 & 0\end{array}\right)\left(\begin{array}{cc} 1 & 0 \\ 0 & -1\end{array}\right)\left(\begin{array}{c} 1 \\ 0\end{array}\right) = 1 \nonumber 
  \end{eqnarray}
  The second term of equation \eqref{LmunuWmunu_pol} is therefore to be interpreted as an enhanced (or reduced, depending on the sign of the coefficient of $\chi'^\dag \sigma_z\chi$) probability for the scattered proton to have spin-up along its momentum direction, which coincides with the momentum transfer.
\item[Transverse Polarization] For a final state with spin-up along the $x$-axis, which is parallel to the reaction plane but perpendicular to the momentum transfer, the appropriate expectation values are:
  \begin{eqnarray}
    \left<\sigma_x \right> = \chi'^\dag(\hat{x},+) \sigma_x \chi'(\hat{x},+) &=& \left(\begin{array}{cc} \frac{1}{\sqrt{2}} & \frac{1}{\sqrt{2}}\end{array}\right)\left(\begin{array}{cc} 0 & 1 \\ 1 & 0\end{array}\right)\left(\begin{array}{c} \frac{1}{\sqrt{2}} \\ \frac{1}{\sqrt{2}}\end{array}\right) = 1 \nonumber \\
      \left<\sigma_z\right> = \chi'^\dag(\hat{x},+) \sigma_z \chi'(\hat{x},+) &=& \left(\begin{array}{cc} \frac{1}{\sqrt{2}} & \frac{1}{\sqrt{2}}\end{array}\right)\left(\begin{array}{cc} 1 & 0 \\ 0 & -1\end{array}\right)\left(\begin{array}{c} \frac{1}{\sqrt{2}} \\ \frac{1}{\sqrt{2}} \end{array}\right) = 0 \nonumber 
  \end{eqnarray}
  The first term of \eqref{LmunuWmunu_pol} should thus be understood as an enhanced/reduced probability for the scattered proton to have spin-up along the direction perpendicular to the momentum transfer, but parallel to the reaction plane. 
  \item[Normal Polarization] Finally, for a state with spin-up along the $y$ axis, which is normal to the scattering plane, both terms vanish:
    \begin{eqnarray}
      \left<\sigma_x\right> = \chi'^\dag(\hat{y},+) \sigma_x \chi'(\hat{y},+) &=& \left(\begin{array}{cc} \frac{1}{\sqrt{2}} & -\frac{i}{\sqrt{2}}\end{array}\right)\left(\begin{array}{cc} 0 & 1 \\ 1 & 0\end{array}\right)\left(\begin{array}{c} \frac{1}{\sqrt{2}} \\ \frac{i}{\sqrt{2}}\end{array}\right) = 0 \nonumber \\
	\left<\sigma_z\right> = \chi'^\dag(\hat{y},+) \sigma_z \chi'(\hat{y},+) &=& \left(\begin{array}{cc} \frac{1}{\sqrt{2}} & -\frac{i}{\sqrt{2}}\end{array}\right)\left(\begin{array}{cc} 1 & 0 \\ 0 & -1\end{array}\right)\left(\begin{array}{c} \frac{1}{\sqrt{2}} \\ \frac{i}{\sqrt{2}} \end{array}\right) = 0 \nonumber
    \end{eqnarray}
    The probability for the scattered proton to have spin-up along the $y$ axis, normal to the scattering plane, is unchanged. 
\end{description}
The same exercise can be repeated for the spin-down eigenstates, which gives the same results with opposite sign. This means that the total scattering cross section, which is summed over final spin states, is unchanged. The results are summarized thusly:
\begin{eqnarray}
  \label{BreitAmplPol}
  L^e_{\mu \nu,A} W^{\mu \nu,A}_N(\mathbf{h}_N = \hat{z}) &=& h\frac{Q^4}{\sin{\frac{\theta_B}{2}}} G_M^2 \nonumber \\
  L^e_{\mu \nu,A} W^{\mu \nu,A}_N(\mathbf{h}_N = \hat{x}) &=& -2hM Q^3 G_E G_M \cot{\frac{\theta_B}{2}} \\
  L^e_{\mu \nu,A} W^{\mu \nu,A}_N(\mathbf{h}_N = \hat{y}) &=& 0 \nonumber
\end{eqnarray}
This completes the derivation of the squared polarized scattering amplitude in the Breit frame. 

A few remarks on the physical interpretation of equations \eqref{BreitAmplPol} are in order. The squared amplitude $L_{\mu\nu}W^{\mu \nu}$, up to dimensionless coupling constants and a factor $Q^{-4}$ for the virtual photon propagator (see \eqref{tensordef}), is the probability amplitude for scattering into the specified final state. In the unpolarized case, the probability of scattering at a specific momentum transfer $Q^2$ and virtual photon polarization $\epsilon$ was derived, regardless of the spin state of either the incoming or the outgoing particles. On the other hand, equations \eqref{BreitAmplPol} represent the change in probability \emph{relative} to the unpolarized case when the spin states of the incident electron and the scattered nucleon are specified.

It is important not to confuse the different probabilistic concepts at work here, and to use conceptually precise language to interpret the meaning of equations \eqref{BreitAmplPol}. In any single reaction, the individual electron and the individual nucleon taking part in the collision are each in some actual spin state which is a linear superposition of spin-up and spin-down eigenstates with respect to an arbitrary quantization axis, and in fact, the quantization axis for either particle can be chosen so that it is in a pure spin-up state with respect to that direction\footnote{The spin state may be changing with time, but a quantization axis can always be chosen for which this statement is instantaneously true.}. For the electron, the natural choice of axis is its momentum direction, and any single electron will be in a quantum state that is a superposition of positive and negative helicity states. Consider a single event in which the incident electron is in a pure positive helicity state\footnote{If the electron is in a pure negative helicity state, then the above results can be read the same way, but with opposite sign, $h \rightarrow -h$. Even if the electron is in an admixture of positive and negative helicity states, it just means that the spin vector of the electron has a nonzero transverse component, and the factor $h$ in $L_{\mu \nu}$ is modulated by the cosine of the angle between the electron's spin and its momentum.}. Then $h=1$ and equation \eqref{BreitAmplPol} can be read as follows:
\begin{enumerate}
\item The probability that the scattered nucleon is in a state with spin-up along its direction of motion is enhanced by $\frac{Q^4}{\sin{\frac{\theta_B}{2}}} G_M^2$ relative to the unpolarized case. The probability that the nucleon is in a spin-down state along the same direction is reduced by the same amount.
\item The probability that the scattered nucleon is in a state with spin up transverse to its motion in the scattering plane is reduced by $2M Q^3 G_E G_M \cot{\frac{\theta_B}{2}}$ relative to the unpolarized case and the probability of a spin-down state along the same direction is enhanced by the same amount.
\item The probability that the scattered nucleon is in a state with spin up along the direction normal to the scattering plane is unchanged relative to the unpolarized case.
\end{enumerate}
In an actual experiment, the spin state of any individual nucleon or electron at the space-time instant of the collision is unknown. Instead, there is a mixed ensemble of many electrons which are spin-polarized, that is, there is a direction in space along which the electron spins are preferentially aligned, with more spin-up than spin-down. Hopefully that direction is as close to longitudinal as possible. There is also a mixed ensemble of initial-state nucleon spins which are unpolarized, meaning that while any single nucleon spin points in some direction, there is no preferred direction in space--the target nucleon's spin is equally likely to point in any direction. The probability amplitudes \eqref{BreitAmplPol} imply that if the electron beam is longitudinally polarized, the statistical ensemble of scattered nucleons will acquire a net polarization, in other words, a preferred direction in space along which more scattered nucleons will have spin-up than spin-down. This acquired polarization is said to be transferred from the electron to the scattered nucleon. 

Equations \eqref{BreitAmplPol} imply that the transferred polarization has longitudinal and transverse components in the scattering plane, and zero component normal to the scattering plane. The components of the transferred polarization are obtained by taking the ratio of the polarized term \eqref{BreitAmplPol} to the unpolarized term \eqref{SymmTerm}. Adopting a self-explanatory notation for the components, one can write
\begin{eqnarray}
  P_l &=& \frac{\left(L_{\mu \nu,A}W^{\mu \nu,A}\right)_{\mathbf{h}_N = \hat{z}}}{\left(L_{\mu \nu,S}W^{\mu \nu,S}\right)} \label{Pldef} \\
  P_t &=& \frac{\left(L_{\mu \nu,A}W^{\mu \nu,A}\right)_{\mathbf{h}_N = \hat{x}}}{\left(L_{\mu \nu,S}W^{\mu \nu,S}\right)} \label{Ptdef} \\
  P_n &=& 0
\end{eqnarray}
which is true in the coordinate system of the Breit frame, in which the $\hat{z}$ direction is \emph{defined} as the direction of the momentum transfer and the $\hat{x}$ direction is \emph{defined} as the in-plane transverse coordinate. In a generic coordinate system, one should rename $(\hat{x},\hat{y},\hat{z})$ to $(\hat{t},\hat{n},\hat{l})$ to avoid confusion.

Of particular interest is the ratio of transverse to longitudinal transferred polarization, because it directly measures the electric-to-magnetic form factor ratio: 
\begin{eqnarray}
  \frac{P_t}{P_l} &=& -\frac{G_E}{G_M} \frac{2M}{Q}\cos{\frac{\theta_B}{2}} \label{ptoverplfirst}
\end{eqnarray}
This result leads to distinct advantages of polarization transfer experiments over Rosenbluth separation experiments in measuring the electric form factor at high momentum transfer.
\subsection{Back to the Lab Frame}
\paragraph{}
The polarization components, defined as ratios of tensor contractions $L_{\mu\nu}W^{\mu\nu}$, are Lorentz-invariant physical observables. The results \eqref{Ptdef} and \eqref{Pldef} thus also apply in the lab frame. It is preferable to express $P_t$, $P_l$, and their ratio in terms of lab-frame kinematic variables. Equation \eqref{BreitLabAngles} can be rearranged in order to rewrite the kinematic factor in \eqref{ptoverplfirst} in terms of lab-frame quantities:
\begin{eqnarray}
  \frac{\cot^2{\frac{\theta_e}{2}}}{1+\tau} &=& \frac{\cos^2{\frac{\theta_B}{2}}}{1-\cos^2{\frac{\theta_B}{2}}} \nonumber \\
  \frac{1}{\cos^2{\frac{\theta_B}{2}}} &=& 1 + (1+\tau)\tan^2{\frac{\theta_e}{2}} \\
  &=& \frac{1}{\cos^2{\frac{\theta_e}{2}}}\left[1 + \tau \sin^2{\frac{\theta_e}{2}} \right] \nonumber \\
  &=& \tan^2{\frac{\theta_e}{2}}\left[\csc^2{\frac{\theta_e}{2}} + \tau \right] \label{kinfactor1}
\end{eqnarray}
Recalling some general properties of elastic scattering, \eqref{kinfactor1} simplifies to:
\begin{eqnarray}
  Q^2 &=& 2M\nu = 2M(E_e-E'_e) \nonumber \\
  \tau &=& \frac{Q^2}{4M^2} = \frac{\nu}{2M} \nonumber \\
  \frac{E_e}{E'_e} &=& 1 + \frac{2E_e}{M}\sin^2{\frac{\theta_e}{2}} \nonumber \\
  \sin^2{\frac{\theta_e}{2}} &=& \frac{M}{2E_e}\frac{E_e-E'_e}{E'_e} \nonumber \\
  \Rightarrow \csc^2{\frac{\theta_e}{2}} + \tau &=& \frac{2E_e E'_e}{M(E_e-E'_e)} + \frac{E_e-E'_e}{2M} \nonumber \\
  &=& \frac{4E_eE'_e + (E_e-E'_e)^2}{2M\nu} \nonumber \\
  &=& \frac{(E_e+E'_e)^2}{Q^2} \nonumber \\
  \Rightarrow \frac{1}{\cos^2{\frac{\theta_B}{2}}} &=& \tan^2{\frac{\theta_e}{2}}\frac{(E_e+E'_e)^2}{Q^2} \nonumber \\
  \frac{1}{\cos{\frac{\theta_B}{2}}} &=& \tan{\frac{\theta_e}{2}}\frac{E_e+E'_e}{Q} \label{secthetabfinal}
\end{eqnarray}
Substituting \eqref{secthetabfinal} into \eqref{ptoverplfirst} and rearranging gives a simple formula for the ratio of the electric and magnetic form factors in terms of the ratio of transferred polarization components:
\begin{equation}
  \frac{G_E}{G_M} = -\frac{P_t}{P_l} \frac{E_e+E'_e}{2M}\tan{\frac{\theta_e}{2}} \label{FFratiomasterformula}
\end{equation}

\begin{figure}[h]
  \begin{center}
    \includegraphics[width=.6\textwidth]{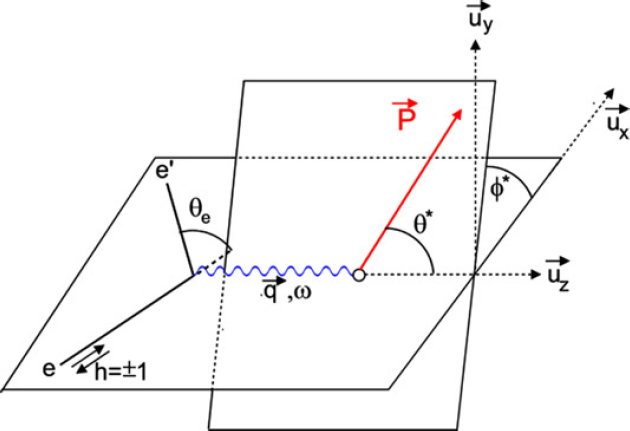}
    \caption{\label{beamtargetasymillustration} Definition of angles in \eqref{beamtgtasymformula}}. Reproduced with permission from \cite{PerdrisatPunjabiVanderhaegen2007}.
  \end{center}
\end{figure}

The final step is to derive the individual transferred polarization components $P_l$ \eqref{Pldef} and $P_t$ \eqref{Ptdef} in the lab frame. The symmetric(unpolarized) term $L_{\mu\nu,S}W^{\mu\nu,S}$ is given by \eqref{SymmTerm}. Recalling the definition of the reduced cross section \eqref{sigmardef}, the unpolarized amplitude is given by:
\begin{eqnarray}
  \left(L^e_{\mu \nu,S} W^{\mu \nu,S}_N\right) &=& 2Q^2 M^2 \left[\frac{G_E^2 + \tau G_M^2}{1+\tau}\cot^2{\frac{\theta_e}{2}} + 2\tau G_M^2\right] \nonumber \\
  &=& 2Q^2 M^2 \cot^2{\frac{\theta_e}{2}} \sigma_r \label{LWunpol}
\end{eqnarray}
so that the longitudinal component is
\begin{eqnarray}
  \sigma_r P_l &=& h \frac{Q^2}{2M^2 \sin{\frac{\theta_B}{2}} \cot^2{\frac{\theta_e}{2}}} G_M^2 \nonumber \\ 
  &=& h \frac{2\tau}{(1+\tau)\sin{\frac{\theta_B}{2}}\cot^2{\frac{\theta_B}{2}}} G_M^2 \nonumber \\
  &=& h \frac{2\tau}{(1+\tau)\cos{\frac{\theta_B}{2}}\cot{\frac{\theta_B}{2}}} G_M^2 \nonumber \\
  &=& 2h \frac{\tau}{1+\tau}\tan^2{\frac{\theta_e}{2}}\frac{E_e+E'_e}{Q} \sqrt{1+\tau} G_M^2 \nonumber \\
  &=& h \sqrt{\frac{\tau}{1+\tau}}\tan^2{\frac{\theta_e}{2}}\frac{E_e+E'_e}{M} G_M^2 
\end{eqnarray}
and the transverse component is
\begin{eqnarray}
  \sigma_r P_t &=& - 2h \frac{Q}{2M} \frac{\cot{\frac{\theta_B}{2}}}{\cot^2{\frac{\theta_e}{2}}} G_E G_M \nonumber \\
  &=& -2h \sqrt{\frac{\tau}{1+\tau}} \tan{\frac{\theta_e}{2}} G_E G_M
\end{eqnarray}
Another popular convention in the literature uses a slightly different definition of the reduced cross section, which differs from \eqref{sigmardef} by a factor of $1+\tau$. In this convention, the reduced cross section is defined as $\sigma_r = G_E^2 + \frac{\tau}{\epsilon}G_M^2$, whereas in the definition \eqref{sigmardef}, the reduced cross section equals $\frac{1}{1+\tau}\left(G_E^2+\frac{\tau}{\epsilon}G_M^2\right)$. Under the alternative definition, the transferred polarization components become 
\begin{eqnarray}
  I_0 P_l &=& h \sqrt{\tau(1+\tau)}\tan^2{\frac{\theta_e}{2}}\frac{E_e+E'_e}{M} G_M^2 \label{Pl} \\
  I_0 P_t &=& -2h \sqrt{\tau(1+\tau)} \tan{\frac{\theta_e}{2}} G_E G_M \label{Pt} \\ 
  I_0 P_n &=& 0 \\
  I_0 &\equiv& G_E^2 + \frac{\tau}{\epsilon}G_M^2 
\end{eqnarray}
This concludes the derivation of the components of the transferred polarization in elastic electron-nucleon scattering in the Born approximation. The key results are \eqref{FFratiomasterformula}, \eqref{Pl}, and \eqref{Pt}. They were first derived in the late 1960s (\cite{AkhiezerRekalo1}, \cite{AkhiezerRekalo2}), and again in the early 1980s (\cite{ArnoldCarlsonGross}). Published in the early days of polarized electron beam technology, \cite{ArnoldCarlsonGross} described a specific program of polarization transfer experiments to measure the proton and neutron electric form factors and the deuteron charge and quadrupole form factors. Although the derivation presented here focused exclusively on polarization transfer experiments, there is another class of double-polarization experiment that deserves mention. Instead of measuring the transferred polarization to an unpolarized target, one can measure the cross-section asymmetry between $+$ and $-$ electron helicity states in elastic scattering on a polarized nucleon target, without measuring the polarizations of the outgoing particles. This technique is sensitive to the form factor ratio in a similar fashion. Rather than proceed with another lengthy derivation along the same lines as the one just given, the relevant formula is quoted below. The asymmetry in elastic scattering between positive and negative electron helicity states is equal to $A_{meas} = P_{beam} P_{target} A_{phys}$, where $A_{phys}$ is given by\cite{Dombey1969,DonnellyRaskin86,PerdrisatPunjabiVanderhaegen2007}
\begin{eqnarray}
  A_{phys} &=& -\frac{2\sqrt{\tau(1+\tau)}\tan{\frac{\theta_e}{2}}}{\frac{G_E^2}{G_M^2} + \frac{\tau}{\epsilon}}\left[ \sin{\theta^*}\cos{\phi^*} \frac{G_E}{G_M} \right. \nonumber \\ 
    & & \left. + \sqrt{\tau\left[1+(1+\tau)\tan^2{\frac{\theta_e}{2}}\right]} \cos{\theta^*} \right] \label{beamtgtasymformula}
\end{eqnarray}
The angles $\theta^*$ and $\phi^*$ are, respectively, the polar and azimuthal angles of the target polarization vector $\mathbf{P}$ with respect to the direction of the momentum transfer $\mathbf{q}$. The azimuthal angle is measured from the reaction plane defined by $\mathbf{k} \times \mathbf{k'}$ toward the plane defined by $\mathbf{q} \times \mathbf{P}$ as shown in figure \ref{beamtargetasymillustration}.
In order to measure the electric form factor using this technique, the optimal orientation of the target polarization is perpendicular to the momentum transfer and parallel to the reaction plane.
\section{Existing Nucleon Form Factor Data}
\paragraph{}
A recent review paper \cite{PerdrisatPunjabiVanderhaegen2007} contains an exhaustive compilation of previous experimental data on nucleon form factors, from Hofstadter's Nobel Prize experiments to the present day. In order to fully characterize the nucleon's electromagnetic structure, it must be studied in both isospin states, the proton and the neutron. The proton can be studied directly using hydrogen targets. The neutron, on the other hand, happens to be unstable; it decays weakly by the reaction $n \rightarrow p + e^- + \bar{\nu}_e$ with a lifetime $\tau = 885.7 \pm 0.8$ seconds\cite{PDG2008}, so there are no free neutron targets available for electron scattering experiments. Since stable neutrons only exist in nuclei, the neutron must be studied indirectly using light, weakly bound nuclei such as deuterium ($^2H$) and Helium-3 ($^3He$). It is for this reason that the proton form factors are known much more precisely and over a greater $Q^2$ range than the neutron form factors.
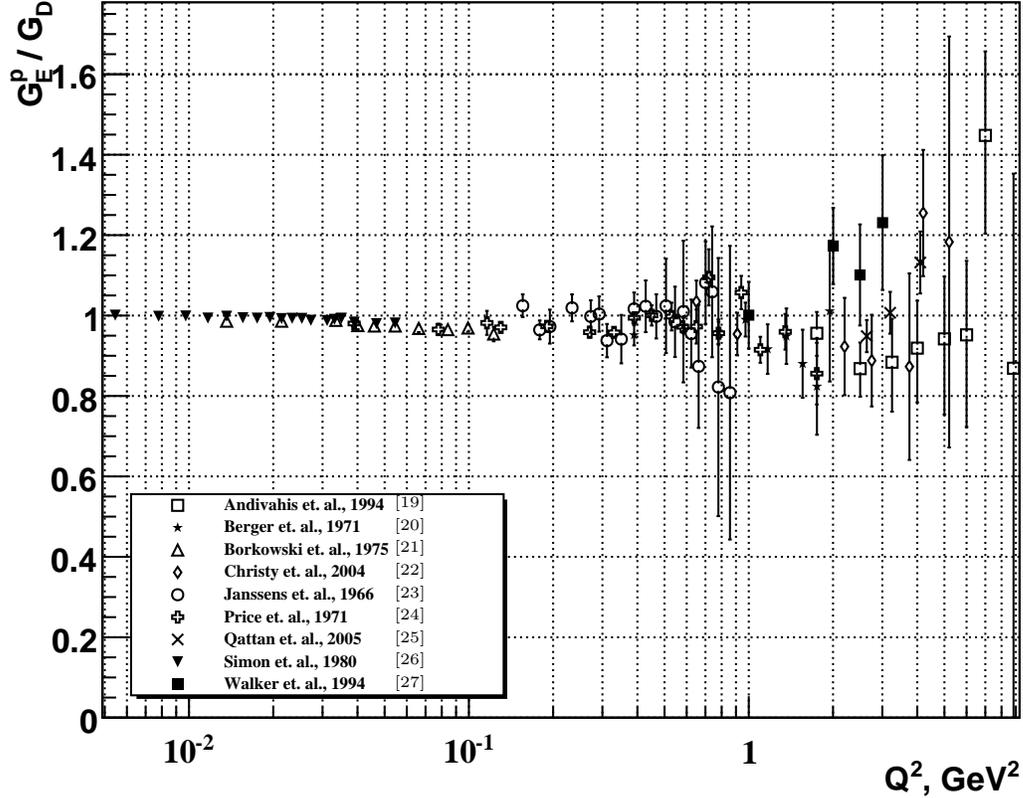
\begin{figure}[h]
  \begin{center}
    \setlength{\unitlength}{1.0\textwidth}
    \begin{picture}(1.0,0.75)
      \put(0,0){\includegraphics[angle=90,width=1.0\textwidth]{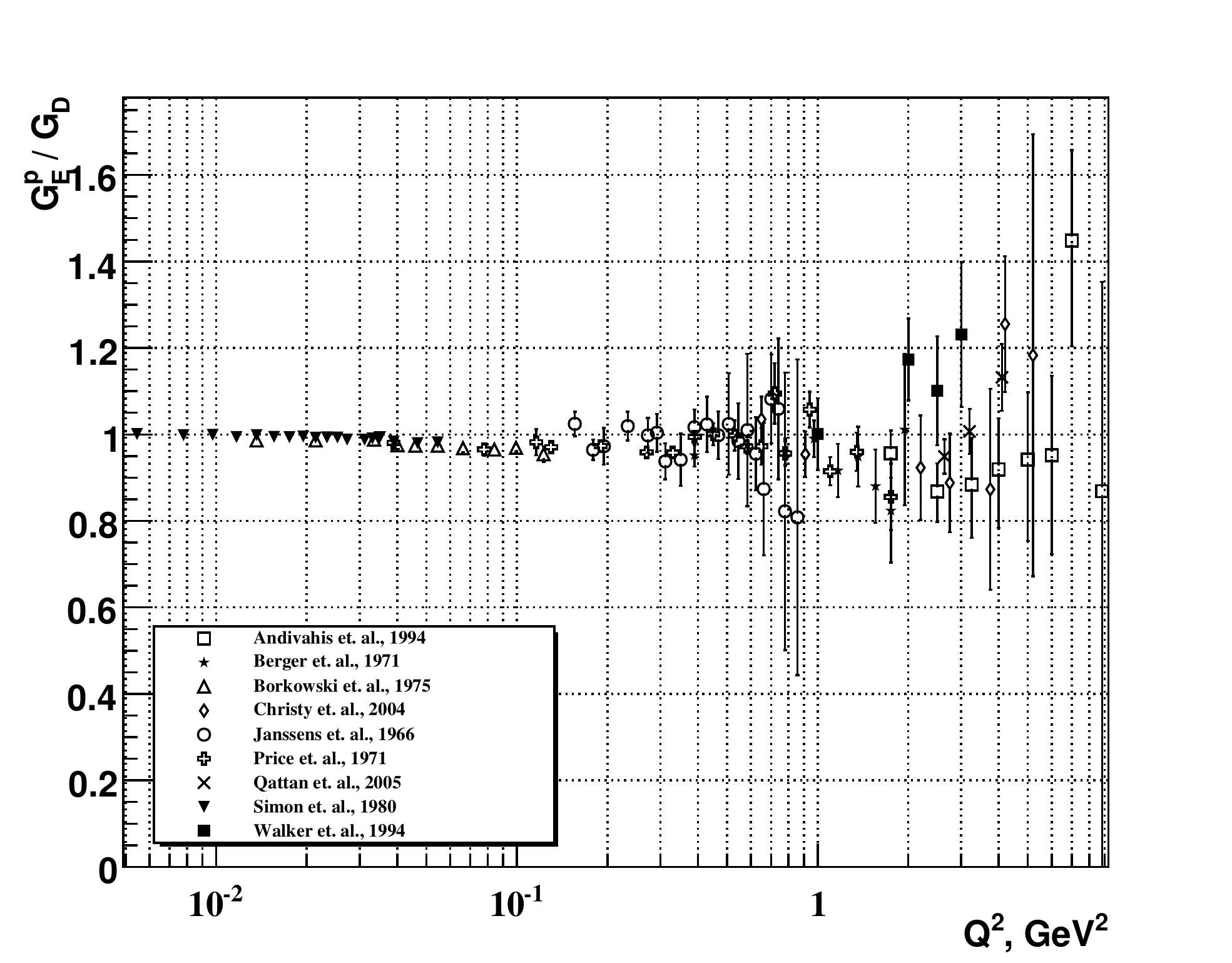}}
      \put(.355,.2615){\tiny \cite{Andi94}}
      \put(.355,.2419){\tiny \cite{Berger71}}
      \put(.355,.2224){\tiny \cite{Bork75}}
      \put(.355,.2028){\tiny \cite{Christy04}}
      \put(.355,.1833){\tiny \cite{Janssens66}}
      \put(.355,.1637){\tiny \cite{Price71}}
      \put(.355,.1441){\tiny \cite{Qattan05}}
      \put(.355,.1246){\tiny \cite{Simon80}}
      \put(.355,.1050){\tiny \cite{Walker94}}
    \end{picture}
    \caption{\label{GEpRosenData} Proton electric form factor data obtained by the Rosenbluth separation technique, normalized to the dipole form factor $G_D(Q^2) = (1+Q^2/\Lambda^2)^{-2}$, $\Lambda^2 = .71\ GeV^2$.}
  \end{center}
\end{figure}
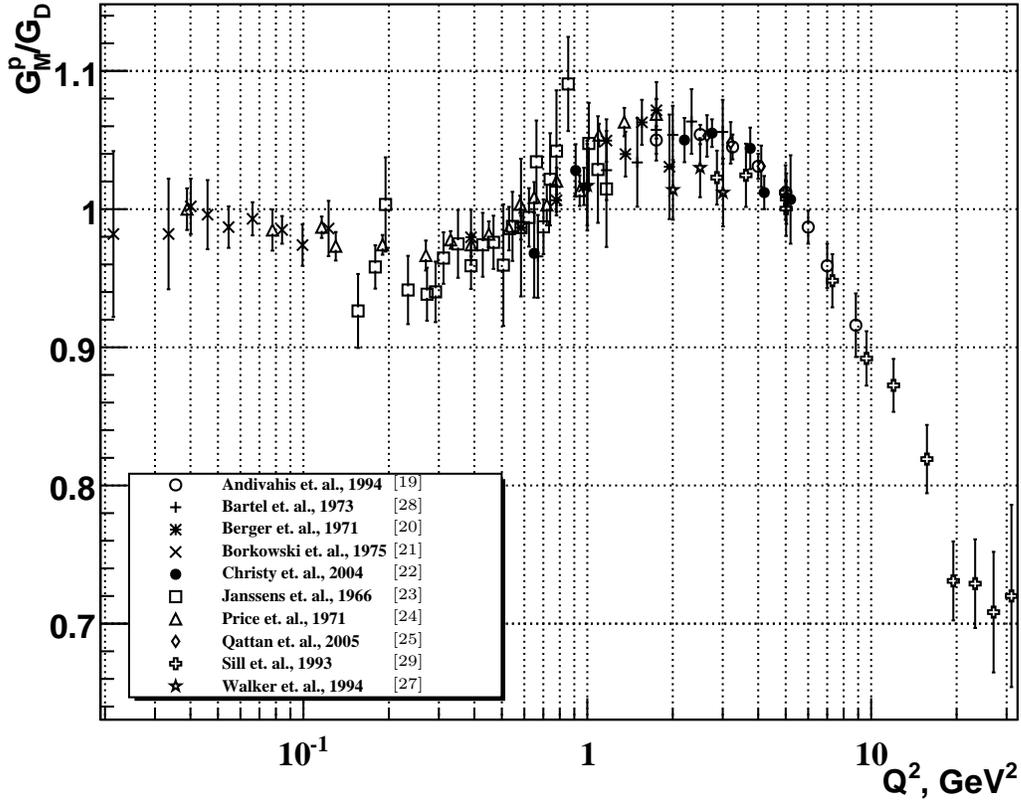
\begin{figure}[h]
  \begin{center}
    \setlength{\unitlength}{1.0\textwidth}
    \begin{picture}(1.0,0.75)      
      \put(0,0){\includegraphics[angle=90,width=1.0\textwidth]{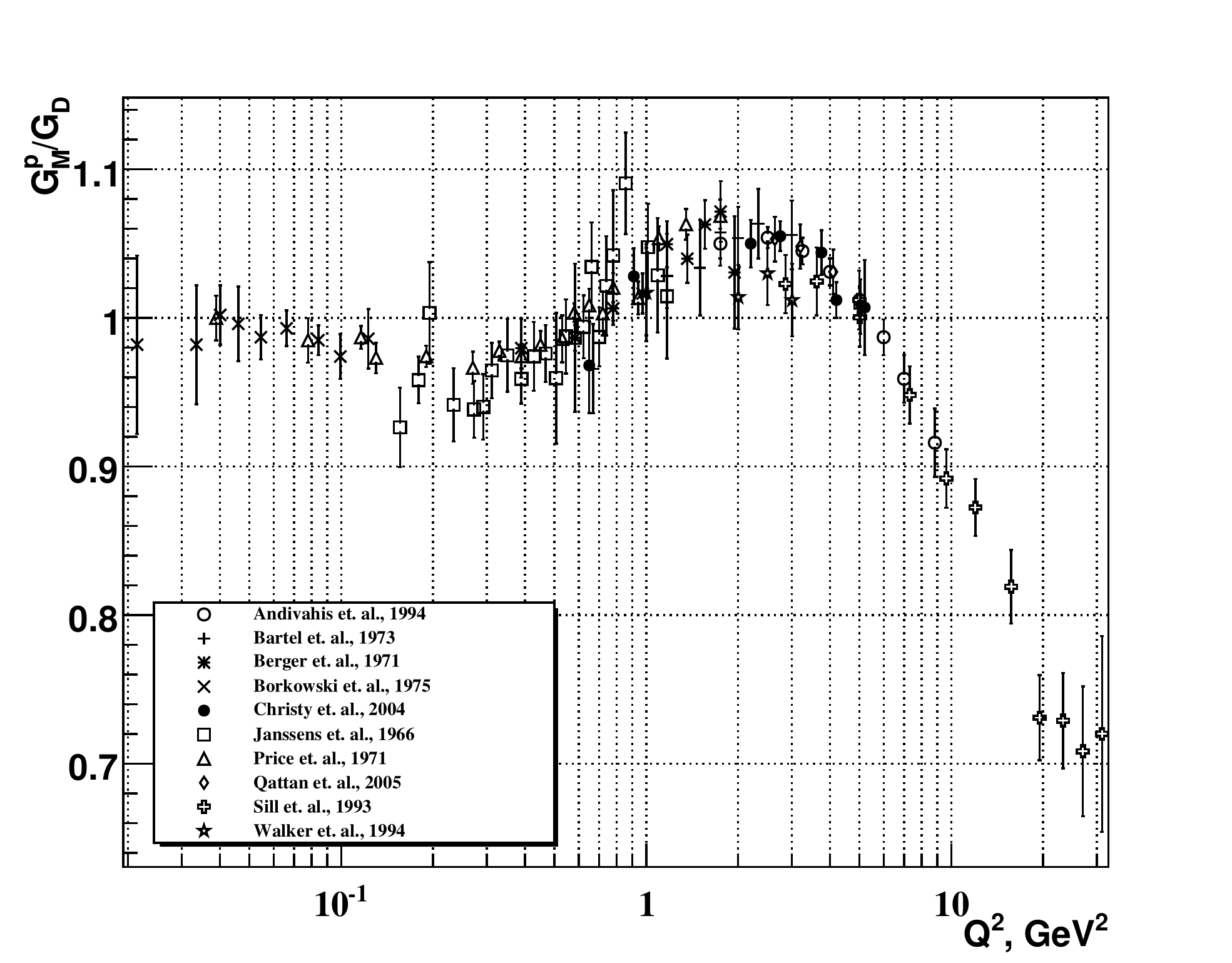}}
      \put(.355,.2811){\tiny \cite{Andi94}}
      \put(.355,.2615){\tiny \cite{Bartel73}}
      \put(.355,.2419){\tiny \cite{Berger71}}
      \put(.355,.2224){\tiny \cite{Bork75}}
      \put(.355,.2028){\tiny \cite{Christy04}}
      \put(.355,.1833){\tiny \cite{Janssens66}}
      \put(.355,.1637){\tiny \cite{Price71}}
      \put(.355,.1441){\tiny \cite{Qattan05}}
      \put(.355,.1246){\tiny \cite{Sill93}}
      \put(.355,.1050){\tiny \cite{Walker94}}
    \end{picture}
    \caption{\label{GMpRosenData} Proton magnetic form factor data obtained from Rosenbluth separation/cross section data, normalized to the dipole form factor $G_D(Q^2) = (1+Q^2/\Lambda^2)^{-2}$, $\Lambda^2 = .71\ GeV^2$. }
  \end{center}
\end{figure}

For relatively low $Q^2$ values, below about $2.0$ GeV$^2$, it is found that the $Q^2$-dependence of the electric and magnetic form factors of the proton, and the magnetic form factor of the neutron, are well described by a dipole form
\begin{eqnarray}
  G_D(Q^2) &=& (1+\frac{Q^2}{\Lambda^2})^{-2} \\
  G_E^p &=& G_D \nonumber \\
  G_M^p &=& \mu_p G_D \nonumber \\
  G_M^n &=& \mu_n G_D \nonumber
\end{eqnarray}
with $\Lambda^2 = .71\ GeV^2$. Figure \ref{GEpRosenData} shows a representative sample of $G_E^p$ data obtained by the Rosenbluth separation technique spanning close to three orders of magnitude in $Q^2$, normalized to $G_D$ to remove the dominant $Q^2$ dependence. From very low $Q^2$ up to about $1-2\ GeV^2$, the agreement of $G_E^p$ with the dipole form factor is excellent. At higher $Q^2$, the precision of the data rapidly gets worse, and although most of the data is consistent with $G_D$, the electric form factor is clearly not very well known from Rosenbluth separation at high $Q^2$. Recalling the Rosenbluth formula \eqref{RosenbluthFormula} for the elastic electron-proton scattering cross section in the Born approximation, the reason for this immediately becomes clear. The magnetic form factor is multiplied by a factor $\tau$; the electric form factor isn't. 

At higher $Q^2$, the contribution of the magnetic form factor begins to dominate the total cross section. The fact that $G_M^p$ is a factor of $\mu_p \approx 2.793$ larger than the electric form factor at $Q^2=0$ shifts the onset of $G_M^p$-dominance of the cross section to even lower $Q^2$. For example, assuming $G_E^p = G_M^p / \mu_p$, then at $Q^2=2.0\ GeV^2$, the magnetic contribution to the Born cross section is already about 82\% for a ``small'' electron scattering angle of 15$^\circ$ and 95\% for a ``large'' electron scattering angle of 90$^\circ$. At $5\ GeV^2$, the magnetic contribution is 92\% at $\theta_e = 15^\circ$ and 98.5\% at $\theta_e = 90^\circ$. Clearly, starting at one to several $GeV^2$ in $Q^2$, the magnetic dominance of the cross section makes Rosenbluth separations prohibitively difficult. Even assuming the exact validity of the one-photon exchange approximation, one must measure the cross section with very high precision over a wide range of $\epsilon$ and have very good control of point-to-point systematic uncertainties and overall normalization uncertainties in the experimental setup.

In fact, effects beyond the Born approximation cannot be neglected in a Rosenbluth separation experiment at high $Q^2$. Radiative corrections to the cross section must be calculated to relate the cross section measured in an experiment to the Born-level cross section which measures the form factors. At leading order in $\alpha$, the radiative corrections to the cross section include both virtual terms, such as one-loop vacuum polarization and vertex corrections, and electron self-energy corrections, and the radiation of real photons; i.e., Bremsstrahlung. The Bremsstrahlung corrections are further divided into external Bremsstrahlung, in which the incident and scattered particles radiate due to interactions with the material they traverse before reaching the detectors (before and after the primary scattering), and internal Bremsstrahlung, in which the incident or scattered electron radiates a real photon in the field of the nucleon participating in the scattering and vice versa. 

The difference between the external and internal Bremsstrahlung correction is that the internal correction is coherent with the Born-level scattering amplitude, meaning the amplitudes interfere, whereas the external correction is incoherent and factorizes from the Born-level process. The virtual corrections depend on $Q^2$, but are independent of $\epsilon$. They do, however, modify the value of $(G_M^p)^2$ as an overall $Q^2$ dependent correction to the cross section. Bremsstrahlung corrections, on the other hand, are energy (and therefore $\epsilon$) dependent, and change the value of $Q^2$. In general, the radiative corrections to the cross section in a Rosenbluth experiment are strongly $\epsilon$-dependent, and the slope of the Rosenbluth plot can change dramatically in going from uncorrected to corrected cross sections. The accuracy with which $G_E^2$ can be determined in a Rosenbluth separation experiment at high $Q^2$ depends critically on the accuracy of the radiative correction, as discussed in \cite{PerdrisatPunjabiVanderhaegen2007}.

Figure \ref{GMpRosenData} shows a representative sample of existing measurements of $G_M^p$ from Rosenbluth separation and/or elastic-$ep$ cross section data. $G_M^p$ is known to higher $Q^2$ than $G_E^p$--the data extend to $Q^2 \approx 30\ GeV^2$. In the highest-$Q^2$ measurements \cite{Sill93}, the magnetic form factor was extracted from a single cross section measurement assuming $G_E^p = G_M^p / \mu_p$ as opposed to a Rosenbluth separation. The cross section is too small at this $Q^2$ to perform a meaningful Rosenbluth separation, given the energy and luminosity capabilities of electron accelerators existing today (or in 1993, when the experiments \cite{Sill93} were published). Although the assumption of form factor scaling is not valid above $Q^2 \approx 1-2\ GeV^2$, the error introduced is approximately as small as the electric contribution to the cross section, i.e., less than one percent. Conversely, at very low $Q^2$, the dominant contribution to the cross section comes from $(G_E^p)^2$, and the magnetic form factor data are sparse below $Q^2 \approx .03\ GeV^2$ while $G_E^p$ is known all the way down to $Q^2 \approx .005\ GeV^2$. 

If the form factors are interpreted as Fourier transforms of its charge and magnetization distributions, and these distributions are assumed to be spherically symmetric, then the dipole form factor corresponds to a charge distribution with an exponential radial dependence:
\begin{equation}
  \rho(r) = \frac{\Lambda^3}{8\pi} e^{-\Lambda r} \label{expdist}
\end{equation}
The charge density in \eqref{expdist} is normalized so that its integral over all space equals 1, the total charge of the proton. Taking the Fourier transform of \eqref{expdist} leads to the dipole form factor as follows:
\begin{eqnarray}
  G(Q) &=& \frac{\Lambda^3}{4} \int_0^\infty e^{-\Lambda r} r^2 dr \int_{-1}^1 e^{iQr\cos{\theta}}d\cos{\theta} \nonumber \\
  G(Q) &=& \frac{\Lambda^3}{4iQ} \int_0^\infty r e^{-\Lambda r}\left[e^{iQr}-e^{-iQr}\right] dr \nonumber \\
  G(Q) &=& \frac{\Lambda^3}{4iQ} \left[\left.r e^{-\Lambda r}\left(\frac{e^{iQr}}{-\Lambda+iQ}+\frac{e^{-iQr}}{\Lambda+iQ}\right)\right|^{r=\infty}_{r=0} \right. \nonumber \\
    & & \left. - \int_0^\infty dr e^{-\Lambda r}\left(\frac{e^{iQr}}{-\Lambda+iQ}+\frac{e^{-iQr}}{\Lambda+iQ}\right)\right] \nonumber \\
  &=& \frac{\Lambda^3}{4iQ} \left[\frac{e^{(-\Lambda+iQ)r}}{(-\Lambda+iQ)^2} - \frac{e^{(-\Lambda-iQ)r}}{(\Lambda+iQ)^2}\right]^{r=\infty}_{r=0} \nonumber \\
  &=& \frac{\Lambda^3}{4iQ}\left[\frac{4i\Lambda Q}{(\Lambda^2+Q^2)^2}\right] = (1+\frac{Q^2}{\Lambda^2})^{-2} \nonumber \\
  &=& G_D(Q^2)
\end{eqnarray}
with $\Lambda^{-1}$ as the characteristic ``size'' of the proton. Noting the agreement of the data with the dipole form factor at sufficiently low $Q^2$, the r.m.s. proton charge radius can be estimated as
\begin{eqnarray}
  \left<r^2\right> &=& -6 \left(\frac{d}{dQ^2} G_E(Q^2)\right)_{Q^2=0} = \left.12(1+\frac{Q^2}{\Lambda^2})^{-3}\Lambda^{-2}\right|_{Q^2=0} \nonumber \\
  \Rightarrow \left<r^2\right> &=& 12\Lambda^{-2} \approx 16.9\ GeV^{-2} = 0.68\ fm^2 \nonumber \\
  \sqrt{\left<r^2\right>} &\approx& 0.82\ fm
\end{eqnarray}
Of course, the Fourier transform interpretation of the form factors is naive and ignores the substantial relativistic effects that invalidate such a simple and intuitive picture of their physical meaning.

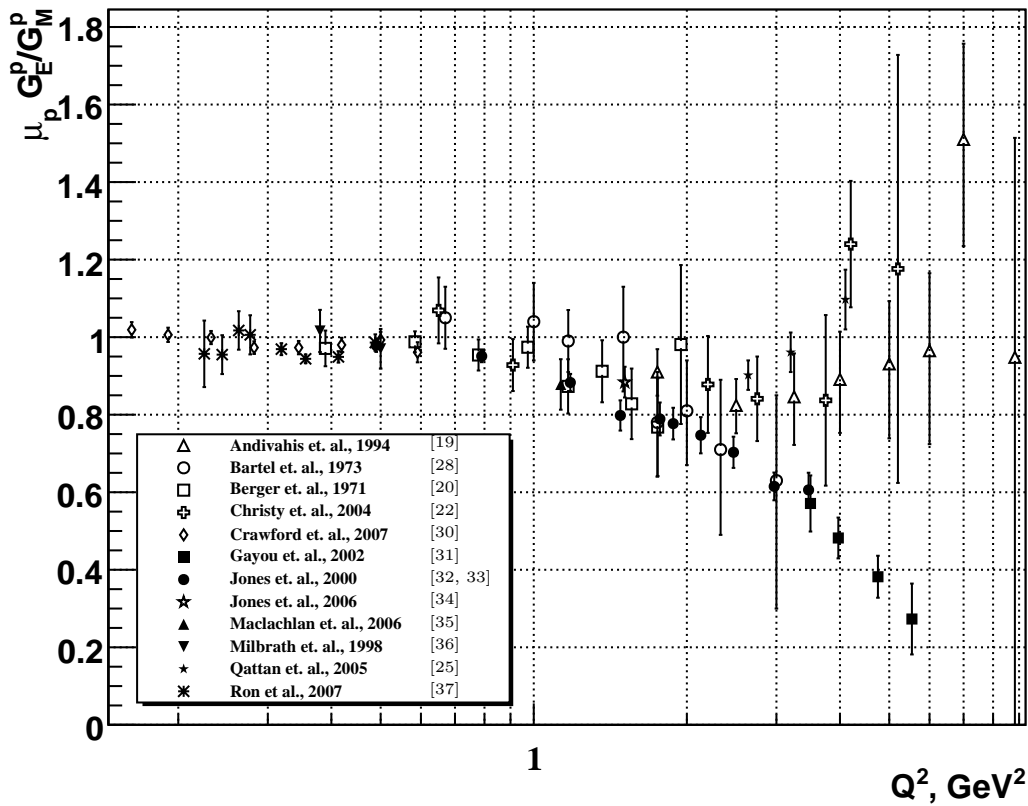
\begin{figure}[h]
  \begin{center}
    \setlength{\unitlength}{1.0\textwidth}
    \begin{picture}(1.0,.75)
      \put(0,0){\includegraphics[angle=90,width=1.0\textwidth]{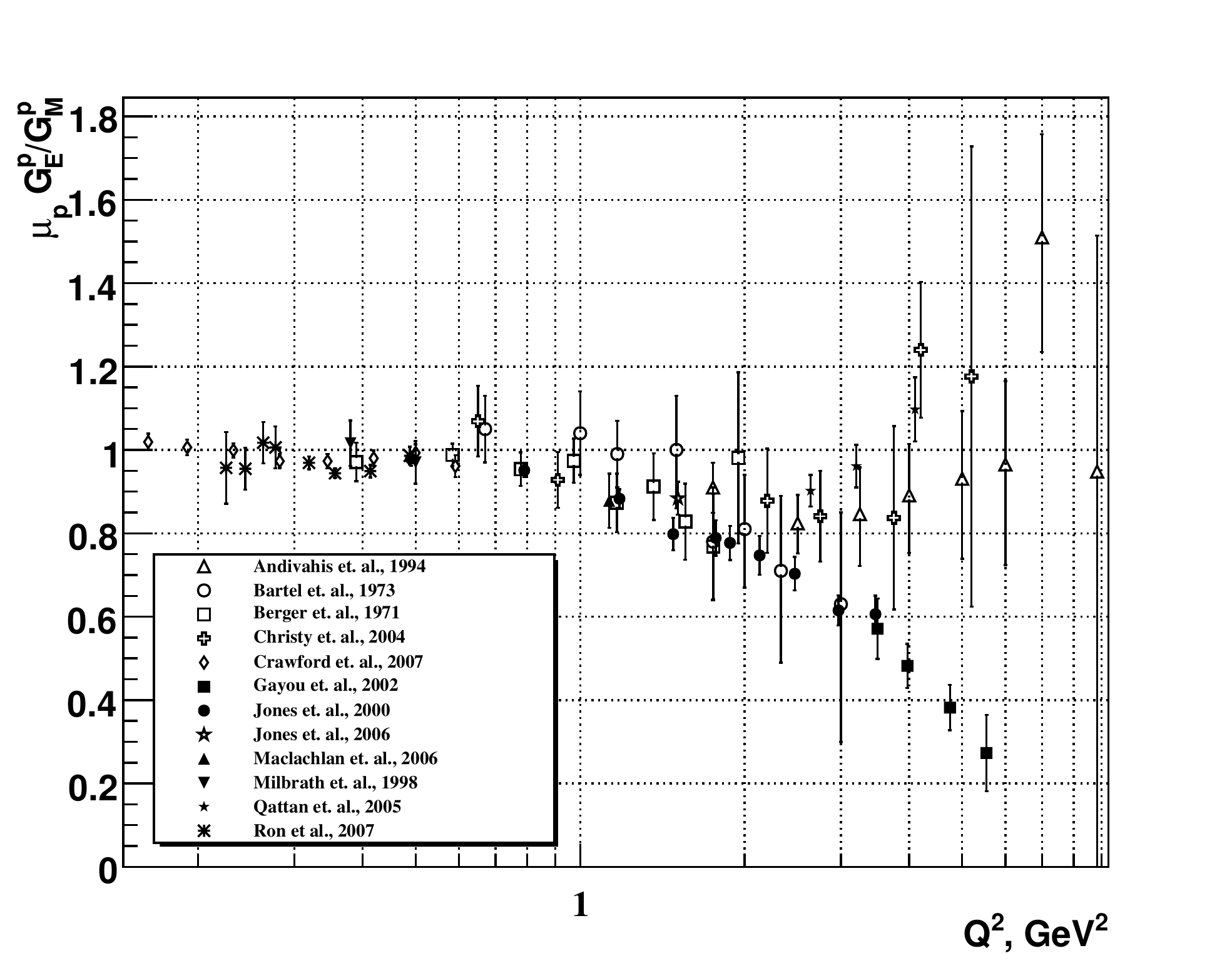}}
      \put(.38,.3202){\tiny \cite{Andi94}}
      \put(.38,.3007){\tiny \cite{Bartel73}}
      \put(.38,.2811){\tiny \cite{Berger71}}
      \put(.38,.2615){\tiny \cite{Christy04}}
      \put(.38,.2419){\tiny \cite{Crawford07}}
      \put(.38,.2224){\tiny \cite{Gayou02}}
      \put(.38,.2028){\tiny \cite{Jones00,Punjabi05}}
      \put(.38,.1833){\tiny \cite{Jones06}}
      \put(.38,.1637){\tiny \cite{Maclachlan06}}
      \put(.38,.1441){\tiny \cite{Milbrath98}}
      \put(.38,.1246){\tiny \cite{Qattan05}}
      \put(.38,.1050){\tiny \cite{Ron07}}
    \end{picture}
    \caption{\label{Ralldata} The proton form factor ratio $\mu_p G_E^p/G_M^p$ from cross section and polarization experiments.}
  \end{center}
\end{figure}

The ability to perform double-polarization experiments allows a precise determination of the proton electric form factor to higher $Q^2$ than is generally possible with Rosenbluth separations. As \eqref{FFratiomasterformula} and \eqref{beamtgtasymformula} show, the polarization observables, unlike the scattering cross section, are sensitive to the interference between the electric and magnetic contributions to the scattering amplitude, making such experiments competitive with and in fact superior to cross section measurements in determining the proton and neutron electric form factors at high $Q^2$, where the magnetic form factor completely takes over the cross section. Figure \ref{Ralldata} shows a representative sample of data obtained from both cross section and polarization experiments on the proton form factor ratio $\mu_p G_E^p/G_M^p$.  
Experiments \cite{Andi94,Bartel73,Berger71,Christy04,Qattan05} were all Rosenbluth separation experiments. Experiments \cite{Gayou02,Punjabi05,Maclachlan06,Milbrath98,Ron07} used recoil polarization, while \cite{Crawford07,Jones06} used polarized targets and measured the beam-target asymmetry \eqref{beamtgtasymformula}. The polarization transfer experiments at high $Q^2$ \cite{Punjabi05,Gayou02} revealed with high precision a strong deviation from the empirical scaling law $\mu_p G_E^p/G_M^p = 1$ and from all existing Rosenbluth separation data at similar $Q^2$. This ``crisis'' precipitated renewed interest in nucleon form factors and intense experimental and theoretical efforts to understand the discrepancy and the form factors themselves that continues today. It is worth remarking that there is little theoretical justification for form factor scaling; it is simply a rough experimental fact for $Q^2<1\ GeV^2$. In fact, the very definitions of the Sachs form factors, $G_E = F_1 - \tau F_2$ and $G_M = F_1 + F_2$, suggest that $G_E / G_M$ should decrease with increasing $Q^2$ and possibly become negative\footnote{The high-$Q^2$ asymptotic behavior of the Sachs form factor ratio should eventually approach the leading asymptotic behavior expected from perturbative QCD. As will be discussed later, simple valence-quark counting rules suggest the ratio of Pauli and Dirac form factors should scale as $Q^2 F_2/F_1 \xrightarrow[Q^2 \rightarrow \infty]{} constant$. This would imply that $\mu_p G_E^p/G_M^p \xrightarrow[Q^2 \rightarrow \infty]{} constant$} (under certain assumptions on the behavior of $F_1$ and $F_2$). In response to the surprising results \cite{Punjabi05,Gayou02}, a new, high-precision Rosenbluth separation experiment\cite{Qattan05} was conducted to discern whether a problem existed with earlier cross section-based experiments in this $Q^2$ region that could be revealed by a more precise experiment. This ``Super-Rosenbluth'' experiment was different from previous Rosenbluth separation experiments in that the scattered proton was detected instead of the scattered electron. Detecting the scattered proton in an elastic $ep$ cross section measurement has several inherent advantages over detecting the scattered electron. 
\begin{itemize}
\item The $\epsilon$ dependence of the proton cross section $d\sigma/d\Omega_p$ is much weaker than the $\epsilon$ dependence of the electron cross section $d\sigma/d\Omega_e$. At low $\epsilon$ values in particular, the Jacobian of the reaction, defined as the ratio of the electron and proton solid angles $d\Omega_e/d\Omega_p$, grows quite large. 
\item The proton momentum is constant at fixed $Q^2$, whereas the electron momentum varies strongly with $\epsilon$.
\item The $\epsilon$-dependence of radiative corrections to the cross section is smaller when the proton is detected. 
\item The $\epsilon$-dependence of the effect of offsets in beam energy and/or scattering angle on the extracted cross section is smaller when the proton is detected.
\end{itemize}
These advantages greatly reduce the systematic uncertainties in a Rosenbluth separation experiment relative to experiments in which the electron is detected, allowing a more precise separation of $G_E^2$ and $G_M^2$. The results of \cite{Qattan05} were consistent with previous Rosenbluth separation experiments, appearing to rule out an undiscovered systematic error in those experiments, and establishing an even stronger disagreement between the cross section and polarization data at high $Q^2$. Presently, significant theoretical and experimental efforts are being devoted to understanding the discrepancy in terms of physics beyond the Born approximation and the standard radiative correction procedures upon which most of the published cross section data are based. 

It is thought that the discrepancy can be largely accounted for by the effect of the two-photon-exchange(TPEX) process, in which both photons are ``hard''. This process is generally neglected in the standard radiative correction procedures. Since the calculation of the TPEX process in elastic $eN$ scattering is necessarily sensitive to the structure of the nucleon through the virtual intermediate hadronic state, it is inherently model dependent and cannot be calculated exactly, in contrast to the case where one of the two photons is ``soft'', which is part of the standard radiative corrections and is well understood, because its dominant infrared part can be factorized in the observables. The process is generally thought to affect the cross section by at most several percent, but in a strongly $\epsilon$ dependent way. Guichon and Vanderhaeghen\cite{GuichonVanderhaeghenTPEX} showed how a TPEX effect of this size could affect the outcome of a Rosenbluth separation drastically while only affecting the result of a polarization transfer experiment at the few percent level. This dramatically illustrates the difficulty of extracting a $\epsilon G_E^2$ term whose contribution to the cross section is similar in relative importance to incompletely understood effects beyond the Born approximation and standard radiative corrections. The enhanced sensitivity to $G_E$ of the recoil polarization method and the diminished relative importance of radiative corrections and TPEX in the determination of $G_E$ through the ratio of polarizations makes it the superior technique to measure this form factor at high momentum transfers.

\begin{figure}
  \begin{center}
    \setlength{\unitlength}{1.0\textwidth}
    \begin{picture}(1.0,.75)
      \put(0,0){\includegraphics[angle=90,width=1.0\textwidth]{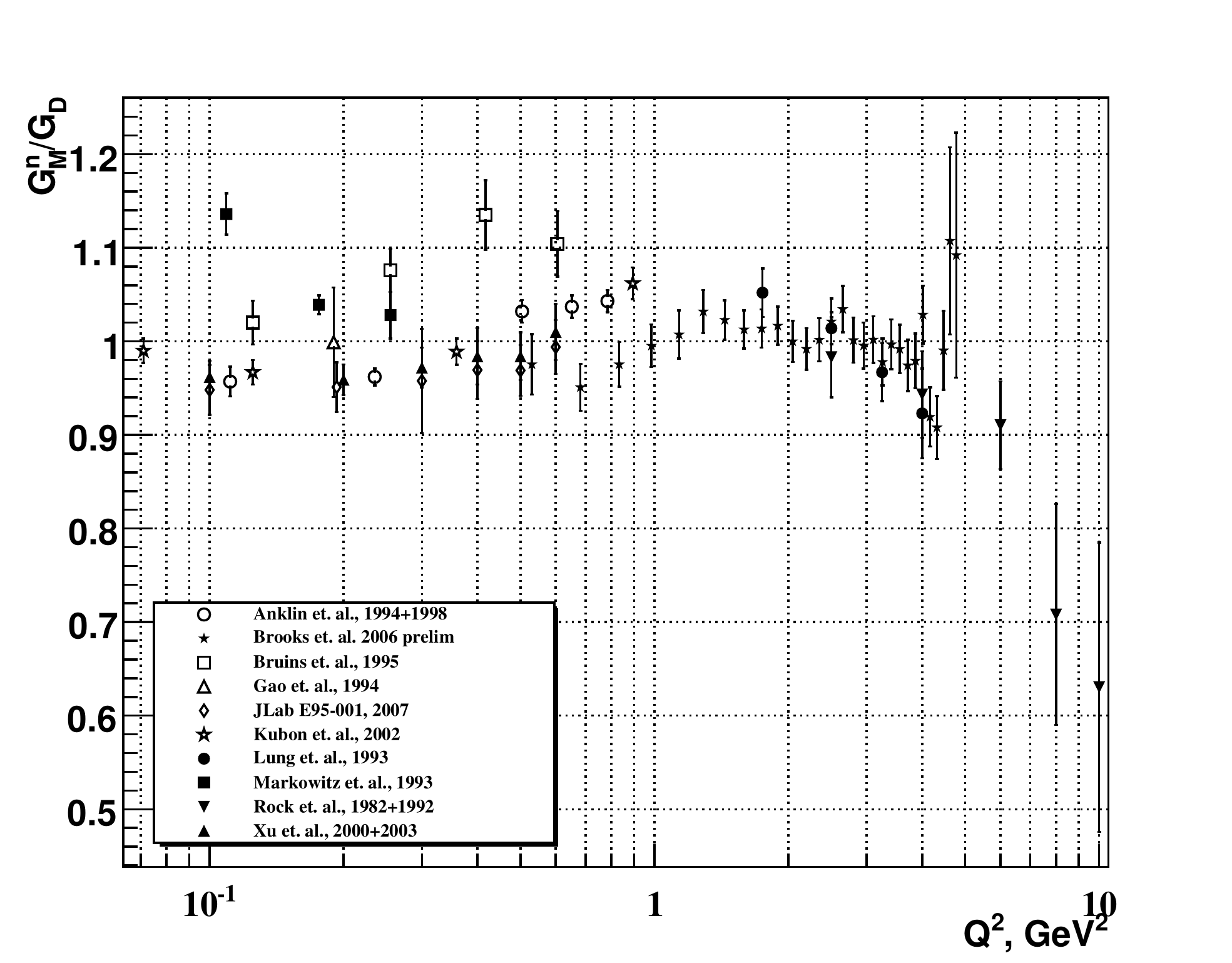}}
      \put(.38,.2811){\tiny \cite{Anklin94,Anklin98}} 
      \put(.38,.2615){\tiny \cite{Brooks06}}
      \put(.38,.2419){\tiny \cite{Bruins95}}
      \put(.38,.2224){\tiny \cite{Gao94}}
      \put(.38,.2028){\tiny \cite{JLab07}}
      \put(.38,.1833){\tiny \cite{Kubon02}}
      \put(.38,.1637){\tiny \cite{Lung93}}
      \put(.38,.1441){\tiny \cite{Markowitz93}}
      \put(.38,.1246){\tiny \cite{Rock82,Rock92}}
      \put(.38,.1050){\tiny \cite{Xu00,Xu03}}
    \end{picture}
    \caption{\label{GMnAllData} Recent neutron magnetic form factor data from cross section and polarization experiments, normalized to the dipole form factor $G_D = (1 + Q^2/\Lambda^2)^{-2}$, where $\Lambda^2 = .71\ GeV^2$.}
  \end{center}
\end{figure}
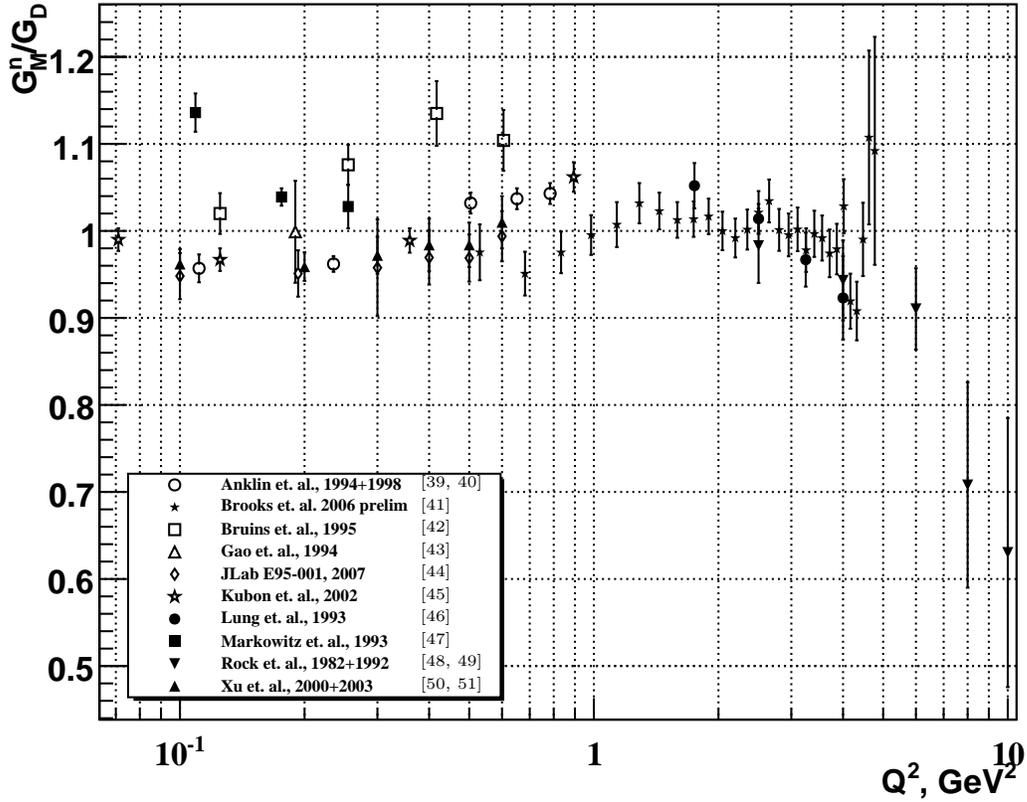

The neutron form factors are not as well known as the proton form factors. They have been measured in both cross section and polarization experiments on deuterium and $^3He$. Extraction of the free neutron elastic form factors from electron scattering experiments on these nuclei requires theoretical models to correct the results for the fact that the neutron in the initial state is bound in a nucleus. Figure \ref{GMnAllData} shows recent results for the neutron magnetic form factor $G_M^n$ from both cross section and polarization experiments, normalized to $G_D$. 
The experiments \cite{Anklin94,Anklin98,Brooks06,Bruins95,Kubon02,Rock82,Rock92} measured simultaneously the cross sections for the reactions $^2H(e,e'n)$ and $^2H(e,e'p)$ in quasi-elastic kinematics. The ratio of the neutron and proton cross sections $\sigma_n/\sigma_p$ on deuterium, combined with knowledge of the free elastic $ep$ scattering cross section, allows one to extract the neutron magnetic form factor in a way that minimizes the dependence of the extraction on the specific deuteron model used. The experiments \cite{Lung93,Markowitz93} both measured the quasi-elastic $^2H(e,e'n)$ scattering cross section but did not measure the ratio to $^2H(e,e'p)$. The experiments \cite{Gao94,JLab07,Xu00,Xu03} used the double-polarization beam-target asymmetry technique with the (inclusive) reaction $^3\vec{He}(\vec{e},e')$ to extract $G_M^n$. Extractions of the neutron magnetic form factor from cross section measurements generally assume $G_E^n = 0$ or use values of $G_E^n$ measured in other experiments. Full Rosenbluth separations of the neutron electric and magnetic form factors are very difficult because the overall neutrality of the neutron fixes $G_E^n(0) = 0$, and $G_M^n$ dominates the cross section at high $Q^2$, as in the proton case. In addition to the ``small''-ness of $G_E^n$, the nuclear model dependence of the extraction further complicates Rosenbluth separation experiments on the neutron.
\begin{figure}
  \begin{center}
    \setlength{\unitlength}{1.0\textwidth}
    \begin{picture}(1.0,.75)
      \put(0,0){\includegraphics[angle=90,width=1.0\textwidth]{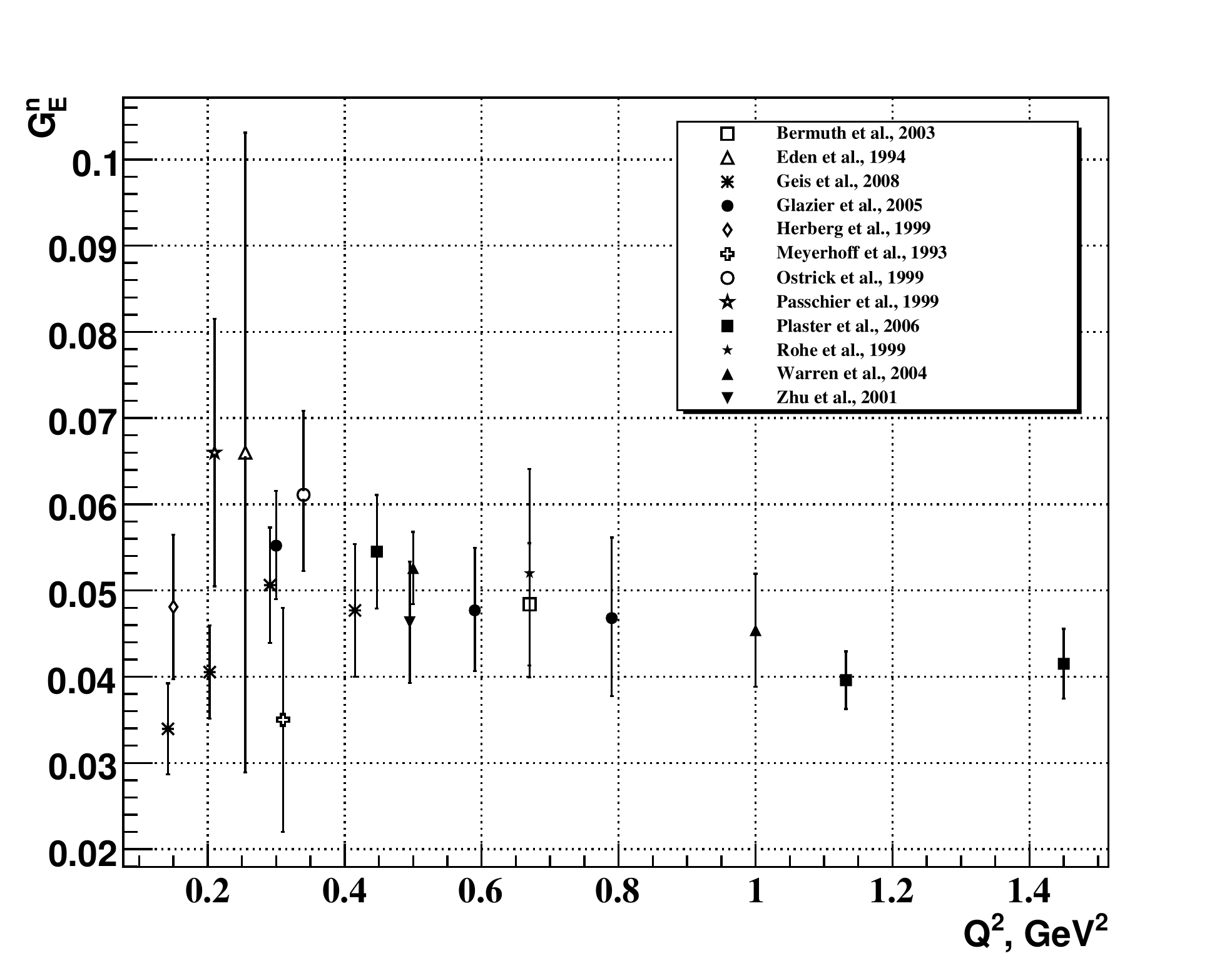}}
      \put(.8,.6722){\tiny \cite{Bermuth03}}
      \put(.8,.6526){\tiny \cite{Eden94}}
      \put(.8,.6331){\tiny \cite{Geis08}}
      \put(.8,.6135){\tiny \cite{Glazier05}}
      \put(.8,.5939){\tiny \cite{Herberg99}}
      \put(.8,.5744){\tiny \cite{Meyerhoff93}}
      \put(.8,.5548){\tiny \cite{Ostrick99}}
      \put(.8,.5353){\tiny \cite{Passchier99}}
      \put(.8,.5157){\tiny \cite{Plaster06}}
      \put(.8,.4961){\tiny \cite{Rohe99}}
      \put(.8,.4766){\tiny \cite{Warren04}}
      \put(.8,.457){\tiny \cite{Zhu01}}
    \end{picture}
    \caption{\label{GEnAllData} Neutron electric form factor data from double-polarization experiments on $^2H$ and $^3He$.}
  \end{center}
\end{figure}
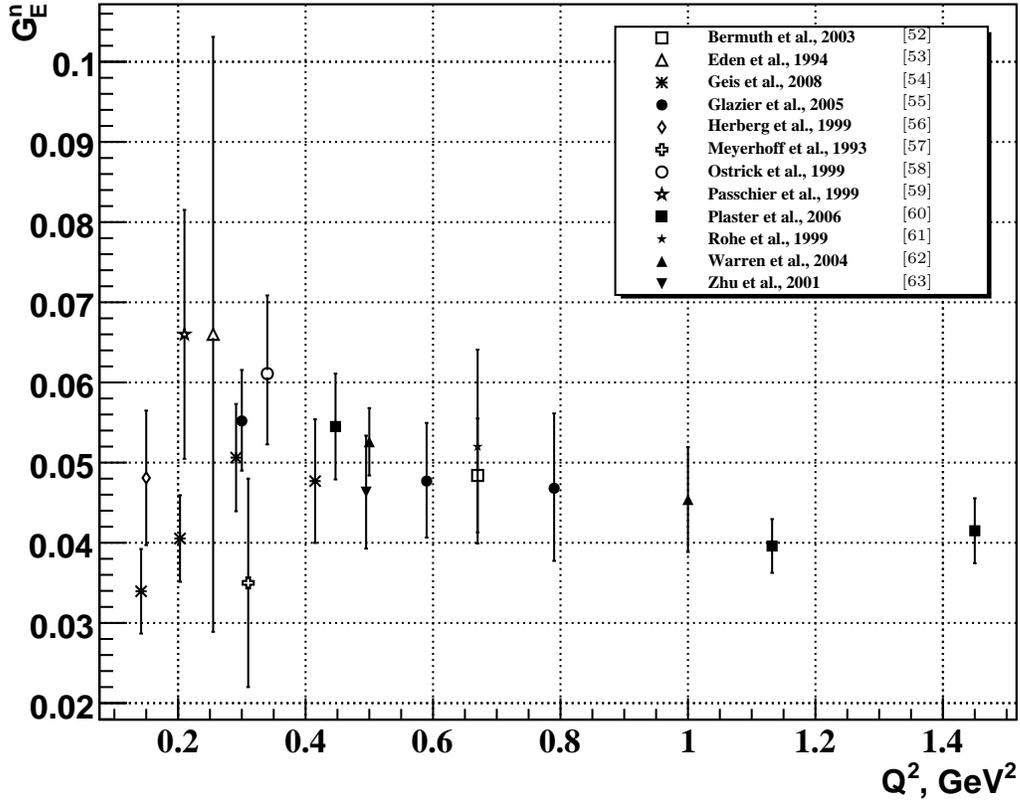
The neutron electric form factor is the most poorly known of the four nucleon form factors. Figure \ref{GEnAllData} shows recent results on $G_E^n$ obtained from double-polarization experiments on $^2H$ and $^3He$. The data shown include beam-target asymmetry measurements in the exclusive reactions $^3\vec{He}(\vec{e},e'n)$ \cite{Bermuth03,Meyerhoff93,Rohe99} and $^2\vec{H}(\vec{e},e'n)^1H$ \cite{Geis08,Passchier99,Warren04,Zhu01} and polarization transfer measurements using the reaction $^2H(\vec{e},e'\vec{n})$ \cite{Eden94,Glazier05,Herberg99,Ostrick99,Plaster06}. A full discussion of the technical and theoretical challenges involved in measuring the neutron form factors is somewhat beyond the scope of this thesis, which is concerned with the proton form factors. The interested reader can consult the review \cite{PerdrisatPunjabiVanderhaegen2007} and the references in figures \ref{GMnAllData} and \ref{GEnAllData}. 

To summarize, the preceding chapter motivated the use of elastic electron-nucleon scattering to study the structure of the nucleon. The electromagnetic interaction between the electron and the nucleon is governed by QED, and first order perturbation theory in $\alpha$ is a very good approximation to the true physical process. Working in the Born approximation, results were derived for both the scattering cross section and polarization transfer observables in terms of electric ($G_E$) and magnetic ($G_M$) form factors which fully characterize the effect of the nucleon's electromagnetic structure on the reaction. The formula for the beam-target asymmetry in polarized target experiments was also presented. Existing proton and neutron form factor data from cross section and polarization experiments were presented and discussed. 

\chapter{Physics of Nucleon Form Factors}
\paragraph{}
In this chapter, the present theoretical understanding of nucleon form factors is addressed, emphasizing the insight that can be gained by measurements at high momentum transfer. As long as calculations of the structure and dynamics of the nucleon from first principles in QCD remain elusive, approximations and phenomenological models of the nucleon continue to provide important insight into their behavior.
\section{Charge and Magnetization Distributions}
\paragraph{}
As shown in appendix \ref{BreitCurrentAppendix}, the nucleon current in the Breit frame has a simple form in terms of the Sachs form factors. The timelike component of the current operator is proportional to $G_E(Q^2)$, while the spacelike three-vector component is proportional to $G_M(Q^2)$. It is tempting to regard $G_E$ and $G_M$ as the Fourier transforms of the charge and magnetization densities of the nucleon in the Breit frame. Though this interpretation is technically correct, it is not terribly meaningful, since the Breit frame corresponds to a different Lorentz boost from the nucleon rest frame for each value of $Q^2$. The traditional density interpretation of the form factors is in fact only valid in the strictly non-relativistic limit $Q^2 \ll M_N^2$, in which the recoil of the nucleon is negligible and the process can be viewed as the scattering of electrons from a static charge distribution as in chapter 1. In this limit, the Breit frame and the lab frame (approximately) coincide. 

Kelly \cite{KellyChargeDist} derived a prescription for relating the Sachs form factors to the rest frame charge and magnetization densities taking relativity into account by defining intrinsic form factors $\tilde{\rho}(k)$ as Fourier-Bessel transforms of the rest frame charge and magnetization densities, and relating the non-relativistic wavenumber $k^2$ of the intrinsic form factor to the Lorentz-invariant four-momentum transfer $Q^2$ through a boost from the Breit frame to the rest frame. The boost factor $\gamma = \sqrt{1+\tau}$ was derived in the discussion of the kinematics of the Breit frame in section \ref{Breitframesection}. The wavenumber is given by $k^2 = \frac{Q^2}{1+\tau}$. This aspect of the prescription is trivial. However, the appropriate relationship between the intrinsic form factor $\tilde{\rho}(k)$ and the Sachs form factor $G(Q^2)$ is inherently model-dependent, since the Lorentz boost for a composite object such as the proton depends on the interactions among the constituent quarks. 

In a number of different models, the relationship can be written as $\tilde{\rho}(k) = G(Q^2)(1+\tau)^\lambda$, where the model dependence is contained in the exponent $\lambda$. Following Mitra and Kumari \cite{MitraKumari}, Kelly chose the prescription $\lambda = 2$ for both the electric and magnetic form factors, which automatically leads to the asymptotic form factor scaling behavior predicted by perturbative QCD at large $Q^2$, and proceeded to fit the proton and neutron electric and magnetic form factor data to obtain the coefficients of an expansion of the rest frame densities in terms of a complete set of radial basis functions, which minimizes the model dependence of the fitted densities. 
\begin{figure}[h]
  \begin{center}
    \includegraphics[width=.5\textwidth]{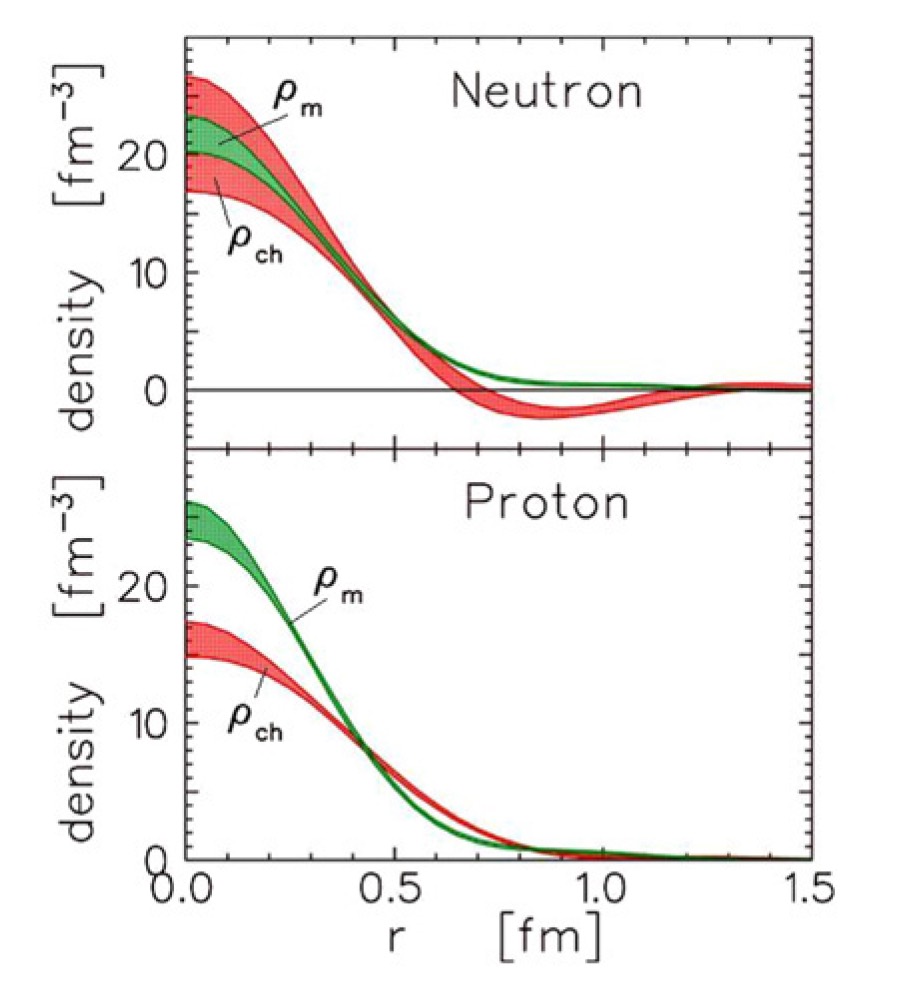}
  \end{center}
  \caption{\label{Kellydensity} Radial proton and neutron charge and magnetization densities obtained from proton and neutron form factor data by Kelly. The neutron charge density has been magnified by a factor of 6 to illustrate its similarity to the magnetization density. The uncertainty bands include statistical uncertainties in the form factor data and incompleteness errors.}
\end{figure}
Figure \ref{Kellydensity} shows the proton and neutron charge and magnetization densities obtained by \cite{KellyChargeDist}, with uncertainty bands indicating the statistical errors in the form factor data and the incompleteness errors due to the finite $Q^2$ range of available data\footnote{Reproduced with permission from \cite{PerdrisatPunjabiVanderhaegen2007}.}. The neutron charge distribution features a positive core surrounded by a negative exterior, which is thought to be a signature of the pion cloud of the nucleon, in which a neutron is pictured as a proton core dressed by a $\pi^-$ cloud. The most notable feature of the proton densities is the broader shape of the charge density relative to the magnetization density, reflecting the precise recoil polarization data which established that $G_E^p$ falls faster than $G_M^p$ at large $Q^2$.

\section{Dispersion Relations and Vector Meson Dominance}
\paragraph{}
Based on the principles of unitarity and analyticity of the form factors considered as functions of $q^2$ in the complex plane, dispersion relations linking the spacelike momentum transfers $q^2 \le 0$ accessible in electron scattering to the timelike region ($q^2 > 0$) accessible in the crossed-channel reaction $e^+e^- \rightarrow N\bar{N}$ can be derived using Cauchy's integral formula:
\begin{eqnarray}
  F(q^2) &=& \frac{1}{\pi} \int_{t_0 = 4m_\pi^2}^{\infty} \frac{ \mbox{Im} F(t) }{(t - q^2)}dt
\end{eqnarray}
Along the positive real axis, the form factor acquires an imaginary part corresponding to the mass spectrum of intermediate virtual hadronic states through which the timelike virtual photon couples to the $N\bar{N}$ final state, the lightest of which is two pions, as indicated by the cutoff of the dispersion integral at $t_0 = 4m_\pi^2$. In dispersion theory analyses of the nucleon form factors, it is customary to work with the isoscalar and isovector Dirac and Pauli form factors instead of the usual proton and neutron Sachs form factors. As their names suggest, they are defined as linear combinations of the proton and neutron form factors of definite isospin: 
\begin{eqnarray}
  F_{1,2}^{iv}(q^2) &\equiv& F_{1,2}^p(q^2) - F_{1,2}^n(q^2) \\
  F_{1,2}^{is}(q^2) &\equiv& F_{1,2}^p(q^2) + F_{1,2}^n(q^2) \\
  F_{1,2}^p &=& \frac{1}{2}\left(F_{1,2}^{is} + F_{1,2}^{iv}\right) \\
  F_{1,2}^n &=& \frac{1}{2}\left(F_{1,2}^{is} - F_{1,2}^{iv}\right)
\end{eqnarray}
These linear combinations of the proton and neutron form factors are the most convenient for dispersion analysis since the virtual hadronic states which contribute to the dispersion integrals can be classified according to their isospin properties. For example, the cutoff at $t_0 = 4m_\pi^2$ applies to the isovector form factors only. For the isoscalar form factors, the lightest hadronic state that can contribute is three pions, which modifies the cutoff to $t_0 = 9 m_\pi^2$. 

Of particular importance for low to intermediate values of $Q^2$ are the three lightest vector mesons $\rho(770)$, $\omega(782)$, and $\phi(1020)$, which carry the same spin (1) and parity (negative) as the photon. The basic quark model flavor wavefunctions of these mesons are given by\cite{PDG2008}
\begin{eqnarray}
  \rho^0 &=& \sqrt{\frac{1}{2}}\left(u\bar{u} - d\bar{d}\right), I^G(J^{PC})=1^+(1^{--}) \\
  \omega &=& \sqrt{\frac{1}{2}}\left(u\bar{u} + d\bar{d}\right), I^G(J^{PC})=0^-(1^{--}) \\
  \phi &=& s \bar{s}, I^G(J^{PC})=0^-(1^{--})
\end{eqnarray}
The coupling of the electromagnetic current operator to these vector meson states is determined by their partial decay width to $e^+ e^-$, which is known experimentally. Based on the quantum numbers of these mesons, the $\rho$ meson should contribute significantly to the isovector form factor at low $Q^2$, whereas the $\omega$ should contribute to the isoscalar form factor. The contribution of the $\phi$ is largely suppressed by the Zweig/OZI rule. Neglecting finite-width effects, the contribution of the lowest-lying vector meson poles to the dispersion integrals takes the form 
\begin{eqnarray}
  F_V(q^2) &=& G_{VN} \frac{m_V^2}{m_V^2 - q^2}F_{VN}(q^2) = G_{VN} \frac{m_V^2}{m_V^2 + Q^2}F_{VN}(q^2)
\end{eqnarray}
where $G_{VN}$ is a meson-nucleon coupling constant to be determined experimentally, and $F_{VN}$ is an intrinsic meson-nucleon form factor. In the basic Vector Meson Dominance (VMD) model, the isovector form factor is determined by $G_{\rho N}$ and the isoscalar form factor is determined by $G_{\omega N}$. The meson-nucleon form factor $F_{VN}$ is universal; i.e., it assumes the same form for the $\rho$ and the $\omega$, and is commonly assumed in the literature to be of monopole form:
\begin{eqnarray}
  F_V(q^2) &=& \frac{1}{1-\frac{q^2}{\Lambda^2}} = \frac{1}{1+\frac{Q^2}{\Lambda^2}}
\end{eqnarray}

In 1973, a VMD-based model by Iachello et al. \cite{Iachello1973} was among the earliest to predict a decrease of the proton $G_E/G_M$ ratio for $Q^2 \ge 1\ GeV^2$, which was in rough agreement with the not-yet-available recoil polarization data from Jefferson Lab{\cite{Jones00,Punjabi05,Gayou02}. In this model, as few as three adjustable parameters were used to fit the form factor data available at the time. In addition to the basic structure outlined above, the authors considered the effect of the finite width of the $\rho$ meson, allowed for a direct photon-nucleon Dirac coupling in $F_1$ in addition to the vector meson pole terms, and considered several alternative functional forms of the intrinsic form factor. 

In 1985, Gari and Kr\"{u}mpelmann\cite{GK1985} presented a model in which a smooth transition from the VMD picture expected to hold at low $Q^2$ and the asymptotic form factor behavior expected at high $Q^2$ in perturbative QCD (see section \ref{pQCDsection}) was built in to the parametrization of the intrinsic form factor, characterized by two different scale parameters $\Lambda_1$ and $\Lambda_2$ governing the transition from meson-baryon dynamics to quark-gluon (pQCD) dynamics in the form factor behavior. This model was further extended in the early 1990s \cite{GK1992,GK1992E} to include the contribution of the $\phi$ meson using an independent parametrization designed to conform to the constraints imposed on the $\phi NN$ coupling imposed by the OZI rule, and the $\phi$ meson contribution was found to have a significant effect on the neutron electric form factor in particular. 

In 2001, Lomon\cite{Lomon2001} extended the model of \cite{GK1985,GK1992,GK1992E} to include the effect of the $\rho$ meson width by replacing the $\rho$ meson pole term with the $\rho$ contribution obtained in the dispersion relation analysis of \cite{MergellMeissnerDrechsel}, and by adding new $\rho'(1450)$ and $\omega'(1420)$ pole terms, obtaining reasonable fits to all four nucleon form factors. Lomon's fit was soon updated\cite{Lomon2002} to incorporate the high-$Q^2$ $G_E^p/G_M^p$ recoil polarization data from Hall A at Jefferson Lab\cite{Gayou02}. In light of the qualitative agreement of the recoil polarization data for $G_E^p/G_M^p$ with \cite{Iachello1973}, Iachello and Bijker\cite{Iachello2004} published an extended version of the 1973 VMD model modified to include a direct coupling in the isovector Dirac form factor as a model for an intrinsic three-quark structure, with the goal of bringing the model into agreement with the data for the neutron electric form factor $G_E^n$ while preserving the successfully predicted decrease of $G_E^p/G_M^p$ with $Q^2$.

VMD models are a special case of more general dispersion-theoretical analysis of the nucleon form factors, in which the spectral function $\frac{\mbox{Im} F(t)}{\pi}$ is constructed from experimental data on the coupling of the nucleon and the electromagnetic current to all possible intermediate hadronic states contributing to the isoscalar and isovector dispersion integrals. H\"{o}hler et al.\cite{Hohler} performed such an analysis using pion-nucleon scattering data and pion form factor data to derive the two-pion continuum contribution to the isovector spectral function, which is found to be an important contribution in addition to the $\rho$ resonance. The analysis was updated in the 1990s\cite{MergellMeissnerDrechsel} to include timelike proton form factor data and again in 2004 to include new precise $G_E^n$ data\cite{HammerMeissner2004}. A more recent dispersion relation analysis of the nucleon form factors by Belushkin et al. \cite{Belushkin2007} added the $K\bar{K}$ and $\rho \pi$ continuum contributions to the isoscalar spectral function along with the $2\pi$ continuum based on the latest available pion form factor data. 

In summary, dispersion-relation analyses and the closely related VMD models of the nucleon form factors provide important insight into the nucleon structure. Most of these models involve a number of adjustable parameters to be fitted to experimental data, and most manage to describe all of the available spacelike and timelike form factor data with reasonable accuracy. As additional constraints on the form factors and the spectral functions become available from experimental data, the number and range of free parameters should decrease. If data of infinite precision were available on all possible reaction channels over an infinite range of momentum transfers, then dispersion relation analysis would provide a crucial check on the internal consistency of the data sets. Since the experimental data are finite in terms of both $Q^2$ coverage and the reaction channels that are measured with reasonable accuracy, dispersion analyses of increasing sophistication can be used to make increasingly strong predictions of observables in the $Q^2$ range where data do not yet exist. Increasing the $Q^2$ coverage and precision of the nucleon form factor data constrains the spectral functions. Similarly, improving the available data base for the relevant hadronic reaction channels contributing to the spectral functions constrains the nucleon form factors in a more-or-less model-independent fashion. Although strict Vector Meson Dominance of the form factor behavior is only expected to hold for relatively low momentum transfers, the fit results for models which incorporate the transition to perturbative QCD shed important light on the evolution with momentum transfer of the relative importance of ``soft'', non-perturbative meson-baryon dynamics and ``hard'', perturbative quark-gluon dynamics.

\section{Constituent Quark Models}
\paragraph{}
In chapter 1, the success of the non-relativistic constituent quark model in baryon spectroscopy was described in terms of the spin-flavor structure of three-quark states from which the spin-1/2 octet and spin-3/2 decuplet baryon states are constructed. To move beyond static properties to predictions of dynamical properties such as form factors, a prediction for the wave function of the quarks in the nucleon is required. The class of nucleon models collectively referred to as constituent quark models involves treating baryons as bound states of three quarks moving in a confining potential, with the nucleon emerging as the ground state of this system. 

In non-relativistic constituent quark models(CQM), the quarks are treated as massive, quasi-particle effective degrees of freedom. A famous example is the Isgur-Karl model\cite{IsgurKarl}, in which the quarks are confined by a long-range harmonic oscillator potential, supplemented by a short-range one-gluon-exchange quark-quark interaction which leads to the color hyperfine interaction which successfully accounts for the mass splittings between the octet and decuplet baryons. This model also predicts a small $D$-state probability for the nucleon ground state, implying a slightly non-spherical charge density and a non-zero electric quadrupole moment of the nucleon charge distribution, which can be accessed indirectly by measuring the $N\rightarrow \Delta$ electric quadrupole ($E2$) and Coulomb quadrupole ($C2$) amplitudes. 

Since the elementary quarks of QCD are light compared to both the confinement scale $\Lambda_{QCD}$ and the nucleon mass, taking relativity into account is important for calculating dynamic quantities such as form factors even at low momentum transfers. The inclusion of relativity in the Hamiltonian formalism of quantum mechanics was explored by Dirac\cite{DiracRelativisticDynamics}, who presented three forms of relativistic dynamics which differ according to the subset of the dynamical variables which are ``kinematical''; i.e., interaction-independent, and those which are dynamical; i.e., quantities that depend on the interactions among the constituents. In general, there are ten dynamical variables (generators of the Poincar\'{e} group) corresponding to four space-time translations, three spatial rotations, and three boosts. 
\begin{description}
\item[Instant form] In the instant form, the dynamical generators are the energy and the three boost operators. The instant form has the advantage that rotations are kinematical, so that constructing states of definite angular momentum is straightforward.
\item[Point form] In point-form dynamics, boosts and rotations are kinematical, and all four components of the four-momentum are dynamical. 
\item[Light-front form] In light-front dynamics, seven of the dynamical variables are kinematical, which is the largest possible number. The dynamical variables are one component of the four momentum operator and two transverse rotations. 
\end{description}
Light-front dynamics are advantageous for the calculation of nucleon form factors because the boost operator for the quark wavefunctions is independent of the details of the interactions among the quarks, effectively separating the center-of-mass motion of the nucleon from the internal motion of its constituents. Its main drawback is that the construction of states with definite total angular momentum depends on the interactions among the quarks.

All relativistic constituent quark model calculations of the nucleon form factors involve the calculation of matrix elements of the electromagnetic current operator between nucleon states defined by an ansatz for the quark wavefunction. Schlumpf\cite{SchlumpfCQM,SchlumpfFF} adopted a wavefunction with a power law dependence in terms of the quark internal momentum variables. This model has only two free parameters, the constituent quark mass $m=263$ MeV and the confinement scale $\alpha=607$ MeV, determined by fitting the magnetic moments and semileptonic decays of the baryon octet, and gives a reasonable description of the existing nucleon form factor data. This wavefunction is of the form 
\begin{eqnarray}
  \phi(M) &=& \frac{N}{(M^2+\alpha^2)^{3.5}} \\
  M &\equiv& \sum_i \sqrt{q_i^2+m^2}
\end{eqnarray}
where $M$ is a function of the internal quark momenta and masses and $N$ is a normalization constant. 

Frank, Jennings, and Miller\cite{FrankJenningsMiller,MillerFrank} demonstrated that the light-front wave function in Schlumpf's constituent quark model leads to a violation of hadron helicity conservation at high energies, leading to a slower falloff of $F_2^p/F_1^p$ than $1/Q^2$, as observed in the recoil polarization data for $G_E/G_M$. In particular, the ratio $QF_2/F_1$ is expected to be roughly constant for $Q^2$ between 2 and 20 GeV$^2$, a result which is borne out by the recoil polarization data between 2.0 and 5.6 GeV$^2$. In all of these models, the importance of relativity is paramount. When working in light-front dynamics in particular, boosting the spin state of the nucleon wave function from the rest frame, in which the spin-flavor state is constructed in the same way as in the non-relativistic quark model, to the light-front (i.e.; the infinite momentum frame) causes the light-cone spinors to undergo Melosh rotations\cite{MeloshRotation} which mix different spin states, introducing non-trivial spin dependence in the light-cone wavefunction, which is responsible for observable effects such as the predicted scaling of $QF_2/F_1$. The Melosh rotation takes the form 
\begin{eqnarray}
  <\lambda'| \mathcal{R}_M(\xi, q_\bot, m, M)|\lambda> &=& \left[\frac{m+\xi M - i\mathbf{\sigma}\cdot (\mathbf{n} \times \mathbf{q})}{\sqrt{(m+\xi M)^2 + q_\bot^2}}\right]_{\lambda' \lambda} \label{MeloshRot}
\end{eqnarray}
where $m$ is the constituent quark mass, $M$ is the mass operator which depends on the internal quark momenta, and $\xi$ is a kinematic factor which also depends on light cone quantities (see, e.g. \cite{SchlumpfCQM} for definitions.). In the model of Frank, Jennings and Miller\cite{FrankJenningsMiller,MillerFrank}, the strong decrease of $G_E^p/G_M^p$ with $Q^2$ emerges as an important consequence of the spin-dependent relativistic effects on the light-front quark wavefunctions embodied by the Melosh rotation \eqref{MeloshRot} which is qualitatively confirmed by the recoil polarization data \cite{Jones00,Gayou02,Punjabi05}. This model tends to predict a somewhat faster falloff of $G_E^p/G_M^p$ than that exhibited by the data, and also predicts a zero crossing of $G_E^p/G_M^p$ near $Q^2 = 6$ GeV$^2$. The results of this experiment provide a severe test of this prediction. 

Despite the relative phenomenological success of nucleon models based entirely on constituent quarks, the shortcomings of these models are well known. In particular, they do not generally satisfy the symmetry properties of the QCD Lagrangian, particularly chiral symmetry. The elementary $u$ and $d$ quarks are nearly massless, and in the limit of exactly massless $u$ and $d$ quarks, the QCD Lagrangian exhibits chiral symmetry; i.e., it is invariant under $SU(2)_L \times SU(2)_R$ rotations of left and right-handed quarks; i.e., quark chirality is conserved in this limit. In nature, the lightest pseudoscalar mesons (pions) appear as the Goldstone bosons of the spontaneously broken chiral symmetry of the QCD Lagrangian. The nonzero masses of the pions observed in nature are acquired through the explicit chiral symmetry breaking of the non-zero $u$ and $d$ quark masses. 

As the lightest hadrons, pions play a particularly important role in the long-distance structure of the nucleon. Miller\cite{MillerLFCBM} added the effects of the pion cloud of the nucleon to the relativistic constituent quark model(rCQM) of \cite{FrankJenningsMiller} by calculating one-loop diagrams involving virtual pions, characterizing the probability that a nucleon fluctuates into a nucleon-virtual pion pair while interacting with the electromagnetic field of a virtual photon, which may couple to either the charged pion or the nucleon, as shown in figure \ref{pionclouddiagrams}. 
\begin{figure}
  \begin{center}
    \includegraphics[width=.7\textwidth]{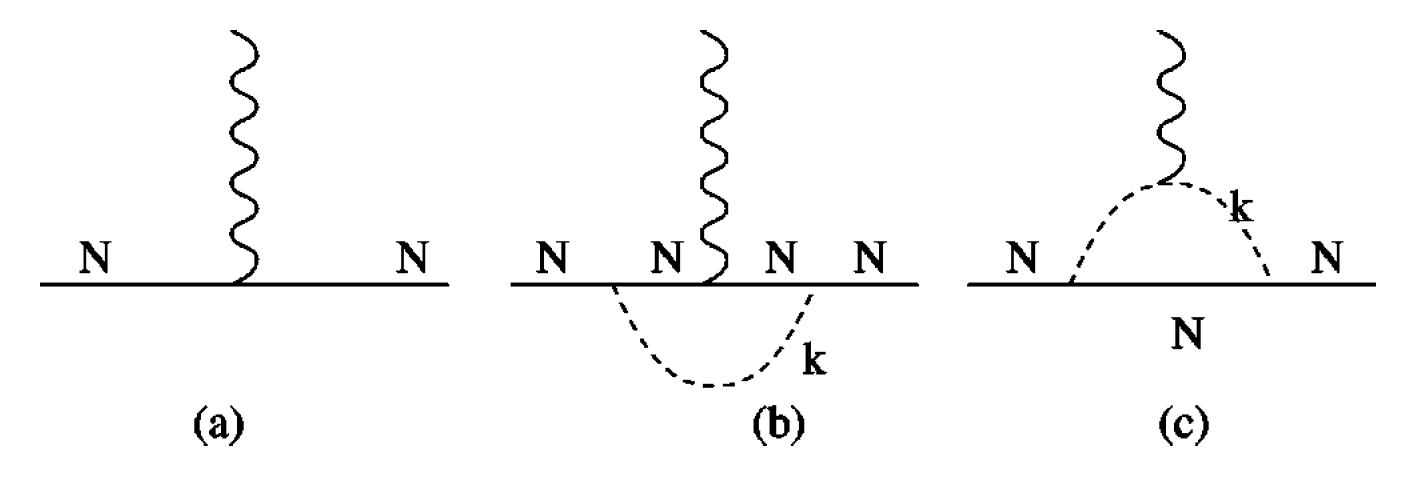}
  \end{center}
  \caption{\label{pionclouddiagrams} One-loop virtual pion diagrams in the Light-Front Cloudy Bag Model.}
\end{figure}
The calculation of these diagrams requires bare virtual photon-nucleon form factors, which are calculated within the rCQM, and relativistic pion-nucleon form factors with an assumed form used by \cite{Zoller1992,Holtmann1996}, and is called the Light-Front Cloudy Bag Model. The pion cloud effects within this model are found to yield important contributions at low momentum transfer, particularly for the neutron electric form factor, which is not well reproduced by relativistic constituent quarks alone. On the other hand, quarks are found to dominate at large momentum transfers.
\section{Form Factors and Perturbative QCD}
\label{pQCDsection}
\paragraph{}
The discussion of perturbative QCD (pQCD) and the nucleon form factors begins with the property of asymptotic freedom in QCD, the theory of the strong interactions. The running coupling constant of QCD is given to leading logarithmic order in the renormalization scale $\mu$ by (see \cite{StructNucleon}, page 82)
\begin{eqnarray}
  \alpha_s(\mu) &=& \frac{4\pi}{\beta_0 \ln\left(\frac{\mu^2}{\Lambda^2}\right)} \label{QCDrunningcoupling} \\
  \beta_0 &=& 11 - \frac{2}{3} n_f
\end{eqnarray}
where $n_f$ is the number of quark flavors that can appear in the $q\bar{q}$ ``vacuum polarization'' loop correction to the gluon propagator. In the Standard Model and certainly at any energy scale currently accessible to experiment, the number of flavors does not exceed six, and the coefficient multiplying $\ln\left(\mu^2/\Lambda^2\right)$ in the denominator is positive, implying that $\alpha_s(\mu^2) \xrightarrow[\mu^2 \rightarrow \infty]{} 0$. The QCD scale $\Lambda$ is the only free parameter of the theory, and determines the energy scale at which the theory becomes strongly coupled and a perturbative expansion in powers of $\alpha_s$ is no longer meaningful. Based on information about $\alpha_s$ from a variety of experiments including deep inelastic scattering, the widths of the $Z$ boson and the $\tau$ lepton, and $e^+e^-$ annihilation data at various energies, the Particle Data Group\cite{PDG2008} quotes a value of $\alpha_s(\mu=M_Z) = 0.1176 \pm 0.002$ for the QCD coupling constant at the Z pole. The results in this thesis are measurements at $Q$ of $2-3\ GeV$, where $\alpha_s \approx 0.3$ based on the running of $\alpha_s$ from high energies where it is measured with reasonable precision. Even if the behavior of the nucleon as measured by form factors is assumed to be calculable entirely within a pQCD framework, the expansion is not likely to converge very rapidly for such a value of $\alpha_s$. In order to calculate the form factors with reasonable precision would require the calculation to several orders beyond the leading order in $\alpha_s$ of all possible Feynman graphs contributing to elastic $eN$ scattering, a non-trivial calculation to say the least for this hard, exclusive process.

Despite these obvious difficulties, this experiment takes place at sufficiently high momentum transfer that quark-gluon dynamics is likely to play an important, if not dominant role in the form factor behavior. Perturbative QCD makes a prediction for the behavior of the nucleon form factors for asymptotically large $Q^2$. 
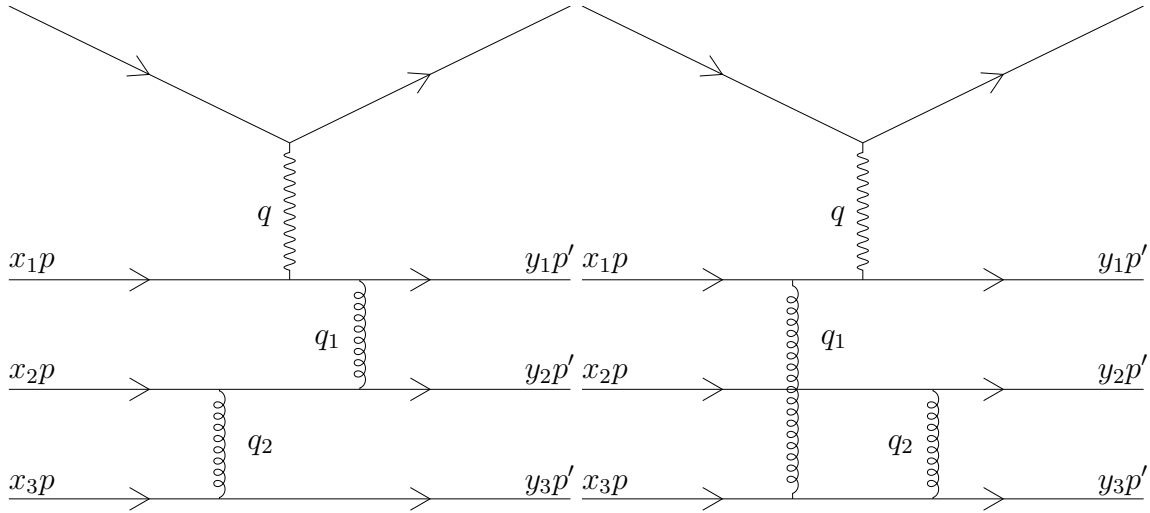
\begin{figure}[h]
  \begin{center}
    \setlength{\unitlength}{1.0in}
    \begin{picture}(6.0,3.0)
      \put(0,0){\includegraphics[angle=90,width=.49\textwidth]{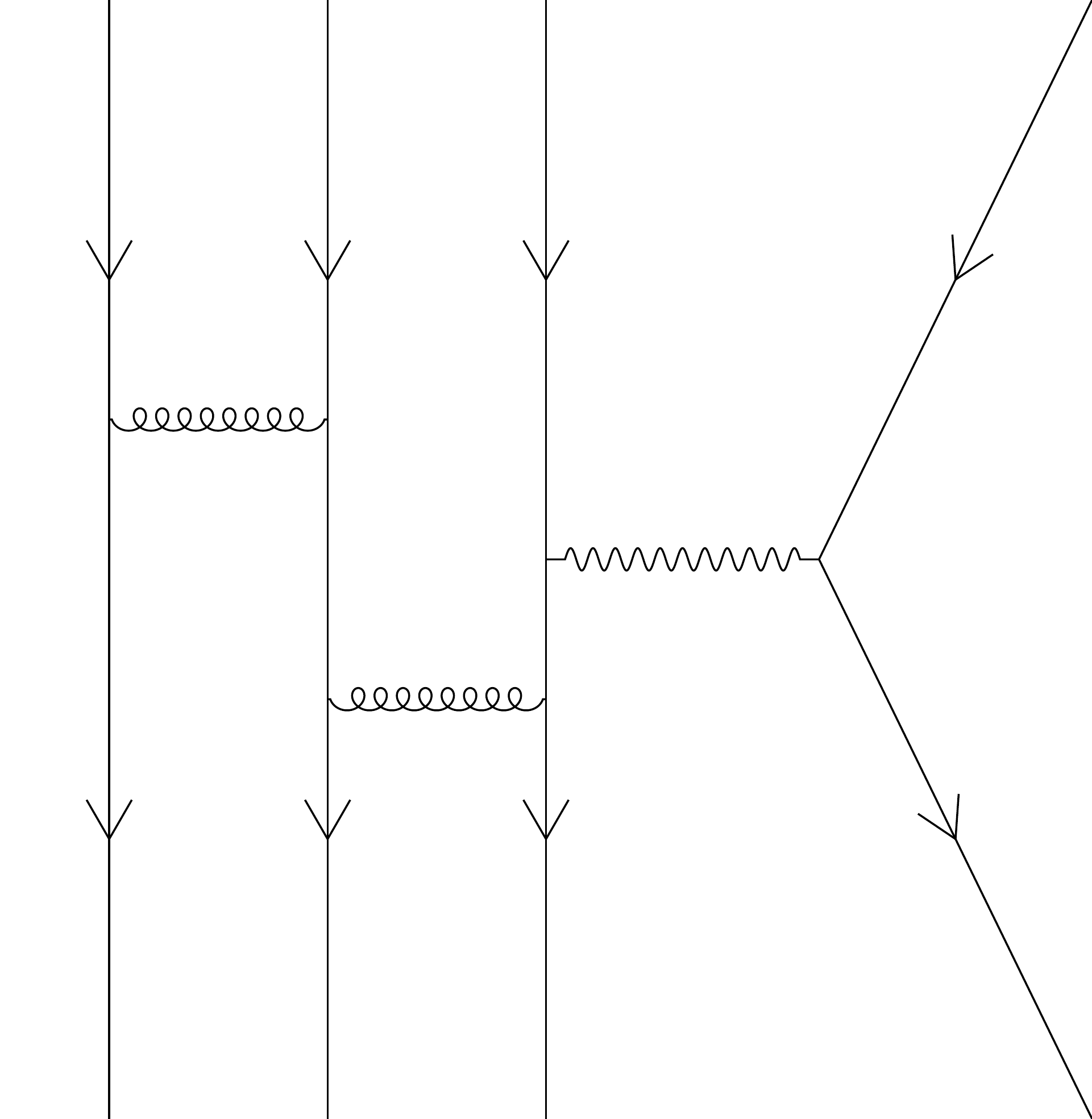}}
      \put(3.0,0){\includegraphics[angle=90,width=.49\textwidth]{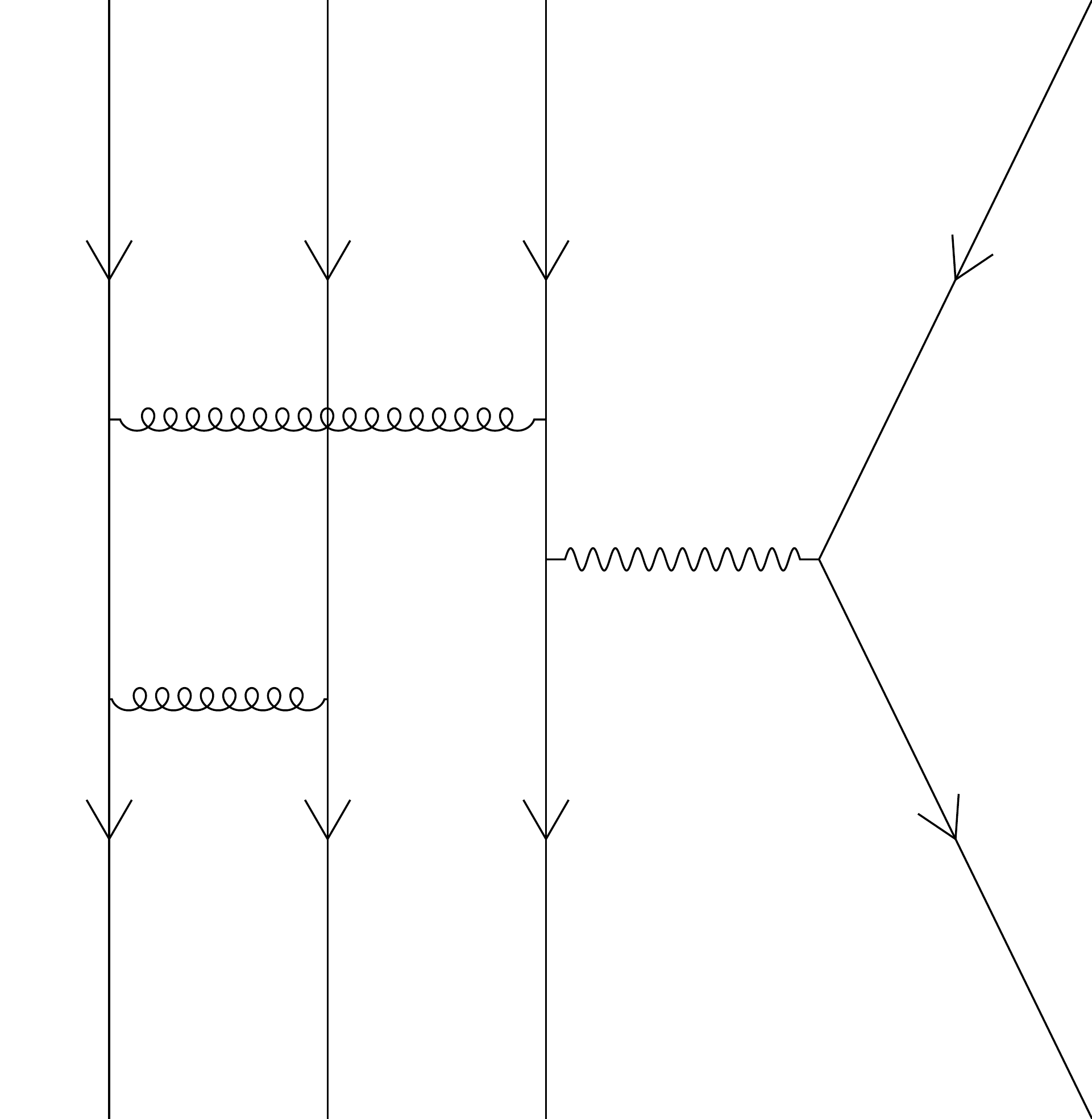}}
      \put(0.0,0.34){$x_3 p$}
      \put(0.0,0.92){$x_2 p$}
      \put(0.0,1.5){$x_1 p$}
      \put(3.0,0.34){$x_3 p$}
      \put(3.0,0.92){$x_2 p$}
      \put(3.0,1.5){$x_1 p$}
      \put(2.7,0.34){$y_3 p'$}
      \put(2.7,0.92){$y_2 p'$}
      \put(2.7,1.5){$y_1 p'$}
      \put(5.7,0.34){$y_3 p'$}
      \put(5.7,0.92){$y_2 p'$}
      \put(5.7,1.5){$y_1 p'$}
      \put(1.3,1.75){$q$}
      \put(1.6,1.10){$q_1$}
      \put(1.25,0.55){$q_2$}
      \put(4.3,1.75){$q$}
      \put(4.25,1.10){$q_1$}
      \put(4.6,0.55){$q_2$}
      \end{picture}
  \end{center}
  \caption{\label{pQCDgraphs} Example feynman diagrams contributing to $eN\rightarrow eN$ at leading order in $\alpha_s$ in pQCD. The total $\mathcal{O} \alpha_s^2$ contribution includes all possible permutations and time-orderings of the two required gluon exchanges.}
\end{figure}
In a reference frame in which the nucleon is moving with infinite momentum, it can be viewed as a system of three massless, weakly-interacting quarks moving collinearly with the nucleon, each carrying a light-front momentum fraction $x_{i,\ (i=1,2,3)}$ where the $x_i$ are required to add up to the total nucleon momentum $x_1+x_2+x_3 = 1$. The transverse momenta of the quarks are assumed small, $\left|\mathbf{k}_{\bot,i}\right| \approx \Lambda \ll Q$ and are neglected. In order for elastic scattering to take place when one of the quarks is struck by a virtual photon with large momentum $Q$, the struck quark must interact with the ``spectator'' quarks such that the three quarks remain collinear after the collision. In perturbative QCD, the quarks interact via single-gluon exchange to leading order in $\alpha_s$. A minimum of two hard\footnote{If one of the exchanged gluons is ``soft'', then the coupling $\alpha_s$ for that gluon exchange increases, perhaps to the point where perturbative QCD no longer applies. However, ``soft'' gluon exchange corresponds to long-distance physics. If either of the ``hard'' ($Q_i^2 \gg \Lambda^2$) gluon exchanges fails to occur, the nucleon is much more likely to break up than to remain in its $qqq$ ground state after the collision. This can be understood in terms of confinement. If one of the quarks is struck by a very hard virtual photon and fails to share the imparted momentum among the two spectator quarks, it will move rapidly away from the original three-quark center of mass. As the attractive color potential between the struck quark and the inert spectator quarks increases with distance, it becomes more energetically favorable for a $q\bar{q}$ pair to pop out of the vacuum, leaving two or more colorless hadrons in the final state. In other words, the struck quark \emph{fragments} into secondary hadrons. Another possibility is the excitation of nucleon resonances such as the $\Delta$. In either case, the reaction is no longer elastic scattering, which is why at very high momentum transfers corresponding to very short distances, elastic scattering cannot occur without a minimum of two \emph{hard} gluon exchanges sharing the transferred momentum among the three quarks. Formally speaking, the arguments leading to the dimensional scaling law rest on certain assumptions about the ultraviolet and infrared behavior of the bound-state nucleon wave function--namely, each of the hadronic constituents must carry a finite fraction of the hadron momentum and no internal mass scale may be present. These assumptions are naturally satisfied in renormalizable field theories such as QCD\cite{BrodskyFarrar1975} for sufficiently large momentum transfers.} gluon exchanges among the three quarks is required for this exclusive process to occur. A subset of the possible two-gluon exchanges is shown in figure \ref{pQCDgraphs}. The virtual photon carries the momentum transfer $Q^2$, while the two gluons carry momentum transfers $q_1^2$ and $q_2^2$, respectively, where $Q^2, Q_{1,2}^2 \gg \Lambda^2$.

It was shown in \cite{BrodskyLepage1979} that the nucleon form factor in this high $Q^2$ limit can be written in the factorized form 
\begin{eqnarray}
  F(Q^2) &=& \int_0^1 dx \int_0^1 dy \Phi^*(y) T(x,y,Q^2) \Phi(x)
\end{eqnarray}
where $dx \equiv dx_1 dx_2 dx_3 \delta(1-x_1-x_2-x_3)$ and $dy \equiv dy_1 dy_2 dy_3 \delta(1-y_1-y_2-y_3)$ are the momentum fractions of the quarks in the initial and final nucleons, respectively, and the sums of the quark momenta are restricted by the $\delta$-functions to add up to the respective nucleon momenta. $T(x,y,Q^2)$ is the transition operator for the ``hard'' scattering process, and $\Phi(x)$, $\Phi^*(y)$ are the light-front quark distribution amplitudes (DAs) integrated over transverse momenta of the quarks in the initial and final nucleons. The perturbative physics is contained in $T$, while the non-perturbative information about the nucleon's ground state wavefunction is contained in the DAs $\Phi$, which are universal. The form of $T$ is obtained by calculating the Feynman diagrams in figure \ref{pQCDgraphs} (and permutations thereof). The leading asymptotic $Q^2$ dependence of the form factor is contained in $T$.

The basic $Q^2$ dependence of $T$ can be guessed by recognizing that each gluon exchange contributes a factor of $\alpha_s(q_i^2)$ for the two $qqg$ vertices, and a factor of $1/q_i^2$ for the gluon propagator:
\begin{eqnarray}
  T(x,y,Q^2) &\propto& \frac{\alpha_s(q_1^2)\alpha_s(q_2^2)}{q_1^2q_2^2} f(x,y) \propto Q^{-4}
\end{eqnarray}
The gluon momenta $q_1^2$ and $q_2^2$ are proportional to $Q^2$. The multiplicative factor $f$, and the proportionality between $q_{1,2}^2$ and $Q^2$, are functions of the quark momentum fractions $x$ and $y$ determined by the arrangement of the gluon and photon lines in the Feynman diagram under consideration. The leading asymptotic $Q^2$ dependence of the form factor contained in $T$ is clearly $F(Q^2) \propto \frac{1}{Q^4}$. This result for the nucleon form factor is a special case of the dimensional scaling laws for large momentum transfer processes derived in \cite{BrodskyFarrar1975}. The leading $Q^{-4}$ dependence applies to the helicity-conserving Dirac form factor. On the other hand, the Pauli form factor $F_2$ characterizes the nucleon spin-flip amplitude. In the limit of massless quarks, which is certainly approximately satisfied in nucleons composed of light $u$ and $d$ quarks at high energies, quark helicity is conserved in interactions mediated by vector fields such as photons and gluons. Therefore, $F_2$ is suppressed by a factor $~m^2/Q^2$ relative to $F_1$, where $m$ is an effective quark mass. This leads to the prediction $F_2(Q^2) \propto 1/Q^6,Q^2 \rightarrow \infty$. The ratio $F_2/F_1$, which is a simple function of the ratio $G_E/G_M$ measured by this experiment, is expected to scale as $Q^2F_2/F_1 \xrightarrow[Q^2\rightarrow \infty]{}\ constant$.
\begin{figure}[h]
  \begin{center}
    \includegraphics[angle=90,width=.98\textwidth]{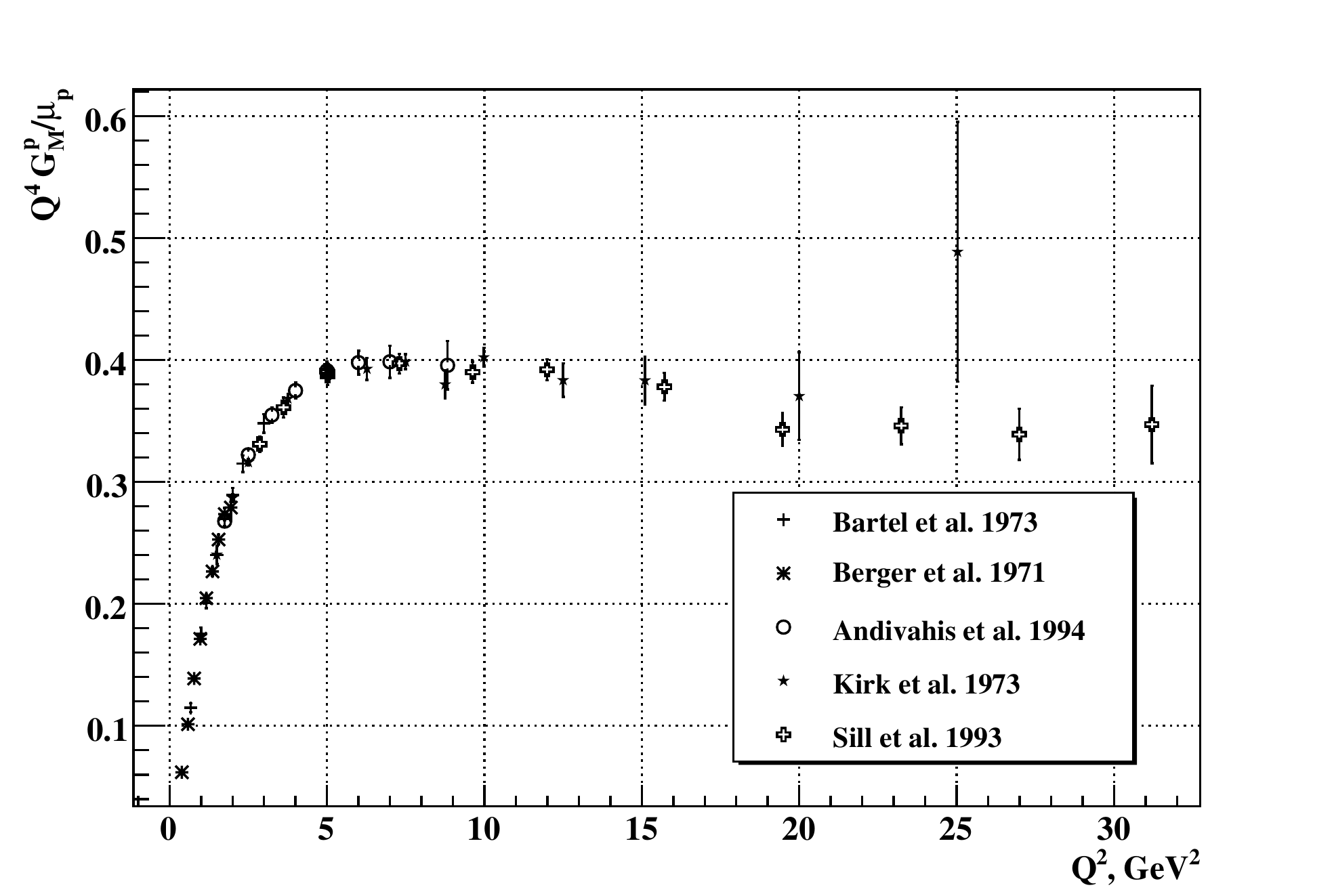}
  \end{center}
  \caption{\label{Q4GMpfig} $Q^4G_M^p/\mu_p$ at high $Q^2$. See text for references.}
\end{figure}
The proton magnetic form factor has been measured to higher $Q^2$ values than any of the other nucleon form factors. Figure \ref{Q4GMpfig} shows the data extending to $Q^2$ above 30 GeV$^2$, plotted as $Q^4G_M^p(Q^2)/\mu_p$. Clearly $G_M^p$ at least approximately satisfies the pQCD scaling expectation starting at $Q^2$ of 5 to 10 GeV$^2$. The data shown in the figure are from \cite{Bartel73,Berger71,Andi94,Kirk1973,Sill93}. 

Since the bare (current) masses of the quarks in the nucleon are negligible in comparison to the mass of the nucleon and to $Q^2$, it has been argued by Belitsky et al.\cite{BelitskyJiYuan2003} that the quark mass contribution to the helicity-flip amplitude is also negligible, and instead the dominant mechanism for nucleon spin flip in QCD is quark orbital angular momentum. In \cite{BelitskyJiYuan2003}, the authors derived the asymptotic $Q^2$ dependence of the Pauli form factor in terms of the amplitude for one of the quarks in either the initial or final state to carry one unit of orbital angular momentum. The leading order contribution to $F_2$ has a $1/Q^6$ dependence with a coefficient depending on the light-cone nucleon wavefunctions at leading (twist-3) and subleading (twist-4) twist. In contrast to the pQCD analysis of $F_1$, in this calculation, the quark transverse momenta $\mathbf{k}_\bot$ were considered to first order in $\mathbf{k}_\bot^2/Q^2$ in order to allow for orbital angular momentum of the constituents. Logarithmic singularities arising in the integration of the nucleon DAs over momentum fractions in this approach contribute an extra logarithmic $Q^2$ dependence of $F_2$. Of particular interest where the results presented in this thesis are concerned is that these considerations lead to a modified scaling behavior for the ratio $F_2/F_1$ at large $Q^2$. Instead of $Q^2F_2 \propto F_1$, Belitsky et al.\cite{BelitskyJiYuan2003} find
\begin{eqnarray}
  \frac{Q^2}{\log^2\left(\frac{Q^2}{\Lambda^2}\right)} F_2 &\propto& F_1
\end{eqnarray}
at large momentum transfer, where $\Lambda$ is loosely related to the QCD scale parameter $\Lambda_{QCD}$. The recoil polarization data\cite{Jones00,Punjabi05,Gayou02} for $F_2^p/F_1^p$ are compatible with such a scaling starting at surprisingly low $Q^2$ for a surprisingly wide range of $\Lambda$ values (see figure \ref{pQCDscalingF2F1}). As this experiment extends the database of $F_2^p/F_1^p$ to higher $Q^2$, it will be interesting to determine the extent to which this scaling continues to be satisfied. The ratio of Pauli and Dirac form factors $F_2/F_1$ is given in terms of Sachs form factors by
\begin{eqnarray}
  F_1 &=& \frac{G_E+\tau G_M}{1+\tau} \\
  F_2 &=& \frac{G_M-G_E}{1+\tau} \\
  \frac{F_2}{F_1} &=& \frac{1-\frac{G_E}{G_M}}{\tau + \frac{G_E}{G_M}}
\end{eqnarray}
\begin{figure}[h]
  \begin{center}
    \includegraphics[angle=90,width=.98\textwidth]{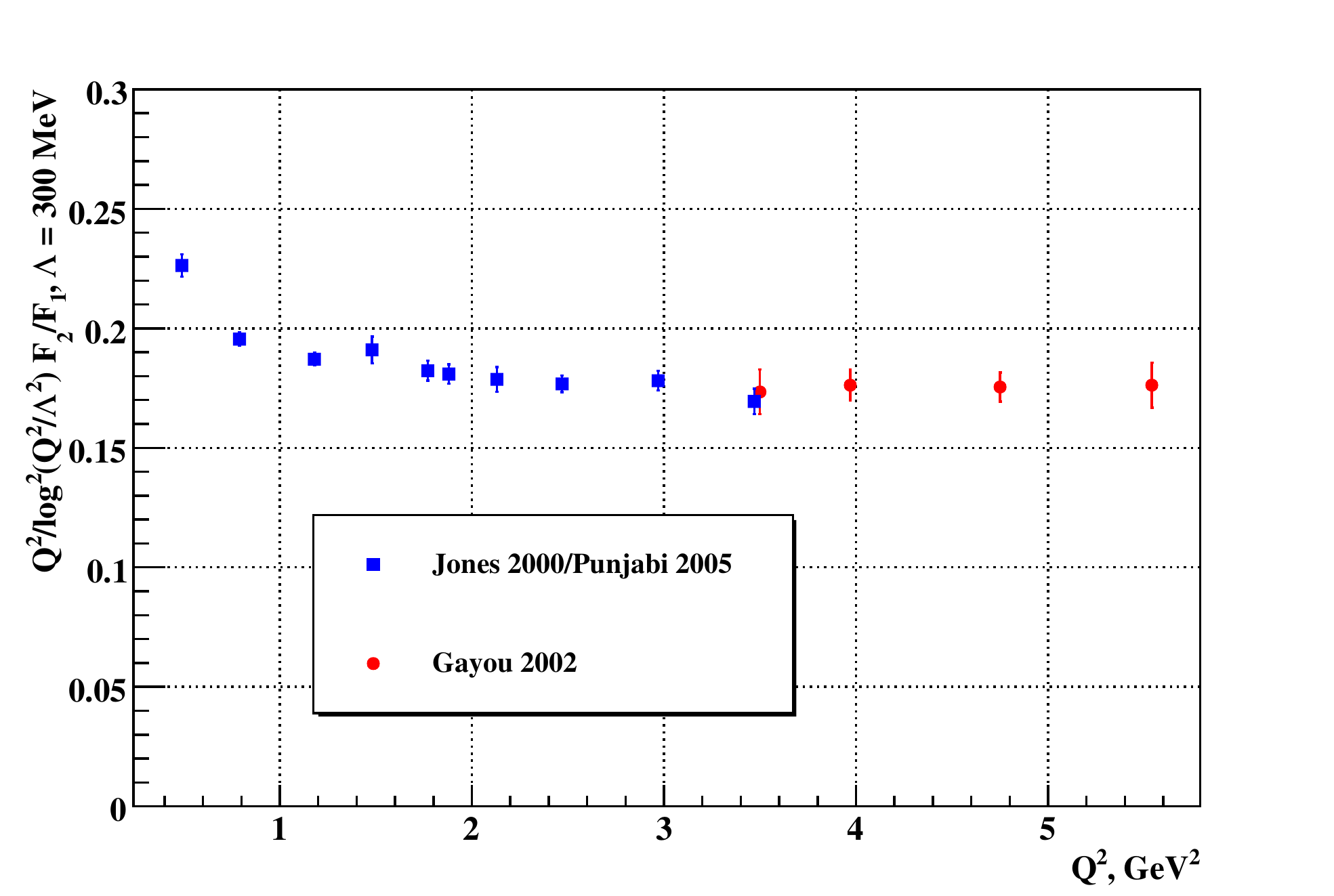}
  \end{center}
  \caption{\label{pQCDscalingF2F1} Jefferson Lab Hall A recoil polarization data plotted as $\frac{Q^2}{\log^2\left(\frac{Q^2}{\Lambda^2}\right)}\frac{F_2}{F_1}$, as a function of $Q^2$, with an arbitrary choice of $\Lambda = 300$ MeV. See text for references.}
\end{figure}

\section{Form Factors and Generalized Parton Distributions}
\paragraph{}
The preceding discussion of the insight into the nucleon form factors gained from perturbative QCD was but an example of the more general method of probing the short-distance structure of hadrons using ``hard'' probes; i.e., high momentum-transfer processes. Another classic example is deep inelastic scattering (DIS), in which inclusive electron scattering from nucleons and nuclei for large values of $Q^2$ and the invariant mass $W$ of the recoiling hadronic system probes the quark structure of the target. Perhaps the most striking success of pQCD, as alluded to previously, is its prediction of the $Q^2$ evolution of the proton structure function $F_2^p$ (not to be confused with the identically named Pauli form factor $F_2$). 

A more recent development is the potential for hard exclusive processes such as deeply virtual Compton Scattering (DVCS) and hard exclusive meson production to open up new frontiers in the understanding of nucleon structure\cite{HardExclusiveReview2001}. This effort has been driven by the recent theoretical development of Generalized Parton Distributions (GPDs). The proof of QCD factorization theorems \cite{JiFactorization,CollinsFactorization,RadyushkinFactorization} for hard exclusive reactions allows the factorization of the amplitude for these processes into a quark-level subprocess which can be calculated perturbatively and a universal, non-perturbative generalized parton distribution function which contains process-independent nucleon structure information. 
\begin{figure}
  \begin{center}
    \includegraphics[width=0.6\textwidth]{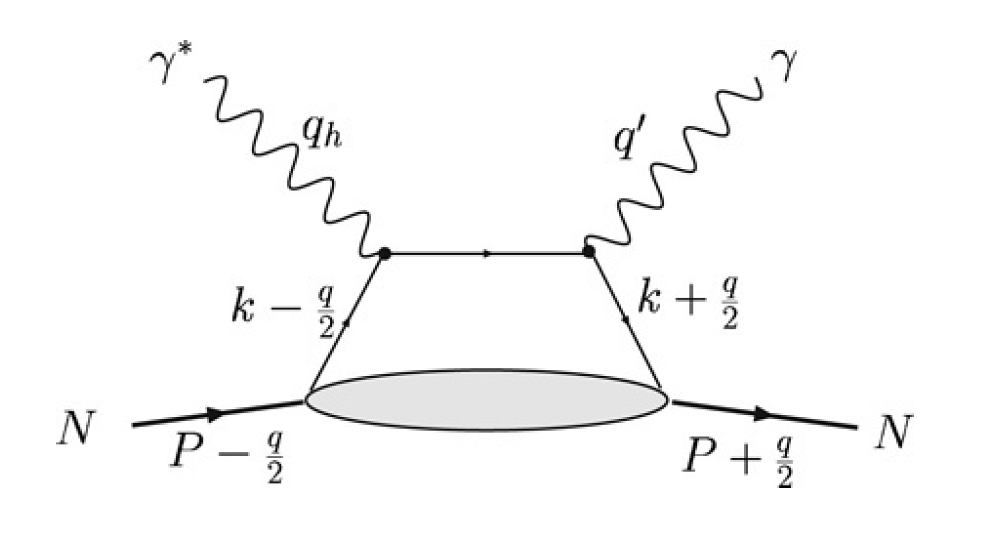}
  \end{center}
  \caption{\label{handbagfig} Handbag diagram for DVCS, illustrating the factorization into the perturbative quark-level subprocess $\gamma^* q \rightarrow \gamma q$, and the universal nucleon structure (pictured as the lower ``blob'') responsible for the presence of the quark line in the upper part of the diagram. The blob is parametrized in terms of GPDs.}
\end{figure}
Figure \ref{handbagfig} illustrates the factorization for the DVCS reaction ($\gamma^*(q_h) + N(p) \rightarrow \gamma(q') + N(p')$) through the ``handbag'' mechanism. In this exclusive reaction, a hard virtual photon (momentum $q_h$) Compton-scatters from an individual quark inside the nucleon, a pure QED process, leaving a real photon of momentum $q'$ and a nucleon of momentum $p' = p+q$ in the final state (note that $q \ne q_h$ is the total momentum transferred to the nucleon in the process). The amplitude for the process is integrated over all possible momenta $k$ of the struck quark. The reaction kinematics are characterized by the light-cone momentum fraction $x$ of the struck quark defined as $k^+ \equiv x\bar{P}^+$, $\bar{P} \equiv (p+p')/2$, the ``skewness'' $\xi$ defined as $q^+ \equiv -2\xi P^+$, and $t \equiv -q^2$, the squared momentum transfer to the nucleon. The amplitude is a function of these three variables, and is given at leading twist in terms of four chiral-even structure functions called $H^q(x,\xi,t), E^q(x,\xi,t), \tilde{H}^q(x,\xi,t), \tilde{E}^q(x,\xi,t)$, each defined for a given quark flavor $q$ and pictured as the blob in the lower part of the diagram in figure \ref{handbagfig}.  These functions are known as the generalized parton distributions (GPDs)\cite{JiGPDPRL1997,JiGPDPRD1997,RadyushkinGPD}. They encode new information about nucleon structure to be gained by studying exclusive reactions which cannot be obtained from inclusive reactions such as DIS. The functions $H, E, \tilde{H}, \tilde{E}$ characterize respectively the vector, tensor, axial vector, and pseudoscalar transition amplitudes.

GPDs provide a consistent synthesis of nucleon structure data from a variety of different classes of reactions. In the forward limit, they reduce to the usual parton distribution functions measured in DIS\cite{HardExclusiveReview2001}:
\begin{eqnarray}
  H^q(x,0,0) &=& \left\{\begin{array}{cc} q(x), &  x > 0 \\ -\bar{q}(-x), & x < 0\end{array} \right. \\
  \tilde{H}^q(x,0,0) &=&  \left\{\begin{array}{cc} \Delta q(x), &  x > 0 \\ \Delta \bar{q}(-x), & x < 0 \end{array} \right.
\end{eqnarray}
By integrating over $x$ at any $\xi$, the elastic nucleon form factors are recovered: 
\begin{eqnarray}
  \int_{-1}^{1} dx H^q(x,\xi,t) &=& F_1^q(t) \\
  \int_{-1}^{1} dx E^q(x,\xi,t) &=& F_2^q(t) \label{GPDmoments}
\end{eqnarray}
Given a model for the GPDs, a prediction for the nucleon form factors can be derived. Guidal et al.\cite{GuidalGPD2005} use a modified Regge ansatz for the GPDs at zero skewness. In order to reduce the integration region for the GPD moments to $0<x<1$, the authors define nonforward parton densities $\mathcal{H}^q(x,t) \equiv H^q(x,0,t) + H^q(-x,0,t)$ (and similarly for $\mathcal{E}^q(x,t)$) such that 
\begin{eqnarray}
  \int_0^1 \mathcal{H}^q(x,t) &=& F_1^q(t) \\ 
  \int_0^1 \mathcal{E}^q(x,t) &=& F_2^q(t)
\end{eqnarray}
for quark flavor $q=u,d$. The proton form factors are given in terms of the form factors for valence quark flavors $u$ and $d$ by $F_{1,2}^p = e_u F_{1,2}^u + e_d F_{1,2}^d$, and the neutron form factors are obtained from the proton by interchanging $u \leftrightarrow d$; i.e., $F_{1,2}^1n = e_u F_{1,2}^d + e_d F_{1,2}^u$. Under this construction, the Dirac form factor automatically satisfies the required normalization condition at $t=0$, namely $F_1^p(0)=1$ and $F_1^n(0)=0$. The modified Regge ansatz of \cite{GuidalGPD2005} relevant to the form factor behavior at large $t = -Q^2$ is parametrized as follows:
\begin{eqnarray}
  \mathcal{H}^q_{R2}(x,t) &=& q_v(x)x^{-\alpha'(1-x)t} \\
  \mathcal{E}^q_{R2}(x,t) &=& \frac{\kappa_q}{N_q}(1-x)^{\eta_q}\mathcal{H}^q(x,t)
\end{eqnarray}
where $\alpha'$ and $\eta_q,\ q=u,d$ are three adjustable parameters which can be fitted to the form factor data, and the normalization factors $\kappa_q/N_q$ guarantee the required $q^2 \rightarrow 0$ behavior $F_2^{p,n}(0) = \kappa_{p,n}$. $q_v(x)$ is the valence unpolarized parton distribution function (PDF) for quark flavor $q$, which is taken from the MRST2002 global NNLO fit\cite{MRST2002}, evolved to a scale $\mu^2=1\ GeV^2$. Given this relatively simple ansatz for the functional form of the GPDs and only three adjustable parameters, the model of \cite{GuidalGPD2005} achieves a remarkably good agreement with experiment for all four nucleon form factors in the entire $Q^2$ range over which they are known. Of particular relevance to the outcome of this experiment is the prediction of a zero crossing of $G_E^p/G_M^p$ at approximately $9\ GeV^2$.  

The connection between form factors and GPDs is a powerful one, as form factor data provide powerful constraints on the $x$ dependence of the GPDs through the moments \eqref{GPDmoments}. GPDs have been related to the total angular momentum $J^q$ carried by quark flavor $q$ in the nucleon \cite{JiGPDPRL1997} through Ji's angular momentum sum rule:
\begin{eqnarray}
  2J^q &=& \int_{-1}^{1} dx x\left\{H^q(x,0,0)+E^q(x,0,0)\right\} \label{Jisumrule}
\end{eqnarray}
In the model of \cite{GuidalGPD2005}, the behavior of the ratio $F_2^p/F_1^p$ determines the behavior of the GPD $E$ as $x \rightarrow 1$, allowing an evaluation of the sum rule \eqref{Jisumrule}. The $F_2^p/F_1^p$ measurements of this experiment to higher $Q^2$ will further constrain this GPD, thus improving the understanding of the spin structure of the proton.

Another promising application of GPDs is the concept of nucleon tomography, a modern version of the traditional interpretation of the form factors as Fourier transforms of the Breit frame charge and magnetization densities. The two-dimensional Fourier transform of GPD $H^q$ with respect to $t$ has been shown to yield the transverse quark density in impact parameter space in the infinite momentum frame (IMF)\cite{BurkardtIJMPA} as a function of longitudinal momentum fraction $x$:
\begin{eqnarray}
  q(x,\mathbf{b}) &=& \int \frac{d^2 \mathbf{q}}{(2\pi)^2} e^{i\mathbf{q}\cdot \mathbf{b}}H^q(x,t=-\mathbf{q}^2)
\end{eqnarray}
Integrating this distribution over all $x$ and summing over quark flavors, one obtains the model-independent IMF transverse charge density $\rho(b)$\cite{MillerGPDchargedensity2007} equal to the two-dimensional Fourier transform of the Dirac form factor $F_1$:
\begin{eqnarray}
  \rho(b) &=& \sum_q e_q \int dx q(x,\mathbf{b}) = \int \frac{d^2 \mathbf{q}}{(2\pi)^2} F_1(Q^2=\mathbf{q}^2) e^{i\mathbf{q}\cdot \mathbf{b}} \label{IMFrhob}
\end{eqnarray}
In contrast to the result of Kelly's analysis of the form factors in terms of three-dimensional Fourier transforms of the charge and magnetization densities with relativistic corrections to go from the Breit frame to the rest frame, which finds a neutron charge density with a positive core and a negative exterior, Miller's analysis\cite{MillerGPDchargedensity2007} of the IMF charge density in impact parameter space using \eqref{IMFrhob} results in a negative central charge density for the neutron, with a positive component at intermediate distances ($\approx$0.5-1.0 fm), and another sign change at large (1-2 fm) distances. 

Diehl et al.\cite{DiehlGPDtomography2005} presented a somewhat more complicated parametrization of the GPDs $H$, $\tilde{H}$, and $E$, and fitted their model to nucleon Dirac, Pauli, and isovector axial form factor $F_A(q^2)$ data. Using their fit results, they constructed tomography plots of unpolarized and (transversely) polarized valence $u$ and $d$ quark distributions in the two-dimensional impact parameter plane for various longitudinal momentum fractions $x$. An interesting property of the polarized transverse quark densities, which are related to Fourier transforms of $E^q$ with respect to $t$, is that for polarization along the $b_x$ direction, the densities are shifted along the $b_y$ axis. The acquired shifts for the $u$ and $d$ quark densities are in opposite directions. The (valence) polarized densities defined in \cite{DiehlGPDtomography2005} are interpreted as the probability density to find a quark with momentum fraction $x$ and impact parameter $\mathbf{b}$ in a nucleon polarized along the $b_x$ direction, less the probability to find an antiquark. 

In summary, the development of GPDs has provided the opportunity for a consistent synthesis of nucleon structure data from a wide variety of reactions in terms of universal, non-perturbative, generalized structure functions which can potentially be measured directly in hard exclusive reactions such as DVCS. They connect the inclusive regime of deep inelastic scattering and PDFs to the exclusive regime of elastic scattering and form factors through constraints on their limiting forward behavior and their first $x$ moments. Through sum rules relating the $x$ moments of GPDs to the total angular momentum carried by quarks in the nucleon, GPDs provide insight into its spin structure. Finally, GPDs enable nucleon tomography in impact parameter space as a function of longitudinal momentum fraction $x$. By integrating over quark momentum, model-independent transverse charge densities in the IMF are obtained as two-dimensional Fourier transforms of the nucleon Dirac form factor $F_1$. The nucleon form factors are an important input to this highly promising field of research, and the measurements of $G_E^p/G_M^p$ (or $F_2^p/F_1^p$) reported in this thesis in particular will help constrain the $x$ dependence of the tensor GPD $E$. 

\section{Summary}
\paragraph{}
This chapter presented an overview of the physics of nucleon form factors with an emphasis on high momentum transfers. Several important topics have been omitted, including a thorough discussion of the pion cloud, the insight gained at low momentum transfers from chiral perturbation theory, and lattice QCD calculations of the nucleon form factors. Since these subjects are more relevant to the understanding of the low momentum transfer behavior of nucleon form factors, they were not discussed in this thesis. The reader is referred to the review paper \cite{PerdrisatPunjabiVanderhaegen2007} for extensive discussions of these and additional topics, and references to the relevant literature. 

\chapter{Description of the Experiment}
\label{chapter3ref}
\paragraph{} Experiment E04-108 collected data from October 2007 to June 2008\footnote{E04-108 ran consecutively with experiment E04-019.} in experimental hall C at Jefferson Lab\footnote{Also known as the Thomas Jefferson National Accelerator Facility}, in Newport News, Virginia. Polarized electrons were excited by circularly polarized laser light from a semiconductor photocathode and accelerated to energies as high as 5.714 GeV by the superconducting radio-frequency resonant cavities of the CEBAF accelerator. After acceleration, the electrons were delivered to experimental Hall C, where they collided with a liquid hydrogen target. Polarized scattered protons were detected in a magnetic spectrometer called the High Momentum Spectrometer (HMS). A Focal Plane Polarimeter (FPP) consisting of blocks of CH$_2$ followed by tracking detectors measured the polarization of these protons. A large solid-angle electromagnetic calorimeter (BigCal) detected the scattered electrons in coincidence with the scattered protons in order to suppress substantial inelastic backgrounds otherwise present in the single-arm proton spectrum at high $Q^2$. In the sections that follow the experimental apparatus will be described in some detail.
\section{Kinematics of Experiments E04-108 and E04-019}
\paragraph{}
Table \ref{kintable1} shows the kinematics of experiments E04-108 and E04-019. The former experiment has as its goal to extend the knowledge of $G_E^p/G_M^p$ as measured in polarization transfer experiments to the highest possible $Q^2$ that can be reached at CEBAF with the highest available beam energy. Experiment E04-019 has the goal of measuring the $\epsilon$-dependence at $Q^2 = 2.5\ GeV^2$ of $G_E^p/G_M^p$ using the same apparatus and method in order to search for signatures of two-photon exchange effects. Experiment E04-108 is the subject of this thesis. 
\begin{table}[h]
  \begin{center}
  \begin{tabular}{|c|c|c|c|c|c|c|}
    \hline
    $Q^2$, GeV$^2$ & $\varepsilon$ & $E_{beam}$, GeV & $\theta_p,\ ^\circ$ & $p_p$, GeV & $E_e$, GeV & $\theta_e, ^\circ$ \\ \hline
    2.5 & 0.154 & 1.873 & 14.495 & 2.0676 & 0.532 & 105.2 \\ \hline
    2.5 & 0.633 & 2.847 & 30.985 & 2.0676 & 1.51 & 44.9 \\ \hline
    2.5 & 0.789 & 3.680 & 36.10 & 2.0676 & 2.37 & 30.8 \\ \hline
    5.2 & 0.377 & 4.053 & 17.94 & 3.5887 & 1.27 & 60.3 \\ \hline
    6.8 & 0.507 & 5.714 & 19.10 & 4.4644 & 2.10 & 44.2 \\ \hline
    8.5 & 0.236 & 5.714 & 11.6 & 5.407 & 1.16 & 69.0 \\ \hline
  \end{tabular}
  \caption{\label{kintable1} Kinematics of experiments E04-108 and E04-019}
  \end{center}
\end{table}

The choice of kinematics is motivated by several considerations which will be discussed below. In all cases, the goal of measuring to the highest possible $Q^2$ was weighed against the increasing difficulty of the recoil polarization technique with increasing $Q^2$ due to the falling elastic scattering cross section which reduces the number of events that can be collected in a fixed amount of beam time, the decreasing analyzing power of the $p + CH_2 \rightarrow X$ reaction at large proton momenta, and the increasing uncertainty associated with the calculation of the precession of the proton spin in the magnets of the HMS. 
\section{The Continuous Electron Beam Accelerator Facility}
\paragraph{}
Before being named Jefferson Lab, the electron accelerator was originally named CEBAF, an acronym for Continuous Electron Beam Accelerator Facility. It consists of two parallel linear accelerators, each capable of approximately 600 MeV of acceleration. Each linac uses superconducting RF-resonant Niobium cavities cooled to well below their transition temperatures by superfluid Helium at $\approx$ 2 K. The use of superconducting RF cavities eliminates power losses to ohmic heating characteristic of room temperature, normal-conducting cavities, allowing the accelerator to operate at roughly 1/3 of the power consumption that would otherwise be required.

The CEBAF linacs are connected by nine recirculating arcs of magnets, with five at the north end and four at the south end. With this ``racetrack'' design, the electron beam can be accelerated in up to five passes through both linacs, for a maximum energy of approximately 6 GeV before extraction and delivery to the three experimental halls. Figure \ref{CEBAFdrawing} shows a schematic layout of the CEBAF accelerator.
\begin{figure}[h]
  \begin{center}
    \includegraphics[width=.9\textwidth]{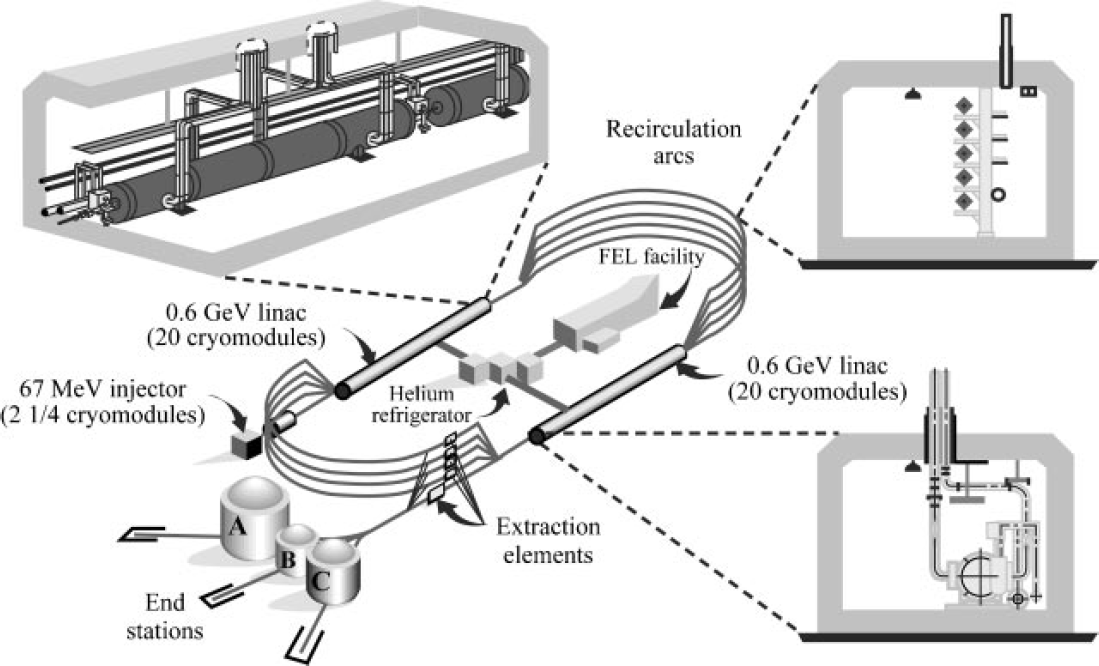}
    \caption{\label{CEBAFdrawing} Schematic of the CEBAF accelerator}
  \end{center}
\end{figure}
\subsection{Polarized Electron Production}
\paragraph{}
The production and acceleration of the electron beam starts with the polarized electron source. Electrons are excited from a GaAs photocathode using circularly polarized laser light. A monolayer of Cs$_2$O on the GaAs results in a surface with negative electron affinity, shifting the vacuum level below the conduction band, facilitating the release of photoelectrons excited across the semiconductor bandgap into the vacuum. The polarization of the electrons takes place by optical pumping between the $P_{3/2}$ valence-band level and the $S_{1/2}$ conduction-band level in GaAs. In bulk GaAs, the four spin substates of the P$_{3/2}$ level are degenerate. For right(left)-circularly polarized photons, the $\Delta m_j=+(-)1$ transition into the conduction band is three times more likely to occur from the degenerate $m_j = -(+) \frac{3}{2}$ substate of the valence $P_{3/2}$ level to the $-(+) \frac{1}{2}$ conduction $S_{1/2}$ substate than from the $-(+) \frac{1}{2}$ valence $P_{3/2}$ substate to the $+(-) \frac{1}{2}$ conduction $S_{1/2}$ substate, owing simply to the Clebsch-Gordan angular momentum coupling coefficient between the initial and excited states, resulting in a theoretical limit of 50\% polarized electrons. However, the polarization can be increased above this limit by lifting the degeneracy of the $P_{3/2}$ states. If the material properties of the cathode are altered in such a way that the degeneracy is broken, then the theoretical limit becomes 100 \% since all electrons are excited into the same polarization state for a given photon polarization.

In practice, lifting of the degeneracy can be accomplished by applying a mechanical strain to the GaAs. One way to provide the strain is by growing the GaAs cathodes on a substrate of GaAsP, which has a different lattice constant. With a single layer of strained GaAs, polarizations of ~75\% are routinely achieved. With so called ``superlattice'' GaAs photocathodes, consisting of a series of alternating thin (several nm) layers of GaAs and GaAsP, polarizations approaching 90\% are routinely achieved, with higher quantum efficiency (QE) than is typical of a single strained layer of GaAs. Jefferson Lab's polarized source uses a superlattice GaAs photocathode to deliver 85\% polarized electron beams with a QE near 1 \%. 

The laser light used to produce electrons from the cathode is provided by three gain switched diode lasers, one for each experimental hall. Each laser is pulsed at a frequency of 499 MHz, and the three lasers are phase shifted relative to each other by 120$^\circ$. Each laser pulse produces a single bunch of electrons, and the combined train of electron bunches has a frequency of 1497 MHz, equal to the fundamental resonant frequency of the RF accelerating cavities in the linacs. In each sequence of three electron bunches, one bunch is destined for each experimental hall. Circular polarization of the laser light is achieved by use of a Pockels Cell, which consists of a birefringent crystal whose birefringence depends linearly on the applied electric field. By varying the electric field applied to the crystal, the polarization components of the light can be phase-shifted relative to each other, allowing control of the orientation of the polarization vector of the laser light passing through it. For this and most other polarized beam experiments at Jefferson Lab to date, the Pockels Cell was used to reverse the polarization of the laser light between left and right circular polarization at a frequency of 30 Hz. Taking data in both positive and negative helicity states enables the cancellation of helicity-independent instrumental asymmetries in the proton polarimeter so that an unambiguous measurement of the physical polarization transfer components can be obtained which is independent of the instrumental asymmetry. The rapid reversal of the beam polarization guarantees equal numbers of events in each polarization state by canceling out slow fluctuations in beam current and target density which affect the luminosity. Additionally, insertable and rotatable half-wave plates are available which may be used by experiments to passively reverse the polarization of the laser light. These tools are particularly important for the program of parity violation experiments at Jefferson Lab which aim to measure asymmetries of a few $ppm$ or smaller. Such experiments are quite sensitive to any small helicity-dependent differences in beam properties, and periodically inserting and retracting the half-wave plate can help to cancel or at least correct for such effects by taking data in both states. This experiment, being largely insensitive to such small differences, did not request any changes to the state of either half-wave plate. A more detailed overview of polarized electron beam technology with references to the scientific literature on the subject is available in \cite{StutzmanPhysToday08}.
\subsection{Acceleration and Beam Delivery to Hall C}
\paragraph{}
The photoemitted electrons are launched into the injector by a 100 kV DC electron gun. The photocathode and the electron gun are housed in an ultra-high-vacuum enclosure ($10^{-11}$ to $10^{-12}$ Torr). Such a high vacuum is crucial to prolonging the lifetime of the photocathode, as ionized atoms from the residual gas in the source enclosure accelerate backward and collide with the cathode, degrading its quantum efficiency. The injector itself accelerates electrons by up to 67 MeV in preparation for entry into the north linac. All of the superconducting RF cavities in the CEBAF accelerator are housed within cryomodules. Each cryomodule consists of eight RF cavities.  Each linac contains 20 cryomodules. The injector consists of $2\tfrac{1}{4}$ cryomodules (18 cavities). A cryomodule is a large cryostat which additionally contains all the necessary support structure for the accelerating cavities, including but not limited to
\begin{itemize}
\item An outer vacuum vessel 
\item Thermal radiation shielding
\item Magnetic shielding
\item A welded helium vessel enclosing each cavity pair to keep the cavities at 2 Kelvin.
\item Superinsulation blankets
\item Waveguides to supply RF power to the cavities
\item Mechanical ``tuners'' apply strains which slightly stretch or compress the cavities to optimize their resonant properties and performance. 
\item Feedthroughs and instrumentation
\end{itemize}
Each cavity is separately powered by a 5 kW maximum power RF klystron operated at 1497 MHz.
Figure \ref{cavitypair} shows one of the four cavity pairs that make up a cryomodule.
\begin{figure}[h]
  \begin{center}
    \includegraphics[width=1.00\textwidth]{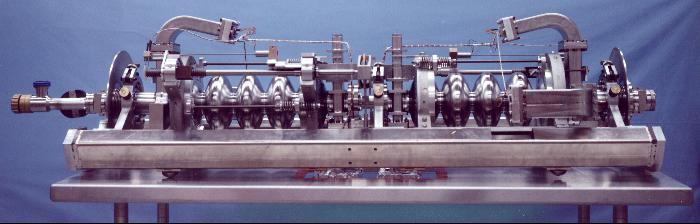}
    \caption{\label{cavitypair} A pair of the standard five-cell Niobium accelerating cavities used in CEBAF.}
  \end{center}
\end{figure}
By the time the electron beam reaches the main linac, it is sufficiently relativistic to stay in phase with the RF field on up to five passes through both linacs, literally riding the crest of an electromagnetic wave to higher energy while moving at a constant speed near the speed of light. At any given time, up to five electron beams of different energies are essentially sitting on top of each other in the linac. At the end of each linac, a series of dipole magnets separates the beam electrons into their component momenta and steers them into one of the recirculating arcs. The arcs consist of conventional room-temperature magnets, with dipoles for steering and quadrupoles for focusing. Each successive arc has a larger field integral in order to steer higher momentum electrons along a path with the same radius of curvature. At the exit of each recirculating arc is an identical series of magnets with opposite polarity to recombine the electrons from the five arc beamlines back into a common beam pipe for another pass through the linac. 

There are five arcs connecting the end of the north linac to the beginning of the south linac, and four arcs connecting the end of the south linac to the beginning of the north linac, for a maximum of five passes through both linacs. The beam destined for any given hall can be extracted from the main racetrack after any number of passes between one and five by RF separator magnets, which also operate at 499 MHz and can be phased so that the beams for Halls A, B, and C are separately extracted after the desired number of passes. After extraction from the main beamline at the appropriate pass, the beams enter the beam switch yard, where, depending on their RF phase, they are steered into the beamline leading to the appropriate experimental hall. This design allows all three experimental halls to run simultaneously with great flexibility in beam properties including polarization, energy, and current, with some constraints described below.

The main constraint on simultaneously delivering different beam energies to three experimental halls is that the energy of each linac is fixed, so that while any given hall can choose a desired number of passes, the energy is restricted to multiples of the linac energy which is common to all three halls. For intensity, the available beam current is unrestricted except that the total current delivered to all three halls simultaneously is limited to approximately 180 $\mu A$. The polarization that can be delivered to three halls simultaneously is limited for a given combination of energies by the precession of the electron spin in the recirculating arc beam lines and the transport lines from the accelerator to the experimental halls. The precession is proportional to the electron's relativistic $\gamma$-factor ($\gamma = E_e / m_e$), its (small) anomalous magnetic moment, and the total bend angle of its trajectory in each pass through the arc and hall transport beam lines. The orientation of the beam polarization is rotated by a Wien filter before injection into the accelerator to optimize the delivery of longitudinal polarization to the experimental halls demanding polarized beam. The complicated dependence of the final polarization orientation in each hall on the Wien filter setting, the number of passes through the accelerator, and the energy per pass actually constrains the combination of energies at which polarized beam can be simultaneously delivered to multiple halls to certain ``magic'' combinations. During part of experiment E04-108, Hall A also needed polarized beam, so a compromise was made in the Wien filter setting to optimize polarization to Halls A and C in which the final longitudinal polarization in Hall C was ~95\% of the full beam polarization at the injector. This compromise resulted in a small increase in the statistical uncertainty on the form factor ratio. Even though the final result for the form factor ratio is independent of precise knowledge of the beam polarization, the statistical error depends on the beam polarization roughly as $h^{-2}$.

In addition to reduced energy costs, the great advantage of superconducting RF technology is that it enables continuous wave (CW) operation of the accelerator at much higher currents than would be possible using normal-conducting cavities. Experiments that use CEBAF almost always run with CW, 100\% duty factor electron beam, for which average current is equal to peak current. The benefit of a 100\% duty factor electron beam is that, for a given average luminosity, ``clean'' events in the experimental detectors are relatively more likely than for a pulsed beam with a low duty factor and a higher peak luminosity, which increases the probability of multiple collisions per beam bunch, multiple particles in the detectors, and increased difficulty in correctly reconstructing individual scattering events. The ability to consistently deliver CW electron beams with high energy, intensity and polarization make CEBAF arguably the leading electron scattering facility in the world for nuclear and particle physics experiments. More information about the CEBAF accelerator along with references can be found in \cite{LeemannCEBAF}.
\subsection{Hall C Beamline Components/Diagnostics}
\paragraph{}
Careful, constant monitoring and tight control of the properties of the electron beam entering Hall C are needed to carry out experiments successfully and safely. The beam current and position are non-intrusively monitored by several redundant devices which enter into feedback circuits used to control, respectively, the beam's intensity and its orbit through the many magnetic elements of both linacs, the recirculating arcs, and the Hall C beamline. In a cross section experiment, an accurate calibration of the current-monitoring devices is needed in order to know the absolute luminosity of the experiment. For a polarization experiment in which an asymmetry is to be measured which does not depend on the absolute luminosity, precise knowledge of the beam current at all times is less important than maximizing the total number of scattering events recorded, which in practice means running at the highest possible luminosity for which clean reconstruction of events in the detectors is still possible. The beam position measurement devices are used by the accelerator operators to tune up the various beam-line elements for stable delivery of beam with the desired properties. In contrast to the beam current and position, there are several important beam properties that must be monitored periodically that involve disruptive measurements for which data taking for the experiment must be interrupted. Beam properties falling into this category include energy, polarization, and the spot size of the beam.
\subsubsection{Beam Position Measurement}
\paragraph{}
The beam position is monitored continuously by devices aptly named Beam Position Monitors (BPMs). Each beam position monitor consists of a resonant cavity with a fundamental frequency equal to that of the accelerator and the Hall C beam. This cavity contains four antennae, rotated by 45 degrees relative to the vertical and horizontal axes in order to minimize synchrotron radiation damage. The difference-sum ratio of the amplitudes of the signals picked up by antennae on opposite sides of the beam is proportional to the distance between the beam and the midpoint of the two antennae. The center of gravity of the four antenna signals measures relative changes in the offset of the beam from its ideal trajectory. The BPMs are used as feedback in steering the beam. The three BPMs closest to the target are monitored closely by the experiments to ensure stability of the beam position on target.

The BPM signals are read out by sampling ADCs which perform an integration over a 20 $\mu$s gate. These ADCs are read out at 60 Hz, and the first 25 data points are averaged and presented once per second by EPICS (Experimental Physics Industrial Control System), the system used by the accelerator and end stations for slow control and monitoring of accelerator and experiment parameters. The same signals are also sent to the Hall C data acquisition electronics and read out by CODA (CEBAF Online Data Acquisition), the standard data acquisition software used by Jefferson Lab, for every event. The BPM information for a particular event is not to be understood as the exact beam position on target for that event, as the signals are not synchronized with the event data itself and the actual position on target is changing rapidly due to the fast raster system which will be described below. In practice, an average beam position is calculated using a rolling average of BPM information over a specified number of previous events, the appropriate choice of which depends on the experiment data rate, and this average beam position is further corrected for each event using the fast raster signals. Since the average beam position on target is usually quite stable over the course of a single CODA run, it is sometimes more practical to simply ignore the event-by-event BPM information and fix the average beam position as a parameter of the analysis, and to use the raster signals to measure the change in beam position relative to the ``fixed'' average position. Once the BPMs are properly calibrated, the BPM signals measure the beam position with an overall accuracy of approximately $\pm 1.0$ mm, and relative changes in beam position are measured with a precision of $\approx 0.2$ mm. More detailed information on the Hall C BPMs is available in \cite{GueyeHallCBPM}.

A more precise and accurate determination of the beam position and profile is obtained using the superharp system. Each superharp consists of a set of two vertical wires and one horizontal wire strung on a moveable frame. These wires can be scanned across a low current beam to measure its profile and absolute position. The signals induced on the wires as they are scanned across the beam are digitized by an ADC and correlated with the wire positions as recorded by an encoder equipped with absolute position readout electronics. Since a harp scan interferes destructively with the electron beam, data taking must be interrupted to perform the measurement. The position accuracy of a single superharp beam profile measurement is better than 20 $\mu$m \cite{YanSuperHarp}. In addition to measuring the beam profile and providing a reference coordinate against which the BPMs can be calibrated, the superharp system is used as part of the beam energy measurement in the Hall C arc as discussed below.  
\subsubsection{Beam Current Measurement}
\paragraph{}
There are three devices used to measure the beam current in Hall C. The first two devices, used to monitor the beam current in real time, are cylindrical cavities designed to resonate in the transverse magnetic mode $TM_{010}$ at the same frequency as the accelerator RF, and are called BCM1 and BCM2 respectively. The advantage of choosing the $TM_{010}$ mode is that the output power is relatively insensitive to the beam position inside the cavity when the beam is close to the cavity's longitudinal axis. When the beam passes through these cavities, this mode is excited and antennae placed inside the cavities are used to convert the RF power of the excited resonance, which is proportional to the square of the beam current, to an analog voltage signal. The electronics used for this conversion are different for BCM1 and BCM2, resulting in slightly different performance characteristics, but for both BCMs, the output voltages are processed by a preamplifier/level-shifter followed by a voltage to frequency converter, before finally being sent to a scaler which is read out every two seconds by EPICS. 

The output power of a resonant cavity depends most strongly on the difference between its resonant frequency and the frequency of the excitation. Since the cavity's resonant frequency is determined mainly by its size and shape, and since the cavity can expand and contract in response to changes in temperature, the gain of the BCM cavities is quite sensitive to temperature. For this reason, the BCM cavities are kept thermally insulated at a constant temperature of 43.3 $^\circ$C. More details on the BCM cavities and their operation can be found in \cite{GNhallcBCM,CAhallcBCM}. 

The third device used to measure the beam current is called an Unser monitor, which is a parametric current transformer\cite{CAhallcBCM,UnserAIP}. The feature of the Unser monitor which makes it useful for beam current measurements is that it has an extremely stable gain and can thus be used as an absolute standard against which to calibrate the BCM cavities, which can experience slow gain drifts over time. On the other hand, the Unser suffers from an unstable zero offset which can drift significantly over short time scales, making it unsuitable for current monitoring in real time. To use the Unser monitor to calibrate the BCMs, alternating runs are taken with no beam in the cavities and with beam of various currents, in order to establish the zero offset and the gain, respectively. For a cross section measurement, careful calibration of the BCMs must be performed periodically in order to minimize the uncertainty on the total charge collected by the experiment. For this experiment, however, since the result did not depend on the total charge delivered to the experiment, no dedicated BCM calibration was performed. Only one rough calibration was performed to verify the integrity of the BCM signals near the beginning of the run.
\subsubsection{Beam Energy Measurement}
\label{secbeamenergymeasurement}
\paragraph{}
To measure the energy of the electron beam in Hall C, the dipole magnets of the Hall C arc transport line are used as a spectrometer. There are eight identical dipoles in the arc\footnote{Additionally, there are twelve quadrupoles, eight sextupoles and eight pairs of beam corrector magnets for transport.}. Pairs of superharps at the entrance and exit of the arc precisely measure the beam position and angles before and after the arc, and an additional pair of superharps at the midpoint of the arc contributes a third measurement of the trajectory and determines its curvature. In order to measure the beam momentum, all elements of the beamline except the dipoles are turned off, and the current in the magnets is varied to steer the beam onto the central trajectory of the arc beamline. For the central trajectory, the total deflection angle of the beam in the arc is 34.3$^\circ$. The momentum is determined from the required current settings, using the precise knowledge of the field integral of the arc dipoles as a function of current:
\begin{equation}
  p_{beam} = \frac{e}{\theta_{arc}} \int B dl
\end{equation}
The accuracy of this method of measuring the beam energy is determined by the accuracy of the positions measured by the superharps and by the knowledge of the field integral in the arc magnets as a function of current. One of the arc dipoles has been precisely field-mapped as a function of current, and is used to calibrate the remaining dipoles, which are assumed to have identical field maps to the reference dipole. The precision on the beam momentum determined by this technique is $\frac{\delta p}{p} \approx 5 \times 10^{-4}$. More details on the technique can be found in \cite{YanArcEnergy}. For this experiment, the reaction under study is elastic electron-proton scattering, in which both outgoing particles are detected in coincidence. By detecting both particles, the kinematics of the reaction are fully determined, and can be used as a measurement of the beam energy. However, the accuracy of using elastic $ep$ scattering to measure the beam energy is limited by the accuracy with which the proton and electron kinematics are reconstructed, and is sensitive to unknown or poorly known offsets in the experimental setup. Since the arc measurement provides precise knowledge of the beam energy, one such measurement was performed for each different beam energy sent to the experiment\footnote{No dedicated arc measurement was performed for data taking during a period in December 2007 during which the nominal beam energy was 3.548 GeV as monitored by EPICS.}, and the measured energy was used to help calibrate the various unknown small offsets in the experiment.  
\begin{table}[h]
  \begin{center}
    \begin{tabular}{|c|c|c|c|}
      \hline Date & $Q^2$, GeV$^2$ & Number of passes & $E_{arc}, MeV$ \\ \hline
      11/19/2007 & 5.2 & 5 & 4052.34 $\pm$ 1.38 \\ \hline 
      11/28/2007 & 2.5 & 3 & 1873.02 $\pm$ 1.09 \\ \hline
      12/11/2007 & 2.5 & 4 & 2847.16 $\pm$ 1.19 \\ \hline 
      1/6/2008   & 2.5 & 4 & 3680.23 $\pm$ 1.31 \\ \hline
      1/23/2008  & 2.5 & 2 & 1868.13 $\pm$ 1.09 \\ \hline
      4/6/2008   & 8.5 & 5 & 5717.32 $\pm$ 1.64 \\ \hline
    \end{tabular}
    \caption{\label{ArcEnergyTable} Arc beam energy measurements taken during experiments E04-108 and E04-019}
  \end{center}
\end{table}

Table \ref{ArcEnergyTable} shows the arc energy measurements performed during the two experiments. In addition to the dedicated arc energy measurements, the beam momentum was monitored continuously through EPICS. The BPM and arc magnet setting information used in the feedback system which stabilizes the beam energy and position is also used to monitor relative fluctuations in beam energy with similar relative precision to the arc measurement. However, in absolute terms, the continuously-monitored EPICS beam energy is less accurate than the arc energy, because the beam position determined by the BPMs is less accurate than that determined by the superharps. The same beam energy and pass configuration was used for the two highest $Q^2$ kinematics at 8.5 and 6.8 GeV$^2$. A shift in the beam energy from 5.717 GeV to 5.712 GeV was detected in the EPICS beam energy monitoring three days after the arc measurement was performed. This shift was determined to be real, although no additional arc measurements were taken. For most of the data taken at the highest beam energy, the lower value of 5.712 GeV is used in the analysis. Once again, since the elastic electron-proton scattering reaction under study serves as an independent check of the beam energy, and since the actual result of the experiment is quite insensitive to small changes in beam energy at the $10^{-3}$ level typical of the observed fluctuations, further arc measurements were deemed unnecessary. It is worth remarking that the beam energy spread is typically less than $5 \times 10^{-5}$, i.e., an order of magnitude smaller than the uncertainty in its absolute determination, and is monitored non-invasively during accelerator operations through the use of synchrotron light interferometry\cite{SLIenergyspread}.
\subsubsection{Beam Polarization Measurement}
\paragraph{}
To measure the beam polarization in Hall C, the pure QED process of double polarized M\"{o}ller scattering is used. The cross section for the reaction $\vec{e} + \vec{e} \rightarrow e + e$ is precisely calculable in QED, and is given in the center-of-mass (CM) frame by\cite{MollerNIM} 
\begin{equation}
  \frac{d\sigma}{d\Omega} = \frac{d\sigma_0}{d\Omega}\left[1+P_t^\parallel P_b^\parallel A_{zz}(\theta)\right]
\end{equation}
where $\frac{d\sigma_0}{d\Omega}$ is the unpolarized cross section for the same process, $P_b^\parallel$ and $P_t^\parallel$ are, respectively, the polarizations of the ``beam'' and ``target'' electrons parallel to the center-of-mass momentum of the incident electron, and $A_{zz}(\theta)$ is called the analyzing power of the reaction and depends on the CM scattering angle $\theta$ as follows:
\begin{equation}
  A_{zz}(\theta) = -\sin^2 \theta \frac{8 - \sin^2 \theta}{(4-\sin^2\theta)^2} \label{MollerAzz}
\end{equation}
From \eqref{MollerAzz}, it is clear that the analyzing power is maximized for electrons scattered by 90$^\circ$ in the center of mass frame. In order to exploit M\"{o}ller scattering to measure the electron beam polarization, a source of polarized target electrons with known polarization is required. In addition, scattered electrons must be detected. In the Hall C M\"{o}ller polarimeter, the electron beam is scattered on a pure iron foil, magnetized to saturation by a 4 Tesla field produced by a superconducting split-coil solenoid. This approach allows accurate knowledge of the polarization of the electrons in the iron. Pairs of electrons scattered at or near 90$^\circ$ in the center-of-mass are detected in coincidence in order to eliminate backgrounds from other processes, such as Mott scattering from the iron nuclei. A system of two quadrupole magnets deflects the scattered electrons to larger angles and provides for analysis of their energy. A system of moveable collimators allows selection of a narrow range of scattering angles around 90$^\circ$ (CM). 

The scattered electrons that pass through the system of quadrupole fields and collimators are detected by lead-glass total absorption shower counters. There are two such counters, one for each scattered electron, each equipped with a photomultiplier tube to collect the Cerenkov light emitted by the electrons and positrons in the shower of the primary electron. For the actual beam polarization measurement, what is measured is the asymmetry in the coincidence counting rate between the two beam helicity states. For this purpose, the analog signals of the two PMTs are amplified, clipped, and converted to 5 and 10 ns wide logic pulses by fast discriminators. The detector package also includes scintillator hodoscopes placed in front of the shower counters, which provide information on the position of the detected electrons. The hodoscopes are not used during the actual beam polarization measurement, but they serve to check the alignment of the various system components during the setup of the measurement. This polarimeter can measure the beam polarization with $<$1\% statistical uncertainty in a short period of time and $\approx$0.5\% systematic uncertainty. However, since the measurement interferes destructively with the beam, it cannot be used to monitor the polarization continuously and data taking must be periodically interrupted to perform M\"{o}ller measurements. 
\begin{figure}[h]
  \begin{center}
    \includegraphics[width=0.7\textwidth]{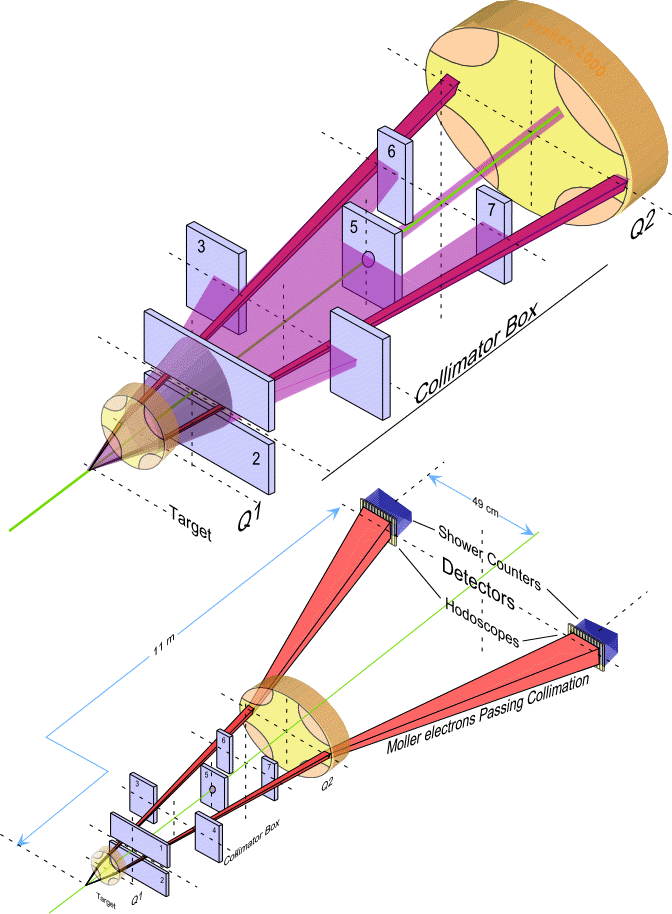}
    \caption{\label{mollerfig} Schematic of Hall C M\"{o}ller Polarimeter}
  \end{center}
\end{figure}

Figure \ref{mollerfig} shows a schematic layout of the Hall C polarimeter, including the collimator system, the two quadrupoles and the detector package. More details on the Hall C polarimeter and the technique of M\"{o}ller polarimetry can be found in \cite{MollerNIM}. In this experiment, since both of the polarization components to be measured are proportional to the beam polarization, one need not know the beam polarization to extract the form factor ratio $G_E^p/G_M^p$. However, the polarization was periodically measured to insure that it was high enough to achieve a reasonable statistical precision on the final result, and, as will be discussed in \ref{FPPsection}, knowledge of the beam polarization allows a calibration of the analyzing power of the proton polarimeter. Knowledge of the beam polarization is also required to extract the transverse and longitudinal transferred polarization components separately.
\begin{table}[p]
  \begin{center}
    \begin{tabular}{|c|c|c|c|c|}
      \hline Date & $Q^2$, GeV$^2$ & $E_{beam}$, GeV & Wien Angle, $^\circ$ & Beam polarization, \% \\ \hline
      11/9/2007  & 5.2 & 4.052 & 37.5 & -79.68  $\pm$ 0.35 \\ \hline
      11/16/2007 & 5.2 & 4.052 & 37.5 & 79.06  $\pm$ 0.31 \\ \hline
      11/28/2007 & 2.5 & 1.873 & 12.4 & -85.14 $\pm$ 0.33 \\ \hline
      11/30/2007 & 2.5 & 1.873 & 12.4 & -85.98 $\pm$ 0.34 \\ \hline
      12/3/2007  & 2.5 & 1.873 & 12.4 & -86.38 $\pm$ 0.30 \\ \hline
      12/5/2007  & 2.5 & 1.873 & 12.4 & -85.65 $\pm$ 0.32 \\ \hline
      12/11/2007 & 2.5 & 2.847 & 86.0 & -83.22 $\pm$ 0.27 \\ \hline
      12/12/2007 & 2.5 & 2.847 & 86.0 & -82.04 $\pm$ 0.30 \\ \hline
      12/12/2007 & 2.5 & 2.847 & 86.0 & -84.12 $\pm$ 0.26 \\ \hline
      12/14/2007 & 2.5 & 2.847 & 86.0 & -84.09 $\pm$ 0.33 \\ \hline
      12/16/2007 & 2.5 & 2.847 & 86.0 & -84.03 $\pm$ 0.31 \\ \hline
      12/16/2007 & 2.5 & 2.847 & 86.0 & -84.82 $\pm$ 0.26 \\ \hline
      12/18/2007 & 2.5 & 3.548 &  6.2 & 85.83 $\pm$ 0.29 \\ \hline
      1/6/2008 & 2.5  & 3.680 &  14.8 & -85.41 $\pm$ 0.29 \\ \hline
      1/9/2008 & 2.5  & 3.680 &  14.8 & -85.71 $\pm$ 0.31 \\ \hline
      1/11/2008 & 2.5 & 3.680 &  14.8 & -84.20 $\pm$ 0.26 \\ \hline
      1/18/2008 & 2.5 & 1.868 &  64.0 & -86.16 $\pm$ 0.28 \\ \hline
      1/21/2008 & 2.5 & 1.868 &  64.0 & -85.69 $\pm$ 0.26 \\ \hline
      1/23/2008 & 2.5 & 1.868 &  64.0 & -84.85 $\pm$ 0.27 \\ \hline
      1/23/2008 & 2.5 & 1.868 &  64.0 & -85.92 $\pm$ 0.32 \\ \hline
      4/6/2008  & 8.5 & 5.717 & -40.2 & 80.10 $\pm$ 0.32 \\ \hline
      4/14/2008 & 8.5 & 5.712 & -40.2 & 78.64 $\pm$ 0.32 \\ \hline
      4/21/2008 & 8.5 & 5.712 & -40.2 & 81.26 $\pm$ 0.43 \\ \hline
      4/28/2008 & 8.5 & 5.712 & -40.2 & 80.35 $\pm$ 0.35 \\ \hline
      5/5/2008 & 8.5  & 5.712 & -57.2 & 85.04 $\pm$ 0.51 \\ \hline
      5/5/2008 & 8.5  & 5.712 & -47.0 & 80.87 $\pm$ 0.33 \\ \hline 
      5/12/2008 & 8.5 & 5.712 & -13.5 & 79.27 $\pm$ 0.42 \\ \hline
      5/12/2008 & 8.5 & 5.712 & -23.7 & 82.13 $\pm$ 0.52 \\ \hline
      5/12/2008 & 8.5 & 5.712 & -3.7  & 71.85 $\pm$ 0.50 \\ \hline
      5/12/2008 & 8.5 & 5.712 & -33.7 & 83.44 $\pm$ 0.52 \\ \hline
      5/19/2008 & 8.5 & 5.712 & -13.5 & 80.14 $\pm$ 0.45 \\ \hline
      5/27/2008 & 8.5 & 5.712 & -13.5 & 80.37 $\pm$ 0.80 \\ \hline
      6/8/2008 & 6.8 & 5.712 & -13.5 & 77.29 $\pm$ 0.69 \\ \hline
    \end{tabular}
    \caption{\label{MollerTable} M\"{o}ller measurements of the beam polarization during experiments E04-108 and E04-019.}
  \end{center}
\end{table}

Table \ref{MollerTable} shows the results of all beam polarization measurements performed during the E04-108 and E04-019 experiments. The actual polarization of the beam at the injector is not the same as the polarization delivered to the experimental halls, because of the precession of the electron spin in the recirculating arcs of the CEBAF accelerator and the Hall C arc owing to its small anomalous magnetic moment. This precession is a complicated function of the number of passes through the accelerator, the linac energy, and the Wien filter setting. The Wien filter, which consists of crossed DC electric and magnetic fields with adjustable strength and orientation, is used to rotate the electron spin at the injector to an initial orientation which is optimized for the delivery of maximum longitudinal polarization to one or more experimental halls after precession in the magnetic beam transport elements. During the E04-019 experiment, the beam polarization was measured more frequently than during the production of the high-$Q^2$ data of the E04-108 experiment. Since it is a high-precision experiment looking for very small effects of TPEX, frequent monitoring of the beam polarization was needed to insure that it was optimal, and to measure the separated longitudinal and transverse polarization components, which are needed to extract the two-photon exchange amplitude discussed in chapter 2. Although the nominal beam energy was the same throughout the high-$Q^2$ data taking of April-June 2008, there were a number of subtle changes in the accelerator configuration dictated by the demands of running Halls A and C with polarized beam at 1 and 5 passes, respectively. There were several changes of the Wien angle setting and even a slight change in the energy balance between the two linacs, resulting in slight changes in the polarization received by the experiment. The typical polarization of about 80\% during this period, which is lower than the maximum of about 85-86\%, was a compromise that allowed Hall A to run a polarized beam experiment simultaneously.
\section{Experimental Targets}
\paragraph{}
The target system used for this experiment consists of several different solid targets and a three-loop cryogenic target system for liquid hydrogen. The solid targets consist of thin foils of Carbon and/or Aluminum used for spectrometer optics calibrations and to measure the contribution of the walls of the cryotarget cell to the experiment background. All of the targets are mounted on a ladder for motion and positioning in the path of the electron beam. The target ladder is enclosed in an evacuated scattering chamber. The scattering chamber is essentially an aluminum cylinder with 2'' thick walls and an internal diameter of 41''. On each side of the scattering chamber are thin aluminum windows through which scattered particles can escape. Table \ref{scatchamberwindowdimensions} shows the dimensions of the windows on each side of the beam. 
\begin{table}[h]
  \begin{center}
    \begin{tabular}{|c|c|c|c|c|}
      \hline
      & $\theta_{min},^\circ$ & $\theta_{max},^\circ$ & height, in. & nominal thickness, in. \\ \hline
      HMS (proton) & 5.5 & 103 & 8 & 0.016 \\ \hline 
      SOS/BigCal (electron) & 24 & 101 & 15 & 0.020 \\ \hline 
    \end{tabular}
  \end{center}
  \caption{\label{scatchamberwindowdimensions} Dimensions of scattering chamber exit windows.}
\end{table}
In addition to the exit windows, the scattering chamber also has beam entrance and exit ports, electrical feedthroughs for instrumentation, ports for vacuum pumps, and plumbing for the cryogenic target system which circulates liquid hydrogen. The beam entrance port is connected directly to the evacuated accelerator beamline so that the beam does not interact with any material before the target.
\subsection{Cryotargets}
\paragraph{}
The cryotarget system consists of a number of target cells connected to a recirculating hydrogen loop. In this experiment, hydrogen was the only cryogen used. Hydrogen liquid is forced to circulate through the loop by a fan operating at 60 Hz. The hydrogen is cooled to the desired operating temperature of 19 K by a heat exchanger with Helium coolant supplied at a temperature of roughly 14 K from the End Station Refrigerator (ESR). The coolant is returned at a temperature of approximately 19 K. The flow of Helium in the heat exchanger is controlled by Joule-Thomson valves. The cool hydrogen then flows through the target cell at 19 K, where it is heated by its interaction with the electron beam. Thermometers at the exit of the heat exchanger, and at the entrance and exit of the target cell, monitor the temperature of the circulating hydrogen continuously, and a variable high-power heater is used to keep the temperature of the hydrogen constant at 19 K through a PID feedback cicuit. The high power heater, which is responsible for correcting macroscopic differences in the heat load, is supplemented by low-power heaters which correct microscopic temperature fluctuations due to small changes in beam current to keep the temperature constant to within $\pm .01$ K. When the beam is on, the heater power is reduced to compensate for the increased heat load of the beam, and is increased when the beam is off to replace the lost beam heating. The maximum power of the heater is approximately 800 W, while the maximum beam heat load in this experiment was 500-600 W. Under normal operating conditions, the JT valve controlling the Helium coolant flow is adjusted so that there is approximately 50-100 W of ``reserve'' heater power when the beam is on, to prevent uncontrolled warming of the target in a situation in which the beam heat load is greater than the cooling power of the heat exchanger. When beam is off for a long time, however, the coolant flow is reduced to conserve cooling and heater power.

The power dissipated in the cryotarget by the electron beam is given by
\begin{equation}
  P_{beam} = I_{beam} \rho L \frac{dE}{dx}
\end{equation} 
where $I_{beam}$ is the beam current in $\mu A$, $\rho$ is the density of liquid hydrogen at 19 K in $g\ cm^{-3}$, $L$ is the target thickness in cm, and $dE/dx$ is the beam collisional energy loss per unit target thickness in $MeV cm^2/g$. At the highest electron beam energy used in this experiment, 5.714 GeV, the ionization energy loss in hydrogen is approximately 4.9 MeV cm$^2$/g. At the operating temperature of 19 K, the density of liquid hydrogen is 0.0723 g/cm$^3$. This leads to a beam heat load of approximately 567 W which must be compensated by the high power heater when the beam is off. 
\paragraph{Fast Raster}
The typical spot size of the electron beam in Hall C is well below 100 $\mu m$. Such a tightly focused beam, at intensities typical of these experiments, can cause intense local heating of the hydrogen liquid which can cause undesirable boiling and fluctuations in target density. Additionally, at high enough intensity, such focused beams can easily melt the thin aluminum walls of the cryocell. To protect against these deleterious effects, the beam spot is enlarged by a raster magnet system to a transverse size of typically 2x2 mm$^2$. These magnets are located approximately 25 meters upstream of the target. The fast raster system magnets, one for horizontal deflection and one for vertical deflection, can be driven by a sinusoidal or triangular waveform. For this experiment, the triangular waveform was used, as it leads to a more nearly uniform transverse beam profile. It also has the advantage of allowing a more accurate determination of the instantaneous beam position from the raster signals which are read out by the data acquisition system for each event. The raster magnets are driven at different frequencies in each direction to prevent the signals from forming a closed Lissajous curve. More information on the Hall C fast raster system can be found in \cite{YanHallCraster}. Figure \ref{rasterfig} shows an example of the fast raster profile on target. The raster current signals are sent to an ADC and read out by CODA. The beam deflections are calculated from the ADC signals as follows:
\begin{eqnarray}
  x_{rast} &=& \frac{\alpha_x}{p_{beam}} (\mbox{ADC}_x - \mbox{PED}_x) \nonumber \\
  y_{rast} &=& \frac{\alpha_y}{p_{beam}} (\mbox{ADC}_y - \mbox{PED}_y) \\
\end{eqnarray}
where the $\alpha$'s are calibration constants determined from harp scan data and the pedestals are determined from the data. The ADC used to digitize the raster signals is a LeCroy 1881M charge-integrating ADC module. As far as the gate width is concerned, the raster signal is a constant DC level\footnote{The raster frequency is 20 kHz, while the typical ADC gate width used during readout of an event in this experiment is no more than a few hundred nanoseconds, so that during the integration time the raster magnet undergoes less than 1\% of an oscillation.}. So the CODA readout is essentially a snapshot in time of the beam deflection when the data acquisition was triggered. The cable delay for the raster signals to reach the ADC modules is similar to the delays of other signals coming from the hall, so the raster signals are in fact reasonably well synchronized with the data. In any case, the oscillations of the raster magnets are slow enough compared to the time scale of event formation and triggering of the data acquisition system that the beam deflection calculated from the raster signals can be regarded, to a very good approximation, as equal to the instantaneous beam deflection when the scattering event responsible for the trigger occured. Because the ADC is integrating what amounts to a constant DC offset, the ``pedestal'' for the raster signal is more properly regarded as the average integral of the raster signal over the gate width. Knowledge of the beam position on target is important, particularly in the vertical direction, since the vertical beam position affects the reconstruction of the scattered proton's momentum and the out-of-plane angle of its trajectory. This effect will be discussed in more depth later.
\begin{figure}[h]
  \begin{center}
    \includegraphics[width=0.95\textwidth]{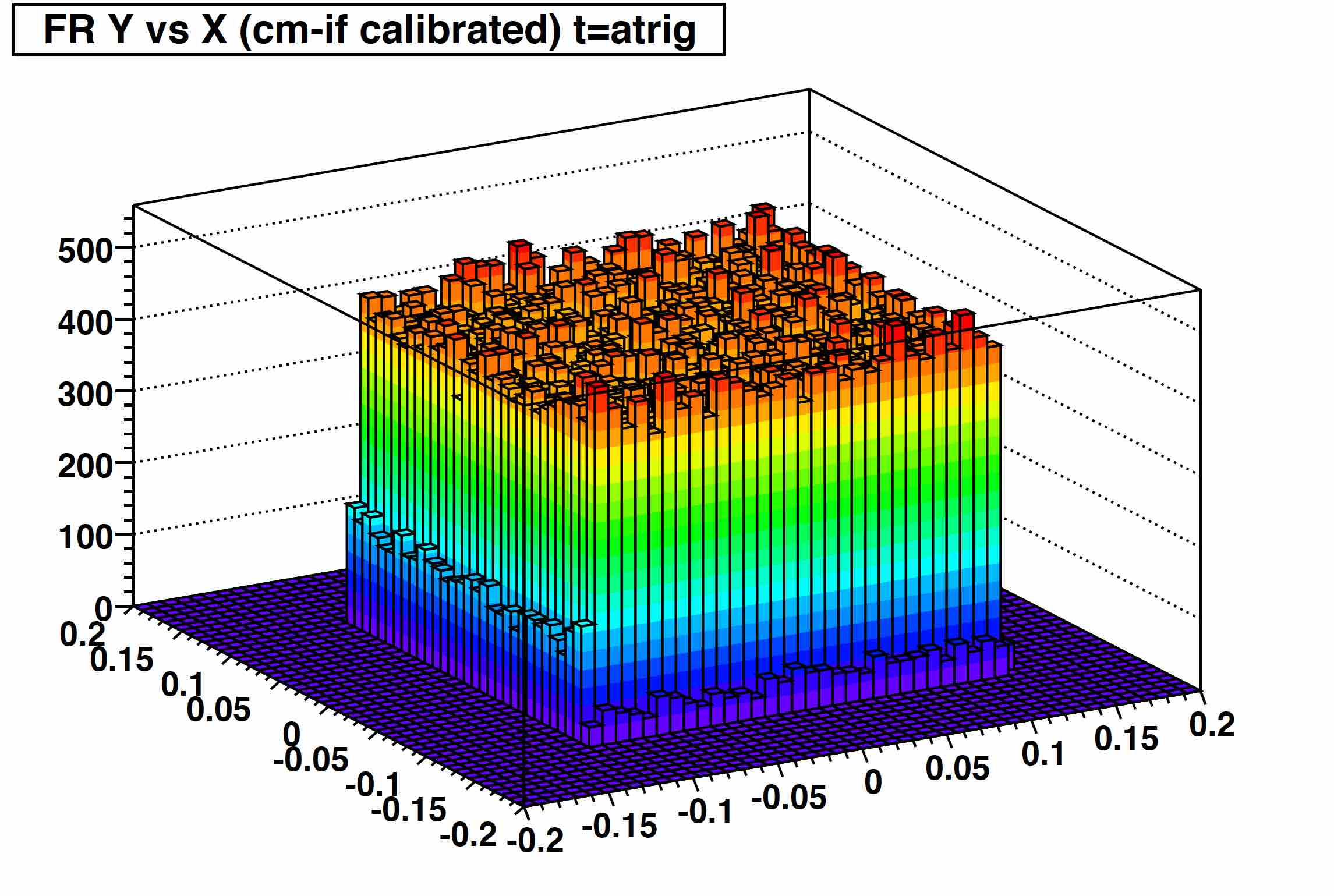}
    \caption{\label{rasterfig} Beam raster pattern used on the cryotarget. The raster position is calculated from the raster current ADC signals read out by CODA.}
  \end{center}
\end{figure} 

Several different configurations of the target cell were used during the experiments E04-108 and E04-019. For the measurements at $Q^2 = 5.2\ GeV^2$, the target used was a 15 cm LH$_2$ cryocell. For all of the other production kinematics of both experiments, a 20 cm cryocell was used. The 20 cm cell was offset by 3.84 cm downstream from the origin along the beamline in order to allow electrons scattered by up to 120 degrees to clear the scattering chamber exit window. Additionally, a 4 cm cell was made available for various studies. Since the goal of these experiments was to measure an asymmetry which is independent of the absolute luminosity, no detailed target thickness or luminosity studies were necessary, so none were performed\footnote{Several quick checks of luminosity were performed by comparing yields from a 4 cm LH$_2$ target and the main 20 cm loop at several values of beam current and raster size. However, these checks were not performed for all kinematics, and only a rough BCM calibration was performed to check the validity of the BCM data.}. However, this meant that no absolute cross sections could be extracted as a byproduct of the data of these experiments without assigning large uncertainties to the target density and thickness. 
\begin{table}[h]
  \begin{center}
    \begin{tabular}{|c|c|c|c|}
      \hline Target position & Entrance window & Exit window & Wall \\
      & thickness (mm) & thickness (mm) & thickness (mm) \\ \hline
       4 cm LH$_2$ loop & $0.127 \pm .005$ & $0.151 \pm .006$ & $0.118 \pm .008$ \\ \hline
      15 cm LH$_2$ loop & $0.115 \pm .001$ & $0.126 \pm .009$ & $0.126 \pm .009$ \\ \hline
      20 cm LH$_2$ loop & $0.122 \pm .005$ & $0.163 \pm .012$ & $0.157 \pm .017$ \\ \hline
    \end{tabular}
    \caption{\label{cryothick} Thicknesses of cryotargets used in E04-108 and E04-019, in cm.}
  \end{center}
\end{table}

Table \ref{cryothick} shows the measured thicknesses of the cell walls and the entrance and exit windows of the three different cryogenic loops used in experiments E04-019 and E04-108. All cryocells were made of Al 7075-T6 aluminum alloy\cite{MeekHallCconfig}. 
\subsection{Dummy/Optics Targets}
\paragraph{} In addition to the production hydrogen targets, there were a number of solid targets available for spectrometer optics calibrations. A single foil of BeO crystal, which exhibits luminescence when irradiated by the electron beam, was used as a beam viewer to verify the initial beam position on target. Single-foil Carbon targets of two different thicknesses and a single-foil of Copper were available for various detector checkouts and systematic studies. Several multi-foil targets were also provided. The so-called dummy targets consisted of pairs of Aluminum foils aligned with the entrance and exit windows of the cryocells. These targets were used both for optics calibrations and to measure the contribution of the cell walls to the experiment background. The 20 cm dummy target has foils at $z = 3.84 \pm 10$ cm, while the 15 cm dummy target has foils at $z = \pm 7.5$ cm. The remaining multi-foil targets were used for spectrometer optics calibrations:
\begin{itemize}
  \item A three-foil Aluminum target with foils located at $z = 0,\pm 7.5$ cm. 
  \item A two-foil Carbon target with foils located at $z = \pm 2$ cm.
  \item A two-foil Aluminum target with foils located at $z = \pm 3.8$ cm.
\end{itemize}
Detailed information on the thickness, material and chemical purity of the solid targets can be found in \cite{MeekHallCconfig}. 
\section{The High Momentum Spectrometer}
\paragraph{} The primary apparatus for these experiments was a superconducting magnetic spectrometer called the High Momentum Spectrometer (hereafter referred to as HMS). Its magnetic system consists of three quadrupole magnets which focus charged particle trajectories and a dipole magnet to momentum-analyze and deflect them into the detector hut. A schematic view of the HMS is shown in figure \ref{HMS_magnet_schematic}.
\begin{figure}[h] 
  \begin{center}
    \includegraphics[width=0.95\textwidth]{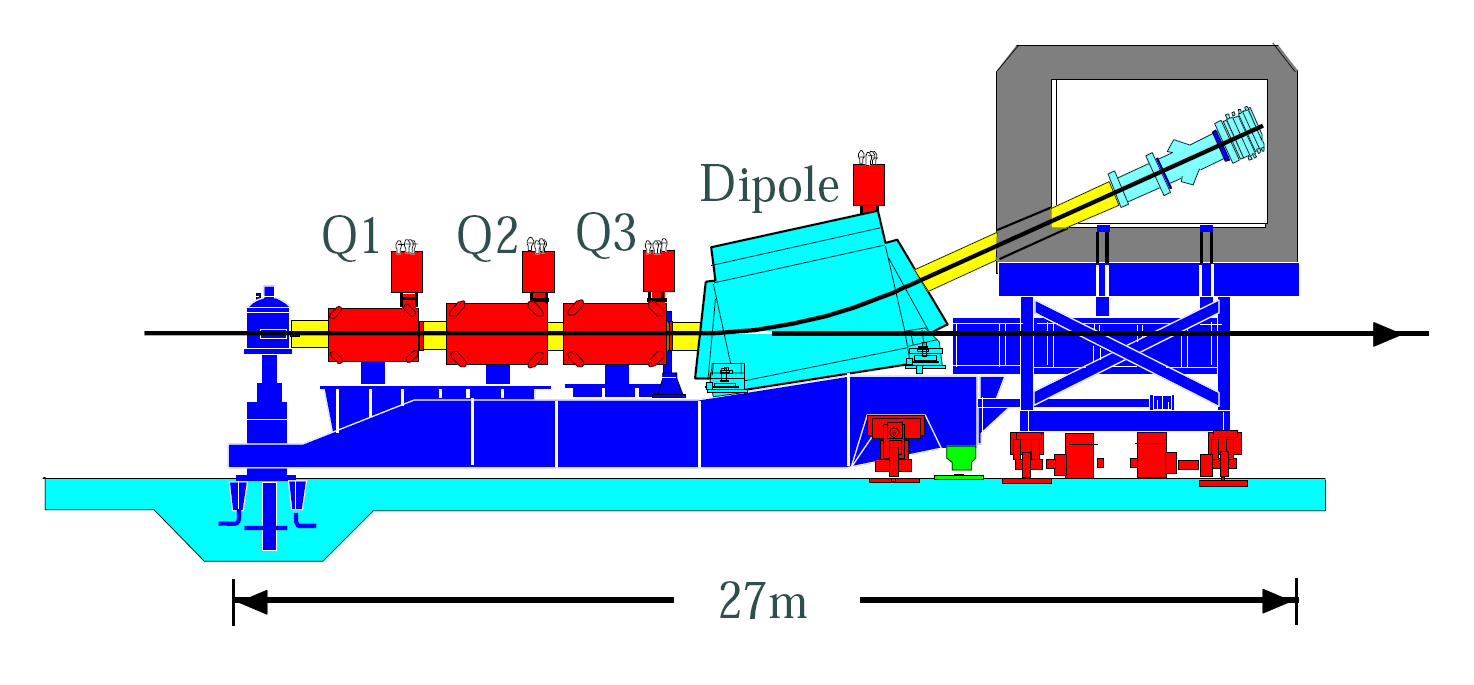}
    \caption{\label{HMS_magnet_schematic} Schematic of the HMS spectrometer.}
  \end{center}
\end{figure}
The HMS detector stack is supported on a common carriage with the magnets so that it remains stationary with respect to the optical axis. The concrete shield hut is supported on a separate carriage. The entire structure rests on concentric rails and can be rotated around the rigid central pivot of the experimental hall. In this experiment, the smallest central angle of the spectrometer was 11.6 degrees, while the largest was 40.5 degrees. The HMS is located on the right side of the beamline as viewed from upstream of the target. The superconducting coils of the magnets are cooled by liquid Helium at 4 K supplied by the ESR. Under normal conditions, including this experiment, the HMS magnets are operated in a point-to-point tune, in which the quadrupoles Q1 and Q3 are focusing in the dispersive direction while Q2 focuses in the non-dispersive direction, resulting in point-to-point focusing in both directions when the dipole is included. In this configuration, the HMS has a large acceptance in momentum, in-plane and out-of-plane angles, and extended target. The magnetic field in the dipole is regulated by an NMR probe, while the quadrupole fields are regulated by current, but also have their fields monitored by Hall probes. The dipole field is stable at the $10^{-5}$ level, while the typical quadrupole current stability is $10^{-4}$. In order to minimize particle losses and resolution degradation due to multiple scattering, and to provide thermal insulation, the entire magnetic length of the HMS is evacuated, from well before the entrance of Q1 and the acceptance-defining collimators to the detector hut just upstream of the first detectors and the location of the optical focal plane of the spectrometer. The air gap between the scattering chamber exit window and the HMS entrance window is only $\approx 15$ cm.

\subsection{Magnets}
\subsubsection{Quadrupoles}
\paragraph{}
The three quadrupole magnets are named Q1, Q2, and Q3 for the order in which scattered particles from the target pass through them. Q2 and Q3 are identical, while Q1 has somewhat smaller dimensions. All three magnets are of cold-Iron superconducting design. Some of the relevant properties of the quadrupole magnets are listed in table \ref{QuadSpecs}.
\begin{table}[h]
  \begin{center}
    \begin{tabular}{|c|c|c|}
      \hline & Q1 & Q2/Q3 \\ \hline
      Max. gradient, G cm$^{-1}$ & 605 & 445 \\ \hline
      ``Good Field'' radius, cm & 22 & 30 \\ \hline
      Max. pole tip field, T & 1.5 & 1.56 \\ \hline
      Radius to pole, cm & 25 & 35 \\ \hline
      Effective length, cm & 189 & 210 \\ \hline
    \end{tabular}
    \caption{\label{QuadSpecs} Basic properties of HMS quadrupole magnets.}
  \end{center}
\end{table}
When increasing the central momentum setting of the HMS, the quadrupoles were typically 'cycled' by ramping to roughly 200 A above their set current and then ramping down to the setpoint. In this way, the set currents were always approached from the same side of the hysteresis loop of the magnets, enhancing reproducibility of the resulting magnetic fields.
\subsubsection{Dipole}
\paragraph{} The HMS dipole is a superconducting magnet with a 25 degree vertical bend for the central ray. Its superconducting coils have a flat racetrack design with no negative curvature. Its poles are flat, giving rise to a highly uniform magnetic field. The width of the gap between the poles is 42 cm. The faces of the poles are inclined at $\pm6 ^\circ$ relative to the normal to the central ray. This inclination gives rise to an ``edge focusing'' effect. Particles of a given momentum entering the dipole at a higher vertical position see a smaller $\int B dl$ and thus undergo smaller deflection than particles moving along the central ray. Similarly, particles entering the dipole at a lower vertical position see a larger $\int B dl$ and undergo a larger deflection than the central ray. The dipole's bend radius is 12.06 m, giving an effective length of 5.26 m for the 25 degree central bend. The momentum dispersion of the HMS is 3.71 cm/\% \cite{HMSresolution}, meaning that a 1\% deviation from the central momentum results in a physical displacement of 3.71 cm from the central ray at the focal plane. This relatively large dispersion gives the HMS excellent momentum resolution. 

Table \ref{MagnetSettings} shows the quadrupole current and dipole field settings for the four different HMS central momentum values used for the production kinematics of this experiment. The setpoints for a given central momentum are proportional to the momentum and are in a constant ratio Q1:Q2:Q3:D.
\begin{table}[h]
  \begin{center}
    \begin{tabular}{|c|c|c|c|c|}
      \hline HMS central momentum, GeV/c & Q1, A & Q2, A & Q3, A & Dipole, T \\ \hline
      2.0676 & 269.3 & 214.1 & 104.3 & 0.568297 \\ \hline
      3.5887 & 467.5 & 371.7 & 180.8 & 0.986386 \\ \hline
      4.4644 & 581.7 & 462.5 & 224.8 & 1.22708 \\ \hline
      5.4070 & 704.7 & 562.8 & 272.2 & 1.48616 \\ \hline
    \end{tabular}
    \caption{\label{MagnetSettings} Quadrupole current and dipole field settings for the standard HMS tune at the different central momentum settings of this experiment.}
  \end{center}
\end{table}
\subsection{Collimators}
\paragraph{}
The HMS is equipped with a system of collimators used for two purposes. First, octagonal collimators of two different sizes are used to define the solid angle acceptance of the HMS. For this experiment, the larger of the two solid-angle defining collimators was used. The so-called pion collimator has an opening that measures 9.150 cm in the horizontal direction and 23.292 cm in the vertical direction, as shown in figure \ref{HMSoctagon}. This collimator is made of densimet (90\% machinable W + 10\% CuNi) with a density of 17 g/cm$^3$, and is 6.35 cm thick. It is flared to match the angular acceptance of the HMS, i.e., its dimensions are 4\% larger at its exit than at its entrance. Its entrance is located at a distance of 166.00 cm from the origin. At this distance, the octagonal collimator subtends a solid angle of 6.74 msr, and the angular acceptance defined by the slit is roughly 70 mrad in the out-of-plane angle and 28 mrad in the in-plane angle. It is worth remarking that for an extended target, a slightly larger range of in-plane angles is accepted since particles coming from anywhere along the target length can pass through the collimator. The collimator is designed to prevent particle losses in the magnetic elements of the spectrometer over a large momentum bite\footnote{While the large collimator is designed to prevent particle losses over a large momentum bite for a point target, and the small collimator is designed to prevent particle losses for an extended target, the goal was to maximize elastic ep statistics, rather than to have a precise understanding of the acceptance. So even with the extended 20 cm LH$_2$ target, the larger collimator was chosen, which meant that there could be increased particle losses at the extremes of the acceptance. However, for most of the kinematics of E04-108, the elastic ep scattering events were concentrated in a fairly narrow momentum bite within $\approx \pm 5\%$ of central momentum.}.
\begin{figure}[htbp]
  \begin{center}
    \includegraphics[angle=90,width=.6\textwidth]{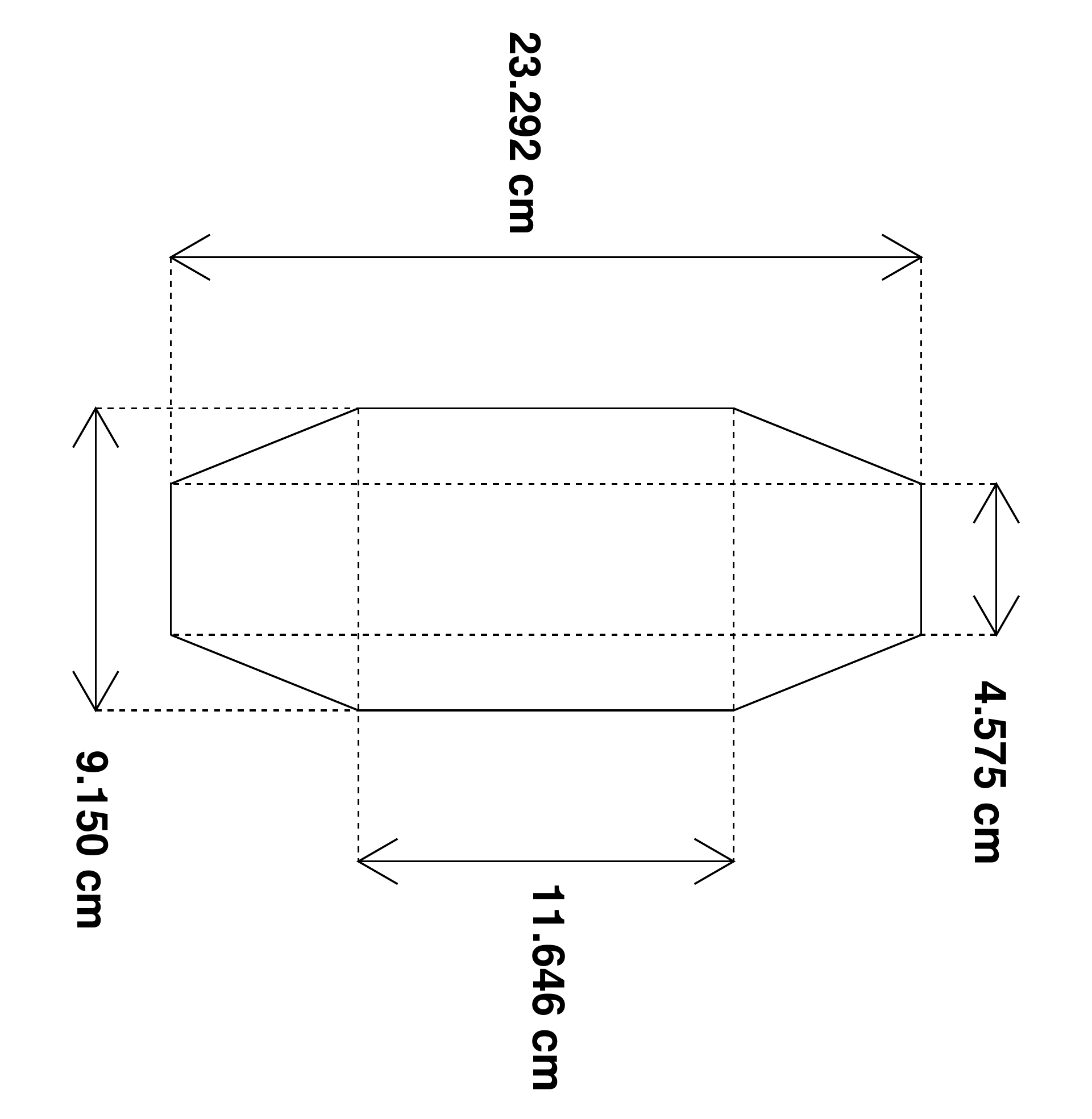}
    \caption{\label{HMSoctagon} HMS collimator dimensions at the collimator entrance.}
  \end{center}
\end{figure}

The second collimator used for this experiment was the sieve slit, which consists of the same material as the other collimators, but is only half as thick (3.175 cm). The sieve slit, as its name suggests, contains a large number of small holes and is used to study the optical properties of the HMS. To use the sieve slit collimator to calibrate the spectrometer optics coefficients, the general procedure is to set the polarity of the HMS magnets to detect electrons, and to measure scattered electrons from a series of thin solid foil targets located at a known position along the beamline. The in-plane and out-of-plane angles are geometrically determined by the ray from the thin target foil to the small sieve slit hole. The holes of the sieve slit are 0.508 cm in diameter, except for the central sieve slit hole, which has a smaller diameter of 0.254 cm and is used to determine the HMS angular resolution. Unlike the aperture of the pion collimator, the sieve holes are not flared. At a distance of 166 cm from the target, the hole radius corresponds to $\pm$1.5 mrad angular acceptance per hole in each direction. The hole spacing is 2.54 cm in the vertical direction and 1.524 cm in the horizontal direction, which corresponds to an angular spacing of 15.3 mrad in the vertical direction and 9.18 mrad in the horizontal direction. Two of the sieve holes are blocked in order to verify the up-down and left-right direction of the reconstructed angles. The outermost rows of sieve holes are located at $\pm$10.16 cm, corresponding to $\pm$61.2 mrad in the dispersive direction, while the outermost columns of sieve holes are located at $\pm$6.10 cm, corresponding to $\pm$ 36.7 mrad in the non-dispersive direction. There are nine rows(columns) of holes in the dispersive(non-dispersive) direction. Note that there are no sieve holes at the vertical extremes of the HMS acceptance ($62 < \left|\theta\right| < 70$ mrad). This means that the optical reconstruction parameters obtained from fitting sieve slit data give relatively poorer resolution when extrapolated into the extreme regions of the acceptance not covered by the sieve slit. More details of the HMS optical properties and reconstruction will be discussed later. 
\subsection{Performance Characteristics}
\paragraph{}
The HMS was designed to have a maximum central momentum of 7.4 GeV/c with moderate momentum, solid-angle and extended target acceptance, and moderate resolution in momentum, angles and vertex position. Table \ref{HMSperformance} summarizes the performance characteristics of the HMS in its standard configuration\footnote{Because these experiments required a special detector and trigger configuration which involved placing a scintillator before the HMS tracking detectors, the actual resolution of the HMS for these experiments was somewhat worse than the values given in table \ref{HMSperformance}, particularly for the reconstructed angles, owing mainly to the multiple scattering introduced by this scintillator plane. The extra trigger plane and its implications for the HMS angular resolution and the analysis of the data will be discussed in depth later.}.
\begin{table}[h]
  \begin{center}
    \begin{tabular}{|l|c|}
      \hline Max. Central Momentum, GeV/c & 7.4 \\ \hline
      Min. Central Momentum, GeV/c & 0.5 \\ \hline
      Momentum Bite, $(p_{max} - p_{min})/p_0$ & 18\% \\ \hline
      Momentum Resolution $\delta p / p$ & $10^{-3}$ \\ \hline
      Solid angle acceptance, msr & 6.74 \\ \hline
      In-plane angle resolution, mrad & 0.8 \\ \hline 
      Out-of-plane angle resolution, mrad & 1.0 \\ \hline
      Useful target length, cm & 10 \\ \hline
      Vertex resolution, mm & 2 \\ \hline
    \end{tabular}
    \caption{\label{HMSperformance} Acceptance and resolution of the HMS in its standard configuration.}
  \end{center}
\end{table}
\subsection{Detector Package}
\paragraph{}
The HMS is equipped with a versatile set of detectors to detect and track charged particles scattered from the target and reconstruct their momenta and trajectories. In the standard configuration, the HMS is equipped with a pair of gas drift chambers for tracking, four planes of scintillator hodoscopes for triggering and timing, gas and aerogel Cerenkov detectors for particle identification, and a lead-glass calorimeter which provides further energy and particle identity information. In these experiments, some parts of the standard detector package had to be removed in order to install the proton polarimeter. Figure \ref{HMSdetectorstack} shows the basic layout of the HMS detector package. 
\begin{figure}[htbp]
  \begin{center}
    \includegraphics[width=.95\textwidth]{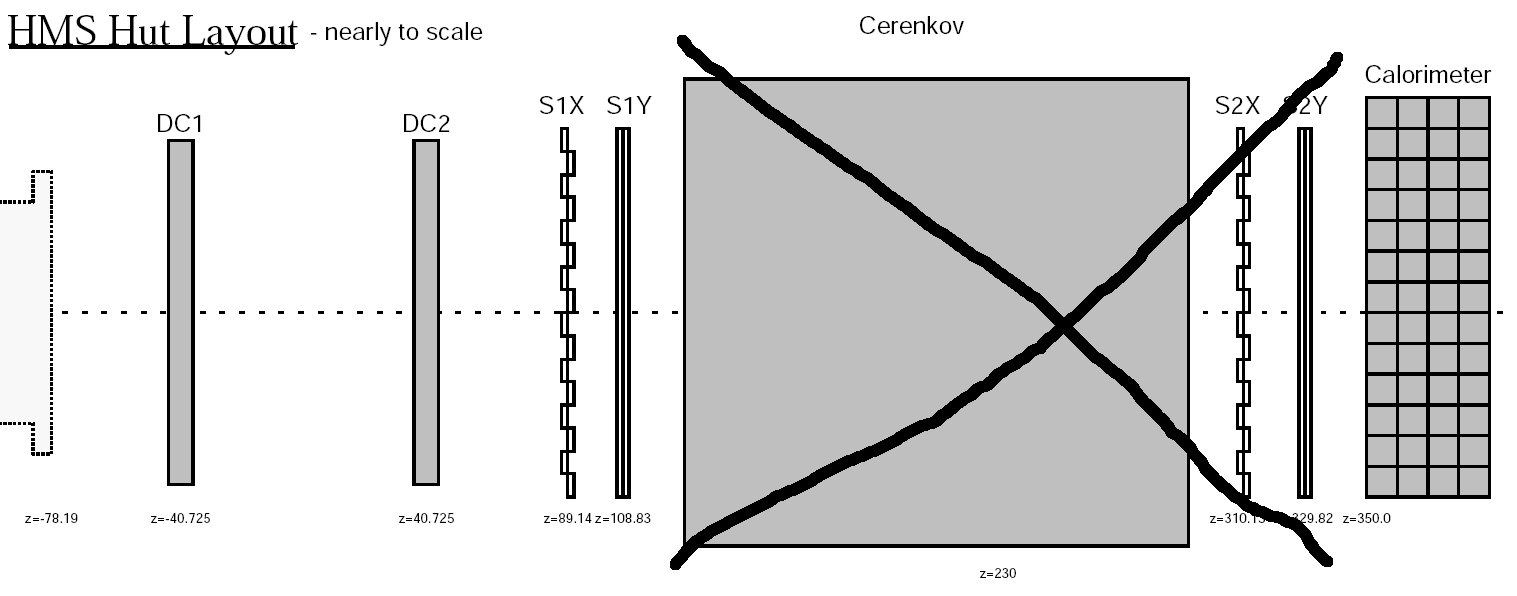}
    \caption{\label{HMSdetectorstack} HMS standard detector stack. DC1 and DC2 are drift chambers. S1X, S1Y, S2X, and S2Y are scintillator hodoscopes. The gas Cerenkov, S2X and S2Y were removed to install the Focal Plane Polarimeter. The exit window of the HMS vacuum system is also shown (far left). The z-coordinate in this drawing is relative to the midpoint between the two drift chambers. Particles coming from the target move from left to right in this picture.}
  \end{center}
\end{figure}
Note that all the detectors in figure \ref{HMSdetectorstack} are oriented perpendicular to the $z$ axis, which coincides with the central ray. 

At this point it is appropriate to define the transport coordinate system used in the analysis. This coordinate system is right-handed, orthogonal and Cartesian. At the target, before entering the spectrometer, the $z$ axis is horizontal and points along the optical axis. The $x$ axis points vertically downward, and the $y$ axis is perpendicular to the $z$ axis in the horizontal plane such that the $(\hat{x},\hat{y},\hat{z})$ axes form a right-handed coordinate system. This coordinate system is fixed relative to the HMS and it rotates with the spectrometer relative to the Hall C coordinate system, and other fixed coordinate systems which will be discussed later. In the HMS detector hut, the $z$ axis is still parallel to the HMS central ray, which is inclined relative to the horizontal by $25^\circ$, corresponding to the vertical bend. The $x$ axis in the detector hut is parallel to the dispersive direction, with positive $x$ pointing in the direction of increasing momentum, i.e., downward. Like the $z$ axis, the $x$ axis is inclined by $25^\circ$ relative to the vertical. The $y$ axis points in the non-dispersive direction and is again oriented so that $(\hat{x},\hat{y},\hat{z})$ is a right-handed, orthogonal coordinate system. Although the $xz$ plane is rotated by 25$^\circ$ in going from the target to the focal plane, the $y$ axes of the target and hut transport coordinate systems coincide. 
\subsubsection{Drift Chambers}
\paragraph{} The drift chambers are the most important part of the HMS detector system. They are used to measure precisely the position and angles of charged particle trajectories at the focal plane. Drift chambers are among the most widely used tracking detectors in nuclear and particle physics. Their moderate cost and excellent resolution make them suitable for a wide range of applications. Drift chambers are gas ionization detectors which achieve high spatial resolution by operating with a combination of gas mixture and electric field which results in saturation of the drift velocity of electrons liberated by ionization. In the saturation region the electron drift velocity is roughly independent of the applied electric field, which in any case will be nearly uniform over the extent of the drift region in a well-designed chamber. By measuring the elapsed time between the initial ionization of the gas by a passing charged particle and the detection of the signal induced by the multiplicative avalanche in the strong $1/r$ electric field gradient in the vicinity of a thin sense wire, the distance between the particle track and the wire can be obtained with accuracies approaching 100 $\mu$m or less. More details on the basic physics underlying drift chamber design and operation can be found in \cite{Leo}, chapter 6. 

The HMS drift chambers consist of a series of planes of parallel wires in a gastight enclosure sealed by thin aluminized Mylar windows. In each wire plane, anode or signal wires made of 25 $\mu m$-diameter gold-plated tungsten alternate with cathode or field wires made of 150 $\mu m$-diameter gold-plated CuBe. The signal wires are maintained at ground potential, while the field wires are maintained at a high negative voltage. The spacing between signal wires is 1 cm. The detection planes of alternating signal and field wires are surrounded by additional field or ``guard'' wires which shape the electric field and define a ``drift cell'' around each sense wire. The basic drift cell in the HMS chambers is rectangular and measures 1.0 cm(horizontal)$\times$0.8 cm(vertical). Each signal wire is surrounded by eight field wires which form a symmetric rectangular cell. The high voltage applied to each field wire is proportional to the distance from that field wire to the signal wire. This high voltage configuration produces equipotential surfaces surrounding the signal wire that are very nearly circular over most of the drift cell, which insures that the drift time measured by a signal wire depends only on the distance of closest approach between the wire and the track that caused the ionization. Three different high voltage settings are required for the eight field wires surrounding a signal wire--one for the corners of the cell, one for the in-plane field wires, and one for the field wires directly above and below the signal wire. The high voltage for the HMS drift chambers, as well as for all the other detectors in the HMS hut, is provided by CAEN power supplies located inside the hut. The power supplies are remotely controlled and monitored by a VME CAEN-net interface through EPICS.

 Each drift chamber contains six parallel planes of wires arranged in the order X, Y, U, V, Y', X' as traversed by incoming particles. The X/X' wires are horizontal and perpendicular to the dispersive direction. The Y/Y' wires are vertical, parallel to the dispersive direction, and perpendicular to the X/X' wires. The U and V wires are oriented at $\pm 15^\circ$ relative to the X/X' wires as shown in figure \ref{HMSdriftchamber_schematic}. The wire planes within each chamber are spaced 1.8 cm apart\footnote{1.8 cm is the vertical distance between signal planes. There are also two additional planes of field wires between signal planes}. The two drift chambers are separated by about $81.45$ cm along the $z$ axis (see figure \ref{HMSdetectorstack}). The large separation in $z$ between the two chambers provides a precise determination of the angles of charged particle trajectories.
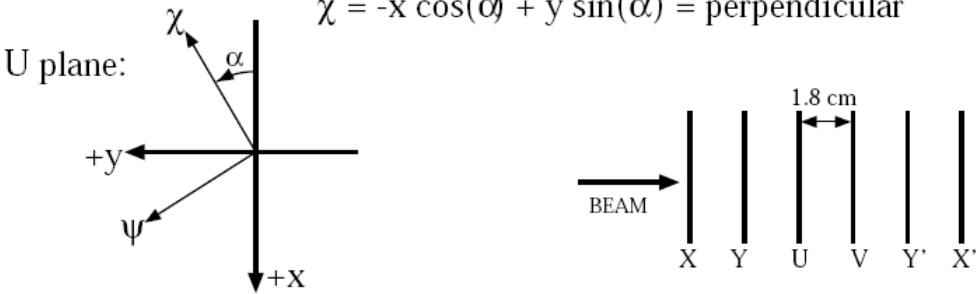
\begin{figure}
  \begin{center}
    \setlength{\unitlength}{1.0in}
    \begin{picture}(6.0,6.0)
      \put(0,0){\includegraphics[height=0.8\textheight]{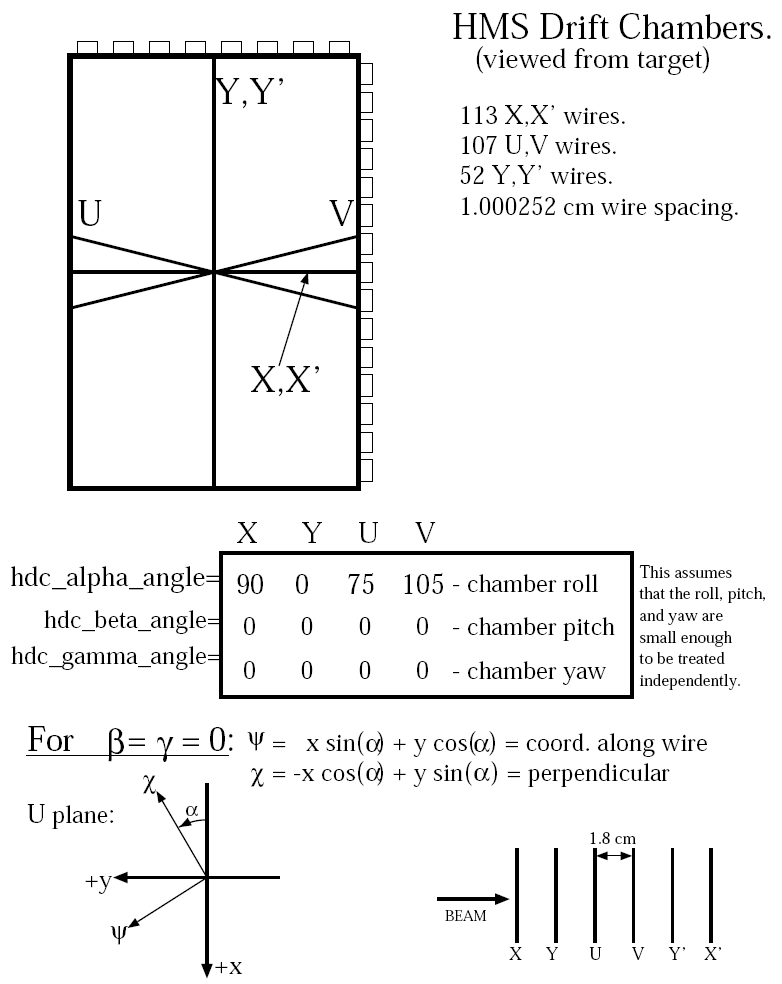}}
      \put(3.5,4.25){\vector(-3,2){0.79}}
      \put(3.55,4.17){Amplifier/Discriminator cards}
    \end{picture}
    \caption{\label{HMSdriftchamber_schematic} Schematic illustration of HMS drift chamber design.}
  \end{center}
\end{figure}
The fact that the U and V wires are much closer to the X/X' wires in orientation than to the Y/Y' wires means that the track position and slope are more precisely determined in the dispersive direction than in the non-dispersive direction. The active area of the drift chambers measures roughly 113 cm in the X (dispersive) direction and 80 cm in the Y (non-dispersive) direction.
 
The gas used by the drift chambers is a 50\%/50\% argon-ethane mixture (by weight), doped with ~1\% isopropyl alcohol by a 0$^\circ$ C bubbler which is used in the flow monitoring system. The role of the argon gas is to provide the primary ionization, while the ethane gas quenches the avalanche near the sense wire and enhances the drift properties of the mixture. The small amount of alcohol prolongs the useful life of the drift chambers, which are operated continuously at high particle rates, by inhibiting the formation of polymers during the recombination of disocciated organic molecules, which can accumulate on the anode and cathode wires and degrade the chambers' performance. During operation the chambers were continuously flushed at rates on the order of a few hundred cm$^3$/min at a pressure just slightly above atmospheric pressure.

The signals induced on the wires are read out at the front end by Nanometrics N-277-L and Lecroy 2735DC amplifier-discriminator (A/D) circuits and transmitted along ECL twisted-pair ribbon cables to multi-hit Lecroy 1877 fastbus TDC modules housed in a fastbus crate inside the HMS hut. Up to sixteen wires are connected to each card. The A/D cards are supplied with DC low voltage to set the discriminator threshold and to power the electronic circuits on the cards. For these experiments, the threshold voltage applied to the A/D cards of the HMS drift chambers was -5.5 V\footnote{The threshold applied to the cards, after accounting for the pre-amplifier electronics and the 1-2 V drop in going from the power supply in the counting house to the A/D cards in Hall C, corresponds to an actual hit threshold of several mV applied to the unamplified raw signal on the anode wire.}. The data acquisition electronics for the drift chambers are located inside the HMS hut for several reasons, mainly to avoid the introduction of lengthy cable delays and the accompanying attenuation and distortion of signals. The TDC modules have a timing resolution of 0.5 ns/count and a 16-bit count range, allowing for the measurement of time intervals up to 32 $\mu$s. 

For the combination of gas mixture and electric field used by the HMS drift chambers, the electron drift velocity is approximately 50 $\mu m/ns$. With a cell size of 1.0 cm, the drift length is in the range $0< d_{drift} < 0.5$ cm, so that the size of the drift timing window is approximately 100 ns. The TDC resolution of 0.5 ns corresponds to a coordinate resolution of 25 $\mu m$, however, this is not the dominant contribution to the spatial resolution. There are also intrinsic fluctuations in the arrival time of the drifting electrons, which of course do not move at a constant speed, but at an effective average velocity which is proportional to the electric field and the mean collision time $\tau$ of electrons drifting in the chamber gas. There is also an intrinsic uncertainty in the absolute positioning of the wires within the chambers and in the absolute positioning of the chambers themselves relative to the HMS optical axis. Finally, there is a contribution to the tracking resolution coming from the fact that while charged particles are assumed to pass through the chambers undeflected and their trajectories are assumed to be straight lines, in reality they undergo multiple scattering in the chamber gas and the mylar windows, and there is even a non-negligible probability of scattering from one or more wires in the course of crossing all twelve wire planes.

The TDCs for the HMS drift chambers are operated in common stop mode. For each wire, the signal from a hit wire starts the timer. When fast signals from the HMS hodoscopes generated by a charged particle passing through the detector stack trigger the main data acquisition system to read out an event, a stop signal is sent to the HMS drift chamber TDCs. After cable propagation and electronic delays, the stop signal arrives approximately 2 $\mu s$ after the initial signal formation, comfortably beyond the 100 ns drift timing window, such that the stop signal is always the last signal to arrive, and arrives at a fixed time relative to the fast hodoscope signals. The TDCs are programmed to read out all of the raw hits on all wires within a 4 $\mu s$ timing window, up to eight hits per wire per event. After reconstruction of the hodoscope signals in software, the time at which the particle passed through the detector stack can be determined with a resolution of about $0.3$ ns, providing for the determination of the drift time with sub-nanosecond resolution. More detailed information about the HMS drift chambers can be found in \cite{HMSdriftchamber}. The reconstruction of charged particle tracks in the HMS drift chambers will be discussed in detail later.
\subsubsection{Hodoscopes}
\label{hodosection}
\paragraph{}
The HMS in its standard configuration is equipped with four planes of scintillator bars which, by virtue of their fast response time, provide a fast trigger and precise timing information. Scintillators are materials containing molecules which emit optical photons when excited by passing energetic charged particles. Each bar is made of BC404 plastic scintillator, with UVT lucite light guides and Photonis XP2262 photomultiplier tubes attached at both ends. The light guide couples the flat rectangular shape of the end of the scintillator bar to the circular photocathode of the PMT. The light given off by the scintillation propagates via total internal reflection through the bar and the light guides to the photocathodes of each PMT, where electrons are released into the vacuum by the photoelectric effect and then amplified by the stages of the PMT. The scintillators and light guides are wrapped in one layer of Aluminum foil and one layer of black Tedlar for light-tightness. The foil wrapping reflects scintillation light emitted at angles exceeding the critical angle for total internal reflection, which would otherwise escape. In plastic scintillators such as those used in the HMS, the scintillation process happens on a very fast time scale with a rise time of less than 1 ns and a decay time on the order of 2-3 ns. It is this fast response which gives the HMS hodoscopes a timing resolution of about 300 ps after corrections for propagation time and pulse-height dependence (walk) of the arrival time of the discriminated signal, which is discriminated relative to a fixed threshold. The PMTs are 2 inches in diameter with 12 stages of dynodes, a gain of roughly $3\times 10^7$, a 2 ns rise time, and a 3 ns (FWHM) pulse duration, which is fairly well matched to the response time of the scintillation mechanism. The transit time from the cathode to the first dynode is 30 ns. The bialkali photocathodes are sensitive to wavelengths in a range from 290-650 nm and reach their peak sensitivity at a wavelength of 420 nm.

With four planes of scintillators with a large separation in z, the standard HMS hodoscope configuration can also provide time of flight information which can be used to determine the particle velocity. Combined with the particle momentum reconstructed from the drift chamber tracks, this information could in principle be used to determine the mass and hence the identity of detected particles. Identification of particles by time-of-flight is, however, extremely limited at high momenta and particle velocities.  The first scintillator plane is separated from the last scintillator plane by 2.6 meters in z. For a charged particle of momentum $p$ and mass $m$ moving on a trajectory with slopes $dx/dz$ and $dy/dz$, the time of flight of a particle between two scintillator paddles with a separation $\Delta z$ is 
\begin{eqnarray}
  t &=& \frac{L}{\beta c} = \frac{\Delta z }{ \sqrt{1 + \left(\frac{dx}{dz}\right)^2 + \left(\frac{dy}{dz}\right)^2}} \frac{\sqrt{p^2+m^2}}{p}
\end{eqnarray}
At the highest central momentum of the experiment, 5.4 GeV/c, the time of flight from the first scintillator plane to the last scintillator plane for a perpendicular electron track is 8.7 ns. The flight times for pions and protons at this momentum are only 3 and 100 picoseconds longer, respectively, than the electron time of flight, so while the time of flight measures the velocity with a relative precision of about 3\% (this is to be compared to the relative momentum resolution from the drift chamber tracking of 0.1\%), the hodoscope timing resolution of 300 ps cannot distinguish between pions, protons, and electrons, the most common particles in the spectrometer.

In the FPP configuration, only the first two planes of hodoscopes (S1X and S1Y) remain (see figure \ref{HMSdetectorstack}), so their relative time-of-flight resolution is even worse and their function is further limited to defining the trigger and measuring the start time relative to which the drift times in the wire chambers are determined. The scintillator bars are 1.0 cm thick, 8 cm wide, and 75 (120.5) cm long for the S1X (S1Y) paddles. The light propagation speed in the paddles typically ranges from 13-17 cm/ns. The measured time difference between the phototubes at opposite ends of a paddle serves as a crude measurement of the longitudinal coordinate at which a charged particle passed through the paddle. With a time resolution of 0.3 ns, the coordinate resolution is roughly 5 cm, comparable to the half-width of the paddles. Since each paddle is only 8 cm wide, the transverse coordinate is already more accurately determined simply by asking which paddle was hit than the longitudinal coordinate is determined by the time difference.

There are 16 paddles in the S1X plane, oriented horizontally so that the transverse dimension of the paddle measures the dispersive coordinate and the longitudinal dimension of the paddle measures the non-dispersive coordinate. The paddles are staggered in z by 2.12 cm and interleaved so that they overlap in X by approximately 0.5 cm, guaranteeing full coverage of the active area of the hodoscope. Similarly, there are ten paddles in the S1Y plane, oriented vertically so that their transverse dimension measures the non-dispersive coordinate and their longitudinal dimension measures the dispersive coordinate. Like the S1X paddles, the S1Y paddles are staggered in z and overlap in Y by approximately 0.5 cm so that the active area is also fully covered in Y. When a charged particle passes through the hodoscope plane, the intersection of the S1X and S1Y paddle(s) that fired localizes the trajectory of the particle to a square area of 8$\times$8 cm$^2$ corresponding to the transverse size of the paddles. This localization can actually be somewhat helpful with track reconstruction in the drift chambers at high rates, as combinations of hits that give tracks which do not point at the scintillators that fired can be rejected in favor of tracks that do point at the scintillators.

Because the third and fourth hodoscope planes were removed to make room for the FPP, and because of the limited separation in z between the S1X and S1Y planes, the trigger formed from S1X and S1Y alone is not very restrictive of the angles of tracks traversing the chambers. As discussed in section \ref{triggersection}, for the higher $Q^2$, lower $\varepsilon$ kinematics of the experiment, the trigger rate from S1X and S1Y alone, even in coincidence with the electron arm (BigCal), was about an order of magnitude too high for the capabilities of the data acquisition system. To address this rate problem, an additional plane of trigger scintillators was installed upstream of the first drift chamber in the HMS, in the space between the HMS vacuum exit window and DC1. This plane, christened ``S0'', consists of two paddles of 1 cm thick plastic scintillator with transverse dimensions of 12 $\times$ 15 square inches. This trigger plane was designed to restrict the HMS trigger to tracks that fire both S0 and S1 in coincidence, and to narrow the active area of the detectors to the region where elastically scattered protons are focused. The coincidence requirement between S0 and S1 decreases the probability of triggers in S1 caused by low-energy ambient radiation, noise and other uninteresting signals in favor of good events with real charged particle tracks coming from the target through the HMS and going through both drift chambers. One of the paddles, called ``S0X2'', is centered on the optical axis of the HMS, and covers approximately the region of the elastic peak. The second paddle, called ``S0X1'', is positioned at smaller X, or lower momentum, and is intended to cover the momentum region of the elastic radiative tail and inelastic scattering. 

The ``S0'' detector reduced the trigger rate to reasonable levels for all kinematics. The S0 scintillator paddles were optically coupled to Photonis XP2020 PMTs via wavelength-shifter bars as shown in figure \ref{S0schematic}, which shows a schematic illustration of the S0 design and a photo of the fully-assembled S0 detector on the bench in the Experimental Equipment Laboratory (EEL) building at Jefferson Lab shortly before its installation in the HMS. The XP2020 is a fast PMT with a 2''-diameter photocathode with very similar characteristics to the XP2262 tubes used in the S1 plane. The design of S0 results in highly efficient detection and triggering, however, the geometry of its scintillator paddles and optical coupling to PMTs is such that its timing resolution is significantly worse than S1, and although its signals are read out by both ADCs and TDCs, the information was not used in the calculation of the start time for the drift chambers. Furthermore, the S1 signals are delayed relative to the S0 signals in the trigger electronics so that the S0-S1 coincidence trigger is always formed by the arrival of the S1 signal, which has a faster time response.
\begin{figure}
  \begin{center}
    \includegraphics[width=0.4\textwidth]{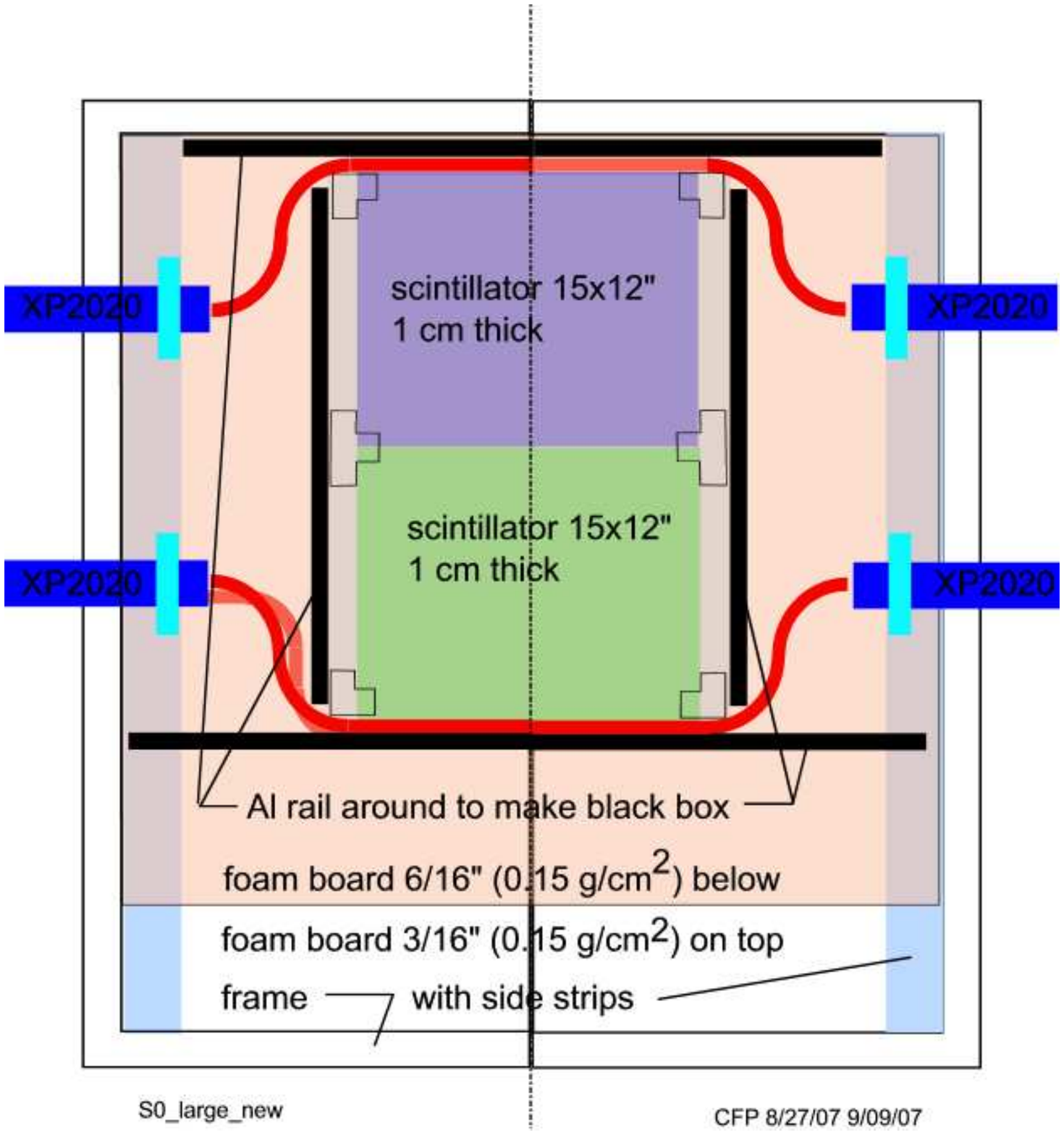}
    \includegraphics[width=0.4\textwidth]{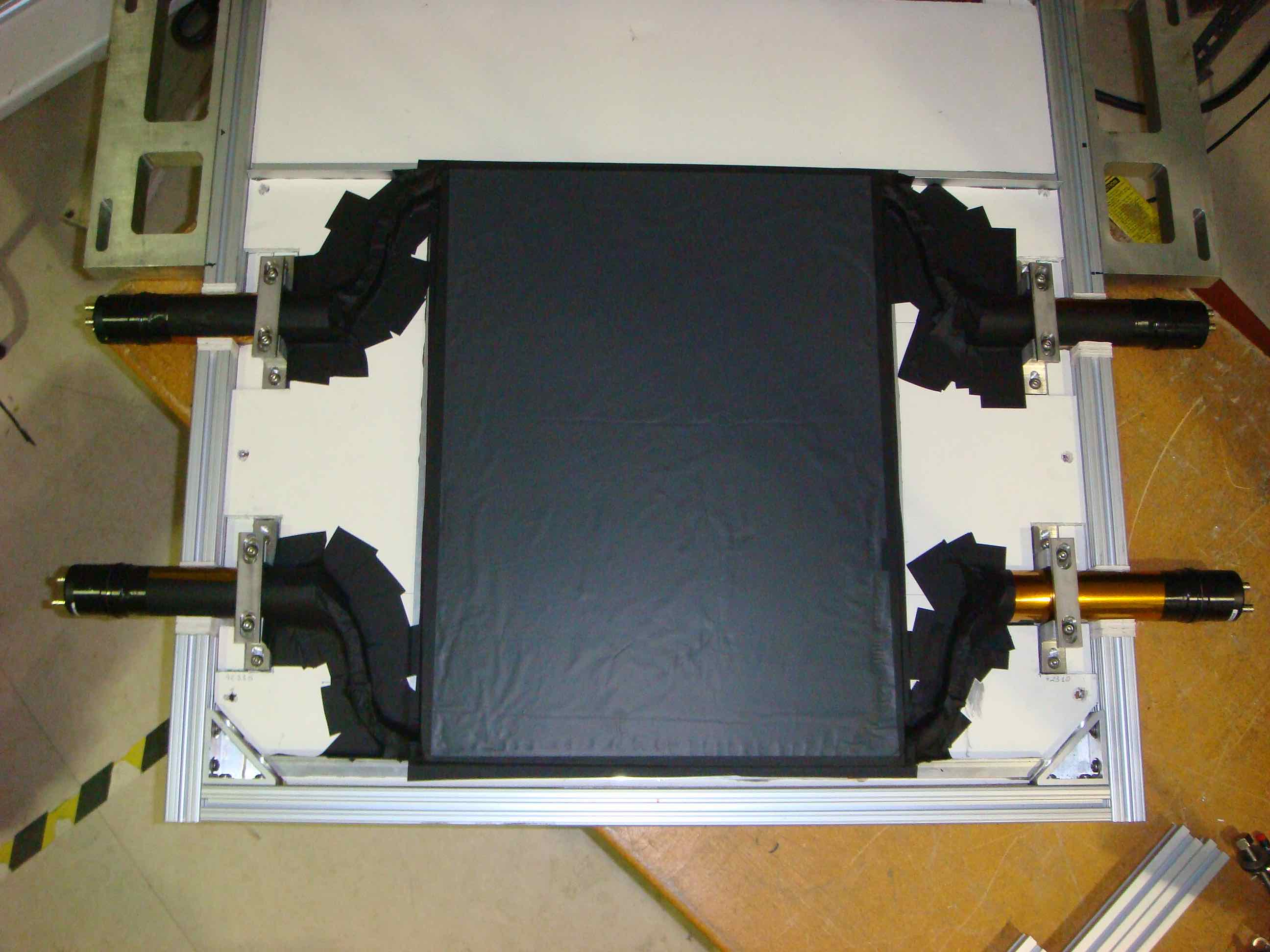}
    \caption{\label{S0schematic} Schematic of the S0 detector design (left), and a photo of S0 assembled in the EEL building at Jefferson Lab (right). }
  \end{center}
\end{figure}

The high voltages for the PMTs for both S1 and S0 were provided by the same CAEN power supplies located in the HMS hut that were used to power the HMS and FPP drift chambers. The hodoscope high voltages were also controlled and monitored through the same interface. In order to obtain uniform signal, timing and efficiency characteristics across all hodoscope PMTs, the hodoscopes were gain matched using a $^{60}$Co $\gamma$-ray source. The PMT voltages were adjusted so that the position of the Compton edge in the gamma-ray spectrum was constant and large enough to give a high trigger efficiency for protons. 

The signals from S1 and S0 were split by a BNC tee at the base of each PMT. One copy of the signal was sent to the custom trigger electronics installed in the HMS hut specifically for these experiments, which is discussed in section \ref{triggersection}. The other copy of the signal was sent to patch panels in the HMS hut along 30 feet of RG58 signal cable. From this patch panel, the signals were sent upstairs to the Hall C counting house on approximately 450 feet of RG8 cable to another patch panel, through which they were connected to yet another splitter. One copy of the signal was sent through approximately 400 ns of cable delay before being sent to charge-integrating LeCroy 1881M Fastbus ADCs for pulse-height readout. The other copy was sent through leading-edge discriminators to VME scalers for count rate monitoring and LeCroy 1872a Fastbus high-resolution TDC modules for timing readout. The 1872a is a 12-bit TDC with a time resolution of approximately 25 ps/TDC channel and a full-scale range of 100 ns. 

In contrast to the drift chamber TDCs which are read out in common stop mode, the hodoscope TDCs are operated in common-start mode. The start signal for the TDCs is provided by the trigger supervisor when the decision is made to read out an event. The combined pulse-height and time information from the PMTs at both ends of each struck paddle is used to reconstruct the time at which the particle passed through the hut. Because the timing of the logic pulses is determined by the point at which the signal exceeds a fixed threshold, larger signals arrive earlier than smaller signals, all else equal. By retaining pulse-height information, a walk correction can be applied to the data which improves the time resolution. The calibration procedure is described in section \ref{hodotimeanalysissection}.

Introducing the S0 detector was not without its drawbacks. Placing a position-insensitive detector upstream of the tracking detectors introduces non-negligible multiple scattering \emph{before} the trajectory of the particle is measured, thus degrading the overall resolution of the HMS. Because of the large dispersion of the HMS, the momentum is predominantly determined by the position of the detected track. Since multiple scattering mainly introduces small angular errors, the momentum resolution is less adversely affected by multiple scattering in S0 than the angular resolution, for which the effect of angular errors is magnified by the optical properties of the HMS. In a gaussian approximation to the angular distribution of multiple scattering, the width of the plane angle distribution is given to a good approximation by \cite{PDG2008} 
\begin{equation}
  \theta_0 = \frac{13.6\ MeV}{\beta c p} z \sqrt{\frac{x}{X_0}}\left[1 + .038\ln (\frac{x}{X_0}) \right] \label{multscatangle}
\end{equation}
where $z$ is the charge of the incident particle (in units of $e$) and $x/X_0$ is the scatterer thickness in units of radiation length. For 1 cm of plastic scintillator, this multiple scattering angle is roughly 1.8 mrad/$\beta cp$ in GeV. At a $Q^2$ of 2.5 GeV$^2$, the proton momentum is about 2.1 GeV/c, so $\theta_0 \approx 1\ mrad$. At the highest central momentum setting of 5.4 GeV/c, $\theta_0 \approx 0.34\ mrad$. 

In addition to the error on the reconstructed angles at the focal plane, there is an error in the reconstructed position at the focal plane due to the projection of the angular error at S0 to the focal plane of the HMS. S0 is located 57 cm upstream of the optical focal plane of the HMS, so the position error introduced is roughly $\theta_0 \times 57\ cm = .57$ mm/mrad. One can get a rough idea of the error magnification at the target by considering the effect due to the first-order HMS optics coefficients only. At this point it is appropriate to introduce the notation used for the variables which describe reconstructed tracks at the focal plane and at the target. The focal plane trajectory is defined by its slopes in the $x$ and $y$ directions and its coordinates at the focal plane, $z=0$: $x_{fp}$, $y_{fp}$, $x'_{fp}\equiv (dx/dz)_{fp}$, and $y'_{fp}\equiv (dy/dz)_{fp}$. The proton trajectory at the target is characterized by its coordinates and angles, and additionally its deviation from the HMS central momentum: $x_{tar}$, $y_{tar}$, $x'_{tar} \equiv (dx/dz)_{tar}$, $y'_{tar} \equiv (dy/dz)_{tar}$, and $\delta \equiv (p-p_0)/p_0$. The target coordinates $x_{tar}$ and $y_{tar}$ are measured in the $z_{tar} = 0$ plane, which is the vertical plane which intersects the origin and faces the HMS head-on\footnote{The $z=0$ plane in spectrometer coordinates is rotated by an angle equal to the HMS central angle $\Theta_{HMS}$ relative to the fixed target coordinate system in which the $z$ axis points downstream along the beam direction.}. The first-order optics coefficients giving the primary contribution to the resolution effect of multiple scattering in S0 are shown in equations \eqref{S0HMSoptics}. 
\begin{eqnarray}
  \left<x'_{tar} \right|\left. x'_{fp} \right> &=& -3.02 \nonumber \\ 
  \left<y'_{tar}\right|\left.y'_{fp}\right> &=& -2.17 \nonumber \\
  \left<\delta\right|\left.x'_{fp}\right> &=& .013 \%/mrad \nonumber \\
  \left<\delta\right|\left. x_{fp}\right> &=& .034 \%/mm \label{S0HMSoptics}
\end{eqnarray}
Table \ref{S0resolutioneffects} shows, for the different central momentum settings used in the experiment, the multiple-scattering angle $\theta_0$ calculated from \eqref{multscatangle}, and the resultant smearing of the resolution in $x'_{tar}$, $y'_{tar}$, and $\delta$.
\begin{table}[h]
  \begin{center}
    \begin{tabular}{|c|c|c|c|c|}
      \hline $p_0$, GeV/c & $\theta_0$, mrad & $\Delta x'_{tar}$, mrad & $\Delta y'_{tar}$, mrad & $\delta p/p$, \% \\ \hline
      2.0676 & 0.965 & 2.9 & 2.1 & 0.023 \\ \hline
      3.5887 & 0.523 & 1.6 & 1.1 & 0.012 \\ \hline 
      4.4644 & 0.416 & 1.3 & 0.9 & 0.0097 \\ \hline
      5.4070 & 0.341 & 1.0 & 0.7 & 0.0080 \\ \hline
    \end{tabular}
    \caption{\label{S0resolutioneffects} The lowest order effect on the HMS angular and momentum resolution due to multiple scattering in the S0 trigger scintillators.}
  \end{center}
\end{table}
For all four central momentum settings, the momentum resolution is least adversely affected. The worst-case momentum smearing is about .02\%, a factor of 5 smaller than the 0.1\% nominal resolution. Therefore, the effect of S0 on the momentum resolution can be regarded as negligible. On the other hand, the effect of S0 on the angular resolution is quite drastic. In the worst case situation at 2.07 GeV/c, the in-plane and out-of-plane angles are smeared by 2.1 and 3 mrad respectively, which is roughly a factor of three worse than the nominal resolution. Even at the highest momentum, the S0 smearing of the angular resolution is approximately equal to the nominal resolution, making the total angular resolution about a factor of $\sqrt{2}$ worse. The angular resolution of the HMS affects the calculation of the proton spin precession in the HMS magnets, because the reconstructed angles are an input to the calculation. However, multiple scattering in $S0$ only affects the angular resolution and does not introduce systematic error in the reconstructed angles.
\section{The Focal Plane Polarimeter} \label{FPPsection}
\paragraph{} In order to measure the recoil proton polarization, a new proton polarimeter was designed, built and installed in the HMS which exploits the spin-orbit coupling in the scattering of protons by hydrogen and carbon nuclei. Given an incident beam of polarized protons, the spin-orbit force causes an azimuthal asymmetry in the angular distribution of scattered protons. The orbital angular momentum operator is given by $\mathbf{L} = \mathbf{r} \times \mathbf{p}$, where $\mathbf{r}$ is the spatial coordinate of the incident proton and $\mathbf{p}$ is the incident proton momentum. As far as $\mathbf{L}$ is concerned, $\mathbf{r}$ is simply the impact parameter of the collision. The spin-orbit coupling is of the form $\mathbf{L} \cdot \mathbf{S}$, where $\mathbf{S}$ is the proton polarization vector. Since $\mathbf{L} \cdot \mathbf{p}=0$ by definition, the spin orbit force is zero for longitudinal polarization of the incident proton. 

Since measuring this component is a non-negotiable requirement of the experiment, it appears at first glance as if using nuclear scattering for polarimetry will not work. However, the proton's large anomalous magnetic moment causes its spin to undergo significant precession relative to its momentum in the HMS magnets, so that the longitudinal polarization at the target can be rotated into transverse and/or normal polarization at the focal plane\footnote{Indeed, if the proton anomalous magnetic moment were zero (up to small QED corrections), we would not be attempting to measure its internal structure because it would be a ``point'' particle like the electron.} depending on its momentum and trajectory. Although the precession of the proton spin is necessary to perform this experiment in the manner described in this thesis, it must be calculated accurately in order to extract the polarization transfer observables $P_t$ and $P_l$ from the measured polarizations at the focal plane, and it turns out that the uncertainty in this calculation is the dominant source of systematic uncertainty in the final result. The spin-orbit term in the potential can be written as $V_{LS} = U_{LS}(r) \mathbf{L} \cdot \mathbf{S}$. The deflecting force arising from this potential is $\mathbf{F}_{LS} = -\nabla V_{LS}$. 
\begin{eqnarray}
  \mathbf{F}_{LS} &=& -\nabla \left(U_{LS}(r) (\mathbf{r} \times \mathbf{p})\cdot \mathbf{S} \right) \nonumber \\
  &=& -\hat{r} \frac{d U_{LS}(r)}{d r} \mathbf{L}\cdot \mathbf{S} - U_{LS}(r) \nabla \left(\left(\mathbf{r}\times\mathbf{p}\right)\cdot \mathbf{S}\right) \nonumber \\
  \mathbf{L} \cdot \mathbf{S} &=& (\mathbf{r} \times \mathbf{p}) \cdot \mathbf{S} = \epsilon_{ijk} r_i p_j S_k = \mathbf{r} \cdot (\mathbf{p} \times \mathbf{S}) \nonumber \\
  \nabla_i ( (\mathbf{r} \times \mathbf{p}) \cdot \mathbf{S} ) &=& \partial_i (\epsilon_{jkl} r_k p_l S_j) \nonumber \\
  &=& \delta_{ik} \epsilon_{jkl} p_l S_j = (\mathbf{p}\times \mathbf{S})_i \nonumber \\
  \Rightarrow \nabla (\mathbf{L} \cdot \mathbf{S}) &=& \mathbf{p} \times \mathbf{S} \nonumber \\
  \mathbf{F}_{LS} &=& -\hat{r} \frac{d U_{LS}(r)}{d r} \mathbf{r} \cdot (\mathbf{p} \times \mathbf{S}) - U_{LS}(r) \mathbf{p} \times \mathbf{S} \label{spinorbitforce}  
\end{eqnarray}
From the point of view of polarimetry, one is less interested in the detailed character of the deflecting force than in how it gives rise to an azimuthal asymmetry. Equation \eqref{spinorbitforce} makes this clear. There are two terms in the potential gradient. The first term, with a magnitude equal to the product of the radial derivative of $U_{LS}$ and the spin-orbit coupling $\mathbf{L}\cdot \mathbf{S}$, is proportional to the inner product of the impact parameter $\mathbf{r}$ with $\mathbf{p}\times \mathbf{S}$. Momentarily ignoring the quantum-mechanical operator nature of $\mathbf{r}$, $\mathbf{p}$, and $\mathbf{S}$ and thinking in classical terms, the deflection force arising from the first term in \eqref{spinorbitforce}, while always directed radially, has a magnitude proportional to the cosine of the angle between the impact parameter and $\mathbf{p}\times \mathbf{S}$, so that deflection occurs preferentially along the direction parallel (or anti-parallel, depending on the sign of $\frac{d U_{LS}(r)}{d r}$) to $\mathbf{p}\times \mathbf{S}$. The second term, which involves the gradient of $\mathbf{L}$, only reinforces this behavior as it is always directed along $\mathbf{p}\times \mathbf{S}$. 

Consider the scattering of a proton of momentum $p$ by an analyzer nucleus at polar and azimuthal angles $\vartheta$ and $\varphi$. Define a coordinate system in which the $z$ axis lies along the incident proton trajectory. Define the $x$ and $y$ axes in the usual way so that $xyz$ forms a right-handed Cartesian coordinate system, and define the azimuthal scattering angle $\varphi$ as the angle measured from the positive $x$ axis toward the positive $y$ axis. The relation of this local coordinate system to the transport coordinate system used in the analysis is discussed in section \ref{fpptrackingsection}. In these coordinates, the incident and scattered proton trajectories are given by $\mathbf{p} = (0,0,p)$ and $\mathbf{p'} = (p'\sin \vartheta \cos \varphi, p'\sin \vartheta \sin \varphi, p'\cos \vartheta)$. Since the proton is preferentially deflected along $-\mathbf{p} \times \mathbf{S} = (pS_y, -pS_x, 0)$\footnote{There is an overall sign uncertainty in the deflection direction depending on whether the spin-orbit attraction is attractive ($U_{LS}<0$) or repulsive ($U_{LS}>0$). This sign ambiguity is unimportant for the extraction of $G_E^p/G_M^p$ since the overall sign cancels in the ratio. Even without a detailed knowledge of the form of the spin-orbit interaction, one can nonetheless determine the sign of the deflection, since the polarization direction of the scattered proton is known from the ratio of form factors and the sign of the beam polarization obtained from M\"{o}ller measurements.}, one can define an angle $\varphi_0$ given by $\tan \varphi_0 = -\frac{S_x}{S_y}$, such that the probability of scattering at an angle $\varphi$ is proportional to $\cos(\varphi - \varphi_0) = \cos \varphi \cos \varphi_0 + \sin \varphi \sin \varphi_0 \propto S_y \cos \varphi - S_x \sin \varphi$. The angular distribution of scattered protons can be written in the following general form:
\begin{equation}
  f(p,\vartheta,\varphi) = \frac{\varepsilon(p,\vartheta)}{2\pi}\left(1 + A_y(p,\vartheta) S_y \cos \varphi - A_y(p,\vartheta) S_x \sin \varphi \right) \label{asym_phys}
\end{equation}
The quantity $\varepsilon(p,\vartheta)$ is called the \emph{efficiency} of the polarimeter and contains the detailed momentum and polar-angle dependence of the nuclear scattering cross section. The quantity $A_y(p,\vartheta)$, called the \emph{analyzing power} of the reaction, describes the proportionality between the polarization of the incident proton and the size of the azimuthal asymmetry. The analyzing power also depends on the proton momentum and polar scattering angle and is determined by the detailed structure of the spin-orbit interaction and its contribution to the total scattering cross section relative to other components of the nuclear force. For a given incident momentum $p$, in order that $f$ be a proper probability distribution, it must satisfy the normalization condition
\begin{eqnarray}
  \int f d\Omega &=& 1 = \int_0^\pi \left(\int_0^{2\pi} f(p,\vartheta,\varphi)d\varphi\right)\sin \vartheta d\vartheta \nonumber \\
  \Rightarrow \int_0^\pi \varepsilon(p,\vartheta) \sin \vartheta d\vartheta &=& 1
\end{eqnarray}
The asymmetry terms vanish in the integration over $\varphi$ since $\int_0^{2\pi} \sin \varphi d\varphi = \int_0^{2\pi} \cos \varphi d\varphi = 0$. To summarize, equation \eqref{asym_phys} shows how the proton polarization components at the focal plane can be measured by measuring the angular distribution of protons scattered by nuclei in some thickness of analyzer material and illustrates the general principle of using nuclear scattering for proton polarimetry\footnote{The same method also works for neutron polarimetry, as the nuclear force is charge-independent; however, since neutrons are generally harder to detect and track than protons, neutron polarimetry is considerably more difficult from a practical point of view.}. 
\subsection{$CH_2$ Analyzer}
\paragraph{}
Several requirements and practical constraints guided the design of a polarimeter for the HMS. Proton polarization measurements can be rather time consuming since the product of polarimeter efficiency and analyzing power which enters the overall figure of merit can be quite small, requiring large numbers of incident protons to obtain the polarization with a reasonable statistical uncertainty. The major design decisions include the choice of analyzer material and thickness, the size of the active area of both the analyzer and the detector, and the type of detector required to obtain the needed angular resolution. The choice of analyzer material is basically driven by the maximum analyzing power that can be obtained while keeping the construction and operating costs reasonable. The choice of analyzer thickness is driven by optimization of the polarimeter efficiency, the fraction of incident protons that undergo ``useful'' scattering. Increasing the analyzer thickness increases the number of scattered protons, but also increases the probability of multiple scattering, absorption and other undesired reactions so that eventually the figure of merit no longer increases with increasing thickness and may actually decrease at very large thickness.

The material chosen as an analyzer for the HMS FPP is CH$_2$ (polyethylene). The choice reflects a compromise between choosing nuclei with high analyzing power and designing a polarimeter that can be built and operated safely and at a reasonable cost. While liquid hydrogen would represent the ideal in terms of analyzing power\cite{Ay_pp}, the cost and non-trivial safety issues involved in installing and operating a large tank of $LH_2$ and the required cryogenic system in the HMS ruled out this option. A calibration measurement was carried out in 2005 at the Joint Institute for Nuclear Research (JINR) in Dubna, Russia to measure the analyzing power of the inclusive reaction $\vec{p} + CH_2 \rightarrow $ one charged particle $+ X$ at proton momenta up to 5.3 GeV/c\cite{Ay_pCH2}, demonstrating sufficient analyzing power to carry out experiment E04-108 in the amount of beam time approved. Among the findings of \cite{Ay_pCH2} was that the analyzing power of CH$_2$ in the region of proton momenta of interest for experiments E04-108 and E04-019 is well described as a function of the incident proton momentum and the transverse momentum defined as $p_t \equiv p \sin \vartheta$ (with $p$ corrected for energy loss in the analyzer up to the interaction point) by the following parametrization: 
\begin{equation}
  A_y(p_t, \left<p_p\right>) = \frac{\sum_{i=1}^4 d_i p_t^i}{\left<p_p\right>}
\end{equation}
with the same coefficients $d_i$ describing the angular distribution regardless of incident momentum, and the dependence on the (acceptance-averaged) incident momentum simply given by $\left<p\right>^{-1}$. \cite{Ay_pCH2} also found that increasing the thickness of CH$_2$ above the nuclear collision length and increasing the angular acceptance in $p_t$ beyond about 0.7 GeV/c do not significantly increase the figure of merit of the polarimeter. The data also showed that the analyzing power of CH$_2$ is about 12\% higher than the analyzing power of Carbon, making it preferable to Carbon as an analyzer material. Some of the relevant properties of polyethylene are shown in table \ref{CH2properties}\cite{PDG2008}:
\begin{table}[h]
  \begin{center}
    \begin{tabular}{|c|c|}
      \hline 
      Density & 0.890 g cm$^{-3}$ \\ \hline
      Nuclear collision length & 56.1 g/cm$^2$ \\ \hline
      $dE/dx$ (minimum ionization) & 2.079 MeV cm$^2$/g \\ \hline
    \end{tabular}
    \caption{\label{CH2properties} Relevant properties of CH$_2$.}
  \end{center}
\end{table}
For the HMS FPP, a double-polarimeter design was adopted in which two 60-cm(53.4 g/cm$^2$)-thick blocks of CH$_2$ were used as analyzers, and tracking detectors (drift chambers, see below) were placed after each block to measure the angular distribution of scattered protons. The thickness of each block is approximately 95\% of a nuclear collision length. There was no need to use thicker analyzers since further increases in analyzer thickness do not further improve the integral efficiency and hence the figure of merit of the FPP as discussed above. The double polarimeter, on the other hand, improves the efficiency where a single polarimeter with twice the thickness of analyzer cannot by essentially taking two snapshots of the angular distribution, one after half the analyzer thickness and again after the full analyzer thickness. Using the information from the first polarimeter, the angular distribution in the second polarimeter can be separated into events which scatter only in the second analyzer but not in the first, and events which scatter in both analyzers.

The analyzer blocks are made of several thick sheets of material held together and surrounded by an aluminum frame. The blocks are split vertically at their midpoint, as is the frame, in order to allow retraction of the analyzer material from the active area of the detectors. This allowed for dedicated ``straight-through'' data taking runs with no analyzer material in front of the detectors in order to help calibrate the FPP drift chambers and, most importantly, to fine-tune the alignment of the chambers in software. The edges of the retractable halves of the analyzer blocks were not flat, but instead designed with an overlapping step, which prevented leakage through the seam between the two halves when the doors were closed. The insertion/retraction of the FPP was accomplished via a low-tech manual crank mechanism, which meant that opening/closing the doors of the FPP required entry into Hall C and the HMS hut. The weight of the analyzer blocks was such that a separate support structure was built to hold the analyzer blocks independently of the detector support frame. This insured that the detectors could not move when inserting/retracting the analyzer doors.

Figure \ref{FPPschematic} illustrates the basic design and dimensions of the FPP.
\begin{figure}[h]
  \begin{center}
    \includegraphics{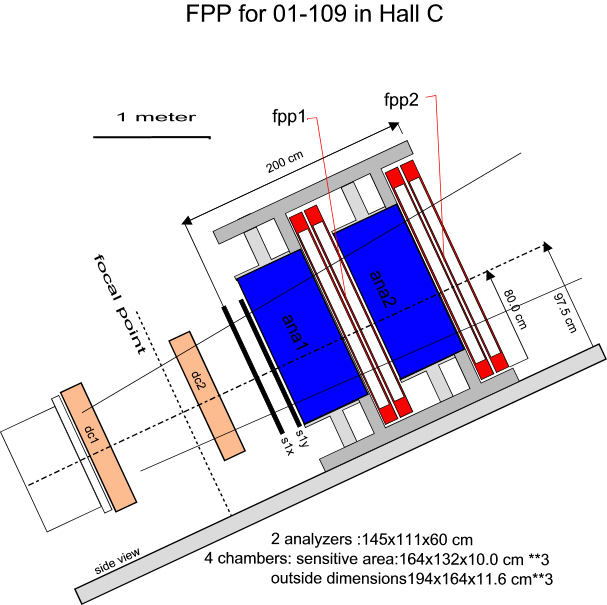}
    \caption{\label{FPPschematic} Schematic of HMS FPP design.}
  \end{center}
\end{figure}
The desired size of the FPP is basically dictated by the desired range of scattering angles and the size of the envelope of elastically scattered protons focused into the HMS hut. The HMS's point-to-point focusing properties and restricted angular acceptance tend to concentrate elastically scattered protons into a narrow range of angles at the focal plane. On the other hand, the HMS's large momentum dispersion and extended target acceptance tend to spread out the position envelope. The central momentum of the HMS was chosen to correspond to the central angle of the HMS for elastic scattering at each given beam energy. This means that the image of the elastic peak was at the center of the acceptance. The size of the elastic envelope changes with the central angle of the HMS owing to its fixed angular acceptance and the kinematic correlation between the proton's scattering angle and its momentum. From \eqref{labfourmomenta}, the formula for the proton's momentum in terms of its scattering angle follows from energy and momentum conservation:
\begin{eqnarray}
  E_e'^2 \sin^2 \theta_e &=& p_p^2 \sin^2 \theta_p \nonumber \\
  E'_e \cos \theta_e + p_p \cos \theta_p &=& E_e \nonumber \\
  E'_e &=& E_e + M_p - \sqrt{p_p^2+M_p^2} \nonumber
\end{eqnarray}
Rearranging the above gives:
\begin{eqnarray}
  E_e'^2 - p_p^2 \sin^2\theta_p &=& E_e'^2 \cos^2 \theta_e \nonumber \\
  &=& \left(E_e - p_p \cos \theta_p\right)^2 \nonumber \\
  E_e^2+p_p^2 - 2E_e p_p \cos \theta_p &=& \left(E_e+M_p-\sqrt{p_p^2+M_p^2}\right)^2 \nonumber \\
  (E_e+M_p)\sqrt{p_p^2+M_p^2} &=& M_p(E_e+M_p) + E_e p_p \cos \theta_p \nonumber \\
  (E_e+M_p)^2p_p^2 &=& E_e^2p_p^2 \cos^2\theta_p + 2M_pE_e(E_e+M_p)p_p\cos\theta_p \nonumber \\
  2M_pE_e(E_e+M_p)p_p\cos\theta_p &=& p_p^2\left(M_p^2+2M_pE_e+E_e^2\sin^2\theta_p\right) \nonumber \\
  p_p &=& \frac{2M_pE_e(E_e+M_p)\cos\theta_p}{M_p^2+2M_pE_e+E_e^2\sin^2\theta_p} \label{pel_hthe}
\end{eqnarray}
Given the in-plane angular acceptance of the HMS of $\approx \pm30$ mrad $=\pm1.7^\circ$, the momentum acceptance for elastic scattering events can be predicted from \eqref{pel_hthe}, which shows that larger beam energies and larger scattering angles tend to increase the momentum acceptance for elastic events while smaller beam energies and smaller scattering angles tend to reduce the momentum acceptance. Table \ref{dptable} shows the typical variation of the momentum and $x_{fp}$ acceptance, calculated from \eqref{pel_hthe} and the nominal 3.71 cm/\% momentum dispersion, for the various kinematics of the experiments.
\begin{table}[h]
  \begin{center}
    \begin{tabular}{|c|c|c|c|c|}
      \hline
      $p_0$, GeV/c & $\theta_0$, $^\circ$ & $E_{beam}$, GeV/c & $(p_{max}-p_{min})/p_0$, \% & $x^{fp}_{max}-x^{fp}_{min}$, cm  \\ \hline 
      2.0676 & 14.5 & 1.868 & 3.9 & 14.5 \\ \hline 
      2.0676 & 31.0 & 2.847 & 9.2 & 34.0 \\ \hline
      2.0676 & 36.1 & 3.680 & 11.0 & 40.8 \\ \hline
      3.5887 & 17.9 & 4.052 & 8.0 & 29.7 \\ \hline
      4.4644 & 19.1 & 5.711 & 10.5 & 39.0 \\ \hline
      5.4070 & 11.6 & 5.712 & 7.5 & 27.8 \\ \hline
    \end{tabular}
    \caption{\label{dptable} Approximate momentum and position acceptances at the focal plane for elastically scattered protons corresponding to the HMS angular acceptance.}
  \end{center}
\end{table}
The smallest momentum acceptance is just $\pm2$\%, while the largest momentum acceptance is about $\pm5.5$\%, corresponding to image sizes at the focal plane of 15 cm and 41 cm, respectively. In order to have a full $p_t$ acceptance of $0.7$ GeV/c, the required angular acceptance is given by $\sin \vartheta_{max} = 0.7\ (GeV/c)/p_0$. At the lowest central momentum setting of $2.07$ GeV/c, this corresponds to an angle $\vartheta_{max}=20^\circ$. The FPP detector acceptance therefore had to be big enough to contain particles scattered by up to 20 degrees in any direction, up to 60 cm upstream of the drift chambers, over the full range of angles and positions of incident protons. In practice this meant building the drift chambers and analyzers as large as practically possible. The analyzer blocks are 145 cm long in the dispersive ($x$) direction and 111 cm long in the non-dispersive ($y$) direction. The drift chambers were designed slightly larger than the analyzers at 164 cm in $x$ and 132 cm in $y$. 
\subsection{Drift Chambers}
\paragraph{}
The detection apparatus for the FPP consists of four drift chambers, with two independent chambers positioned after each analyzer to measure the track(s) of the scattered particle(s). Drift chambers have sufficiently high spatial/angular resolution for all the kinematics of these experiments. The angular resolution of the polarimeter becomes more important at high proton momenta, because the angular distribution of the analyzing power becomes concentrated at smaller angles. The basic resolution requirement is that the angular resolution be smaller than the width of the multiple-Coulomb-scattering peak, so that events from the Coulomb peak, which have no analyzing power, will not be spread out by the angular resolution to larger angles, where real nuclear scattering events with useful analyzing power reside. The angular resolution of the FPP drift chambers is approximately 1-2 milliradians as discussed in section \ref{fpptrackingsection}. The full width of the Coulomb peak for the 5.4 GeV/c central momentum setting turns out to be roughly 0.6 degrees or 10 milliradians, so there is no significant broadening of the Coulomb peak due to the angular resolution of the FPP. 

Each drift chamber contains three planes of signal wires made of 30 $\mu$m-diameter gold-plated Tungsten strung at a tension of 70 g. The spacing between signal wires in each plane is 2 cm. Between the signal wires are alternating field wires of 100 $\mu$m-diameter Beryllium+Bronze alloy strung at a tension of 150 g. Surrounding the detection layers of alternating signal and field wires are planes of cathode wires spaced 0.3 cm apart at 0.8 cm above and below the detection layers, forming a drift cell measuring 2.0 cm (horizontal) $\times$ 1.6 cm (vertical). The FPP drift cell has the same proportions as the HMS drift cell, but is twice as large. The wires in the cathode planes are made of the same material as the in-layer field wires, but are thinner at 80 $\mu$m in diameter and strung at a lower tension of 120 g. Because of the small spacing of the field wires in the cathode planes compared to the in-layer spacing of alternating signal and field wires, the electric field created by these wires behaves almost like that of a plane at constant potential, with only small spatial oscillations of the electric field coming from the fact that these planes consist of a finite number of wires with finite spacing all held at the same potential.

The signal wire planes have three different wire orientations as shown in figure \ref{fppchamberlayers}. The wires in the first plane (in order of increasing $z$) make an angle of $+45^\circ$ relative to the $x$ axis and thus measure the coordinate along the $-45^\circ$-line; i.e., the line $y=-x$. The wires in the second plane are perpendicular to the $x$ axis and measure the $x$ coordinate. The wires in the third plane make an angle of $-45^\circ$ relative to the $x$ axis and measure the coordinate along the line $y=x$. All four chambers are identical and have planes stacked in the same order so that the stacking order $-45^\circ$, $0^\circ$, and $+45^\circ$ (in terms of the measured coordinate, not the wire orientation) is repeated for both chambers within each pair. Finally, 30 $\mu$m thick aluminized mylar windows covering the entire active area form the gas enclosure for each drift chamber. Each pair of chambers is held in place rigidly by spacer blocks in order that the relative positioning of the two chambers within a pair is fixed and reproducible. The drift chambers are attached to the spacer blocks by a series of bolts penetrating through the whole chamber. The pair is then attached to the detector support frame by tracks machined onto the spacer blocks. 
\begin{figure}
  \begin{center}
    \includegraphics[height=0.5\textheight]{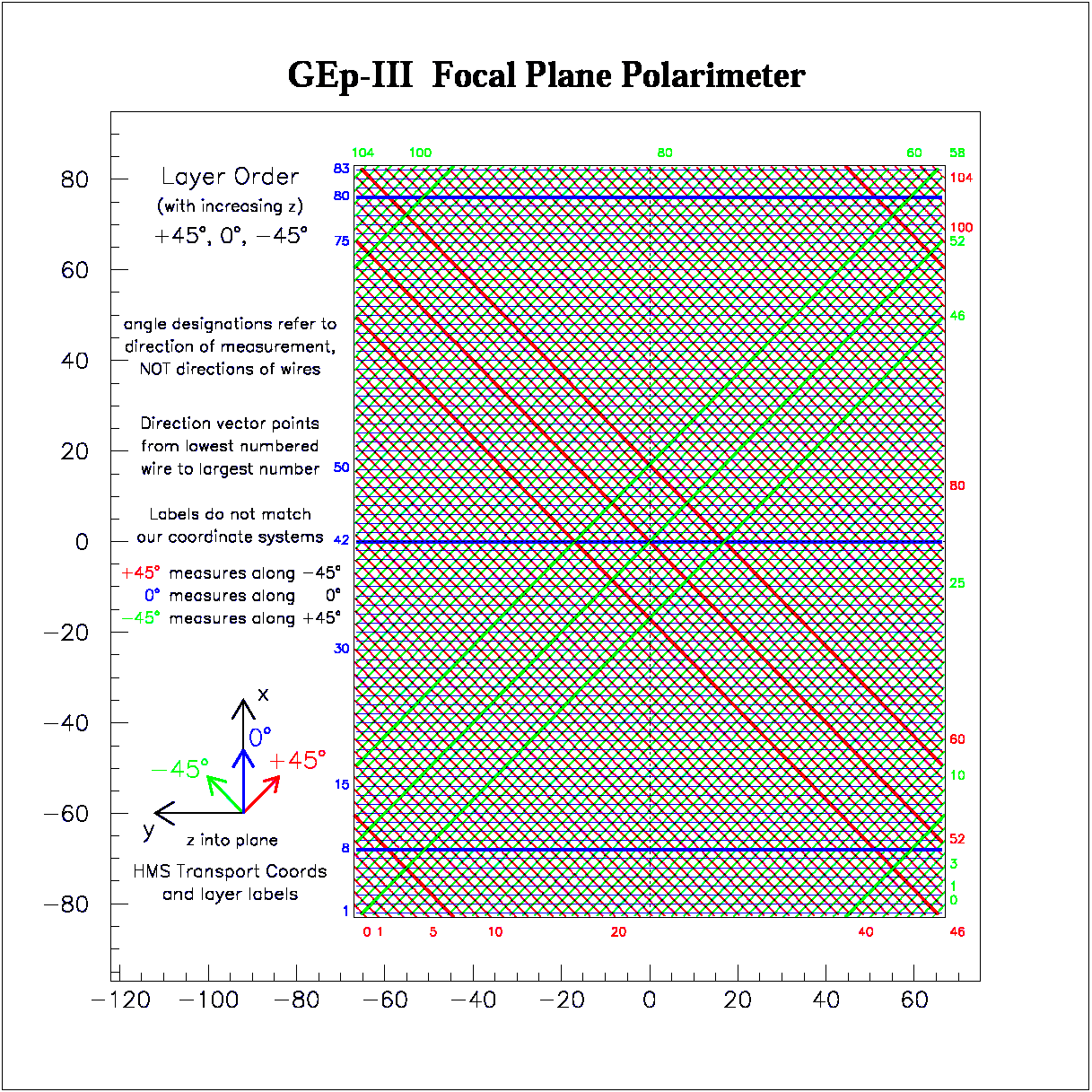}
    \caption{\label{fppchamberlayers} Wire layout for FPP drift chambers. Contrary to the note in the picture, the coordinate axes shown do not coincide with the HMS transport coordinates, unless we imagine that we are hanging upside-down from the ceiling and looking upstream along the negative $z$ axis. This contradicts the note ``z into plane'' in the figure. The proper layer order is still +45$^\circ$(red), 0$^\circ$(blue), -45$^\circ$(green) along the +$z$ axis.}
  \end{center}
\end{figure}

The FPP drift chambers were connected to the same gas mixing system used by the HMS drift chambers and were supplied with the same gas mixture of 50\%/50\% argon/ethane by weight. They were also operated at a similar gas pressure and flow rate. Because the same gas system was used, the FPP drift chambers exhibited drift properties very similar to the HMS drift chambers. Since the FPP drift chambers have a $\pm 1$ cm drift cell, as opposed to the $\pm 0.5$ cm cell size of the HMS drift chambers, the drift time window was about twice as large in the FPP chambers; i.e., $\approx$200 ns. The high voltage power for the FPP drift chambers was provided by the standard CAEN power supplies used for the hodoscopes and the HMS drift chambers. The chambers were designed so that separate high voltages could be applied to the in-layer field wires and the cathode-plane wires; however, during the experiment, the same high voltage of approximately -2400 V was applied to both sets of wires. The signal wires were maintained at ground potential. 

The FPP signals were read out by amplifier/discriminator cards connected to the individual wires. Each A/D card used by the FPP connected to eight signal wires. The low voltage levels needed to power the amplifier circuits and set the discriminator threshold were provided by custom Acopian power supply units. The threshold voltage applied to the cards was -3.0 V during most of the experiment. The plateau curve of hit rate as a function of the applied high voltage was determined using this threshold value as shown in figure \ref{FPPplateau}. 
\begin{figure}[h]
  \begin{center}
    \includegraphics[height=0.5\textheight]{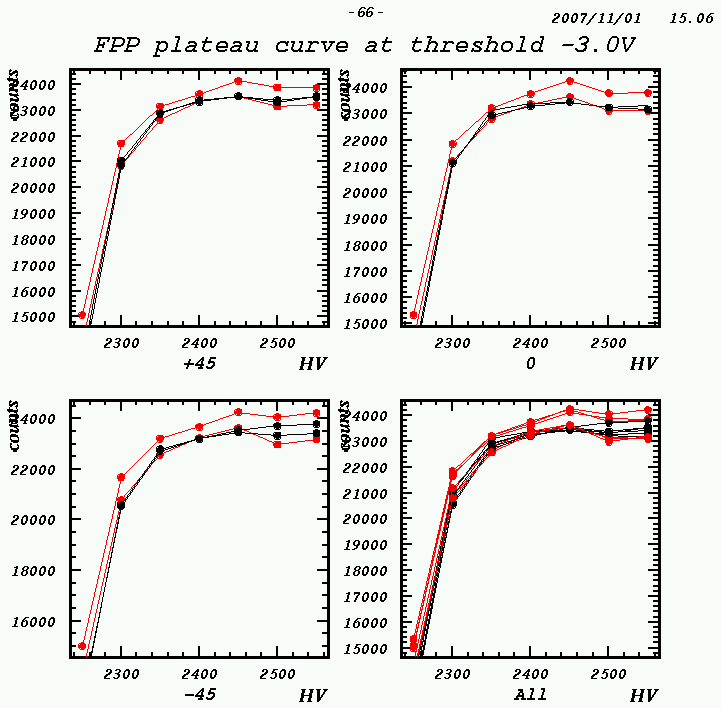}
    \caption{\label{FPPplateau} Plateau curve for the FPP drift chambers. Maximum efficiency is reached when the hit rate saturates as a function of the applied high voltage.}
  \end{center}
\end{figure}
When the hit rate saturates as a function of high voltage for a given threshold, the maximal firing efficiency of the wires is reached. The amplified, discriminated signal outputs in the form of ECL logic levels were transmitted on twisted-pair ribbon cable to the digitizing electronics for readout. Guard rails were installed around the sides of the chambers to protect the A/D cards and to guide the ribbon cables. 

Two different types of electronics were used to read out the FPP drift chambers during the experiments. During the early phase of the experiment, from October 2007 to February 2008, VME-based F1 TDC modules were used. Two VME crates were needed to hold enough modules to read out all the wires in all four FPP chambers. During the second phase of the experiment, from April to early June 2008, an additional Fastbus crate was installed in the HMS hut and the FPP signals were instead read out using LeCroy 1877 TDCs, the same kind used by the HMS drift chambers. The reason the data acquisition was switched from VME to Fastbus had to do with the high crash frequency of the VME crates during the production run of the low-$\varepsilon$ kinematics of E04-019, with the HMS positioned at an angle of 14.5$^\circ$. In that position, the VME crates suffered frequent crashes which interrupted the data acquisition and somewhat negatively impacted the quality of the data. Although the reason for the crashes was never fully understood, their frequency was strongly correlated with the HMS being positioned at forward angles, placing the detector hut close to the beam dump. Even though the HMS detectors are very well protected from excessive radiation levels by the concrete shield hut, the radiation levels in the hut were higher at forward angles. It is not clear whether the crashes were induced by radiation or by the high hit rates prevalent at forward angles. In either case, the Fastbus crate used to read out the HMS drift chambers suffered from no such crashes. Therefore, in preparation for taking data with the HMS at an even smaller angle of 11.6 degrees\footnote{Although at a higher beam energy of 5.71 GeV}, the data acquisition for the FPP drift chambers was switched over to Fastbus. As expected, the frequency of crashes of the FPP data acquisition system was much lower using the Fastbus TDCs than the VME TDCs; however, no direct comparison was made of the crash rate of the FPP data acquisition with Fastbus and VME under identical experimental conditions, so no rigorous conclusions could be drawn pertaining to the relative merits of the two technologies in the high-luminosity/rate/radiation environment present in Hall C during these experiments. 

The VME F1 TDCs are not like standard TDCs which start counting from zero at the arrival of a start signal until the arrival of the STOP signal at a later time. Instead, the F1 TDCs are free-running, meaning that they simply start counting when the data acquisition is enabled and keep counting, recording the count value every time a signal arrives and rolling over when the full-scale count range is reached. To determine the time of each hit relative to the time at which the particle passed through the drift chambers required in addition that the HMS trigger signal be recorded by the F1 TDCs as well. Otherwise, it would not have been possible to determine the relative timing of the hits, as the recorded times were absolute times with an unknown zero offset. The relative timing of the hits could only be determined with respect to other signals recorded by the same crate, since all the TDC modules in the same crate were synchronized with respect to a single clock. The FPP data acquisition system was configured so that the trigger signal always arrived later than the hit signal, making the correction of the relative hit time for rollover of the TDC count straightforward. The determination of the drift time using the F1 TDC information is discussed in appendix \ref{F1decode}.

For the data taken using Fastbus TDCs, the method for determining the drift time was the same as for the HMS drift chambers. Common stop mode was used, with the same 4 $\mu$s time window/8 hit maximum for readout. The count resolution was still 0.5 ns. The F1 TDCs also have a 16-bit count range, but with a smaller count resolution of approximately 125 ps, meaning that the rollover of the TDC count occurs roughly every 8 $\mu$s. In the FPP setup, as with the HMS, the trigger signal arrives no more than 2 $\mu$s later than the hit signals, avoiding the possibility of the TDC count rolling over to zero more than once between the arrival of the hit signals and the arrival of the trigger signal.
\begin{figure}[h]
  \begin{center}
    \includegraphics[width=.9\textwidth]{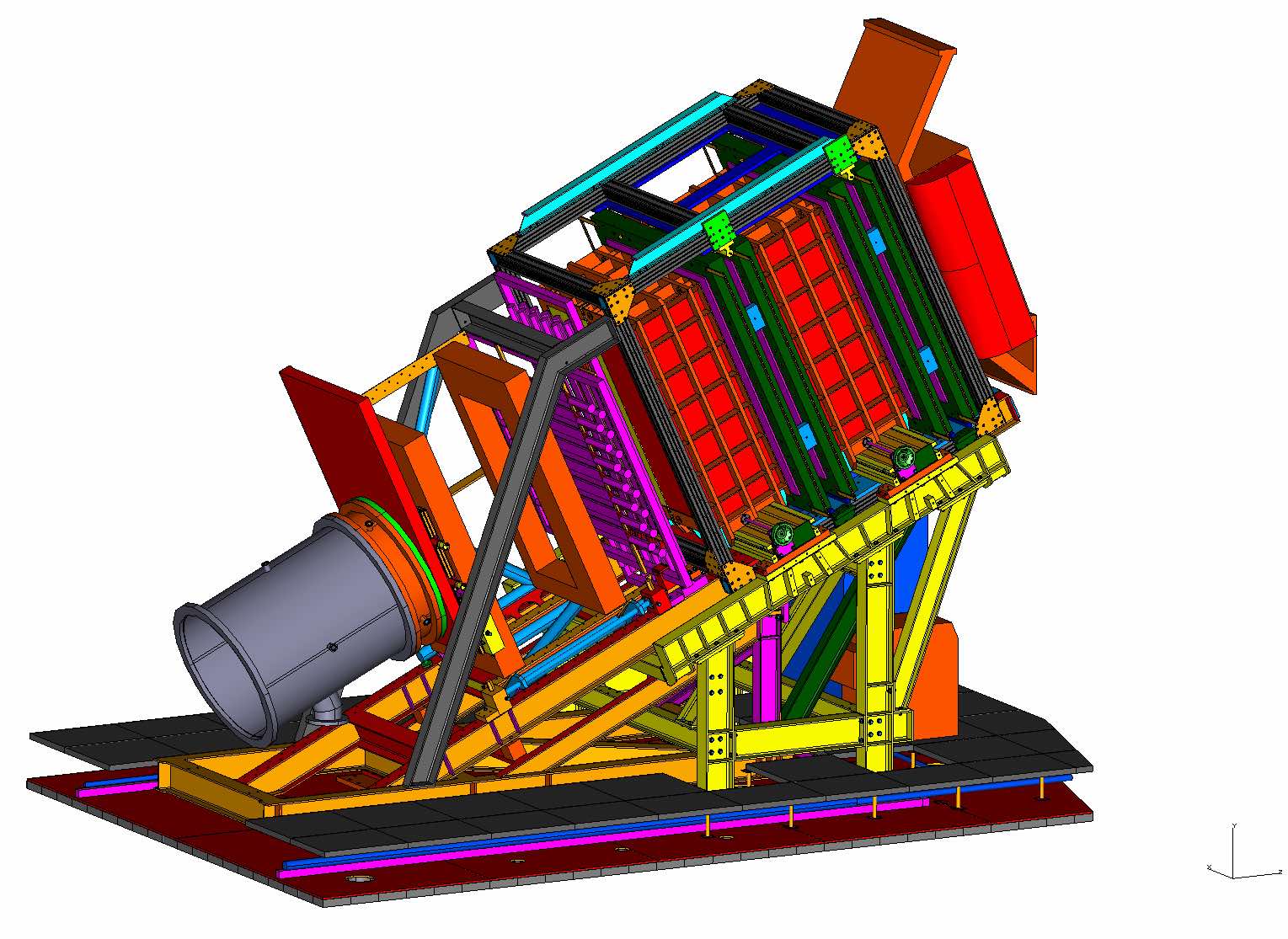}
    \caption{\label{HMSFPPfig} Design drawing of the entire HMS detector package including drift chambers and hodoscopes, the S0 trigger plane, and the FPP.}
  \end{center}
\end{figure}

In summary, the Focal Plane Polarimeter for the HMS was designed to measure the polarization of protons with momenta from 2.0 up to 5.4 GeV/c and beyond. CH$_2$ was chosen as the analyzer material. Measurements of the analyzing power of the reaction $p+CH_2 \rightarrow X$ at Dubna\cite{Ay_pCH2} showed that the overall figure of merit of the polarimeter does not increase when the thickness of analyzer is increased beyond the nuclear collision length of CH$_2$. With this result in mind, the HMS FPP was designed as a double polarimeter with two analyzers, each approximately 1 $\lambda_T$ thick and followed by pairs of drift chambers to measure the angular distribution of scattered protons. The analyzers and the drift chambers were designed to be large enough to have an angular acceptance with full $2\pi$ azimuthal coverage for transverse momenta $p_T = p\sin \vartheta$ up to 0.7 GeV/c, beyond which the polarimeter figure of merit essentially saturates. The double polarimeter configuration, by taking snapshots of the angular distribution after half the total thickness of analyzer and after the full thickness, allows a significant improvement in polarimeter efficiency which could not otherwise have been achieved by, e.g., doubling the analyzer thickness in a single polarimeter setup. Figure \ref{HMSFPPfig} shows a design drawing of the entire HMS detector package including the FPP with its independent support structure. The details of reconstructing particle tracks and angular distributions in the FPP are discussed in section \ref{fpptrackingsection}.
\section{BigCal Electromagnetic Calorimeter}
\label{BigCalSection}
\paragraph{}
At high energies and momentum transfers, a number of inelastic reaction channels can overlap with elastic electron-proton scattering within the acceptance and resolution of the HMS. In order to avoid or minimize complications and uncertainties involved in extracting the polarization of a sample of events with significant background contamination, it is highly desirable to detect the scattered electron in coincidence with the scattered proton in order to suppress inelastic backgrounds to as low a level as possible. One possible solution would have been to use the existing SOS spectrometer in Hall C to detect electrons. However, this would have meant an unacceptable loss of elastic $ep$ statistics at high $Q^2$, because the Jacobian of the reaction, defined as the ratio of the electron and proton solid angles at the chosen kinematics, grows large at high energies and large electron scattering angles\footnote{or, equivalently, low values of $\varepsilon$}. This means that to cover the full range of scattered electron angles corresponding to the solid-angle acceptance of the HMS at a given kinematics requires significantly larger solid-angle coverage than the fixed angular acceptance of the SOS can provide. The Jacobian in elastic $ep$ scattering is defined as follows: 
\begin{eqnarray}
  J &\equiv& \left|\frac{d \Omega_e}{d\Omega_p}\right| \nonumber \\
  d\Omega_e &=& \sin \theta_e d\theta_e d\phi_e \nonumber \\
  d\Omega_p &=& \sin \theta_p d\theta_p d\phi_p \nonumber \\
  J &=& \left|\frac{\sin \theta_e}{\sin \theta_p} \frac{d\theta_e}{d\theta_p} \frac{d\phi_e}{d\phi_p}\right| \nonumber \\
  d\phi_e &=& d\phi_p \nonumber \\
  \frac{\sin \theta_e}{\sin \theta_p} &=& \frac{p_p}{E'_e} \nonumber \\
  \Rightarrow J &=& \frac{p_p}{E'_e} \frac{d\theta_e}{d\theta_p} 
\end{eqnarray}
It is already apparent from the factor $p_p/E'_e$ that the Jacobian of the reaction increases rapidly with $Q^2$, as the proton momentum increases and the scattered electron energy decreases. The derivative of the scattered electron angle with respect to the scattered proton angle turns out to be a more complicated expression. Using implicit differentiation, the expression for $d\theta_e/d\theta_p$ becomes  
\begin{eqnarray}
  E'_e \sin \theta_e &=& p_p \sin \theta_p \nonumber \\
  \left(\sin \theta_e \frac{dE'_e}{d\theta_e} + E'_e \cos \theta_e\right)d\theta_e &=& \left(\sin \theta_p \frac{dp_p}{d\theta_p} + p_p \cos \theta_p\right) d\theta_p \nonumber \\
  \frac{d\theta_e}{d\theta_p} &=& \frac{\sin \theta_p \frac{dp_p}{d\theta_p} + p_p \cos \theta_p}{\sin \theta_e \frac{dE'_e}{d\theta_e} + E'_e \cos \theta_e} \label{dethdpth} 
\end{eqnarray}
The derivative of the scattered electron energy with respect to its scattering angle, $dE'_e/d\theta_e$, is given by
\begin{eqnarray}
  \frac{dE'_e}{d\theta_e} &=& \frac{d}{d\theta_e}\left(\frac{E_e}{1+\frac{E_e}{M_p}(1-\cos \theta_e)}\right) \nonumber \\
  &=& -\frac{E_e^2}{M_p\left(1+\frac{E_e}{M_p}(1-\cos \theta_e)\right)^2}\sin \theta_e \nonumber \\
  &=& -\frac{E_e'^2}{M_p}\sin \theta_e 
\end{eqnarray}
The derivative of the proton momentum with respect to the proton scattering angle, $dp_p/d\theta_p$, is given by
\begin{eqnarray}
  \frac{dp_p}{d\theta_p} &=& \frac{d}{d\theta_p}\left(\frac{2M_pE_e(E_e+M_p)\cos\theta_p}{M_p^2+2M_pE_e+E_e^2\sin^2\theta_p}\right) \nonumber \\
  &=& -p_p \tan \theta_p - p_p\left(\frac{2E_e^2 \sin \theta_p \cos \theta_p}{M_p^2+2M_pE_e+E_e^2\sin^2\theta_p}\right) \nonumber \\
  &=& -p_p \tan \theta_p - p_p^2 \sin \theta_p \frac{E_e}{M_p(E_e+M_p)}
\end{eqnarray}
Substitution of the expressions for the derivatives $dE'_e/d\theta_e$ and $dp_p/d\theta_p$ into \eqref{dethdpth} gives
\begin{eqnarray}
  \frac{d\theta_e}{d\theta_p} &=& \frac{p_p}{E'_e}\left[\frac{\cos \theta_p - \tan \theta_p \sin \theta_p - \sin^2 \theta_p \frac{p_pE_e}{M_p(E_e+M_p)}}{\cos \theta_e - \frac{E'_e}{M_p}\sin^2\theta_e}\right] \\
  \Rightarrow J &=& \frac{p_p^2}{E_e'^2}\left|\frac{\cos \theta_p - \tan \theta_p \sin \theta_p - \sin^2 \theta_p \frac{p_pE_e}{M_p(E_e+M_p)}}{\cos \theta_e - \frac{E'_e}{M_p}\sin^2\theta_e}\right| \label{Jacobian}
\end{eqnarray}
Equation \eqref{Jacobian} shows how the Jacobian of the reaction grows with $Q^2$, with an overall factor $p_p^2/E_e'^2$ multiplying a complicated expression involving $\theta_p$, $\theta_e$, the proton mass, the beam energy $E_e$, and the scattered electron energy $E'_e$. Table \ref{JacobianTable} shows the result of \eqref{Jacobian} evaluated at the central kinematic variables for each setting.
\begin{table}[h]
  \begin{center}
    \begin{tabular}{|c|c|c|c|c|c|c|}
      \hline $E_{e}$, GeV/c & $p_p$, GeV/c & $\theta_p$, $^\circ$ & $E'_e$, GeV/c &$\theta_e$, $^\circ$ & $J$ & $\Delta \Omega_e$, msr \\ \hline
      1.87 & 2.0676 & 14.5 & 0.53 & 105.2 & 15.7 & 106 \\ \hline
      2.85 & 2.0676 & 31.0 & 1.51 & 44.9 & 2.20 & 14.8 \\ \hline
      3.68 & 2.0676 & 36.1 & 2.37 & 30.8 & 0.925 & 6.23  \\ \hline
      4.05 & 3.5887 & 17.9 & 1.27 & 60.3 & 8.36 & 56.3 \\ \hline 
      5.71 & 4.4644 & 19.1 & 2.10 & 44.2 & 4.77 & 32.1 \\ \hline 
      5.71 & 5.4070 & 11.6 & 1.16 & 69.0 & 22.0 & 148 \\ \hline 
    \end{tabular}
    \caption{\label{JacobianTable} The Jacobian, defined as $J \equiv \frac{d\Omega_e}{d\Omega_p}$, evaluated from \eqref{Jacobian}, at the central values of all the relevant kinematic variables for experiments E04-108 and E04-019. $\Delta \Omega_e \equiv J \Delta \Omega_p$ is the required electron solid angle coverage corresponding to the 6.74 msr acceptance of the HMS for the scattered proton.}
  \end{center}
\end{table}
It must be noted that $J$ as defined by \eqref{Jacobian} is essentially the derivative of the electron solid angle with respect to the proton solid angle, and it represents the infinitesimal change in the electron angle induced by a given infinitesimal change in the proton angle. For the kinematics with large $J$ in particular, $J$ varies substantially over the range of scattering angles encompassed by the solid angle $J \Delta \Omega_p$ so that it is not strictly true that $\Delta \Omega_e = J \Delta \Omega_p$ for finite $\Delta \Omega_{p/e}$. Nonetheless, the values in table \ref{JacobianTable} show that a large acceptance electron detector is needed for a coincidence experiment. For all but the smallest-$\theta_e$ kinematics, the Jacobian is at least 2, and at the highest $Q^2$ kinematics, nearly 150 msr of solid-angle coverage is needed for the electron arm.

In addition to the acceptance matching requirement discussed above, the design goal for the electron detector was to have resolution comparable to or exceeding the resolution of the HMS in either the electron scattering angles, the electron energy/momentum, or both. Since the elastic scattering reaction is overdetermined kinematically, it was not necessary for the electron detector to measure both the scattering angles and the energy with high precision. Instead, having high resolution of one quantity or the other was sufficient to achieve a clean separation between elastic and inelastic events.

The type of electron detector chosen was a lead-glass electromagnetic calorimeter. The properties of lead-glass shown in table \ref{TF1properties} lend themselves naturally to calorimetry. It has a high density, a high index of refraction, and a relatively small radiation length. 
\begin{table}[h]
  \begin{center}
    \begin{tabular}{|c|c|}
      \hline $n$ & 1.6522 \\ \hline 
      $\rho$, $g/cm^3$ & 3.86 \\ \hline 
      $X_0$, cm & 2.74 \\ \hline 
      $R_M$, cm & 4.70 \\ \hline 
      $E_c$, MeV & 15 \\ \hline 
    \end{tabular}
    \caption{\label{TF1properties} Properties of TF1-0 lead glass relevant to electromagnetic calorimetry.}
  \end{center}
\end{table}
It is also highly transparent, making it an efficient collector of photons emitted by showering particles. The calorimeter, called BigCal, was assembled from 1,744 type TF1-0 lead-glass bars. The glass for BigCal came from two sources. First, 1,024 blocks of dimension 3.8$\times$3.8$\times$45.0 cm$^3$ were contributed by the Institute for High Energy Physics (IHEP) in Protvino, Russia. These blocks were stacked in a 32$\times$32 square array on the bottom of the calorimeter, with the 3.8$\times$3.8 cm$^2$ ends facing the target. The remaining blocks came from a calorimeter that had been used to study  real Compton scattering (RCS) on the proton in Hall A at JLab, but originally came from the Yerevan Physics Institute in Armenia. These blocks were made of the same kind of lead-glass, but had slightly different dimensions at 4.0$\times$4.0$\times$40 cm$^3$. These  ``RCS'' bars were stacked in a 30 (horizontal) $\times$ 24(vertical) array on top of the Protvino bars, again with the small-area (4$\times$4 cm$^2$) ends facing the target. The fully assembled calorimeter was segmented in 56 rows and 32(30) columns for rows 1-32(33-56) as shown in figure \ref{BigCal_trigger_schematic}. 
\begin{figure}[h]
  \begin{center}
    \includegraphics[width=.6\textwidth]{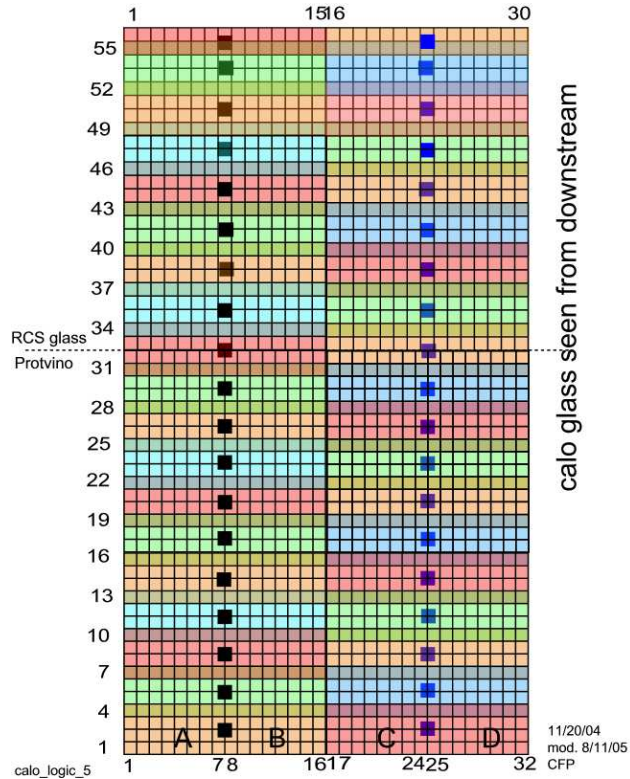}
    \caption{\label{BigCal_trigger_schematic} The 1,744 lead-glass blocks of BigCal. The different colors indicate the groupings of channels for the trigger, each marked by a black or blue square. The groups of 64 overlap by one row vertically, increasing the trigger efficiency. There are 38 total groups.} 
  \end{center}
\end{figure}

The area of the calorimeter facing the target is roughly $122\times218$ cm$^2$. This is to be compared with the horizontal and vertical angular acceptances of the HMS of approximately $\pm$40 mrad(horizontal)\footnote{Although the collimator-defined angular acceptance is only $\pm$30 mrad for a point target, a greater range of angles is accepted for an extended target, for which particles scattered upstream(downstream) of the origin can pass through the collimator at smaller(larger) angles than particles scattered from the origin.} $\times \pm$70 mrad(vertical). The solid-angle coverage of BigCal is approximately $\Delta \Omega_e = A_{cal}/r_{cal}^2$, where $A_{cal}$ is the area of the calorimeter facing the target and $r_{cal}$ is the distance from the target (the origin), to the surface of BigCal. The placement of BigCal in Hall C was determined by the acceptance matching requirement shown in table \ref{JacobianTable}. The distances required by acceptance matching are compared to the actual distances used in the experiment in table \ref{BigCalDistanceTable}. 
\begin{table}[h]
  \begin{center}
    \begin{tabular}{|c|c|c|c|}
      \hline $\theta_e$, $^\circ$ & $J$ & $r_{cal}(J)$, cm & $r_{cal}^{actual}$, cm \\ \hline
      105.2 & 15.7 & 500.0 & 493.2 \\ \hline
      44.9 & 2.2 & 1336 & 1200 \\ \hline
      30.8 & .925 & 2060 & 1108 \\ \hline
      60.3 & 8.36 & 685.3 & 605.0 \\ \hline
      44.2 & 4.77 & 907.2 & 608.2 \\ \hline
      69.0 & 22.0 & 422.4 & 430.4 \\ \hline
    \end{tabular}
    \caption{\label{BigCalDistanceTable} Distance from the origin to the surface of BigCal required by acceptance matching, compared with the actual distance used in the experiment. Obstacles and space constraints in Hall C limited the possible angles and distances at which BigCal could be placed.}
  \end{center}
\end{table}
A number of physical obstacles in Hall C restricted the possible locations of BigCal. Among those obstacles were the rails for the SOS carriage, an elevated concrete platform covering a fraction of the floor of Hall C, and a variety of infrastructure and equipment located throughout Hall C. The ability to place BigCal at large distances from the target was also restricted by the finite length of signal and high voltage cables used to power and instrument the detector, combined with the location of the data acquisition electronics and high-voltage power supplies. Additional constraints include the actual physical size of Hall C, the size of the support platform for BigCal and its front-end electronics, and the maximum radius at which the overhead crane in Hall C can safely lift and lower an object the size and weight of BigCal. Fortunately, it is mostly harmless to place BigCal closer to the target than the acceptance-matching distance. It is only when BigCal is placed further from the target than the acceptance-matching distance that a fraction of elastically scattered electrons fail to hit the calorimeter and go undetected. 

The thickness of the lead-glass blocks in BigCal is approximately 15 radiation lengths. This is enough material to fully stop electrons with energies up to 10 GeV. Therefore, all elastically scattered electrons in this experiment deposit their full energy in BigCal. An energetic electron striking the calorimeter initiates an electromagnetic cascade,  a process wherein the primary electron produces Bremsstrahlung photons as it loses energy by radiation, photons which in turn produce $e^+e^-$ pairs, which in turn radiate more photons, and so on until the initial electron, and all secondary particles produced in the cascade, lose enough energy to fall below the critical energy, after which they lose energy predominantly through ionization and are eventually absorbed. The primary electron and the secondary pair-produced electrons and positrons move at close to the speed of light, and faster than $c/n$, where $n$ is the index of refraction of lead-glass. They therefore emit Cerenkov radiation at optical wavelengths, and it is this Cerenkov light which is collected and transduced into electrical signals.  Electromagnetic showers generate large numbers of optical photons. The signal measured by a given PMT is proportional to the number of photoelectrons emitted by the cathode, which is the convolution of the wavelength spectrum of emitted Cerenkov photons with the spectral sensitivity of the cathode, and for lead-glass instrumented with PMTs with conventional bialkali photocathodes, typical yields are about 1,000 photoelectrons per GeV of shower energy.

The individual lead-glass bars are optically isolated from each other by an aluminized mylar wrapping, insuring that the light radiated in each bar is contained within that bar. At the end of each bar is a Russian FEU84-12 stage ``venetian blind'' photomultiplier tube, optically coupled to the glass through a 5 mm-thick Si-pad ``cookie''. The PMTs and attached cookies are held in 2'' thick moveable aluminum cross bars, with each cross bar holding 4 rows of PMTs. In the lower half of BigCal, each cross bar holds 128 PMTs, while in the upper half, each cross bar holds 120 PMTs. The base of each PMT is attached to the crossbar via threaded rods screwed into the cross bar, pressing the PMTs and the cookies firmly against the glass. Each of the 56 rows of PMTs has a patch board which connects the externally supplied high voltage power to the voltage divider circuits on the PMT bases, and also connects the signal output of the PMT base to cables which transmit the signals to the front-end electronics. The glass, PMTs, cross bars, and patch panels are all contained within a black box. The signal and high voltage cables that connect the patch panels to the outside world enter/exit the black box through labyrinth openings.

On the front of the calorimeter is a .5-inch thick aluminum holding plate with 1,744 .25-inch-diameter holes drilled in front of each block to allow viewing of the glass and to provide an opening for light produced by the source for the gain monitoring system to reach the blocks. The light source consisted of a .5''-thick lucite plate installed in front of the aluminum holding plate, illuminated from the sides by a single LED through a fiber-optic splitter, with outputs coupled to the lucite through connectors plugged into small holes distributed uniformly up and down the sides of the plate. The lucite is supported by an aluminum frame which includes a .25''-thick wall in front of the lucite. Finally, at the very front of the detector is a series of four removable 1''-thick aluminum absorber plates. These plates were used to shield BigCal from low-energy photons in order to mitigate radiation damage to the lead-glass in the high-radiation environment present in Hall C. While the absorber is effective at reducing radiation damage, it degrades the energy resolution of the detector significantly. On the other hand, it has an almost negligible impact on the position resolution. The full 4-inch absorber thickness was used for all kinematics of the experiment except for the $\varepsilon=.15$, $\theta_e = 105^\circ$ setting, where only one of the four plates was used. In these kinematics, the central scattered electron energy is only about 530 MeV. At large scattering angles, radiation levels are lower, so radiation damage is not as serious a problem for the operation of the calorimeter, allowing a thinner absorber to be used. At such a low electron energy, the full absorber thickness would have unacceptably degraded the energy resolution, and, by reducing the overall signal size and resolution, would have forced the use of a lower trigger threshold, which would have reduced the efficiency and increased the trigger rate from non-elastic reactions. 

The analog signals from the PMTs begin a long journey that ends at LeCroy 1881M Fastbus charge-integrating ADCs. First, the signals are sent to specialized NIM summing modules. Each module has two groups of eight inputs, and four outputs, each equal to the analog sum of the eight inputs amplified by a factor of 4.2. One of the four outputs is inverted. Each summing module also has sixteen outputs in the back which are 4.2X-amplified copies of the 16 individual input signals. These outputs are connected by 34-conductor flat cables to patch panels which connect to 100-meter long signal cables which transmit the signals to the electronics platform, where they are patched to twisted pair ribbon cable which finally connects the analog signals to the ADCs for readout. The signals are grouped so that each sum of eight signals corresponds to a group of eight blocks in the same row (see figure \ref{BigCal_trigger_schematic}). There are four such groups per row. In the Protvino (lower) half, each row has 32 blocks so the grouping is quite natural. The groups are referred to as xxA(1-8), xxB(9-16), xxC(17-24), and xxD(25-32), where xx is the row number. In the RCS (upper) half of the calorimeter, since there are only 30 blocks per row, two groups in each row have only seven blocks. The grouping in the RCS blocks is xxA(1-8), xxB(9-15), xxC(16-23), and xxD(24-30), where xx is again the row number which runs from 33-56. The sums of eight are referred to as the ``first-level'' sums for reasons related to the BigCal trigger. One of the outputs of each first-level summing module connects to discriminators which output a logic pulse when the analog input exceeds a fixed threshold. The discriminator output is sent on 50-meter signal cable to the electronics platform for timing readout using LeCroy 1877 model Fastbus TDCs and rate monitoring using VME scalers. The other output of each first-level sum goes to another identical summing unit, where it is combined in groups of eight with other first-level sums to form ``second-level'' sums of 64 channels, which are used in the trigger system, which will be discussed in detail in section \ref{triggersection}.

The width of the ADC gate over which the BigCal signals were integrated ranged from 150 ns at the kinematics with the lowest scattered electron energy to 250 ns at the kinematics with the highest scattered electron energy. This wide integration gate made the suppression of noise and the proper termination of signals particularly important as all the electronic noise in the system was integrated over the full gate width, widening the ADC pedestal and degrading the resolution for small signals in particular.

The high voltage power for the BigCal PMTs was provided by LeCroy and CAEN high voltage crates. Six LeCroy crates were located on the electronics platform with the two FastBus crates used to read out BigCal. These crates provided 1,024 individual channels of high voltage and were used to power the PMTs coupled to the bottom (Protvino) blocks. These power supplies were controlled and monitored using custom slow-control software specially designed for this kind of crate. The remaining 720 channels of BigCal were powered by CAEN high voltage supplies located upstairs in the ``G0 cage'' on the second floor of the counting house. The CAEN power supplies were remotely controlled and monitored using the same CAEN-net interface as those in the HMS hut.

The choice of a lead-glass electromagnetic calorimeter for the electron detector reflects a choice in favor of high coordinate and angular resolution at the expense of moderate to poor energy resolution--certainly much worse than a magnetic spectrometer. The energy resolution of an electromagnetic calorimeter can be expressed as $\sigma_E/E = a / \sqrt{E} \oplus b \oplus c / E $\cite{PDG2008}. The term $a$ characterizes the intrinsic statistical fluctuations in electromagnetic shower development, and is usually taken to be 5\% for a lead-glass calorimeter with 100\% sampling fraction. The sampling fraction for BigCal is actually less than 100\% since roughly one radiation length of inactive material (the aluminum absorber) was placed in front of BigCal, making $a$ proportionally worse. The constant term $b$ includes calibration uncertainties, the effects of radiation damage (significant for BigCal), and detector non-uniformities. The $c$ term contains the effect of summing electronic noise over all the channels included in the shower. The Moliere radius of lead-glass, which characterizes the transverse size of electromagnetic showers, is 4.7 cm, which is to be compared with the $\approx$4 cm transverse size of the blocks. Approximately 90\% of the shower is contained within one $R_M$, and $\approx$99\% of the shower is contained within 3.5 Moliere radii. This means that electron showers are typically spread out in 3$\times$3 to 5$\times$5-block clusters. By calculating the center of gravity of the cluster of signals in a shower, its coordinate can be reconstructed with a resolution significantly better than the canonical $L/\sqrt{12}\approx 1.1(1.15)$ cm resolution provided by the granularity of the 3.8(4.0) cm blocks\footnote{To see how the $L/\sqrt{12}$ ``resolution'' appears, assume that the block size $L$ is large enough that the entire shower is contained in one block. Alternatively, assume that the shower coordinate is assigned to the center of the block with the largest signal. Assume a uniform distribution of shower coordinates within a block. Then the R.M.S. error in the reconstructed coordinate is $\sqrt{\left<\Delta_x^2\right>} = \sqrt{1/L \int_{-L/2}^{L/2} dx x^2} = L/\sqrt{12}$}. The ideal coordinate resolution of BigCal, which was obtained from a Monte Carlo simulation\cite{IHEP_leadglass} including the effect of the absorber but neglecting the effects of radiation damage, ranges from about 7.5 mm at the lowest scattered electron energy of 530 MeV to about 3.5 mm at the highest electron energy of 2.37 GeV. 

Because the target is extended, the BigCal coordinate measurement does not on its own measure the electron scattering angle. To measure the scattering angle, one must also know the position of the interaction vertex, which is reconstructed by the HMS. The HMS reconstructs $y_{tar}$ with $\approx$2 mm resolution. The desired quantity is the position of the interaction vertex along the beamline, $z_{vertex} \approx y_{tar} / \sin \Theta_{HMS}$, where $\Theta_{HMS}$ is the central angle of the HMS. The electron angle resolution can be estimated by considering scattering in the horizontal plane only. Assume BigCal is located at an angle $\Theta_{cal}$ at a distance $R_{cal}$ from the origin, and define $x_{cal}$ as the horizontal coordinate along the surface of BigCal. Additionally define a ``Hall'' coordinate system in which the $z_{Hall}$ axis points downstream along the beamline, and the $x_{Hall}$ axis is horizontal and points toward BigCal. Furthermore, assume that the positive $x_{cal}$ axis points toward large angles. Then, the ``Hall'' coordinates of the electron corresponding to the surface coordinate measured at BigCal are given by
\begin{eqnarray}
  x_{Hall} &=& x_{cal} \cos \Theta_{cal} + R_{cal} \sin \Theta_{cal} \nonumber \\
  z_{Hall} &=& -x_{cal} \sin \Theta_{cal} + R_{cal} \cos \Theta_{cal} 
\end{eqnarray}
The electron scattering angle is defined by the ray from the reconstructed interaction vertex to the measured coordinates at BigCal $\Delta \mathbf{r} \equiv \mathbf{r}_{cal} - \mathbf{r}_{vertex}$:
\begin{eqnarray}
  \cos \theta_e &=& \frac{z_{Hall} - z_{vertex}}{\sqrt{x_{Hall}^2 + (z_{Hall}-z_{vertex})^2}} \equiv \frac{\Delta z}{L} \approx \frac{\Delta z}{R_{cal}}\\
  \left(\sin \theta_e d\theta_e\right)^2 &=& \left(\frac{\partial \cos \theta_e}{\partial x_{cal}}\right)^2dx_{cal}^2 + \left(\frac{\partial \cos \theta_e}{\partial z_{vertex}}\right)^2dz_{vertex}^2 \label{BigCalres1}
\end{eqnarray}
To determine the dominant contribution to the angular resolution of BigCal, the electron flight path length $L$ in the denominator can be approximated by $R_{cal}$. Then equation \eqref{BigCalres1} becomes
\begin{eqnarray}
  \sin^2\theta_e d\theta_e^2 &=& \frac{1}{R_{cal}^2} \left(\sin^2\Theta_{cal} dx_{cal}^2 + dz_{vertex}^2\right)
\end{eqnarray}
For the central electron angle $\theta_e = \Theta_{cal}$ and the electron angle resolution becomes:
\begin{eqnarray}
  d\theta_e &=& \frac{1}{R_{cal}}\sqrt{dx_{cal}^2 + \frac{dy_{tar}^2}{\sin^2 \Theta_{cal} \sin^2 \Theta_{HMS}}} \label{BigCalres2}
\end{eqnarray}
Equation \eqref{BigCalres2} shows that there is an angle-independent contribution to the electron angle resolution $dx_{cal}/R_{cal}$ coming from the shower coordinate resolution, and an additional angle-dependent contribution coming from the resolution of the interaction vertex reconstructed by the HMS. Table \ref{eth_res_table} shows the variation of $d\theta_e$ for all the kinematics.
\begin{table}[h]
  \begin{center}
    \begin{tabular}{|c|c|c|c|c|c|}
      \hline $E'_e$, GeV & $\Theta_{cal}$, $^\circ$ & $\Theta_{HMS}$, $^\circ$ & $R_{cal}$, cm & $dx_{cal}$, cm & $d\theta_{e}$, mrad \\ \hline
      0.53 & 105.2 & 14.5 & 493.2 & 0.74 & 2.25 \\ \hline 
      1.51 & 44.9 & 31.0 & 1200.0 & 0.44 & 0.59 \\ \hline
      2.37 & 30.8 & 36.1 & 1108.0 & 0.35 & 0.68 \\ \hline
      1.27 & 60.3 & 17.9 & 605.0 & 0.48 & 1.47 \\ \hline 
      2.10 & 44.2 & 19.1 & 608.2 & 0.37 & 1.57 \\ \hline 
      1.16 & 69.0 & 11.6 & 430.4 & 0.50 & 2.73 \\ \hline
    \end{tabular}
    \caption{\label{eth_res_table} Kinematical dependence of electron scattering angle resolution. The coordinate resolution of BigCal is estimated by $dx_{cal} \approx 0.54\ cm / \sqrt{E'_e}$, with $E'_e$ in GeV, motivated by the results of a detailed GEANT Monte Carlo simulation of the BigCal detector response to electromagnetic showers. The HMS $y_{target}$ resolution is (conservatively) estimated at 2 mm. Multiple scattering of the electron in the air between BigCal and the target is neglected. Multiple scattering will make the resolution $d\theta_e$ worse, especially at large $R_{cal}.$ }
  \end{center}
\end{table}
The coordinate resolution of BigCal was studied in depth using a Monte Carlo simulation of BigCal developed at Protvino\cite{IHEP_leadglass}. The simulation is based on the GEANT3 software libraries and was used to examine the development of electromagnetic showers in the lead-glass of BigCal in depth. It includes measurements of the wavelength dependence of the index of refraction and the absorption coefficent of the glass, and also includes the reflective Mylar wrapping of the lead-glass bars, the photocathodes of the PMTs, and all materials in front of the glass. For each simulated shower, the code traces every Cerenkov photon to the cathode of the PMT, where it causes photoemission of an electron with a probability governed by the quantum efficiency of the cathode. For table \ref{eth_res_table}, the coordinate resolution with the full absorber in place was estimated as $dx_{cal} = 5.4\ \mbox{mm} / \sqrt{E'_e}$, with $E'_e$ in GeV, based on the results of the simulation. The electron angle resolution ranges from 0.6 mrad to 2.7 mrad. Multiple scattering of the electron in the air between the target and BigCal is neglected in these estimates. It is non-negligible compared to the quoted resolution for the large-$R_{cal}$ kinematics in particular. Additionally, radiation damage to the glass accumulated over the course of the experiment worsened the coordinate resolution only slightly even as it dramatically degraded the energy resolution. 

The energy resolution of BigCal was substantially worse than the nominal $5\%/\sqrt{E}$ for two reasons. First, the absorber in front of BigCal, which is approximately 1 $X_0$ thick, reduced the sampling fraction, the ratio of active absorber (lead-glass) to the total amount of material (lead-glass + absorber). Second, significant radiation damage to the glass accumulated over the course of the experiment made the constant term in $\sigma_E/E$ much larger than the typical 1\%. Although the glass was partially ``cured'' using UV annealing during the two-month accelerator maintenance shutdown in February-March 2008, it did not regain full transparency in the limited amount of curing time and the energy resolution suffered accordingly during the April-June 2008 run period.

The main result of the above discussion is that the coordinate resolution of BigCal translates into excellent angular resolution, comparable to that of the HMS. When the Jacobian of the reaction is taken into account, the angular resolution of BigCal actually turns out to be far better than needed for most of the kinematics (those with large $J$), and perfectly adequate for the small-$J$ kinematics, for which inelastic backgrounds are much lower in any case. In other words, when the electron angles measured by BigCal are compared to those expected from elastic kinematics of the detected proton, the resolution of the difference is dominated by the HMS resolution when the Jacobian is large.

In summary, the kinematics of elastic $ep$ scattering at high $Q^2$ result in a large Jacobian, $d\Omega_e/d\Omega_p$. In order to perform a coincidence experiment, a large-acceptance electron detector is required to match the electron acceptance to the proton acceptance defined by the HMS collimator. Since the two-body reaction kinematics are overdetermined, simultaneously precise measurements of the electron energy and angles were unnecessary. The lead-glass electromagnetic calorimeter BigCal provided the needed acceptance with moderate-to-poor energy resolution but sufficient angular resolution that the cleanliness of the separation between elastic and inelastic events was limited only by the resolution of the HMS itself 
\section{Trigger and Data Acquisition}
\label{triggersection}
\paragraph{}
The design of the data acquisition system for the experiment was primarily driven by the electronics requirements of BigCal. All 1,744 PMT signals were read out by ADCs. Each LeCroy 1881M ADC module has 64 channels, so 28 modules totaling 1,792 channels were used, leaving 48 spare channels. This number of modules required two Fastbus crates. Additionally, the first-level sum of 8 signals were discriminated and sent to TDCs for timing readout. This required 224 channels of TDCs. Each LeCroy 1877 module had 96 channels, so three modules were needed. Finally, the ``second-level'' analog sums of 64 were sent to ADCs and also discriminated and sent to TDCs for readout, requiring 38 additional channels of ADCs and TDCs. This required one more 1881M ADC module, but no more TDC modules since 64 channels in the third 1877 module were unused. Both fastbus crates were located in Hall C on a platform with multiple electronics racks. The main data acquisition electronics and other important equipment were also installed on this platform, as will be discussed below. The platform was located on the opposite side of the beam from BigCal, where more space was available. In order to shield the sensitive electronics from the high radiation levels present when the electron beam is in Hall C, a large concrete bunker was built around the electronics platform. The bunker completely surrounded the sides of the electronics platform facing the beam dump and the target, except for small openings serving as feedthroughs for cables, and a wall was built on the far side of the platform in order to support the roof of the bunker. A large opening was left in the side of the bunker furthest from the beam dump to allow personnel access and improve ventilation. Cables with insufficient length to wind all the way around to the large opening, including the 50 meter cables carrying the timing signals from BigCal and the shorter fast cables carrying the trigger signals from BigCal and the HMS, were fed through the small openings on the front (facing the target) and side (facing the beam dump) walls. 
\subsection{HMS Proton Trigger}
\paragraph{}
Signals from the HMS hodoscopes were used to form a single-arm proton trigger using a number of NIM modules in the HMS hut. The standard HMS trigger electronics are located upstairs in the counting house. Since only two planes of hodoscopes and S0 were involved in the HMS trigger, and since other standard HMS detectors had been removed, the trigger requirements for this experiment were quite different from those of the standard HMS configuration, and would have required substantial modifications of the intricate maze of connections in the Hall C counting house electronics room, with enormous potential for mistakes and unintended consequences. However, this was not the primary motivation for building a custom trigger setup--the main reason was timing. Had the main coincidence trigger been located in the counting house, it would not have been possible to send a gate back down to the BigCal ADCs in Hall C in time for the arrival of the BigCal signals. More complicated solutions would have been required for the readout of BigCal. This was the main determinant of the decision to form the coincidence trigger and locate the trigger supervisor in the electronics bunker in Hall C. No such timing restrictions faced any of the other detectors. The HMS and FPP drift chambers and the discriminated first and second-level sum signals from BigCal were read out by TDCs in common-stop mode, and the hodoscope signals for both ADCs and common-start TDCs had enough fixed and adjustable delay on the way to the readout electronics in the Hall C counting house that they could be timed in without any trouble. 
 
The HMS trigger was formed by requiring a coincidence between ``S0'' and ``S1''. Each of the two paddles of S0 has two PMTs. The ``S0'' trigger required both PMTs on either paddle of S0 to fire. A separate trigger type was defined for each paddle of S0. This allowed them to be prescaled separately in the data acquisition system. Having two independent HMS triggers was particularly useful for the $Q^2=8.5$ GeV$^2$ kinematics and the $Q^2=2.5$ GeV$^2$ kinematics at $\epsilon=.15$, both of which involved forward angles of the HMS and high singles rates, especially on the paddle covering the lower-$x$/$\delta$ region where inelastic reactions dominate. For these data points, the entire distribution of elastic events was contained within the higher-$x$/$\delta$ paddle so that the trigger rate contributed by the low-$x$ paddle could be prescaled as needed with a negligible loss of elastic events. The ``S0'' trigger logic is shown in figure \ref{S0trig_diagram}. 
\begin{figure}[h]
  \begin{center}
    \includegraphics[angle=-90,width=.90\textwidth]{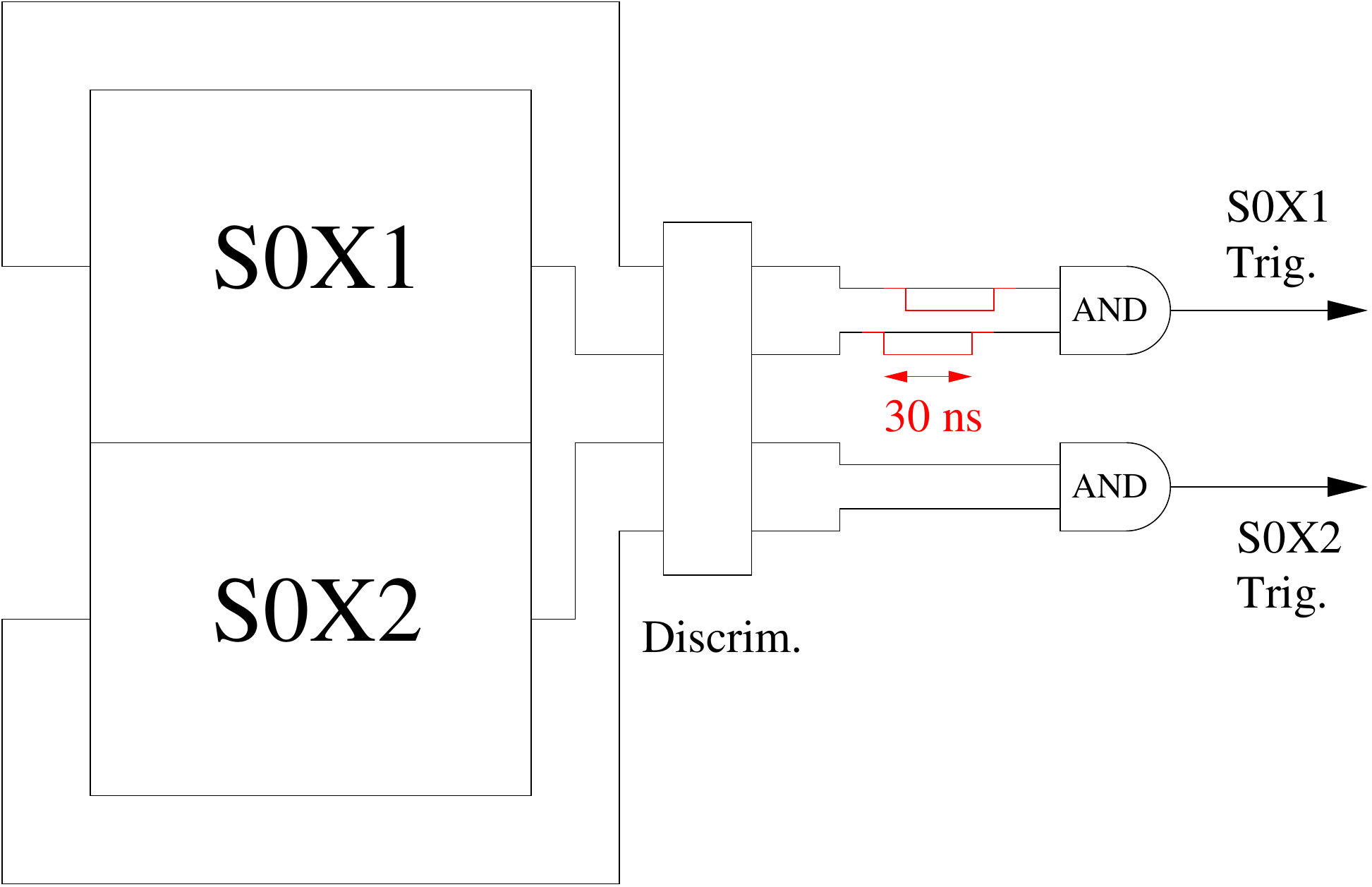}
    \caption{\label{S0trig_diagram} S0 trigger logic diagram. The discriminator output pulses for both S0 and S1 signals are 30 ns wide, defining the coincidence time tolerance for the HMS trigger as illustrated above.}
  \end{center}
\end{figure}
The ``S1'' trigger was formed by requiring both PMTs in at least one paddle from S1X and at least one paddle from S1Y to fire, as shown in figure \ref{S1trig_diagram}. The two HMS triggers were defined by coincidences between S1 and the respective paddles of S0: 
\begin{itemize}
\item HMS1 $\equiv$ S1 AND SOX1
\item HMS2 $\equiv$ S1 AND S0X2
\end{itemize}
This logic is illustrated in figure \ref{COIN_trig_diagram}. Extra delay was added to the signals from the $+$ end of S1X ($-y$ end) so that the S1X+ paddles always determined the timing of the event.
\begin{figure}[h]
  \begin{center}
    \includegraphics[angle=-90,width=.90\textwidth]{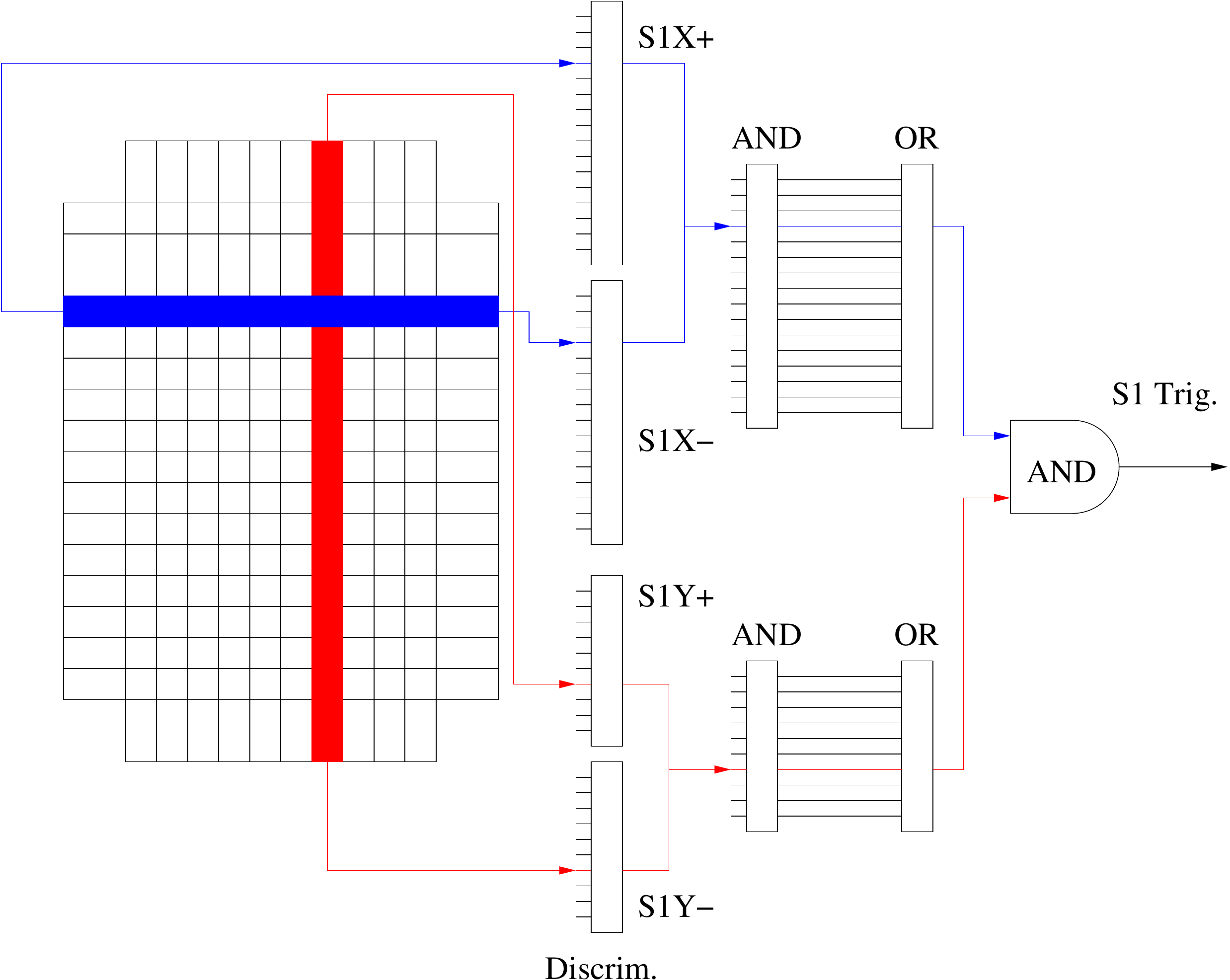}
    \caption{\label{S1trig_diagram} S1 trigger logic diagram.}
  \end{center}
\end{figure}

The HMS trigger electronics used fixed-threshold discriminators with a threshold of 40 mV and a 30 ns output pulse width. The output pulse width determines the coincidence time tolerance in the trigger between the two ends of each paddle, between S1X and S1Y, and between the S0 and S1 triggers. The $\pm$30 ns tolerance was wide enough to allow for all possible differences in light propagation time, walk, and cable and electronic delays within the hodoscope system and trigger electronics.
\subsection{BigCal Electron Trigger}
\paragraph{}
The trigger for BigCal was formed from the ``second-level'' sums of 64 shown in figure \ref{BigCal_trigger_schematic}. Each summing module at the first level has three useful outputs, one of which is sent to a discriminator to produce the 224 timing signals. The overlap scheme in the trigger was implemented by using both of the spare outputs for each first-level sum in the overlap rows 4, 7, 10, 13, \dots. Using identical summing modules, the first-level sums were combined into second-level sums containing up to 64 channels each. The number of summed signals was thus reduced to 38 at the second level. The calorimeter is split into two halves horizontally and 19 overlapping sections vertically. Each group contains four rows and sixteen (or fifteen in the RCS section) columns, except for the last group in each column, each of which only has two rows. The last row of each group is always the first row of the next group, so the channels in these special rows get summed twice, once each in two different groups. The extra identical outputs of the NIM summing modules were quite useful for this purpose. There is no overlap in the horizontal direction. The groups are named according to the scheme $xx$AB/CD, where $xx$ is the row number of the first row included in the group, and AB/CD refers to the labels of the first-level horizontal groups included in each second-level group. For example, group 10AB includes columns 1-16 (groups A and B in each row) in rows 10-13, and group 37CD includes columns 16-30 in rows 37-40. The vertical overlap allows a higher trigger threshold with no loss of efficiency compared to a non-overlapping design. This is easy to understand by simply considering, in the case of no overlap, an electron that impacts the surface of BigCal at the boundary between two trigger groups. Roughly half the shower energy is deposited in each group. This establishes a hard upper limit on the trigger threshold of half the incident energy since any higher threshold would result in efficiency losses at the boundaries of groups. Overlapping groups solve this problem by sharing the rows at the boundaries between groups. The trigger threshold can then be increased above half the incident energy without efficiency loss, except at the vertical boundary between the left and right halves. Increasing the threshold closer to the full incident energy will of course eventually result in a loss of efficiency even with overlap, but the benefit of the overlap scheme is obvious.

Each of the 38 second level sum outputs was sent to one of four sixteen-channel discriminator units, each with a unique remotely programmable threshold (one threshold per-unit, not per-channel), allowing a different threshold to be applied to each of the four quadrants of BigCal. In practice, either the same threshold was applied to all four quadrants, or the left and right halves were operated with slightly different thresholds reflecting the variation of the electron energy with its angle/position at the calorimeter. The discriminator outputs, one for each second-level sum, were then sent to logical fan-in/fan-out units, which had the effect of applying a global OR logic gate to all 38 trigger sums; i.e., if any trigger sum exceeded the threshold, a trigger was generated. All of the summing modules for the first and second-level sums and the trigger were located on the BigCal support platform behind the black box. The output pulses of the discriminators for the second-level sums, the logic FiFo units, and the final OR of the four quadrants to generate the trigger signal to send to the electronics platform were all 50 ns wide. This pulse width defined the time tolerance for the coincidence trigger between BigCal and the HMS.  
\subsection{Coincidence Trigger}
\paragraph{}
Both the HMS and BigCal triggers were sent to the bunker on special fast (~4 ns/m) signal cables which were needed to form a coincidence trigger and send a gate signal to the BigCal ADCs in time for the arrival of the signals from the PMTs. Two coincidence triggers were defined, one for each of the two HMS trigger types. The relative timing of the HMS and BigCal trigger signals was always adjusted so that the BigCal trigger arrived first, and the HMS signal arrived later, defining the timing of each event. 

This timing scheme is illustrated in figure \ref{cointrig_time}. The two histograms on the left show the raw TDC spectrum of the trigger signals from the HMS2 trigger and the BigCal trigger for coincidence events. The trigger signals were digitized in LeCroy 1877 TDCs in the same fastbus crate with the BigCal timing signals. These TDCs have a count resolution of 0.5 ns, so two TDC counts correspond to 1 ns in the raw TDC spectrum. The important feature to notice is that the vast majority of the HMS trigger signals occur at a fixed time, meaning that the HMS trigger sets the timing of the event, including the timing of all the ADC gate and TDC start/stop signals. The constant level of TDC counts at times up to 50 ns earlier than the self-timing peak corresponds to accidental coincidences. Looking at the BigCal trigger TDC spectrum, one sees a lower number of self-timing events. The BigCal self-timing events correspond to accidental coincidences in which an HMS trigger arrives first, and then an uncorrelated BigCal trigger arrives later, giving a false coincidence between the two within the allowed amount of overlap ($\pm$50 ns). At earlier times, the true coincidence peak appears in the BigCal trigger TDC spectrum. The correlation between the HMS and BigCal triggers is shown in the lego plot on the right. Here the relationship is more obvious. The HMS trigger time is plotted on the $y$ axis vs. the BigCal trigger time on the $x$ axis. Most of the events reside at a fixed HMS trigger time with the BigCal trigger time arriving up to 50 ns earlier (Again, since the TDC is operated in common STOP, larger times correspond to earlier times in this plot.). The peak in the BigCal time spectrum along the $x$ axis at a constant HMS time is the true coincidence peak. The small level of events at constant $x$, i.e., constant BigCal time, with the HMS time up to 50 ns earlier, corresponds to accidental coincidence triggers in which the BigCal trigger sets the timing of the event. 
\begin{figure}
  \begin{center}
    \includegraphics[angle=90,width=.45\textwidth]{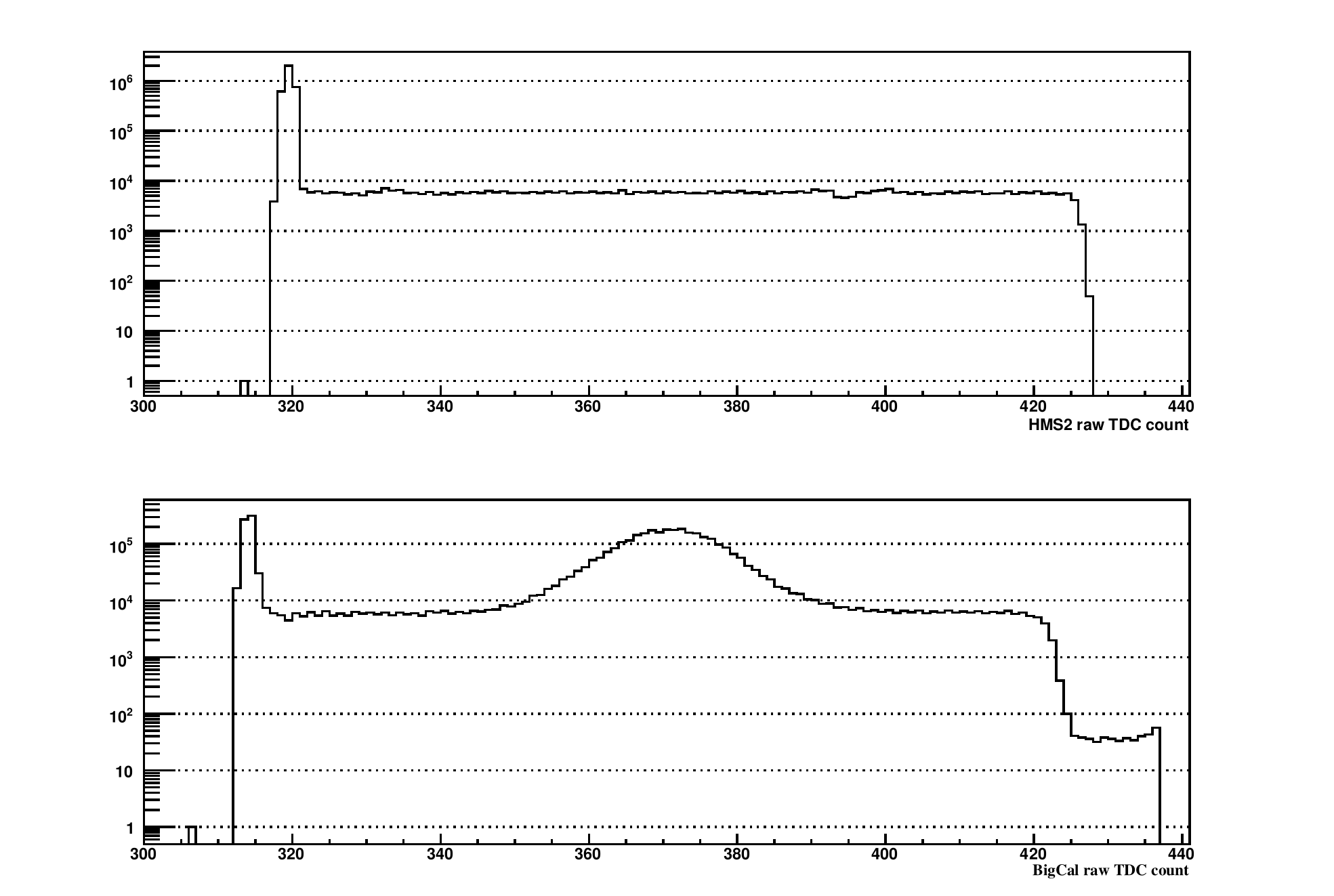}
    \includegraphics[angle=90,width=.45\textwidth]{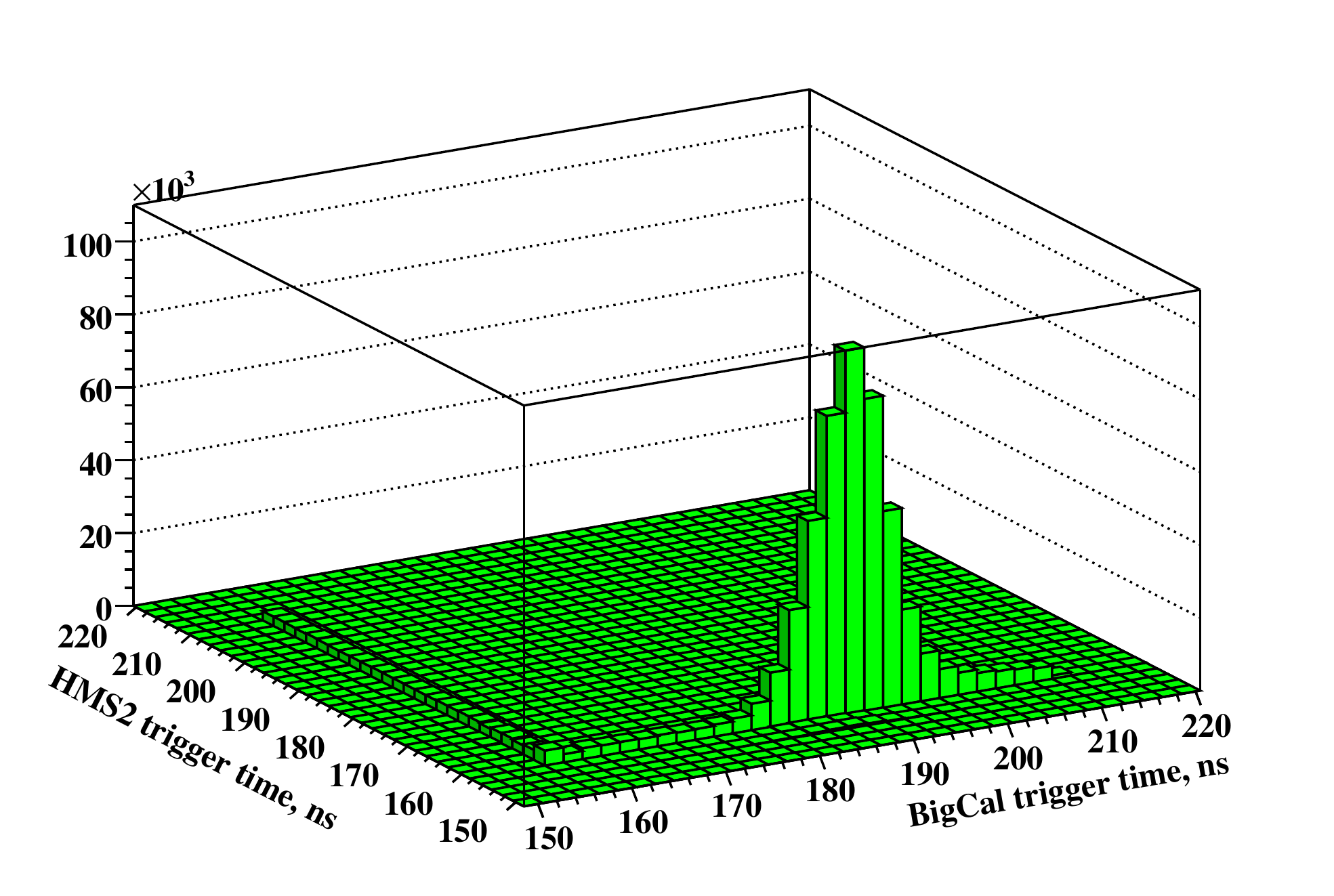}
    \caption{\label{cointrig_time} Raw TDC spectrum (left) of HMS2 (top) and BigCal (bottom) trigger signals. Correlation between HMS2 and BigCal trigger times (right). The TDCs measuring the trigger times were operated in common STOP mode, so larger times correspond to earlier signals.}
  \end{center}
\end{figure}

In total, five different trigger types were defined for production data taking\footnote{The five trigger types mentioned here do not include two additional trigger types used for BigCal. First, a cosmic ray trigger based on the coincidence between scintillators placed above and below BigCal was used to collect muon tracks in BigCal for an initial calibration. Second, a light source trigger was defined to read out events with signals in all 1,744 blocks from the light produced by the gain monitoring system.}, including three single-arm triggers (HMS1, HMS2, and BigCal), and two coincidence triggers (HMS1+BigCal and HMS2+BigCal) as shown in figure \ref{COIN_trig_diagram}. During production data taking, the two coincidence triggers were the most important. The BigCal single-arm trigger was always heavily prescaled because its raw rate was so high it would have overwhelmed the data acquisition system. The two HMS single arm triggers were usually prescaled down to a rate of 1-10 Hz. The ``COIN2'' trigger, coming from the paddle of S0 at the center of the HMS acceptance where elastically scattered protons are focused, was always read out with a prescale factor of 1, meaning all such triggers resulted in reading an event to disk, except when another trigger arrived during the time required to read out the event. The readout time is determined by the size of the event, the number of channels read out, the number of hits, the conversion time for all the digitizing electronics, and the time needed to write the data to disk over the network. A copy of each trigger signal was also sent to VME scalers to monitor the raw trigger rates and the computer deadtime. The ``COIN1'' trigger, which comes from the paddle of S0 covering the inelastic and elastic radiative tail region of the HMS acceptance, was prescaled differently for different kinematics. For the kinematics at larger $\epsilon$, all COIN1 triggers were accepted, since the elastic envelope was spread out over both paddles of S0 (see table \ref{dptable}). On the other hand, for the lowest-$\varepsilon$ kinematics, where nearly all elastic events were contained within the second paddle of S0, the COIN1 triggers were prescaled down to a low rate to reduce the computer deadtime and the number of inelastic events recorded to disk. 
\begin{figure}[h]
  \begin{center}
    \includegraphics[angle=-90,width=.90\textwidth]{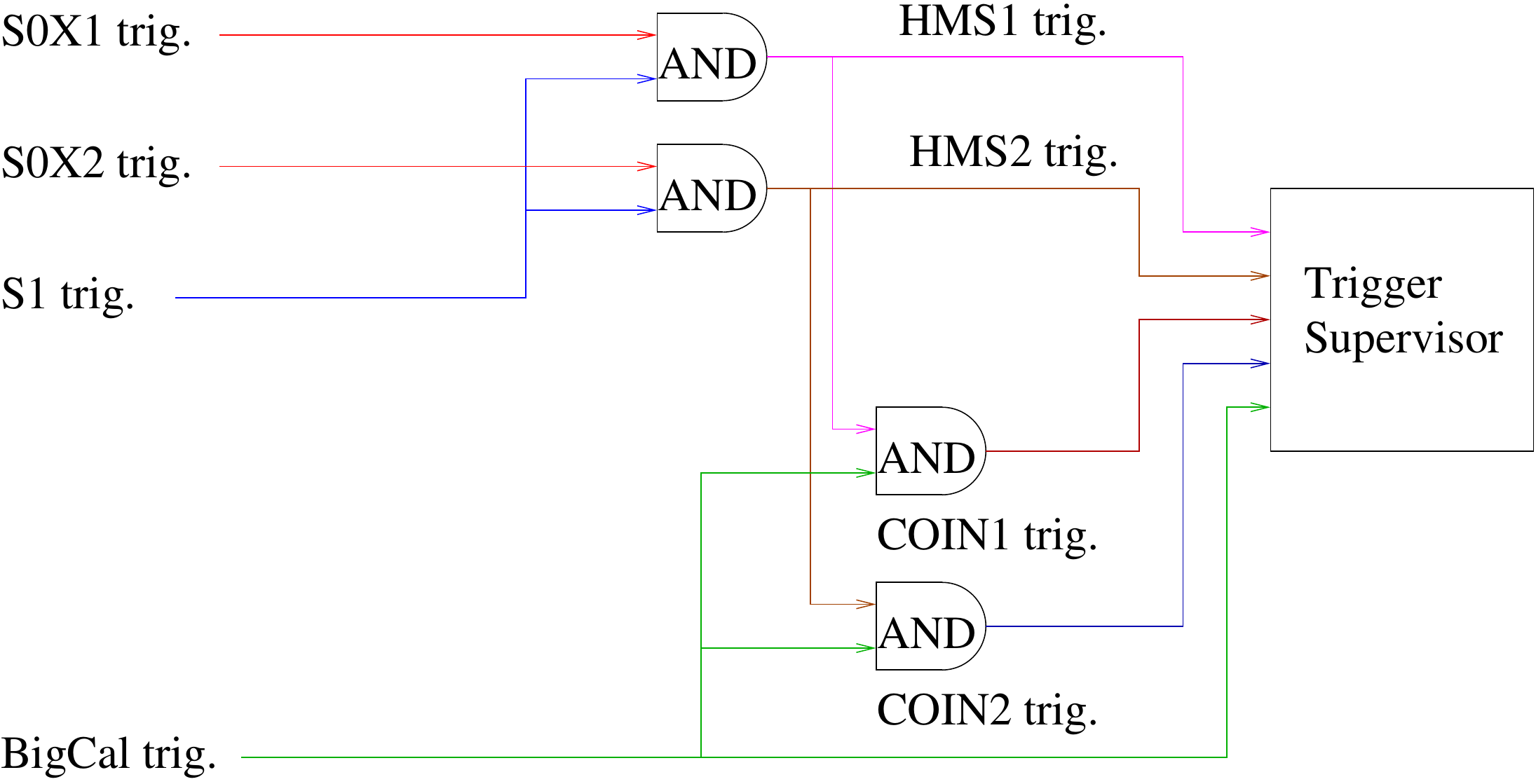}
    \caption{\label{COIN_trig_diagram} Coincidence trigger logic diagram showing the five types of triggers used by the experiment. Each of the five trigger types could be prescaled independently.}
  \end{center}
\end{figure}
\subsection{Helicity Signals}
\paragraph{}
Among the most important signals for experiments using polarized electron beams are the signals indicating the polarization state of the beam. These signals are sent from the Machine Control Center (MCC), which, as suggested by its name, is where the operations of the accelerator are controlled, to the Hall C counting house on fiber-optic cables. The helicity signals are converted to NIM logic levels by a fiber translator unit, from which they are then sent to the data acquisition electronics in several redundant ways. For this experiment, the beam helicity was directly reported and sent with a repeating +-+- structure, changing sign every $1/30^{th}\ s$. Direct reporting means that the sign of the helicity signal corresponds directly to the sign of the beam polarization.

The helicity information is contained in three different signals. First, there is a logic level called $h+$, indicating a positive helicity state. Its absence signals a negative helicity state. Second, there is the so-called ``MPS'' signal, which is a logic pulse of 500 $\mu$s duration, which indicates the ``blank-off'' period during which the helicity state is changing. During the time it takes to reverse the beam polarization, the state of the beam is undefined and the actual beam polarization is unstable. The settling time after each reversal is basically governed by the time required for the Pockels cell voltage to stabilize at the new setting. The MPS signal is used to indicate this uncertain polarization state, and events arriving during this window are discarded. The duration of the MPS window is actually quite conservative and is in fact longer than the time required for the polarization to stabilize after reversal. It also eliminates any jitter/bouncing present during the change of state of the logic level. Since the polarization changes 30 times per second, the MPS window results in a loss of 1.5\% of events. The third signal, the so-called ``quartet'' signal, is a logic pulse arriving at 1/4 the frequency of the polarization reversal signals. Although the quartet signal was put into the data stream via one of the latched inputs of the trigger supervisor, the information was neither needed nor used since this experiment opted for direct helicity reporting. 

The helicity signals taken from the fiber translator were put into the data stream in two places. First, separate $h+$ and $h-$ logic signals were defined. The $h+$ signal was defined as the logical AND of $h+$ and the absence (NOT) of MPS. The $h-$ signal was defined as the logical AND of NOT $h+$ and NOT MPS. These NIM logic levels were then sent to NIM/ECL level translators on the way to ADCs for readout. Since the signals were DC logic levels, their presence in the ADC readout is signaled by saturation of the ADC count. The integration of a large DC level by the ADC always resulted in reading out the full count range. In the absence of a signal, only the pedestal is read out. In the offline data analysis, the presence of h+(-) AND the absence of h-(+) was required in order to assign a beam helicity state of h+(-). Otherwise, a helicity of zero was assigned, indicating an unstable/changing polarization state during readout of the event. 

The $h+$, MPS, and quartet signals were also sent to the electronics platform in Hall C. Using NIM modules located on the platform, redundant $h+$ and $h-$ signals were formed with the same definition as the counting house signals. These signals were read out using two of the latched inputs of the trigger supervisor. In the offline analysis, as in the case of the ADC signals, the combination of h+(-) AND NOT  h-(+) was required in order to form h+(-); otherwise, a zero value was assigned. Putting the helicity information in the data stream in two different places in redundant ways guaranteed that the beam polarization state was known with high confidence. In the final analysis, agreement between the trigger supervisor helicity and the ADC helicity was not required, since a small percentage of all data runs had unreliable ADC information. Except for the runs with bad ADC information, the trigger supervisor and the ADC agreed on the beam helicity state for roughly 99.9\% of events.
\begin{figure}[h]
  \begin{center}
    \includegraphics[angle=90,width=.95\textwidth]{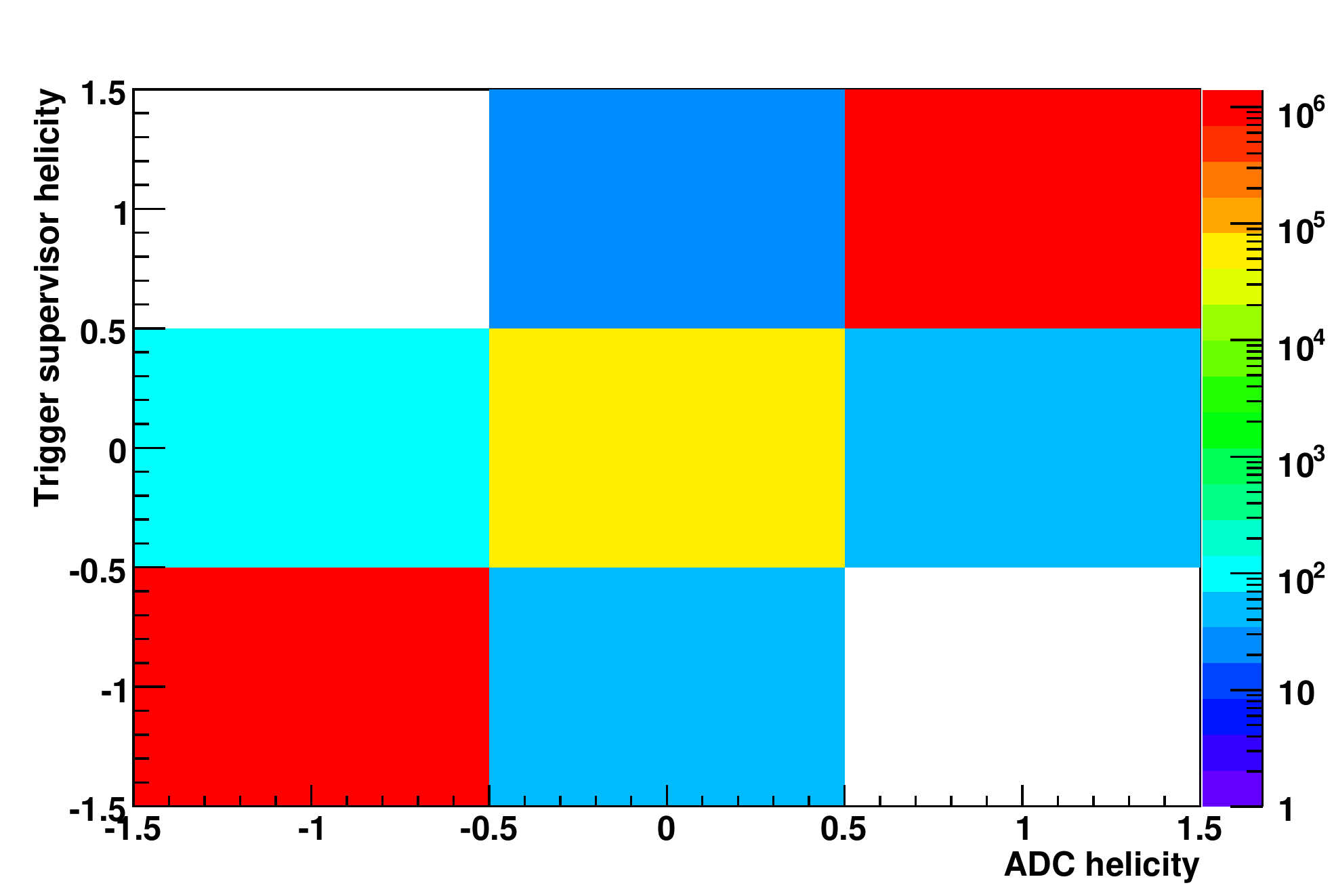}
  \end{center}
  \caption{\label{Helicity_comp} Correlation between the trigger supervisor helicity signal and the counting house ADC helicity signal for CODA run 69842. For this run, the two signals agree for roughly 99.99\% of events. The average beam current for this run was approximately 67 $\mu$A.}
\end{figure}

The mechanism for disagreement between the two signals is the slight difference between the times at which they are recorded. The helicity state at the trigger supervisor is latched at the instant an event trigger is formed. This signal is delayed by the length of cable required to send it from the counting house to Hall C, which is on the order of several hundred ns. The trigger supervisor records what the state of the fiber-translator helicity signal was at a time $\Delta t$ \emph{before} the event readout was triggered. On the other hand, the counting house helicity signal is undelayed except for short cable and electronic delays between the fiber translator and the ADC. But it is read out only when the gate arrives from the trigger supervisor, which is $\approx \Delta t$ \emph{after} the event readout is triggered. Therefore, the trigger supervisor and the ADC measure the beam helicity state at two different times separated by $\approx 2 \Delta t$ on the order of one microsecond. This time difference is still quite small compared to the 30 Hz reversal rate and the 500 $\mu$s blank-off period, but there is still a non-negligible chance of disagreement between the two signals, particularly given the possibility of noise giving an incorrect readout in the ADC. Figure \ref{Helicity_comp} shows the correlation between the trigger supervisor and ADC helicity signals, plotted on a logarithmic color scale. The color scale shows the overwhelming agreement between the signals, with more than 99.99\% of events lying on the diagonal. The fraction of events with ``zero'' helicity is 1.5\% as expected. Because the MPS window is 500 $\mu$s, and the time difference between the TS and ADC helicity signals is of order 1 $\mu$s, it should be physically impossible for one signal to give a positive helicity and the other signal to give a negative helicity. In this example, there are no events in which such a combination of signals occurs. In some runs, however, a small number of events appeared in the (1,-1) and (-1,1) helicity bins. Such events indicate a problem with one of the signals, and were most likely caused by noise in the ADC signals.

In the final analysis, the trigger supervisor helicity was used regardless of the ADC information, which mainly served as an independent on-line check of the validity of the signals. To the extent that the fiber-translator helicity signal represents the true beam polarization state at the instant of the reaction responsible for triggering an event readout, the comparable delays involved in sending the helicity information to Hall C and in sending the trigger signals from the detectors to the electronics platform puts the trigger supervisor helicity signals closer in time to the real event than the ADC signals. However, in the overwhelming majority of events, the polarization state does not change during the entire time from scattering in the target, to the formation of raw signals in the detectors, to the trigger arriving at the trigger supervisor, to the arrival of the gate at the counting house ADCs, meaning that the beam helicity state is totally unambiguously determined in very nearly 100\% of events. Events arriving during the MPS window are assigned a polarization state of zero and are eventually discarded in the offline analysis.
\subsection{Pedestal (Pulser) Trigger}
\paragraph{}
At the beginning of each data run, 1000 events from random triggers generated by a pulser are recorded in order to measure the centroids and widths of the pedestals of every ADC channel in the experiment. This is accomplished through the ``enable 1''(EN1) and ``GO'' output bits of the trigger supervisor which control the state of the run. The GO signal enables the inputs of the trigger supervisor when a run is started. The EN1 signal is used to disable all triggers except the pedestal trigger until 1000 pedestal events are acquired. This behavior is enforced by wiring the external trigger electronics so that the pulser trigger is ANDed with the combination GO AND NOT EN1 and all other triggers are ANDed with the combination GO AND EN1. Once the first 1000 events are acquired, the trigger supervisor is programmed to turn on the EN1 bit which blocks pedestal triggers and enables all other triggers. 

Another difference between the pedestal events and physics events is that whereas physics events are read out with sparsification, pedestal events are read out without sparsification. Sparsification means that the only ADC channels read out are those with ADC count values above a threshold which is programmed into the readout controllers (ROCS) in each Fastbus or VME crate. Additionally, it means that only those TDC channels in which a stop signal is present are read out. This reduces the event size and readout time and hence the computer deadtime. During the pedestal events, on the other hand, sparsification is turned off, meaning every ADC channel is read out, which allows measurement of the pedestal positions and widths in every channel. 

After the acquisition of the pedestal events, sparsification is turned on and the thresholds loaded into the ROCs are used to speed up the data acquisition for physics events. The ADC information from the pedestal events can then be analyzed to monitor the pedestal positions and widths. The desired readout threshold for the ADCs is typically calculated to be 2.5 pedestal widths above the pedestal position. The actual readout thresholds programmed into the ADCs by CODA are put into the data stream in special ``GO info'' events at the start of each run. These thresholds are then compared for every run to the desired threshold calculated from the analysis of the pedestal events and a warning message is printed if the desired threshold differs from the actual threshold by more than two $\sigma$. Separate text files containing the desired thresholds, one for each Fastbus crate with ADCs, are generated for each run. This file is in the proper format so that it can be optionally loaded into CODA at the start of the next run simply by replacing the threshold file read in by CODA with the new threshold file. Changes in the pedestal position, width, and desired threshold occur with changes in the timing of trigger and gate signals relative to the arrival time of the analog signals to be digitized, changes in detector high voltage, changes in the gate width, and changes in the experiment kinematics which result in different particle rates and/or signal sizes.

The pedestal events are an excellent feature of the Hall C data acquisition setup, since they allow the automatic calculation of new pedestal positions, which are then subtracted from the data to get the true signal size, and new pedestal widths, which can be used to calculate software thresholds that can optionally be used for additional suppression of the pedestal in the data analysis. The software threshold option is particularly useful for the relatively harmless situation in which the readout threshold is set too low. The ability to compare the readout threshold to the desired threshold for every run also protects against the quite harmful situation in which the readout threshold is set too high, resulting in a loss/artificial suppression of data.
\subsection{Data Acquisition}
\subsubsection{Electronics: Trigger Supervisor and Readout Controllers}
\paragraph{}
The primary element of the data acquisition system is the Trigger Supervisor (TS). Detailed descriptions of the JLab trigger supervisor system can be found in \cite{TS_manual,TS_proceedings}. In this section a brief description of its functionality pertaining to this experiment is provided. The TS is used to control the state of the data run and the readout of all the various branches of the data acquisition system when a run is in progress. It is also used to control the pre-scaling of the various trigger types. The pre-scale factors are programmed into the TS by CODA whenever a run is started, and are used to reduce the number of triggers of a given type accepted. For example, a prescale factor of 10 means that only one out of every 10 triggers is accepted. 

The TS is located on the electronics platform in the bunker in Hall C in a VME crate along with a number of scaler modules used to monitor the hit rates for the first and second-level sums in BigCal, the various triggers, and the beam current monitoring devices. It is daisy-chained to the various branches of the data acquisition system through branch cables. There are three main branches. The first branch, located on the electronics platform, includes the two FastBus crates used to read out BigCal. The second branch, located in the Hall C counting house, includes one FastBus crate and one VME crate. The FastBus crate contains ADCs and TDCs which are used to read out the HMS hodoscope and calorimeter signals, and the BPM and raster signals. The VME crate contains scalers which monitor the hit rate on the HMS hodoscopes. The third branch, located in the HMS hut, includes the crates used to read out the HMS and FPP drift chambers. As discussed above, this branch of the DAQ system used two VME crates with F1 TDCs during the first phase of the experiment, and a single Fastbus crate during the second phase of the experiment for the FPP drift chambers. The HMS drift chambers were always read out by a single Fastbus crate. 

Each individual crate is controlled by a single-board CPU called a readout controller or ROC for short. The ROCs are connected to the Hall C network via either ethernet or fiber-link. Each crate's ROC collects the data from all ADC, TDC and/or scaler modules in its crate into ``banks'' in memory and attaches header information such as the identifiying information of the ROC and the bank length in data words, followed by the individual data words. Each ROC's data bank becomes a fragment of a physics event that is eventually assembled together with all the other ROC banks into a complete physics event by CODA as discussed below. The trigger supervisor itself is also a ROC and has its own data bank which enters the data stream. The trigger supervisor ROC bank contains the status of its 12 input bits, including 8 pre-scalable trigger inputs and 4 additional latched inputs which were used for the helicity signals in this experiment. When special scaler read events are triggered, the data from the scaler modules located in the same VME crate is also read into the TS ROC bank. The trigger supervisor communicates with the ROCs and controls their readout functions via twisted-pair ribbon cables (the branch cables). It can support up to 32 ROCs on four branches (up to 8 ROCs on each branch). Each branch has an 8-deep FIFO buffer memory.

When the trigger supervisor is in a state of readiness to accept new triggers and receives a signal on one of its trigger inputs, it generates a signal called a level 1 (L1) accept, which is fanned out to various locations to form gate, start, and stop signals for the various ADCs and TDCs. On the electronics platform, it goes to a gate generator which creates the gate for the BigCal ADCs. It is also used to stop the TDCs in the second BigCal fastbus crate, which include timing signals for the first and second-level sums of BigCal and the HMS1, HMS2, and BigCal triggers. It is also sent to the upstairs counting house, where it is used to start the TDCs for the HMS hodoscope signals and to gate the hodoscope and calorimeter ADCs. Finally, it is sent to the HMS hut where it is used to stop the TDCs for the HMS and FPP drift chambers. Whenever each crate receives a gate/start/stop signal, an event is created in the memory buffers of the modules. Each crate's ROC, controlled by the trigger supervisor, serves as the interface between the individual digitizing modules and the outside world. 

Upon accepting a trigger which passes the prescale circuits and generating a level 1 accept, the trigger supervisor locks up and enters a latched state. A 4-bit readout code, permitting up to 16 independent readout functions, is calculated from the 12-bit trigger latch pattern in a user-programmable fashion. This readout code is sent along 4 data lines on the branch cable. If none of the branch buffers is full, the TS drops the L1 accept signal and is ready to accept new triggers. If, on the other hand, any of the branch buffers is full, the TS holds its state until space is available in all buffers. Upon loading the readout code into the branch buffer a ``strobe'' signal is sent to all the ROCs on each branch telling the ROCs to read out their event fragments according to the readout code. Once the ROCs are finished processing their data, they send an acknowledgement (``ack'') signal back to the trigger supervisor. Each ROC has its own dedicated ``ack'' line on the branch cable. When the trigger supervisor receives the ``ack'' signal from all ROCs on the branch, the readout code data lines are reset, the buffer counter is decremented, and the strobe signal is dropped. Each ROC drops its ``ack'' signal upon detecting the absence of the strobe signal.

In order to accommodate front-end modules with no buffering capability, the trigger supervisor supports an unbuffered data acquisition mode, in which all branch buffers are programmed to have a buffer depth of one. In this mode, no new triggers can be accepted until all the ROCs have finished processing their event data. Although all of the ADC and TDC modules used in this experiment were capable of buffering, the unbuffered data acquisition mode was used during the vast majority of the production data collection. When running in buffered mode, careful synchronization between all the ROCs and the trigger supervisor is necessary to prevent catastrophic data loss in a situation in which the individual event fragments from the ROCs do not correspond to the same physical trigger. The main benefit of using buffered mode DAQ is the reduction in computer deadtime obtained by allowing the triggering of new events before the acquisition of previous events is completed. However, the rates on the main coincidence trigger were low enough across all kinematics that the computer deadtime was acceptable even in the unbuffered mode, which is both simpler to implement and less fraught with potential hazard. Given the size and complexity of the non-standard data acquisition requirements of this experiment, the unbuffered mode was safer and easier to use. But it is worth noting that the choice of unbuffered mode was not merely one of convenience. At one point during the run, memory issues in the F1 TDC modules in one of the VME crates used to read out the FPP chambers caused them to fail to decrement their event counters after readout. Detailed investigations showed that even though the modules in fact had only one event per trigger, they erroneously showed extra events. This issue prevented buffered mode data acquisition for as long as it persisted.
\subsubsection{Computer Deadtime and the S0 Trigger}
\paragraph{}
The DAQ or computer deadtime is the fraction of time the data acquisition system cannot accept new triggers because the data acquisition computer is busy reading out the previous event. It is a function of the rate of triggered events and the time required to read out an event. The readout time is determined by several factors, including the data conversion time in the digitizing modules, the time required by the ROCs to read the data from the modules into their memory, and the time required for the ROCs to send the data out over the network. The latter requirement, which depends on the available network bandwidth and the volume of data being sent out, is by far the most time consuming. The data conversion time for 1881M ADCs is on the order of several tens of $\mu$s, while the time required to send the data for all hit channels out over the network is on the order of several hundreds of microseconds. The total event processing time, which takes several hundred microseconds, sets an absolute upper limit on the data acquisition rate, whether running in buffered or unbuffered mode, which was typically 3-5 kHz for the configuration of this experiment. The computer deadtime was monitored during the experiment by comparing the number of (prescaled) incident triggers to the number of triggers read out. Both numbers are counted by scalers. The readout time can be inferred from the computer deadtime and the event rate. It depends on the average event size, which can be greatly reduced through sparsification, the bandwidth of the Hall C network, and the available memory, processing, and read/write speed of the computer on which the data acquisition software processes are hosted.
\begin{table}[h]
  \begin{center}
    \begin{tabular}{|c|c|c|c|} %
      \hline $Q^2(GeV^2)/\epsilon/I_{beam}$ & HMS1/ps1 & HMS2/ps2 & BigCal/ps3 \\ \hline
      5.2/.377/74.2 & 33.6/99999 & 11.3/99999 & 7044.8/99999 \\ \hline
      2.5/.154/80.5 & 1303.0/99999 & 67.8/99999 & 10638.4/99999 \\ \hline
      2.5/.633/100.0 & Disabled & 467.7/99999 & 1775.3/99999 \\ \hline 
      2.5/.789/84.0 & Disabled & 240.0/99999 & 2990.3/99999 \\ \hline
      6.8/.507/73.3 & 7.6/2000 & 4.7/2000 & 32640.4/99999 \\ \hline
      8.5/.236/83.1 & 407.7/1000 & 17.7/500 &  39319.4/99999 \\ \hline
    \end{tabular}
    \begin{tabular}{|c|c|c|c|}
      \hline $Q^2(GeV^2)/\epsilon/I_{beam}$ & COIN1/ps4 & COIN2/ps5 & DAQ Deadtime \\ \hline
      5.2/.377/74.2 & 7.2/10 & 1.9/1 & 15.9\% \\ \hline
      2.5/.154/80.5 & 168.9/100 & 20.4/1 & 21.9\% \\ \hline
      2.5/.633/100.0 & Disabled & 21.7/1 & 25.5\% \\ \hline
      2.5/.789/84.0 & Disabled & 19.9/1 & 18.9\% \\ \hline 
      6.8/.507/73.3 & 2.5/1 & 1.6/1 & 7.8\% \\ \hline
      8.5/.236/83.1 & 158.0/2000 & 7.3/1 & 11.9\% \\ \hline
    \end{tabular}
    \caption{\label{TrigRatesDeadtimes} Typical trigger rates/prescale factors and computer deadtimes observed during the experiment at different kinematics. Each table entry corresponds to one data run. Rates are given in Hz/$\mu$A. The examples in which the HMS1 and COIN1 triggers are marked as ``disabled'' refer to an alternate trigger configuration which was used for kinematics for which the S1-only trigger rate was low enough that the S0 trigger was unnecessary. Though S0 remained in the detector stack for these runs, only one HMS trigger (HMS2) was defined which used the S1 trigger logic described above. The COIN2 trigger was defined as S1 AND BigCal for these kinematics.}
  \end{center}
\end{table}

Table \ref{TrigRatesDeadtimes} shows a sampling of the trigger rates and computer deadtimes observed across the various kinematics of the experiment when running at or near the full beam current. During the data taking of the two highest-$\varepsilon$ kinematics, a problem with the discriminator for the S0 trigger reduced the rate and efficiency of that trigger. Additionally, there were problems with the signals coming from S0 itself. Though these problems were later fixed, the decision was made at the time to take data with an HMS trigger which bypassed S0 and used only the S1 trigger logic shown in figure \ref{S1trig_diagram} for the HMS single arm trigger. Since the trigger rates in BigCal (see table \ref{TrigRatesDeadtimes}) were lower for these kinematics, which involved large distances to BigCal, high electron energies and high trigger thresholds, the accidental coincidence rate between the S1-only trigger and BigCal was manageable. 

One immediately notices upon close examination of table \ref{TrigRatesDeadtimes}, that the HMS single-arm trigger rates were much higher with an S1-only trigger. At an angle of 35.4 degrees, the S1-only trigger rate was 240 Hz/$\mu$A, and at 31 degrees, that rate increased to almost 500 Hz/$\mu$A. S1-only trigger data were only taken at three different HMS angle settings, which include those listed in table \ref{TrigRatesDeadtimes}, and the data taken with an HMS central angle of 36.1 degrees, for which the S1-only HMS trigger rate was 220 Hz/$\mu$A. All three of these HMS angle settings had the same central momentum setting of 2.0676 GeV/c. Since the singles rate in the HMS depends on the beam energy, HMS central momentum, and HMS central angle, one cannot directly extrapolate from those three data points to predict what the S1-only trigger rate would have been for all the other kinematics. But an educated guess can be made using the S1 scaler data. Although there were scalers in the data stream for each of the individual paddles of S1X and S1Y for logical combinations of the paddles related to the \emph{standard} HMS trigger, there was no dedicated scaler channel counting the frequency of the logical combination of S1 signals used in the trigger for this experiment, with the exception of the runs when the trigger was defined by S1 only.

Despite the lack of a dedicated S1 trigger scaler for the kinematics where the S0 AND S1 trigger was used, the S1-only trigger rate could be estimated from the S1 scalers by noting the fact that for the three different kinematic settings where the S1-only trigger was used, the ``hS1'' scaler rate was observed to be proportional to the S1 trigger rate, to a very good approximation. The hS1 signal is an exclusive OR of the ``S1X'' and ``S1Y'' signals in the upstairs counting house, which are not quite the same as the S1X and S1Y signals in the custom hut trigger. The difference is that while the hut signal requires hits on PMTs on both ends of the same paddle, the counting house signal requires at least one hit on both sides, but removes the requirement that the two hits come from the same paddle, in the interest of reducing the amount of electronics needed.
\begin{table}[h]
  \begin{center}
    \begin{tabular}{|c|c|c|c|c|}
      \hline
      $Q^2/\varepsilon$ & hS1 & est. S1 trig. & BigCal & Accidental(est.), Hz $@$ 80 $\mu$A \\ \hline
      5.2/.377 & 3278 & 207 & 7045 & 907 \\ \hline
      2.5/.154 & 12335 & 778 & 10643 & 5064 \\ \hline 
      6.7/.507 & 4678 & 295 & 32640 & 5416 \\ \hline
      8.5/.236$^a$ & 10044 & 634 & 53231 & 17534 \\ \hline
      8.5/.236$^b$ & 6372 & 402 & 39319 & 8663 \\ \hline
    \end{tabular}
  \end{center}
  \caption{\label{RandomRateTable} Estimated accidental coincidence rate between BigCal and the HMS had an S1-only trigger been used, for the kinematics which used S0 in the trigger. All rates are given in Hz/$\mu$A except the accidental rate which is given in Hz at 80 $\mu$A, which was the typical beam current during the experiment. The two different runs shown at 8.5 GeV$^2$ were taken with (a) all sixteen paddles of S1X turned on, near the beginning of data taking, and (b) with eight paddles of S1X turned off in order to reduce the rate. The four paddles closest to each end of the acceptance were turned off since the middle eight paddles were sufficient to cover the full elastic envelope. The reduction in the BigCal rate from (a) to (b) comes mainly from the effects of radiation damage and slight adjustments in the trigger threshold.}
\end{table}

The accidental coincidence rate between two signals, produced randomly at average rates $N_1$ and $N_2$, within a coincidence time window $\Delta t$, is governed by Poisson statistics. Given the presence of the first signal, the probability that a random hit from the second signal arrives within $\pm\Delta t/2$ is given by $P(2|1) = \sum_{n=1}^\infty (N_2 \Delta t)^n e^{-N_2\Delta t} / n! = e^{-N_2\Delta t} (e^{N_2 \Delta t} - 1) = 1 - e^{-N_2\Delta t}$. Similarly, given the presence of signal 2, the probability of a random hit from signal 1 during $\Delta t$ is $P(1|2) = 1 - e^{-N_1 \Delta t}$. To get the random coincidence rate, since the probabilities $P(1)P(2|1)$ and $P(2)P(1|2)$ describe non-mutually-exclusive events, the probability for random coincidences must be written as $P(1\cap 2) = P(1)P(2|1) + P(2)P(1|2) - P(1)P(2) = N_1\Delta t(1-e^{-N_2\Delta t}) + N_2\Delta t(1-e^{-N_1\Delta t}) - N_1 N_2 (\Delta t)^2$. Thus, in terms of rates, $dN_{random}/dt = N_1(1-e^{-N_2\Delta t})+N_2(1-e^{-N_1\Delta t})-N_1N_2\Delta t$. For $N_1\Delta t, N_2\Delta t \ll 1$, this becomes $1-e^{-N_{1,2}\Delta t} \approx N_{1,2} \Delta t$, so that the random coincidence rate reduces to $dN_{random}/dt \approx N_1 N_2 \Delta t$.

Table \ref{RandomRateTable} shows why the S0 trigger scintillator was absolutely necessary for these experiments. The ``hS1'' column shows the actual scaler rate, in Hz/$\mu$A, of the ``hS1'' exclusive OR signal observed during the experiment. The second ``S1 trig.'' column shows the expected S1-only trigger rate, assuming that said rate is directly proportional to the hS1 rate. The BigCal column shows the actual BigCal scaler rate observed in the experiment, in Hz/$\mu$A. The final column shows the expected accidental coincidence rate between the S1-only and BigCal triggers, using the full expression for the rate rather than the approximation $N_1N_2 \Delta t$, since the BigCal rate is large enough that at 80 $\mu$A, the condition $N\Delta t \ll 1$ is not satisfied, and the difference between the full expression and the approximate expression is non-negligible. These estimates show that the introduction of S0 was absolutely necessary, since the rate of random coincidences alone with an S1 trigger would have met or exceeded the maximum data acquisition rate in Hall C even under an optimistic assumption of 5 kHz maximum event rate. Reading out random coincidences with a computer deadtime near 100\% would have meant a loss of most if not all elastic ep statistics. Hence, a tighter trigger was required, not optional.  

To conclude the digression into the trigger challenges facing this experiment, it is worth remarking that several alternatives to placing new scintillators upstream of the tracking chambers existed. One alternative was to place the scintillators between the two drift chambers, near the focal point of the spectrometer. This option would have had the benefit of reducing the optical magnification of the angular errors introduced by multiple scattering, but at the expense of reducing the lever arm for the trigger and possibly reducing the track reconstruction efficiency and accuracy, to the extent that significant multiple scattering in between the two drift chambers would make a straight-line approximation to the track less accurate\footnote{Also, neither drift chamber reconstructs the trajectory angles with sufficient precision on its own to reconstruct the multiple scattering angles in the scintillator between the two chambers.}. Nevertheless, this alternative would have resulted in better resolution of the trajectory angles at the target, but may not have provided the needed rate reduction. 

A second alternative would have been to include the HMS shower counter in the trigger with no additional scintillators. The advantage of this approach is that the best possible tracking resolution would have been preserved. The obvious drawback is that a calorimeter trigger would have been quite inefficent, for two reasons. First, since the signals in the lead-glass calorimeter are based on the collection of Cerenkov light, and since protons are relatively heavy, the calorimeter signals for protons would have been small, with large statistical fluctuations, leading to either relatively inefficient triggering or thresholds too low to provide the needed rate reduction. Second, many protons that undergo useful scattering in the first polarimeter would not hit the calorimeter and therefore be lost. A calorimeter trigger probably would have resulted in an unacceptable loss of efficiency given the design of the FPP and the proton momenta and scattering angles involved in this experiment. On the other hand, such a trigger might become useful and acceptable at higher energies, since the angular distribution of the efficiency and analyzing power for scattering in CH$_2$ becomes concentrated at increasingly forward angles, and since the proton signal in lead-glass would increase at higher momenta. Indeed, given the increasing importance of the angular and momentum resolution at high energies to maintaining the systematic uncertainty in the precession calculation and the inelastic background contamination at acceptable levels, a hadronic-calorimeter based trigger may very well become the best available option in future recoil polarization experiments. 
\subsubsection{CODA Software}
\paragraph{}
CEBAF Online Data Acquisition (CODA) is a system of software libraries used to program and provide real time control of the data acquisition system\cite{CODA_proc}. The CODA libraries handle all parameters of the run, including prescale factors, ADC thresholds, the number of pedestal triggers required before accepting physics events, which ROCs and which modules from the crates they control are to be read out and in what fashion, the buffering depth, the programmable parameters of all the various ADC and TDC modules, and so on\dots CODA also provides a graphical user interface called RunControl for starting and ending data runs and easily transitioning between different data acquisition configurations. Among the most important programs in the CODA libraries is the event builder (EB), which runs on the host computer and, when a run is in progress, assembles the incoming event fragments (ROC banks) from the different branches of the DAQ system into full physics events in a standard format, attaching important header information. Once the event is built, it is placed in a buffer where it can then be accessed by other processes or written to disk. In this experiment, the events were written straight to disk, and the data were monitored for quality and analyzed by a separate code, the Hall C ENGINE.

There are four broad classes of events defined in the Hall C CODA setup. \textbf{Status events}, inserted into the data stream at the beginning and end of run, contain basic parameters of the run, including prescale factors, the ADC thresholds that were programmed into the modules, and some slow control variables monitored through EPICs. \textbf{Physics events} correspond to experiment triggers, and include all of the detector and beamline\footnote{The BPM and raster magnet signals are recorded for every physics event by ADCs in the fastbus crate in the counting house.} information recorded by ADCs and TDCs when a physics trigger causes these modules to be read out. \textbf{Scaler read} events, forced into the data stream by CODA once every two seconds, cause all the scaler modules to be read out. These include the BigCal, trigger, and beam current scalers in the VME crate containing the trigger supervisor in the bunker in Hall C, and the hodoscope scalers in the VME crate in the Hall C counting house. Finally, \textbf{EPICS events}, inserted by CODA every thirty seconds, are used to record slow control information, such as spectrometer magnet settings, beam, target, and accelerator parameters, and other important information. 
\begin{figure}[h]
  \begin{center}
    \includegraphics[angle=-90,width=.95\textwidth]{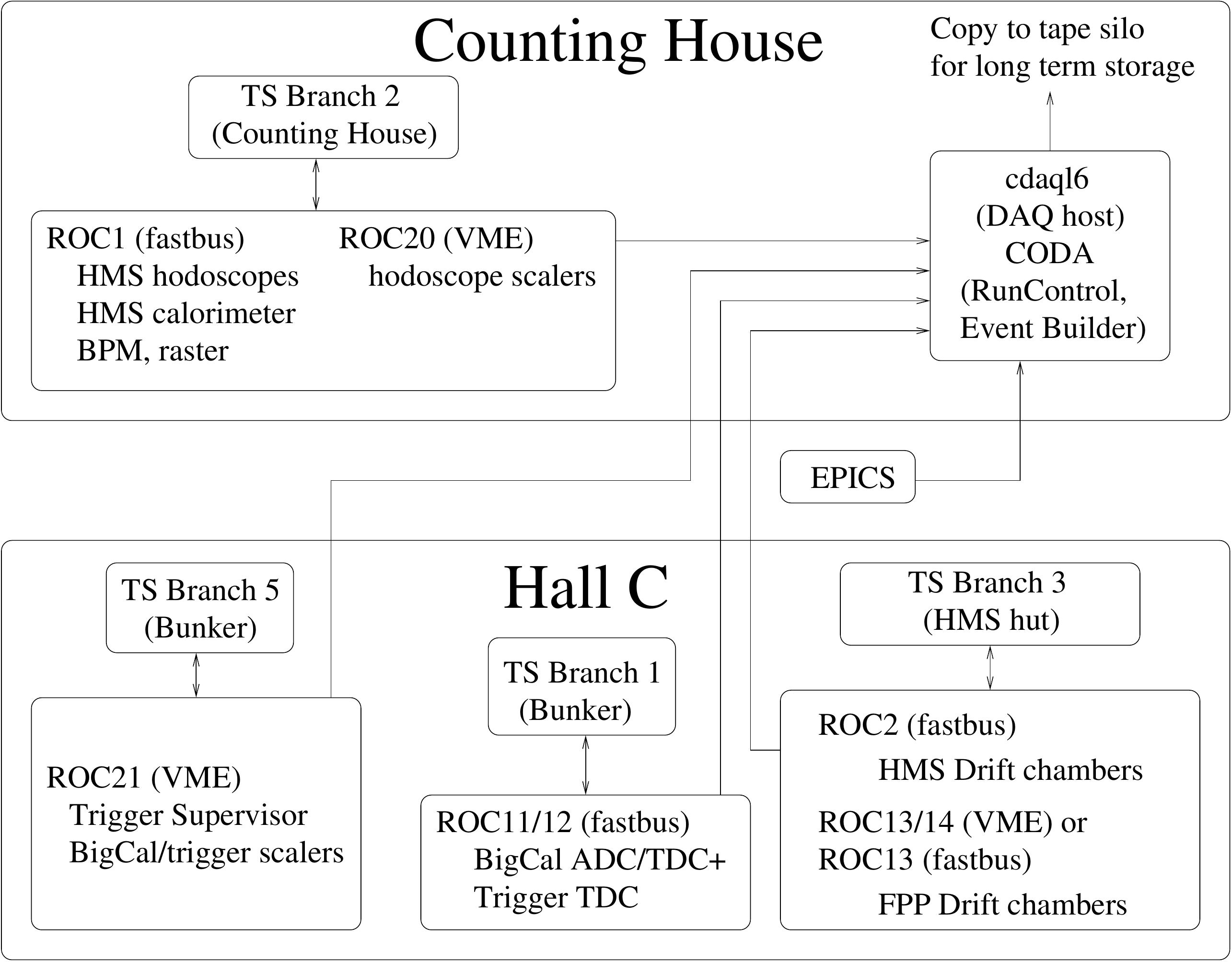}
    \caption{\label{DAQschematic} Schematic layout of $G_E^p$ Hall C DAQ configuration. Each crate of digitizing modules is controlled by a ROC, which communicates with the TS via branch cables. The TS accepts a trigger if and only if it is no longer busy reading out the previous event. When a trigger is accepted, a L1 accept signal goes out to external circuitry to generate ADC gates and TDC start and/or stop signals. Then, the TS signals the ROCs to read out their data via the branch cables, holding a busy state until all ROCs signal they are finished processing their data, at which point the TS is ready to accept new triggers. The event fragments from each ROC go out over the Hall C network to the Event Builder running on the host computer, which assembles the fragments into a complete physics event and passes it to other processes which write the data to disk. Eventually, the data is copied to tape for long-term storage.}
  \end{center}
\end{figure}

Figure \ref{DAQschematic} summarizes the layout of the special data acquisition configuration for this experiment. The data acquisition computer cdaql6 hosts the CODA run control and event builder processes, and writes the data directly to disk for efficient acquisition and short term storage. Because of the high volume of incoming data from the experiment, the hard disk on cdaql6 periodically became full. A cron job running on cdaql6 automatically copied the oldest data files on disk to the tape silos of the JLab Mass Storage System (MSS), then deleted them from the disk, only after verifying their successful transfer to tape, to keep enough free space for the incoming data at all times. The typical event size was roughly 2 KB per event. With typical event rates on the order of 1 kHz, the data rate from the experiment was roughly 2 MB/s, a total data throughput which was comfortably within the capabilities of the modern network and disk hardware in hall C.

\chapter{Data Analysis}
\paragraph{}
The analysis of the data consisted of three major tasks. The first task was the reconstruction of individual scattering events. After decoding the raw data from the HMS and BigCal, the trajectories and momenta of the scattered particles were reconstructed, as well as the position of the interaction vertex along the length of the target. The scattering angles of the proton in the CH$_2$ analyzer of the FPP were also reconstructed. The second task was the separation of elastic and inelastic events, which involved studying kinematic correlations, including the correlation between the detected proton's momentum and its scattering angle, and the correlation between the electron scattering angle and the proton momentum. The third task was the extraction of the polarization transfer components $P_t$ and $P_l$ from the observed angular distribution in the FPP, which involved first measuring the asymmetry and then calculating the proton spin precession in the HMS magnets for each event.
\section{Event Reconstruction}
\paragraph{}
In the following section, the decoding of the raw data, the calibration of the various detectors, and the reconstruction of relevant kinematic variables and FPP scattering angles for each event is discussed.
\subsection{Event Decoding}
\paragraph{}
A CODA event in the raw data file consists of an array of 32-bit integer data words. Each event starts with a header which includes information such as the size of the event and the event type. The CODA event type is a function of the trigger latch pattern. Special event types were defined for status events, EPICS events and scaler read events. There were also four relevant physics event types. 
\begin{table}
  \begin{center}
    \begin{tabular}{|c|c|c|}
      \hline Trigger Type & TS input(s) & CODA event type \\ \hline 
      HMS1 single & 1 & 1 \\ \hline 
      HMS2 single & 2 & 1 \\ \hline 
      BigCal single & 3 & 5 \\ \hline 
      COIN1 trigger & 4 & 6 \\ \hline 
      COIN2 trigger & 5 & 6 \\ \hline 
      Pedestal trigger & 8 & 4 \\ \hline
    \end{tabular}
    \caption{\label{CODAeventtypes} Trigger types and CODA event types.}
  \end{center}
\end{table}
Table \ref{CODAeventtypes} shows the event type assigned to the various trigger types by CODA. Both HMS triggers were assigned an event type of 1, while both coincidence triggers were assigned an event type of 6. The BigCal singles trigger was assigned to event type 5 and the pedestal events were assigned to event type 4. The ENGINE is the standard analysis code for Hall C, which handles the decoding of the raw event information and the reconstruction of events. The code behaves differently depending on the event type. The first task performed by the engine is the parsing of a master configuration file containing basic information about the run such as the path to the raw data file, parameters which enable or disable the analysis of different event types, and pointers to database files containing histogram definitions, test/cut definitions, and miscellaneous parameters of the analysis such as detector calibration constants and the HMS transport coefficients used to reconstruct the proton momentum, angles, and vertex position from the track reconstructed by the drift chambers.

The configuration file for the ENGINE also defines the so-called ``detector map'' file. This file contains the correspondence between fastbus or VME channel addresses and physical detector locations. The ENGINE parses the map file looking for keywords defining specific parameters needed by the decoding routines. Whenever the combination \emph{keyword}=\emph{value} is found, the quantity referred to by \emph{keyword} assumes the value \emph{value} until the next time a line is found which specifies a new value for \emph{keyword}. The following keywords are defined:
\begin{description}
  \item[ROC] Specifies ROC number for all following entries until the next ``ROC='' line is found. Each ROC corresponds to one crate of digitizing electronics.
  \item[slot] Specifies the current slot within the crate referred to by ROC. Each slot contains one digitizing module.
  \item[Nsubadd] Specifies the number of channels or subaddresses per module. 
  \item[MASK] Specifies the hex mask needed to extract the actual signal (digitized time or amplitude) from the data word. This mask depends on the type of module under consideration and how it is programmed.
  \item[BSUB] Specifies the least-significant bit of the ``channel'' or subaddress field within the data word.
  \item[detector] Specifies the ``detector ID'' to which the channels are to be assigned. Each detector in the experiment was assigned a hard-coded ID parameter within the ENGINE.
\end{description}

All other lines in the map file consist of three or four integers separated by commas. These lines are interpreted as the channel, ``plane'', ``wire'' and ``type'' of an individual detector signal. The channel refers to the subaddress of that signal within the digitizing module. The ``plane'', ``wire'', and ``type'' definitions refer to physical detector locations, allowing a software mapping of detector channels in up to three dimensions, with the restriction that the ``type'' variable is only allowed to vary from 0 to 3, and the number of ``types'' per detector ID is only allowed to be 1, 2, or 4. For example, for the HMS drift chambers, ``plane'' and ``wire'' refer to actual wires within actual planes, with only one signal type which defaults to zero. In BigCal, ``plane'' refers to row number and ``wire'' refers to the column number of individual lead-glass bars. Only one signal ``type'' was allowed for BigCal as well. The HMS hodoscopes employed the ``type'' variable in such a way that types 0, 1, 2, and 3 referred to the ADC signal of the PMTs on the ``+''(type 0) and ``-''(type 1) ends, and the TDC signals for the PMTs on the ``+''(type 2) and ``-''(type 3) ends, respectively. The ``plane'' number for the HMS hodoscopes ran from 1 to 4 and referred to S1X(1), S1Y(2), S2X(3) and S2Y(4), while the ``wire'' number referred to each individual scintillator bar within a plane, which ran from 1 to 16 for the x planes and 1 to 10 for the y planes. Preparing a correct detector map file was the main requirement for the decoding of events. The subroutines in the ENGINE then used the information in this file to place the decoded data into the appropriate ``hit'' arrays for processing by the detector reconstruction subroutines.
\subsection{Hodoscope Timing Analysis}
\label{hodotimeanalysissection}
\paragraph{}
The discussion of event reconstruction begins with the calibration of the HMS hodoscopes. For this experiment, the main purpose of the hodoscopes was to measure the time at which a particle passed through the detector stack in order to set the zero time for the drift chambers of both the HMS and the FPP. The amplitude of the signal from each PMT was digitized by integrating ADCs. The timing of the signal was measured by high-resolution common-start TDCs as described above. To reconstruct the timing of the event from the raw ADC and TDC information, a number of corrections are applied. First, the raw TDC value is converted to time in nanoseconds with an assumed conversion factor of 25.9 picoseconds per TDC count. This value for the time scale of the TDC was based on calibration measurements using a time interval generator checked against the accelerator RF performed during the initial commissioning of the HMS\cite{ArringtonThesis}. This conversion factor represents the average of all the individually calibrated modules, with channel-to-channel variations within a single module possible at the percent level and variations between modules of up to 6\%. Since these variations were small compared to the time resolution of the scintillators themselves, using the average conversion factor was sufficient. The ``raw'' time is given by $t_{raw} = 25.9\ ps \times TDC$.

The next correction to the scintillator time was to subtract an overall zero offset, determined for each channel, to account for variations in cable delay between channels. The goal was to determine the time of each scintillator hit relative to an arbitrary reference time, which in turn defines the start time for the drift chamber tracking. The absolute signal propagation time from the PMTs to the TDCs was irrelevant, except in making sure the signal arrived within the gate sent from the trigger supervisor. Therefore, in the calibration procedure which determines all the calibration parameters for all PMTs simultaneously, one of the PMTs is arbitrarily chosen to have an offset of zero, and the offsets for all the other PMTs are calibrated relative to this reference PMT. 

Since the PMT signals are discriminated against a fixed threshold, significant time walk or pulse-height dependence of the signal timing is possible. The time measured by the TDC is the time at which the leading edge of the signal exceeds the fixed threshold. Since the pulse has a characteristic shape with a finite rise time, larger signals will exceed the threshold earlier than smaller signals. Thus, a correction to the time for the pulse height is made of the form
\begin{equation}
  t_{corr,walk} = t_{raw} - \frac{w}{\sqrt{ADC}}
\end{equation}
where the constant $w$ is determined for each individual PMT in the calibration procedure. Figure \ref{hodo_walk} shows an example of the pulse-height dependence of $t_{raw}$. 
\begin{figure}
  \begin{center}
    \includegraphics[angle=90,width=.99\textwidth]{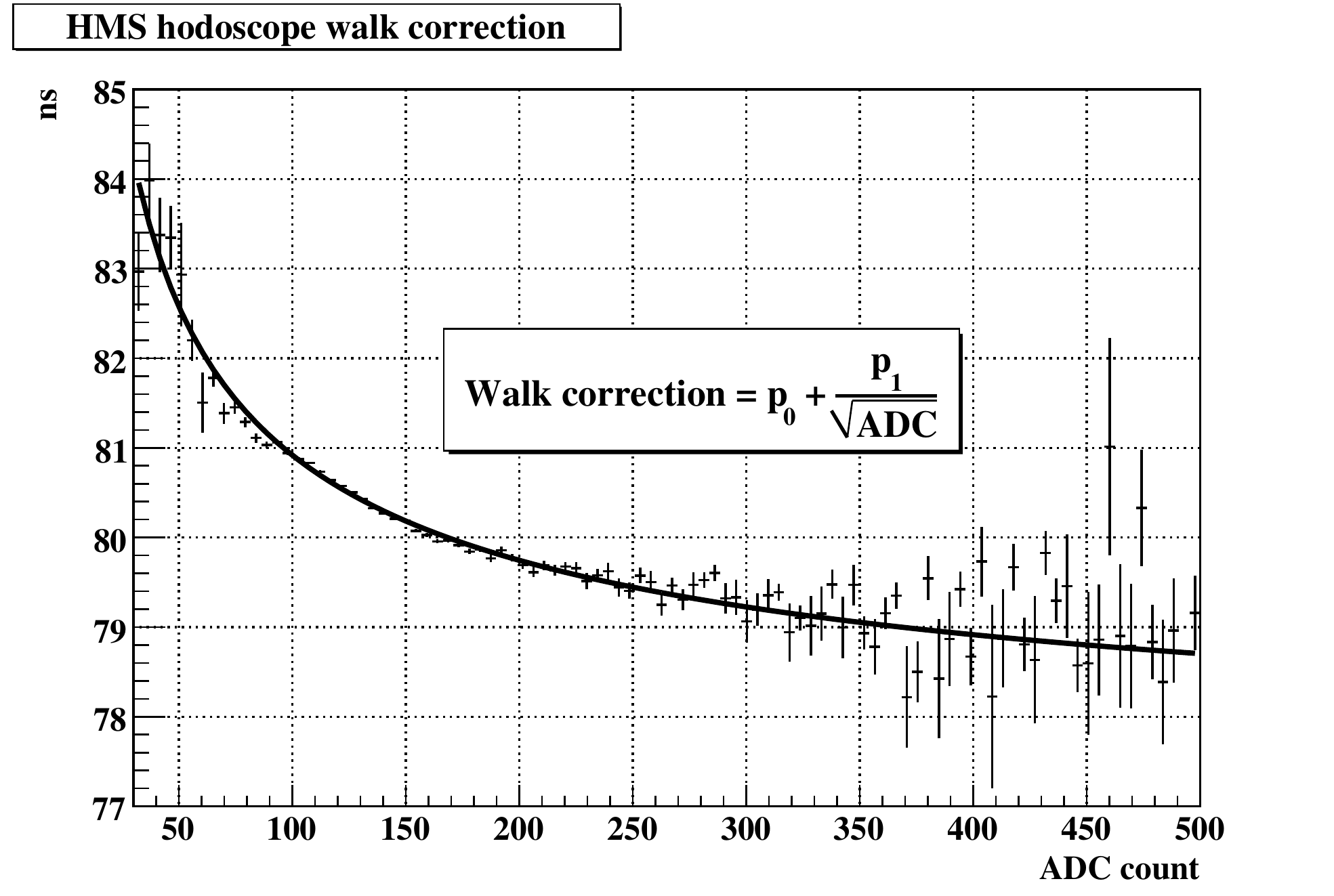}
    \caption{\label{hodo_walk} Profile histogram of $t_{raw}$ as a function of the pulse height measured by the ADC for the PMT at the + end of S1X paddle 8, with fitted walk correction shown.}
  \end{center}
\end{figure}

The next correction to the timing signals was to correct for the light propagation time in the scintillator bar. Each PMT was assigned a ``speed'' parameter related to the effective average light propagation speed in the scintillator. Given a light velocity $v_{eff}$, bar length $L$, and a particle passing through the bar at a distance $d$ from the midpoint of the bar, with positive $d$ pointing toward the $+$ end, the expected time difference between the two PMT signals is given by 
\begin{eqnarray}
  t_+ - t_- &=& \left(\frac{L}{2v_{eff}} - \frac{d}{v_{eff}}\right) - \left(\frac{L}{2v_{eff}} + \frac{d}{v_{eff}}\right) \\
  &=& -\frac{2d}{v_{eff}} \\ 
  \Rightarrow d &=& \frac{1}{2}v_{eff} \left(t_- - t_+\right)
\end{eqnarray}
Thus, the zero-offset-subtracted, walk-corrected time difference between the + and - PMTs measures the longitudinal coordinate at which the particle passed through the paddle. Once this approximate coordinate is known, the propagation time for each PMT is subtracted to calculate an average corrected time for the scintillator.
\begin{eqnarray}
  t_{+,corr} &=& t_+ - \frac{1}{v_{eff,+}}\left(\frac{L}{2}-d\right) \\
  t_{-,corr} &=& t_- - \frac{1}{v_{eff,-}}\left(\frac{L}{2}+d\right) \\
  t_{avg.,corr} &=& \frac{1}{2}\left(t_{+,corr} + t_{-,corr}\right)
\end{eqnarray}
Figure \ref{S1prop} shows the time difference between the + PMT and the - PMT due to light propagation delay for paddle S1X8 (left), and paddle S1Y5 (right), as a function of the distance to the + PMT from the intersection point of the particle track with the paddle, determined by projecting the reconstructed drift chamber track to the paddle.
\begin{figure}
  \begin{center}
    \includegraphics[angle=90,width=.48\textwidth]{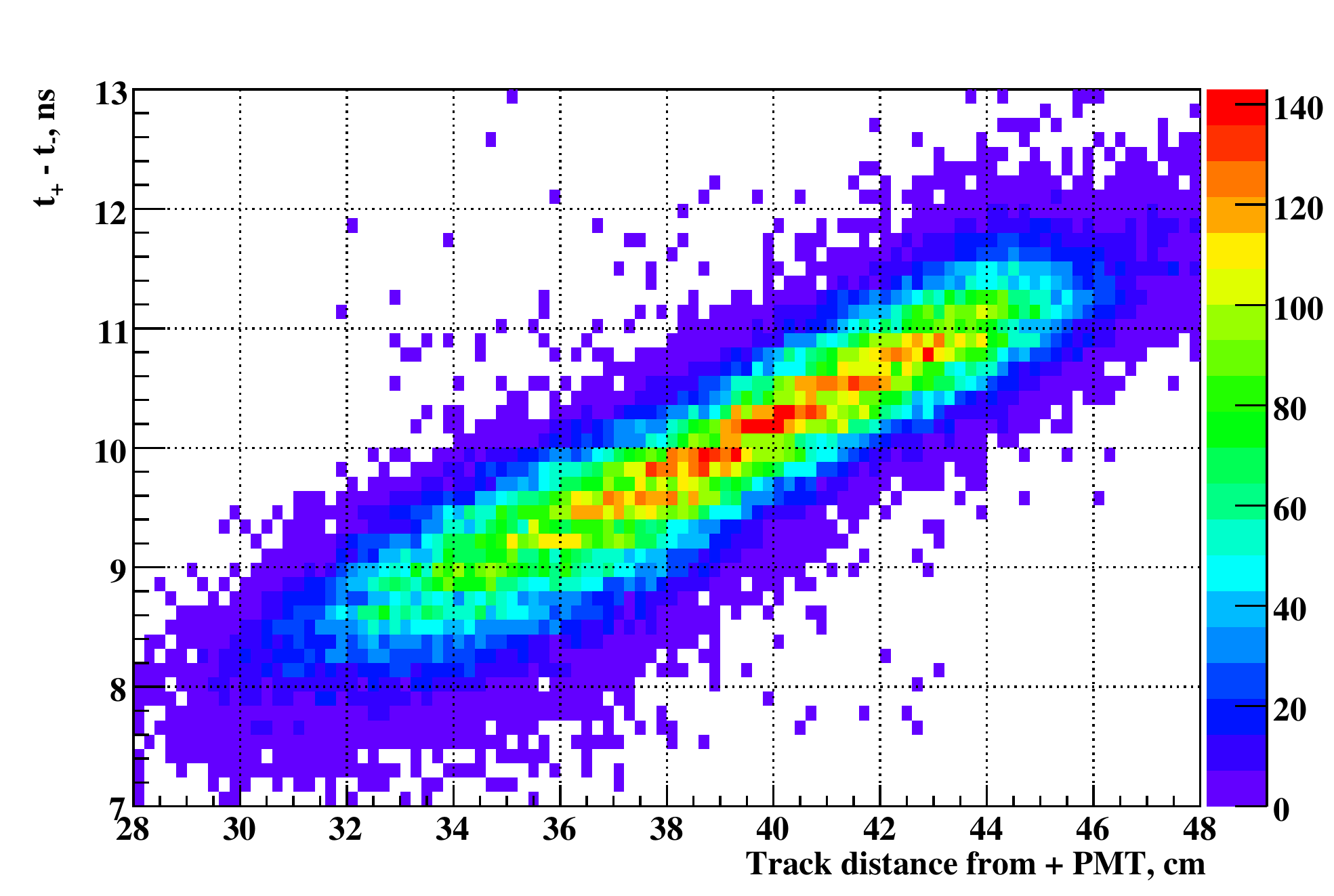}
    \includegraphics[angle=90,width=.48\textwidth]{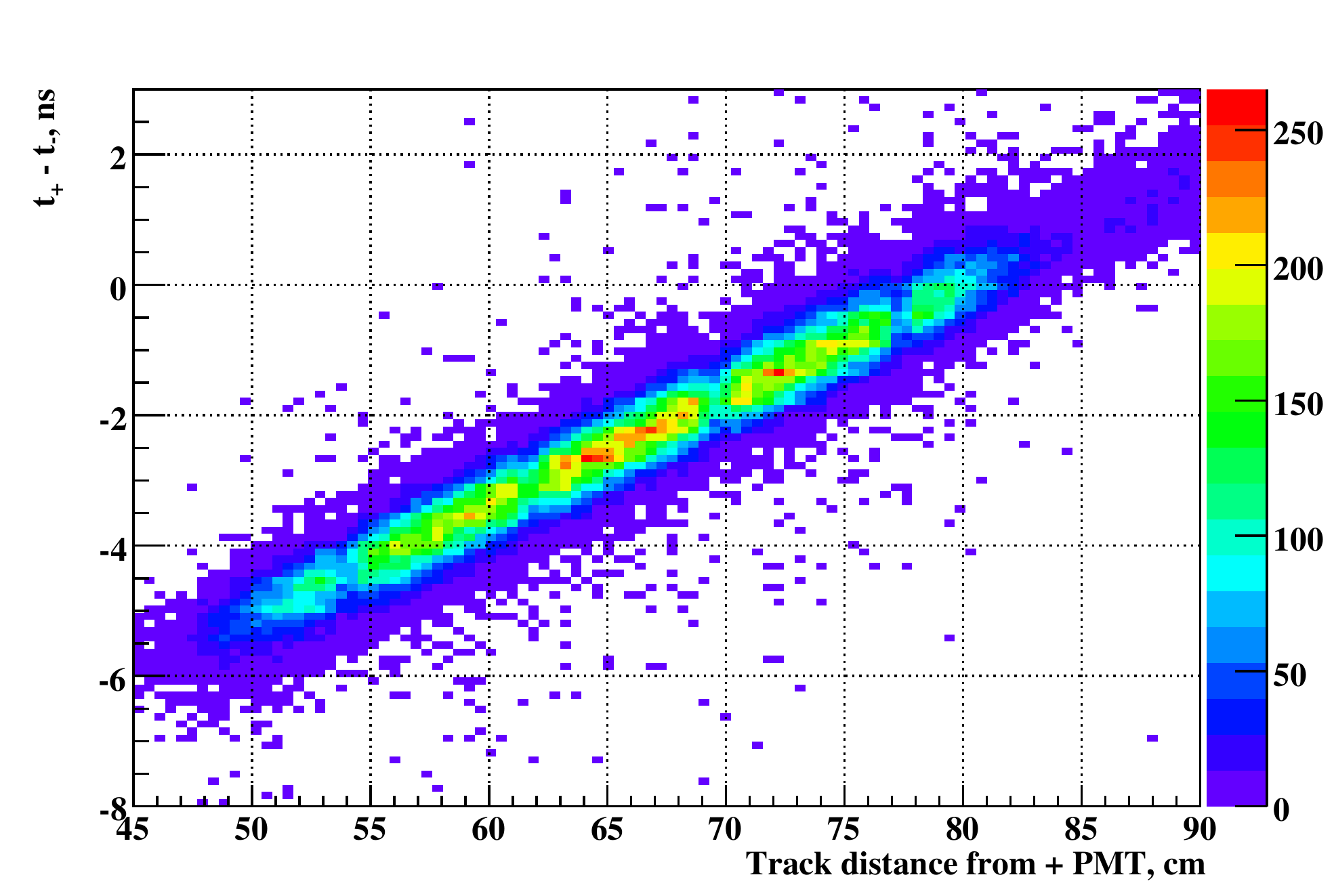}
    \caption{\label{S1prop} Time difference between + and - ends of the same paddle due to propagation delay, as a function of the propagation path length from the particle impact point to the + PMT, for paddle S1X8 (left) and S1Y5 (right). The data in this plot had not yet been corrected for the zero offset between the two signals.}
  \end{center}
\end{figure}
Even though the light propagation speed should be the same for PMTs at opposite ends of the same paddle, slight differences in the effective propagation speed are possible due to e.g. slightly different light-guide characteristics. Therefore, each PMT is assigned its own $v_{eff}$ which is not required to be the same as the $v_{eff}$ of the tube at the opposite end of the same paddle and is varied independently in the calibration procedure. In practice the differences between $v_{eff,+}$ and $v_{eff,-}$ for a given paddle are quite small, and paddle-to-paddle $v_{eff}$ variations are no more than 20\%. Finally, in order to determine the start time for the drift chamber tracking, the corrected times are projected back to the HMS focal plane ($z=0$). In the first stage of the code, since the drift chamber track is not yet known, the trajectory slopes $x'$ and $y'$ are unknown. However, deviations from $x'=y'=0$ for true tracks are always small, so it is a reasonable first approximation to assume $x'=y'=0$. Since the particle momentum is also not yet known, particles are assumed to have the central momentum of the spectrometer. The mass of the desired particle is an input parameter of the code. In this case, the proton mass was used. The focal plane time corresponding to the corrected scintillator time is given by 
\begin{eqnarray}
  t_{fp} = t_{corr.,avg.} - \frac{z}{\beta(p_0)c}
\end{eqnarray}
where $z$ is the position of the scintillator element relative to the focal plane $z=0$ and $\beta(p_0)c$ is the velocity of a proton with momentum equal to the HMS central momentum.

In order to obtain the best set of parameters, the standard HMS time-of-flight calibration code written by Peter Bosted was used. This code fits all three parameters (zero offset, walk correction coefficient, and effective propagation speed) for all PMTs simultaneously. It uses the HMS drift chamber tracks rather than the +/- time difference to more precisely measure the impact coordinate of the particle, and uses the reconstructed spectrometer momentum and the measured trajectory angles to determine, respectively, the proton's velocity and the length of its flight path between the various elements, in order to calculate the time-of-flight correction to each scintillator time. Using all of this information and the raw ADC and TDC information, the code solves a system of linear equations in the parameters which minimizes the sum of squared corrected-time differences between PMT hits on the same track:
\begin{eqnarray}
  \chi^2 &=& \sum_{i=1}^{N_{event}} \sum_{j=1}^{N^{(i)}_{hit}-1}\sum_{k=j+1}^{N^{(i)}_{hit}}\left[\left(t^{(i)}_{raw,j} - \frac{z_j}{\beta(p_i)c} - t_{0,j} - \frac{d^{(i)}_{prop,j}}{v_{eff,j}} - \frac{w_j}{\sqrt{ADC^{(i)}_j}} \right) - \right. \nonumber \\
    & & \left. \left(t^{(i)}_{raw,k} - \frac{z_k}{\beta(p_i)c} - t_{0,k} - \frac{d^{(i)}_{prop,k}}{v_{eff,k}} - \frac{w_k}{\sqrt{ADC^{(i)}_k}} \right)\right]^2
\end{eqnarray}
The calibration code runs separately from the ENGINE, and requires a special input file prepared by the ENGINE, requiring typically one thousand to ten thousand events per PMT to be calibrated. The number of parameters is three per PMT ($t_0$, $1/v_{eff}$, and $w_j$) for a total of 156 parameters. After running the calibration code, obtaining the best fit parameters, and re-analyzing the data, the final timing resolution of the HMS hodoscopes, averaged over all PMTs in S1X and S1Y, is approximately 250 picoseconds. 
\begin{figure}
  \begin{center}
    \includegraphics[angle=90,width=.95\textwidth]{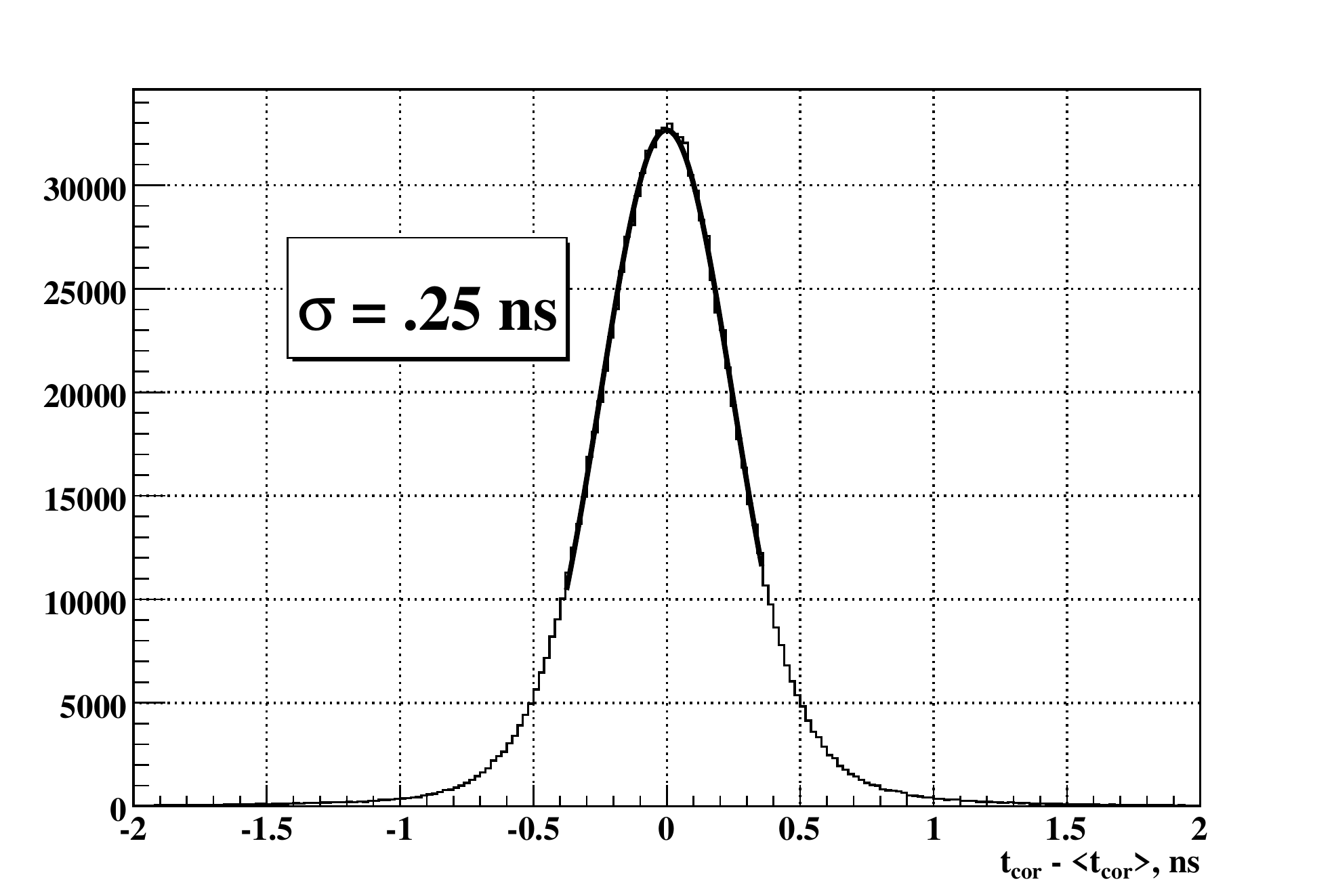}
  \end{center}
  \caption{\label{finalhodores} Difference between corrected time and the average corrected time of all \emph{other} hits on a track. This quantity measures the hodoscope timing resolution, with all corrections applied, after calibration.}
\end{figure}
The time resolution achieved is shown in figure \ref{finalhodores}, which shows a histogram of the difference between the corrected focal-plane time of each hit and the average corrected focal-plane time of all the other hits on the track, not including the hit in question. Excluding the hit under consideration from the average time means that it does not contribute to the determination of the average time, which would make the resolution appear better than it really is.

The timing window for hodoscope hits is 100 ns wide. Even within this relatively short window, it is possible for extra hits not related to the trigger to leak in. Recalling the discussion of trigger rates in the previous chapter, the S1 exclusive-OR rate during the experiment was on the order of several hundred kHz and sometimes approaching 1 MHz. In the worst case of 1 MHz, there is a roughly 10\% probability for an extra PMT hit to arrive during the timing window. In order to choose the right hits, then, a filtering algorithm was written into the hodoscope timing analysis. This algorithm creates a histogram with 200 time bins between 0 and 100 ns, with bins starting at $t_{min}=0,0.5,1.0,1.5,...$ ns. A time tolerance parameter adjustable by the user sets the bin width. Then, in each bin, the code counts the number of hits with corrected times satisfying $t_{min}<t_{cor}<t_{min} + \delta$, where $\delta$ is the tolerance parameter. Note that the time tolerance parameter is typically set to be significantly greater than 0.5 ns, so that the time bins have significant overlap. Then, the time bin with the greatest number of hits is chosen, and all hits with corrected times outside this bin are thrown out. This procedure helps to select the hits from the PMTs that caused the trigger and filter out random hits that accidentally arrived during the 100-ns window. For this experiment, given the 250 ps timing resolution, a conservative tolerance of 5 ns was used to filter out bad hits. This filtering algorithm reduces the effective timing window for the hodoscope PMT signals by a factor of 20, greatly reducing the probability of using unwanted background hits.

A final remark on the hodoscope timing is to mention that the S0 ADC and TDC signals were not used in the start time determination, since the timing resolution of these signals was almost an order of magnitude worse than the S1 timing resolution. However, it was still useful to have them in the data stream to measure the efficiency of the S0 PMTs when they were not part of the trigger.
\subsection{HMS Drift Chamber Track Reconstruction}
\paragraph{}
The next step in the HMS reconstruction is to reconstruct the proton trajectory in the HMS drift chambers. The track reconstruction proceeds by first converting the raw TDC information for all wires with a hit into rough drift time information. The drift chambers are read out by TDCs with 0.5 ns count resolution operating in common stop mode, in a 2 $\mu$s window with up to eight hits per wire per event allowed. In the raw TDC spectrum, a low-level background of random hits populates the full time window. A significant fraction of these hits are thrown out by a loose cut on the raw TDC value surrounding the good hits from the track which caused the trigger, which form a prominent peak in the spectrum roughly 100 ns wide. An initial drift time value is assigned to the remaining hits as follows:
\begin{eqnarray}
  t_{drift} = -t_{raw} - t_{start} + t_0 \label{HMSroughdrift}
\end{eqnarray}
In equation \eqref{HMSroughdrift}, $-t_{raw} = -0.5\ \mbox{ns} \times TDC_{raw}$ is the measured hit time, with a minus sign attached because of the common stop. $t_{start}$ is the ``start time'' reconstructed from the hodoscopes, equal to the average corrected focal plane time of all the PMTs in the 5 ns time-tolerance bin with the most PMT hits. $t_0$ is an overall time offset used to align the drift time spectrum so that the good hits lie roughly in the window between 0 and 100 ns. For this experiment, a separate $t_0$ offset was calculated for every wire, and updated periodically for configuration and kinematics changes which resulted in slight timing changes.

After applying the loose TDC cut, the process of pattern recognition begins. First, the total number of hits per chamber passing the TDC cut must lie between user-adjustable lower and upper limits for pattern recognition to proceed. For this experiment, since accurate track reconstruction was more important than tracking efficiency, the minimum number of hits per chamber was set to five. The maximum number of hits was set to 35, to prevent pattern recognition from proceeding for events with a small probability of success requiring the consideration of geometrically increasing numbers of hit combinations. Within each chamber, the tracking algorithm considers all possible pairs of ``non-parallel'' wires with hits, and records the intersection points of their wire positions in the $xy$ plane, without considering their drift distance information. Pairs of wires whose orientation angles differ by less than 17.5 degrees are not considered. This means that the U and V wires in the HMS drift chambers are \emph{not} paired with the X wires, but \emph{are} paired with each other. Since most tracks move nearly perpendicular to the wire planes, and since the separation in $z$ between planes within a chamber is small, it is adequate for pattern-recognition purposes to imagine that all six planes in each wire chamber are located at the same $z$, and consider the intersection points between wires as points in a two-dimensional plane. After recording all pairs of hits on non-parallel wires, the code compares each pair of hits to all other pairs of hits in that chamber, comparing the squared distance between the two intersection points to a prescribed maximum separation. 
\begin{equation}
  r^2 = (x_1-x_2)^2+(y_1-y_2)^2 < r^2_{max}
\end{equation}
In the analysis, $r^2_{max}=2$ cm$^2$ was used. Every combination of pairs of hits passing this squared-distance criterion is added to an array of valid combinations or ``combos''. The first time a combination of two unique pairs of non-intersecting wires is found whose intersection points are close enough together to satisfy the squared-distance criterion, a ``space point'' is formed which includes all unique hits in the combo; i.e., all four hits in the two pairs, unless any wires are shared between the two pairs. Space point coordinates $x$ and $y$ are defined as the average position of the intersection points of the two pairs that make up the combo. After the first space point is formed, all additional combos are compared to all previously existing space points. If the squared distance between the coordinates of the combo under consideration and the coordinates of a previously existing space point is smaller than $r^2_{max}$, any hits in the current combo not already included in the existing space point are added to it. Each ``combo'' is only allowed to be part of one space point, meaning that the first time a combo is added to an existing space point, it is not compared to any subsequent space points. If the combo under consideration is more than $r^2_{max}$ away from all previously existing space points, a new space point is created with this combo as a seed, and all subsequent combos are tested against the new space point.

Once all hits in the chamber are grouped into space points in the manner above, the position of the track is roughly known. Using this information, small corrections to the drift time are made for signal propagation time in the wires based on the distance along each wire from the track position to its readout card. Some space points may have more hits than needed to fit a track. In particular, if two adjacent wires in the same plane fired, they will end up on the same space point. When fitting tracks to hits, it is preferable to use only one hit per wire plane. A special ``cloning'' routine examines all the found space points for multiple hits in the same plane, and creates new space points with all possible combinations of one wire per plane within an existing space point, with the exception that the total number of space points is not allowed to exceed a hard-coded upper limit of 20 per drift chamber, and the number of planes with multiple hits in a single space point is not allowed to exceed 3. If any space point still contains multiple hits in the same plane after the cloning procedure is carried out, because of e.g. 4 or more planes in a space point having multiple hits, then the single hit in that plane is chosen which has the shortest drift time. The last task of the pattern recognition routine is to throw away all space points for which either the total number of hits is less than five or the total number of ``combos'' contributing to the space point is less than 4.

The next task of the HMS tracking algorithm is to use the (now propagation-corrected) drift times to fit track ``stubs'' to each of the individual space points. Since most tracks are moving nearly perpendicular to the wire chambers, it is reasonable to assume that the distance measured by the drift time is the in-plane distance from the track to the wire, even though in reality the drift time measures the distance to the point of closest approach between the track and the wire, which may be slightly out-of-plane. To convert from drift times to drift distances, it is assumed that the drift distance is a monotonically increasing function of the drift time $d_{drift} = d(t_{drift})$. Since the size of a drift cell in the HMS is small compared to the envelope of tracks populating the active area of the chambers, the change in the relative number of incident tracks within any given drift cell can be assumed to be small. It is a very good approximation to assume a uniform distribution of drift distances within a cell. Even given small cell-to-cell variations in the relative non-uniformity of the distribution of incident tracks as a function of drift distance, when averaged over all the drift cells in a plane of wires and/or all planes within the chambers, the assumption of uniform drift distance is quite robust. This assumption greatly simplifies the time-to-distance conversion, since it is straightforward to map the observed drift time spectrum onto a uniform distance distribution. The drift time distribution $n(t_{drift})$ is defined as the probability density of events as a function of drift time within the allowed time window.
\begin{eqnarray}
  \int_{t_{min}}^{t_{max}} n(t) dt &=& 1
\end{eqnarray}
For a uniform distribution of events, then, the drift distance for a given drift time is simply given by the integral of the observed drift time spectrum up to the measured $t_{drift}$:
\begin{eqnarray}
  \frac{1}{d_{max}}\int_0^{d_{drift}} dx = \frac{d_{drift}}{d_{max}} &=& \int_{t_{min}}^{t_{drift}} n(t) dt \nonumber \\
  d_{drift}(t_{drift}) &=& d_{max} \int_{t_{min}}^{t_{drift}} n(t) dt \label{driftmap_eq}
\end{eqnarray}
\begin{figure}
  \begin{center}
    \includegraphics[angle=90,width=.49\textwidth]{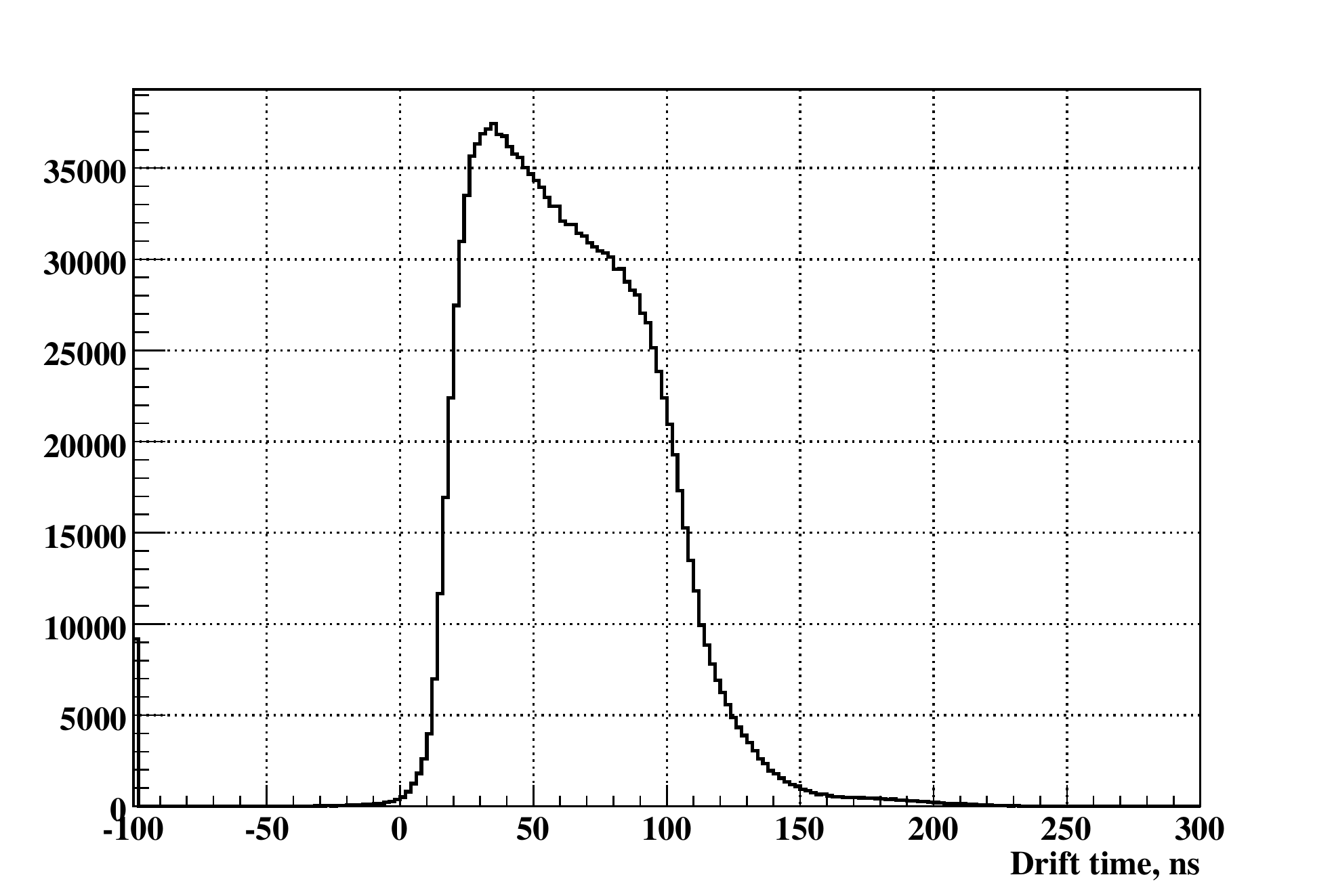}
    \includegraphics[angle=90,width=.49\textwidth]{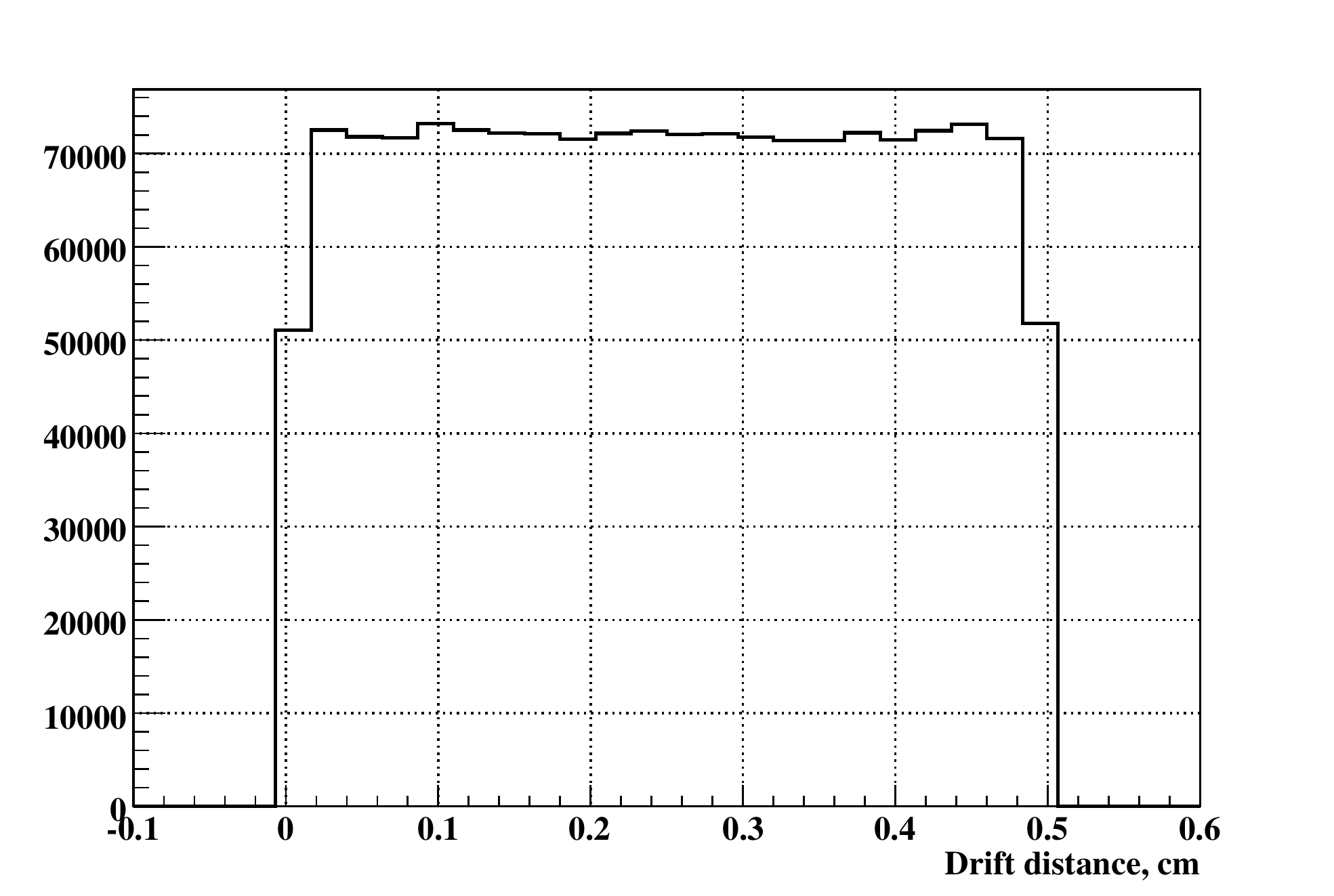}
  \end{center}
  \caption{\label{HMSdriftmap} Example drift time spectrum for the HMS drift chambers (left) and calculated drift distance spectrum (right).}
\end{figure}
Figure \ref{HMSdriftmap} shows an example drift time spectrum and the resultant drift distance spectrum. Using the time spectrum, a time-to-distance map is generated by calculating the integral \eqref{driftmap_eq} for each of 138 2 ns-wide time bins starting at $t_{min}=-24$ ns up to $t_{max}=250$ ns. This map is then included in the parameter database of the ENGINE. The drift distance is then calculated from the drift time for each hit by linearly interpolating within time bins in the drift map file.

Once the drift distance is determined, however, the track still isn't known. Only the magnitude of the distance is determined, not its sign. The track could have passed on either side of the wire at the same distance. The appropriate drift sign of the hits in a space point is determined by fitting test track stubs to the possible left-right combinations of the hits within a point. A small-angle approximation is used to determine the sign of the hits in the Y planes if they are both present on the track. Because the tracks are very nearly perpendicular to the chambers $\left|y'\right| \leq \pm .03$, and the positions of the wires in the Y' plane are staggered by one half of a cell size or 0.5 cm relative to the Y plane, it is quite reasonable to choose the left-right assignment for these wires that makes the track go between the two wires. This reduces the number of left-right combinations that have to be tested from $2^6=64$ to $2^4=16$ and speeds up the tracking code. Since the X and X' planes are farther apart and the spread of good tracks in $x'$ is about twice as large as in $y'$, no small-angle approximation is used for these planes. The small-angle approximation also assumes that $y'=0$ when performing the ``stub'' fits, so only $x$, $y$, and $x'$ are fitted. The left-right combination of hits is chosen that gives the best $\chi^2$ for the stub fit. The coordinate of the hit along the direction measured by its wire is then $w_{hit} = w_{pos} \pm d_{drift}$ where $w_{pos}$ is the wire position. Additionally, since it is possible for some hits to appear on multiple space points, the best drift sign of each hit is remembered for each space point on which it appears, to be recalled later for full track fitting.

After performing the stub fits to determine the best left-right combination, the two chambers are considered together. The code compares all possible combinations of one space point from each chamber, linking space points together to form tracks in an attempt to find the best possible combination of hits to form the full track. The track stubs fitted to the hits in each space point are projected to the focal plane. The slope $x'$ and the coordinates $x_{fp}$ and $y_{fp}$ are compared. Since $y'$ is assumed to be zero in the stub fits, it is ignored. If a stub from chamber 1 agrees with a stub from chamber 2 in all three parameters to within user-defined tolerances\footnote{Given the $y'=0$ assumption in the stub fits, the $y$ tolerance parameter for stub linking basically defines the maximum allowed $y'$ for the full track. In this analysis, a tolerance of 4 cm was used, restricting the possible values of $y'$ to $\pm.05$, which is fairly conservative given the smaller spread in $y'$ of true tracks.}, the stubs are joined to form a track. A single space point from one chamber can be used in multiple tracks if there are multiple space points in the other chamber with which it agrees. Even if only a single particle passed through the chambers, it is still possible to reconstruct multiple tracks if, for example, a track leaves hits in two adjacent wires such that two space points are formed that share all but one of their hits. Events such as this are much more common than true multi-particle events, but with single-particle rates approaching 1 MHz, there is still a non-negligible chance of a second particle going through the chambers within the $\approx 100$ ns good drift time window defined by the track responsible for the trigger. The hits left by such a track would be significantly out of time with the start time defined by the hodoscopes, but could still wreak havoc during the pattern recognition, since each valid ``combo'' can only appear in one space point, making it easy to either mix hits from different tracks in the same space point or to fail to build space points containing the correct combination(s) of hits. This is particularly true for kinematics in which tracks are focused onto a narrow region of the focal plane, which unfortunately coincide with high single-particle rates, since both phenomena are associated with forward central angles of the HMS. 

Once all possible ``stub links'' have been performed, a number of candidate tracks are defined, each one corresponding to a different combination of space points and therefore a different combination of hits. Using the left-right combinations of the hits determined from the stub fitting and the measured drift distances, a straight line is fit to all 10-12 hits\footnote{Recall that a minimum of five hits per space point is required, and only one hit per plane is used in each space point, so that the total number of hits on a track in our analysis can only vary between 10 and 12.} on the track, determining the best fit parameters $(x_{fp},y_{fp},x'_{fp},y'_{fp})$ which define the trajectory. Along the way, the reduced $\chi^2$ of the track and the residuals of each hit on the track are computed in order to check the quality of the tracking.

In the analysis, it is assumed that only one particle went through the drift chambers per event. Although this single-particle assumption is not strictly true, only one track is chosen as the best track for each event for further analysis. After all candidate tracks have been reconstructed, a number of subsequent procedures are carried out for each track before choosing the best track. For this experiment, just two additional subroutines were carried out before choosing the best track. First, each candidate track is transported back to the target using the HMS optical reconstruction coefficients. The trajectory angles $x'_{tar}$ and $y'_{tar}$, the momentum deviation $\delta$, and the vertex position $y_{tar}$ are reconstructed. Then, each track is projected to the hodoscope planes, and the hodoscope timing analysis is revisited with reconstructed tracks in hand. The number of PMT hits on the scintillators pointed to by each track is counted, and the timing of those hits is compared to the $t_{start}$ value determined by the initial hodoscope analysis and used in the tracking algorithm. The hit times are recalculated using the track information. By projecting the drift chamber tracks to the scintillators, the position at which the particle crossed the scintillators is more precisely determined than by the time difference between the + and - PMTs, and since the momentum and angles of the trajectory are known, the time-of-flight of the particle from the focal plane to the scintillators is also more precisely known. A ``time at focal plane'' is determined for each track based on the average corrected focal plane time of all hits on scintillators pointed to by that track. Once the transport of tracks back to the target and the improved scintillator reconstruction is complete, the ``best'' track is selected by a ``pruning'' algorithm. 

The pruning algorithm works by subjecting all candidate tracks to a series of tests. At first, all tracks are assumed to be valid. For each test considered, if any of the remaining tracks passes that test, all tracks failing that test are discarded. In this way, the code is guaranteed to choose at least one track from among all the candidates, even if all tracks fail one or more of the tests. For this kind of track selection algorithm, the order in which the tests are considered matters somewhat. In this analysis, the following tests were considered in the order in which they are listed: 
\begin{itemize}
\item $\left|x'_{tar}\right|\leq 100$ mrad 
\item $\left|y'_{tar}\right|\leq 50$ mrad
\item $\left|\delta\right| \leq 9$\% 
\item $\left|y_{tar}\right|\leq 10$ cm
\item $N_{PMT}$ on track $\geq$ 3
\item $\left|t_{fp}-t_{start,0}\right|\leq10$ ns\footnote{$t_{start,0}$ is the user-defined ``central'' start time parameter, which corresponds to the position of the peak in the start time distribution corresponding to the HMS self-timing in the coincidence trigger.}
\end{itemize}
The test definitions for the pruning algorithm are very permissive, as they are designed only to throw away tracks with reconstructed vertex variables far outside the useful acceptance of the HMS. Applying tight prune tests in these variables can have the undesired effect of shifting unwanted tracks of marginal quality into the acceptance. The tests on the number of hodoscope PMTs on the track and the focal plane time of those hits is designed to bias the track selection in favor of the track which caused the trigger. If more than one track passes all the prune tests, the track from among the remaining tracks which has the lowest $\chi^2$ is chosen as the best track.

Figure \ref{HMSdisplay1} shows an example of the wire hit pattern of a high-quality track in the HMS drift chambers. This event display highlights only the wires with hits that end up on the best chosen track. In this event, only one track was found. Figure \ref{HMSdisplay2} shows, for the same event, the projection of the track along the direction measured by each of the four wire orientations, with a line showing the fitted track, and markers showing the wire positions, the hit positions reconstructed from the measured drift times, and the track position at each plane.
\begin{figure}
  \begin{center}
    \includegraphics[angle=90,height=.4\textheight]{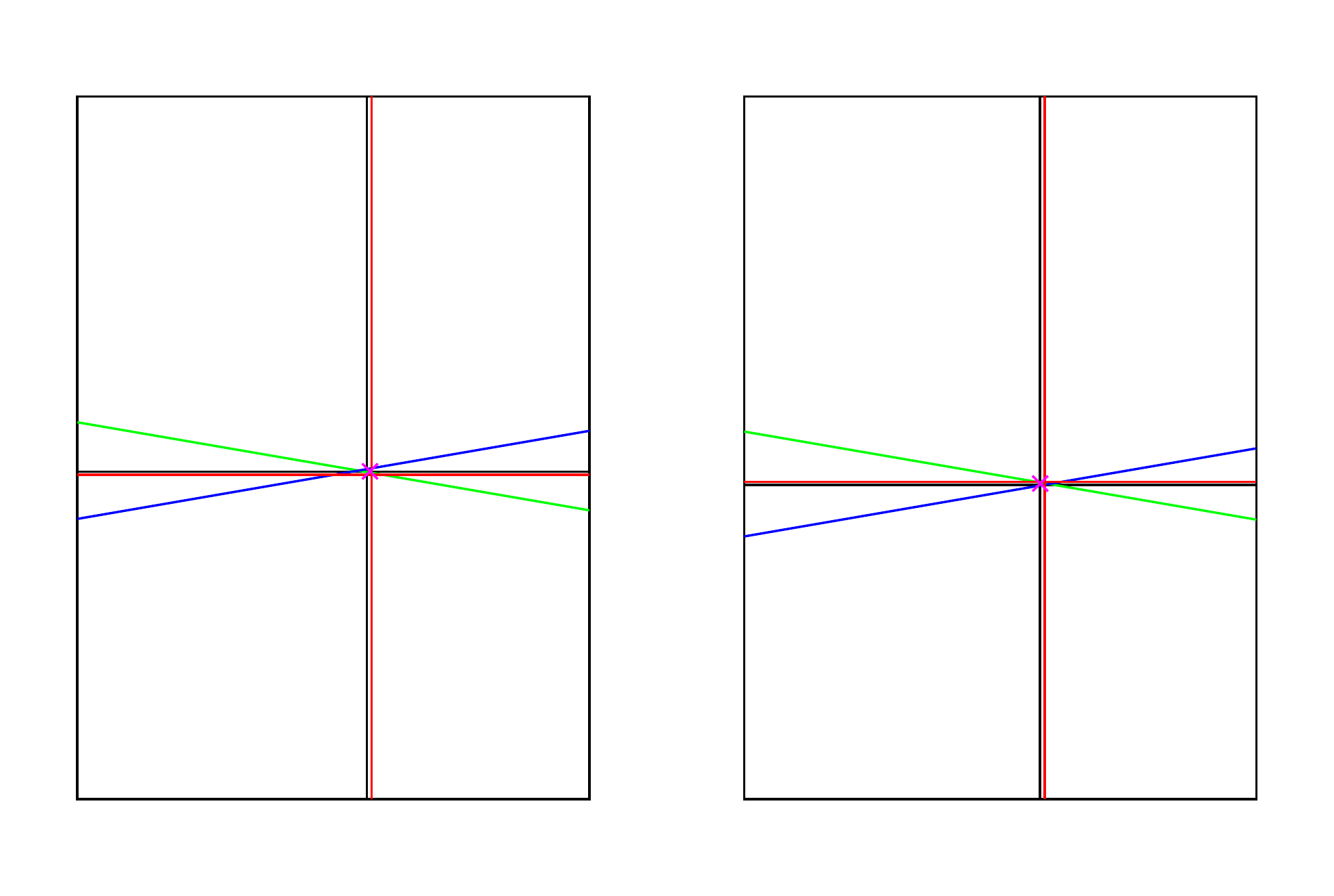}
  \end{center}
  \caption{\label{HMSdisplay1}Example event display of results of HMS tracking. Hit wires in chamber 1 (above, left) and chamber 2 (above, right), with X(black) and X'(red) horizontal wires, Y(black) and Y'(red) vertical wires, and U(green) and V(blue) wires. The projection of the fitted track to the middle of each chamber is marked by the magenta $x$.}
\end{figure}
\begin{figure}
  \begin{center}
    \includegraphics[angle=90,width=.99\textwidth]{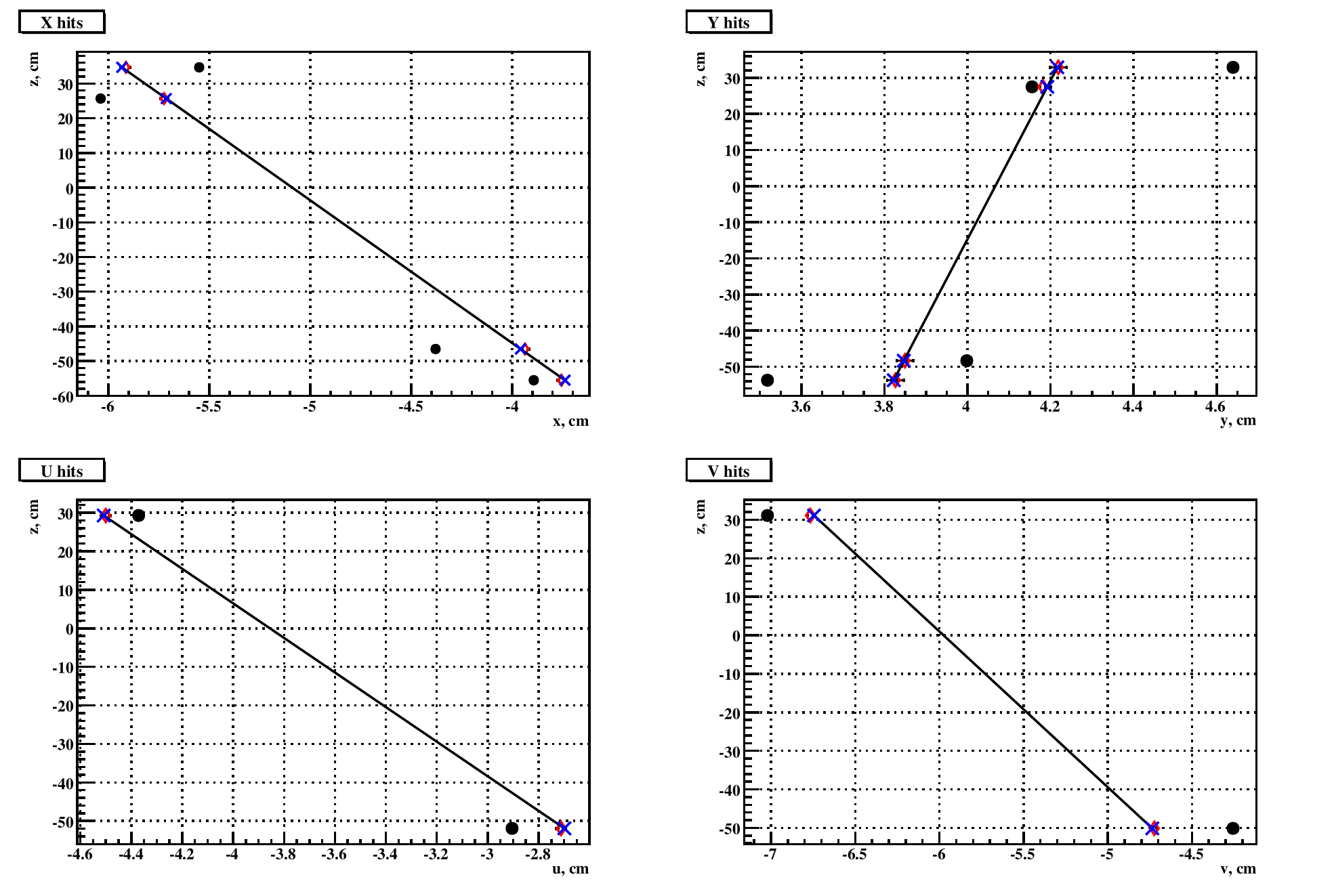}
  \end{center}
  \caption{\label{HMSdisplay2}  Track projections along the $x$(top left), $y$(top right), $u$(bottom left), and $v$(bottom right) axes, along with wire positions (black circles), in-plane hit positions (red diamonds with 200 $\mu$m error bars) reconstructed from drift times, and fitted track coordinates at each plane (blue $\times$'s.). Note that the horizontal scale in each plot is much smaller than the vertical scale, which corresponds to the distance between the two drift chambers.}
\end{figure}
\begin{figure}
  \begin{center}
    \includegraphics[angle=90,width=.99\textwidth]{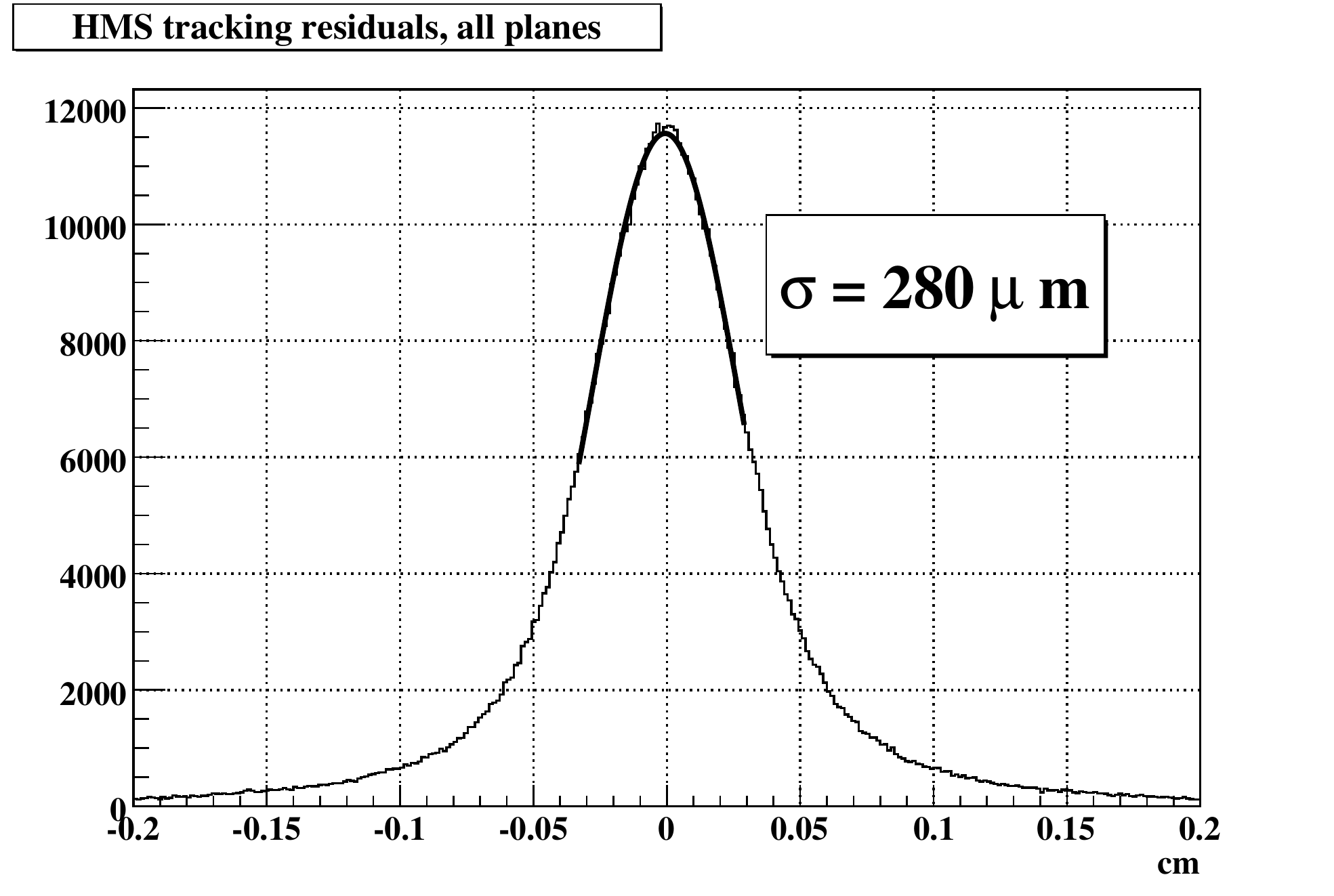}
  \end{center}
  \caption{\label{HMSres} Tracking residuals in the HMS drift chambers, averaged over all planes, with no $\chi^2$ cut applied. Less than 2\% of events lie outside the $\pm2$ mm range of this histogram.}
\end{figure}

Figure \ref{HMSres} shows a histogram of the tracking residuals in the HMS drift chambers after fully calibrating the hodoscope timing (which determines the start time relative to which the drift times are measured) and the time-to-distance maps. The final resolution achieved is approximately 280 $\mu$m (1 $\sigma$), which is the result averaged over all wire planes\footnote{There are slight plane-to-plane variations in the width of the tracking residuals.}. 

Since the $U$ and $V$ planes are oriented only 15$^\circ$ away from the horizontal, they can be thought of as measuring the $x$ coordinate. The $Y$ planes dominate the measurement of the $y$ coordinate. Therefore, there are four planes in each chamber measuring $x$, and two planes in each chamber measuring $y$, giving a per-drift-chamber spatial resolution of 140 $\mu$m (200 $\mu$m) in $x$($y$). The resolution of the focal plane track slopes is 0.24 mrad (0.35 mrad) in $x'_{fp}$($y'_{fp}$). Recalling equations \eqref{S0HMSoptics}, the approximate contribution of the drift chamber resolution to the resolution of the trajectory angles at the target from the lowest-order HMS optics coefficients is 0.72 mrad in $x'_{tar}$ and 0.76 mrad in $y'_{tar}$. Multiple scattering in the air between the HMS vacuum exit window and the first drift chamber, the air between the drift chambers, and the drift chambers themselves make the actual angular resolution somewhat worse, and, as discussed in section \ref{hodosection}, the introduction of S0 before the first drift chamber adds a multiple scattering contribution which dominates the angular resolution.
\subsection{HMS Optics}\label{opticssection}
\paragraph{}
Once the proton trajectory at the focal plane is reconstructed from the pattern of drift chamber hits and measured drift times, the next step is to reconstruct its momentum, trajectory angles and position at the target. The optical properties of the HMS are very well known after more than a decade of data taking and repeated calibration measurements. The state of the particle coming from a reaction in the target, for purposes of transport through the HMS magnets, is characterized by its position at $z_{spec}=0$, $(x_{tar}, y_{tar})$, the angles of its trajectory, $(x'_{tar}, y'_{tar})$, and its deviation from the central momentum of the HMS, $\delta \equiv (p-p_0)/p_0$. The correspondence between these ``target'' coordinates in the five-dimensional $(x_{tar}, y_{tar}, x'_{tar}, y'_{tar}, \delta)$ phase space and the four focal-plane trajectory coordinates $(x_{fp}, y_{fp}, x'_{fp}, y'_{fp})$ is one-to-one when one of the target coordinates is fixed. 

For the analysis of HMS data, one is faced with solving the inverse problem of reconstructing the target coordinates from the focal plane coordinates. Since only four parameters of the track are measured at the focal plane, it is generally not possible to reconstruct all five target coordinates. However, since the beam impinges on the target at an almost fixed vertical position which is known on a per-event basis from the raster signals, it is adequate as a first approximation to set $x_{tar}$, which is the vertical position at $z_{spec}=0$ of the scattered particle, equal to the vertical beam position, and then solve for the other four target coordinates $y_{tar}$, $x'_{tar}$, $y'_{tar}$, and $\delta$ assuming this value of $x_{tar}$. For thin targets, this approximation is quite accurate. For extended targets such as the 20 cm cryotarget used in this experiment, however, significant corrections arise for particles scattered from different points along the length of the target. 

Quite generally, the solution for the reconstruction coefficients can be expanded in a Taylor-series up to arbitrary order:
\begin{eqnarray}
  (x',y',y,\delta)_{tar} &=& \sum_{\alpha,\beta,\mu,\nu,\lambda=0}^{\alpha+\beta+\mu+\nu+\lambda \leq n} C_{(x',y',y,\delta)}^{\alpha \beta \mu \nu \lambda}(x_{fp})^\alpha (y_{fp})^\beta (x_{fp}')^\mu (y_{fp}')^\nu (x_{tar})^\lambda \label{HMSrecon_expansion}
\end{eqnarray}
The order $n$ of the expansion is defined so that only terms for which the sum of the five exponents does not exceed $n$ are included in the expansion. The coefficients $C$ are determined in an iterative fitting procedure \cite{CMOP} from a more or less reasonable starting set of coefficients determined by a model of the spectrometer. COSY\cite{COSY}, a differential-algebra based code for the modeling of charged-particle optics and other applications, was used as the starting model for the HMS. The fitting of the expansion coefficients is carried out up to sixth order. To calibrate the spectrometer reconstruction coefficients, a series of dedicated optics runs using the sieve slit collimator and thin multi-foil targets was taken. For these runs, the S0 detector was removed. Another, smaller trigger scintillator was placed between the two drift chambers near the optical focal point of the spectrometer in order to achieve better angular resolution than in the default trigger configuration. In addition to the sieve-slit, multi-foil target runs, a set of ``delta scan'' data consisting of a series of elastic-$ep$ runs using the 20-cm hydrogen target were taken. For these runs, the HMS central angle was varied at a fixed central momentum in order to scan the envelope of elastic-$ep$ scattering events across the acceptance of the HMS. Since the proton scattering angle and momentum are correlated, different regions of the HMS acceptance in $\delta$ were populated at different HMS central angles. Additionally, since the measured scattering angle determines the proton momentum through two-body elastic kinematics, the delta-scan data could be used to fit the reconstruction matrix elements for $\delta$.

Each of the four reconstructed target variables has an independent polynomial expansion, so the four sets of coefficients are optimized separately. For the $y_{tar}$ and angle optimization, the multi-foil/sieve-slit data is used. A number of thin solid target foils at a known, surveyed $z$ position along the beamline provide point targets at known, fixed $y_{tar}$. Additionally, the sieve slit collimator blocks all tracks but those passing through narrow holes at known, surveyed positions. Having tracks originating from point targets at known positions, passing through small holes at known positions, provides a set of events populating nearly the full acceptance of the HMS in $y_{tar}$, $x'_{tar}$, and $y'_{tar}$, with well-defined rays of known $x'_{tar}$, $y'_{tar}$, and $y_{tar}$ that can be compared to the values reconstructed from the expansion \eqref{HMSrecon_expansion} in the optimization process.
\begin{figure}[h]
  \begin{center}
    \includegraphics[angle=90,width=.99\textwidth]{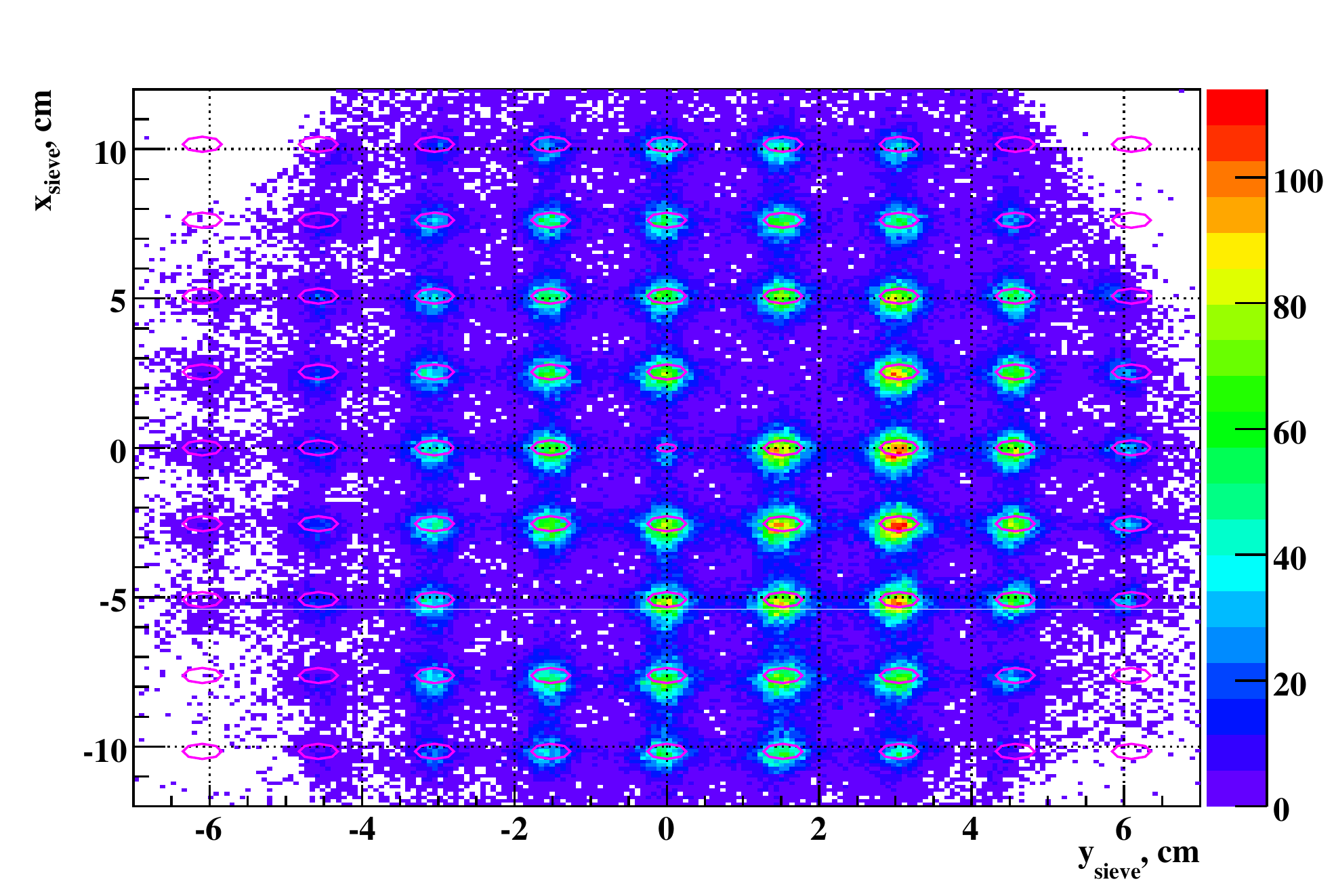}
  \end{center}
  \caption{\label{xysieve} Reconstructed HMS tracks projected to the sieve slit. Magenta ellipses are the sieve holes. In these runs, the HMS was set for negative polarity to detect electrons at a central momentum of 2.4 GeV/c and a central angle of 22.0 degrees. Data are from a two-foil Aluminum target with foils at $z=0.95\pm3.8$ cm, a two-foil Carbon target with foils at $z=0.95\pm2.0$ cm, and a three-foil Aluminum target with foils located at $z=0.95\pm7.5$ cm and at $z=0.95$ cm. A cut was applied on the HMS calorimeter energy to select electrons.}
\end{figure}

Figure \ref{xysieve} shows the projection of reconstructed tracks to the sieve slit, for data taken with the HMS set at negative polarity in order to detect electrons, at a central momentum of 2.4 GeV/c and a central angle of 22.0 degrees. The data in the plot come from three different multi-foil targets\footnote{A global offset of +0.95 cm from the ideal $z$ position was applied to all the optics targets, reflecting the result of a survey of the target positions\cite{JonesPrivateComm1}.}:
\begin{itemize}
\item The three-foil Aluminum target, with foils located at $z=0.95 \pm 7.5$ cm and at $z=0.95$ cm. 
\item The two-foil Carbon target, with foils located at $z=0.95\pm2.0$ cm.
\item The two-foil Aluminum target, with foils located at $z=0.95\pm3.8$ cm.
\end{itemize}
The data plotted in figure \ref{xysieve} were reconstructed after optimization, using the same reconstruction coefficients that were used in the final analysis. For the optimization of $x'_{tar}$, a special procedure was carried out which took into account the dependence of $x'_{tar}$ on $x_{tar}$ for a given $(x,y,x',y')_{fp}$. At $y_{tar}=z_{vertex}=0$, $x_{tar}$ is exactly equal to the vertical beam position\footnote{The beam position from the BPMs and the raster signals is in a different coordinate system in which $+y$ points vertically upward, so in fact $x_{tar}=-y_{beam}$}$^,$\footnote{$z_{vertex}$ is defined as the position of the interaction vertex along the target/beamline.}. However, since the beam intersects the ray of the scattered proton trajectory at $z_{vertex}\neq 0$, for non-zero $y_{tar}$/$z_{vertex}$, the track must be projected back to $z_{spec.}=0$:
\begin{eqnarray}
  x_{tar} = -y_{beam} - x'_{tar} z_{vertex} \cos \Theta_{HMS} \label{xtardef}
\end{eqnarray}

In this experiment, the hydrogen target extends over an asymmetric range in $z_{vertex}$ of $-6.16\ \mbox{cm}\leq z_{vertex} \leq 13.84\ \mbox{cm}$. For rays at the far end of the target and near the maximum $|x'_{tar}|$, significant corrections to $x_{tar}$ arise. For example, with the HMS positioned at an angle of 11.6 degrees, a particle scattered at $z_{vertex}=13$ cm and $x'_{tar}=60$ mrad experiences a correction to $x_{tar}$ of approximately 7.6 mm, which is much larger than the rastered beam spot size of 2 mm and introduces significant corrections to the reconstruction of both the momentum and $x'_{tar}$ of the scattered particle. Although this represents the worst-case correction for this experiment, it absolutely cannot be neglected in the reconstruction of $x'_{tar}$ and $\delta$. 

It is easy to see qualitatively how the vertical beam position affects the momentum reconstruction. Consider a particle detected at a position $x_{fp}$ in the HMS. Suppose the beam position on the target is actually $y_{beam}=-x_{tar}=+3$ mm, but the momentum is reconstructed assuming $y_{beam}=-x_{tar}=0$. By assuming that the particle starts at a higher $x$, a smaller vertical deflection is assumed than actually took place, which corresponds to a smaller total bend angle, and a higher momentum than the true momentum. By similar logic, if the true beam position is $y_{beam}=-3$ mm, than the reconstructed momentum will be lower than the true momentum for an assumed $y_{beam}=0$. A similar effect operates for $x'_{tar}$.

$x_{tar}$-dependent reconstruction coefficients were calculated up to sixth order within the COSY model of the HMS. The lowest order reconstruction matrix elements involving $x_{tar}$ are $\left<\delta|x_{tar}\right> = 0.077\ \%/mm$ and $\left<x'_{tar}|x_{tar}\right>=1.1\ mrad/mm$. Immediately one sees the importance of $x_{tar}$ for extended targets. With corrections to $x_{tar}=-y_{beam}$ approaching 8 mm for rays near the extremes of vertex position and $x'_{tar}$, and especially at small $\Theta_{HMS}$, the correction to the momentum can be as large as 0.6\%, and the correction to $x'_{tar}$ can be as large as 9 milliradians, as compared to the 0.1\% momentum resolution and $\approx$1 mrad angular resolution of the HMS\footnote{The $x_{tar}$ dependence of $y_{tar}$ and $y'_{tar}$, on the other hand, is much smaller.}. In previous optimizations of the HMS reconstruction coefficients, the $x_{tar}$ correction had been neglected, since previous experiments had not used targets as long as the cryotarget of this experiment and smaller corrections were involved. When using the reconstruction coefficients in the analysis, the value of $x_{tar}$ is not \emph{a priori} known, so the full correction cannot be applied. It is approximated by $-y_{beam}$. Once $x'_{tar}$, $y'_{tar}$ and $y_{tar}$ are roughly known from the first iteration of the reconstruction, one can calculate $x_{tar}$ and perform an arbitrary number of additional iterations of the reconstruction to improve $x'_{tar}$ and $\delta$ an arbitrary number of times. One additional iteration is enough in practice to correct the reconstructed quantities to a level well below the intrinsic resolution of the HMS and certainly below the resolution needed by this experiment.

The $\delta$ matrix elements involving $x_{tar}$ were found to significantly improve the momentum reconstruction with no further optimization needed. On the other hand, the reconstruction of $x'_{tar}$ became worse in the second iteration of reconstruction when the existing coefficients were used. In order for a second iteration to improve the reconstruction of $x'_{tar}$, the expansion coefficients had to be refitted taking into account the $x_{tar}$ corrections for foils located at large $|z|$. In the fitting procedure, all matrix elements involving non-zero powers of $x_{tar}$ were fixed; only those elements with no $x_{tar}$ dependence were allowed to vary. The fitting procedure was performed three times. In each subsequent fit, the matrix elements of the previous fit were used as the starting point for the next fit. The following plots demonstrate the improvement in the reconstruction of $x'_{tar}$. 
\begin{figure}
  \begin{center}
    \includegraphics[width=.49\textwidth]{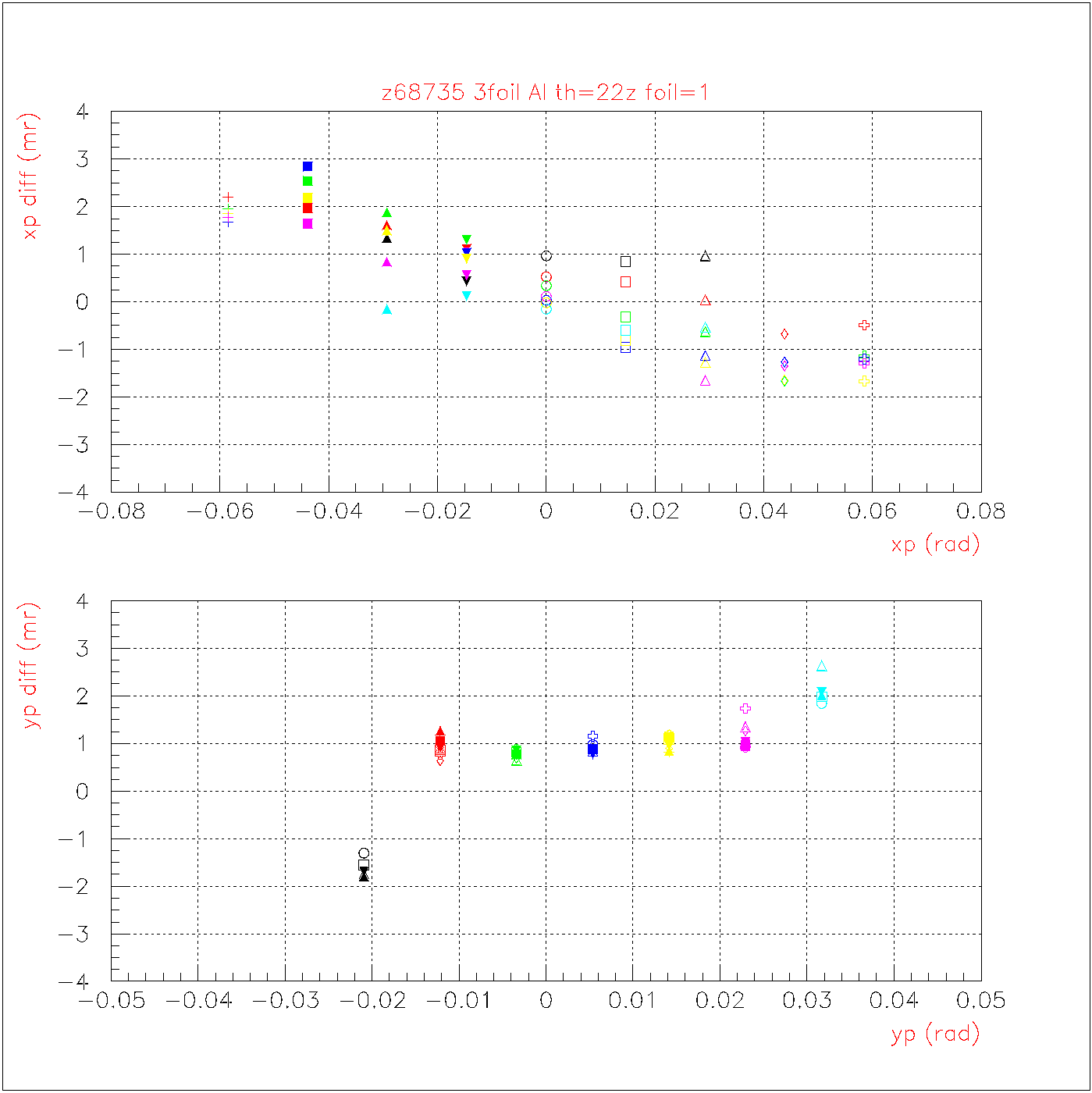}
    \includegraphics[width=.49\textwidth]{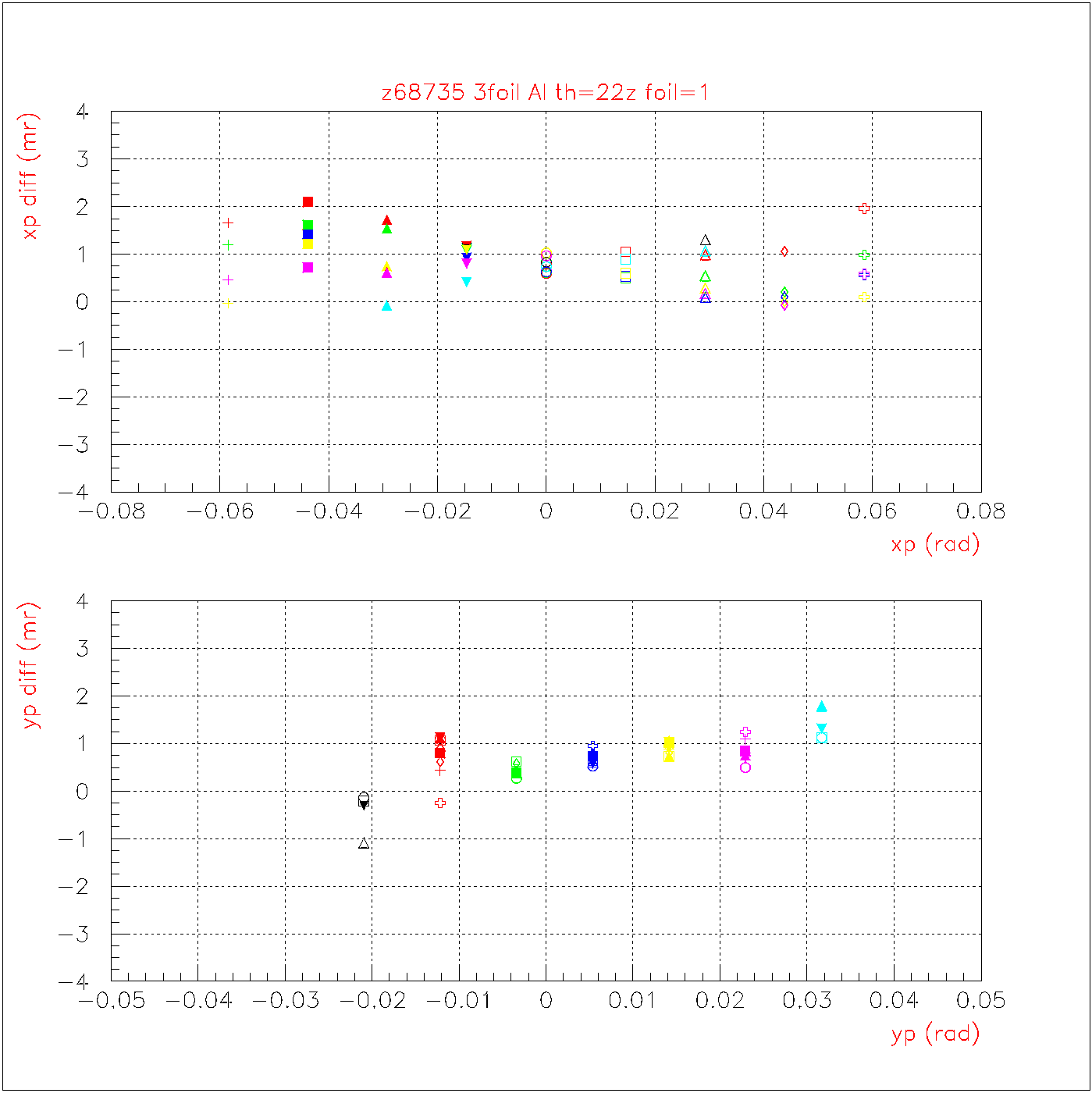}
  \end{center}
  \caption{\label{improved_foil1} Difference between reconstructed and known angles for foil 1, located at $z_{vertex}=-6.55$ cm. Top plot shows $x'_{recon.}-x'_{true}$ vs. $x'_{true}$, bottom plot shows $y'_{recon}-y'_{true}$ vs. $y'_{true}$. Left: reconstruction using \textbf{old} coefficients, fitted without $x_{tar}$ correction. Right: reconstruction using \textbf{new} coefficients, fitted with $x_{tar}$ correction.}
\end{figure}
\begin{figure}
  \begin{center}
    \includegraphics[width=.49\textwidth]{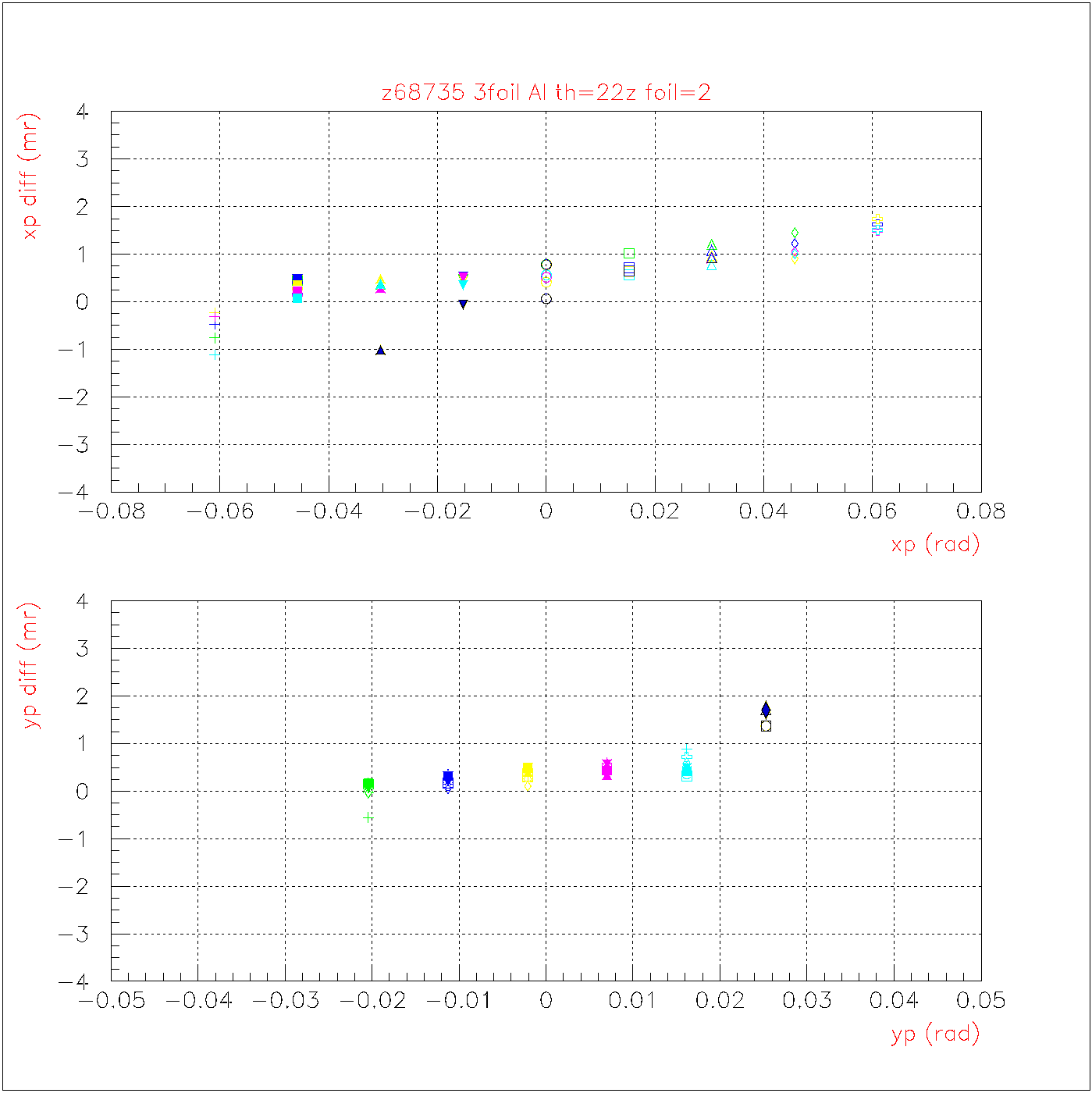}
    \includegraphics[width=.49\textwidth]{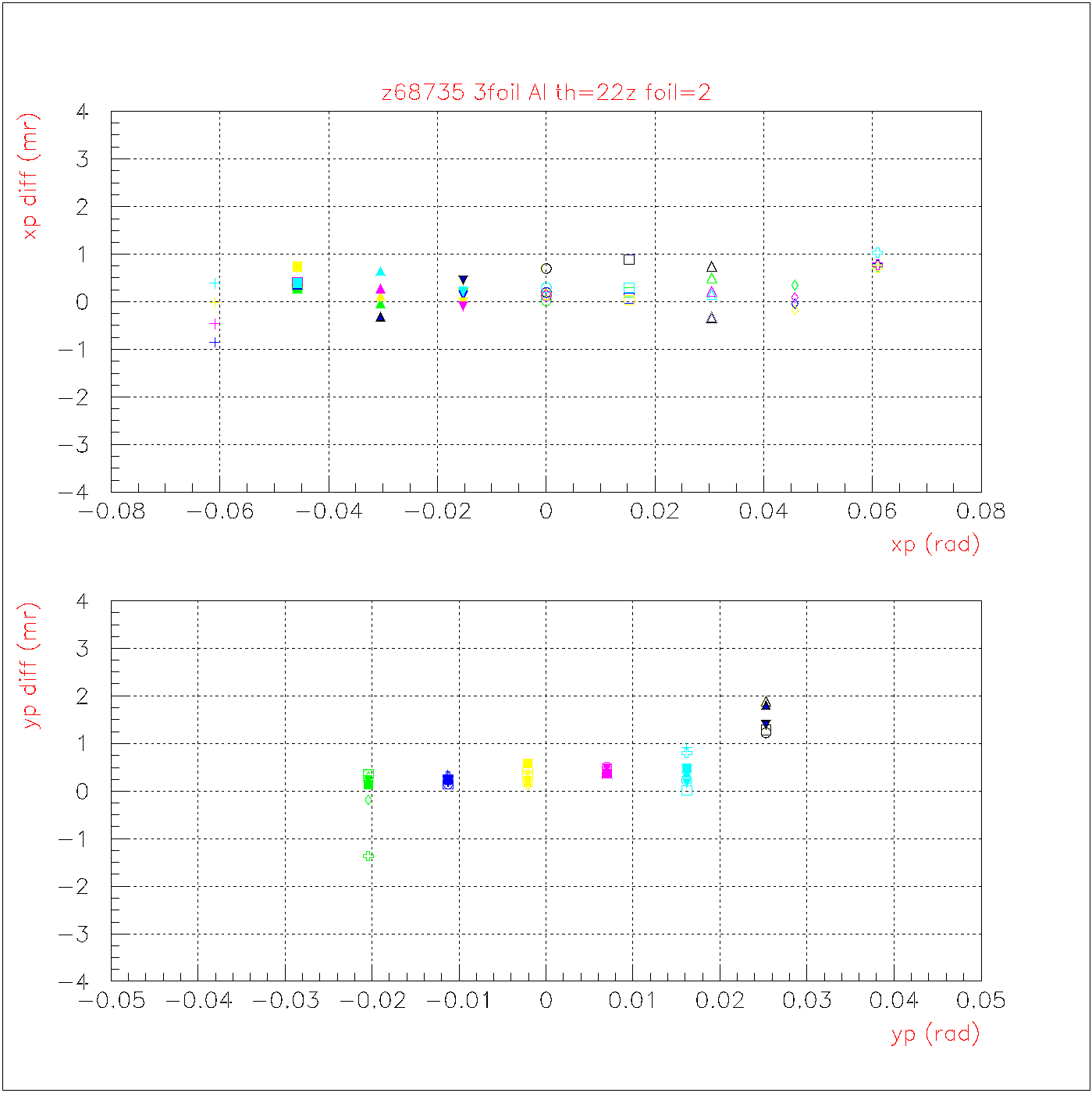}
  \end{center}
  \caption{\label{improved_foil2} Difference between reconstructed and known angles for foil 2, located at $z_{vertex}=0.95$ cm. Top plot shows $x'_{recon.}-x'_{true}$ vs. $x'_{true}$, bottom plot shows $y'_{recon}-y'_{true}$ vs. $y'_{true}$. Left: reconstruction using \textbf{old} coefficients, fitted without $x_{tar}$ correction. Right: reconstruction using \textbf{new} coefficients, fitted with $x_{tar}$ correction.}
\end{figure}
\begin{figure}
  \begin{center}
    \includegraphics[width=.49\textwidth]{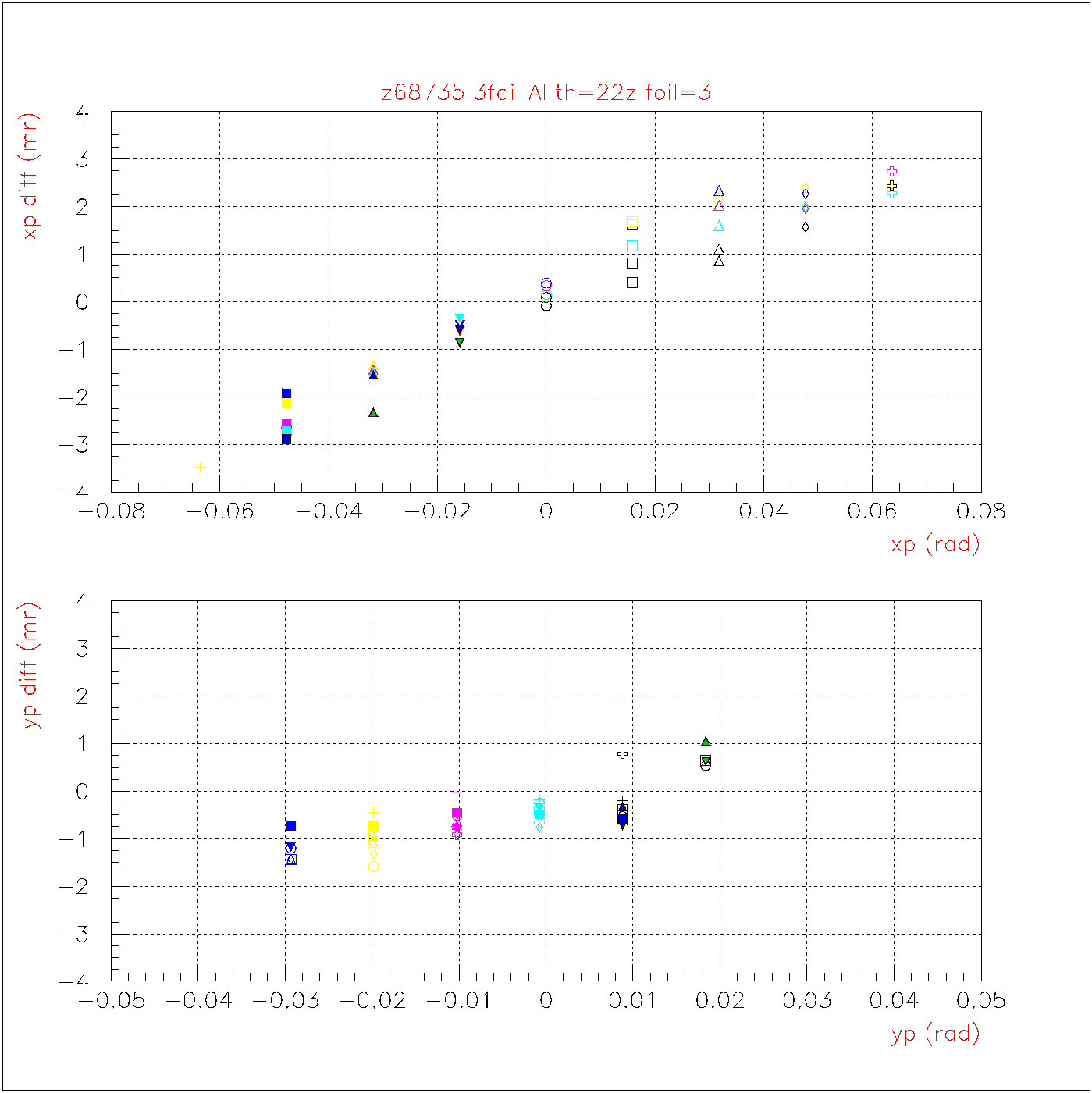}
    \includegraphics[width=.49\textwidth]{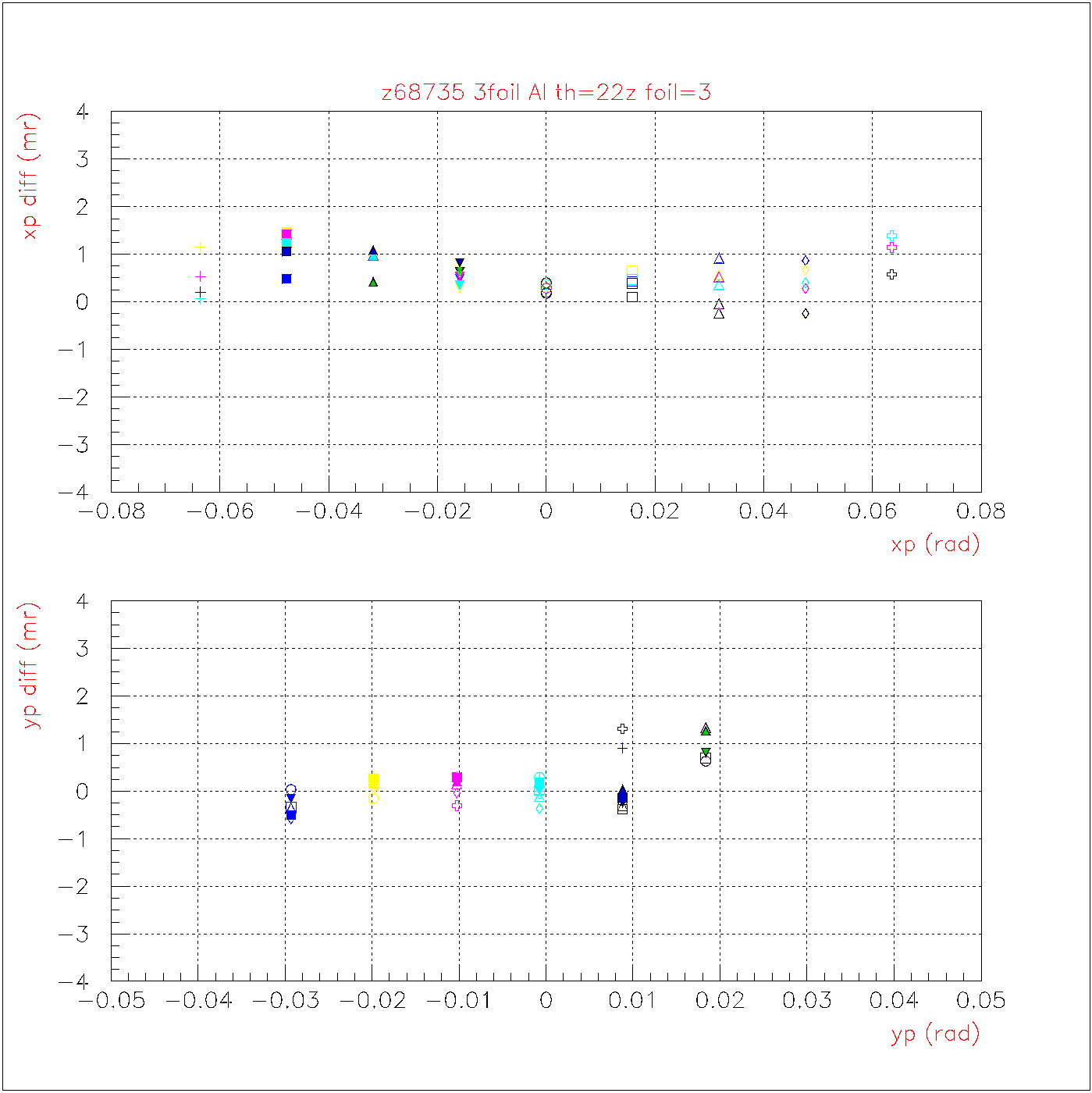}
  \end{center}
  \caption{\label{improved_foil3} Difference between reconstructed and known angles for foil 3, located at $z_{vertex}=8.45$ cm. Top: $x'_{recon.}-x'_{true}$ vs. $x'_{true}$. Bottom: $y'_{recon}-y'_{true}$ vs. $y'_{true}$. Left: reconstruction using \textbf{old} coefficients, fitted without $x_{tar}$ correction. Right: reconstruction using \textbf{new} coefficients, fitted with $x_{tar}$ correction.}
\end{figure}
\begin{figure}
  \begin{center}
    \includegraphics[angle=90,width=.99\textwidth]{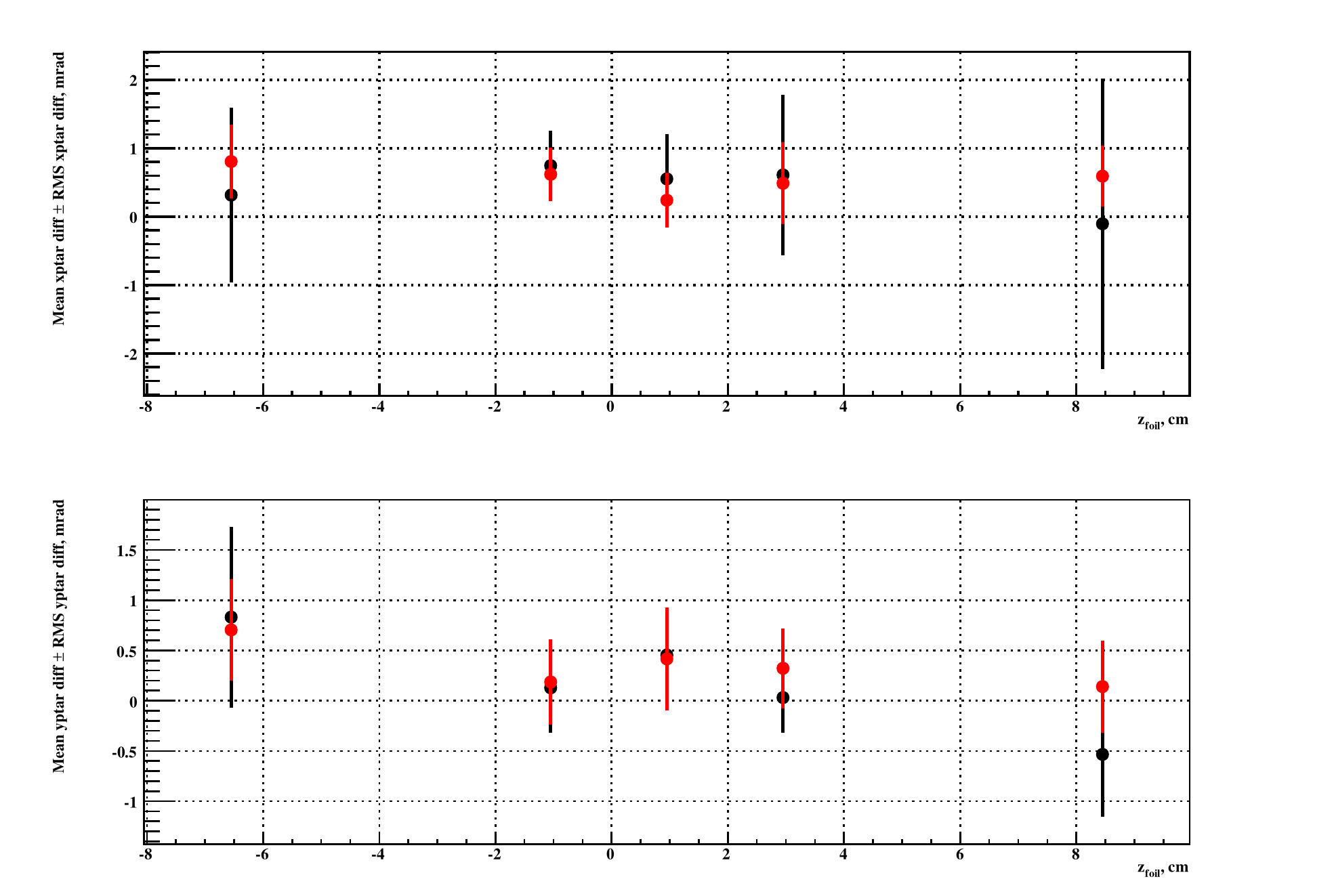}
  \end{center}
  \caption{\label{avgdiffs_vs_z} Average difference (over all sieve holes for a given foil) between reconstructed and known angles as a function of $z_{foil}$. Data points are average differences in mrad, with ``error bars'' representing the R.M.S. hole-to-hole differences in mrad. Black points are reconstructed using the old coefficients. Red points are reconstructed using the new coefficients.}
\end{figure}

At a given HMS central angle $\Theta_{HMS}$, a particle scattered from a thin foil located at $z_{vertex} = z_{foil}$ and passing through a sieve hole centered at $(x,y)=(x_{hole},y_{hole})$ has 
\begin{eqnarray}
  y'_{true} = \frac{y_{hole} - z_{foil}\sin \Theta_{HMS}}{D - z_{foil}\cos \Theta_{HMS}}
\end{eqnarray}
where $D$ is the distance from the origin to the sieve slit collimator. Similarly, $y_{tar}$ is determined from $y'_{true}$ and $z_{foil}$ by 
\begin{eqnarray}
  y_{true} = z_{foil} \left( \sin \Theta_{HMS} - y'_{true} \cos \Theta_{HMS}\right)
\end{eqnarray}
and $x'_{tar}$ is determined from 
\begin{eqnarray}
  x'_{true} = \frac{x_{hole} + y_{beam}}{D - z_{foil}\cos \Theta_{HMS}}
\end{eqnarray}
Figures \ref{improved_foil1}--\ref{improved_foil3} show the improvement made possible by including the $x_{tar}$ correction in the fitting of $x'_{tar}$. Each plot shows the difference $x'_{tar}-x'_{true}$(top) and $y'_{tar}-y'_{true}$(bottom) between reconstructed and true angles as a function of $x'_{true}$ and $y'_{true}$, respectively, in milliradians. Each marker type corresponds to a different sieve slit row (constant $x$), while each marker color corresponds to a different sieve slit column (constant $y$). Each sieve hole is represented by a unique marker type and color. The plots on the left in each figure show the data reconstructed using the old HMS matrix elements, while the plots on the right in each figure show the data reconstructed using the new, re-optimized HMS matrix elements. In each plot, if the reconstruction coefficients worked perfectly well, and all of the following quantities were known exactly
\begin{itemize}
  \item The positions of the sieve holes relative to the HMS optical axis. 
  \item The absolute positions and angles of the beam on target. 
  \item The absolute position of the target foils. 
  \item The distance from the target foils to the collimator. 
  \item The pointing angles of the spectrometer.
\end{itemize} 
, then all points would lie along the $x'(y')_{diff}=0$ axis. For the $x'_{tar}$ reconstruction coefficients, the improvement is obvious and unambiguous. For the foils at $z=-6.55$ cm and at $z=+8.45$ cm, the $x'$ difference reconstructed from the old matrix elements exhibited a strong slope, changing by up to 6 mrad over the full $x'_{tar}$ acceptance, symptomatic of the neglect of the $x_{tar}$ correction. After optimization, the slopes are largely absent. For $y'_{tar}$, re-optimization yields slight improvement in slopes that were already quite small, with $y'_{tar}$ differences no more than 2 mrad for any single sieve hole. 

Figure \ref{avgdiffs_vs_z} shows the $x'$ and $y'$ differences as a function of $z_{foil}$ for the five foils used in the optimization, which included the Aluminum foils in figures \ref{improved_foil1}--\ref{improved_foil3}, and the two-foil Carbon target, with foils located at $z_{foil}=-1.05$ and $2.95$ cm, respectively. The data are plotted as a function of $z_{foil}$ and averaged over all sieve holes for each foil. The ``error bars'' are not uncertainties but RMS hole-to-hole deviations from the mean. The black points show the results using the old optics coefficients, while the red points show the results using the new optics coefficients. For both $x'$ and $y'$, the optimization is observed to flatten the $z$ dependence of the deviation from the ``true'' value and reduce its RMS spread, particularly for the foils at large $|z|$.
\begin{figure}
  \begin{center}
    \includegraphics[angle=90,width=.95\textwidth]{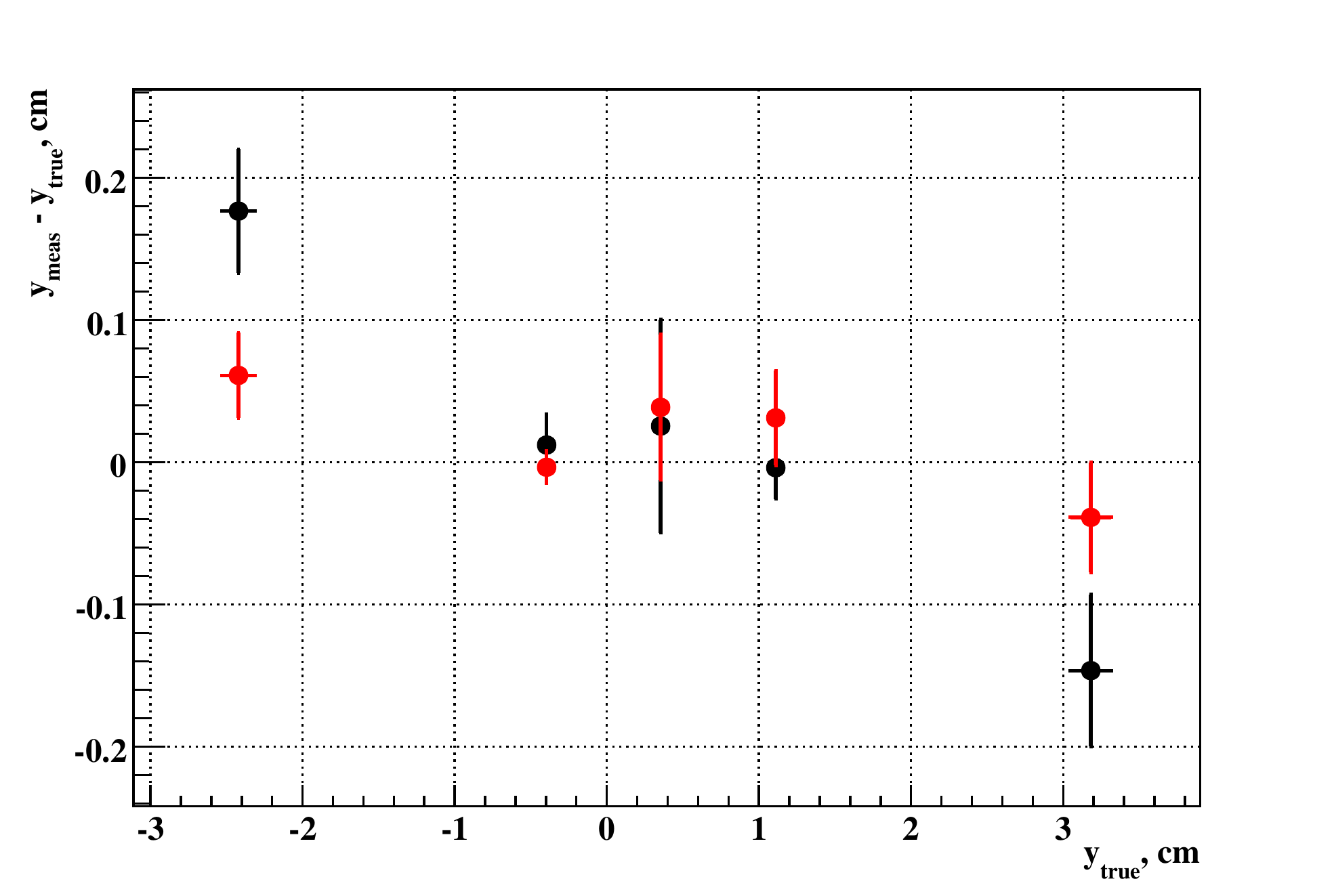}
  \end{center}
  \caption{\label{ytar_diffs} Mean (data points), and R.M.S. (``error bars'') difference between reconstructed $y_{tar}$ and $y_{true}$, before (black) and after (red) optimization.}
\end{figure}

The reconstruction of the position of the interaction vertex in the transverse spectrometer coordinate; i.e., $y_{tar}$, was also optimized. Figure \ref{ytar_diffs} shows the improvement. As in figure \ref{avgdiffs_vs_z}, the data points represent the difference between $y_{measured}$ and $y_{true}$ as a function of $y_{true}$, averaged over all sieve columns for a particular target foil. The vertical ``error bars'' represent the R.M.S. spread in $y_{diff}$ over all sieve columns for a given foil, while the horizontal bars represent the R.M.S. difference between $y_{true}$ for a given sieve column and the $y_{true}$ value averaged over all sieve columns for a given foil. The slight variations of $y_{true}$ as a function of sieve column simply reflect the fact that each sieve column represents a different $y'_{true}$ for a given foil position, and to get $y_{true}$, one must project from $z_{spec} = z_{foil}\cos \Theta_{HMS}$ back to $z_{spec} = 0$.
\begin{figure}
  \begin{center}
    \includegraphics[width=.9\textwidth]{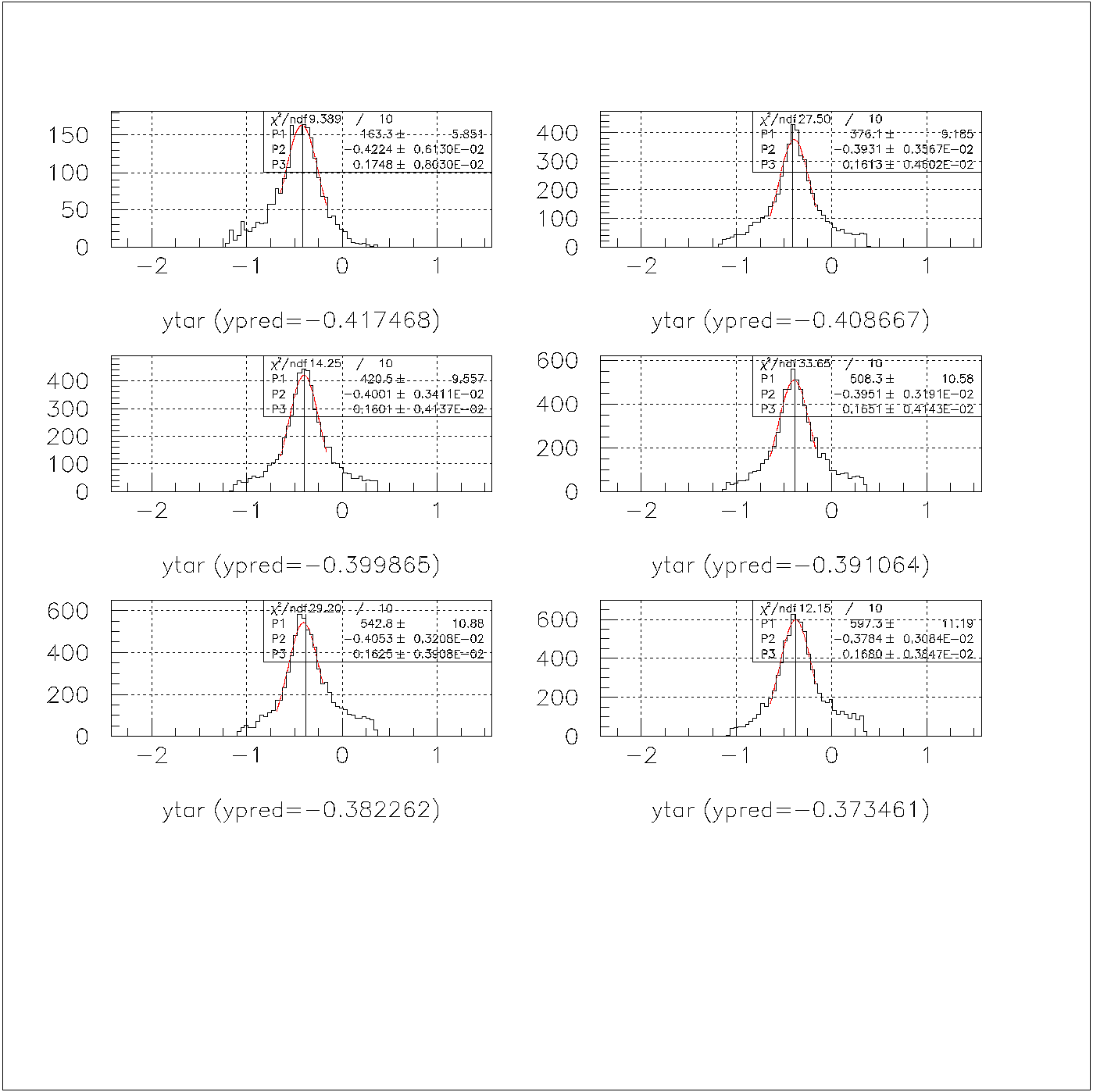}
  \end{center}
  \caption{\label{ytar_res} Resolution in $y_{tar}$. Each histogram represents the $y_{tar}$ distribution for a different sieve slit column for the carbon foil located at $z_{foil}=-1.05$ cm. ``ypred'' is what has been referred to as $y_{true}$; i.e., the $y_{tar}$ position of the foil projected to $z_{spec}=0$.}
\end{figure}
Figure \ref{ytar_res} shows the resolution achieved in $y_{tar}$. Each histogram in the picture shows the reconstructed $y_{tar}$ distribution for a different sieve slit column for the carbon foil located at $z_{foil}=-1.05$ cm. The $1\sigma$ resolution in $y_{tar}$ is approximately 1.6 millimeters. Similar results were obtained for all other foils after optimization. The first-order optical coupling of $y_{tar}$ to the focal-plane trajectory angles is relatively weak. $y_{tar}$ is most sensitive to the position of the trajectory at the focal plane. Therefore, like $\delta$, it is minimally affected by multiple scattering in S0. Even though the data in figure \ref{ytar_res} were taken with no S0, the observed resolution in $y_{tar}$ with S0 in the detector stack from data taken with the ``dummy'' target was approximately 1.7 millimeters at $p_0=5.4$ GeV/c. At lower momenta, the resolution was slightly worse.

To check the quality of the momentum reconstruction, elastic $ep$ scattering from hydrogen was used. From the reconstructed scattering angle of the proton, $\theta_p$, and the known beam energy corrected for energy loss in the target, the proton momentum is determined by equation \eqref{pel_hthe}.  The data were taken with a beam energy $E_{beam} = 4.109$ GeV, with the HMS central momentum set to 2.02 GeV/c. BigCal was placed at an angle of approximately 25.8 degrees at a distance of approximately 8.82 meters from the target\footnote{The position and orientation of BigCal were not surveyed at this setting so the numbers could not be independently confirmed, only re-aligned in software according to the location of the elastic peak.}. Elastic scattering data were obtained at central angles of 40.5$^\circ$, 39.5$^\circ$, 38.5$^\circ$, 37.5$^\circ$, 36.5$^\circ$, and 36.0$^\circ$. The angular acceptance of the HMS is roughly $\pm$1.8$^\circ$, so scanning across a 4.5$^\circ$ range in $\theta_p$ is sufficient to populate the full acceptance of the HMS with elastically scattered protons. Another advantage of having taken this data at relatively large $\theta_p$ is that a large range of $y_{tar}$ is populated as well, since $y_{tar} \approx z_{vertex} \sin \Theta_{HMS}$. In this way, the validity of the momentum reconstruction was checked over an effective target length in $y_{tar}$ that was larger than that of all the other kinematics of both experiments which used smaller $\Theta_{HMS}$ with the same target length along $z$.

\begin{figure}[h]
  \begin{center}
    \includegraphics[width=.99\textwidth]{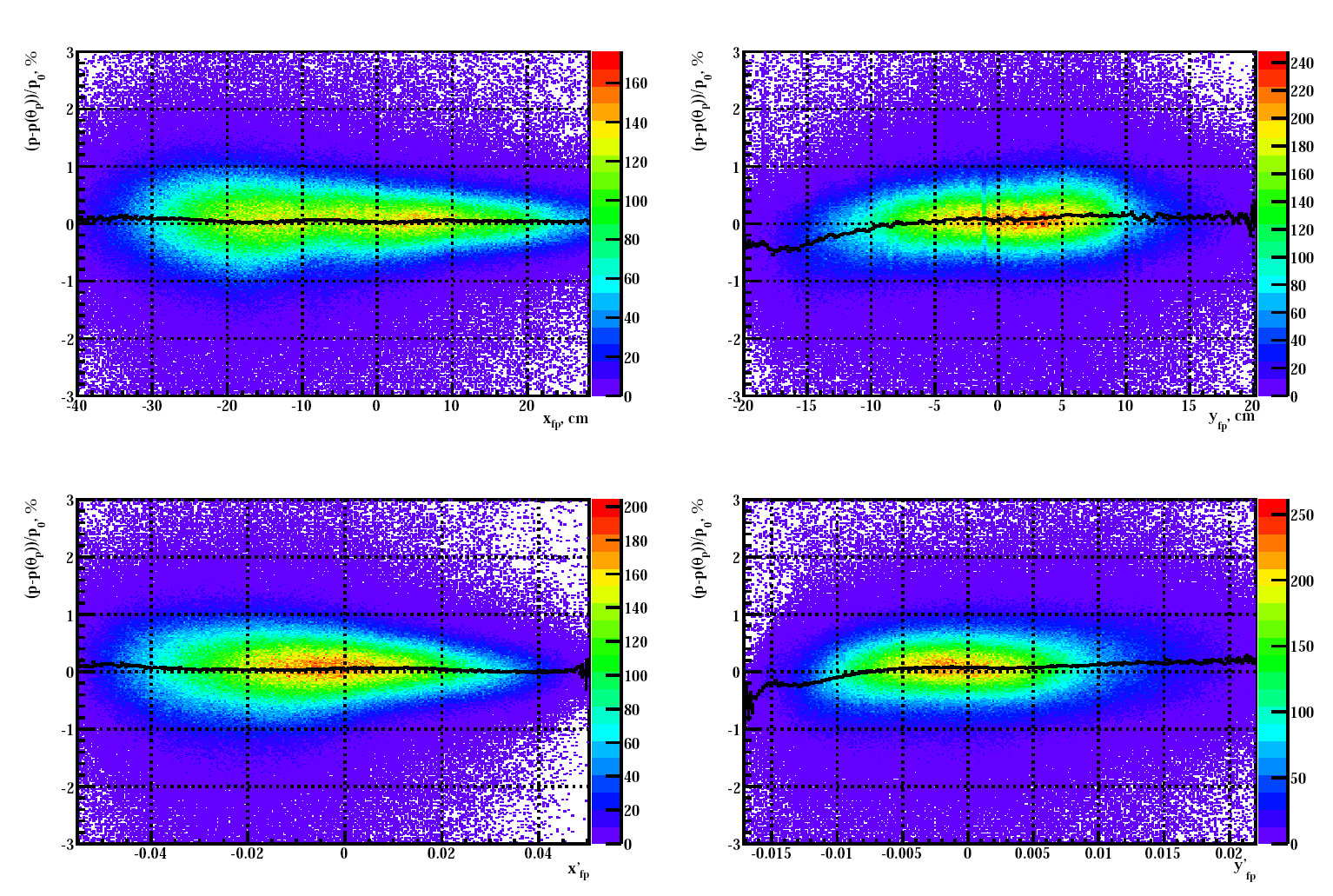}
  \end{center}
  \caption{\label{pdiff_vs_fp} Difference between reconstructed proton momentum and expected momentum from the reconstructed proton angle as a function of $x_{fp}$ (top left), $y_{fp}$ (top right), $x'_{fp}$ (bottom left), and $y'_{fp}$ (bottom right). The momentum difference is expressed as a percentage of central momentum. The black points represent the mean momentum difference in each bin.}
\end{figure}

Figure \ref{pdiff_vs_fp} shows the difference between the reconstructed proton momentum and the momentum of an elastically scattered proton at the reconstructed proton angle $\theta_p$, as a function of the focal plane trajectory parameters $x_{fp}$, $y_{fp}$, $x'_{fp}$, and $y'_{fp}$, expressed as a percentage of central momentum. The black curves in each plot show the mean momentum difference in each bin and illustrate the excellent quality of the reconstruction. As a function of $x_{fp}$ and $x'_{fp}$, the quantities which primarily determine $\delta$, there is no significant variation of $p-p(\theta_p)$ across the full acceptance. As a function of $y_{fp}$ and $y'_{fp}$, however, some significant variations appear at large, negative values of $y_{fp}$ and $y'_{fp}$. These variations come mainly from errors in the reconstruction of $\theta_p$ for events coming from extreme positions along the length of the target, since the HMS reconstruction coefficients are not very well behaved when extrapolated outside the range in $y_{tar}$ and $y'_{tar}$ of the data used in the fitting procedure, as illustrated in figure \ref{pdiff_vs_target}.

\begin{figure}[h]
  \begin{center}
    \includegraphics[width=.99\textwidth]{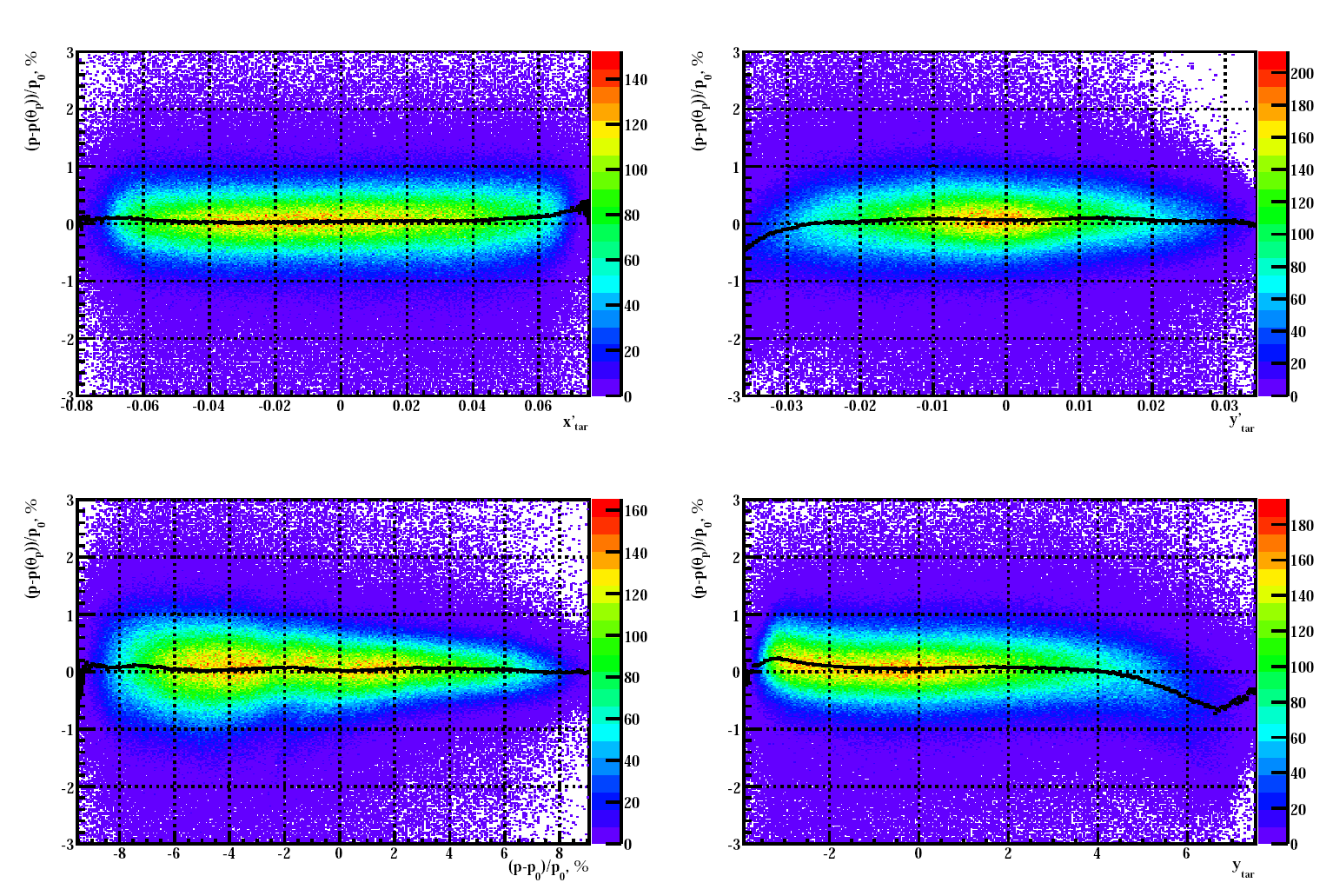}
  \end{center}
  \caption{\label{pdiff_vs_target} Difference between reconstructed proton momentum and the momentum of an elastically scattered proton at the reconstructed $\theta_p$, as a function of the reconstructed target variables $x'_{tar}$ (top left), $y'_{tar}$ (top right), $\delta$ (bottom left), and $y_{tar}$ (bottom right).}
\end{figure}

Figure \ref{pdiff_vs_target} shows the same momentum difference as a function of the reconstructed target variables $x'_{tar}$, $y'_{tar}$, $\delta$, and $y_{tar}$. The mean momentum difference shows significant deviation from a constant for $y_{tar} < -2.5$ cm and $y_{tar} > 4$ cm, and also for $y'_{tar} < -28$ mrad and $y'_{tar} > 32$ mrad, reflecting the decreasing quality of the reconstruction coefficients outside the range of the data used in the fit (recall figures \ref{improved_foil1}--\ref{ytar_diffs}). Given the absence of any significant correlation between the reconstructed momentum difference and the quantities $x_{fp}$ and $x'_{fp}$ which are dominant in determining the reconstructed $\delta$, and also the absence of significant correlations between $p-p(\theta_p)$ and reconstructed $x'_{tar}$ and $\delta$, it was concluded that the residual correlations between $p-p(\theta_p)$ and $y_{tar}$/$y'_{tar}$ were mainly due to errors in the reconstruction of $\theta_p$, which is primarily determined by $y'_{tar}$. Further optimization of the reconstruction matrix elements for $\delta$ was deemed unnecessary based on the above results. It is worth remarking that the correlations between $p-p(\theta_p)$ and $y_{tar}$, $y'_{tar}$, $y_{fp}$ and $y'_{fp}$ were significantly reduced by the optimization described above, and that the pictures shown in figure \ref{pdiff_vs_fp} and figure \ref{pdiff_vs_target} were significantly worse prior to the optimization. The final reconstruction matrix elements obtained from fitting $y_{tar}$, $y'_{tar}$, and $x'_{tar}$ including secondary $x_{tar}$ corrections provided the most accurate possible reconstruction over the largest possible acceptance that could be obtained from the optics calibration data taken in this experiment. Further improvements would have been possible given more optics data on multi-foil targets at larger $\Theta_{HMS}$ and/or foils located at larger $\left|z_{foil}\right|$; however, the optimization, which took place at $\Theta_{HMS}=22^\circ$, was sufficient to cover (nearly) the full acceptance in $y_{tar}$ for all but two kinematic settings, at $\theta_p \approx 31.0^\circ$, $\varepsilon \approx .63$ and $\theta_p \approx 36.1^\circ$, $\varepsilon \approx .79$ for $Q^2=2.5$ GeV$^2$. The $\delta$-scan data showed that the momentum reconstruction was already well optimized for $\left|\delta\right|<9$\%, a $\delta$ acceptance significantly larger than the range populated by elastically scattered protons for any of the kinematics (see table \ref{dptable}). In the final analysis, the results obtained using hard cuts around the ``safe'' regions in $x'_{tar}$, $y'_{tar}$ and $y_{tar}$ were compared to the results obtained using less restrictive cuts to exclude possible systematic error arising from regions where the optics are less well calibrated.
\subsection{FPP Drift Chamber Track Reconstruction}
\label{fpptrackingsection}
\paragraph{}
The next step in event reconstruction is to track protons (and other charged particles) scattered in the CH$_2$ blocks of the FPP. The tracking algorithm for the FPP drift chambers works somewhat differently than that of the HMS drift chambers. The HMS tracking algorithm is designed for speed, with pattern recognition and ``stub'' fitting proceeding separately for each chamber before considering both chambers together. The fact that all tracks of interest in the HMS drift chambers are very nearly perpendicular to the wire planes further simplifies the tracking procedure. Because of the relatively large number of wire planes (12) in the HMS drift chambers, this approach is preferable to a ``brute force'' tracking algorithm which considers all possible combinations of one hit per plane and performs full drift-based tracking to each combination. The FPP drift chambers, on the other hand, have only six wire planes with which to define a track, so a more exhaustive consideration of possible hit combinations is possible without requiring prohibitive CPU time. Furthermore, tracks of interest in the FPP drift chambers cover a much wider range of angles than in the HMS drift chambers. 

As in the HMS drift chamber tracking, the first step in the FPP tracking is to convert raw TDC values to rough drift times. This is done via equation \eqref{HMSroughdrift}, with a separate $t_0$ offset defined for each wire to align all the drift times in a window from 0 to approximately 200 ns. A loose cut is applied on the rough drift time to partially suppress noise and random hits not associated with the trigger. For the data taken with Fastbus TDCs, the raw hit time $t_{raw}$ is simply given by the count resolution (0.5 ns) times the raw TDC value. For the data taken with the VME TDCs, special care must be taken to identify events where the free-running counters roll back to zero between the arrival of the drift chamber hits and the arrival of the stop signal, as detailed in appendix \ref{F1decode}. Once the raw hit times are known and loose cuts are applied, the process of pattern recognition and track reconstruction begins. 

The pattern recognition procedure for the FPP drift chambers starts by grouping individual wire hits into ``logical hits'' consisting of clusters of hits on up to three adjacent wires in a given plane. By grouping adjacent hits, which probably come from the same track, the total number of hit combinations is reduced considerably. If more than three adjacent wires have hits, only the first three wires are considered for that logical hit, and subsequent wires are ignored. After the raw hits are grouped into logical hits, the algorithm examines all possible combinations of one logical hit per plane. On the first iteration, the tracking algorithm requires all six planes to fire, but if one plane fails to fire, it will also test five-plane combinations. If more than one plane fails to fire, the algorithm gives up as there will be insufficient hits to fully define a track.

For all combinations of one logical hit per plane, the code fits a ``simple'' track (straight line) to the positions of the hit wires, ignoring the drift distance information. A $\chi^2$ test is defined as 
\begin{equation}
  \chi^2 \equiv \sum_{i=1}^{N_{hit}} \left(\frac{w_{track} - w_{wire}}{\sigma_{wire}}\right)^2 
\end{equation}
where $w_{track}$ is the track projection along the direction measured by the wire, $w_{wire}$ is the position of the wire center, and $\sigma_{wire}$ is a resolution parameter defined as the wire spacing $d$ divided by $\sqrt{12}$. If only one wire per plane is considered, then the maximum allowed track-wire distance should be 1 cm or $d/2$. Thus, the maximum contribution of a single wire to $\chi^2$ is $\left(\frac{d/2}{d/\sqrt{12}}\right)^2 = 3$. For six planes, there are two degrees of freedom in the fit, so the maximum possible $\chi^2$ for a ``good'' track is $\chi^2_{max}/n.d.f =  3\times6 / 2 = 9$. In order to handle the case where a logical hit contains two or three adjacent hits in the same plane, the code fits an initial track with all wires included and then projects that track to each plane to calculate the residual for each hit wire in that plane. The closest wire to the initial track in each plane is chosen and a second fit is performed with only one wire per plane, and a cut is applied to the $\chi^2$ of the second track to decide whether the combination of logical hits under consideration should be considered as a candidate for drift-based tracking.

For every combination of one logical hit per plane passing the $\chi^2$ cut of the fit to wire positions only, the next step is to consider the drift time information and perform full track reconstruction. Since the rough track coordinates are known from the fits to wire positions at this stage of the analysis, the drift times are corrected for the delay due to the finite signal propagation speed on the wires based on the distance along the wire from the point where the track crossed the wire plane to the readout card. An example of a corrected drift time spectrum is shown in figure \ref{fppdrifttime}. 
\begin{figure}[h]
  \begin{center}
    \subfigure[]{\label{fppdrifttime}\includegraphics[angle=90,width=.49\textwidth]{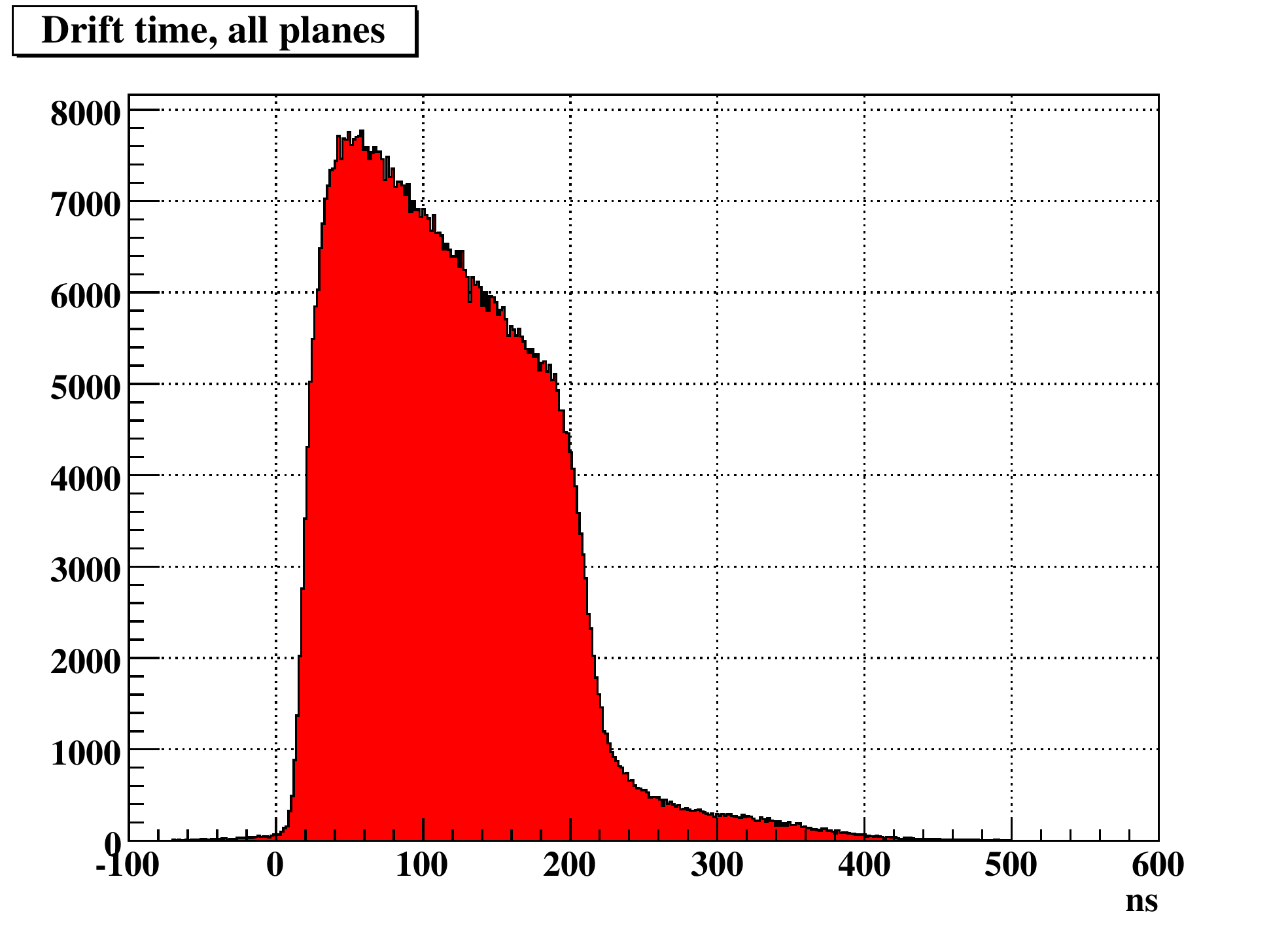}}
    \subfigure[]{\label{fppdriftdist}\includegraphics[angle=90,width=.49\textwidth]{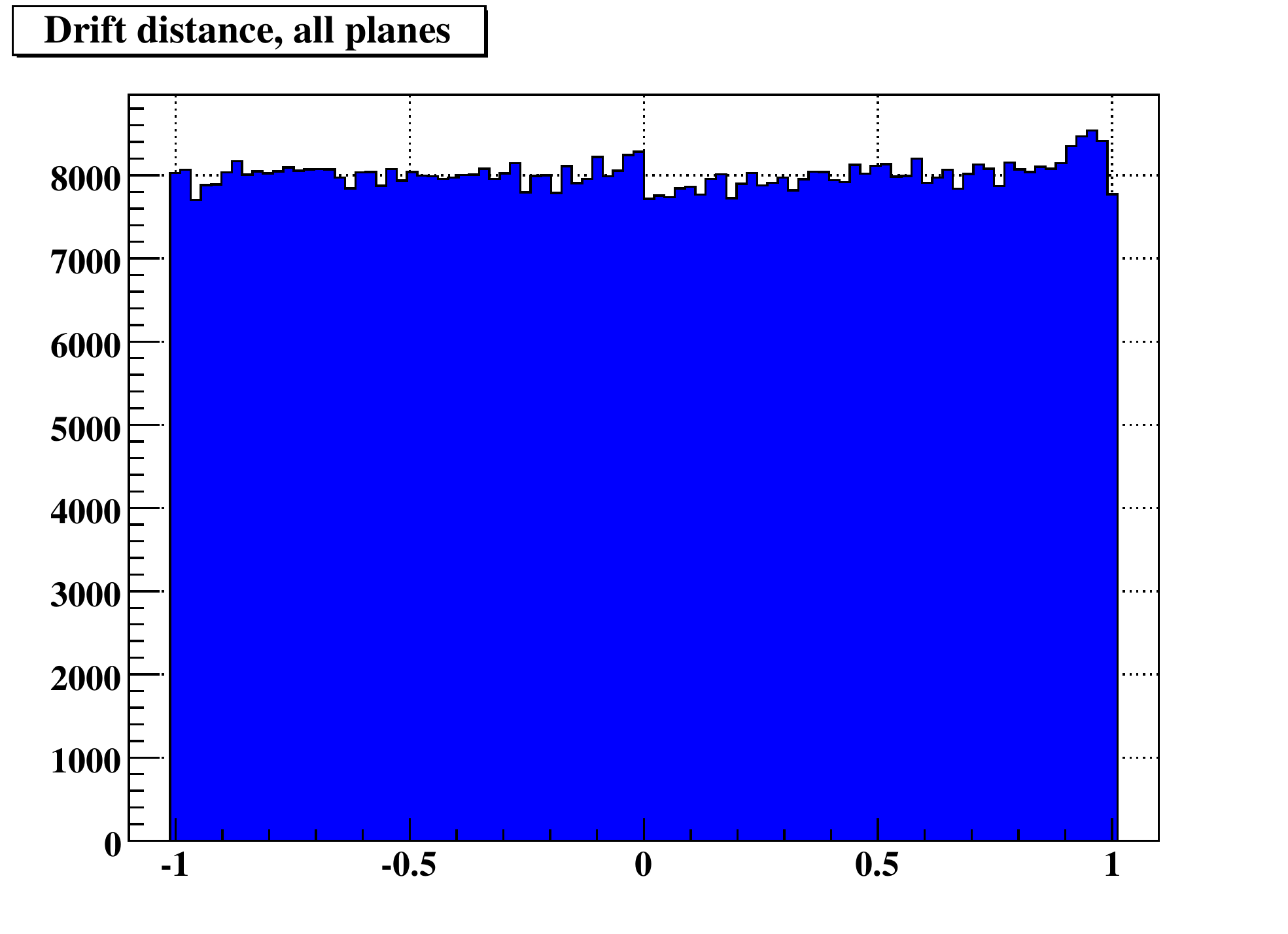}}
  \end{center}
  \caption{\label{fppdrifttimedist} FPP drift time spectrum (a) and FPP drift distance spectrum (b) after drift map calculation.}
\end{figure}
The spectrum in figure \ref{fppdrifttime} was obtained at the highest HMS central momentum setting of 5.4 GeV/c. It is qualitatively similar to the HMS drift time spectrum, but approximately twice as wide, with a more pronounced tail at large drift times (from approximately 250-400 ns). The large-time tail in the FPP drift time spectrum arises because of the fact that tracks which scatter at large angles in the CH$_2$ analyzer can cross the wire planes of the FPP at large angles of incidence relative to the normal to the wire planes, passing through the corners of drift cells and giving hits with larger closest-approach distance in regions of highly non-uniform electric field. Despite the presence of this tail, the observed drift time spectrum is translated into a time-to-distance map using a uniform drift distance mapping as in equation \eqref{driftmap_eq}. The resulting drift distance spectrum is shown in figure \ref{fppdriftdist}. 

Unlike figure \ref{HMSdriftmap}, figure \ref{fppdriftdist} also shows the sign of the drift distance determined from fitting. Although tracks may scatter at large angles and cross the wire planes at significantly non-normal incident angles, the drift time calculated from the drift distance was still equated with the in-plane track-wire distance, which ranges from 0 to 1 cm. For the high-$Q^2$ kinematics ($Q^2=$5.2, 6.8, and 8.5 GeV$^2$), the tracks of interest (those with significant analyzing power) have scattering angles $\vartheta$ of no more than $\approx 20^\circ$, such that the error introduced by equating the measured drift distance with the in-plane distance is no more than $\approx 1\ cm/\cos \vartheta - 1 cm \approx 64\ \mu m$. In the example above, the drift time spectrum is restricted to hits on tracks with scattering angles $\vartheta<15^\circ$, which is approximately the upper limit of tracks with significant analyzing power for $Q^2=8.5$ GeV$^2$. The drift map was also calculated using this cut, in order to optimize the track reconstruction for the events of interest. 

By restricting the range of drift distances to $\pm 1$ cm, events passing through the corners of drift cells, which can have drift distances up to $\sqrt{(1.0\ cm)^2 + (0.8\ cm)^2}=1.28$ cm, will be incorrectly mapped to a drift distance of $\approx$ 1 cm. Fortunately, however, events with hits in the corners of a drift cell are very likely to have hits on an adjacent wire with drift times and distances well within the ``safe'' region, and the tracking algorithm is designed to favor these hits in the track fitting procedure. The slight residual non-uniformities in the drift distance spectrum do not significantly affect the quality of reconstructed tracks. The residuals of FPP tracks after calibration of the time-to-distance map are shown in figure \ref{fppresid}.
\begin{figure}[h]
  \begin{center}
    \includegraphics[angle=90,width=.98\textwidth]{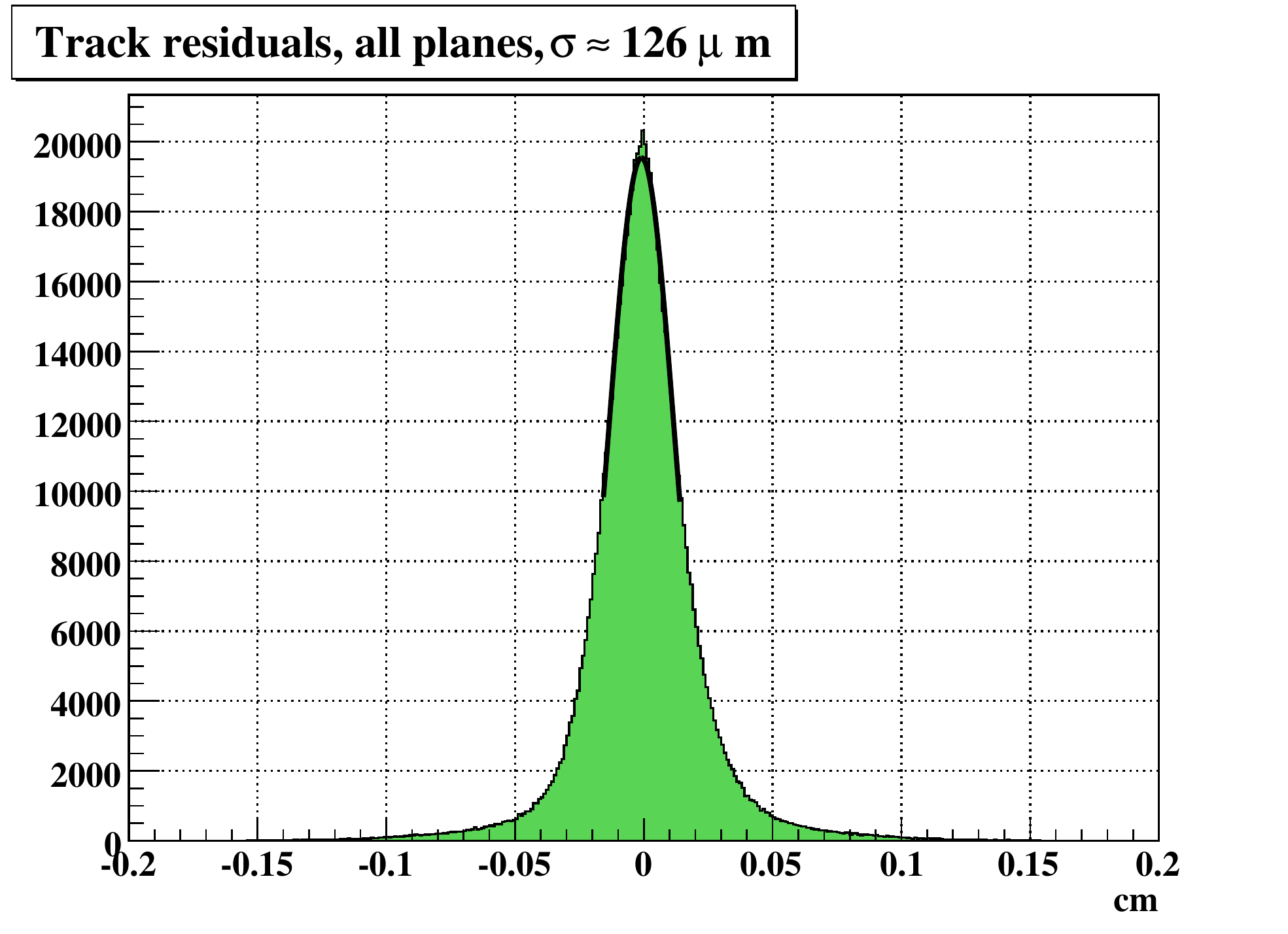}
    \caption{\label{fppresid} Residuals of FPP tracking.}
  \end{center}
\end{figure}
The 1$\sigma$ resolution, fitted to the full-width at half-maximum of the peak, is approximately 126 microns. 

Apart from the fact that the FPP drift chambers were brand-new and relatively leak-proof, the FPP tracking residuals are more than a factor of two smaller than the HMS tracking residuals because of the smaller number of planes on the track and the smaller separation in $z$ between planes. Because the six wire planes provide only one more coordinate measurement than the minimum required to define a track, the fitted track, which is of course strongly influenced by the hits it contains, tends to be closer to those hits, on average, than a track defined by twelve hits as in the HMS drift chambers. The smaller $z$ separation reduces the amount of multiple scattering of the measured track along the path over which it is measured. Each drift chamber in each FPP has one plane measuring only the $x$ coordinate, and two planes measuring the $x$ and $y$ coordinates in equal proportions. With a separation of 21.3 cm between the middle planes of each chamber in a pair, the angular resolution of the FPP drift chambers is approximately 0.37 mrad in $x'$ and 0.42 mrad in $y'$, which is only slightly worse than the angular resolution of the HMS drift chambers. The resolution of the relative angles between tracks measured by the FPP and by the HMS drift chambers also has a significant contribution from multiple scattering in the roughly two cm thickness of scintillator lying between the HMS and FPP drift chambers (see table \ref{S0resolutioneffects}), not to mention the thickness of CH$_2$ analyzer traversed by the proton before and after the primary scattering. 

Because of the relatively high particle flux in the HMS for many of the kinematic settings of the experiment, and because of the large acceptance of the FPP drift chambers, it is often difficult, especially with only six wire planes, to choose the best combination of hits for a track unambiguously and correctly, particularly if only the wire positions are considered. For this reason, all wire combinations passing the $\chi^2$ test of the fit to wire positions are considered as possible candidates for tracks. For each such combination, the drift distances are calculated and the best left-right combination of the hits is determined by fitting a track to all $2^6=64$ possible combinations of drift signs and choosing the combination which gives the smallest $\chi^2$. Then, from among all candidate tracks, the combination of hits which has the smallest $\chi^2$ of drift-based tracking is chosen as the first true track. At this point, all logical hits on the first track are marked as used, and the tracking algorithm starts over, searching for more tracks among the remaining unused hits. A cut on the $\chi^2$ of drift-based tracks is applied such that no hit on the track is allowed to be further from the fitted track than 1.55 millimeters, or just over 12$\sigma$. If the $\chi^2$ of the fit exceeds this test, then the hit with the worst contribution to $\chi^2$ is dropped and the fit is repeated until either the track passes the $\chi^2$ test or the number of hits/planes on the track drops below the minimum. 

In the first iteration of pattern recognition/track reconstruction, at least six hits are required with at least one hit per plane. If no combination of wires and drift distances can be found with the required number of hits/planes that passes the $\chi^2$ test, then the hit/plane requirement is dropped to five and the algorithm is repeated. If all six planes fire and one or more multi-hit clusters are present, then hits within multi-hit clusters are always considered for removal before single hits. The tracking algorithm strongly favors tracks with all six wire planes firing over tracks with only five planes, since five-plane tracks are actually under-determined given the FPP drift chamber design, particularly with respect to left-right determination. If the track passes the $\chi^2$ test, with all six planes firing and one or more multi-hit clusters present such that ``extra'' hits might be removed to improve $\chi^2$, it is treated according to the following procedure, with user-customizable behavior through a ``clean track'' $\chi^2$ test:
\begin{itemize}
\item For each two-hit cluster on a track, the sign of the drift distance of both hits is fixed to force the track to go between the two wires. Each three-hit cluster is reduced to two hits by testing both two-hit combinations which include the middle hit, and throwing away the outer hit which gives the worst $\chi^2$ when paired with the middle hit.
\item If the ``clean'' $\chi^2$ test is set smaller than the maximum $\chi^2$ test, then the code will remove the worst hit from within a multi-hit cluster according to its contribution to $\chi^2$, and this procedure will repeat until $\chi^2$ falls below the ``clean'' $\chi^2$ test or the track drops to exactly six hits with one hit per plane. In the final analysis, the ``clean'' $\chi^2$ test was set to zero, resulting in the reduction of all six-plane tracks to one hit per plane.
\item If only five planes fire, then all multiple hits are retained on the track as long as the $\chi^2$ is below the maximum, since these extra hits help constrain the otherwise ambiguous left-right combination. 
\end{itemize}

\begin{figure}[h]
  \begin{center}
    \subfigure[]{\label{fppnsimple}\includegraphics[angle=90,width=.49\textwidth]{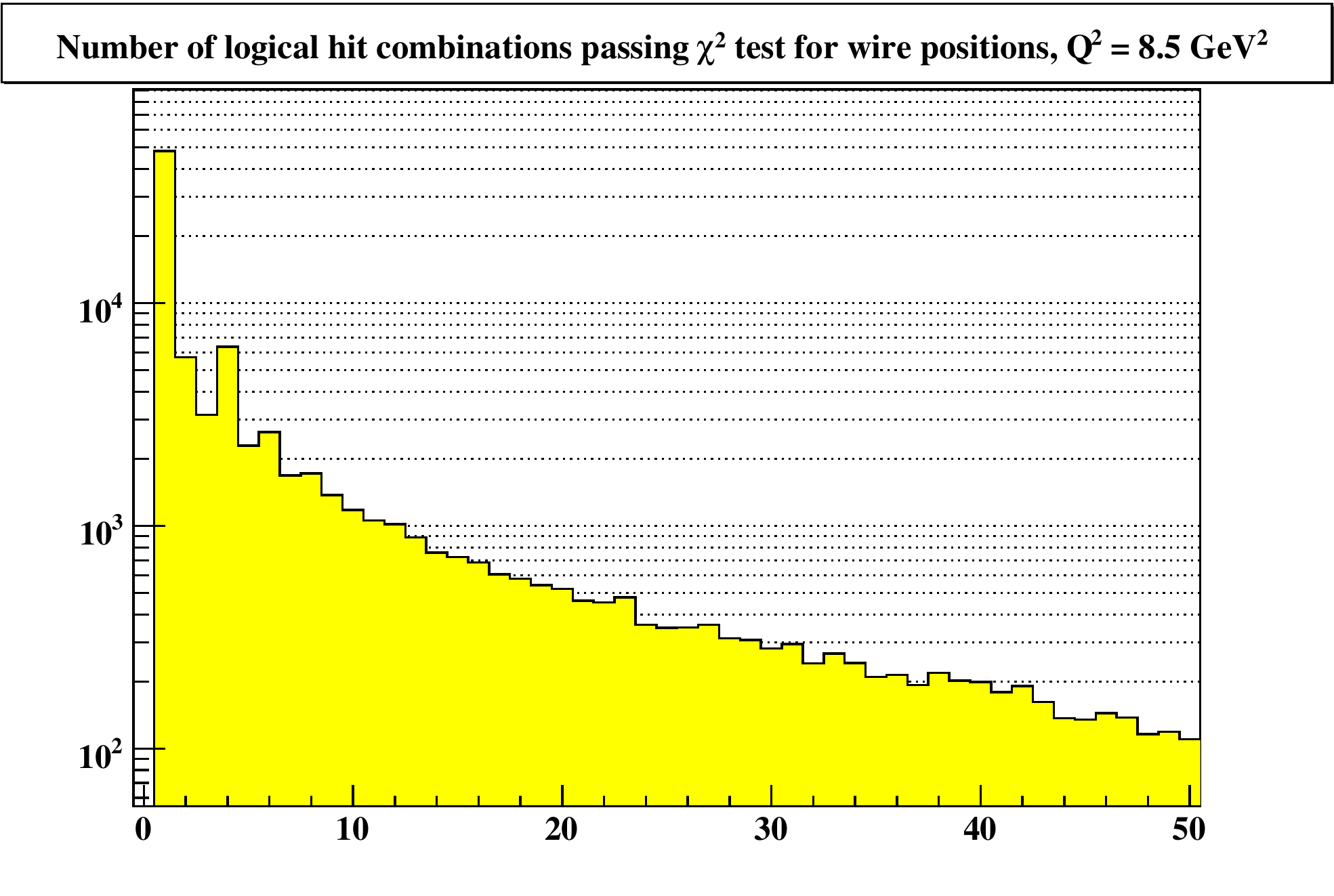}}
    \subfigure[]{\label{fppngooddrift}\includegraphics[angle=90,width=.49\textwidth]{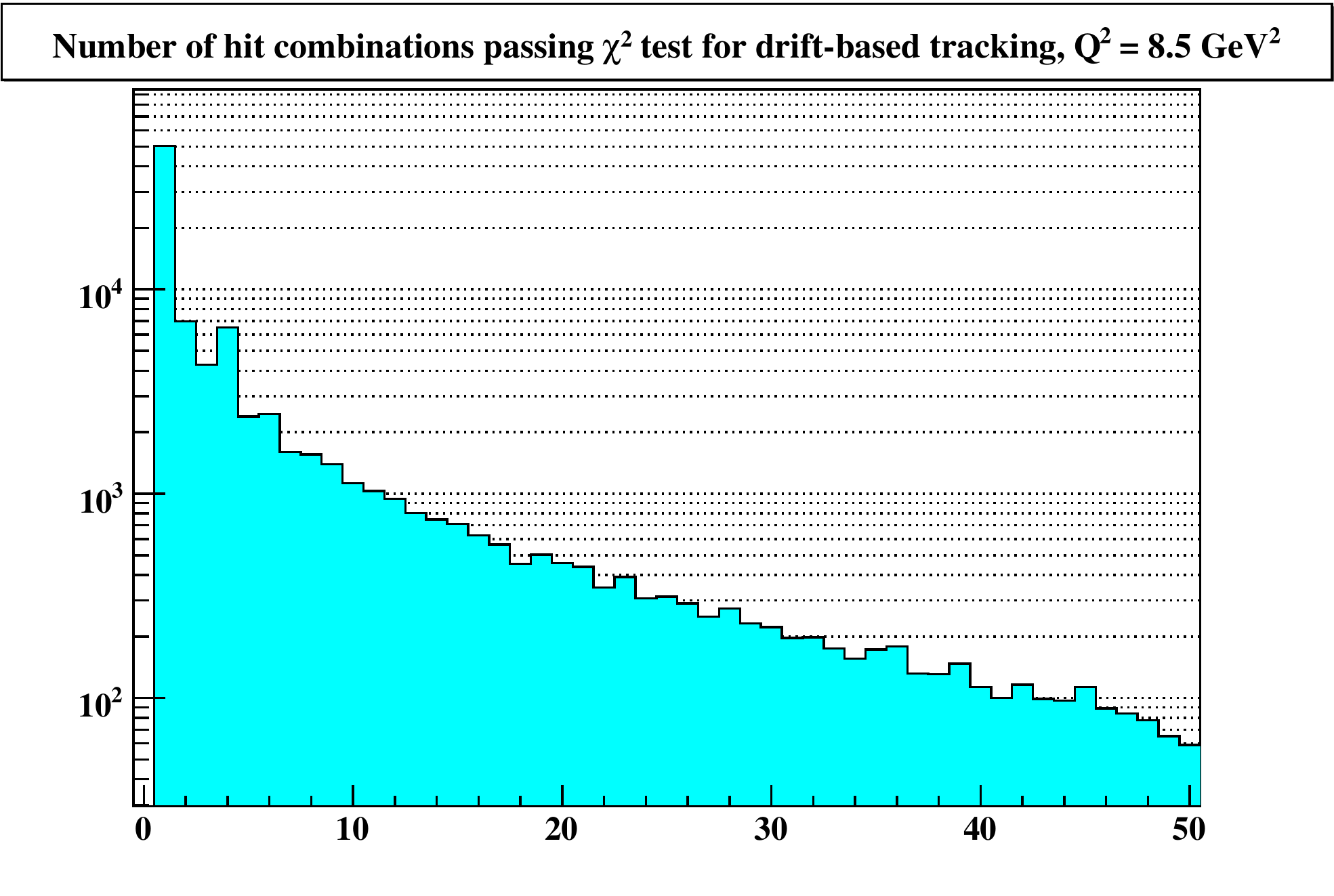}}
  \end{center}
  \caption{\label{fpphitmult} Multiplicity in FPP tracking. The number of wire combinations passing the $\chi^2$ test of the fit to wire positions with no drift information (a), and (b) the number of wire combinations passing the $\chi^2$ test of the fit to wire positions $\pm$ drift distances.}
\end{figure}
Figure \ref{fpphitmult}\footnote{Both histograms in figure \ref{fpphitmult} are for the first set of drift chambers only. The multiplicities in the second set of drift chambers are even higher.} illustrates the problem of the high multiplicity of potentially valid hit combinations per event in the FPP tracking at $Q^2=8.5$ GeV$^2$, for which single-particle rates in the HMS approached 1 MHz. Figure \ref{fppnsimple} shows the number of ``logical hit'' combinations passing the $\chi^2$ cut for ``simple'' tracking, on the first iteration of pattern recognition. Approximately 54\% of events with at least one valid track have exactly one valid wire combination. The other 46\% of events have two or more valid wire combinations. In this situation, it is generally not possible to reliably choose the correct combination of hits for a track without considering the drift information. Figure \ref{fppngooddrift} shows, for the same events as figure \ref{fppnsimple}, the number of hit combinations passing the $\chi^2$ test for drift-based tracking. Even after considering drift, only 55\% of events have just one valid track. Though many hit combinations still pass the $\chi^2$ test with drift taken into account, choosing the combination of hits/drifts with the smallest $\chi^2$ gives a correct track with high probability. 

In a simple Monte-Carlo simulation of events with exactly two real tracks in the FPP drift chambers and a conservative input coordinate resolution of 200 $\mu$m, it was demonstrated that choosing hits based only on the $\chi^2$ of ``simple'' tracking resulted in an \emph{incorrect} choice in approximately 20\% of events, whereas choosing hits based on the $\chi^2$ of drift-based tracking resulted in an incorrect choice in less than 2\% of events. To the extent that the intrinsic coordinate resolution is even better than 200 $\mu$m in the FPP drift chambers, as suggested by figure \ref{fppresid}, then the fraction of events with correctly chosen hit combinations should be even closer to 100\%, when two or fewer real tracks pass through the chambers per event. On the other hand, events with \emph{more} than two real tracks will cause additional confusion of the tracking algorithm, and one expects the efficiency for correctly reconstructing all tracks to decrease further with increasing numbers of real tracks in the chambers per event.

The number of hit combinations passing drift-based tracking roughly follows the number of wire combinations passing simple tracking at relatively small $N$. This reflects the relatively loose maximum $\chi^2$ test applied to the drift-based tracks. At larger $N$, however, the number of ``good drift'' tracks falls more rapidly than the number of good ``simple'' tracks as one should expect. It is important to note that in the case of multiple real tracks in the drift chambers, virtually any hit combination involving the pairing of one localized set of 2-3 hits left by a real track in the first drift chamber with another such set of hits in the second drift chamber is likely to give a valid track, unless the tracks are from uncorrelated events and are significantly out of time with the HMS trigger. If the proton which caused the trigger scatters in the CH$_2$ and produces multiple secondary particles which are detected by the drift chambers in time with the scattered primary, then many ``space point''-like pairings will appear to give reasonable tracks even if the groupings of hits do not come from the same track. This is because of the minimal drift chamber design in which the track coordinates are essentially measured at only two values of $z$, and the fact that, since tracks are allowed to have virtually any slope, the track position in the second chamber is not significantly constrained by the measured coordinates in the first chamber and vice versa. The six-plane design contains an inherent, irreducible ambiguity in the reconstruction of events with multiple tracks, which are rather frequent in the data at $Q^2=8.5$ GeV$^2$, which could be largely eliminated for future experiments by adding a third wire plane for each wire orientation. 

In addition to the FPP chambers' relatively poor ability to reconstruct events with multiple tracks, the six-plane design with symmetric wire orientations has another inherent drawback pertaining to the left-right determination. Since each chamber consists of three planes, close together in $z$, with wire orientations of $\pm45^\circ$ and $90^\circ$ relative to the vertical, tracks passing through regions of the chamber with certain ``magic'' arrangements of wire positions will leave hit patterns for which more than one solution exists for the left-right combination. One of the most obvious examples of this situation is shown in figure \ref{FPPleftrightambig}.
\begin{figure}[h]
  \begin{center}
    \includegraphics[angle=90,width=.6\textwidth]{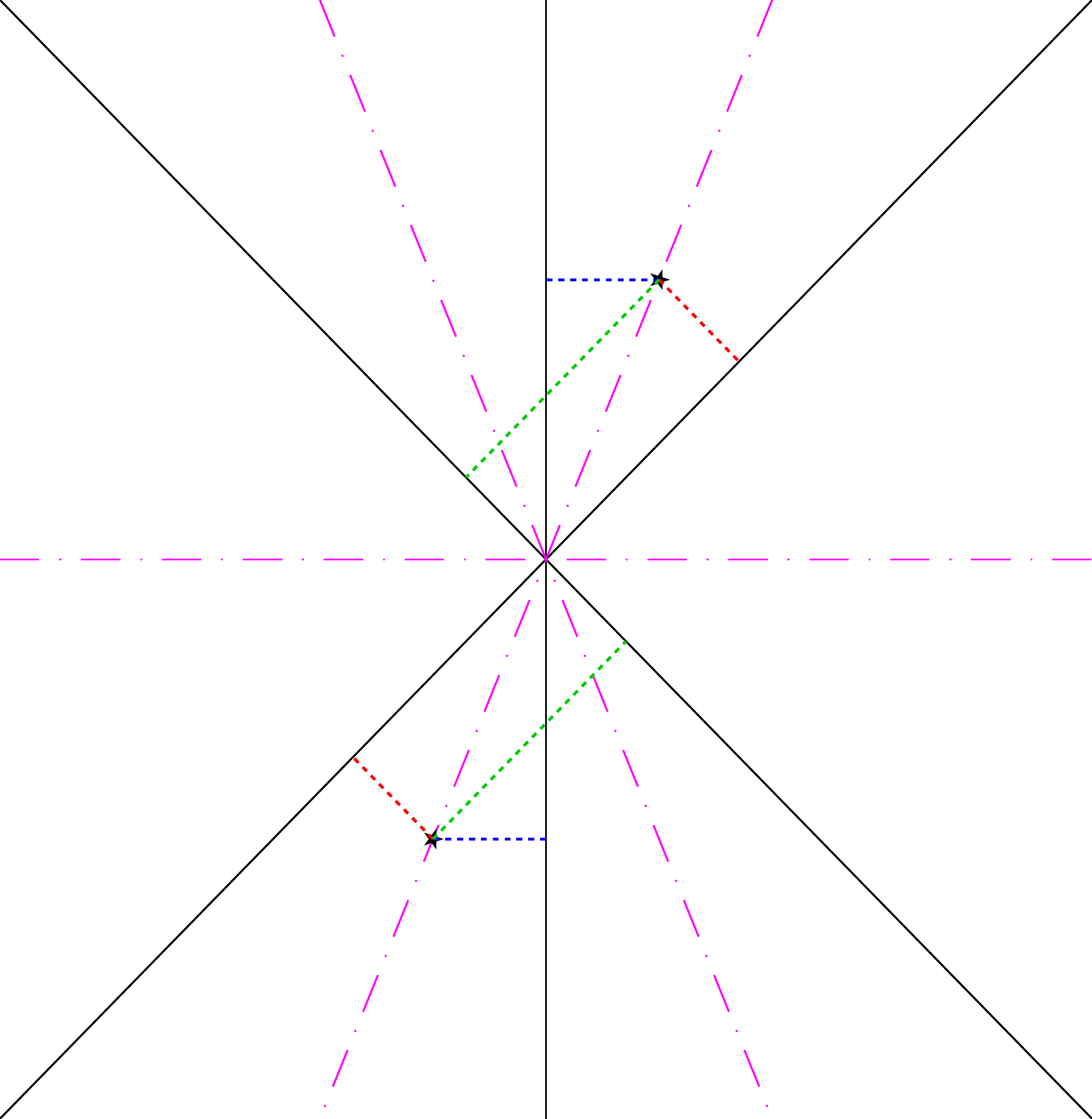}
  \end{center}
  \caption{\label{FPPleftrightambig} Illustration of left-right ambiguity for three wires (black lines) intersecting at the same point, as in the center of the FPP drift chambers. If the track crosses a drift chamber in such a region along any of the lines bisecting two wires (magenta dot-dashed lines), the same combination of drift distances (red/green/blue dashed lines) gives two equally valid solutions for the track crossing point (black stars).}
\end{figure}
The FPP drift chambers were designed so that at the origin (center) of each drift chamber, the three different wire orientations intersect at the same point in the $xy$ plane, namely $(0,0)$. Because of the symmetry of the wire orientations, any track crossing the chamber at near-perpendicular incidence along any of the three lines bisecting the wires as shown in figure \ref{FPPleftrightambig} has a mirror-image solution with identical drift distances on the other side of the origin. Fortunately, since the spacing of the $\pm45^\circ$ wires in the $x$ direction is $\sqrt{2}$ times the nominal wire spacing of 2 cm (because of the equal spacing along the coordinate measured for all wire planes), this pattern never repeats itself exactly (but sometimes approximately). Thus, in small regions of the chamber acceptance, there is an approximately 50\% chance of choosing the wrong drift sign. 

The purpose of the FPP drift chambers is to measure the asymmetry of the angular distribution of tracks scattered in CH$_2$. Both the rate-dependent ``space point pairing'' ambiguity and the rate-independent intrinsic left-right ambiguity are reconstruction inefficiencies which are independent of the beam helicity state. Therefore, they affect only the statistical precision of the result, and not the result itself. Both ambiguities can be somewhat mitigated by looking not only at $\chi^2$, but also the reconstructed scattering parameters relative to the incident proton track in the HMS, before choosing the best combination of hits, as discussed below. The potential benefit of such considerations is, however, limited by the need to avoid potentially negative side effects of biasing the reconstruction of tracks toward favorable scattering conditions, including the possibility of shifting bad tracks into the ``good track'' acceptance and diluting the asymmetry.

\begin{figure}[h]
  \begin{center}
    \subfigure[]{\label{fppntrack1}\includegraphics[angle=90,width=.49\textwidth]{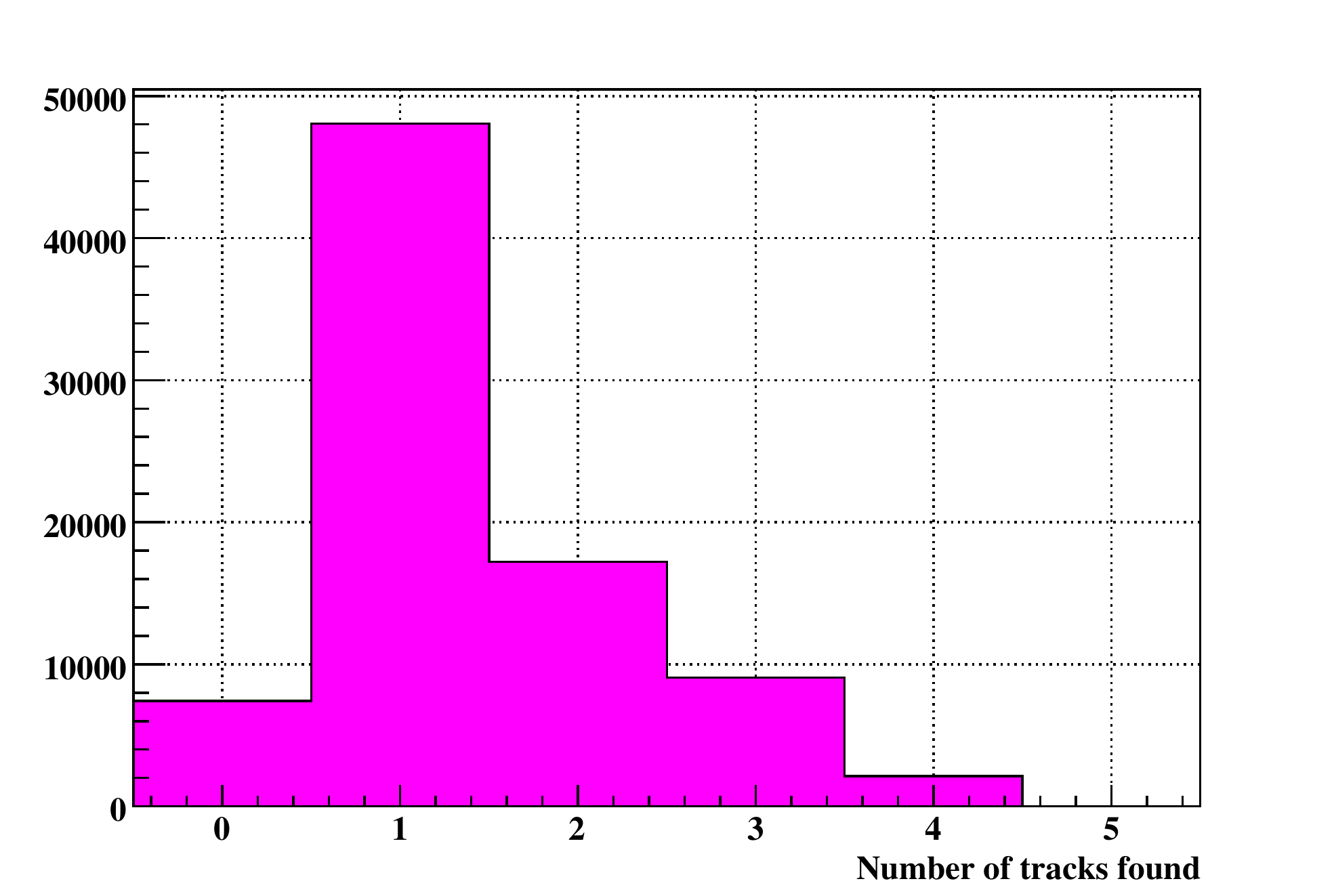}}
    \subfigure[]{\label{fppntrack2}\includegraphics[angle=90,width=.49\textwidth]{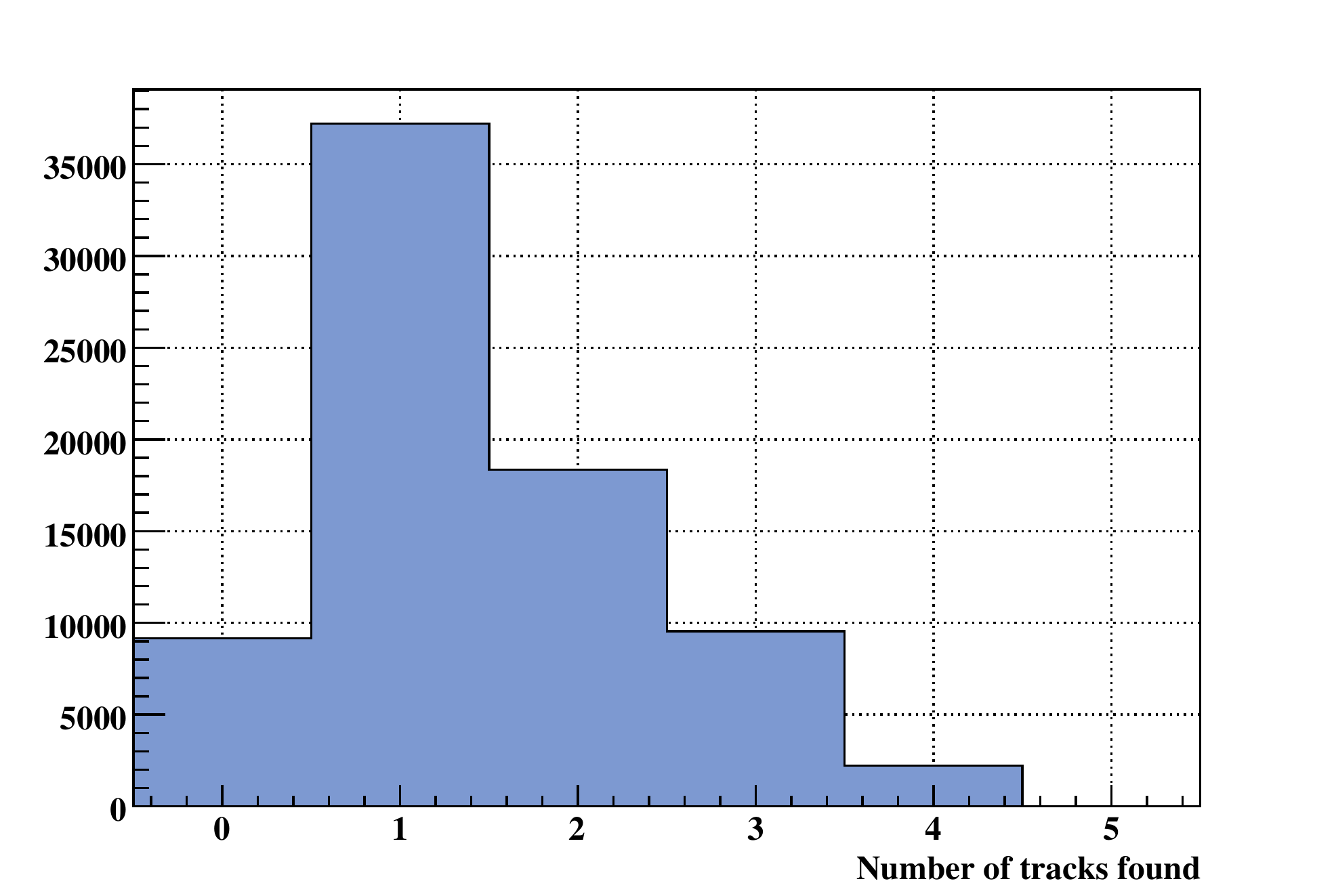}}
  \end{center}
  \caption{\label{FPPntrack} Number of tracks found per event at $Q^2=8.5$ GeV$^2$, in FPP1 (a), for all events in which a good HMS track was found, and FPP2 (b), for all events in which a good HMS track and a good FPP1 track were found.}
\end{figure}

Despite the daunting proliferation of hit combinations producing apparently valid tracks, once the best initial combination (based on $\chi^2$ of the fit to drift distances) is chosen and its hits are removed from consideration, the multiplicity of tracks decreases substantially, showing that many of the candidate hit combinations found on the first iteration of reconstruction share one or more logical hits. Figure \ref{FPPntrack} shows the distribution of the number of tracks reconstructed per event, for the first polarimeter (\ref{fppntrack1}) and the second polarimeter (\ref{fppntrack2}). In figure \ref{fppntrack1}, the presence of a good HMS track is required, with cuts applied on the reconstructed target quantities $x'_{tar}$, $y'_{tar}$, $y_{tar}$, and $\delta$ to guarantee reasonably good events. The number of zeroes in fig. \ref{fppntrack1} is about 8.8\%. Events with no found tracks come partly from reconstruction inefficiencies, but also from the non-negligible probability of 
\begin{enumerate}
\item Scattering by such a large angle in the first block of CH$_2$ as to be outside the acceptance of the first set of drift chambers
\item Charge-exchange/absorption/proton capture reactions in the CH$_2$ producing either no particle or a neutron which goes undetected by the drift chambers.
\end{enumerate}
The percentage of events with a non-zero number of tracks that have exactly one track is 62.8\%, and there are almost no events with more than four tracks. Similarly, for the second polarimeter (fig. \ref{fppntrack2}), requiring a good track in the HMS \emph{and} the first polarimeter, roughly 12.0\% of events have no track, and the percentage of single-track events (55.3\%) is lower among those with at least one track. The higher percentage of zeroes and the higher track multiplicity in the second polarimeter are to be expected, since there is twice as much thickness of CH$_2$ intercepting the flux of incident protons before the second set of drift chambers. As in the case of FPP1, there are significant numbers of 2, 3, and 4-track events with a negligible probability for five or more tracks.

\subsubsection{Scattering Angle/Closest Approach Reconstruction and Best Track Selection}
\paragraph{}
Once track reconstruction is complete, the next step is to reconstruct the parameters of the scattering in CH$_2$. All tracks reconstructed in the first chamber pair are compared to the incident track reconstructed by the HMS drift chambers, and all tracks reconstructed in the second chamber pair are compared to the HMS track, and also any tracks reconstructed in the first polarimeter. Each track is characterized by its $(x,y)$ coordinates at $z=0$, and its slopes $(x'\equiv \frac{dx}{dz},y'\equiv \frac{dy}{dz})$. The first and simplest quantity to reconstruct is the polar scattering angle $\vartheta$. The incident and scattered trajectories are expressed in terms of unit vectors $\hat{n}_{i}$ and $\hat{n}_f$, respectively:
\begin{eqnarray}
  \hat{n}_i = \frac{1}{\sqrt{1+x_i'^2 + y_i'^2}}\left(\begin{array}{c} x'_i \\ y'_i \\ 1 \end{array}\right) &,& \hat{n}_f = \frac{1}{\sqrt{1+x_f'^2 + y_f'^2}}\left(\begin{array}{c} x'_f \\ y'_f \\ 1 \end{array}\right)
\end{eqnarray}
The polar scattering angle $\vartheta$ is then given by 
\begin{eqnarray}
  \cos \vartheta &=& \hat{n}_f \cdot \hat{n}_i
\end{eqnarray}
The next step is to reconstruct the azimuthal scattering angle $\varphi$. The consistent definition of $\varphi$ everywhere in the data analysis is very important to get the right answer for $G_E^p/G_M^p$. $\varphi$ is defined in the local coordinate system defined by the incident trajectory, with $\hat{n}_i$ as the $z$ axis. The $x$ and $y$ axes of this coordinate system are both orthogonal to the $z$ axis, and the definition of $\varphi=0$ is more or less arbitrary, but must be defined exactly the same way everywhere in order to be correct and meaningful. The chosen definition is for the local $y$ axis to always be parallel to the $yz$ plane of the transport coordinate system. Thus, the unit vector in \emph{transport} coordinates of the $y$ axis in the local coordinate system of the incident proton is given by 
\begin{eqnarray}
  \hat{y}_{local} &\equiv& \frac{\hat{n}_i \times \hat{x}_{transport}}{\left|\hat{n}_i \times \hat{x}_{transport}\right|} 
\end{eqnarray}
where $\hat{x}_{transport} \equiv \left(1, 0, 0\right)$ is the $x$ axis of the (fixed) transport coordinate system and $\hat{n}_i$ is the unit vector of the incident (HMS) track, still expressed in transport coordinates. This definition guarantees that the local $y$ axis with respect to which $\varphi$ is measured is always perpendicular to both the incident trajectory and the $x$ axis of the fixed transport coordinates. Finally, the $x$ axis of the local coordinate system is defined such that $(\hat{x}_{local},\hat{y}_{local},\hat{n}_i)$ is right-handed and orthonormal:
\begin{eqnarray}
  \hat{x}_{local} &\equiv& \hat{y}_{local} \times \hat{n}_i \label{localxaxisdef}
\end{eqnarray} 
where again all vectors are expressed in transport coordinates. $\hat{x}_{local}$ defined by equation \eqref{localxaxisdef} is already a unit vector since $\hat{y}_{local}$ and $\hat{n}_i$ are already orthonormal. In the local coordinate system thus defined, the azimuthal scattering angle is defined by
\begin{eqnarray}
  \hat{n}_f &\equiv& \sin\vartheta \cos \varphi \hat{x}_{local} + \sin \vartheta \sin \varphi \hat{y}_{local} + \cos \vartheta \hat{n}_i
\end{eqnarray}
so that 
\begin{eqnarray}
  \tan \varphi &=& \frac{\hat{n}_f \cdot \hat{y}_{local}}{\hat{n}_f \cdot \hat{x}_{local}} \label{fppphidef}
\end{eqnarray}
In the analysis code, the ``atan2(y,x)'' function, which gives the result in the correct quadrant given two signed numbers $y$ and $x$, is used to determine $\varphi$ from \eqref{fppphidef}. This FORTRAN intrinsic function returns a value between $-\pi \le \varphi \le \pi$. In this analysis, $\varphi$ is taken to be the angle measured from the positive $x$ axis toward the positive $y$ axis, with $\varphi=0$ along the $x$ axis. Events with $-\pi \le \varphi < 0$ are shifted by $2\pi$, $\varphi \rightarrow \varphi + 2\pi$ so that the final $\varphi$ value runs from $0$ to $2\pi$. 

The next step is to find the point and distance of closest approach between the incident and scattered trajectories. This is a textbook minimization problem--in this case, the expression for the squared distance between two straight lines in three-dimensional space is to be minimized. Given two lines $(x_1(z_1), y_1(z_1)) = (x_1(0) + x'_1 z_1,y_1(0) + y'_1 z_1)$ and $(x_2(z_2), y_2(z_2)) = (x_2(0) + x'_2 z_2,y_2(0) + y'_2 z_2)$, the squared distance between the two lines at two arbitrarily chosen points $z_1$ and $z_2$ along those lines is given by 
\begin{eqnarray}
  s^2 &=& \left(x_1(z_1)-x_2(z_2)\right)^2 + \left(y_1(z_1)-y_2(z_2)\right)^2 + \left(z_1 - z_2\right)^2 \label{closestapproachdistsquared}
\end{eqnarray}
which is minimized as a function of $z_1$ and $z_2$ by setting the partial derivatives $\partial s^2/\partial z_1$ and $\partial s^2 / \partial z_2$ to zero:
\begin{eqnarray}
  0 &=& \left(x_1(z_1)-x_2(z_2)\right)x'_1 + \left(y_1(z_1)-y_2(z_2)\right)y'_1 + \left(z_1 - z_2\right) \nonumber \\
  0 &=& \left(x_1(z_1)-x_2(z_2)\right)x'_2 + \left(y_1(z_1)-y_2(z_2)\right)y'_2 + \left(z_1 - z_2\right)
\end{eqnarray}
Substituting $x_i(z_i) = x_i(0) + x'_i z_i$ and $y_i(z_i) = y_i(0) + y'_i z_i$ yields the following matrix equations for $z_1$ and $z_2$:
\begin{eqnarray}
  \left(\begin{array}{cc} 1 + x_1'^2 + y_1'^2 & -1 - x_1' x_2' - y_1' y_2' \\ -1 - x'_1 x'_2 - y'_1 y'_2 & 1 + x_2'^2 + y_2'^2 \end{array}\right)\left(\begin{array}{c} z_1 \\ z_2 \end{array} \right) = \nonumber \\
   \left(\begin{array}{c} x'_1(x_2(0) - x_1(0)) + y'_1(y_2(0) - y_1(0)) \\ x'_2(x_1(0)-x_2(0)) + y'_2(y_1(0)-y_2(0))\end{array}\right) 
\end{eqnarray}
which can be re-written in the convenient shorthand 
\begin{eqnarray} 
  \left(\begin{array}{cc} a & -b \\ -b & c \end{array}\right)\left(\begin{array}{c} z_1 \\ z_2 \end{array}\right) &=& \left(\begin{array}{c} -x'_1 \Delta x_0 - y'_1 \Delta y_0 \\ x'_2 \Delta x_0 + y'_2 \Delta y_0 \end{array}\right) \nonumber \\
  \left(\begin{array}{c} z_1 \\ z_2 \end{array}\right) &=& \frac{1}{ac - b^2}\left(\begin{array}{cc} c & b \\ b & a \end{array}\right)\left(\begin{array}{c} -x'_1 \Delta x_0 - y'_1 \Delta y_0 \\ x'_2 \Delta x_0 + y'_2 \Delta y_0 \end{array}\right) \label{z1z2close}
\end{eqnarray}
with 
\begin{eqnarray}
  a &\equiv& 1 + x_1'^2 + y_1'^2 \nonumber \\
  c &\equiv& 1 + x_2'^2 + y_2'^2 \nonumber \\
  b &\equiv& 1 + x'_1 x'_2 + y'_1y'_2 \label{zcloseequationdefs}\\ 
  \Delta x_0 &\equiv& x_1(0) - x_2(0) \nonumber \\
  \Delta y_0 &\equiv& y_1(0) - y_2(0) \nonumber
\end{eqnarray}
From the solutions \eqref{z1z2close} for the parameters $z_1$ and $z_2$ of the point of closest approach between the two lines, the point of closest approach $z_{close}$ is defined as the average $z_{close} \equiv \frac{1}{2}\left(z_1+z_2\right)$, and the distance of closest approach $s_{close} \equiv \sqrt{s^2(z_1,z_2)}$, obtained by substituting the solutions into \eqref{closestapproachdistsquared}.

The scattering is fully defined by the angles $\vartheta$ and $\varphi$ and the closest approach parameters $s_{close}$ and $z_{close}$. In the ideal scattering scenario, the proton scatters elastically from a single carbon or hydrogen nucleus within the CH$_2$, and a single, solitary track is detected in the FPP drift chambers. The reconstructed $s_{close}$ of this track will be zero, up to the smearing effects of multiple scattering in the scintillators and CH$_2$ before and after the scattering, and tracking resolution. The $z_{close}$ of this ideal track should lie within a range corresponding to the thickness of the CH$_2$. If the proton scatters by a non-negligible angle more than once within the analyzer, it is likely that $s_{close}$ will be blown up and $z_{close}$ will no longer necessarily lie inside the analyzer. 

The last parameter of the scattering to be determined is a logical variable called a ``cone test''. The cone test is designed to eliminate false azimuthal asymmetries arising from the rectangular acceptance by requiring that, for the measured scattering angle $\theta$ and vertex position $z_{close}$, the full $2\pi$ azimuthal cone at angle $\vartheta$ surrounding the incident track lies within the acceptance of the drift chambers measuring the scattered track. Given $(x_{fp}, y_{fp})$ and $(x'_{fp},y'_{fp})$ of the incident track, and $\vartheta$, $z_{close}$ of the scattered track, the acceptance is checked by calculating the maximum and minimum $x$ and $y$ coordinates at the position of each chamber plane. The maximum (minimum) $x/y$ coordinate occurs when the proton scatters along the $+(-)x/y$ direction by an angle $\vartheta$. The position of the incident track at the interaction vertex is given by 
\begin{eqnarray}
  x_{close} &=& x_{fp} + x'_{fp} z_{close} \\
  y_{close} &=& y_{fp} + y'_{fp} z_{close}
\end{eqnarray}
If the scattering is purely along the $x(y)$ direction, then the scattered track makes an angle $\arctan x'_{fp}(\arctan y'_{fp}) \pm \vartheta$ with the $x(y)$ axis. Its slopes $x'_{scat}$ and $y'_{scat}$ are given by
\begin{eqnarray}
  x'_{scat} &=& \tan \left(\arctan x'_{fp} \pm \vartheta \right) = \frac{x'_{fp} \pm \tan \vartheta }{1 \mp x'_{fp}\tan \vartheta} \\
  y'_{scat} &=& \tan \left(\arctan y'_{fp} \pm \vartheta \right) = \frac{y'_{fp} \pm \tan \vartheta}{1 \mp y'_{fp} \tan \vartheta}
\end{eqnarray}
Since $\vartheta$ is always positive, as long as $z_{close}$ lies in front of the drift chambers, the maximum and minimum coordinates are given by:
\begin{eqnarray}
  x_{max} &=& x_{close} + \frac{x'_{fp} + \tan \vartheta}{1-x'_{fp}\tan \vartheta}( z_{back} - z_{close} ) \\
  x_{min} &=& x_{close} + \frac{x'_{fp} - \tan \vartheta}{1+x'_{fp}\tan \vartheta}( z_{back} - z_{close} ) \\
  y_{max} &=& y_{close} + \frac{y'_{fp} + \tan \vartheta}{1-y'_{fp}\tan \vartheta}( z_{back} - z_{close} ) \\
  y_{min} &=& y_{close} + \frac{y'_{fp} - \tan \vartheta}{1+y'_{fp}\tan \vartheta}( z_{back} - z_{close} )
\end{eqnarray}
where $z_{back}$ is the $z$ coordinate of the last plane of the drift chambers measuring the track. If the size of the active area of the drift chambers is $L_x$ in the $x$ direction and $L_y$ in the $y$ direction, then the cone test is satisfied if the following conditions are met:
\begin{eqnarray}
  x_{max} &\le& x_0 + \frac{L_x}{2} \\
  x_{min} &\ge& x_0 - \frac{L_x}{2} \\
  y_{max} &\le& y_0 + \frac{L_y}{2} \\ 
  y_{min} &\ge& y_0 - \frac{L_y}{2}
\end{eqnarray}
$(x_0,y_0)$ are the coordinates of the center of the FPP drift chambers relative to the transport coordinate system; i.e., the HMS optical axis.
\begin{figure}[h]
  \begin{center}
    \includegraphics[width=.99\textwidth]{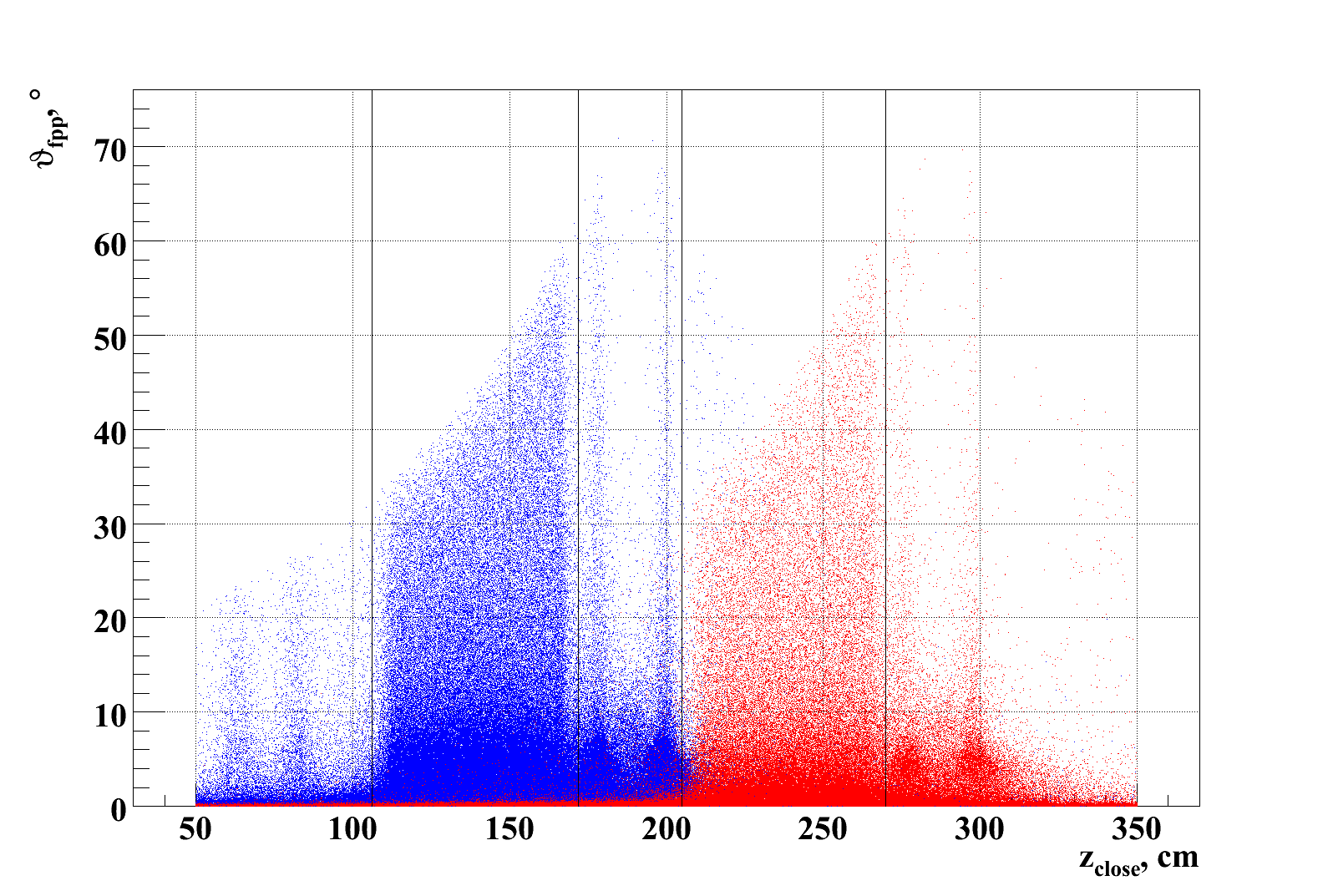}
  \end{center}
  \caption{\label{FPPthetazclose} Correlation between polar scattering angle $\vartheta$ and point of closest approach $z_{close}$ at $Q^2=8.5$ GeV$^2$, for events with exactly one found track in FPP1 (blue) and FPP2 (red), passing the cone test, with $s_{close} \le 3(6)$ cm for FPP1(2). In the second polarimeter, events with small scattering angles in the first polarimeter were selected ($\vartheta_1 \le 1^\circ$). Black lines illustrate the $z_{close}$ cut used in the final analysis.}
\end{figure}

Figure \ref{FPPthetazclose} illustrates the effect of the cone test on the distribution of reconstructed tracks. The polar scattering angle $\vartheta$ in degrees is plotted on the $y$ axis as a function of the reconstructed point of closest approach $z_{close}$. Several prominent features of the spectrum deserve mention. First, the two ``stripes'' around $z\approx 60$ cm and $z\approx 80$ cm are tracks that scatter from the scintillators used in the trigger. Scattering in the $CH_2$ analyzers is clearly visible in the broad stripes from about $106 \le z_{close} \le 172$ cm and $205 \le z_{close} \le 270$ cm. The characteristic ``razor blade'' shape of the distribution comes from the application of the cone test. Events that scatter further upstream of the drift chambers satisfy the conetest for a smaller range of angles than tracks which scatter closer to the chambers. Note that the maximum $\vartheta$ passing the conetest ranges from just over 30$^\circ$ at the front of the CH$_2$ to just over 60$^\circ$ at the end of the CH$_2$ closest to the drift chambers. The solid black vertical lines in the plot illustrate the $z_{close}$ cut applied to the first and second analyzers used in the final analysis. 

The stripes at the locations of the drift chambers come at least in part from scattering within the drift chambers, but are also symptomatic of the difficulties in track reconstruction inherent in the FPP design. One mechanism for the ``collapse'' of $z_{close}$ to the position of the drift chambers is when an ``unscattered'' track passes through the analyzer, but one of its ``space points'', which lies more or less on top of the incident track, is incorrectly paired with another space point from a second track, giving rise to a larger $\vartheta$ with $z_{close}$ at the position of the ``good'' space point. Another mechanism for an event to reconstruct to a value of $z$ inside the drift chambers is the left-right ambiguity discussed above. If the incident track doesn't scatter in the first analyzer, the measured track should have $\vartheta \approx 0$. But if this track passes through the FPP drift chambers such that the condition depicted in figure \ref{FPPleftrightambig} is satisfied, then one (or both) of the space points of the measured track may be significantly displaced from its true position, resulting in a reconstructed track with erroneously large $\vartheta$ and $z_{close}$ inside the FPP drift chambers. This situation may also occur if one of the planes on the track does not fire. In such a situation, the left-right combination of the hits in the chamber with one missing plane may be underdetermined; i.e., several different combinations of drift signs may give roughly equal $\chi^2$.

\begin{figure}[h]
  \begin{center}
    \subfigure[]{\label{FPPthetadist}\includegraphics[angle=90,width=.49\textwidth]{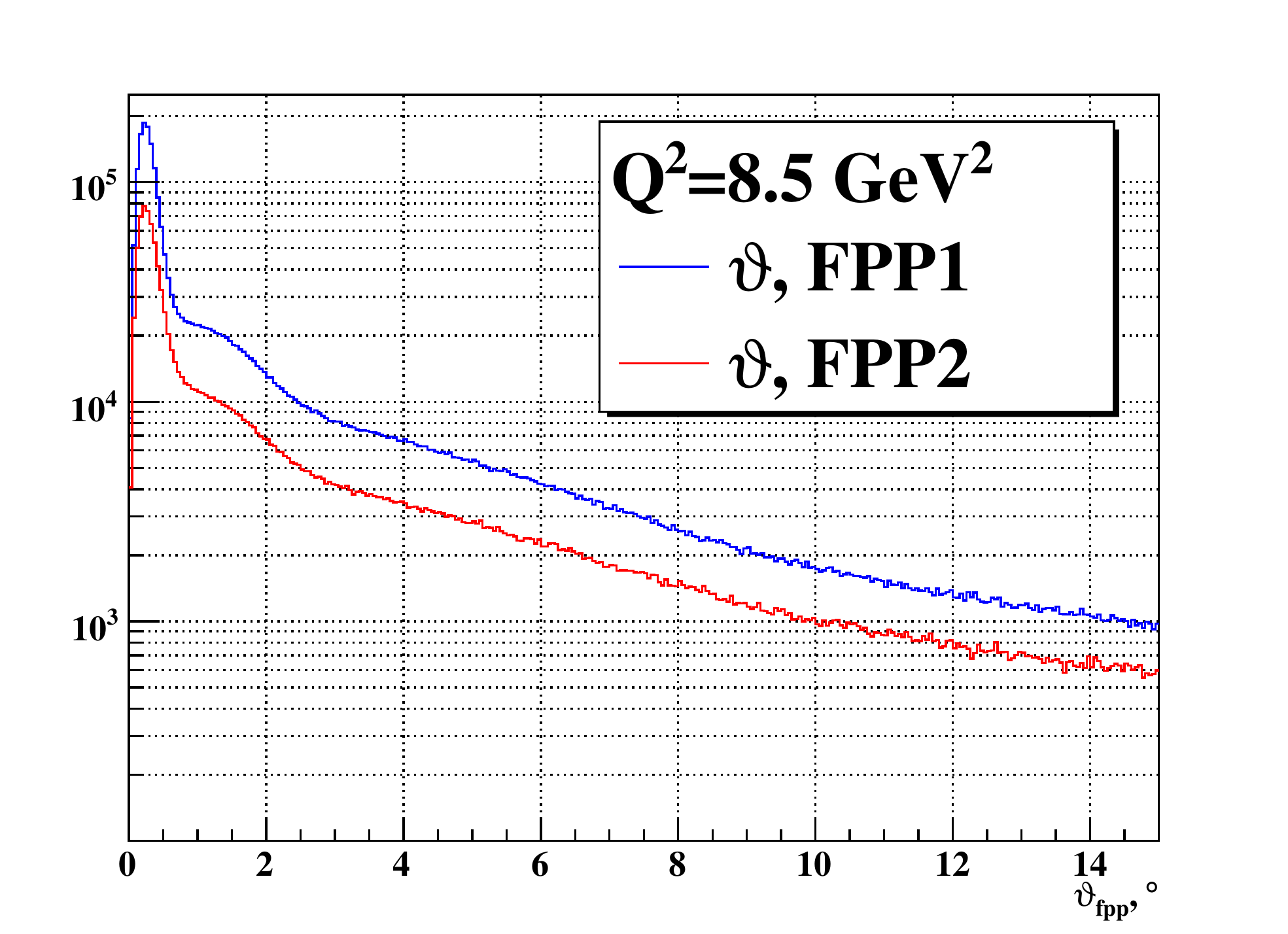}}
    \subfigure[]{\label{FPPsclosedist}\includegraphics[angle=90,width=.49\textwidth]{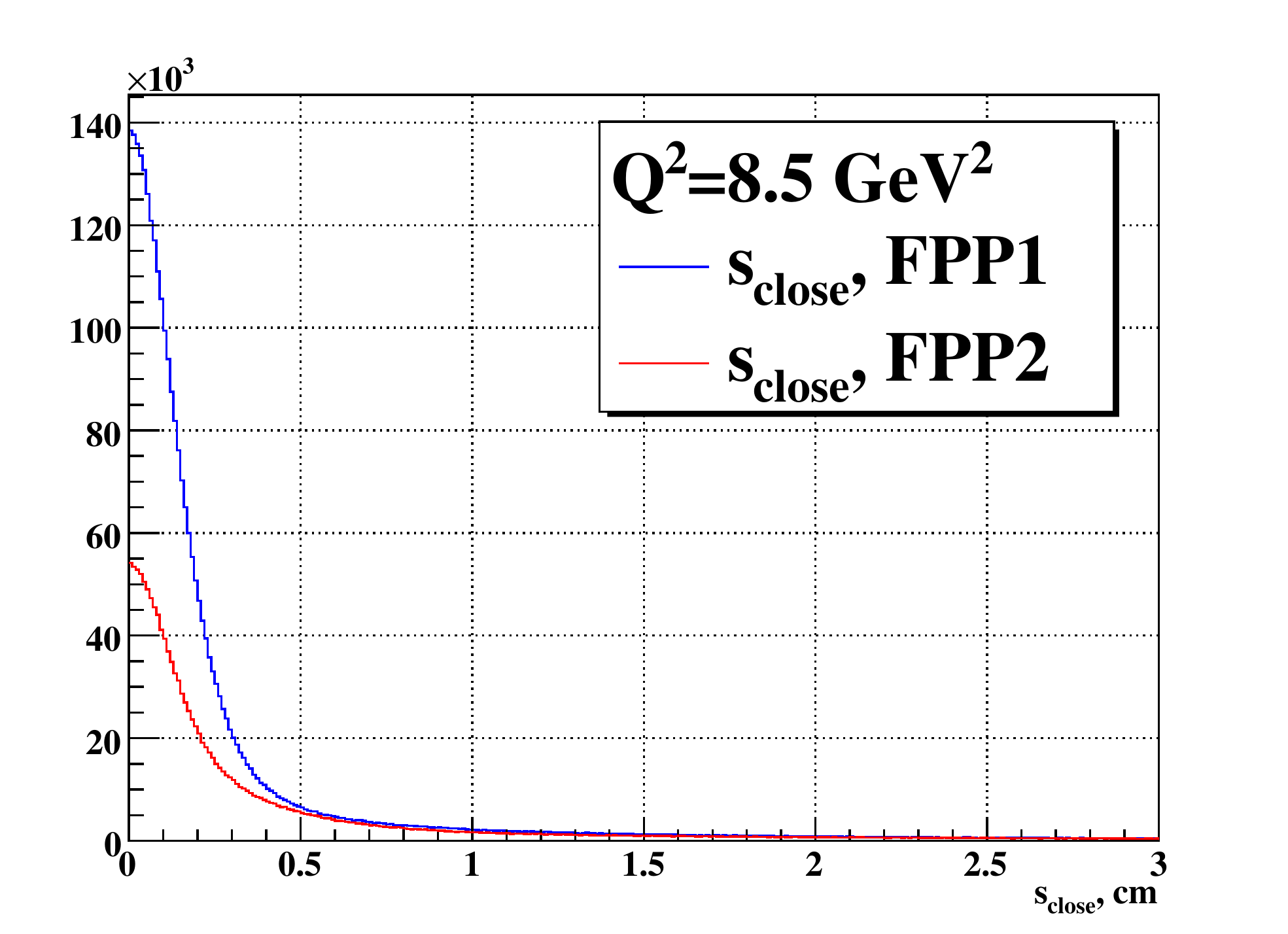}}
  \end{center}
  \caption{\label{FPPthetasclose} $\vartheta_{fpp}$ (a) and $s_{close}$ (b) distributions in both polarimeters at $Q^{2}=8.5$ GeV$^2$, for tracks passing the $z_{close}$ and cone test cuts.}
\end{figure}
Figure \ref{FPPthetasclose} shows the distributions of polar scattering angle $\vartheta_{fpp}$ (\ref{FPPthetadist}) and the distance of closest approach $s_{close}$ (\ref{FPPsclosedist}) at $Q^2=8.5$ GeV$^2$, for events passing the cone test and the $z_{close}$ cuts around the CH$_2$ analyzer illustrated in figure \ref{FPPthetazclose}. The distributions shown are for events with exactly one track found in the polarimeter in question. The peak at small angles in the $\vartheta_{fpp}$ distribution is dominated by Coulomb scattering, and events inside this peak have negligible analyzing power. The dip as $\vartheta\rightarrow 0$ simply reflects the vanishing phase-space volume at $\vartheta=0$ for tracks reconstructed by detectors with finite resolution in $x'$ and $y'$. The $s_{close}$ distribution is peaked at zero, with long non-Gaussian tails at larger distances reflecting the possibility for protons to scatter more than once in the CH$_2$, destroying the $s_{close} \approx 0$ result obtained for tracks that scatter only once. The distribution is wider for the second polarimeter, which is to be expected since the proton must cross twice as much material before it is detected in the second polarimeter. 

In the first polarimeter, the scattering angles and closest approach parameters are always calculated with respect to the HMS track, and if multiple tracks are found, the single track with the smallest polar angle $\vartheta_{fpp}$ is chosen as the ``best'' track for use in the polarization analysis, for the following reasons: 
\begin{itemize}
\item The scattering cross section is larger at small angles. Choosing the track with the smallest scattering angle gives the greatest likelihood of choosing the most interesting particle (the scattered incident proton) as opposed to low-energy secondaries produced by inelastic reactions or random tracks uncorrelated with the trigger. 
\item The analyzing power distribution is heavily concentrated at small angles.
\end{itemize}
In the second polarimeter, the angles may be calculated with respect to either the HMS track or any of the tracks found in the first polarimeter. Making the appropriate choice for each event is a non-trivial matter. In the subset of events in which the incident proton is unscattered; i.e., it is detected in the first polarimeter with a ``small''\footnote{The definition of ``small'' scattering angles corresponds roughly to the width of the Coulomb peak, which depends on the incident proton momentum roughly as $p^{-1}$. For purposes of this analysis, ``Coulomb'' events are defined as events with $p_T \equiv p\sin \vartheta_{fpp} \le 0.07$ GeV/c.} scattering angle, the angles calculated relative to the HMS track are approximately equal to the angles calculated relative to the FPP1 track, and the choice only matters to the extent that either set of angles is more accurately measured than the other. 

For events in which the measured track in FPP1 has a larger scattering angle, it would appear at first glance that one should always calculate the angles in FPP2 relative to the FPP1 track. However, given the not insignificant probability of failed tracking or mistracking of events in the FPP1 drift chambers, it is preferable to revert to the angles calculated relative to the HMS track for events in which the FPP1 track is either unreliable or absent. Since only one track is chosen from each polarimeter for the analysis, a criterion is needed to choose not only the best track in FPP2 to use in the analysis but also the best track from either the HMS or FPP1 to use as the reference track in calculating the scattering angles of the chosen track. For both choices, the smallest-angle approach is used. Each track found in FPP2 is compared to each track found in FPP1 and also to the HMS track, and the combination of (FPP2) track and (HMS or FPP1) reference track resulting in the smallest $\vartheta$ is chosen, regardless of the angle of the FPP1 track relative to the HMS track. This track selection algorithm can be summarized as seeking out the smallest-angle scattering in each polarimeter. It was found to give the best figure of merit from among several possible approaches, and is reflected in the $\vartheta$ (figure \ref{FPPthetadist}) and $s_{close}$ (figure \ref{FPPsclosedist}) distributions. The use of such an algorithm for the second polarimeter is validated by the nearly identical shape of the FPP1 and FPP2 $\vartheta$ distributions shown.

\subsubsection{FPP Alignment}
\label{fppalignmentsection}
\paragraph{}
The result for $G_E^p/G_M^p$ depends critically on the accuracy of the reconstructed scattering angles $\vartheta$ and particularly $\varphi$, which in turn depends on knowing the location and orientation of the FPP drift chambers to a very good accuracy. The accuracy of the alignment of the FPP drift chambers becomes more important at higher momenta, where the angular distribution of the analyzing power is concentrated at more forward angles, since the error in $\varphi$ blows up as $\Delta\varphi \propto \sin^{-1} \vartheta$ as $\vartheta \rightarrow 0$. For example, at $Q^2 = 8.5$ GeV$^2$, the maximum of the analyzing power is found at $\vartheta \approx 4^\circ$. The accuracy of the surveyed chamber positions is not better than approximately $\pm$ 1 millimeter. In order to fine-tune the alignment of the FPP drift chambers in software and reduce the systematic uncertainty in the reconstructed scattering angles, dedicated ``straight-through'' data runs were taken with both CH$_2$ doors open, in order to align the measured FPP tracks with the incident HMS tracks in software. 

The software alignment procedure works by fitting a quadratic, position-dependent correction to each of the four FPP track parameters that minimizes the difference between the slopes and positions of the FPP and HMS tracks. A ``grid'' (essentially a two-dimensional histogram) of each of the track parameter differences $\Delta x$, $\Delta y$, $\Delta x'$, and $\Delta y'$ between the HMS and FPP track slopes and positions\footnote{Positions are evaluated at the $z$ coordinate of the midpoint of the pair of FPP drift chambers being aligned.} is filled from the reconstructed straight-through track data. Each grid consists of 25 bins in $x$ between -50 and +50 centimeters, and 10 bins in $y$ between -20 and +20 centimeters. Within each bin, the average and root-mean-square difference in each of the four track parameters is calculated. The measured $x$ and $y$-dependent offsets are then used to fit a correction of the form
\begin{eqnarray}
  x_{track} &\rightarrow& x_{track} + c_{x,00} + c_{x,10} x + c_{x,01} y + c_{x,20} x^2 + c_{x,02} y^2 + c_{x,11} xy \\
  y_{track} &\rightarrow& y_{track} + c_{y,00} + c_{y,10} x + c_{y,01} y + c_{y,20} x^2 + c_{y,02} y^2 + c_{y,11} xy \\
  x'_{track} &\rightarrow& x'_{track} + c_{x',00} + c_{x',10} x + c_{x',01} y + c_{x',20} x^2 + c_{x',02} y^2 + c_{x',11} xy \\
  y'_{track} &\rightarrow& y'_{track} + c_{y',00} + c_{y',10} x + c_{y',01} y + c_{y',20} x^2 + c_{y',02} y^2 + c_{y',11} xy 
\end{eqnarray}
which is the most general possible correction to the track at quadratic order in $x$ and $y$. Provided the rotations of the FPP drift chambers relative to their ideal orientation (perpendicular to the $z$ axis of the transport coordinate system) are sufficiently small that the $\mathcal{O}3$ and higher terms in the polynomial expansion of said rotations are negligible, fitting the correction above is equivalent to finding the best set of position offsets and rotations, and is significantly simpler in practice. 

Figures \ref{FPP1anglediffs} and \ref{FPP1anglediffcorr} show the quality of alignment of the track slopes achieved using the above procedure. Figure \ref{FPP1dxdiff} shows the difference between the reconstructed $x'$ of the FPP1 track and that of the HMS track for straight-through data after applying the alignment correction described above. Figure \ref{FPP1dydiff} shows the $y'$ difference. 
\begin{figure}[h]
  \begin{center}
    \subfigure[]{\label{FPP1dxdiff}\includegraphics[angle=90,width=.49\textwidth]{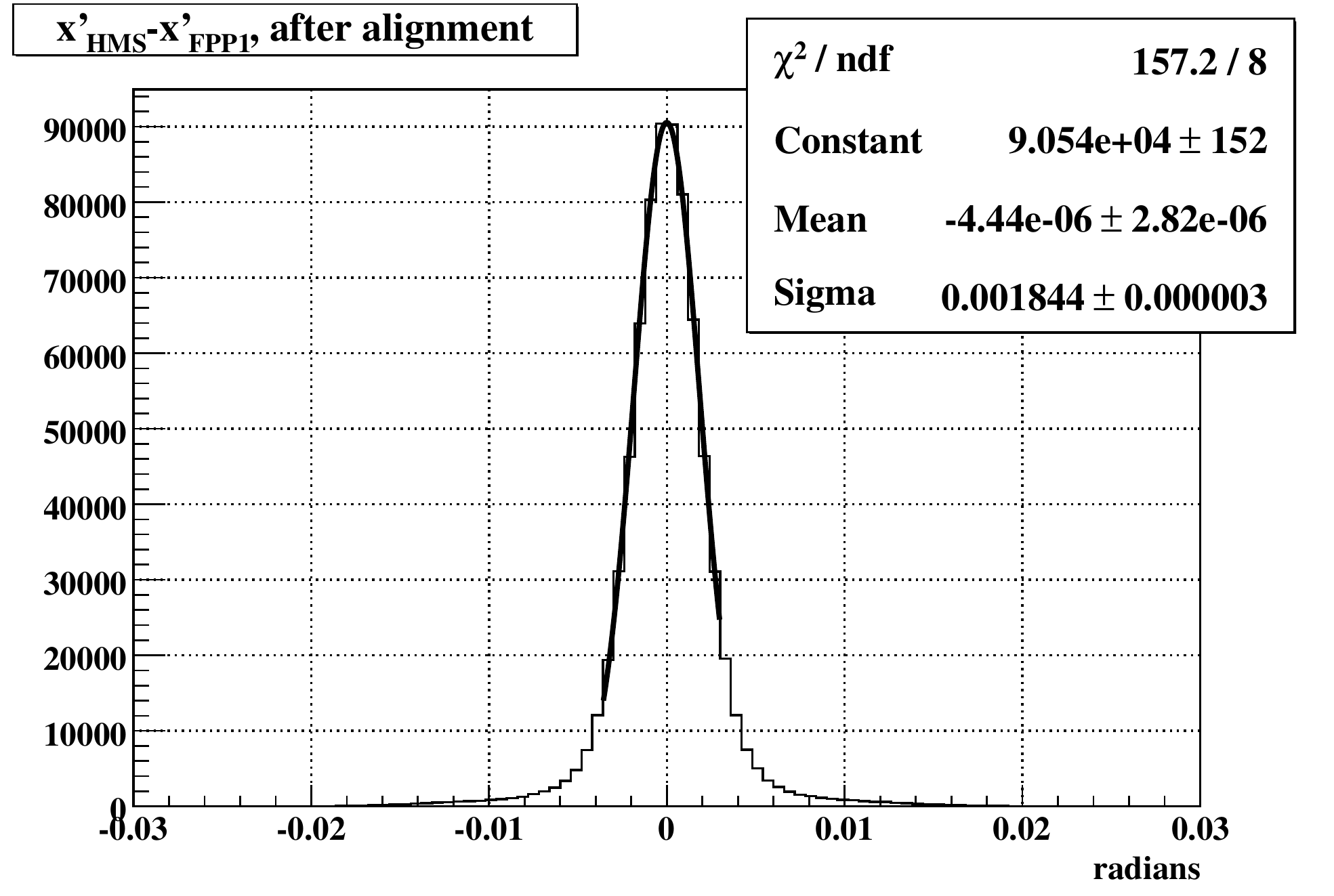}}
    \subfigure[]{\label{FPP1dydiff}\includegraphics[angle=90,width=.49\textwidth]{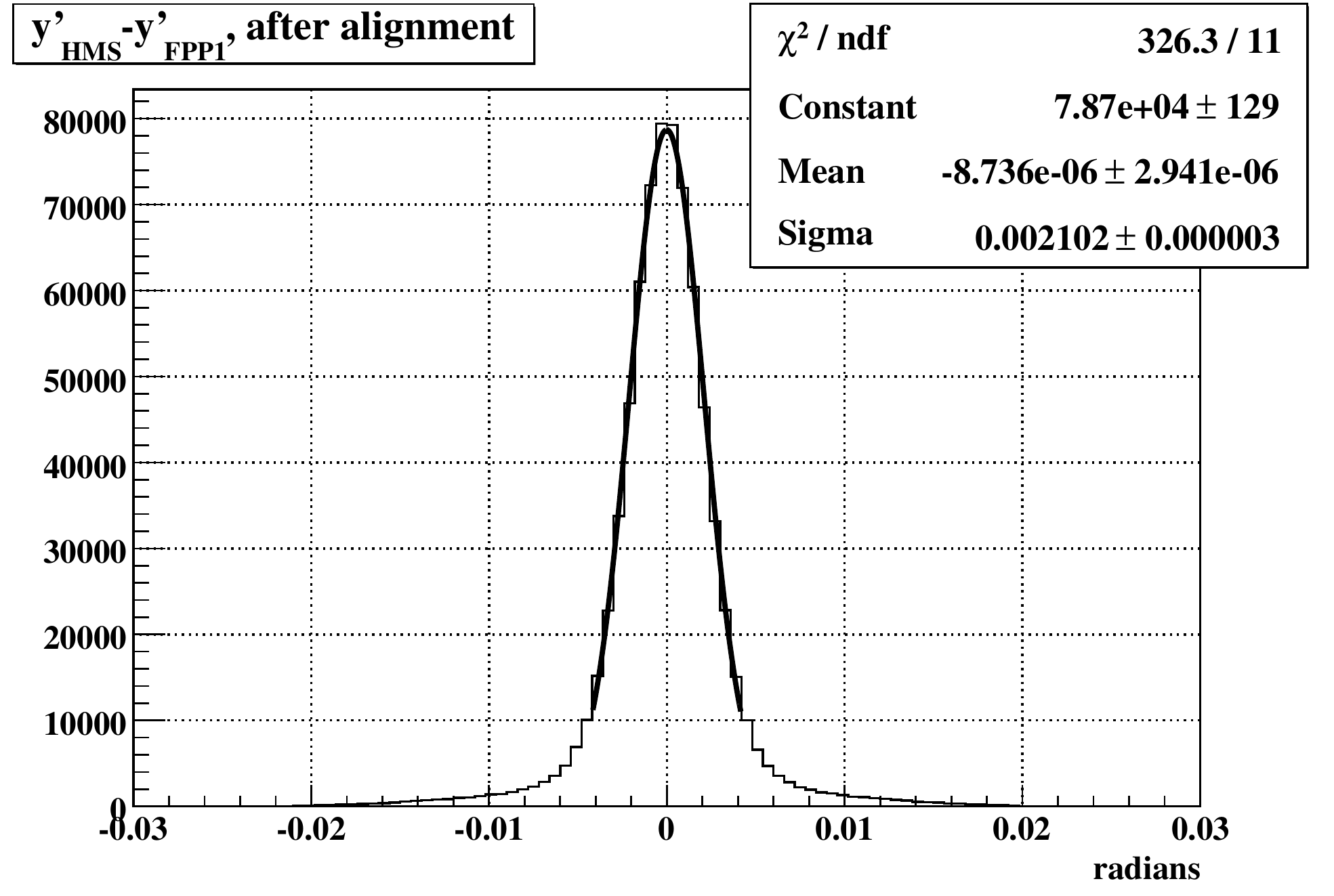}}
  \end{center}
  \caption{\label{FPP1anglediffs} Difference in $x'$ (a) and $y'$ (b) between the FPP1 drift chamber track and the HMS track, after alignment.}
\end{figure}
The mean values of the fits to the difference distributions show that the track slopes are aligned to better than $10^{-5}$. The $\sigma$'s of the distributions provide an estimate of the angular resolution, which turns out to be $\approx$2.3 mrad in $x'$ and $\approx$2.6 mrad in $y'$ for 2 GeV protons. For the 2.2 GeV electrons shown in figure \ref{FPP1anglediffs}, the resolution is slightly better. At this momentum, the $\theta_0$ of the multiple scattering distribution in 2 cm of plastic scintillator for electrons is $\approx$0.8 mrad, representing a significant contribution to the overall angular resolution. The total angular resolution is the combined effect of the intrinsic tracking resolution of the HMS and FPP drift chambers, and multiple scattering of the proton in the scintillators, the drift chambers themselves, and the air inside the HMS hut. The alignment of the track slopes at the $\approx 10^{-5}$ level limits the maximum ($\vartheta$-dependent) systematic uncertainty on $\varphi$ to $\Delta \varphi \le 10^{-5}/\sin(0.6^\circ) \approx 0.95$ milliradians, and usually much smaller. At $\vartheta = 4^\circ$, for instance, $\Delta \varphi = .14$ mrad. In the final analysis, as discussed below, a systematic uncertainty was assigned for a more conservative estimate of the alignment uncertainty of 0.1 mrad in each direction by applying a $\vartheta$-dependent shift to the azimuthal angle $\varphi$ in the asymmetry analysis. The shift in the form factor ratio induced by a shift in the azimuthal angle by $\Delta \varphi = \pm0.14\ \mbox{mrad}/\sin \vartheta$ was then taken to be the contribution of FPP alignment uncertainty to the overall systematic uncertainty in $G_E^p/G_M^p$.
\begin{figure}[h]
  \begin{center}
    \subfigure[]{\label{FPP1dxcorr}\includegraphics[angle=90,width=.49\textwidth]{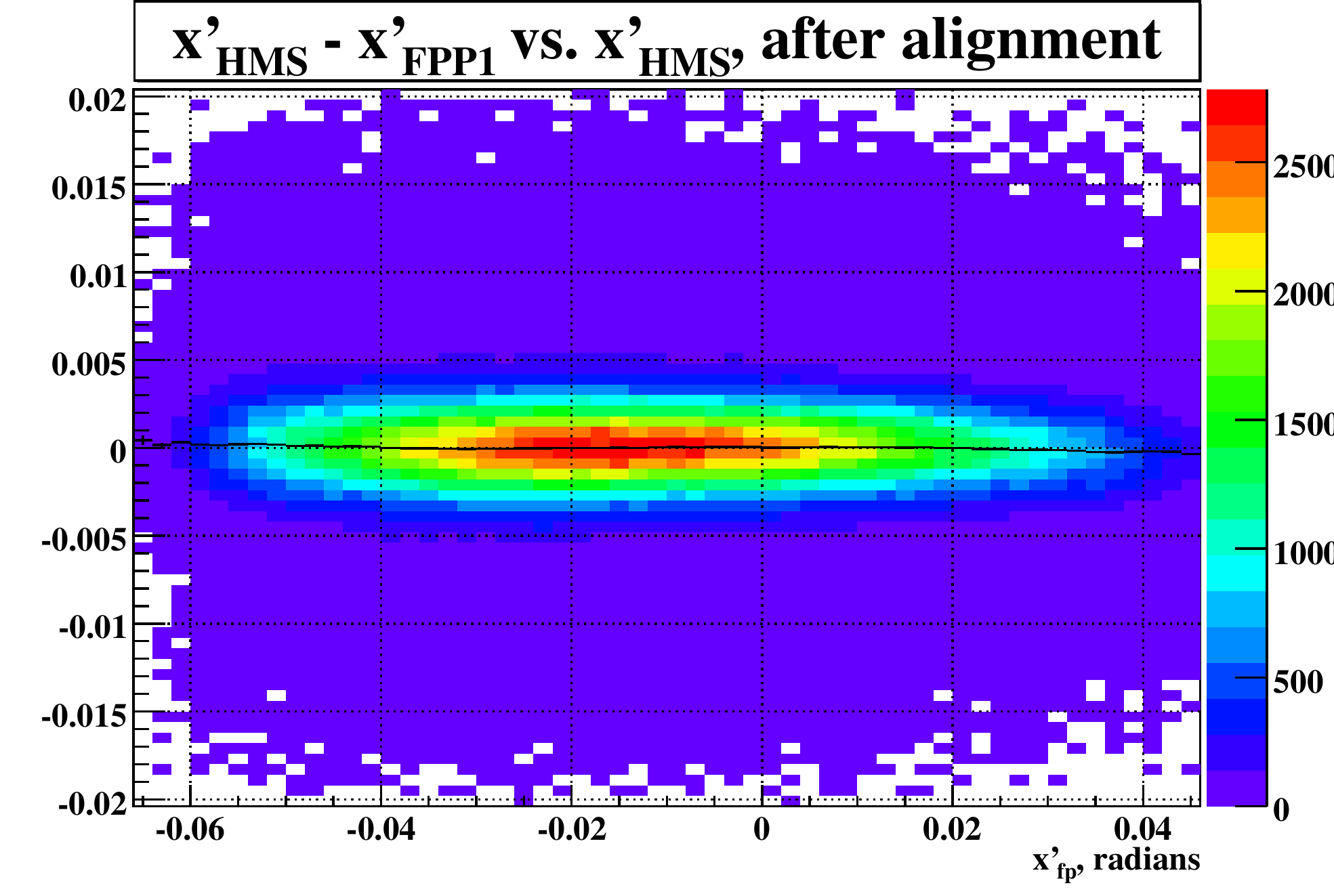}}
    \subfigure[]{\label{FPP1dycorr}\includegraphics[angle=90,width=.49\textwidth]{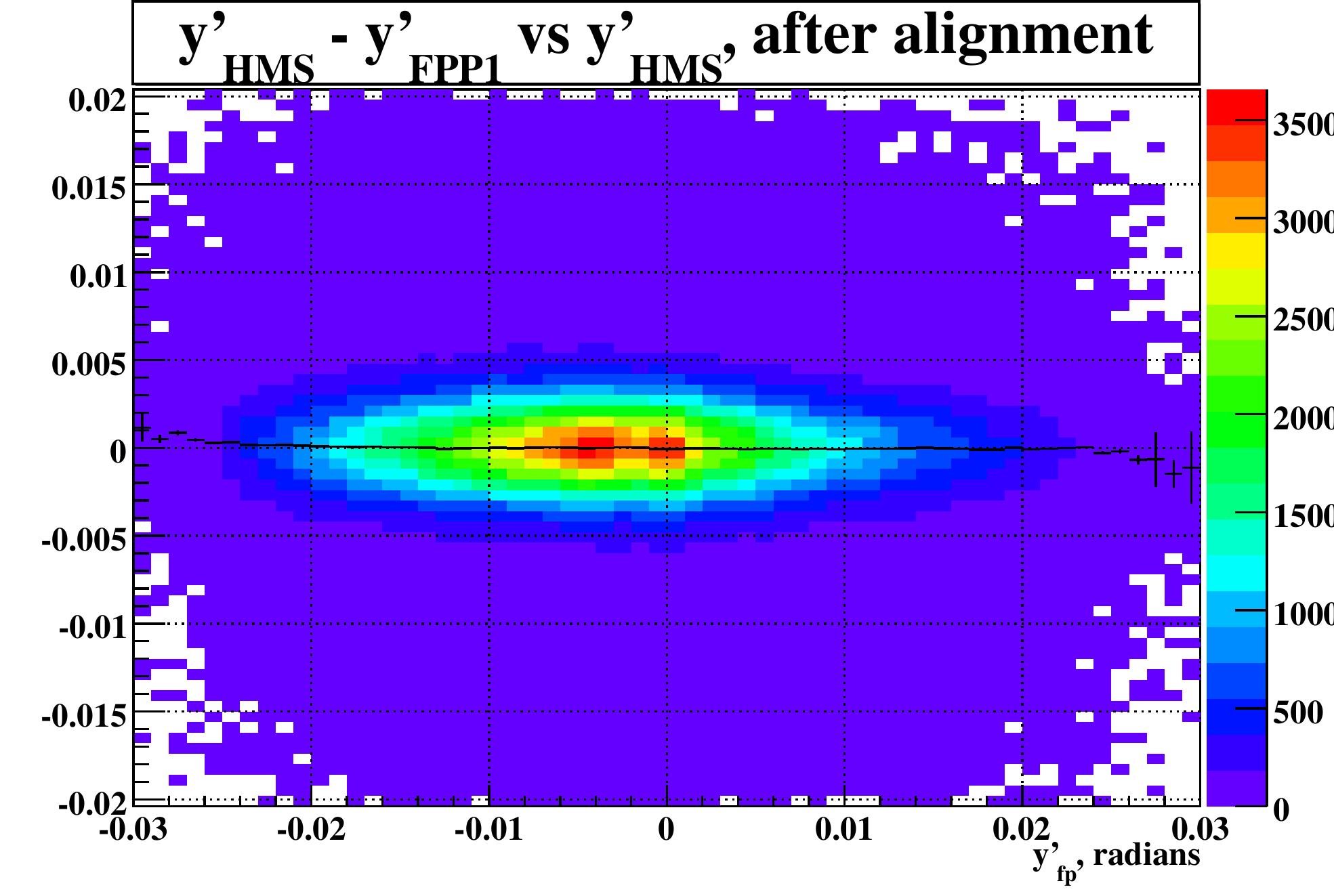}}
  \end{center}
  \caption{\label{FPP1anglediffcorr} $\Delta x'$ vs. $x'_{fp}$ (a) and $\Delta y'$ vs. $y'_{fp}$ from straight-through data for the FPP1 drift chambers after alignment. No significant correlations remain.}
\end{figure}
Figures \ref{FPP1dxcorr} and \ref{FPP1dycorr} show the correlations $\Delta x'$ vs. $x'_{fp}$ and $\Delta y'$ vs. $y'_{fp}$ after the alignment procedure, demonstrating the lack of significant correlations between the angle differences and the measured angles. Results of similar quality were obtained for the track coordinates and slopes in both chamber pairs. The fitted corrections to the track angles were no more than $\approx$1.6 milliradians at the very extremes of the acceptance, and much smaller in the interesting part of the acceptance, indicating not only the applicability of the correction but also the fact that the FPP drift chambers' orientation was reasonably close to ideal.
\subsection{Electron Coordinate and Energy Reconstruction with BigCal}
\paragraph{}
The last step in event reconstruction is the reconstruction of the electron scattering angles and energy using BigCal. The total information for each event includes 
\begin{itemize}
\item Raw ADC values for each PMT signal with a hit above the programmed CODA threshold.
\item Raw TDC values for each ``first level'' sum of 8 signals as described in section \ref{BigCalSection}.
\item Raw ADC and TDC values for each second-level sum of 64 signals, per section \ref{BigCalSection}.
\end{itemize}
The engine searches this data for the cluster of hits corresponding to the scattered electron which caused the trigger and formed a coincidence with the HMS proton trigger. The first task of the engine is to convert raw ADC values to amplitudes. This is accomplished for every run using the pedestal (type 4) events, for which every ADC channel is read out. The CODA ADC threshold for BigCal was typically set to a conservative value of $\approx 1\sigma$ above the pedestal mean, where $\sigma$ is the pedestal width (R.M.S.), to avoid the suppression of good data while still accomplishing significant event size and readout time reduction through zero suppression. In software, the pedestal width obtained from the pulser events was used to calculate a software threshold, typically higher than the ``hardware'' threshold, such that $ADC_{raw} - PED \ge 2.5 \sigma_{PED}$. Only the hits passing this threshold were retained for subsequent analysis.

The second step in the reconstruction of BigCal is the conversion of ADC amplitudes to energies. Each channel of BigCal is assigned a unique calibration constant which varies with time as accumulated radiation damage reduces the light output of each lead-glass bar, and hence, the size of the signal measured by the ADC. For the elastic scattering reaction under study, the scattered electron energy is known from both the measured proton momentum and the measured electron angle. Therefore, the gain of each channel of BigCal could be continuously monitored and recalibrated in software. Each elastic electron shower consists of a cluster of up to $5\times5$ hits, the sum of whose energies equals the incident electron energy:
\begin{equation}
  E_e = \sum_{i\in 5\times5} c_i A_i \label{calibconstdef}
\end{equation}
In equation \eqref{calibconstdef}, $c_i$ is the calibration constant and $A_i$ is the pedestal-subtracted ADC value for each hit in the cluster. During the commissioning of BigCal, dedicated elastic $ep$ scattering data with BigCal positioned far enough away from the target to be fully populated with elastically scattered electrons was used to perform the initial gain matching of the BigCal PMTs. The high voltages of all 1,744 PMTs of BigCal were adjusted to give a signal size of 1,000 ADC channels per GeV of energy deposited by the showering electron. Matching the signal sizes as closely as possible was crucial to insure that channel-to-channel gain variations would not give rise to local trigger inefficiencies for channels with signals too small relative to the fixed trigger threshold and/or extra junk triggers for channels with signals too large compared to the threshold. The calibration constants were found by minimizing the differences between the reconstructed and known electron energies in a sample of elastic $ep$ scattering events:
\begin{eqnarray}
  \chi^2 &=& \sum_{i=1}^{N_{event}} \left[E_{true}^{(i)} - \sum_{j\in 5\times5} c_j^{(i)} A_j^{(i)}\right]^2 \label{BigCalcalibchi2}
\end{eqnarray}
Minimizing the $\chi^2$ defined in equation \eqref{BigCalcalibchi2} as a function of the 1,744 calibration constants $c_j$ involves solving a system of 1,744 linear equations in as many unknowns, which is easily accomplished using standard linear algebra software libraries. During the initial gain matching phase, several iterations of calibration and high-voltage adjustment were required since each PMT has a slightly different characteristic gain curve as a function of applied high voltage. During the experiment, the calibration procedure was repeated periodically to keep the calibration constants up to date and to adjust the PMT high voltage as needed to compensate for the gain losses due to radiation damage.

\begin{figure}[h]
  \begin{center}
    \subfigure[]{\label{bigcalcalib1}\includegraphics[angle=90,width=.49\textwidth]{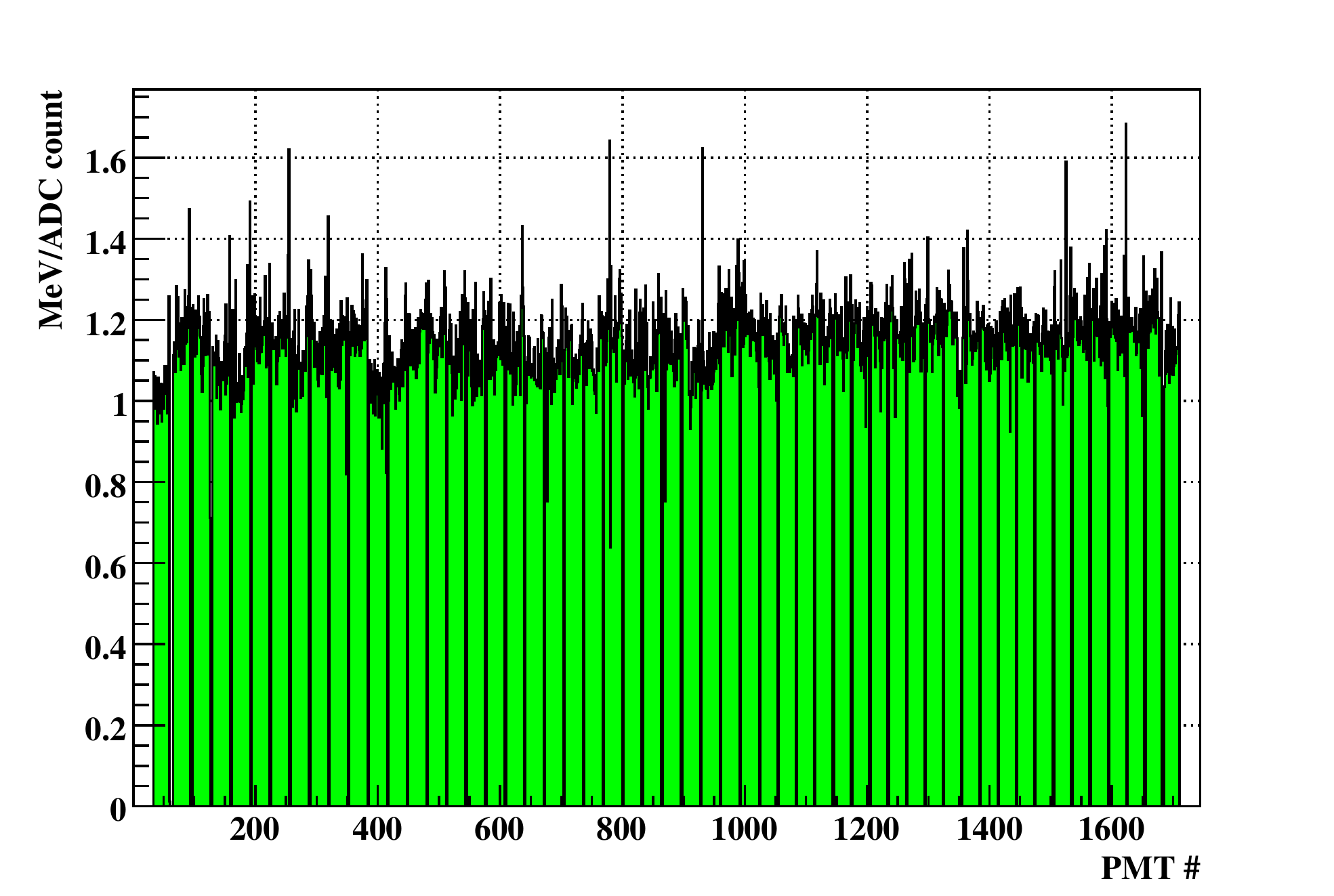}}
    \subfigure[]{\label{bigcalcalib2}\includegraphics[angle=90,width=.49\textwidth]{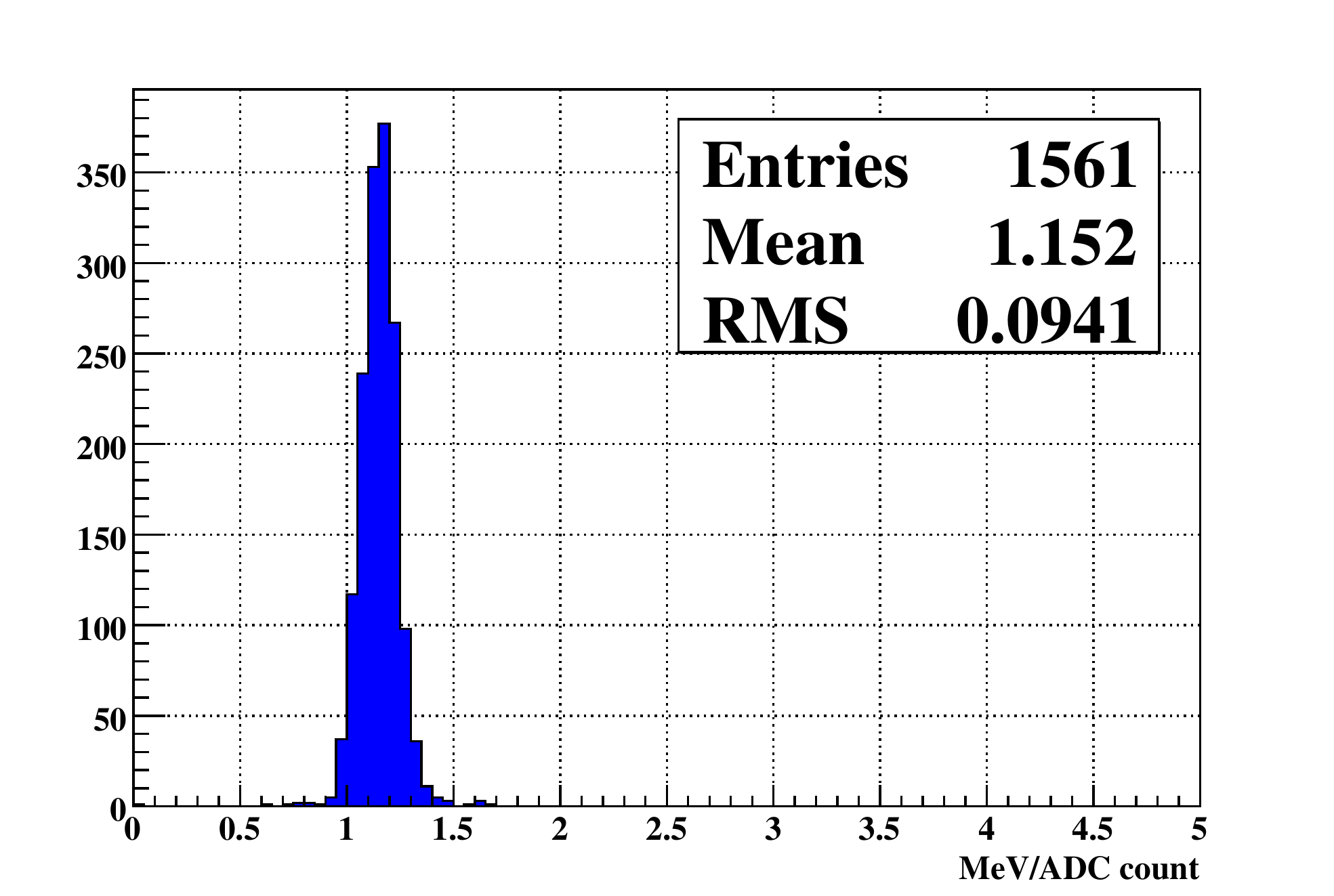}}
  \end{center}
  \caption{\label{BigCalCalibration} BigCal calibration example. (a) Calibration constants by PMT, in MeV per ADC count. (b) Distribution of calibration constants.}
\end{figure}
Figure \ref{BigCalCalibration} shows an example calibration result obtained from $\approx 250000$ elastic $ep$ events in CODA runs 66574-66577 at $Q^2=5.2$ GeV$^2$. The central scattered electron energy for this setting is 1.27 GeV. In this example, BigCal had been operated in beam for approximately two weeks since the initial gain matching/high-voltage adjustment. Figure \ref{bigcalcalib1} shows the calibration constant for each individual PMT, while figure \ref{bigcalcalib2} shows the distribution of calibration constants\footnote{In both plots, the blocks lining the edges of BigCal were omitted, since showers initiated near the edge of BigCal may lose a significant, unknown fraction of their energy which cannot easily be accounted for in the calibration procedure. When the edge blocks are included (not shown), their calibration constants come out systematically higher than those of the blocks at least one row or column away from the edge of BigCal, reflecting the fraction of the shower that is undetected. In the analysis, the calibration constants for the edge blocks were replaced with the average calibration constant of all non-edge blocks.}. The average result in this case was 1.15 MeV per ADC count. The spread in the constants after two weeks of operation in beam was approximately 9\%. 

\begin{figure}[h]
  \begin{center}
    \includegraphics[angle=90,width=.98\textwidth]{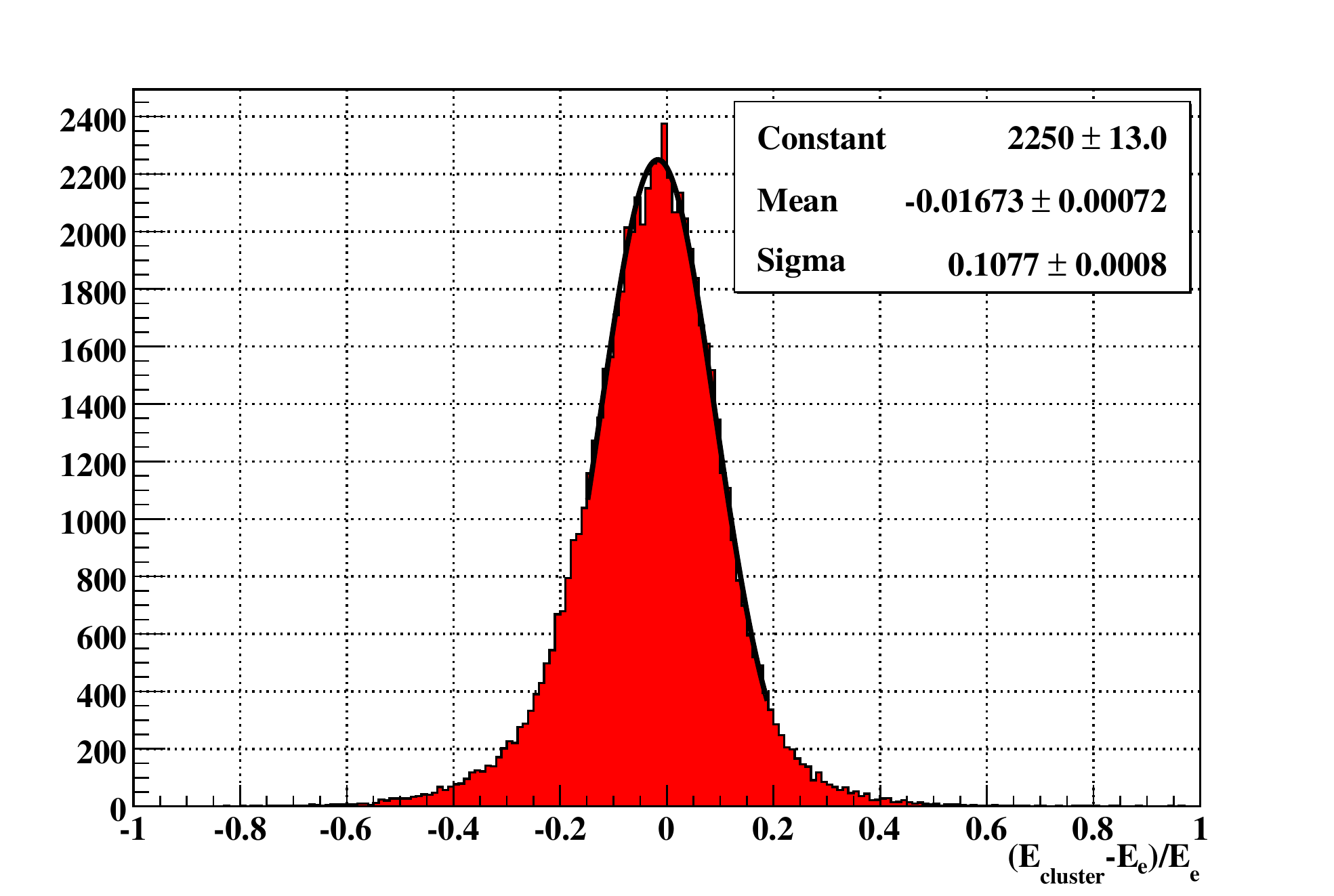}
  \end{center}
  \caption{\label{BigCalEres}Energy resolution of BigCal with 4'' aluminum absorber. $\frac{\sigma_E}{E} = 10.8\%$}
\end{figure}
Figure \ref{BigCalEres} shows the energy resolution of BigCal after calibration, also at $Q^2=5.2$ GeV$^2$, $E_e = 1.27$ GeV. The histogram shows the difference between the energy sum of all hits in the cluster detected in BigCal and the known electron energy calculated from the measured proton momentum and the beam energy, assuming elastic kinematics. Cuts were placed around the elastic peak in $p-p(\theta_p)$ and $p-p(\theta_e)$ in order to obtain a clean sample of elastic events. The resulting energy resolution for these kinematics is $\sigma_E/E = 10.8\%$. The ``ideal'' resolution obtained for these kinematics in the BigCal Monte Carlo simulation, with all four inches of absorber in front of BigCal was $\sigma_E/E = 9.1\%$. The difference between the ideal energy resolution obtained in the simulation and the experimental energy resolution comes from the combined effects of electronic noise, calibration uncertainties, and radiation damage. The resolution in figure \ref{BigCalEres} was obtained approximately two weeks after the start of the experiment. The difference between the ideal and experimental energy resolution only became worse with time as the cumulative effect of radiation damage increased. By the end of the experiment, the energy resolution increased to $\sigma_E/E = 15.4\%$ at $E_e=2.09$ GeV, and the relative gain of BigCal dropped by a factor of nearly 2.5 relative to the beginning of the experiment\footnote{Recall that the absolute gain was maintained at an approximately constant value by periodic high voltage increases.}.

Despite the poor energy resolution, the coordinate resolution of BigCal was only minimally affected by radiation damage, and was more than sufficient to cleanly separate elastic and inelastic events throughout the duration of the experiment. After converting ADC amplitudes to energies, the process of cluster finding proceeds. A relatively simple algorithm is used to find clusters of hits. Starting with the largest-energy hit, the code looks for hits in the four nearest-neighbor blocks around the maximum (above, below, left, and right). The maximum energy hit is stored in the cluster and marked as used so that it is only added to the cluster once. Each nearest-neighbor block with a hit above the ADC threshold is then added to the cluster and marked as used, and the process is repeated for each new hit, adding any unused nearest-neighbor hits above threshold until no more hits are found in the nearest neighbor blocks of any hit in the cluster. Clusters are allowed to expand freely in any direction until a maximum of 25 blocks is reached, at which point cluster growth is simply truncated. In practice, the number of blocks per cluster with a hit above threshold almost never reaches the maximum due to the relatively high software ADC threshold (2.5$\sigma_{PED}$) required to suppress pedestal noise. In terms of the electron energy, this ADC threshold suppresses all blocks containing less than about 2-3\% of the total shower energy\footnote{The effective energy threshold varies slightly across kinematic settings due to changes in the ratio of the pedestal width to the total shower signal. The width of the BigCal pedestal comes from integrating the noise and random background in each channel over the entire gate width. The noise/background level depends strongly on the BigCal singles rate, which is determined by the beam current, the beam energy, and the angle and distance at which BigCal is placed.}.

\begin{figure}[h]
  \begin{center}
    \includegraphics[angle=90,width=.99\textwidth]{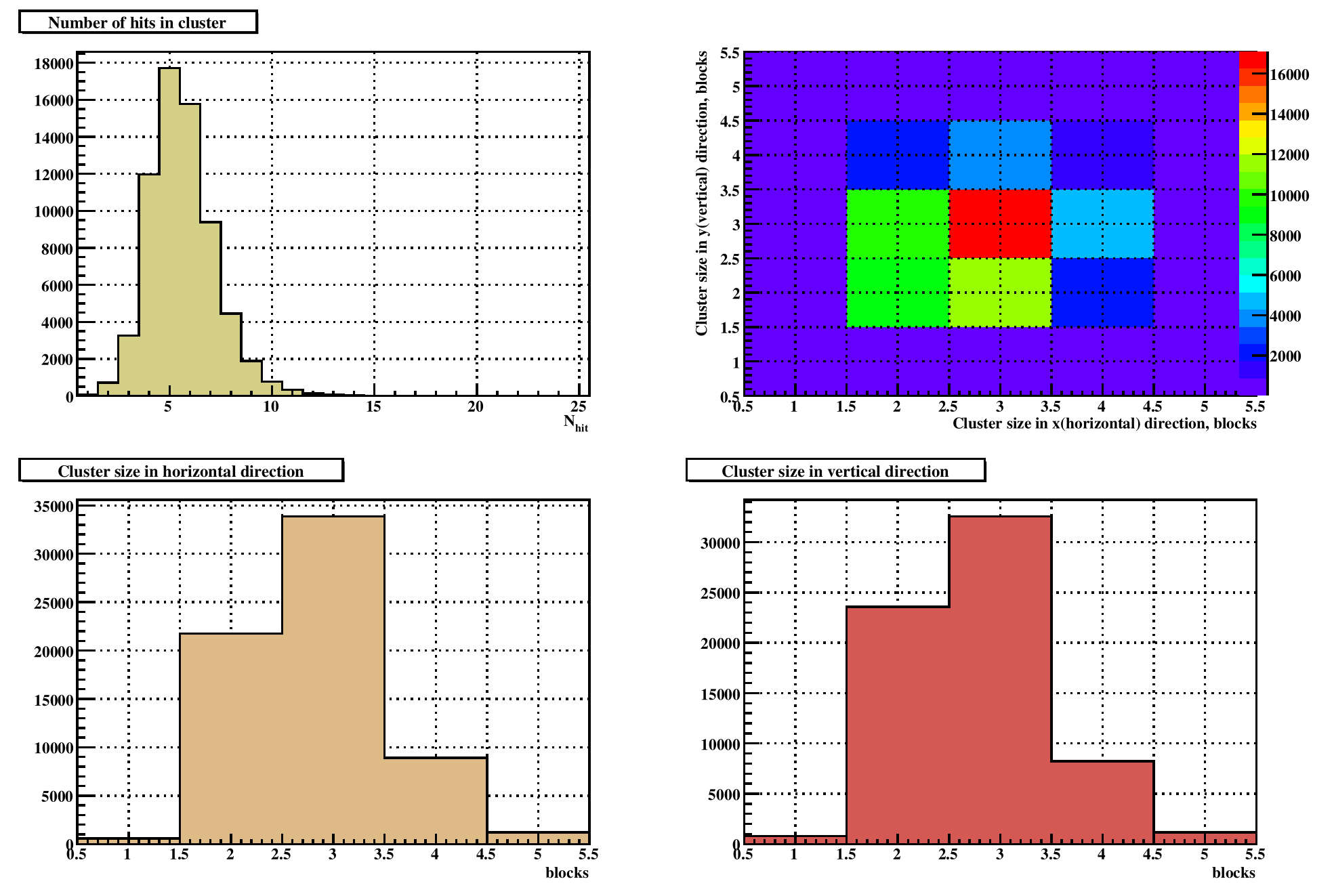}
  \end{center}
  \caption{\label{bigcalclustersize} Size of clusters in BigCal for $Q^2=5.2$ GeV$^2$, $E_e=1.27$ GeV, for events passing elastic kinematic cuts. Number of hits (top left), cluster size in horizontal direction (bottom left), cluster size in vertical direction (bottom right), and their correlation (top right).}
\end{figure}
The behavior of the clustering algorithm was controlled by two additional energy cuts. First, a threshold was applied to the energy of the central maximum. Since each iteration of the cluster finding algorithm seeks out the hit with the maximum energy, whenever the remaining unused hit with the largest energy falls below this threshold, cluster finding stops and no additional clusters are generated. Secondly, a threshold was applied to the total energy of the cluster. Cluster finding stops when no additional clusters with total energy sums exceeding this threshold are found. To optimize the clustering behavior for elastic $ep$ events, a subset of the data was analyzed with very low thresholds. Elastic events were selected using the methods of section \ref{elasticeventselectionsection}, and the distributions of the total cluster energy and the energy of the central maximum were obtained. The cluster sum and central maximum thresholds were both set as high as possible without rejecting significant numbers of elastic events. In practice, this meant that the threshold applied to the total cluster energy was no more than 40\% of the elastic $ep$ energy and the threshold applied to the energy of the central maximum was no more than 15\% of the elastic $ep$ energy.

Figure \ref{bigcalclustersize} illustrates the typical size of hit clusters in BigCal for events passing elastic kinematic cuts at $Q^2=5.2$ GeV$^2$. The most probable cluster size is $3\times3$, while the most probable total number of hits is five. The fraction of clusters with a size of only one block along either direction is less than 2\%, and the fraction of clusters with exactly one block is less than 0.1\%. Many of the clusters of single-block width in one or both directions had their maxima located at the edges of BigCal. Such clusters were allowed in the analysis. More than 98\% of events passing elastic kinematic cuts for this setting have a size of at least $2\times2$. 

The fact that practically all showers initiated by elastically scattered electrons at this energy are shared between multiple lead-glass bars allows for excellent coordinate resolution. For each cluster, an energy-weighted average block position or ``moment'' is calculated as
\begin{eqnarray}
  \left<x\right> &=& \sum_{i\in 3\times3}\frac{E_i}{E_{cluster}}(x_i - x_{max}) \\
  \left<y\right> &=& \sum_{i\in 3\times3}\frac{E_i}{E_{cluster}}(y_i - y_{max})
\end{eqnarray}
where $(x_i,y_i)$ are the coordinates of the center of the $i^{th}$ bar in the cluster, and $(x_{max},y_{max})$ are the coordinates of the center of the maximum-energy block around which the cluster is built. The sum in this case runs over the $3\times3$ grid of blocks centered on the maximum, in contrast to the energy sum, which runs over a $5\times5$ grid. This is in order to prevent small signals at the periphery of the cluster from distorting the calculated central shower coordinate\footnote{The signals in cells outside the $3\times3$ grid centered on the maximum are typically small with large statistical fluctuations and low signal-to-noise ratios, and are disproportionately weighted in the moment sums because of their large distances from the maximum (2 cell sizes or approximately 1.7 Moliere radii). Including these signals in the moment sums does not improve the coordinate resolution and in some cases makes it worse.}.

To convert the measured cluster moments to shower coordinates, a procedure similar to the drift map calculation for the FPP and HMS drift chambers is used. Denoting the block size $d$, the allowed range of shower moments is 
\begin{eqnarray}
  -\frac{d}{2} \le &\left<x\right>& \le \frac{d}{2} \\
  -\frac{d}{2} \le &\left<y\right>& \le \frac{d}{2}
\end{eqnarray}
The measured distribution of moments, shown in figure \ref{bigcalmoments}, is mapped onto a uniform shower coordinate distribution within the central, maximum-energy cell, resulting in the distribution of shower coordinates shown in figure \ref{bigcaldiffs}.
\begin{figure}[h]
  \begin{center}
    \subfigure[$\left<x\right>$]{\label{bigcalxmom}\includegraphics[angle=90,width=.49\textwidth]{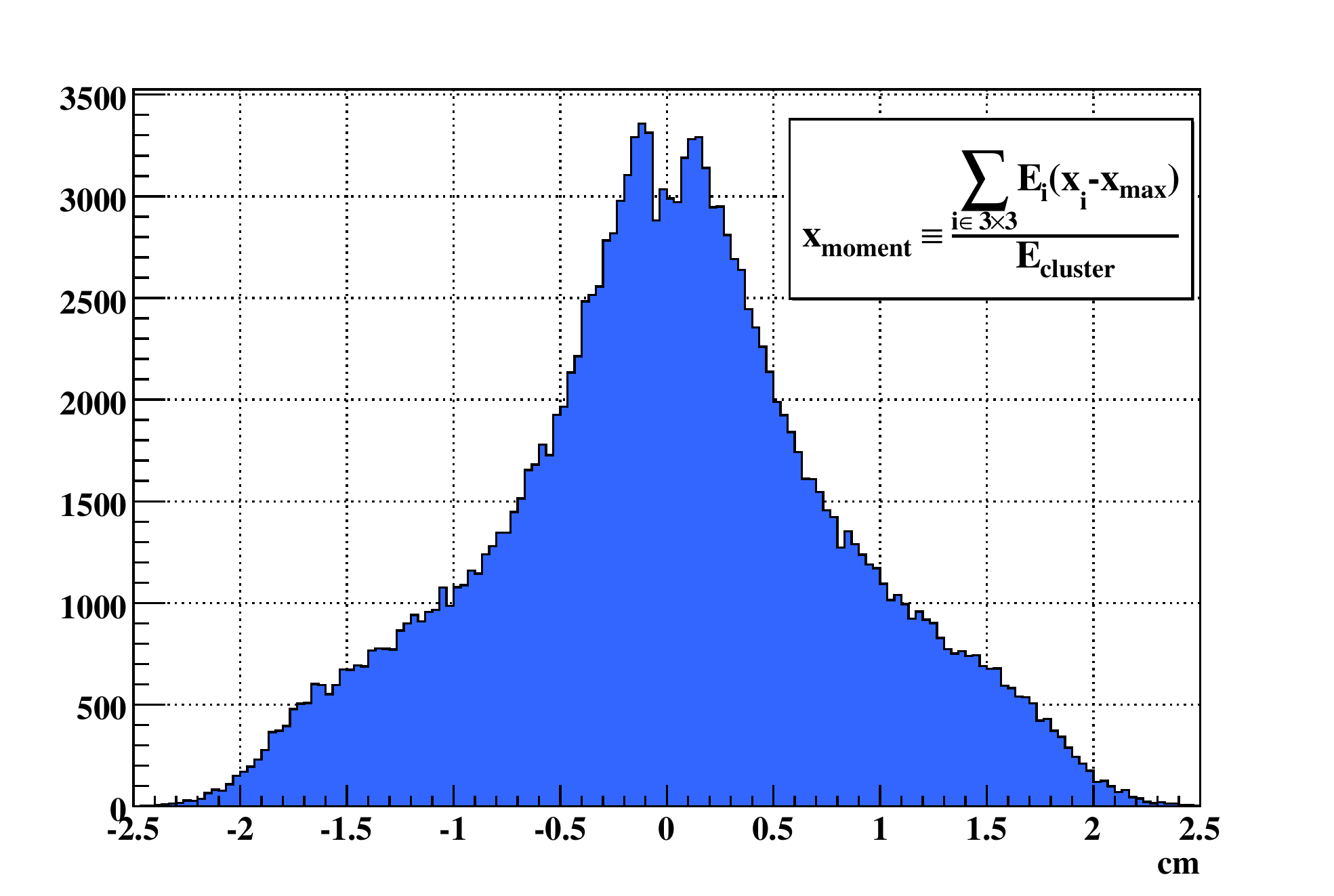}}
    \subfigure[$\left<y\right>$]{\label{bigcalymom}\includegraphics[angle=90,width=.49\textwidth]{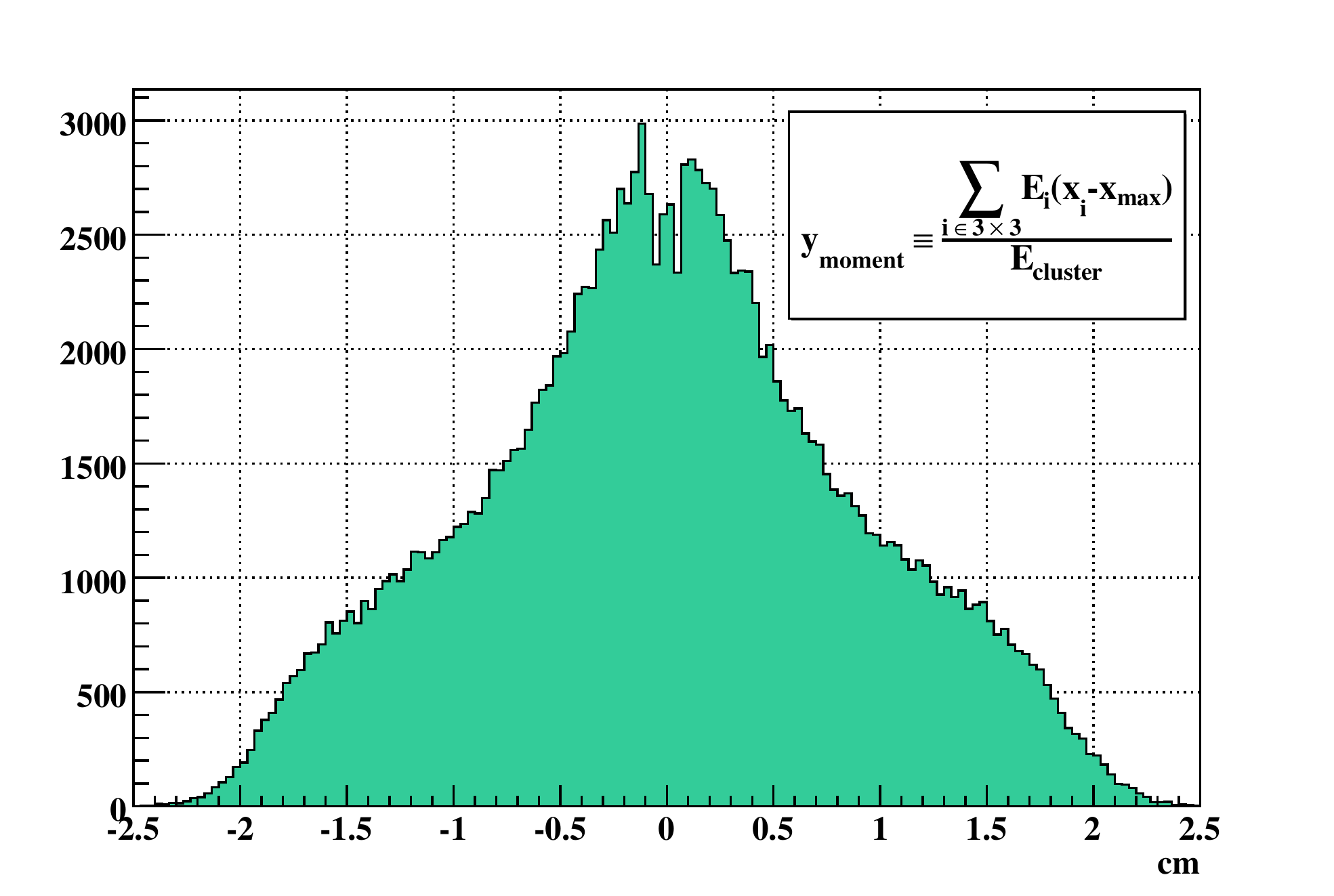}}
  \end{center}
  \caption{\label{bigcalmoments} Cluster moments in BigCal.}
\end{figure}
\begin{figure}[h]
  \begin{center}
    \subfigure[$x_{cluster}-x_{max}$]{\label{bigcalxdiff}\includegraphics[angle=90,width=.49\textwidth]{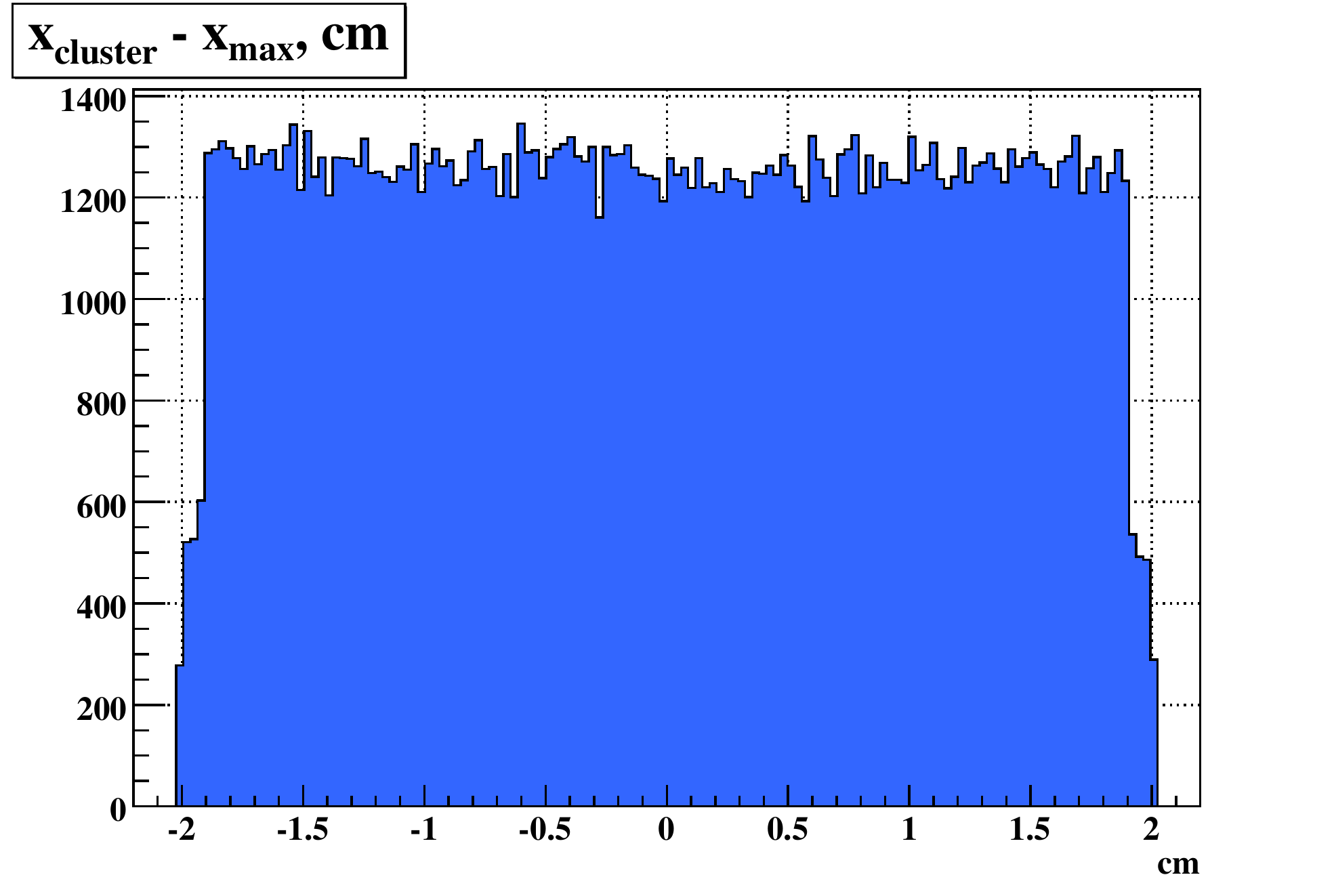}}
    \subfigure[$y_{cluster}-y_{max}$]{\label{bigcalydiff}\includegraphics[angle=90,width=.49\textwidth]{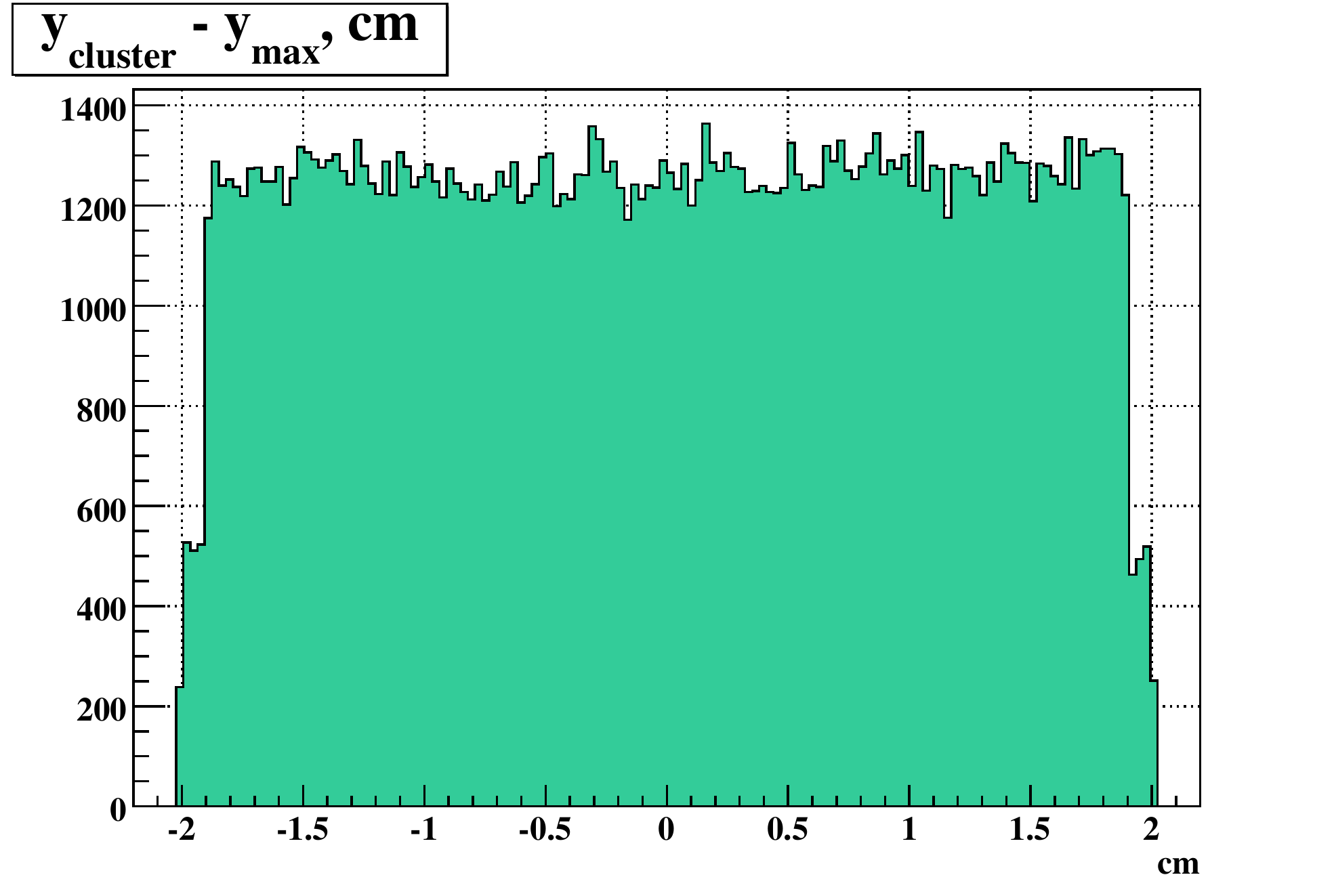}}
  \end{center}
  \caption{\label{bigcaldiffs} Calculated shower coordinate, relative to the center of the maximum-energy cell in the cluster.}
\end{figure}
To perform this mapping, the calorimeter was divided into 28 sections of $8\times8$ blocks, with separate $<x>$ and $<y>$ mappings calculated for each section. Each section had a slightly different distribution of moments, because of the different incident angles of electrons hitting different parts of the calorimeter, which gave rise to different effective shower shapes in each section. The calculated mapping between shower moments and shower coordinates is shown in figure \ref{bigcalscurves}. The characteristic ``S'' shape of the mapping makes intuitive sense. Showers impacting near the center of the maximum-energy block tend to share less energy with neighboring blocks, and also tend to distribute that energy more evenly. The moments of this kind of cluster change more slowly as a function of the impact coordinate, giving rise to the steeper slope of $x(<x>)$ near the block center. At larger $<x>$, more energy is shared with neighboring blocks, and $<x>$ varies more rapidly as a function of the shower impact coordinate. 
\begin{figure}[h]
  \begin{center}
    \subfigure[$x_{clust}-x_{max}$ vs. $<x>$]{\label{bigcalscurvex}\includegraphics[width=.49\textwidth]{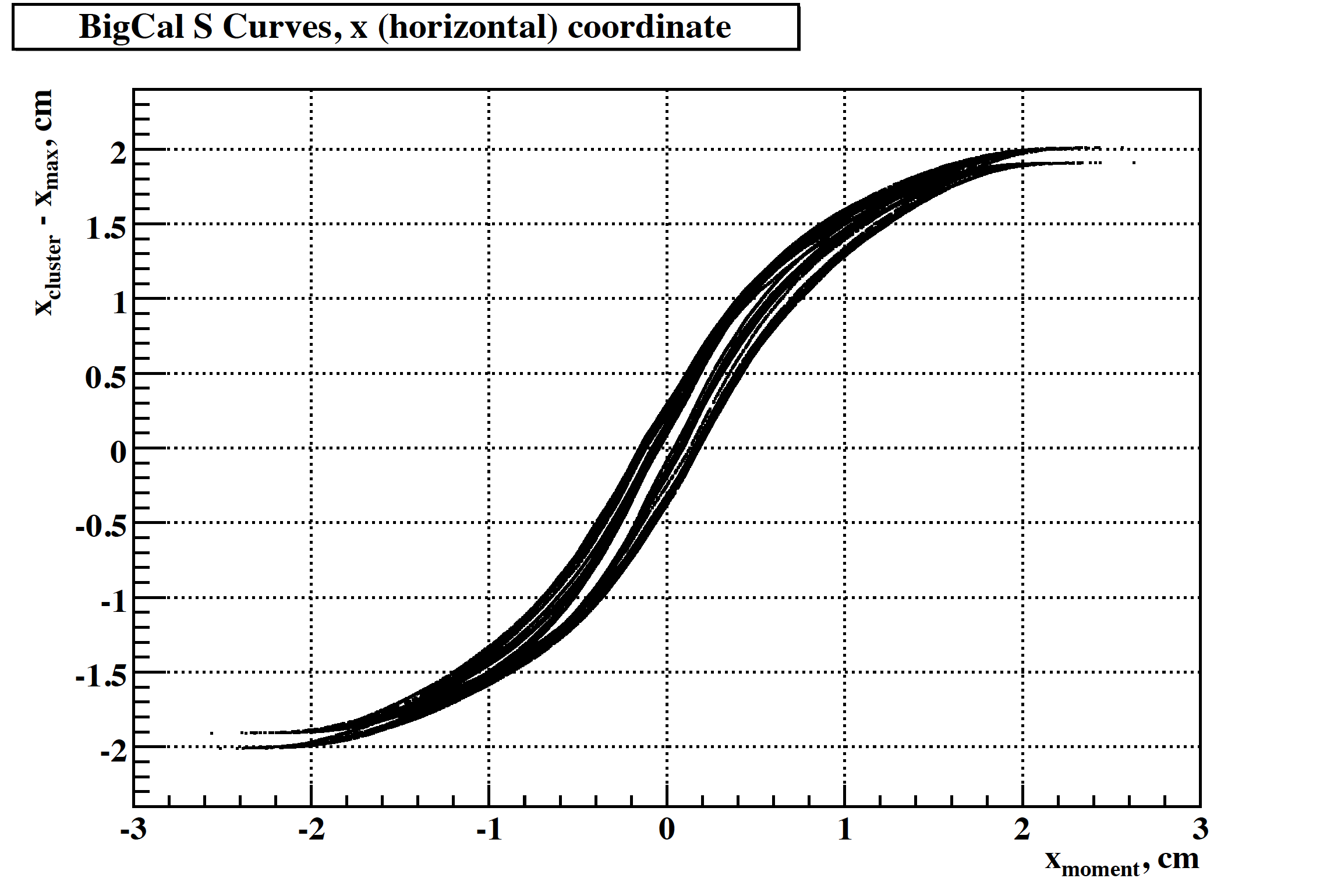}}
    \subfigure[$y_{clust}-y_{max}$ vs. $<y>$]{\label{bigcalscurvey}\includegraphics[width=.49\textwidth]{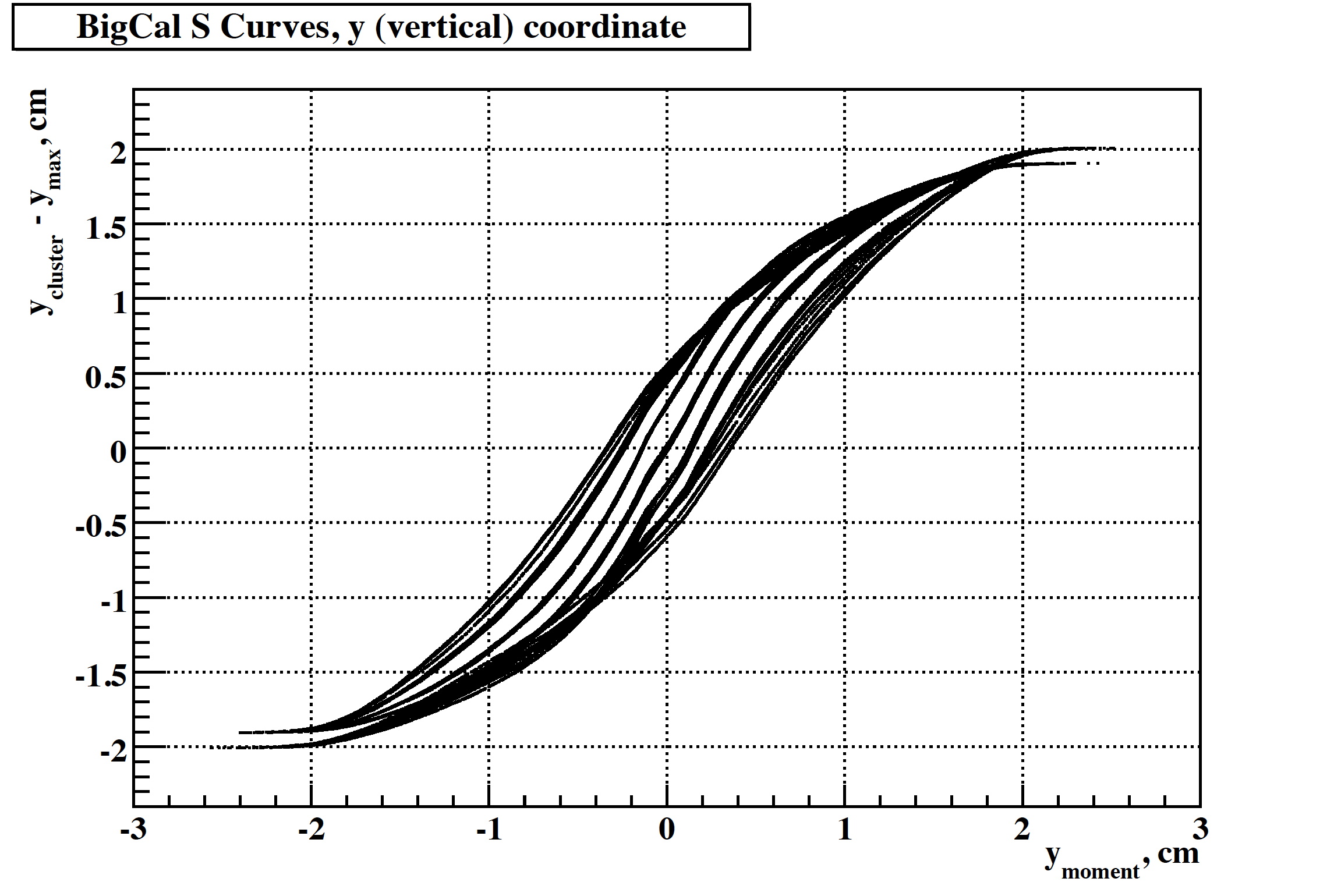}}
  \end{center}
  \caption{\label{bigcalscurves} Calculated mapping between cluster moments and shower coordinates illustrating characteristic ``S'' shape.}
\end{figure}
There are twenty-eight different curves in figures \ref{bigcalscurvex} and \ref{bigcalscurvey}, one $x(<x>)$ curve and one $y(<y>)$ curve for each $8\times8$ section of the calorimeter. The curves corresponding to the lower half of the calorimeter cover a range equal to the size of the Protvino blocks ($-1.9$ cm $\le x_{clust}-x_{max} \le 1.9$ cm), while the curves corresponding to the upper half cover a range equal to the size of the RCS blocks ($-2.0$ cm $\le x_{clust}-x_{max} \le 2.0$ cm). For a uniform distribution of shower coordinates, given the moment distributions $f(<x>)$ and $g(<y>)$, the cluster coordinates are given by
\begin{eqnarray}
  x(<x>) &=& -\frac{d}{2} + \int_{-\frac{d}{2}}^{<x>} f(<x>) d<x> \\
  y(<y>) &=& -\frac{d}{2} + \int_{-\frac{d}{2}}^{<y>} g(<y>) d<y>
\end{eqnarray}

The cluster coordinates thus defined are equated to the point at which the showering particle impacts the surface of BigCal, and are restricted to lie inside the area of the maximum-energy cell. This assumption is not strictly true. For electrons impinging at significantly non-normal incident angles, the shower maximum can be significantly displaced from the surface impact point. The longitudinal depth of the maximum energy deposition in electromagnetic shower development for electron-induced showers is well approximated by\cite{PDG2008}
\begin{equation}
  t_{max} = X_0 \ln \left(\frac{E_e}{E_c}\right) - \frac{1}{2} 
\end{equation}
where $E_e$ is the energy of the primary electron, $X_0$ is the radiation length of lead glass, and $E_c$ is the critical energy. If the incident electron trajectory makes angles $\theta_x$ and $\theta_y$ relative to the normal to the surface of BigCal, then the displacement of $t_{max}$ in $x$ and $y$ from the impact point at the surface is given by 
\begin{eqnarray}
  x_{max} - x_0 &=& t_{max} \sin \theta_x \\
  y_{max} - y_0 &=& t_{max} \sin \theta_y 
\end{eqnarray}
The displacement of the \emph{observed} shower maximum from the surface impact coordinate should be proportional to the displacement of the maximum energy deposition. The constant of proportionality was determined using the BigCal Monte Carlo simulation by comparing the true surface impact coordinates to those reconstructed using the mapping above, and was found to be very close to $\sqrt{\frac{1}{2}}$ for all the kinematics of these experiments. The fact that this constant is independent of the distance from the origin $R$ and the incident electron energy $E_e$ is strong evidence for the validity of the assumption that the displacement of the maximum from the surface coordinate is proportional to $t_{max}$. Therefore, a ``distortion'' correction was applied to each cluster as follows. Under the assumption of a point target; i.e., assuming the scattered electrons start at the origin, the incident angles of a cluster detected at a point $(x,y)$ on the calorimeter are
\begin{eqnarray}
  \sin \theta_x  &=& \frac{x}{\sqrt{x^2+R^2}} \\
  \sin \theta_y  &=& \frac{y}{\sqrt{y^2+R^2}}
\end{eqnarray}
where $R$ is the distance from the origin to the surface of BigCal. The final shower coordinates, corrected for the displacement of the shower maximum from the surface impact coordinates, are therefore given by
\begin{eqnarray}
  x_{shower} &=& x(<x>) - \frac{t_{max}}{\sqrt{2}} \frac{x(<x>)}{\sqrt{(x(<x>))^2+R^2}} \\
  y_{shower} &=& y(<y>) - \frac{t_{max}}{\sqrt{2}} \frac{y(<y>)}{\sqrt{(x(<x>))^2+R^2}}
\end{eqnarray}
The slight variations in the incident angle due to the extended target were neglected in applying the above correction.

Figure \ref{bigcaldistcorr} shows the effect of this correction on the accuracy of the electron coordinate reconstruction. 
\begin{figure}[h]
  \begin{center}
    \subfigure[$x(<x>)-x_{elastic}$ vs. $x(<x>)$]{\label{bigcaldxxnodistcorr}\includegraphics[angle=90,width=.49\textwidth]{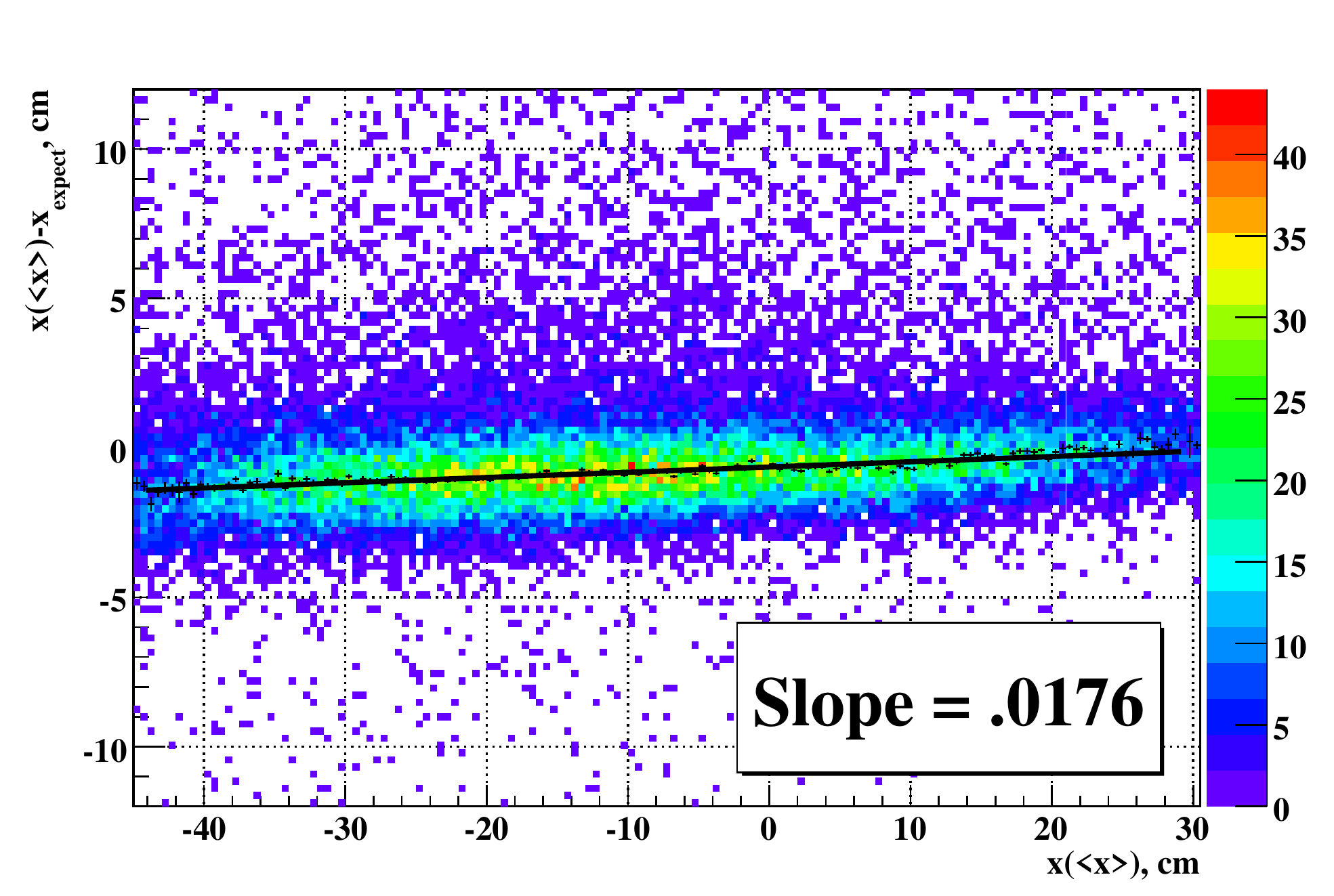}}
    \subfigure[$x_{shower}-x_{elastic}$ vs. $x_{shower}$]{\label{bigcaldxxdistcorr}\includegraphics[angle=90,width=.49\textwidth]{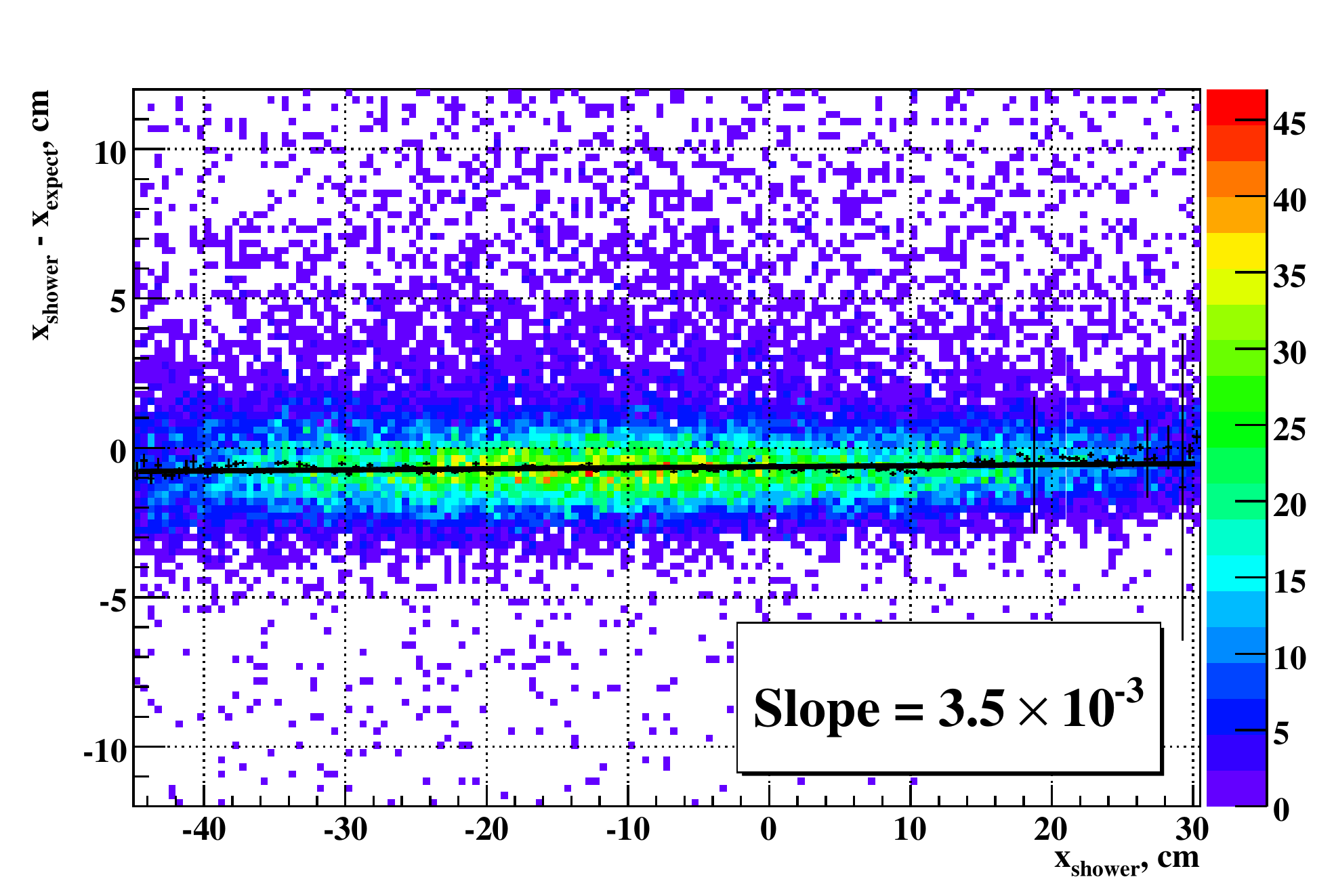}}
  \end{center}
  \caption{\label{bigcaldistcorr} Difference between reconstructed shower coordinate and expected coordinate of elastically scattered electron, as a function of the reconstructed coordinate, (a) before incident-angle correction, and (b) after incident-angle correction.}
\end{figure}
Figure \ref{bigcaldxxnodistcorr} shows the correlation between the uncorrected coordinate $x(<x>)$ on the $x$ axis, and the difference between the uncorrected coordinate and the expected coordinate calculated from elastic kinematics of the detected proton on the $y$ axis. There is an obvious correlation, which is caused by the displacement of the shower maximum discussed above. Figure \ref{bigcaldxxdistcorr} shows how the incident angle correction eliminates this correlation. 

The final coordinate resolution for BigCal obtained using the method described above in the Monte Carlo simulation is $\Delta x \approx 0.54\ \mbox{cm}/\sqrt{E_e}$. Estimating the true coordinate resolution achieved from the data is complicated by the fact that the resolution of the coordinates predicted from elastic kinematics of the measured proton is, in most cases studied in these experiments, comparable to or worse than the intrinsic coordinate resolution of BigCal, such that any estimate of the BigCal resolution relies on and is highly sensitive to estimates of the HMS resolution, which is only approximately known. Since the angular resolution of the HMS is degraded by S0, the measured proton momentum is used to calculate the scattered electron angle:
\begin{eqnarray}
  Q^2 = 2M_p T_p &=& 2E_e E'_e(1 - \cos \theta_e) \\
  T_p &=& \sqrt{p_p^2 + M_p^2} - M_p \\
  \cos \theta_e &=& 1 - \frac{M_pT_p}{E_e(E_e-T_p)} \label{ethexpectformula}
\end{eqnarray}
To calculate the expected electron coordinates requires in addition the position of the interaction vertex $z$ and the azimuthal angle of the scattered proton $\phi_p$, so one cannot completely avoid using the measured proton angles. The $z$ coordinate of the interaction vertex is determined by the intersection point between the beam ray and the projection of the spectrometer ray onto the horizontal plane, and is a function of $y_{tar}$ and $y'_{tar}$:
\begin{equation}
  z_{beam} = y_{tar}\left(\frac{\cos \Theta_{HMS}}{\tan\left(\Theta_{HMS}-\arctan y'_{tar}\right)}+\sin \Theta_{HMS}\right) \label{zbeam_formula}
\end{equation}
where $\Theta_{HMS}$ is the central angle of the HMS. The azimuthal angle of the scattered proton is determined by rotating the proton's measured trajectory, which is in transport coordinates, into Hall C or ``beam'' coordinates:
\begin{eqnarray}
  \hat{x}_{transport} &\equiv& \frac{x'_{tar}}{\sqrt{1+x_{tar}'^2 + y_{tar}'^2}} \\
  \hat{y}_{transport} &\equiv& \frac{y'_{tar}}{\sqrt{1+x_{tar}'^2 + y_{tar}'^2}} \\ 
  \hat{z}_{transport} &\equiv& \frac{1}{\sqrt{1+x_{tar}'^2 + y_{tar}'^2}} \\
  \left(\begin{array}{c} \hat{x} \\ \hat{y} \\ \hat{z} \end{array}\right)_{beam} &=& \left( \begin{array}{ccc} 1 & 0 & 0 \\ 0 & \cos \Theta_{HMS} & -\sin \Theta_{HMS} \\ 0 & \sin \Theta_{HMS} & \cos \Theta_{HMS} \end{array} \right)\left(\begin{array}{c} \hat{x} \\ \hat{y} \\ \hat{z} \end{array}\right)_{transport}
\end{eqnarray}
The ``physics angles'' of the scattered proton, the polar angle $\theta_p$ with respect to the beamline and the azimuthal angle $\phi_p$ are then defined by
\begin{eqnarray}
  \cos \theta_p &=& \hat{z}_{beam} \\
  \tan \phi_p &=& \frac{\hat{y}_{beam}}{\hat{x}_{beam}} \label{phipdef}
\end{eqnarray}
The proton's polar angle $\theta_p$ is not used in the calculation, since $\theta_e$ is already determined by $p_p$. The azimuthal angle of the electron is determined by the requirement of co-planarity; i.e., $\phi_e = \phi_p + \pi$. In the ``beam'' coordinate system, the HMS azimuthal angle is centered at $\phi_p = -\frac{\pi}{2}$, and BigCal is centered at $\phi_e = \frac{\pi}{2}$. The $x$ axis of this coordinate system points vertically downward, and the $y$ axis points toward beam left (toward BigCal). By historical accident, the coordinate system for BigCal was defined differently, such that $x$ is the horizontal coordinate, with positive $x$ pointing in the direction of increasing $\theta_e$, and $y$ is the vertical coordinate, with the $+y$ axis pointing vertically upward. Therefore, an alternate ``beam'' coordinate system is defined which is simply a re-labeling of the coordinate axes, with $y \rightarrow x$ and $x \rightarrow -y$, so that the new $y$ axis points vertically upward and the new $x$ axis points toward beam left. In this coordinate system, the trajectory of the electron is 
\begin{eqnarray}
  \hat{x}_e &=& \sin \theta_e \sin \phi_e \\
  \hat{y}_e &=& -\sin \theta_e \cos \phi_e \\
  \hat{z}_e &=& \cos \theta_e
\end{eqnarray}
The coordinates of the electron at any point $s$ along its trajectory, with $s\equiv 0$ at the interaction point, are given by $\mathbf{r}(s) \equiv (x(s),y(s),z(s))$ defined as
\begin{eqnarray}
  x(s)  &=& x_{beam} + s \hat{x}_e  \\
  y(s)  &=& y_{beam} + s \hat{y}_e  \\
  z(s)  &=& z_{beam} + s \hat{z}_e 
\end{eqnarray}
The electron trajectory intersects the surface of BigCal at $s=s_0$ such that the unit normal vector to the surface of BigCal $\hat{n}$ is orthogonal to the ray from the center of BigCal $\mathbf{R}_0$ to the trajectory vector $\mathbf{r}(s_0)$:
\begin{eqnarray}
  \hat{n} \cdot (\mathbf{r}(s_{0}) - \mathbf{R}_0) &=& 0 \label{intersectionpoint} \\
  \hat{n} &\equiv& \left(\sin \Theta_{cal}, 0, \cos \Theta_{cal} \right) \\
  \mathbf{R}_0 &\equiv& R_{cal} \hat{n}
\end{eqnarray}
Solving \eqref{intersectionpoint} for $s_0$ gives
\begin{eqnarray}
  R_{cal} &=& (x_{beam} + s_0 \hat{x}_e)\sin \Theta_{cal} + (z_{beam} + s_0 \hat{z}_e) \cos \Theta_{cal}   \\
  s_0 &=& \frac{R_{cal} - x_{beam} \sin \Theta_{cal} - z_{beam} \cos \Theta_{cal}}{\hat{x}_e \sin \Theta_{cal} + \hat{z}_e \cos \Theta_{cal}}
\end{eqnarray}
The coordinates measured at BigCal are the coordinates parallel to its surface, with $+x_{cal}$ pointing in the direction of increasing $\theta_e$ and $+y_{cal}$ pointing vertically upward. The expected coordinates from the proton kinematics $(x_{HMS}, y_{HMS})$ are given by a simple rotation of $\mathbf{r}(s_0)$:
\begin{eqnarray}
  x_{HMS} &=& x(s_0) \cos \Theta_{cal} - z(s_0) \sin \Theta_{cal} \\
  y_{HMS} &=& y(s_0)
\end{eqnarray}

Figure \ref{bigcalxres} shows the difference between the measured $x$ coordinate at BigCal and the expected coordinate $x_{HMS}$ at $Q^2=6.7$ GeV$^2$. The data for this kinematic setting were obtained at the end of the experiment, so they represent the worst-case scenario for the degradation of the position resolution due to radiation damage. On the other hand, the central electron energy for this setting was a fairly high 2.09 GeV, and both the coordinate and energy resolution get worse at lower energies. Nonetheless, these data provide a suitable test case to estimate the position resolution of BigCal.
\begin{figure}[h]
  \begin{center}
    \includegraphics[angle=90,width=.99\textwidth]{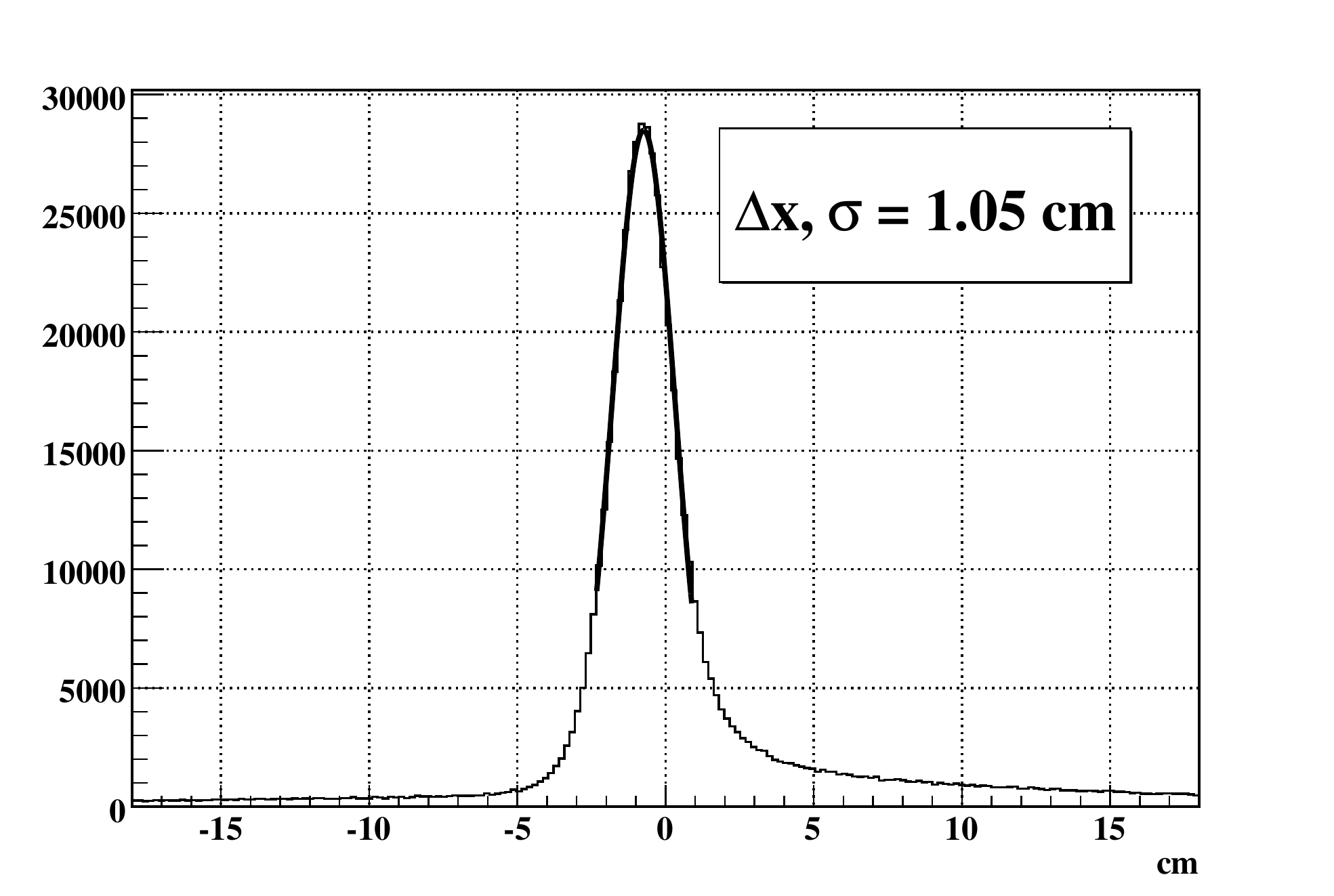}
  \end{center}
  \caption{\label{bigcalxres} $x_{cal}-x_{HMS}$ with $R_{cal}=608.2$ cm, $\Theta_{cal}=44.2^\circ$, $E_{beam}=5.71$ GeV, $E'_e=2.09$ GeV, $\Theta_{HMS}=19.07^\circ$, and $p_0=4.4644$ GeV/c.}
\end{figure}
The resolution of the difference $\Delta x$, which is a combination of the BigCal and HMS resolution, is approximately 1.05 cm. To estimate the contribution from BigCal, one must first estimate the contribution from the HMS. Assuming a momentum resolution of $\delta p/p = 0.1\%$ and $\delta y_{tar} = .17$ cm, the change in the expected electron angle defined by equation \eqref{ethexpectformula} induced by a 0.1\% shift in momentum is $\Delta \theta_e(p \rightarrow p + dp) = 1.4$ mrad for the central kinematics at $Q^2=6.7$ GeV$^2$. The resulting change in $x_{HMS}$ is $\Delta x_{HMS,p} \approx R_{cal} \Delta \theta_e = 0.82$ cm. There is also an error in $x_{HMS}$ due to the error in $z_{beam}$, $\Delta x_{HMS,y} \approx \sin \Theta_{cal} \Delta z_{beam} \approx \frac{\sin \Theta_{cal}}{\sin \Theta_{HMS}} \Delta y_{tar} = 0.36$ cm. The quadrature sum of the momentum and $y_{tar}$ contributions to the resolution $\Delta x_{HMS,tot}$ is $\Delta x_{HMS,tot} = \sqrt{(\Delta x_{HMS,p})^2+(\Delta x_{HMS,y})^2} = 0.90$ cm. The observed resolution in $\Delta x \equiv x_{cal} - x_{HMS}$, which should be zero for elastically scattered electrons, equals the quadrature sum of the HMS and BigCal contributions:
\begin{eqnarray}
  \Delta(x_{cal} - x_{HMS}) &=& \sqrt{ (\Delta x_{cal})^2 + (\Delta x_{HMS,tot})^2 } \\
  \Rightarrow \Delta x_{cal} &=& \sqrt{(\Delta(x_{cal}-x_{HMS}))^2 - (\Delta x_{HMS,tot})^2} = 0.55\ \mbox{cm} \label{bigcalresolutionestimate}
\end{eqnarray}
The result \eqref{bigcalresolutionestimate} of $\Delta x_{cal} = 0.55$ cm with $E'_e=2.09$ GeV was obtained at the end of the experiment when the most radiation damage had been done. It is to be compared with the ``ideal'' resolution from the Monte Carlo of $0.54$ cm$/\sqrt{E_e} = 0.37$ cm. It is a very rough estimate, given the fact that the HMS contribution to the resolution of $\Delta x$ is nearly twice as large, and the result for the BigCal resolution is quite sensitive to the estimated HMS resolution. Nonetheless, it is not too far from the ideal result and significantly better than $d/\sqrt{12}$. One of the key facts of this experiment is that for most of the kinematics, the limiting resolution, as far as the separation of elastic and inelastic events is concerned, is that of the HMS. Among the three high-$Q^2$ data points, $Q^2=6.7$ GeV$^2$ actually has the smallest Jacobian at $J=4.77$, and it is the only point for which the contribution of BigCal to the resolution could actually be estimated. The data points at $Q^2 = 5.2$ GeV$^2$ and $Q^2=8.5$ GeV$^2$ have $J=8.36$ and $J=22.0$ respectively, making the HMS resolution even more dominant in determining e.g. the $\Delta x$ resolution. 

The resolution of the vertical coordinate difference $y_{cal}-y_{HMS}$ is typically 3-5 times worse than the resolution of the horizontal coordinate. That is because whereas $x_{HMS}$ is mainly determined by the proton momentum, which is measured with excellent resolution, the vertical coordinate $y_{HMS}$ is mainly determined by $x'_{tar}$, the resolution of which is blown up by multiple scattering in S0. The vertical coordinate in BigCal is measured with approximately the same precision as the horizontal coordinate, and actually serves as a more precise determination of the proton azimuthal angle for elastic events.

\subsubsection{BigCal Timing}
\paragraph{}
In the preceding sections, the use of the BigCal ADC information alone to reconstruct the shower coordinates and energy was discussed. The timing signals from BigCal, as measured by the TDCs, were also useful in isolating the best clusters, particularly at high rates. As previously discussed, the signals were combined into groups of eight channels for timing purposes in order to save on cabling and electronics. Each first-level sum has an associated TDC signal, and each second-level sum has both an ADC and TDC signal. TDCs for the first and second-level sums were operated in common stop mode. Several corrections were applied to the raw TDC values to convert them to hit times. 

First, the time difference between each first level sum signal and the BigCal trigger was determined. An example time difference spectrum is shown in figure \ref{bigcalhittimediff}. Second, the timing of the BigCal trigger relative to the HMS start time was determined for events in the true coincidence peak of the trigger time distribution (see figure \ref{cointrig_time}). An example of this time difference is shown in figure \ref{bigcaltrigtimediff}.
\begin{figure}[h]
  \begin{center}
    \subfigure[$t_{hit} - t_{trig}$, ns]{\label{bigcalhittimediff}\includegraphics[angle=90,width=.49\textwidth]{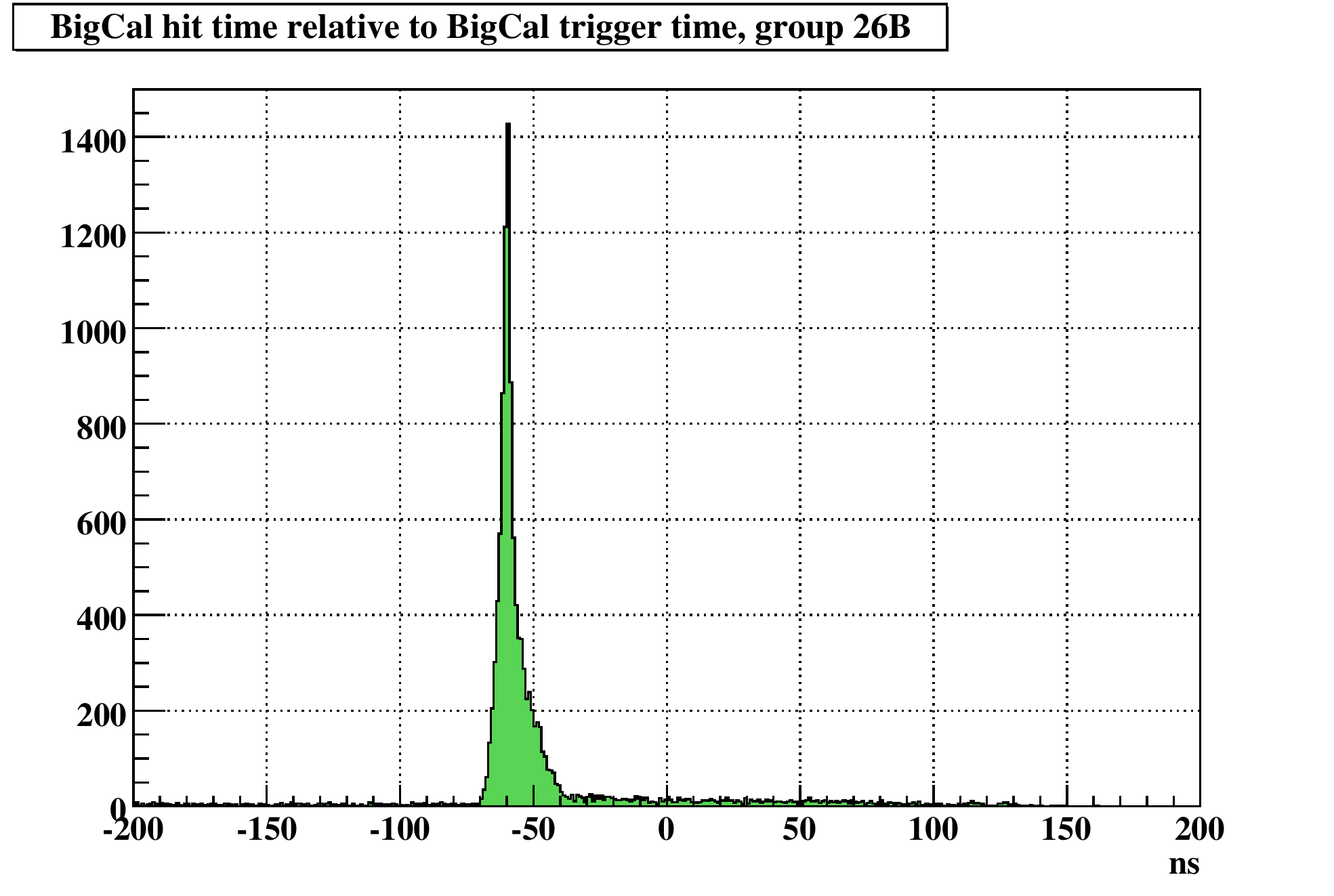}}
    \subfigure[$t_{trig} - t_{start}$, ns]{\label{bigcaltrigtimediff}\includegraphics[angle=90,width=.49\textwidth]{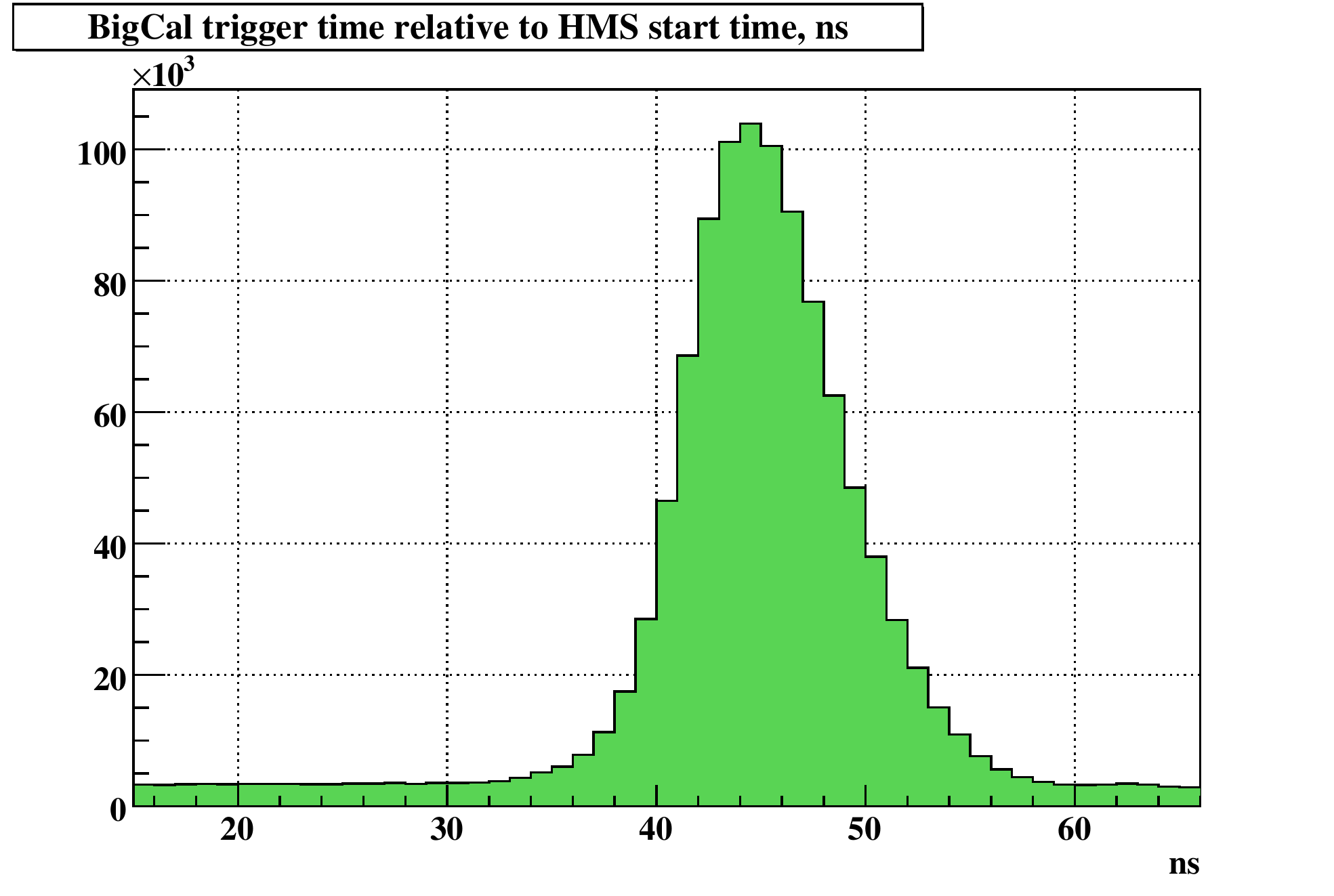}}
  \end{center}
  \caption{\label{bigcaltimes} (a) Time difference between first-level sum 26B and BigCal trigger. (b) Time difference between BigCal trigger and HMS start time (determined by hodoscopes).}
\end{figure}
The two combined offsets thus determined serve to roughly align the timing of the first-level sums with the HMS start time. 

The final correction to the BigCal hit times was a walk correction to account for the pulse-height dependence of the arrival time of the signals, which were discriminated against a fixed threshold of approximately 100 MeV. The walk correction used was of the same form as that used for the HMS hodoscopes. The signal size of the first level sums is not directly measured, but can be calculated by summing the eight individual channels in each group:
\begin{equation}
  t_{corr} = t_{raw} - t_0 - \frac{w}{\sqrt{\sum_{i=1}^{8} (ADC - PED)_i}} \label{bigcalwalkcorr}
\end{equation}
To determine the walk correction parameters $w$ for each channel, profile histograms of the time difference between each BigCal first level sum and the HMS start time as a function of the sum of the eight measured ADC signals in that group were fitted to the functional form \eqref{bigcalwalkcorr}. An example fit result is shown in figure \ref{bigcalwalkcorrfig}. 
\begin{figure}[h]
  \begin{center}
    \subfigure[]{\label{bigcalwalkcorrfig}\includegraphics[angle=90,width=.49\textwidth]{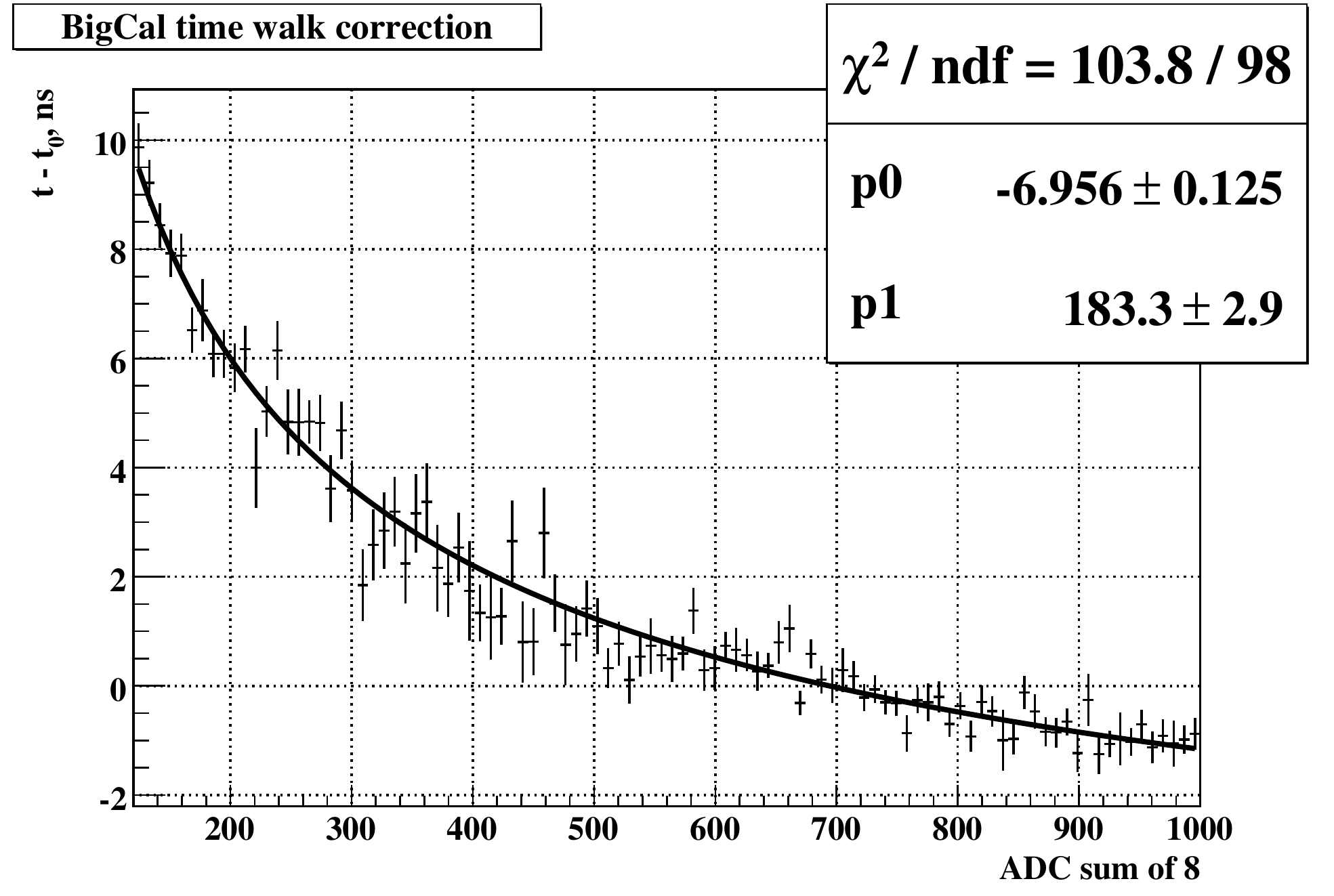}}
    \subfigure[]{\label{bigcaltimeres}\includegraphics[angle=90,width=.49\textwidth]{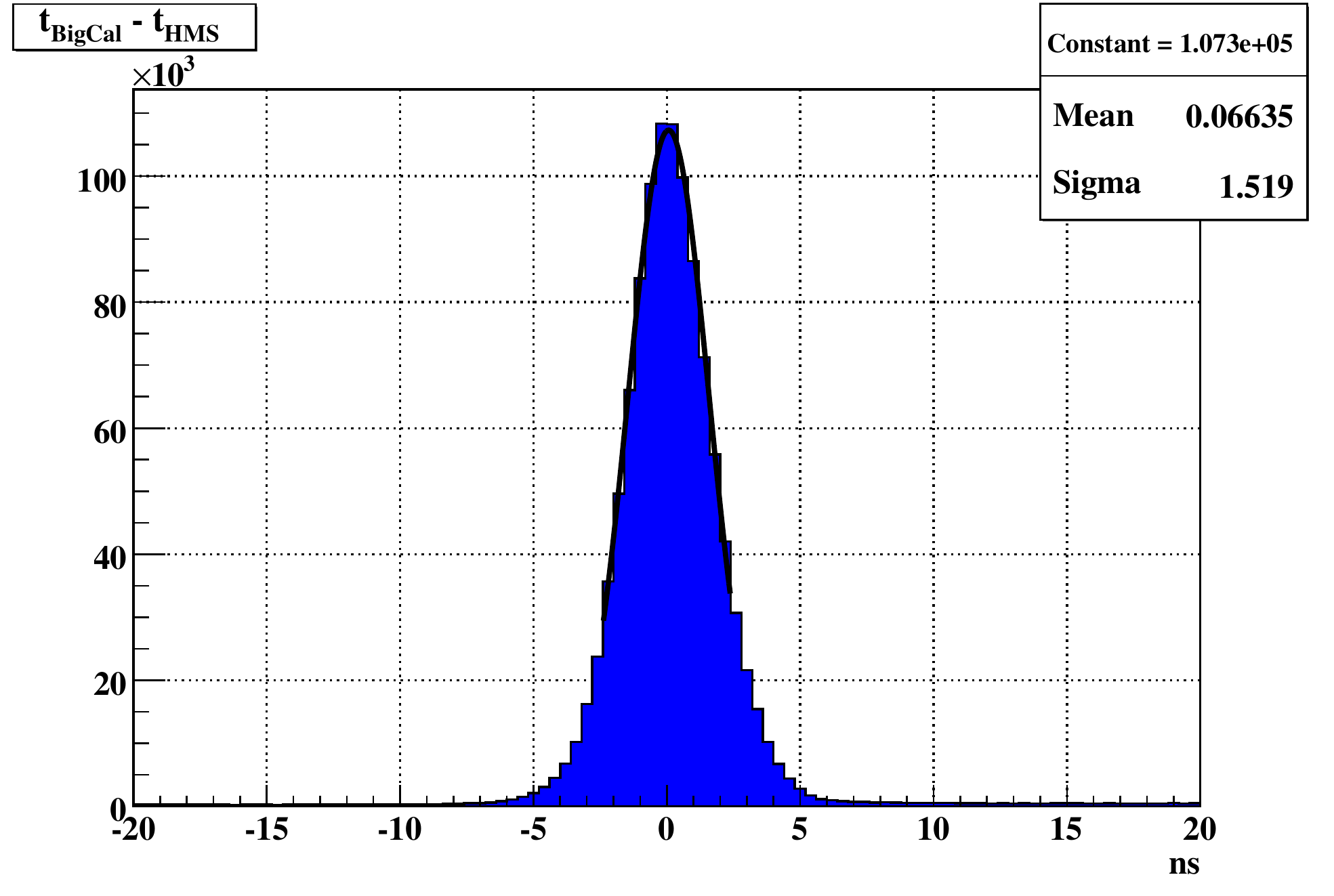}}
  \end{center}
  \caption{\label{bigcaltimefig} (a) Walk correction to BigCal timing signals. (b) Final timing resolution of BigCal at $E_e \approx 2.3$ GeV.}
\end{figure}

The final timing resolution of BigCal after applying the offset and walk correction to each TDC channel ranges from 1.5 to 2 ns, depending on the electron energy. A ``cluster time'' is calculated for each cluster as the energy-weighted average of corrected hit times for all unique TDC hits in the cluster. Figure \ref{bigcaltimeres} shows the resolution achieved for an electron energy of $E_e \approx 2.3$ GeV. A loose cut was placed on this cluster time in the analysis to reject clusters that were out-of-time in favor of clusters with good timing. 

In addition to the first-level timing signals, ADC and TDC signals were recorded for each of the 38 second-level trigger sums. The TDC hits of the second level sums responsible for the presence of the trigger always arrive at a fixed time relative to the BigCal trigger. This is because the discriminators used on the front end to form the BigCal trigger are also used to produce the timing signals for the second-level sums, meaning these signals are only present for channels which exceed the trigger threshold. For this reason, requiring the presence of one or more second-level TDC hits in time with the BigCal trigger and in the trigger logic group(s) associated with a given cluster serves as a very powerful way to select the cluster corresponding to the shower responsible for the presence of a BigCal trigger, which significantly reduces the multiplicity of clusters found per event under high-rate conditions. For this reason, the second level timing signals proved even more useful than the first level timing signals in this important respect.
\section{Elastic Event Selection}
\label{elasticeventselectionsection}
\paragraph{}
After the reconstruction of each event is complete, the next task is to separate elastic and inelastic scattering events. Although it is possible, in principle, to separate elastic and inelastic events based on the reconstructed proton momentum $p_p$ and scattering angle $\theta_p$ alone, the resolution of the HMS is insufficient to achieve a clean separation. In other words, inelastic reactions have significant overlap with elastic scattering within the resolution of the HMS, especially at large $Q^2$ values. By detecting the scattered electron in coincidence and measuring its scattering angles with a resolution comparable to or in some cases exceeding the resolution of the HMS for the variables considered, the cleanest possible separation is achieved. The elastic-inelastic separation provided by BigCal turns out to be most important for the highest-$Q^2$ data point. 
\subsection{Proton $\delta$-$\theta$ Correlation}
\paragraph{}
For a given beam energy $E_e$, the scattering angle $\theta_p$ and momentum $p_p$ of the recoiling proton in elastic scattering are related by equation \eqref{pel_hthe}. The difference between the measured momentum and the momentum predicted by \eqref{pel_hthe} at the measured angle therefore defines the degree of ``inelasticity'' of a given event. This difference is subsequently referred to as $\Delta_p \equiv 100 \times \frac{p-p(\theta_p)}{p_0}$, which is expressed as a percentage of the HMS central momentum for comparison with the nominal resolution of the HMS. Elastic events show up as a prominent peak at $\Delta_p = 0$, with a width determined by the HMS momentum and angular resolution, the beam energy, and the HMS central angle. 
\begin{figure}[h]
  \begin{center}
    \subfigure[$Q^{2}$ = 5.2 GeV$^2$]{\label{pmissp52}\includegraphics[angle=90,width=.49\textwidth]{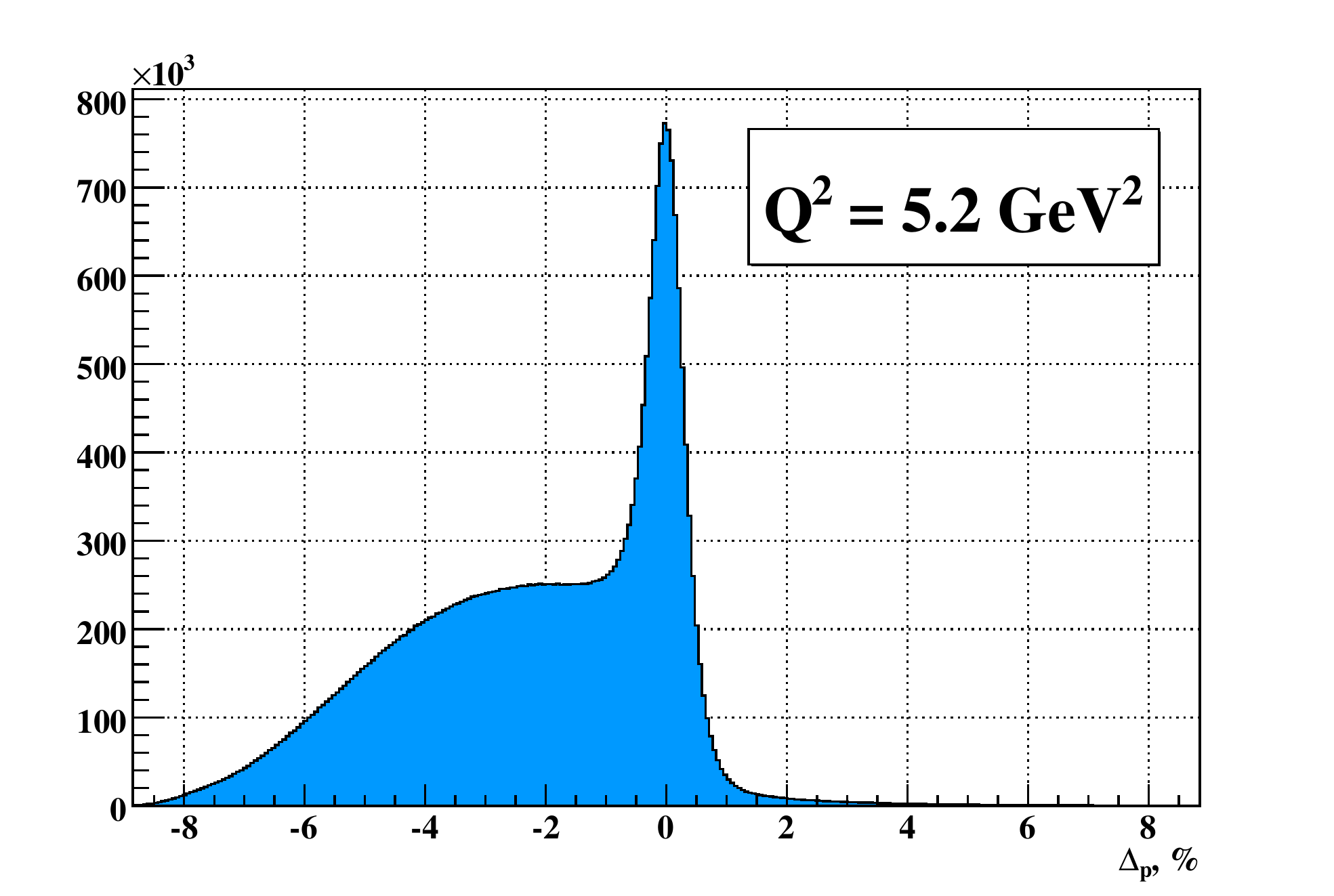}}
    \subfigure[$Q^{2}$ = 6.7 GeV$^2$]{\label{pmissp67}\includegraphics[angle=90,width=.49\textwidth]{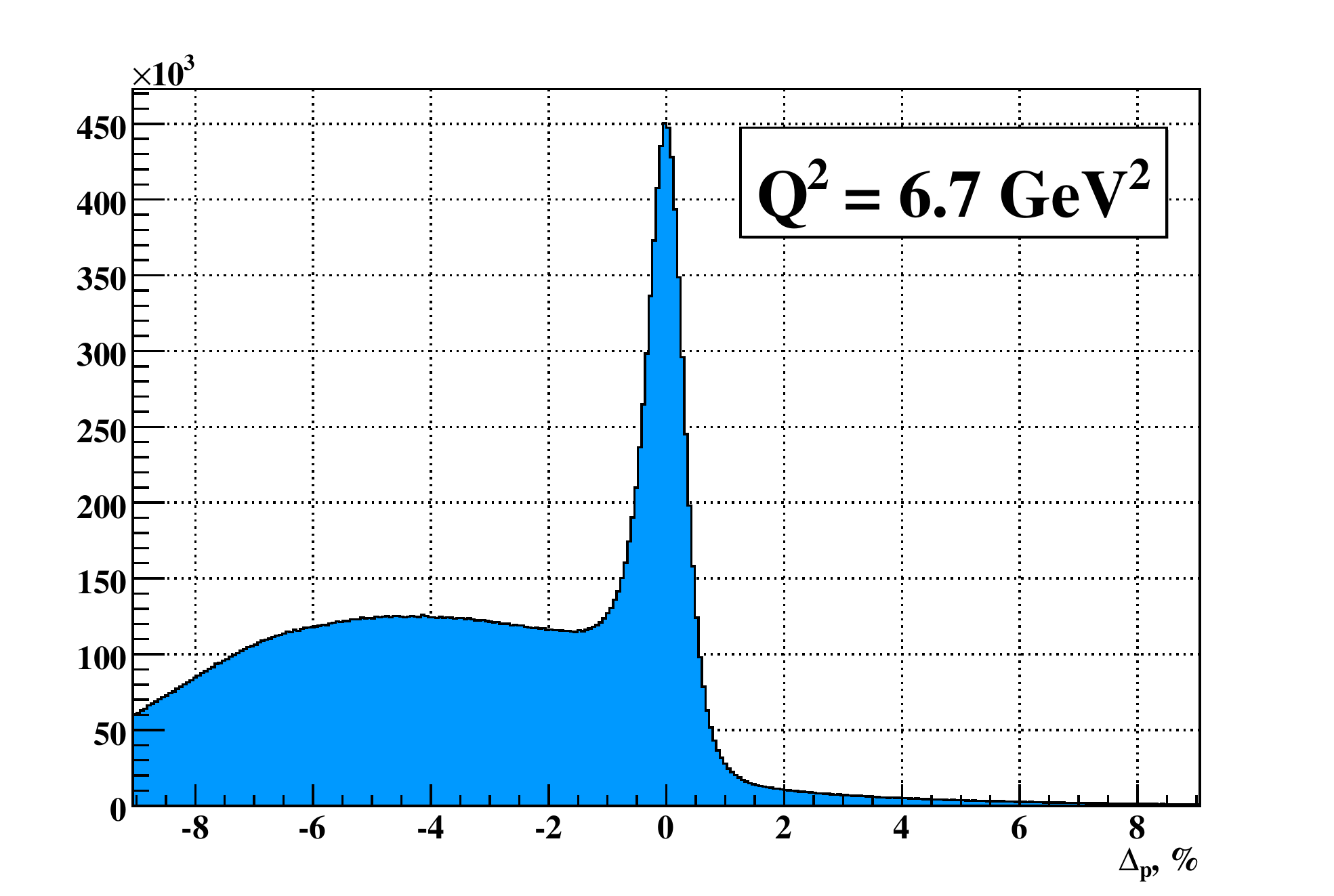}}
  \end{center}
  \caption{\label{pmissp52_67} $\Delta_p$ spectrum for $Q^2=5.2$ GeV$^2$ (a) and $Q^2=6.7$ GeV$^2$ (b), before applying cuts on BigCal.}
\end{figure}
Figure \ref{pmissp52_67} shows the $\Delta_p$ spectrum in the region near the elastic peak for two different kinematic settings. For both of these data points, the elastic peak is clearly visible together with a significant inelastic background, represented by the shoulder on the left side of the peak in each plot. The ``super-elastic'' events on the right side of the peak have momenta exceeding the expected proton momentum for elastic scattering. These events come primarily from quasi-elastic $Al(e,e'p)$ reactions in the target endcaps and contribute a small fraction of the total non-elastic background. 

The ``sub-elastic'' events on the left side of the peak have lower momenta than the expected momentum for elastic scattering. These events are primarily protons from neutral pion photoproduction reactions initiated by hard Bremsstrahlung photons in the 2.3\% radiation-length hydrogen target. When a beam electron radiates a real photon at or near the full beam energy, the exclusive reaction $\gamma + p \rightarrow \pi^0 + p$ has two-body kinematics quite similar to elastic $ep$ scattering, when the radiated photon energy is large compared to the $\pi^0$ mass. The pion immediately decays to two photons $\pi^0 \rightarrow \gamma + \gamma$. In the $\pi^0$ rest frame, the decay photons emerge back-to-back, at a random, uniformly distributed angle, and each photon has an energy exactly equal to half the $\pi^0$ mass. Upon boosting this isotropic, monoenergetic decay photon distribution to the lab frame, in which the pion is produced with a large momentum determined by the two-body kinematics of the reaction, both the angle and energy distributions are folded forward along the pion trajectory, resulting in a significant probability for one or both photons to hit BigCal with enough energy to pass the trigger threshold. Since electron-induced and photon-induced showers give identical signals in BigCal, the only way to distinguish between pion and elastic events is based on the energy and position of the detected particle. In the end, only the position of the detected cluster provided for a meaningful separation between elastic and $\pi^0$ events due to the poor energy resolution of BigCal. 
\begin{table}[h]
  \begin{center}
    \begin{tabular}{|c|c|c|c|}
      \hline $E_{beam}$, GeV & $p_0$, GeV/c & $\theta_p$,$^\circ$ & $p_{ep\rightarrow ep}(\theta_p) - p_{\gamma p \rightarrow \pi^0 p}(\theta_p)$, MeV (\% of $p_0$) \\ \hline
      1.87 & 2.0676 & 14.50 & 12.21 (0.590) \\ \hline
      2.85 & 2.0676 & 31.00 & 9.06 (0.438) \\ \hline
      3.68 & 2.0676 & 36.10 & 7.41 (0.359) \\ \hline
      4.05 & 3.5887 & 17.94 & 9.37 (0.261) \\ \hline
      5.71 & 4.4644 & 19.07 & 8.21 (0.184) \\ \hline 
      5.71 & 5.4070 & 11.59 & 9.53 (0.176) \\ \hline
    \end{tabular}
  \end{center}
  \caption{\label{pdiff_pi0elastic} Momentum difference between elastically scattered protons and protons from $\gamma + p \rightarrow \pi^0 + p$, for $E_\gamma = E_{beam}$, evaluated at the central kinematics of experiments E04-108 and E04-019.}
\end{table}

Table \ref{pdiff_pi0elastic} shows the difference in proton momentum between elastic $ep$ scattering and $\pi^0$ photoproduction, calculated from two-body kinematics at the central HMS angle, for a radiated photon with the full beam energy. The differences are on the order of 10 MeV for all the kinematics of both experiments. As a percentage of the HMS central momentum, the differences range from approximately 0.18\% to 0.6\%. Given the combined angular and momentum resolution of the HMS, the small ``threshold'' for pion photoproduction in terms of $\Delta_p$ allows for significant overlap with elastic scattering for all kinematics. The smallest percentage difference occurs at $Q^2 = 8.5$ GeV$^2$. Unless very tight cuts are applied on the inelastic side of the peak, all kinematics will suffer some contamination from the $\pi^0$ background. The amount of remaining background depends on the relative rates of the two reactions. The pion photoproduction rate involves the convolution of the Bremsstrahlung flux near endpoint ($E_\gamma \approx E_{beam}$) and the photoproduction cross section. 

Another reaction contributing to the non-elastic background is real Compton scattering ($\gamma + p \rightarrow \gamma + p$), which is also initiated by hard Bremsstrahlung photons radiated from the primary beam. Unlike pion photoproduction, the kinematics for Compton scattering at endpoint ($k=E_{beam}$) are identical to elastic $ep$ scattering since the photon is massless, and whereas the random angle distribution of the $\pi^0$ decay photons tends to destroy the kinematic correlation between the particle detected in BigCal and the proton in the HMS, Compton-scattered photons are kinematically correlated with the scattered proton in exactly the same way as elastically scattered electrons. Only the energy difference between the incident Bremsstrahlung photon and the primary electron beam separates Compton scattering from elastic $ep$ scattering. Although it would have been possible to distinguish photons from electrons by placing either a gas Cerenkov detector or a deflecting magnet in front of BigCal, no such measures were adopted for this experiment, making the Compton background essentially irreducible. Generally speaking, the cross section for real Compton scattering is much smaller than for pion photoproduction and is of secondary concern for this analysis. The worst-case kinematic setting in this experiment for both Compton scattering and pion photoproduction backgrounds was $Q^2 = 8.5$ GeV$^2$, where the elastic scattering cross section was small enough that the ratio of background events to signal events within the applied elastic kinematic cuts was significant.

Pion photoproduction and Compton scattering kinematics are characterized by the Mandelstam invariants $s$ and $t$. At $Q^2=8.5$ GeV$^2$, the central kinematics give $s = 11.6$ GeV$^2$ and $-t = Q^2 = 8.5$ GeV$^2$ (for $E_\gamma = E_{beam}$). Pion photoproduction cross section data exist over a wide range in $s$, but the $-t$ range of the data is rather limited. The highest available momentum transfer measured for $\gamma p \rightarrow \pi^0 p$ is approximately 7.5 GeV$^2$ for a photon energy of 5 GeV\cite{Anderson1976}, or $s \approx 10.3$ GeV$^2$. For Compton scattering, high momentum transfer data have recently become available\cite{Danagoulian2007} for $s$ up to 11 GeV$^2$ and $t$ up to 6.5 GeV$^2$. The highest momentum transfer data point for the differential cross section for Compton scattering from \cite{Danagoulian2007} was $d\sigma / dt = 5.6$ pb/GeV$^2$ at $s = 10.9$ GeV$^2$ and $-t = 6.5$ GeV$^2$, whereas for $\pi^0$ photoproduction\cite{Anderson1976}, a result of $d\sigma / dt = 8.4$ nb/GeV$^2$ was obtained at $s=10.3$ GeV$^2$ and $-t=7.5$ GeV$^2$. Though the kinematics of these two data points are not identical to each other or to E04-108 kinematics, one immediately sees that the Compton cross section at the nearest available data point is approximately three orders of magnitude smaller than the photoproduction cross section. The smallness of the Compton cross section relative to the photoproduction cross section was also confirmed in another experiment at lower momentum transfers in which both cross sections were measured simultaneously\cite{Shupe1979} for photon energies from 2 to 6 GeV and momentum transfers up to 4 GeV$^2$, and found ratios $R(s,t) = (d\sigma/dt)_{\gamma p \rightarrow \gamma p}/(d\sigma / dt)_{\gamma p \rightarrow \pi^0 p}$ never exceeding 5\%. 

The authors of \cite{Danagoulian2007} found a scaling behavior of $s^{-8}$ for the Compton scattering cross section at fixed momentum transfer, to be contrasted with the approximate $s^{-7}$ scaling of the photoproduction cross section found in \cite{Anderson1976}. Although it was not possible to actually measure the Compton background contribution from the data of this experiment, a reasonable estimate was obtained by extrapolating from the data of \cite{Danagoulian2007} using the empirical $s^{-8}$ scaling behavior of $d\sigma / dt$, which showed that the Compton scattering background, while irreducible, was safely below the level where it could significantly impact the form factor ratio results.
\begin{figure}[h]
  \begin{center}
    \includegraphics[angle=90,width=.99\textwidth]{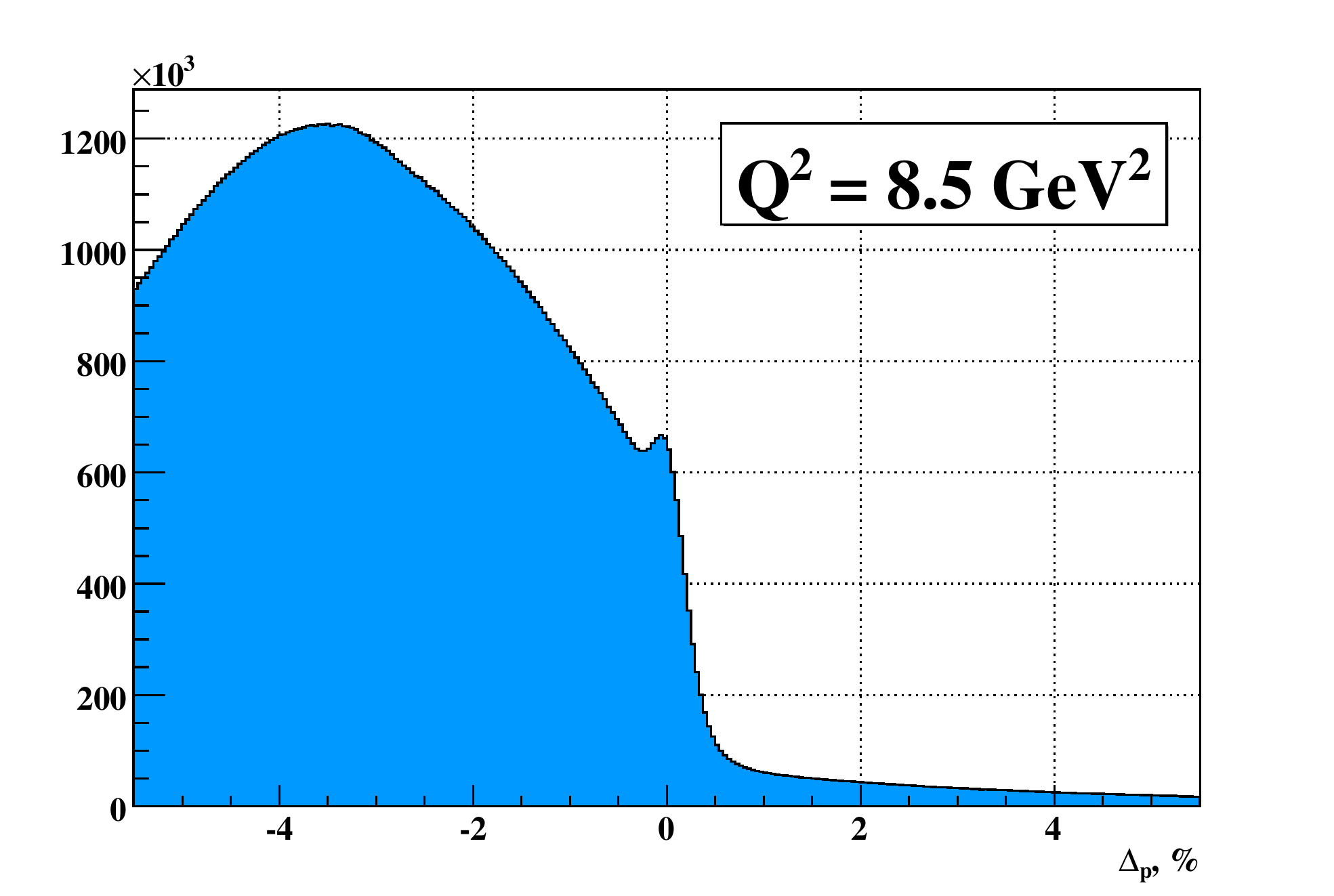}
  \end{center}
  \caption{\label{pmisspraw85} $\Delta_p$ spectrum at $Q^2 = 8.5$ GeV$^2$. No cuts applied on BigCal.}
\end{figure}

The $\pi^0$ background, on the other hand, was large enough to have a significant impact on the analysis of the data at $Q^2 = 8.5$ GeV$^2$. Figure \ref{pmisspraw85} shows the $\Delta_p$ spectrum in the region near the elastic peak for this setting, before applying any cuts on the detected electron in BigCal. In contrast to the data at lower $Q^2$, the backgrounds nearly overwhelm the signal in this case, with the elastic peak barely registering as a small bump at $\Delta_p=0$, on top of a significant $\pi^0$ background. The relatively high level of events in the super-elastic region indicates that there is also a contribution from scattering in the target walls, which is to be expected with the HMS at the relatively forward angle of 11.6$^\circ$. The situation illustrated in figure \ref{pmisspraw85} demonstrates the necessity of using BigCal to detect the scattered electron in order to isolate the elastic events.
\subsection{Electron-Proton Kinematic Correlation}
\paragraph{}
The scattering angles of the electron are measured precisely using the ray from the interaction point reconstructed by the HMS to the position of the detected electron at the surface of BigCal (see table \ref{eth_res_table}). Since the proton momentum is the most precisely measured kinematic variable by the HMS and the electron angle $\theta_e$ is the most precisely measured variable by BigCal, the elastic peak is separated by comparing the measured proton momentum to the expected proton momentum for elastic kinematics of the measured electron angle. The transverse coordinates of the interaction vertex are simply the beam positions measured by the BPM and raster signals. The $z$ coordinate of the interaction vertex is given in terms of $y_{tar}$ and $y'_{tar}$ measured by the HMS by equation \eqref{zbeam_formula}. The $(x,y,z)_{e}$ coordinates of the electron at the surface of BigCal are given in terms of the reconstructed cluster coordinates $(x,y)_{clust}$ and the BigCal distance and angle $R_{cal}$ and $\Theta_{cal}$ by 
\begin{eqnarray}
  x_e &=& x_{clust} \cos \Theta_{cal} + R_{cal} \sin \Theta_{cal} \\ 
  y_e &=& y_{clust} \\
  z_e &=& -x_{clust} \sin \Theta_{cal} + R_{cal} \cos \Theta_{cal}
\end{eqnarray}
The electron scattering angles $\theta_e$ and $\phi_e$ are then calculated as follows
\begin{eqnarray}
  L &=& \sqrt{(x_e - x_{beam})^2 + (y_e - y_{beam})^2 + (z_e - z_{beam})^2} \\
  \cos \theta_e &=& \frac{z_e - z_{beam}}{L} \\
  \tan \phi_e &=& \frac{y_e - y_{beam}}{x_{e}-x_{beam}}
\end{eqnarray}
If the reaction is elastic scattering, the electron angle and the proton momentum are related by 
\begin{eqnarray}
  E'_e &=& \frac{E_e}{1+\frac{E_e}{M_p}(1-\cos \theta_e)} \\
  T_p(\theta_e) &=& E_e - E'_e = \frac{E_e}{1+ \frac{M_p}{E_e (1-\cos \theta_e)}} \\
  p_p(\theta_e) &=& \sqrt{T_p^2 + 2M_pT_p}
\end{eqnarray}
and the electron and proton must be coplanar, i.e., $\phi_e = \phi_p + \pi$. 

A momentum difference $\Delta_e \equiv 100 \times \frac{p_p - p_p(\theta_e)}{p_0}$ is defined using the same convention as $\Delta_p$ defined above, but replacing $p_p(\theta_p)$ with $p_p(\theta_e)$. For all of the high $Q^2$ kinematics, the Jacobian of the reaction is large enough that the HMS momentum resolution dominates the resolution of $\Delta_e$, resulting in a narrower elastic peak in this variable than in $\Delta_p$, and also providing a measure of the HMS momentum resolution. Furthermore, the broad opening angle distribution of the decay photons from $\pi^0$ photoproduction events tends to spread out the $\Delta_e$ distribution of the background and reduce the amount of overlap with the elastic peak. This is in stark contrast to the protons from $\pi^0$ photoproduction events, which have almost zero separation in $\Delta_p$ from elastically scattered protons for endpoint Bremsstrahlung photons. Figure \ref{pmisserawlowerQ2} shows the $\Delta_e$ distributions at $Q^2$ of 5.2 and 6.7 GeV$^2$. Figure \ref{pmisseraw85} shows the $\Delta_e$ spectrum at $Q^2 = 8.5$ GeV$^2$, before applying cuts to any other variables.
\begin{figure}[h]
  \begin{center}
    \subfigure[$Q^2=5.2$ GeV$^2$]{\label{pmisseraw52} \includegraphics[angle=90,width=.48\textwidth]{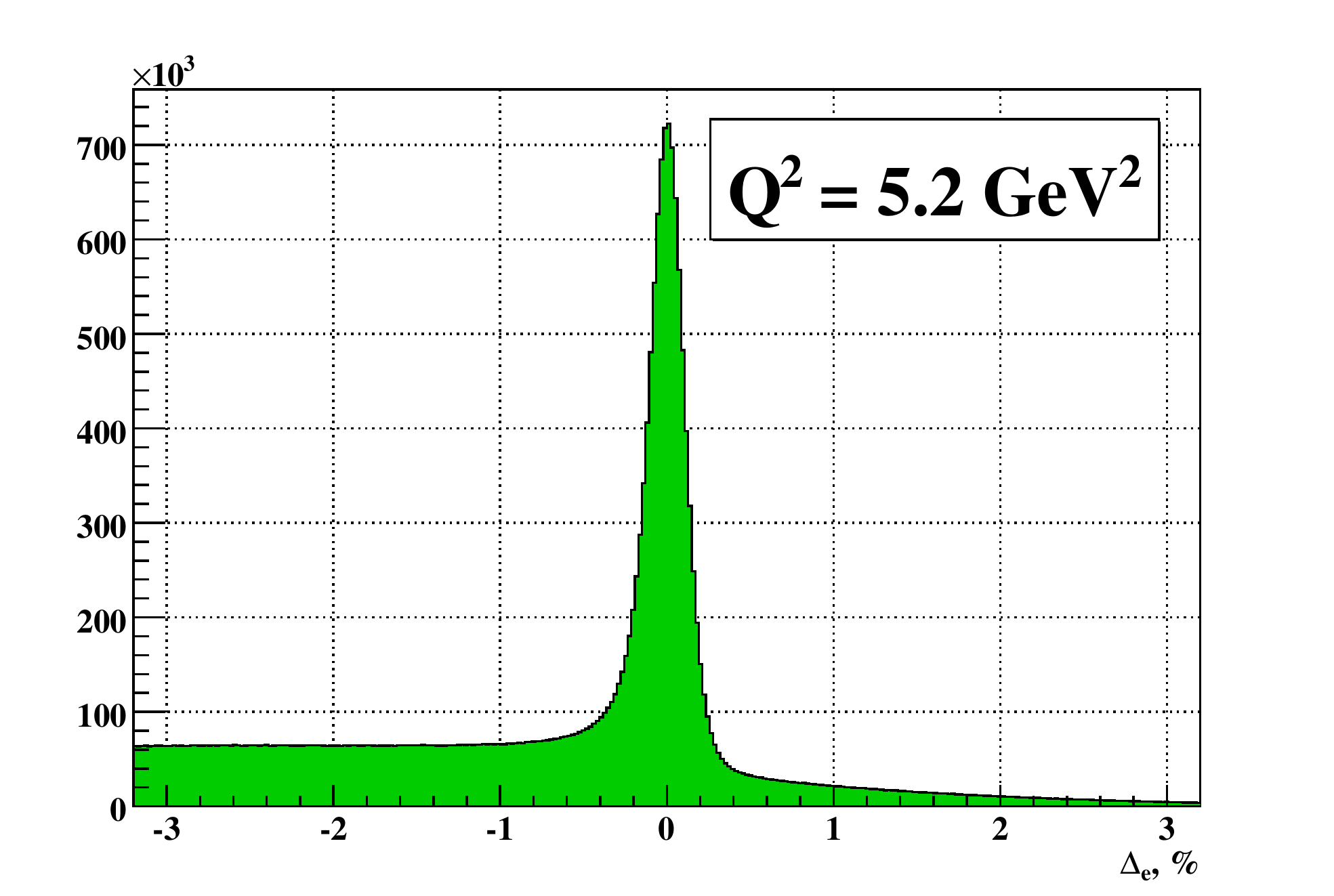}}
    \subfigure[$Q^2=6.7$ GeV$^2$]{\label{pmisseraw67} \includegraphics[angle=90,width=.48\textwidth]{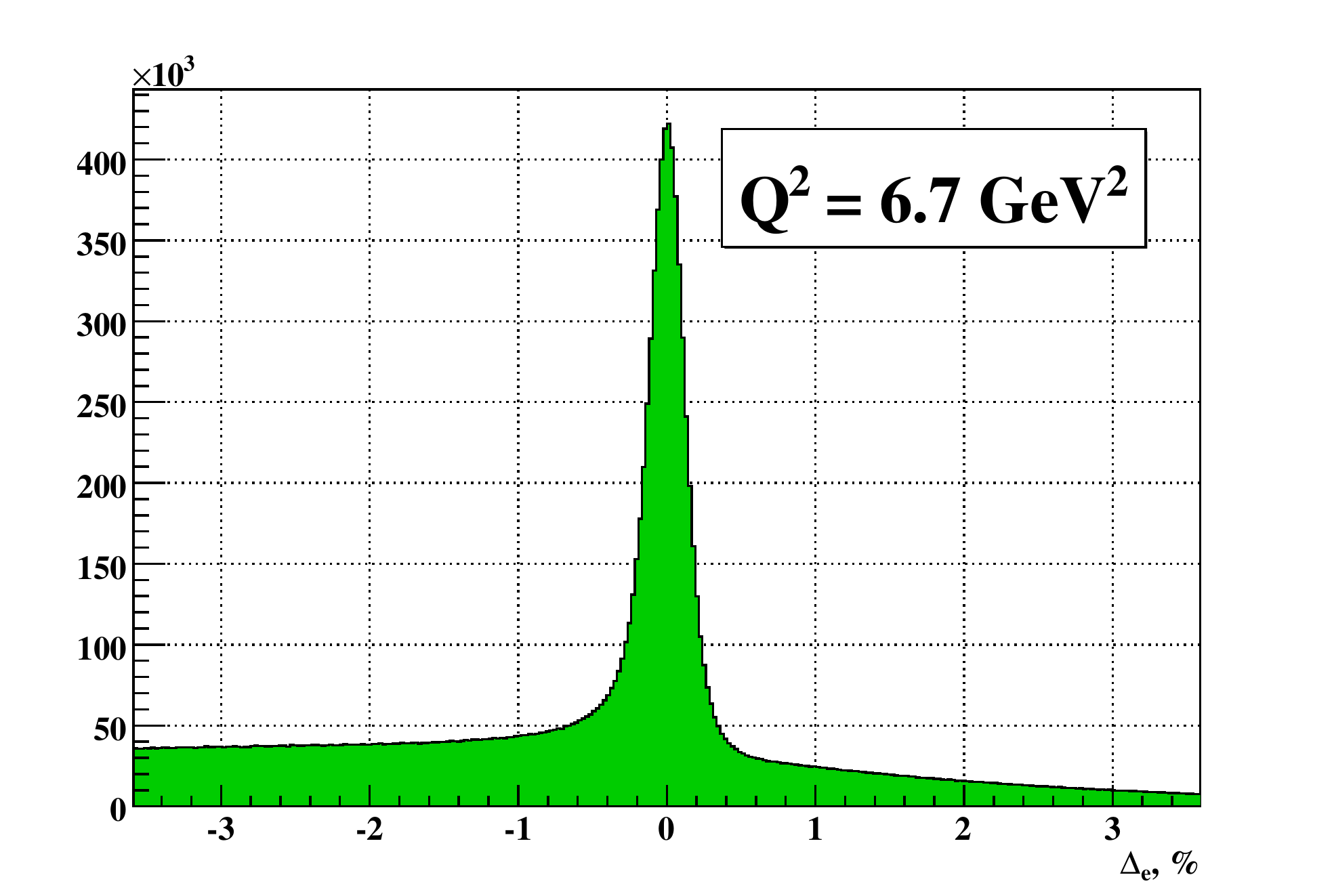}}
  \end{center}
  \caption{\label{pmisserawlowerQ2} $\Delta_e$ distribution at $Q^2=5.2$ and $6.7$ GeV$^2$, before applying any coplanarity cuts.}
\end{figure}

\begin{figure}[h]
  \begin{center}
    \includegraphics[angle=90,width=.99\textwidth]{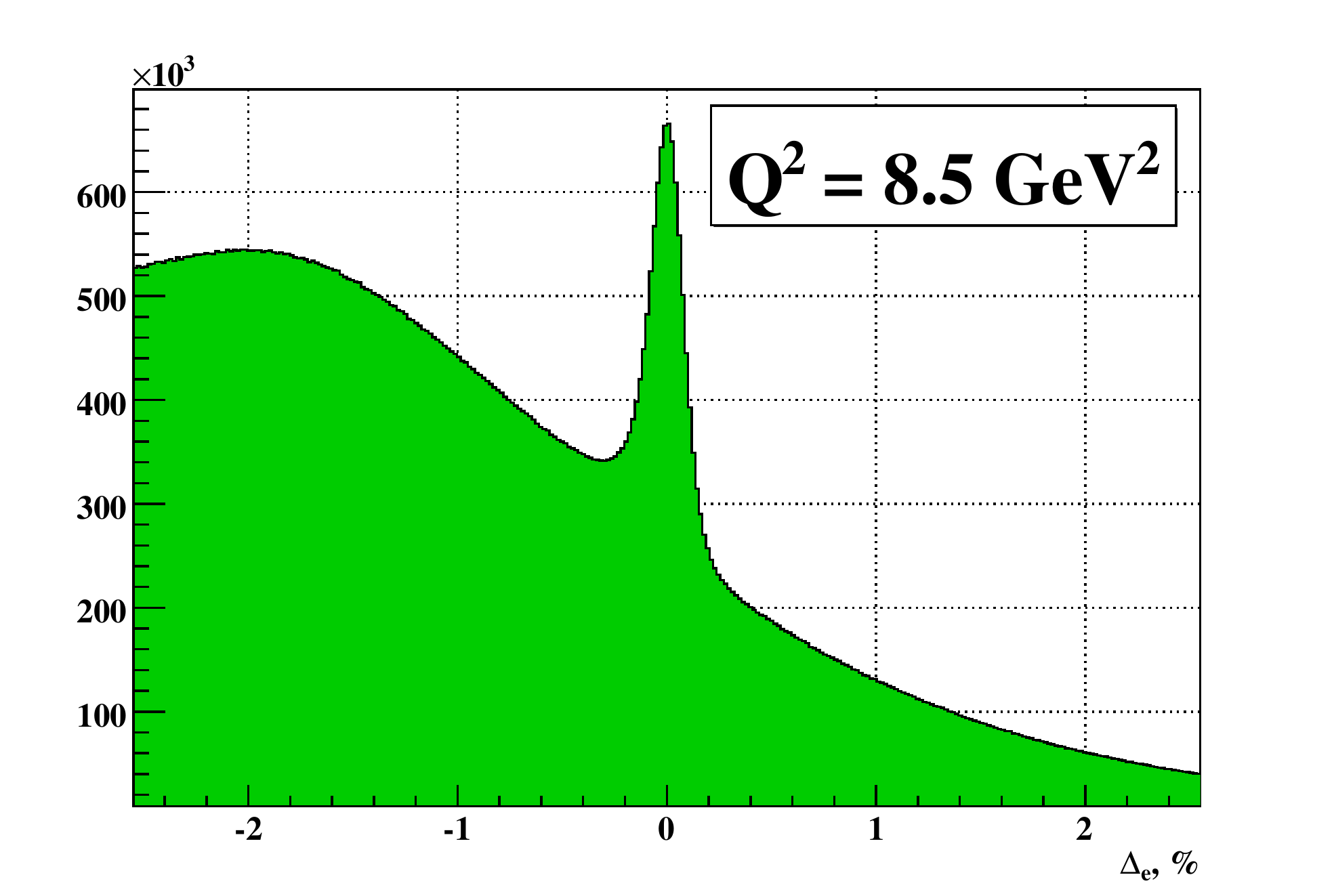}
  \end{center}
  \caption{\label{pmisseraw85} $\Delta_e$ spectrum at $Q^2=8.5$ GeV$^2$. No cuts applied on $\Delta_p$ or coplanarity.}
\end{figure}
The elastic peak is now clearly visible. Though it still rests on top of a formidable $\pi^0$ background, the signal-to-background ratio is much improved compared to the $\Delta_p$ case, even \emph{before} applying any $\Delta_p$ or coplanarity cuts.

The last piece of the puzzle is the coplanarity between the electron detected in BigCal and the proton detected in the HMS. The ``acoplanarity'' is defined as $\Delta \phi \equiv \phi_e - \phi_p - \pi$. The coordinate resolution of BigCal is approximately the same in both directions, such that the out-of-plane angle of the electron is measured with significantly better resolution than that of the proton, which is measured by the HMS with resolution degraded by S0. The proton azimuthal angle was defined in equation \eqref{phipdef}. For the central trajectory ($x'_{tar} \approx y'_{tar} \approx 0$), the resolution in $\phi_p$ is a function of the resolution in $x'_{tar}$, and varies with the HMS central angle as follows:
\begin{eqnarray}
  \phi_p &=& \arctan \left(\frac{y'_{tar} \cos \Theta_{HMS} - \sin \Theta_{HMS}}{x'_{tar}}\right) \\
  \frac{\partial \phi_p}{\partial x'_{tar}} &=& -\frac{y'_{tar} \cos \Theta_{HMS} - \sin \Theta_{HMS}}{x_{tar}'^2 + (y'_{tar} \cos \Theta_{HMS} - \sin \Theta_{HMS})^2} \\
  \Rightarrow d\phi_p &\xrightarrow[x'_{tar},y'_{tar} \rightarrow 0]{}& -\frac{dx'_{tar}}{\sin \Theta_{HMS}}
\end{eqnarray}
\begin{figure}[h]
  \begin{center}
    \subfigure[$Q^2 = 6.7$ GeV$^2$]{\label{dphiraw67}\includegraphics[angle=90,width=.48\textwidth]{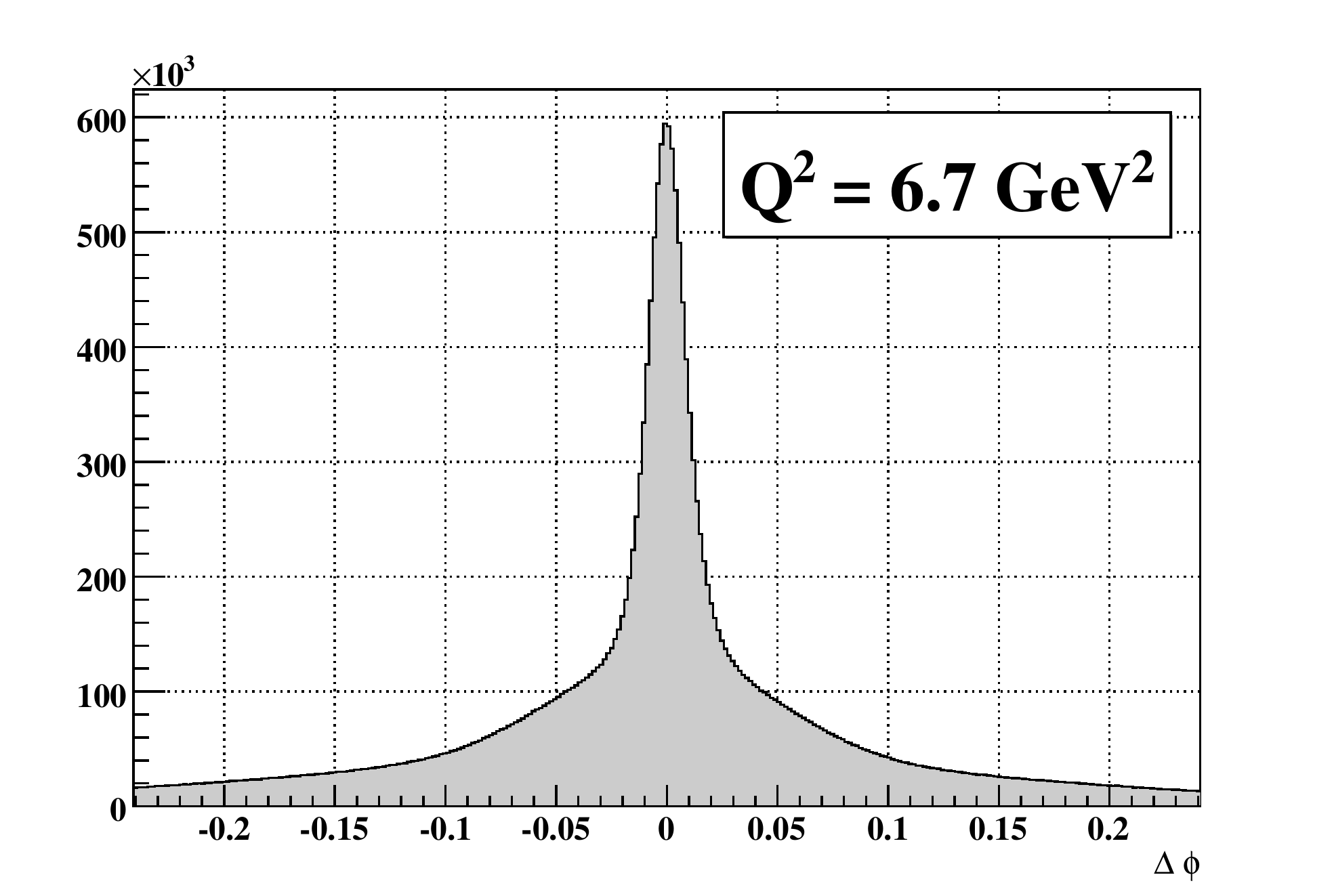}}
    \subfigure[$Q^2 = 8.5$ GeV$^2$]{\label{dphiraw85}\includegraphics[angle=90,width=.48\textwidth]{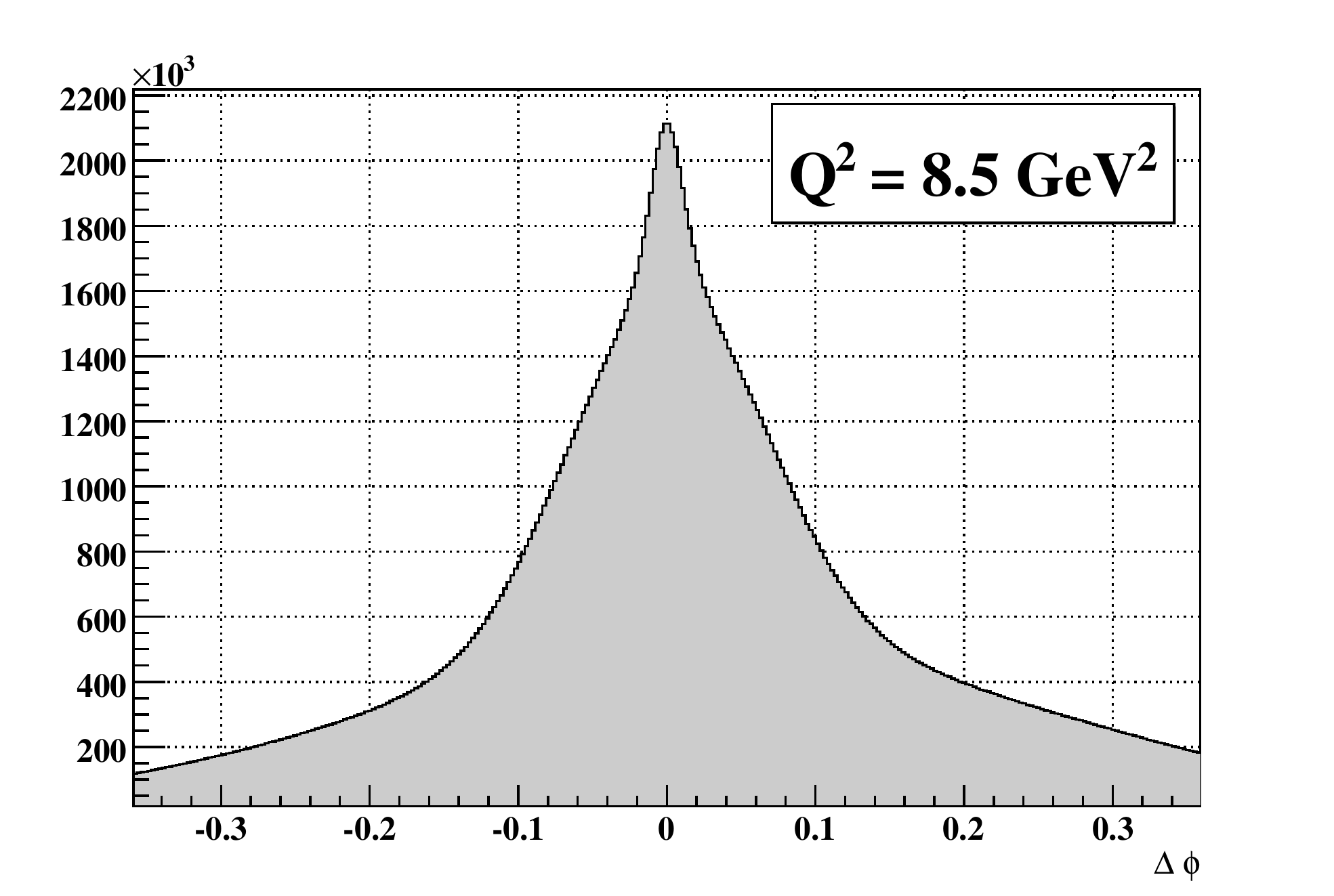}}
  \end{center}
  \caption{\label{dphiraw} $\Delta \phi$ with no cuts applied, for $Q^2 = 6.7$ and 8.5 GeV$^2$.}
\end{figure}
Figure \ref{dphiraw} shows the distribution of $\Delta \phi$, before applying any cuts, at $Q^2$ of 6.7 GeV$^2$ (\ref{dphiraw67}) and 8.5 GeV$^2$ (\ref{dphiraw85}). One immediately notices that the resolution in $\Delta \phi$ is insufficient to provide meaningful elastic-inelastic separation on its own. However, when combined with inelasticity cuts $\Delta_e$ and $\Delta_p$, the coplanarity cut provides significant additional background suppression.

To select elastic events, cuts are applied to the momentum differences $\Delta_e$ and $\Delta_p$, and the azimuthal angle difference $\Delta \phi$. While there are different ways of defining the cuts used to select elastic events, all of them are essentially equivalent to the method outlined above, as the three inelasticity variables defined in the preceding discussion exhaust all possible information about the reaction kinematics obtained from the reconstruction of the HMS and BigCal, except for the measured energy in BigCal. Given the already poor energy resolution of BigCal and the significant probability of large energy losses in the four-inch thick absorber, even a loose missing energy cut did not achieve significant additional background suppression after applying the angular correlation cuts defined above, and was in fact found to reject significant numbers of elastic events. For this reason, no cuts were applied to the cluster energy in BigCal in the final analysis, other than those used in the cluster-finding algorithm discussed in the previous section.

An alternative method of correlating the electron and proton kinematics of particular importance is to compare the detected electron position at BigCal to the expected position calculated from elastic kinematics of the proton detected in the HMS. The details of the calculation were already outlined above in the discussion of the coordinate resolution of BigCal. The coordinate differences are defined as $\Delta x \equiv x_{clust} - x_{HMS}$ and $\Delta y \equiv y_{clust} - y_{HMS}$\footnote{These differences are defined in the BigCal coordinate system, for which $x$ is the horizontal coordinate and $y$ is the vertical coordinate.}. Elastic events are found at $(\Delta x, \Delta y) = (0,0)$ as shown in figure \ref{dxdyraw85}. 

\begin{figure}[h]
  \begin{center}
    \includegraphics[width=.99\textwidth]{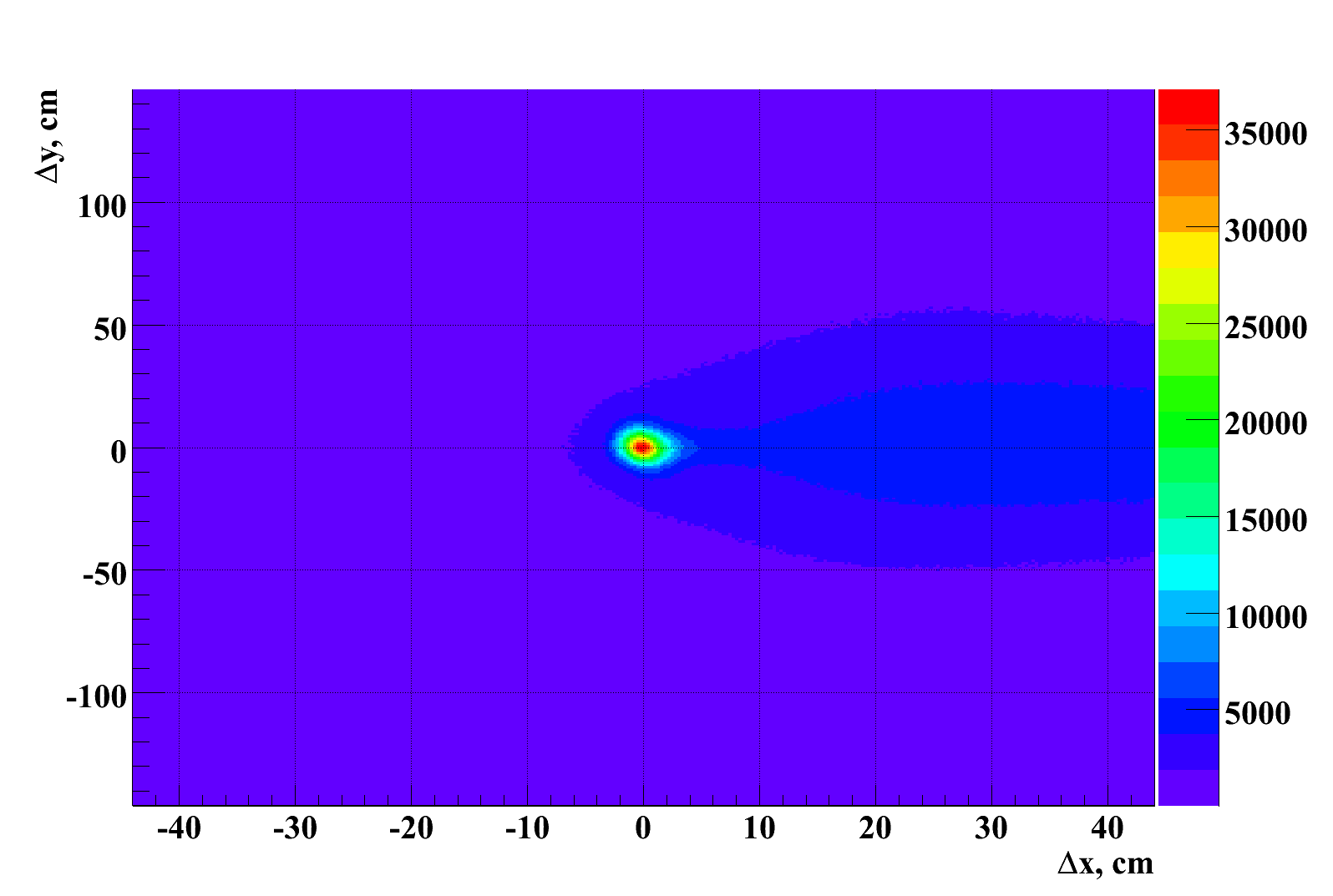}
  \end{center}
  \caption{\label{dxdyraw85} $\Delta y$ versus $\Delta x$, with no $\Delta_p$ cuts applied, for $Q^2=8.5$ GeV$^2$.}
\end{figure}
Figure \ref{dxdyraw85} shows the distribution of $\Delta y$ versus $\Delta x$ for $Q^2 = 8.5$ GeV$^2$. The elastic peak is clearly visible on top of a broadly distributed background of $\pi^0$ decay photons located primarily at positive $\Delta x$, with a $\Delta y$ distribution symmetric about $\Delta y = 0$. In terms of the inelasticity variables defined above, the horizontal coordinate difference $\Delta x$ is roughly equivalent to the momentum difference $\Delta_e$, while the vertical coordinate difference $\Delta y$ is roughly equivalent to the azimuthal angle difference $\Delta \phi$. Events at positive $\Delta x$ correspond to larger deflection angles than the expected $\theta_e$ for elastic scattering and therefore correspond to larger expected proton momenta than the measured proton momentum; i.e., $+\Delta x$ corresponds to $-\Delta_e$.
\begin{figure}[h]
  \begin{center}
    \subfigure[$\Delta x$ projection at $\Delta y = 0$]{\label{xproject}\includegraphics[angle=90,width=.48\textwidth]{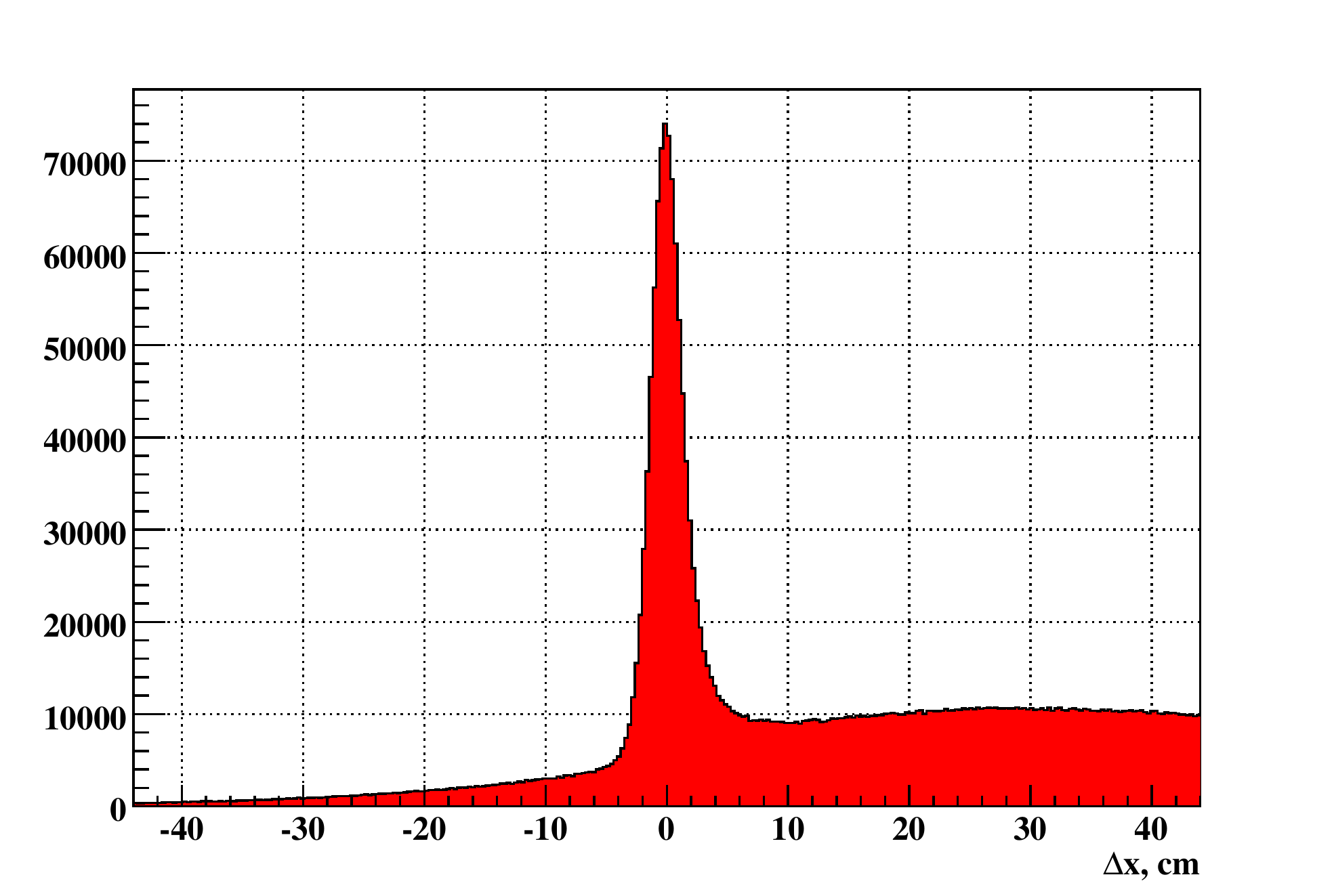}}
    \subfigure[$\Delta y$ projection at $\Delta x = 0$]{\label{yproject}\includegraphics[angle=90,width=.48\textwidth]{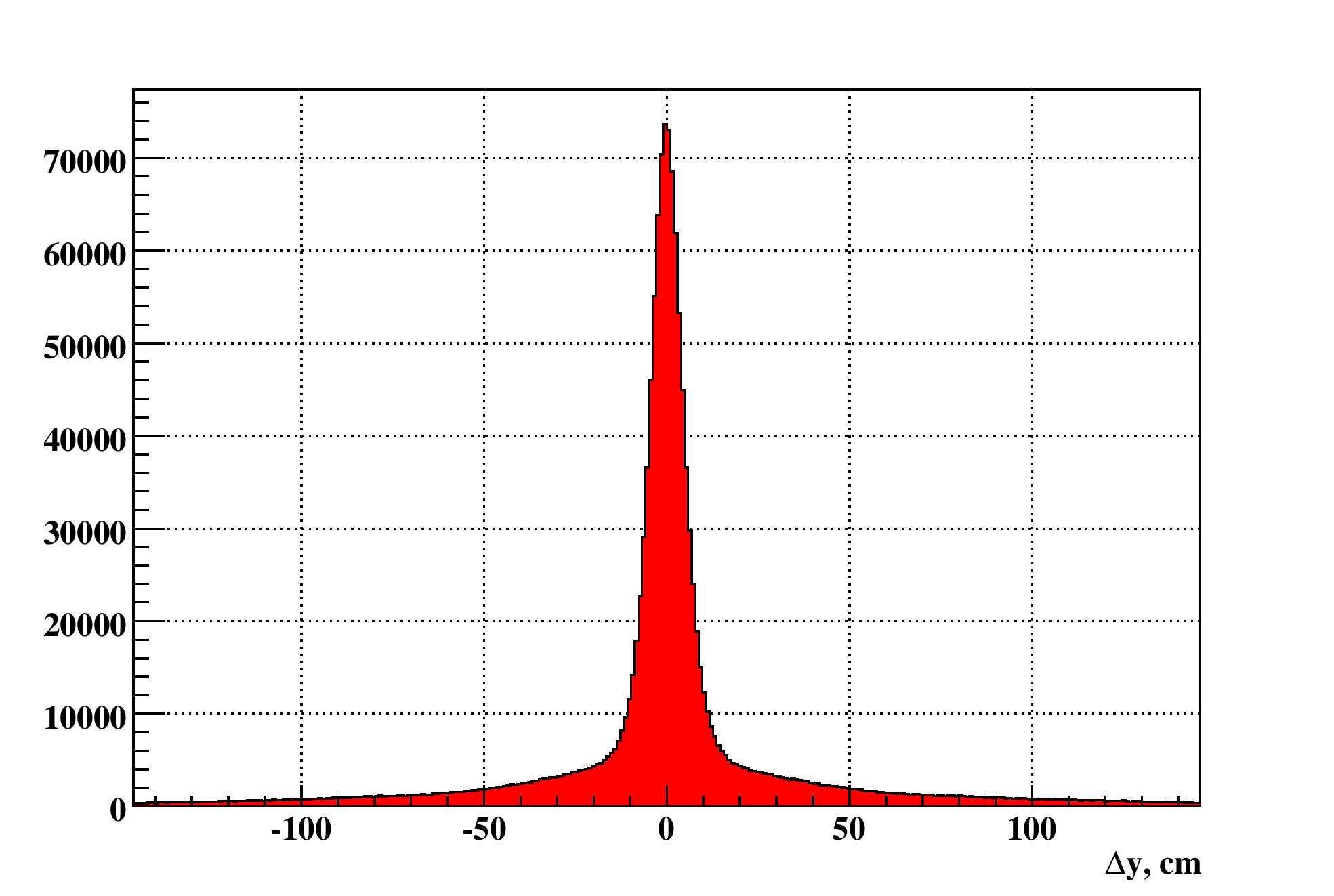}}
  \end{center}
  \caption{\label{dxdyprojections} Projection of the $\Delta y$ vs. $\Delta x$ distribution along the $\Delta x$ axis (a), and along the $\Delta y$ axis (b) for $Q^2 = 8.5$ GeV$^2$.}
\end{figure}
Figure \ref{dxdyprojections} shows the projections of the two-dimensional distribution of figure \ref{dxdyraw85} along the $\Delta x$ axis (\ref{xproject}) and the $\Delta y$ axis (\ref{yproject}). In the two-dimensional correlation plot and its projections, one immediately sees that the level of background events found at the expected scattering angle $\theta_e$ \emph{and} the expected azimuthal angle $\phi_e$ is low, but still significant compared to the number of elastic events. A ``clean'' sample of elastic events is obtained only when cuts are applied to all three inelasticity variables:
\begin{itemize}
\item The proton momentum difference $\Delta_p$
\item The electron momentum difference $\Delta_e$ (or equivalently the horizontal position difference $\Delta x$).
\item The azimuthal angle difference $\Delta \phi$ (or equivalently the vertical position difference $\Delta y$).
\end{itemize}
Comparing figure \ref{yproject} to figure \ref{dphiraw85}, the unambiguous appearance of the elastic peak in the $\Delta y$/$\Delta \phi$ spectrum for events found at the expected $\theta_e$ for elastic scattering is evident. The next section will illustrate the effect of applying these cuts and present estimates of the remaining inelastic background.
\subsection{Elastic Event Selection Cuts}
\label{elasticcutssection}
\paragraph{}
The data were analyzed using both methods of elastic event selection described above. In the first method, cuts were applied to the inelasticity variables $\Delta_e$ and $\Delta_p$, and the acoplanarity $\Delta \phi$. In the second method, which was also used for the final analysis, the $\Delta_e$ and $\Delta \phi$ cuts were replaced by an elliptical cut applied to the $(\Delta x$, $\Delta y)$ distribution:
\begin{eqnarray}
  \sqrt{\left(\frac{\Delta x}{x_{cut}}\right)^2 + \left(\frac{\Delta y}{y_{cut}}\right)^2} &\le& 1 \label{ellipsecutdefinition}
\end{eqnarray}
The second method achieves a better background suppression than the first method because the shape of the cut matches the shape of the elastic peak in the two-dimensional $(\Delta x,\Delta y)$ space, whereas a $(\Delta_e,\Delta \phi)$ cut, which amounts to selecting a section of spherical solid angle centered around the elastic peak, does not. In fact, the latter cut is roughly equivalent to a rectangular cut in the $(\Delta x, \Delta y)$ plot. Compared to a rectangular cut, an elliptical cut throws out events in the corners of the rectangle where the background-to-signal ratio is higher, leading to a cleaner sample of elastic events. 
\begin{figure}[h]
  \begin{center}
    \subfigure[Elliptical $(\Delta x, \Delta y)$ cut.]{\label{ellipsecut85}\includegraphics[width=.48\textwidth]{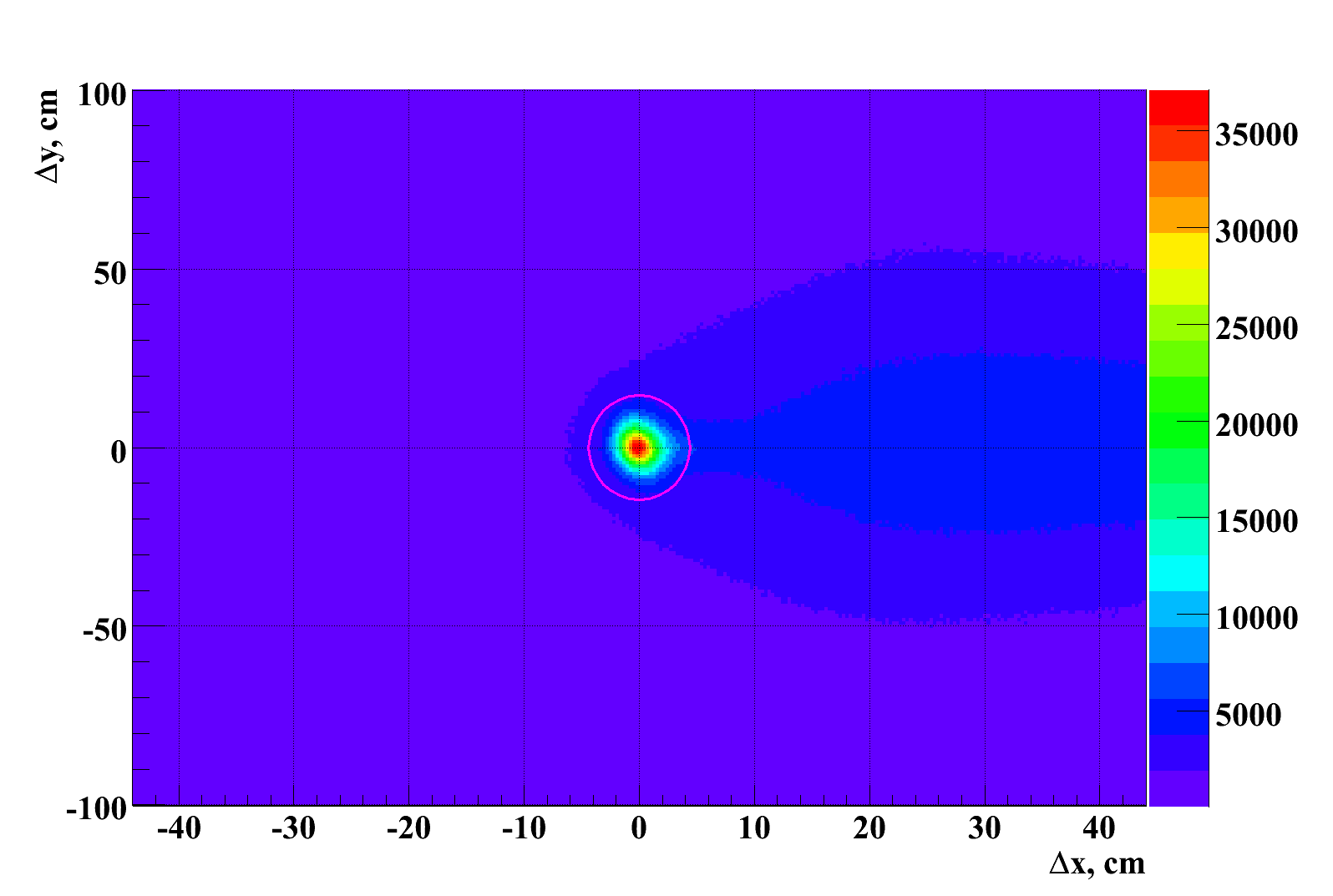}}
    \subfigure[$\Delta_p$ spectrum after applying elliptical cut.]{\label{pmisspcutxy85}\includegraphics[angle=90,width=.48\textwidth]{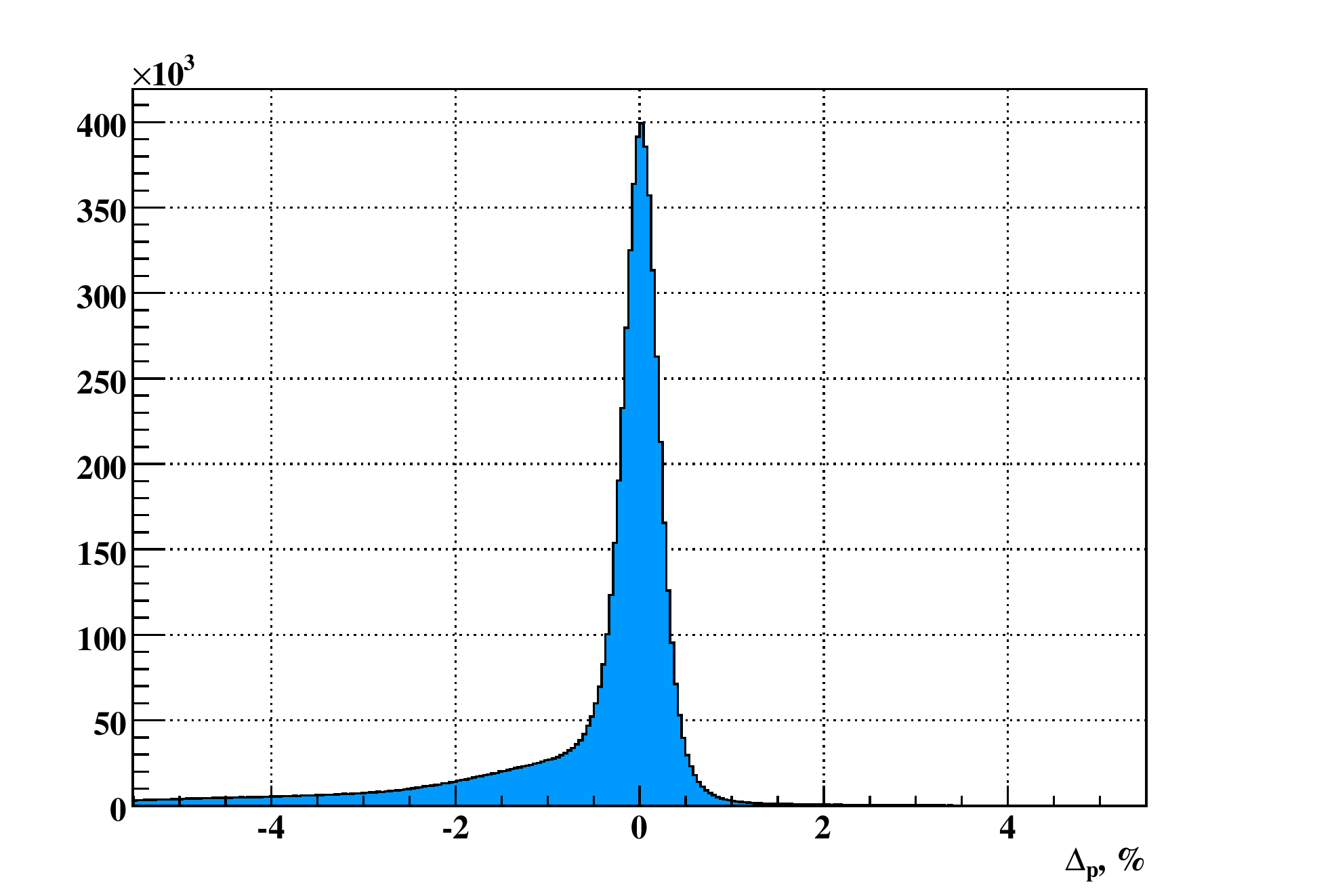}}
  \end{center}
  \caption{\label{ellipse85} Selection of elastic events using elliptical $(\Delta x, \Delta y)$ cut for $Q^2=8.5$ GeV$^2$ (a), and $\Delta_p$ spectrum of events passing elliptical cut.}
\end{figure}
Figure \ref{ellipse85} illustrates the quality of the elastic event selection achieved using the elliptical cut. In figure \ref{ellipsecut85}, an elliptical cut with $(x_{cut}, y_{cut})=(4.4,14.6)$ cm is applied to the $(\Delta x, \Delta y)$ spectrum at $Q^2 = 8.5$ GeV$^2$. The $\Delta_p$ spectrum of events inside the ellipse is shown in figure \ref{pmisspcutxy85}. Compared to the $\Delta_p$ spectrum of all events (figure \ref{pmisspraw85}), a fairly clean selection of elastic events is achieved, with a residual $\pi^0$ background indicated by the tail on the inelastic side of the peak and a near-total suppression of the super-elastic events coming from the target endcaps.

At first glance, it appears as if the events at negative $\Delta_p$ could be part of the elastic radiative tail, but this is not the case. Since $\Delta_p$ and $\Delta_e$ are both calculated from the measured proton momentum and the measured electron/proton angles assuming an incident electron energy equal to the full beam energy, elastic scattering events in which radiation from either the incident beam or the scattered proton occurs will have the effect of lowering the momentum of the scattered proton for a given angle $\theta_p$. This will reduce $\Delta_p$, but it will also reduce $\Delta_e$, so that these events will fail the $\Delta_e$ cut as well. To illustrate how the cuts defined above suppress the elastic radiative tail, the shift in kinematics of elastic scattering caused by the emission of a soft photon of energy $k$ from the incident electron beam was considered. 
\begin{figure}[h]
  \begin{center}
    \includegraphics[angle=90,width=.98\textwidth]{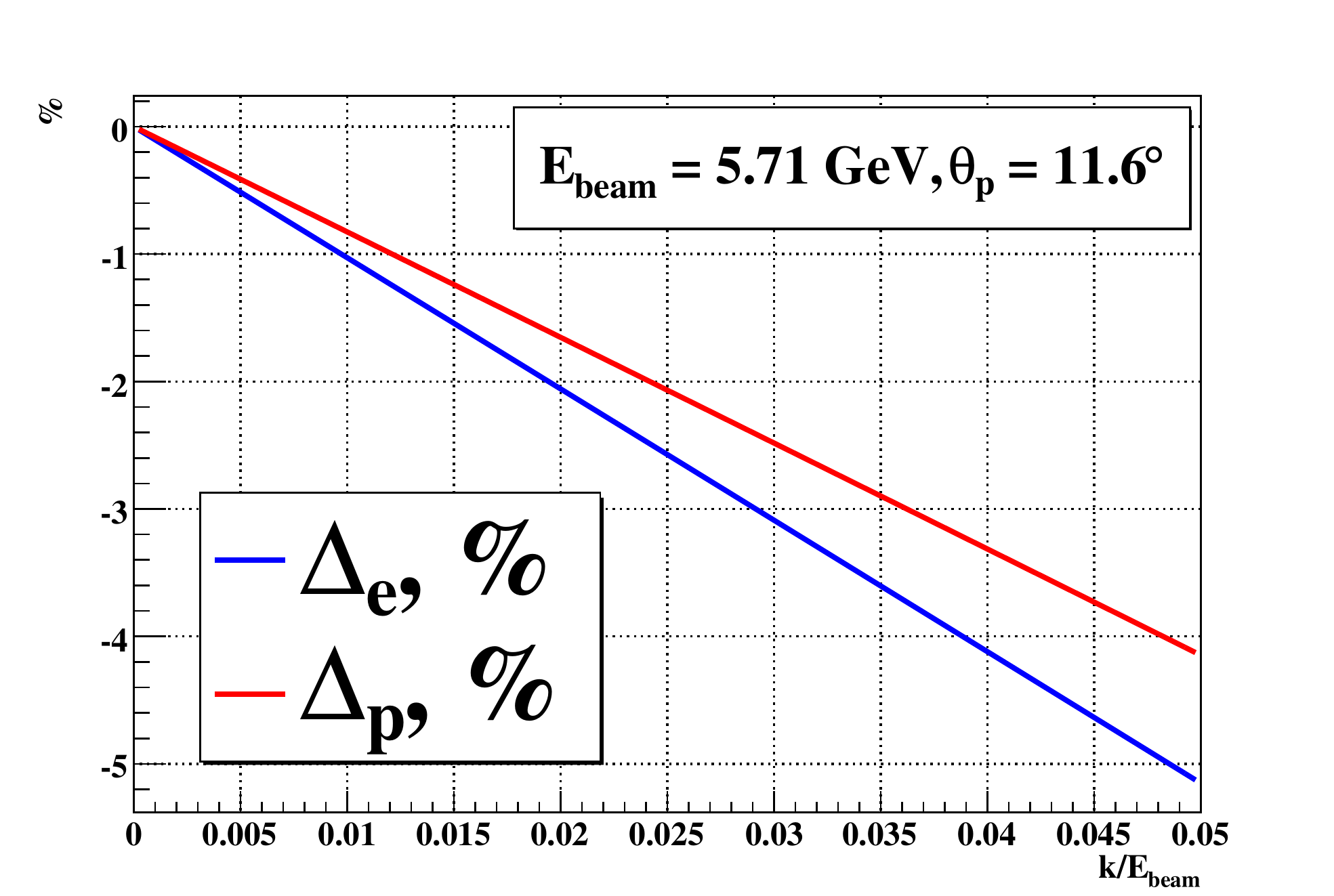}
  \end{center}
  \caption{\label{kinshift} Shift in observed $\Delta_e$ and $\Delta_p$ as a function of the fraction of the beam energy carried away by the (undetected) radiated photon, $Q^2 = 8.5$ GeV$^2$.}
\end{figure}
Assuming a fixed $\theta_p$ and a beam energy given by $E_{rad} = E_{beam} - k$, the true, observed values of the proton momentum $p_p$, the electron scattering angle $\theta_e$, and the electron energy $E'_e$ were calculated for elastic scattering. Then, the inelasticity tests $\Delta_e$ and $\Delta_p$ were calculated from the expected values $p_p(\theta_p)$ and $p_p(\theta_e)$ assuming $k = 0$. Figure \ref{kinshift} shows the shift in both inelasticity variables due to radiation of energy $k$ from the incident beam, as a function of $k/E_{beam}$, at $Q^2=8.5$ GeV$^2$. Both variables decrease with $k/E$ at different rates, with $\Delta_e$ decreasing faster than $\Delta_p$ because of the large Jacobian. An elliptical cut with $\left|\Delta x\right| \le 4.4$ cm is roughly equivalent to a cut of $\left|\Delta_e\right| \le 0.25$\%. For a detector with infinite resolution, a shift in $\Delta_e$ of $-0.25$\% due to radiation from the incident beam corresponds to an even smaller shift in $\Delta_p$ of $-0.20$\%, which is well inside the resolution of the peak (see figure \ref{pmisspcutxy85}). In short, because initial state radiation reduces $\Delta_e$ even faster than $\Delta_p$ for elastic scattering, a tight cut around the elastic peak in $\Delta_e$ has the effect of suppressing the elastic radiative tail at negative $\Delta_p$, implying that the observed tail at (large) negative values of $\Delta_p$ for events passing the elliptical cut comes from inelastic reactions such as $\pi^0$ photoproduction.

In order to suppress these inelastic events, a cut around the elastic peak in $\Delta_p$ is applied. The quality of the background suppression achieved by this cut is evident in figure \ref{xyprojections_cut85}, which shows the projections of the $(\Delta x, \Delta y)$ spectrum after applying a roughly $3\sigma$ cut around the elastic peak in the $\Delta_p$ spectrum.
\begin{figure}[h]
  \begin{center}
    \subfigure[$\Delta x$ projection, $\Delta y = 0$.]{\label{dxcut85}\includegraphics[angle=90,width=.48\textwidth]{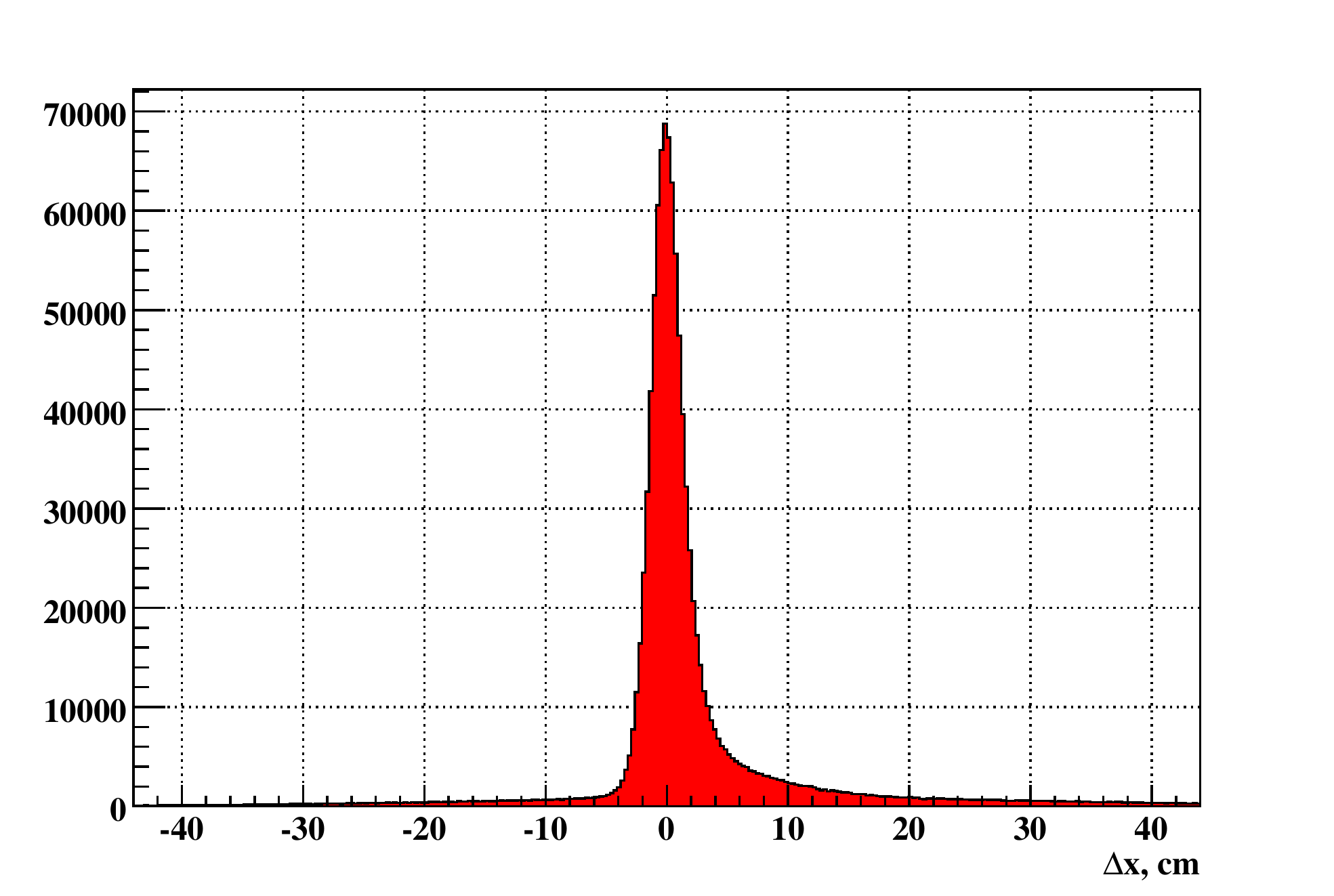}}
    \subfigure[$\Delta y$ projection, $\Delta x = 0$.]{\label{dycut85}\includegraphics[angle=90,width=.48\textwidth]{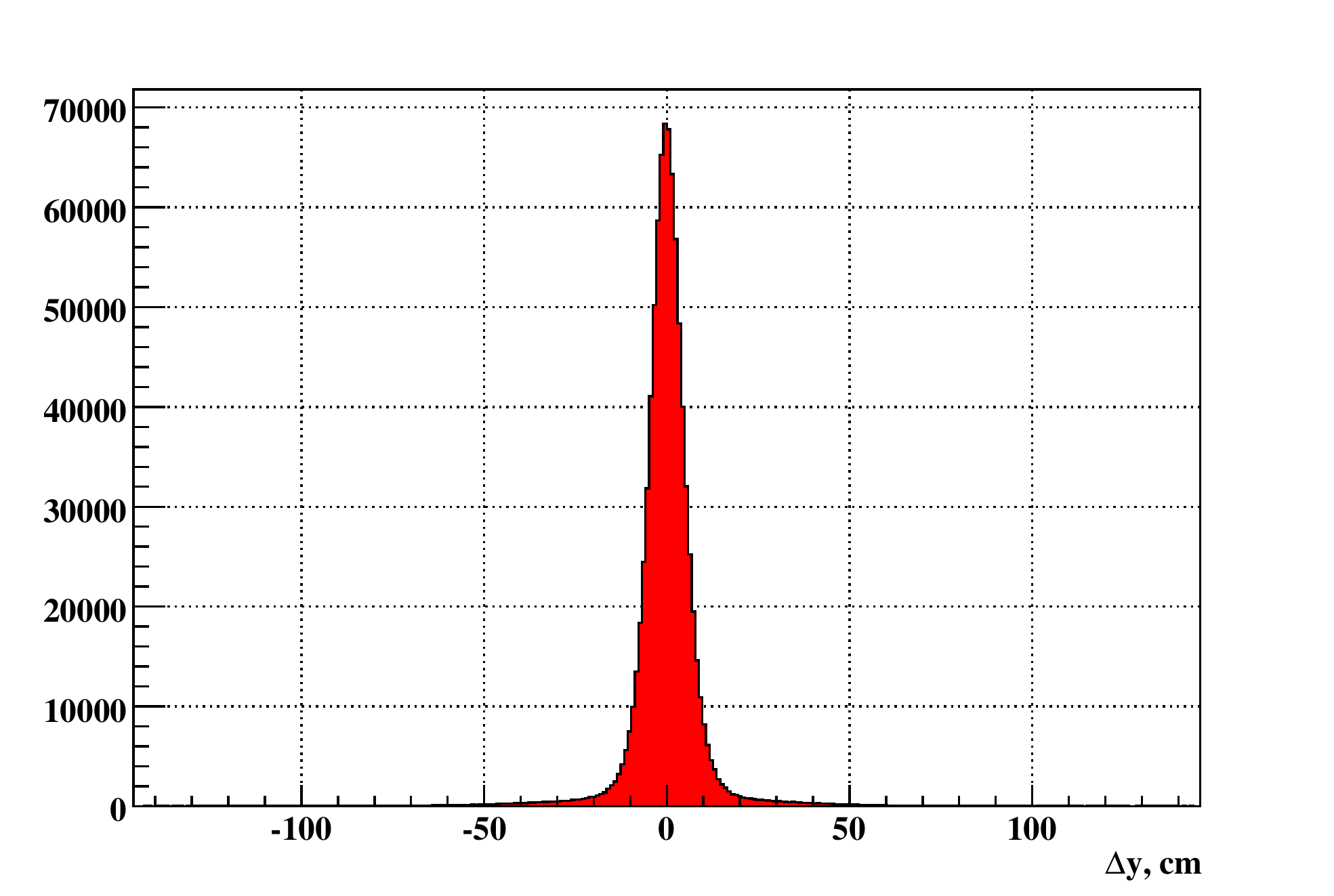}}
  \end{center}
  \caption{\label{xyprojections_cut85} Projection along the $\Delta x(\Delta y)$ axis for $\Delta y(\Delta x) = 0$, $Q^2=8.5$ GeV$^2$, for events with $\left|\Delta_p\right| \le 0.62$\%. }
\end{figure}
Compared to the same projections obtained before applying the $\Delta_p$ cut (figure \ref{dxdyprojections}), the background level is evidently much lower.
\subsection{Inelastic Background Estimate}
\paragraph{}
After applying all elastic event selection cuts, there is still an unavoidable residual background of primarily $\pi^0$ photoproduction events induced by near-endpoint photons overlapping with the elastic peak in $\Delta_p$, with at least one decay photon that happens to hit BigCal at the expected position of an elastically scattered electron. The main determinant of the inelastic contamination within the cuts is the relative rate of elastic scattering and pion photoproduction. At $Q^2=8.5$ GeV$^2$, the $\pi^0$ photoproduction cross section is actually much larger than the elastic $ep$ scattering cross section, leading to the overwhelming background in the single-arm\footnote{Strictly speaking, figure \ref{pmisspraw85} is not a true ``single-arm'' spectrum, but a coincidence spectrum of the proton arm before applying any cuts on the electron arm.} proton $\Delta_p$ spectrum of figure \ref{pmisspraw85}.

Obtaining an estimate of the inelastic contamination is an important component of the data analysis, because even a relatively small contamination can have a significant effect on the extracted form factor ratio, particularly if the polarization of protons from background reactions is substantially different from the polarization of elastic events. Since the elliptical position correlation cut achieves the cleanest elastic/inelastic separation, this section will focus on the method used to estimate the remaining background of this cut.

\subsubsection{Gaussian Extrapolation}
\paragraph{}
The background fraction is estimated directly from the data by parametrizing the signal and background shapes and fitting the two-dimensional $(\Delta x, \Delta y)$ spectrum. The main $\pi^0$ decay photon background is parametrized using a Gaussian shape in two dimensions:
\begin{eqnarray}
  N_{inel}(x, y) &=& N_{inel} \exp \left[-\frac{(x-\mu_{x,inel})^2}{2\sigma_{x,inel}^2}-\frac{(y-\mu_{y,inel})^2}{2\sigma_{y,inel}^2}\right] \label{GaussBGshape}
\end{eqnarray}
The background in the $\Delta x, \Delta y$ plot exhibits tails at very large $\Delta y$ at a very low level. Events in this region, which do not exhibit the same shape as the main $\pi^0$ background, come from other reactions, including the target walls and a small contribution from particles detected in the HMS which do not come from the target through the spectrometer. Particles which scatter from the target or other materials along the beamline but do not scatter into the HMS can enter the detector hut through small exposed gaps in the shielding between the exit of the dipole and the entrance to the shield hut. These particles may then re-scatter from various materials into the active area of the detectors\footnote{It is generally assumed that the background from such ``non-spectrometer'' particles is very small and is highly suppressed by the cuts already applied to the reconstructed target coordinates; therefore, no additional cuts were necessary to suppress this kind of event. On the other hand, such events are much more common when the HMS is positioned at a very forward angle as it was for the $Q^2=8.5$ GeV$^2$ point, and inevitably some small number of these events leaked into the acceptance of the applied cuts.}. In order to achieve a better fit to the whole spectrum, the background parametrization was expanded to include the sum of two Gaussians of the form \eqref{GaussBGshape}.

The elastic peak is parametrized using a more complicated form. A simple Gaussian shape cannot account for the tail at positive $\Delta x$ (figure \ref{dxcut85}), which is assumed to be an effect of the elastic radiative tail as opposed to the inelastic background\footnote{Recall that a tail at positive $\Delta x$ corresponds to a tail at negative $\Delta_e$.}. A reasonable approximation to the shape of the elastic peak is to form the product of a Gaussian distribution in $y$ with the sum of Gaussian and Landau distributions in $x$. The Landau distribution arises in the theory of straggling in the energy loss distribution of charged particles passing through thin absorbers\cite{Leo}. It is defined by the following integral:
\begin{eqnarray}
  \phi(x) &\equiv& \frac{1}{\pi} \int_0^\infty e^{-u \ln u - ux} \sin{\pi u} du \label{landaudef}
\end{eqnarray} 
The integral \eqref{landaudef} was evaluated numerically using standard software libraries. The parametrization of the elastic peak assumes the following form: 
\begin{eqnarray}
  N_{el}(x,y) &=& \left[N_{gauss} \exp \left[-\frac{(x-\mu_{x,el})^2}{2\sigma_{x,el}^2}\right]+N_{landau} \phi\left(\frac{x-\alpha}{\beta}\right)\right] \times \nonumber \\
  & & \exp \left[-\frac{(y-\mu_{y,el})^2}{2\sigma_{y,el}^2}\right]
\end{eqnarray}
The parametrization of the sum of signal and background is then given by
\begin{eqnarray}
  N(\Delta x, \Delta y) &\equiv& N_{el}(\Delta x, \Delta y) + N_{inel}(\Delta x, \Delta y)
\end{eqnarray}
This fit function has eighteen adjustable parameters:
\begin{itemize}
  \item Means $(\mu_x, \mu_y)$, widths $(\sigma_x,\sigma_y)$, and normalization constant $N_{inel}$ for each of the two Gaussians describing the background shape (ten parameters).
  \item Normalization constants for the Gaussian ($N_{gauss}$) and Landau ($N_{Landau}$) contributions to the elastic peak (two parameters).
  \item Means $(\mu_x,\mu_y)$ and widths $(\sigma_x,\sigma_y)$ for the Gaussian parts of the elastic peak parametrization (four parameters). 
  \item Most probable value $\alpha$ and width $\beta$ for the Landau component of the elastic peak parametrization (two parameters).
\end{itemize}

This parametrization was not motivated by any specific physical considerations. The detailed shape of the elastic peak was actually not terribly important, since the goal of fitting the two-dimensional spectrum was merely to separate the signal from the background. Once this goal was accomplished, the two-dimensional background shape resulting from the fit was extrapolated into the cut region, and the ratio of this estimated background to the total number of counts (from the data) inside the ellipse was taken as the inelastic background. The fit results for the elastic peak shape were not used for the background estimate. 
\begin{figure}[h]
  \begin{center}
    \includegraphics[width=.98\textwidth]{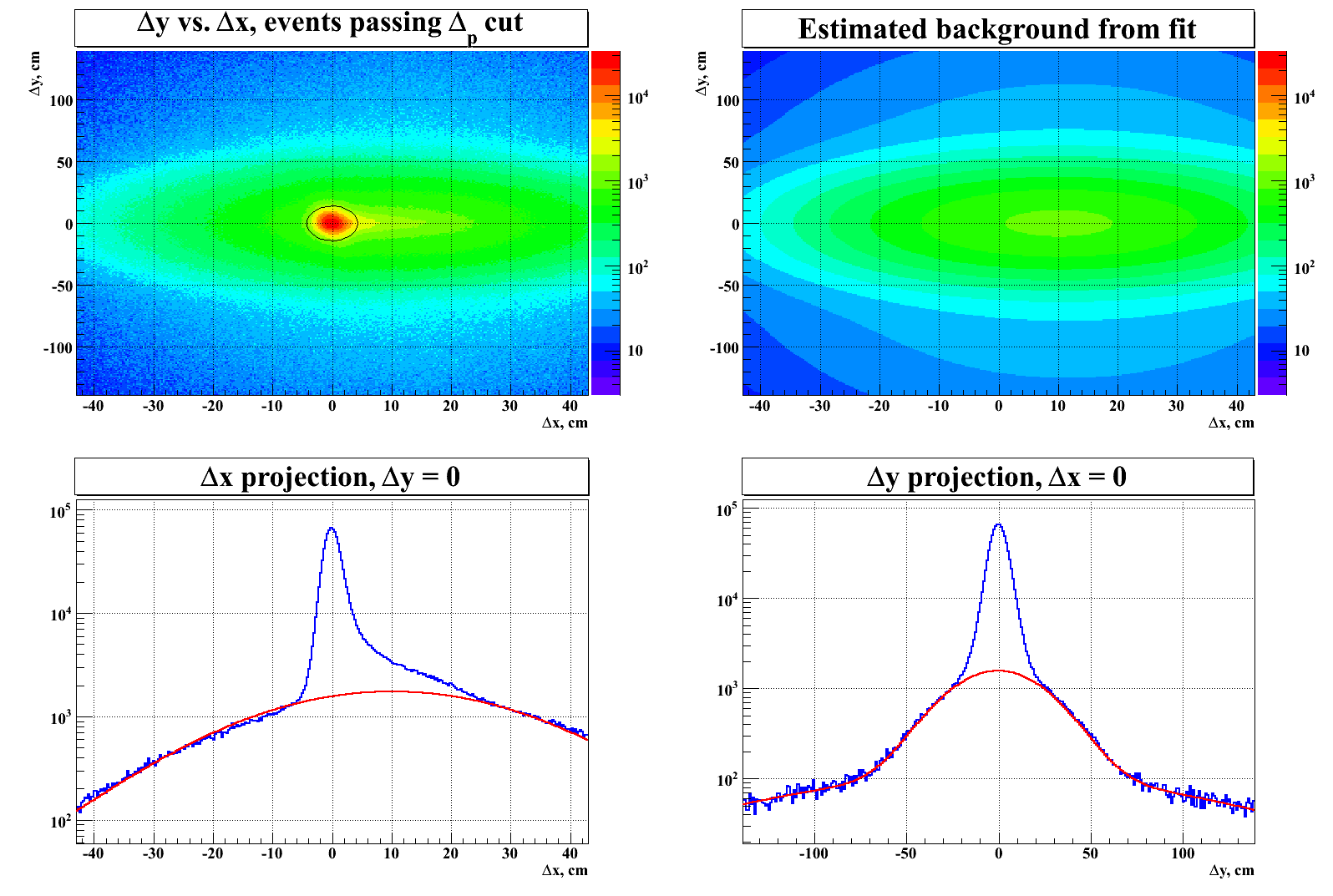}
  \end{center}
  \caption{\label{backgroundellipse85} Inelastic background estimate obtained by fitting the $(\Delta x, \Delta y)$ spectrum, $Q^2 = 8.5$ GeV$^2$. Data (top left), and estimated background (top right). Projection on the $\Delta x$ axis (bottom left) and the $\Delta y$ axis (bottom right) of the data (blue) and estimated background (red).}
\end{figure}

Figure \ref{backgroundellipse85} shows an example result of the fit procedure for $Q^2=8.5$ GeV$^2$. The top left panel shows the $(\Delta x, \Delta y)$ spectrum after applying a cut around the elastic peak in $\Delta_p$. The black ellipse represents the $\Delta x$/$\Delta y$ cut region. The top right panel shows the two-dimensional shape of the estimated background resulting from the fit, plotted on the same color scale as the data. The Gaussian parametrization achieves a very good description of the background shape. The bottom two panels show the projections on the $\Delta x$ axis (left) and the $\Delta y$ axis (right) of the data (blue) and the estimated background (red). The fraction of inelastic background in the data selected by the cut was obtained by integrating the estimated background and the data over the elliptical cut region:
\begin{eqnarray}
  f \equiv \frac{N_{inel}}{N_{data} = N_{inel} + N_{el}} &=& \frac{\int_{-x_{cut}}^{x_{cut}} \int_{-y_{cut}\sqrt{1 - x^2/x_{cut}^2}}^{y_{cut}\sqrt{1-x^2/x_{cut}^2}} N_{inel}(x,y)dy dx}{\int_{-x_{cut}}^{x_{cut}} \int_{-y_{cut}\sqrt{1 - x^2/x_{cut}^2}}^{y_{cut}\sqrt{1-x^2/x_{cut}^2}} N_{data}(x,y)dy dx}
\end{eqnarray}
Table \ref{EllipseBackgroundTable} shows the results of this calculation for $Q^2 = $ 5.2, 6.7, and 8.5 GeV$^2$. 
\begin{table}[h]
  \begin{center}
    \begin{tabular}{|c|c|c|c|c|}
      \hline $Q^2$, GeV$^2$ & $\Delta_p$ cut, \% & $x_{cut}$, cm & $y_{cut}$, cm & $f$, \% \\ \hline 
      5.2 & 0.852 & 4.3 & 15.3 & 1.1 \\ \hline 
      6.7 & 0.900 & 3.1 & 10.2 & 0.8 \\ \hline
      8.5 & 0.600 & 4.3 & 13.9 & 5.6 \\ \hline
    \end{tabular}
    \caption{\label{EllipseBackgroundTable} Estimated background $f$ of $3\sigma$ elliptical position correlation and $\Delta_p$ cuts.}
  \end{center}
\end{table}
The estimates of the inelastic background obtained by this method were combined with the measured polarization of events rejected by the cuts to obtain a correction to the form factor ratio. The uncertainty on this correction is an important source of systematic uncertainty in the result of the experiment, and was estimated by performing the same background estimation procedure using cuts of varying width, and observing the extent to which the background-corrected form factor ratio is invariant with respect to the cuts. 
\begin{table}[h]
  \begin{center}
    \begin{tabular}{|cc|ccc|}
      \hline \multicolumn{2}{|c|}{Background} & \multicolumn{3}{c|}{Ellipse cut width, $\sigma$} \\ 
      \multicolumn{2}{|c|}{fraction, \%} & 3 & 5 & 6 \\ \hline 
      & 3 & 5.5 & 8.3 & 9.6 \\ 
      $\Delta_p$ cut, $\sigma$ & 5 & 9.9 & 14.5 & 16.7 \\
      & 6 & 13.5 & 19.5 & 22.1 \\ \hline
    \end{tabular}
    \caption{\label{DATAbackgroundtable} Cut width dependence of inelastic background contamination estimated using the two-dimensional Gaussian extrapolation of the inelastic background under the elastic peak at $Q^2 = 8.5$ GeV$^2$. Cut widths are given in units of the resolution of the cut variables. The elliptical cut width was only varied in the $\Delta x$; i.e., ``inelasticity'' direction. The cut width in the $\Delta y$ or ``acoplanarity'' direction was fixed at 3$\sigma$ for the estimates presented here. }
  \end{center}
\end{table}
Table \ref{DATAbackgroundtable} shows the variation of the inelastic contamination estimated using the Gaussian extrapolation as a function of the width of the elastic cuts at $Q^2=8.5$ GeV$^2$.
\subsubsection{Monte Carlo Estimate}
\paragraph{}
The above method of estimating the background involves assumptions about the shape of the signal and background in the two-dimensional $(\Delta x, \Delta y)$ space; namely, the shape of the background is assumed to be Gaussian, and the tail at positive $\Delta x$ is attributed to the elastic radiative tail and is assumed to be part of the signal. Neither assumption is exactly satisfied in practice. In order to provide an independent check of this background estimate and to understand the signal and the inelastic background in terms of the underlying physics, a Monte Carlo simulation of the experiment was performed using SIMC, the standard Hall C Monte Carlo package. The adaptation of SIMC to this experimental configuration is described in appendix \ref{SIMCappendix}. The most significant additions to the standard configuration of the Monte Carlo are the presence of BigCal as the electron arm, which is approximated by a rectangular acceptance, and the presence of S0 in the HMS detector hut as part of the trigger and a source of multiple scattering.

Four reactions were simulated, including elastic $ep$ scattering, $\pi^0$ photoproduction, real Compton scattering, and quasi-elastic $(e,e'p)$ from the target endcaps. Since the simulation did not include inefficiencies, only the relative rate of these different reactions was estimated, not the absolute cross section. An important difference between the simulation and the true experimental situation is that in the simulation, the interaction vertex is sampled uniformly along the length of the target, reflecting the assumption of constant target density (and therefore constant luminosity) and a negligible loss of beam intensity along the target length. Deviations from a uniform distribution of events along the length of the target in SIMC reflect the combined acceptance function of the HMS and BigCal. However, the $y_{target}/z_{beam}$ distribution in the experimental data is significantly different from that of the simulation, with fewer events from the downstream end of the target relative to the upstream end of the target. This reduction is primarily caused by a reduction in density along the target length due to localized boiling of the liquid hydrogen in the heat load of the intense electron beam. One other possible cause of this effect not accounted for by the simulation is a $z_{beam}$-dependent reduction in the trigger efficiency of BigCal. Electrons scattering from the downstream end of the target must have a larger average scattering angle $\theta_e$ in order to scatter into the acceptance of the HMS and BigCal. Since larger scattering angles correspond to lower electron energies $E'_e$, which have a higher probability of failing the BigCal trigger threshold, such an effect could be partially responsible for the reduction in luminosity for the downstream end of the target. However, since this threshold was set relatively low throughout the experiment, and certainly well below half the elastic $ep$ energy, it is thought that target boiling is the main cause of the effect.

The dominant background is $\pi^0$ photoproduction, the rate of which is proportional to the Bremsstrahlung flux, which increases approximately linearly along the target length. Since the rate of the main background reaction increases along the target length, and since the simulation does not account for the reduction in luminosity along the target length observed in the data, the simulation tends to overestimate the rate of $\pi^0$ photoproduction relative to elastic scattering in comparison to the data when averaged over the full acceptance. In order to account for this effect, a $z_{beam}$-dependent re-weighting of simulated events was used to map the SIMC $z$ distribution, which only includes acceptance effects, onto that of the data, which reflects the convolution of the experimental acceptance with the $z$-dependent target density reduction, bringing the Monte Carlo background estimate into better agreement with the background estimated directly from the data using the Gaussian extrapolation of the $(\Delta x, \Delta y)$ spectrum. A reduction in target density affects the elastic yield and the photoproduction/Compton yield differently because, whereas the electron beam intensity is unaffected by the reduced target density, the Bremsstrahlung flux \emph{is} affected, since the radiation lengths of the target are the source of this flux. Therefore, while the $ep$ luminosity scales linearly with the target density, the luminosity for Bremsstrahlung-induced reactions scales with the square of the target density, reducing the yield of such reactions relative to electron scattering for a given reduction in target density.
\begin{figure}[h]
  \begin{center}
    \includegraphics[angle=90,width=.98\textwidth]{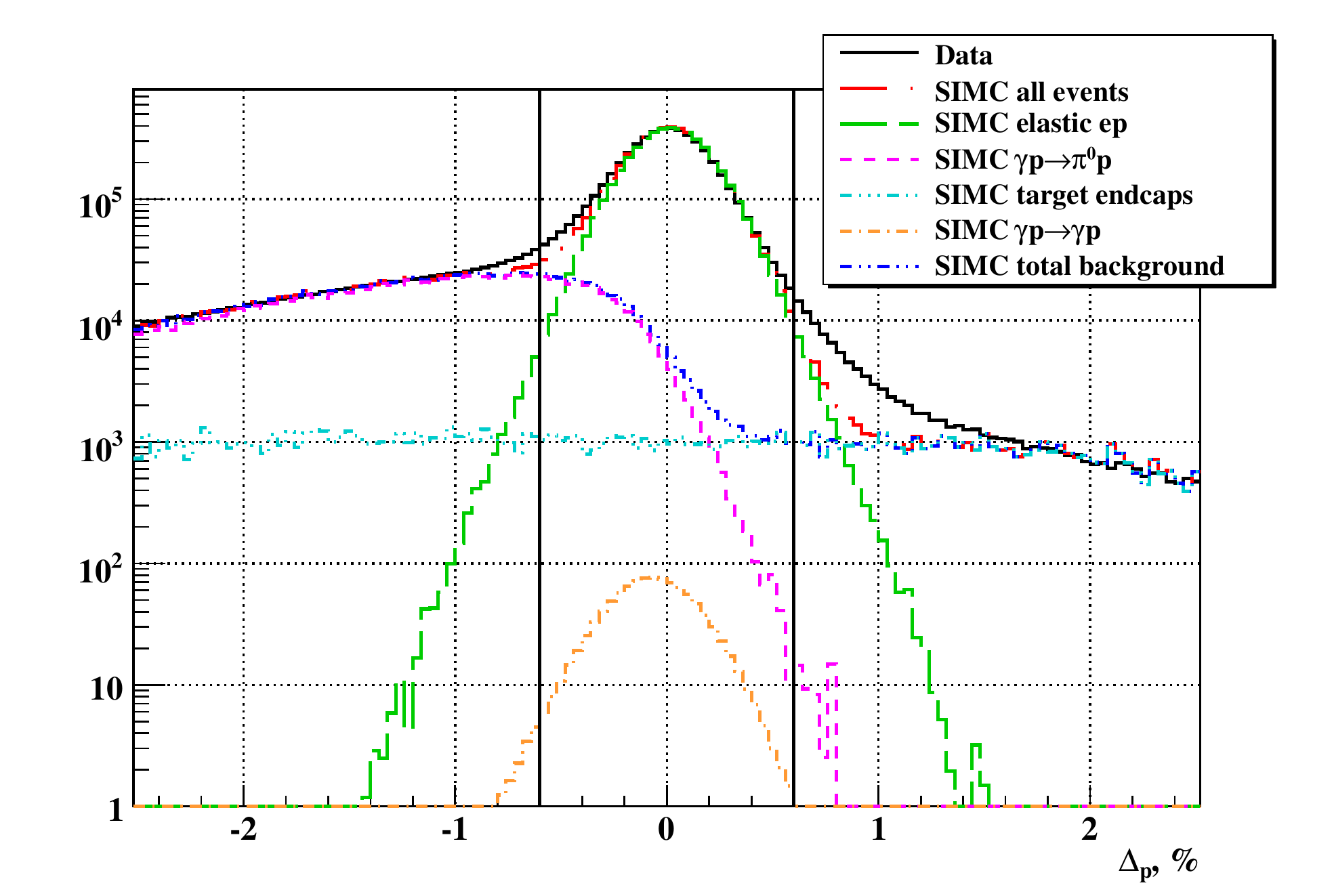}
  \end{center}
  \caption{\label{SIMCdatacomp} SIMC-data comparison of $\Delta_p$ spectra after applying elliptical $(\Delta x,\Delta y)$ cut, $Q^2 = 8.5$ GeV$^2$.}
\end{figure}

Figure \ref{SIMCdatacomp} compares the experimental $\Delta_p$ distribution to the combined $\Delta_p$ distribution of all simulated reactions in the region near the elastic peak. The spectra are obtained after applying the elliptical position correlation cut at BigCal. An overall normalization constant was applied to each reaction to give the best agreement with the data. The target endcap contribution was normalized to the data in the far super-elastic region where it is assumed to be the only reaction that contributes, and then the sum of $\pi^0$ photoproduction, Compton scattering, and the target endcaps was normalized to the data on the far inelastic side of the peak, where the elastic radiative tail has been removed by the position correlation cut and only the background contributes. The black vertical lines indicate a $3\sigma$ cut around the elastic peak in $\Delta_p$. 

The estimated Compton scattering contribution turned out to be much smaller than either the $\pi^0$ contribution or the cell wall contribution, so it could not be meaningfully renormalized on its own. Instead, the Compton spectrum was scaled by the same factor needed to bring the $\pi^0$ contribution into agreement with the data. The data are fairly well reproduced, except for the non-Gaussian tails of the elastic peak observed in the data, which are significantly underestimated by the simulation, even though the resolution of the elastic peak agrees with the data within $\approx \pm 2\sigma$ of the maximum of the peak. The tails are assumed to be elastic events exhibiting non-Gaussian resolution effects not accounted for by the simulation. 
\begin{table}
  \begin{center}
    \begin{tabular}{|cc|ccc|}
      \hline \multicolumn{2}{|c|}{SIMC background} & \multicolumn{3}{c|}{Ellipse cut width, $\sigma$} \\ 
      \multicolumn{2}{|c|}{fraction, \%} & 3 & 5 & 6 \\ \hline 
      & 3 & 5.8 & 9.0 & 10.5 \\ 
      $\Delta_p$ cut, $\sigma$ & 5 & 10.4 & 15.6 & 18.0 \\
      & 6 & 14.1 & 20.8 & 23.7 \\ \hline
    \end{tabular}
    \caption{\label{SIMCbackgroundtable} Cut width dependence of inelastic background contamination estimated by SIMC at $Q^2 = 8.5$ GeV$^2$. Cut widths are given in units of the resolution of the cut variables. The elliptical cut width was varied only in the $\Delta x$ direction for a fixed $\Delta y$ cut of 3$\sigma$.}
  \end{center}
\end{table}

Table \ref{SIMCbackgroundtable} shows the simulated inelastic background contamination for different combinations of cut widths, in units of the resolution of the cut variables\footnote{For the elliptical cut, the vertical and horizontal coordinates have different resolution, and the cut takes the form $\sqrt{\left(\frac{\Delta x}{n\sigma_x}\right)^2+\left(\frac{\Delta y}{3\sigma_y}\right)^2}$, where $n$ is the number given in table \ref{SIMCbackgroundtable}}, at $Q^2=8.5$ GeV$^2$. Compared to table \ref{DATAbackgroundtable}, the Monte Carlo estimate is systematically higher, by a factor ranging from 4 to 10\%, even after renormalizing events as a function of $z_{beam}$. Since the simulation does not fully reproduce the tails of the elastic peak, the elastic peak in the data is inherently less ``sharp'' than the elastic peak of the simulation, reducing its height relative to the background. Simultaneously renormalizing the elastic peak and inelastic background of the simulation to match the $\Delta_p$ spectrum of the data thus reduces the height of the elastic peak relative to the background without widening the peak to match the data, resulting in a slight overestimate of the fraction of events accepted by the cuts that are inelastic. The most important aspect of these results, however, is that they are in good qualitative agreement with the estimates obtained directly from the data, and that the shape of the $\Delta_p$ distribution is well understood in terms of the underlying physics of the signal and the background. The near-total suppression of the elastic radiative tail by the elastic cuts defined above was also confirmed by the simulation. 

In summary, a prescription for estimating the inelastic background directly from the data was presented and the results were compared to a full Monte Carlo simulation of the experiment. The background estimates obtained by the two different methods agreed to within $\pm 10$\%, and the level of disagreement between the two methods can plausibly be attributed to the slight difference in the resolution of the elastic peak between the simulation and the data. In the context of achieving the goals of this analysis, additional fine-tuning of the simulation to achieve exact agreement with the data was unnecessary, since the main purpose of the simulation was to provide an independent check of the background estimate obtained from the data and to demonstrate that the reactions contributing to the inelastic background are well understood. Conservatively assuming a relative uncertainty on the estimated inelastic contamination of $\pm15$\%, and combining this estimate with the measured polarization of the background to obtain the correction to the form factor ratio, the size and uncertainty of the correction for contamination at the level of $\approx6$\% (for $3\sigma$ elastic cuts) at $Q^2 = 8.5$ GeV$^2$ is well below the statistical precision of the measurement, making further improvement of the simulation unnecessary. The irreducible background from real Compton scattering was estimated to be 0.02\% (for $3\sigma$ cuts), which is far below the level where it could significantly affect the form factor ratio. Since the Compton cross section was estimated using a very crude extrapolation from the existing data nearest the $Q^2=8.5$ GeV$^2$ kinematics of this experiment, the relative uncertainty in the estimate is large, but even if the cross section were an order of magnitude larger, it would still be too small to have a significant impact. 

\section{Extraction of Polarization Observables}
\label{polextractsection}
\paragraph{}
To measure the form factor ratio $G_E^p/G_M^p$, the ratio of transverse to longitudinal polarization of the scattered proton must be extracted from the data. To obtain this ratio requires the measurement of the azimuthal asymmetry of protons scattered in the CH$_2$ analyzers of the FPP, and the calculation of the precession of the proton spin through the HMS magnets. The precession depends on the trajectory of the scattered proton as it enters the HMS, since protons of different $(x_{tar},y_{tar},x'_{tar},y'_{tar},\delta)$ see slightly different magnetic fields and experience slightly different deflections depending on their angles, coordinates and momenta. For each event, the precession is calculated from the COSY model of the HMS, and the physical polarization transfer observables $P_t$ and $P_l$ are extracted from the experimental angular distribution using an unbinned maximum likelihood method. 
\subsection{Focal Plane Asymmetry}
\label{fpasymsection}
\paragraph{}
Section \ref{FPPsection} demonstrated how spin-orbit coupling in the nuclear force gives rise to an azimuthal asymmetry in the angular distribution of protons scattered from carbon and hydrogen nuclei in CH$_2$. Equation \eqref{asym_phys} describes the asymmetry in terms of the transverse polarization components of the incident proton. The relative minus sign between the $S_y \cos \varphi$ term and the $S_x \sin \varphi$ term reflects the preferential deflection of the proton along the direction defined by $\mathbf{p} \times \mathbf{S}$. The definition of the direction of the $x$ and $y$ axes with respect to the proton momentum is in fact arbitrary, but the relative sign between the two terms in \eqref{asym_phys} holds for any right-handed coordinate system. 

In an ideal polarimeter, the geometrical acceptance and the detection and tracking efficiency are all independent of the azimuthal angle $\varphi$. In real life, however, a device of the size and complexity of the FPP will inevitably exhibit some instrumental or false asymmetries. These asymmetries will be minimized in a well-designed polarimeter and a well-calibrated data analysis, but they are very difficult to completely eliminate. The most important difference between the physical asymmetries resulting from the polarization of the incident proton and the false asymmetries resulting from the $\varphi$-dependent variations in acceptance, detection efficiency, and tracking efficiency is that the false asymmetries are independent of the beam helicity, while the physical asymmetries change sign upon reversal of the beam polarization\footnote{This is true of the \emph{transferred} polarizations, but not the \emph{induced} polarization, which is zero for elastic scattering in the Born approximation}. The experimental angular distribution including physical and instrumental asymmetries is written as follows:
\begin{eqnarray}
  N^{\pm}(p,\vartheta,\varphi) &=& N_{0}^{\pm} \frac{\varepsilon(p,\vartheta)}{2\pi} \Big[ 1+(a_1 \pm hA_y(p,\vartheta)P_y^{fpp})\cos \varphi + \nonumber \\
    & & (b_1 \mp hA_y(p,\vartheta)P_x^{fpp})\sin \varphi  + \nonumber \\
    & & (a_2 \cos 2\varphi + \ldots) + (b_2 \sin 2\varphi + \ldots) \Big] \label{asym_real}
\end{eqnarray}
In equation \eqref{asym_real}, $N_0^\pm$ is the number of incident protons in the $\pm$ beam helicity state, $h$ is the beam polarization, $\varepsilon$ and $A_y$ are the efficiency and analyzing power defined in \eqref{asym_phys}, $P_x^{fpp}$ and $P_y^{fpp}$ are the polarization components of the incident proton, and $a_1, b_1, a_2, b_2, \ldots$ are false asymmetry terms, which may have higher-order Fourier components than $\cos \varphi$ and $\sin \varphi$.

The rapid 30 Hz reversal of the beam polarization guarantees very nearly equal numbers of events in each helicity state. By forming sum and difference distributions, the physical asymmetries and the false asymmetries can be separated unambiguously. Integrating over the interesting range of $p$ and $\vartheta$ in both the numerator and the denominator yields sum and difference azimuthal angle distributions defined as  
\begin{eqnarray}
  f_{sum}(\varphi) &\equiv& \frac{N^+(\varphi)}{N_0^+} + \frac{N^-(\varphi)}{N_0^-} \nonumber \\
  &=& \frac{1}{2\pi}\big[1+a_1 \cos \varphi + b_1 \sin \varphi + \nonumber \\
    & & a_2 \cos 2\varphi + b_2 \sin 2\varphi + \ldots \big] \label{phisum} \\
  f_{diff}(\varphi) &\equiv& \frac{N^+(\varphi)}{N_0^+} - \frac{N^-(\varphi)}{N_0^-} \nonumber \\
  &=& \frac{h\overline{A_y}}{\pi}\left[P_y^{fpp} \cos \varphi - P_x^{fpp} \sin \varphi \right] \label{phidiff}
\end{eqnarray}
in which the physical asymmetries cancel in the sum distribution \eqref{phisum} and the false asymmetries cancel in the difference distribution \eqref{phidiff}. The cancellation of the false asymmetries by reversing the beam helicity is a very important aspect of the recoil polarization technique because it decouples the extraction of the polarization transfer observables from detailed knowledge and understanding of the false asymmetry. Furthermore, since the induced polarization is zero in elastic scattering, the false asymmetry can be measured using the sum distribution, and any possible effects of a false asymmetry, however small, can in principle be corrected. 

When the numbers of events in each helicity state are equal ($N_0^+ = N_0^-$), the cancellation of the false asymmetry also approximately holds for each helicity state separately. Using the shorthand $\lambda_0(\varphi)$ for the false asymmetry terms, the asymmetries for the $+$ and $-$ helicity states relative to the sum distribution become
\begin{eqnarray}
  \frac{N_+(\varphi)}{N_+(\varphi)+N_{-}(\varphi)} &=& \frac{1}{2}\left[\frac{1+\lambda_0(\varphi) + hA_y(P_y^{fpp} \cos \varphi - P_x^{fpp} \sin \varphi)}{1+\lambda_0(\varphi)}\right] \nonumber \\
  &=& \frac{1}{2} \left[1+\frac{hA_y(P_y^{fpp} \cos \varphi - P_x^{fpp} \sin \varphi)}{1+\lambda_0(\varphi)}\right] \\
  \frac{N_-(\varphi)}{N_+(\varphi)+N_{-}(\varphi)} &=& \frac{1}{2}\left[\frac{1+\lambda_0(\varphi) - hA_y(P_y^{fpp} \cos \varphi - P_x^{fpp} \sin \varphi)}{1+\lambda_0(\varphi)}\right] \nonumber \\
  &=& \frac{1}{2} \left[1-\frac{hA_y(P_y^{fpp} \cos \varphi - P_x^{fpp} \sin \varphi)}{1+\lambda_0(\varphi)}\right]
\end{eqnarray}
in which the false asymmetry terms cancel in the numerator but still appear in the denominator as small $\phi$-dependent modulations of the physical asymmetry, which remains in the numerator. This partial cancellation for separate helicity states is illustrated in figure \ref{asymcancelfig} at $Q^2=8.5$ GeV$^2$, while the difference between the $+$ and $-$ distributions of figure \ref{asymcancelfig} is shown in figure \ref{asymdifffig}.
\begin{figure}[h]
  \begin{center}
    \includegraphics[angle=90,width=0.98\textwidth]{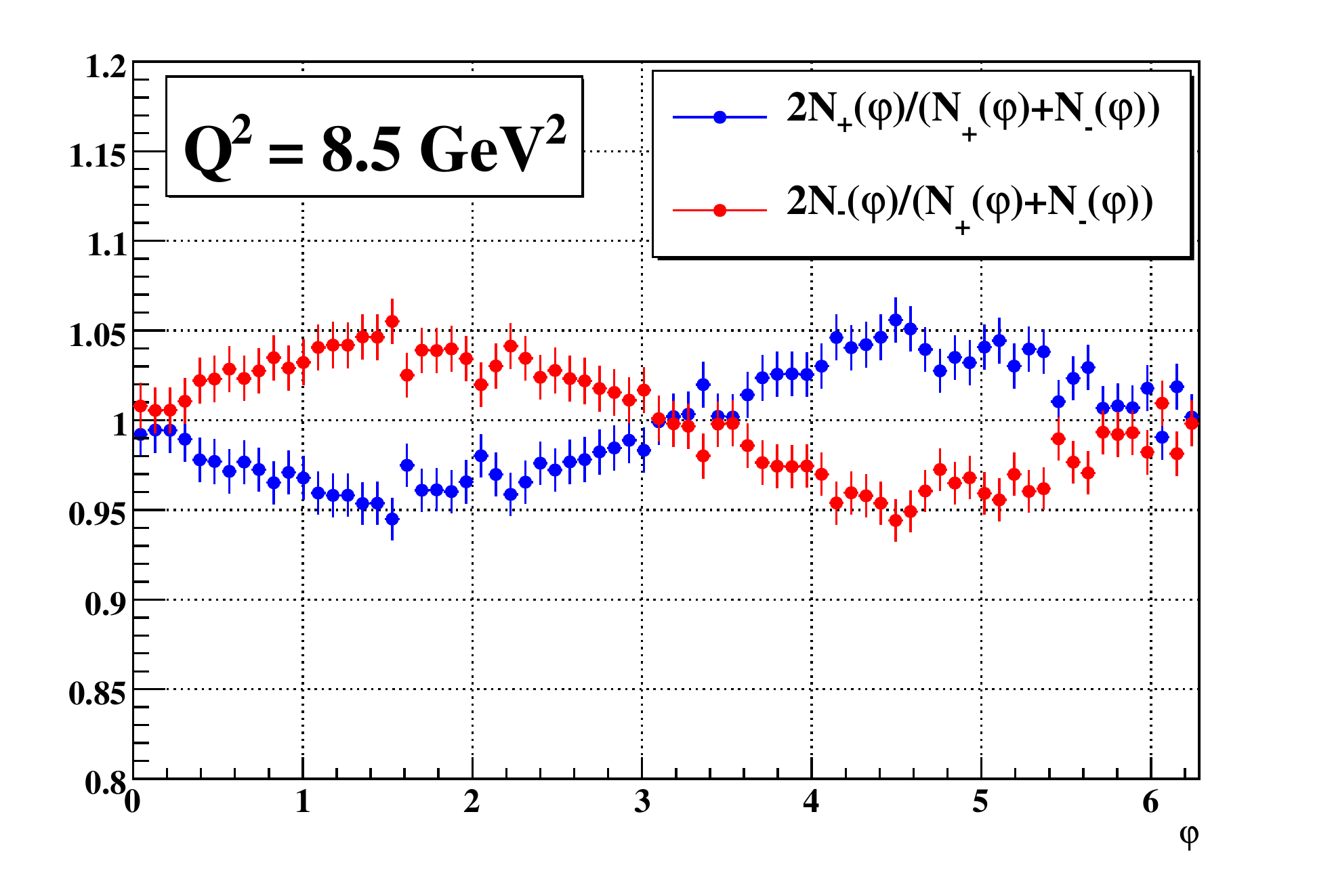}
  \end{center}
  \caption{\label{asymcancelfig} Helicity-dependent asymmetry terms, $Q^2=8.5$ GeV$^2$.}
\end{figure}
\begin{figure}[h]
  \begin{center}
    \includegraphics[angle=90,width=0.98\textwidth]{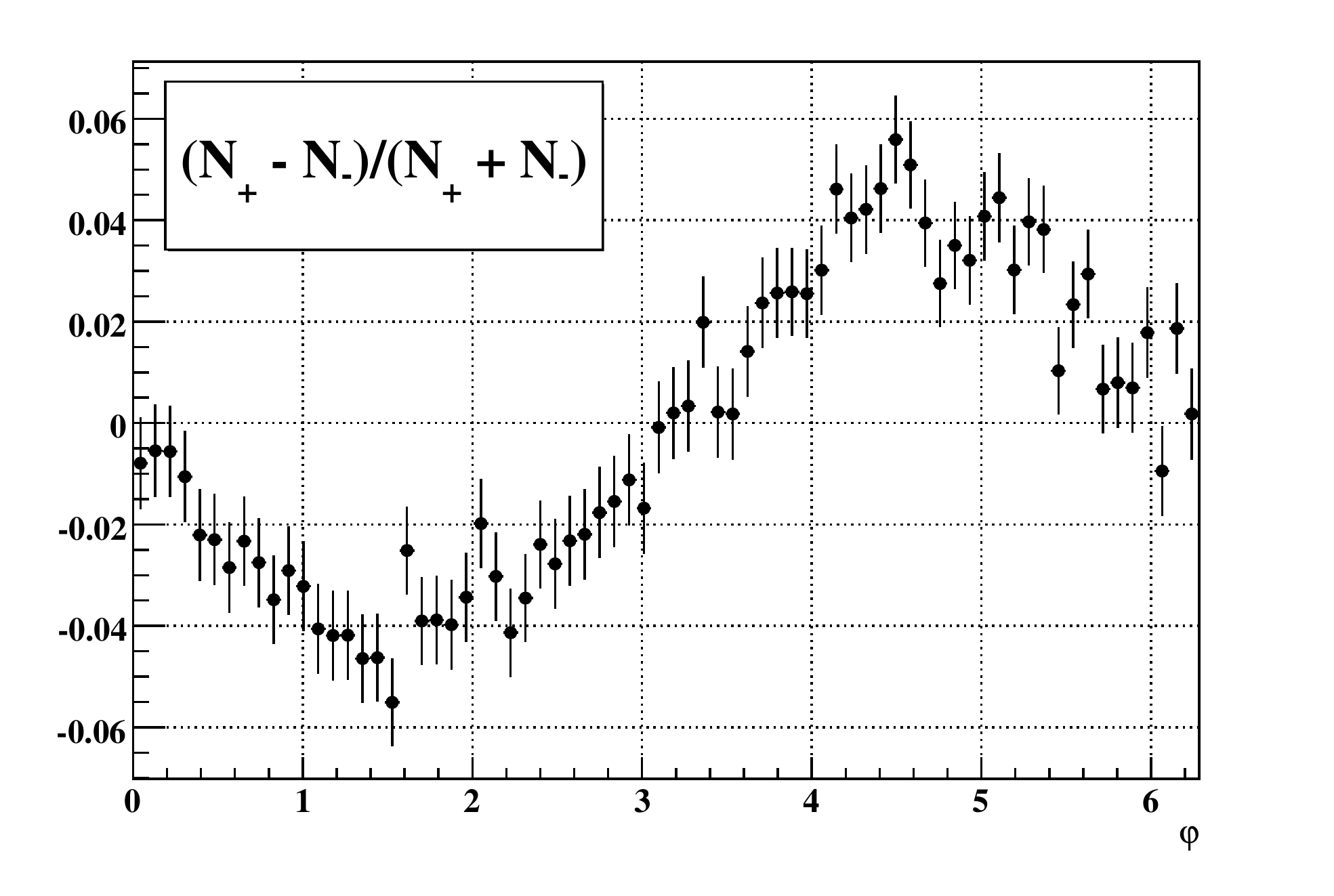}
  \end{center}
  \caption{\label{asymdifffig} Helicity-difference asymmetry, $Q^2=8.5$ GeV$^2$, displayed as $(N_+-N_-)/(N_++N_-)$.}
\end{figure}
The phase shift of the asymmetry determines the ratio of the two transverse polarization components $P_x^{fpp}$ and $P_y^{fpp}$ at the focal plane. This phase shift is related to the form factor ratio through the precession of the proton spin from the target to the focal plane.
\subsubsection{FPP Event Selection Cuts}
\paragraph{}
For each event, a single track is chosen from each polarimeter (drift chamber pair). If multiple tracks are found, the track corresponding to the smallest scattering angle $\vartheta_{fpp}$ is chosen. Additional cuts were applied to reconstructed tracks in the FPP to select the cleanest possible sample of nuclear scattering events for the asymmetry analysis. 
\begin{itemize}
\item A cut was applied to $z_{close}$ to select events which scattered in the CH$_2$ analyzer as illustrated in figure \ref{FPPthetazclose}. A cut of $107\ cm \le z_{close} \le 171\ cm$ was applied to events in the first polarimeter and a cut of $205\ cm \le z_{close} \le 270\ cm$ was applied to events in the second polarimeter.
\item A relatively loose cut was applied to the distance of closest approach $s_{close}$ between the incident and scattered proton tracks. This cut restricts the analysis to events that correspond to a single scattering of the incident particle with reasonably high probability. A cut of $s_{close} \le 3\ cm$ was applied for events in which either the angles in the first polarimeter were measured with respect to the HMS track or the angles in the second polarimeter were measured with respect to the track in the first polarimeter. For events in which the angles in the second polarimeter were measured with respect to the HMS track, a cut of $s_{close} \le 6\ cm$ was applied.
\item All events failing the cone test were rejected in order to eliminate acceptance-related false asymmetries.
\item A momentum-dependent scattering-angle cut was applied in order to suppress events at very large angles with negligible analyzing power and events at very small angles, which have small analyzing power because of their overlap with the Coulomb peak and poor azimuthal angle resolution which diverges as $1/\sin \vartheta$. For a given central momentum $p_0$ of the HMS, a cut of $0.07\ GeV \le p_0 \sin \vartheta_{fpp} \le 1.2\ GeV$ was applied. Although most of the figure-of-merit of the polarimeter is concentrated in the range $0.1\ GeV \le p_T \le 0.7\ GeV$, allowing events at larger and smaller $p_T$ in the analysis does improve the overall figure-of-merit slightly when events are weighted by analyzing power.
\end{itemize}
All of the cuts listed above were applied in the analysis. For each event, the incident proton may scatter in the first analyzer, the second analyzer, neither, or both. When analyzing the asymmetry of both polarimeters combined, all events passing the scattering parameter cuts for the first polarimeter are always counted. In this subset of events, if the chosen track in the second polarimeter also passes the scattering parameter cuts relative to the chosen track in the first polarimeter, the event is counted a second time using the angles of the second track relative to the first, since the two independent scatterings constitute two independent measurements of the polarization of the incident proton\footnote{Counting two polarization-analyzing scattering events for a single incident proton assumes that the polarization of the proton after the first scattering is unchanged. Examination of the asymmetry of the second scattering reveals this assumption to be valid for sufficiently small (but non-zero) angles of the first scattering. In the analysis, the analyzing power for the second scattering was measured as a function of the angle of the first scattering to determine the appropriate cut.}. 

Events which fail the scattering cuts of the first polarimeter fall into two basic categories; correctly-tracked events with $\vartheta$ outside the useful range, and events which are mistracked in the first polarimeter. Events in the first category may be further broken down into small-angle (i.e., Coulomb) and large-angle scattering. Small-angle events as measured in the first polarimeter make up the majority of useful scattering events in the second polarimeter. Large-angle events in the first polarimeter, when properly tracked, are found not to contribute significantly to the figure of merit of the second polarimeter. Events which fail the cuts for FPP1 because of mistracking, though impossible to unambiguously identify, make up a significant fraction of all events. A strong signature for mistracking, as discussed previously, is the reconstruction of $z_{close}$ to a point inside the drift chambers, which occurs quite frequently. These events are rejected from the FPP1 asymmetry analysis. In such cases, the event may still be detected in the second polarimeter with scattering angles and closest approach parameters compatible with scattering in the second analyzer when compared to the HMS track. Such events are assumed to be single-scattering events even if the chosen (smallest-angle) track in the first polarimeter is incompatible with such a scenario due to either mistracking or an incorrect choice made by the track selection algorithm from among multiple tracks, and are counted in the analysis as such.

An important dichotomy exists in the data between single-track and multiple-track events. Events with multiple reconstructed tracks in the polarimeter in question exhibit much lower analyzing power than events with only a single reconstructed track\footnote{A single-track event for the first polarimeter is defined as an event with exactly one reconstructed track in the first chamber pair, regardless of how many tracks were found in the second chamber pair, while a single-track event in the second polarimeter is defined as an event with exactly one reconstructed track in the second chamber pair and either one track or no track found in the first chamber pair. A multiple-track event in the first polarimeter has two or more reconstructed tracks by definition, while a multiple-track event in the second polarimeter has two or more reconstructed tracks in either the first or second chamber pair.}. In fact, the asymmetry of multi-track events is so much smaller than the single-track asymmetry that including multi-track events in the asymmetry analysis actually dilutes the overall figure of merit. Given this situation, a better figure of merit is achieved by analyzing the single-track and multi-track events separately, and combining the results after the fact. Multi-track events contribute a very small fraction of the total figure of merit.
\begin{figure}[h]
  \begin{center}
    \subfigure[Single-track events.]{\label{asym85onetrackonly}\includegraphics[angle=90,width=.49\textwidth]{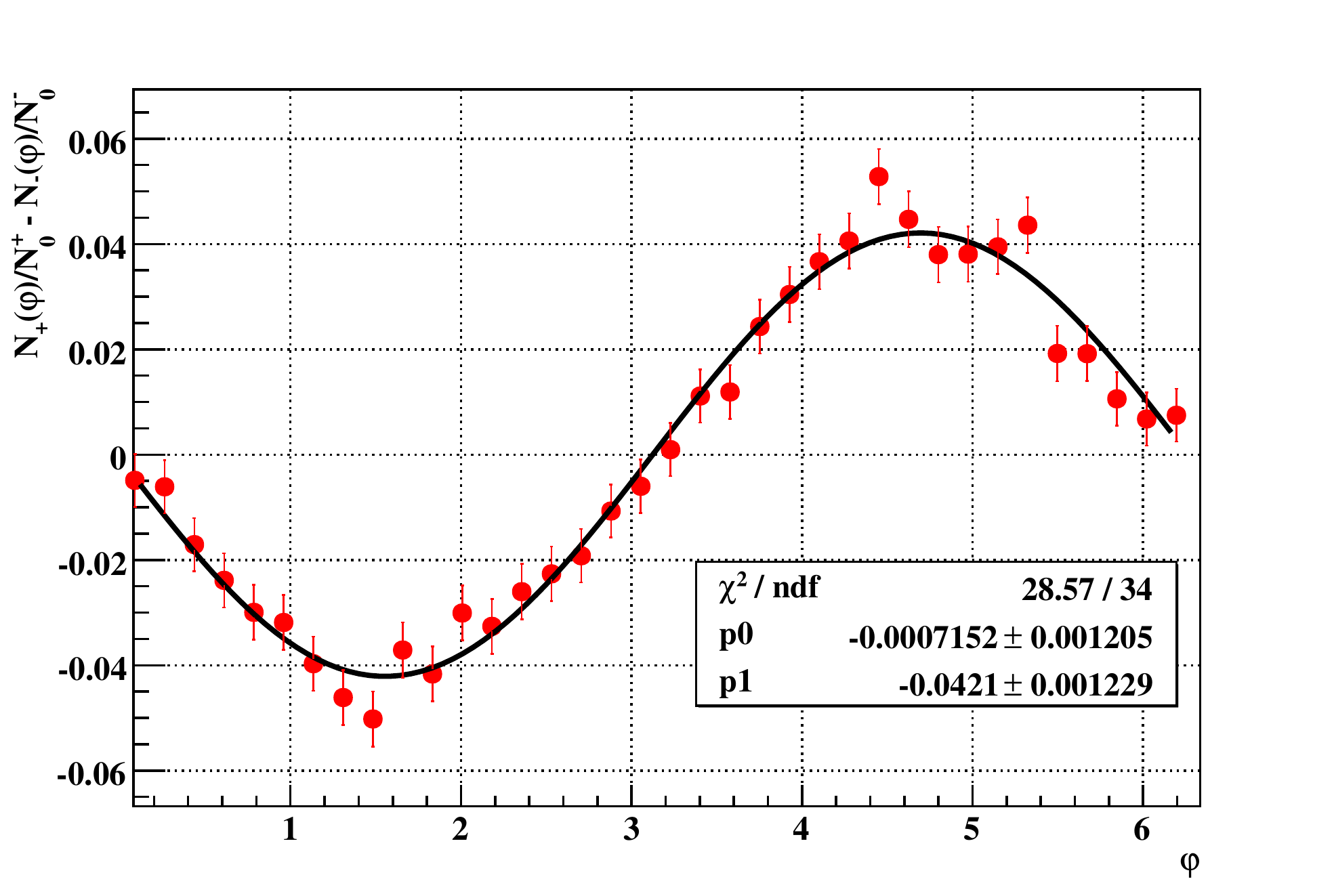}}
    \subfigure[Comparison of single-track, multiple-track, and combined asymmetries.]{\label{asym85multiplicity}\includegraphics[angle=90,width=.49\textwidth]{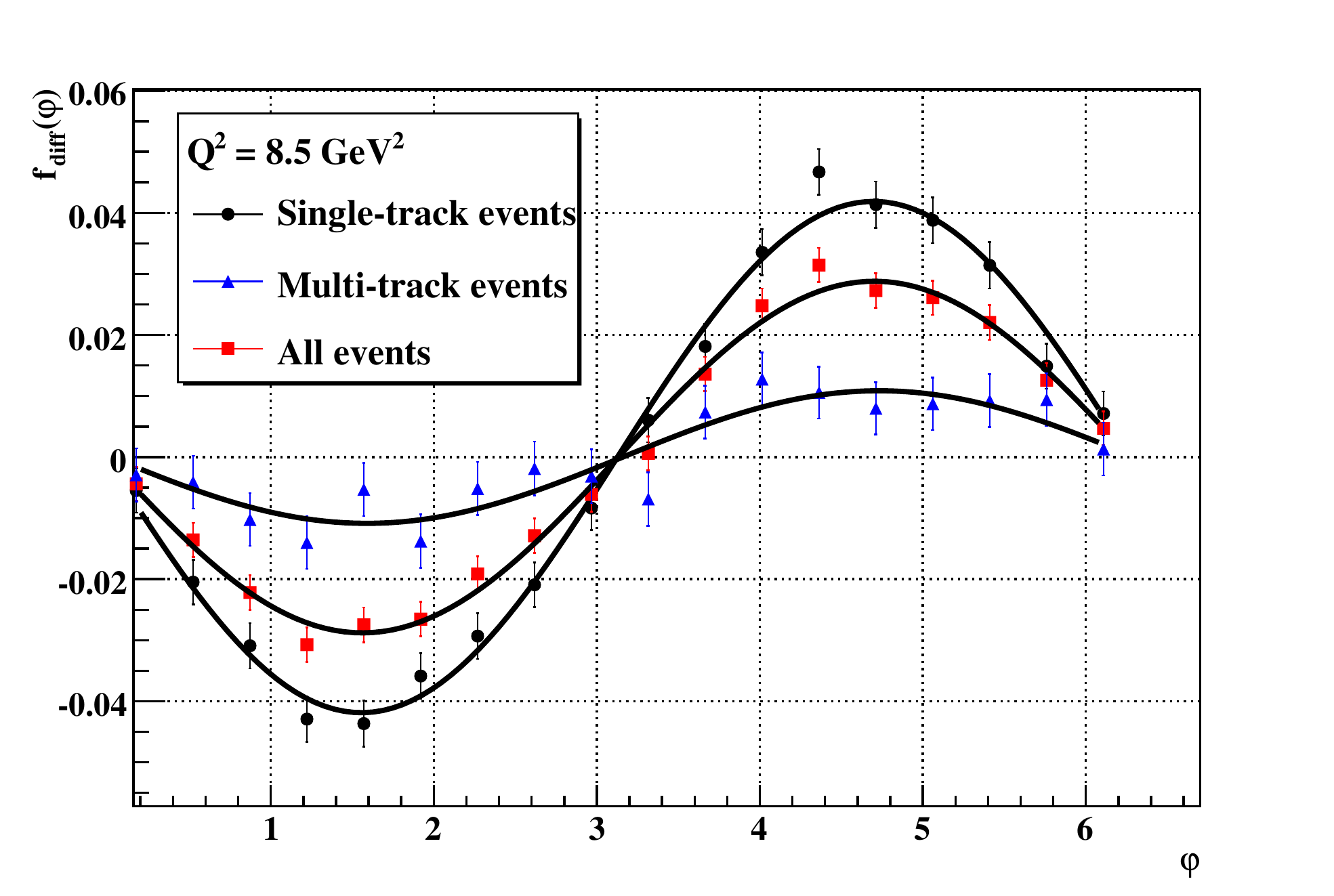}}
  \end{center}
  \caption{\label{asymfig} Helicity-difference asymmetry at the focal plane, $Q^2=8.5$ GeV$^2$.  Fit function is $p_0\cos \varphi + p_1 \sin \varphi$.}
\end{figure}

Figure \ref{asymfig} illustrates the stark difference between the single-track and multiple-track asymmetries at $Q^2 = 8.5$ GeV$^2$. In this case, the multiple-track asymmetry is approximately four times smaller than the single-track asymmetry. Approximately 42\% of all events at this $Q^2$ have multiple tracks. The statistical figure of merit of the asymmetry is proportional to the product of the number of events and the square of the amplitude of the asymmetry; i.e., the analyzing power. In this example, the ratio of the figure of merit of the combined asymmetry to that of the single-track asymmetry becomes
\begin{eqnarray}
  A_{y,eff} &=& \left(f_{single} A_{y,single} + f_{multi} A_{y,multi} \right) \\
  &\approx& 0.725 A_{y,single} \\
  \frac{N_{tot} A_{y,eff}^2}{N_{single}A_{y,single}^2} &\approx& 0.91 
\end{eqnarray}
demonstrating how the dilution of the combined asymmetry outweighs the number of added events, leading to the decision to analyze single and multiple-track events separately. It is worth pointing out that although the amplitude of the multi-track asymmetry is much smaller than that of the single-track asymmetry, the phase shifts of the two asymmetries are compatible, as they should be since they measure the same polarization components. It appears that reactions in the CH$_2$ producing multiple particles simply have lower analyzing power than single-particle reactions such as elastic nuclear scattering. While it is certainly true that track reconstruction difficulties could be partially responsible for the reduced asymmetry of multiple-track events, the fact that the difference in the asymmetry was observed to be largely independent of the various track reconstruction strategies attempted suggests that either
\begin{itemize}
\item The difficulties in correctly reconstructing multiple tracks in drift chambers of this design at single-particle rates typical of experiment E04-108 are insurmountable, or$\ldots$
\item The reduced asymmetry of multiple-track events is a real, irreducible physical effect, reflecting the fact that reactions in CH$_2$ producing multiple charged particles detected in the drift chambers simply have less analyzing power than single-particle reactions such as elastic $\vec{p}+$CH$_2$ scattering.
\end{itemize}
The determination of the exact cause of the single-track/multi-track dichotomy is a rather involved project that goes beyond the scope of this thesis, but may prove to be an important consideration for the design of future experiments requiring proton polarimetry.
\subsubsection{Fourier Analysis}
\paragraph{}
The helicity-dependent asymmetries at the focal plane, which are related to the polarization transfer observables, can be obtained by direct Fourier (i.e., moment) analysis of the helicity difference distribution:
\begin{eqnarray}
  \int_0^{2\pi} f_{diff}(\varphi) \cos \varphi d\varphi &=& h\overline{A_y}P_y^{fpp} \\
  \int_0^{2\pi} f_{diff}(\varphi) \sin \varphi d\varphi &=& -h\overline{A_y}P_x^{fpp}
\end{eqnarray}
The integrals can be estimated by sums over all events:
\begin{eqnarray}
  \int_0^{2\pi} f_{diff}(\varphi) \cos \varphi d\varphi &=& \frac{1}{N_0^{+}} \sum_{i=1}^{N_0^+} \cos \varphi_{i} - \frac{1}{N_0^-}\sum_{i=1}^{N_0^-} \cos \varphi_{i} \\
  \int_0^{2\pi} f_{diff}(\varphi) \sin \varphi d\varphi &=& \frac{1}{N_0^{+}} \sum_{i=1}^{N_0^+} \sin \varphi_{i} - \frac{1}{N_0^-}\sum_{i=1}^{N_0^-} \sin \varphi_{i}
\end{eqnarray}
where the sums are over all events of a given helicity state. The statistical uncertainty in the individual $\cos \varphi$ and $\sin \varphi$ moments is determined by 
\begin{eqnarray}
  \sigma^2\left(\overline{\cos \varphi}\right) &=& \frac{s^2(\cos \varphi)}{N} = \frac{1}{N}\overline{\left(\cos \varphi - \overline{\cos \varphi}\right)^2} \nonumber \\ 
  &=& \frac{1}{N}\left(\frac{1}{N}\sum_i \cos^2 \varphi_i - \left(\frac{1}{N}\sum_i \cos \varphi_i\right)^2 \right) \\
  \sigma^2\left(\overline{\sin \varphi}\right) &=& \frac{s^2(\sin \varphi)}{N} = \frac{1}{N}\overline{\left(\sin \varphi - \overline{\sin \varphi}\right)^2} \nonumber \\ 
  &=& \frac{1}{N}\left(\frac{1}{N}\sum_i \sin^2 \varphi_i - \left(\frac{1}{N}\sum_i \sin \varphi_i\right)^2 \right)
\end{eqnarray}
leading to uncertainties in the focal plane asymmetries given by
\begin{eqnarray}
  \sigma \left(h\overline{A_y}P_y^{fpp}\right) &=& \sqrt{\sigma^2_+\left(\overline{\cos \varphi}\right) + \sigma^2_- \left(\overline{\cos \varphi}\right)} \\
  \sigma \left(h\overline{A_y}P_x^{fpp}\right) &=& \sqrt{\sigma^2_+\left(\overline{\sin \varphi}\right) + \sigma^2_- \left(\overline{\sin \varphi}\right)}
\end{eqnarray}
where $\sigma_\pm$ is the uncertainty in the given moment for the $\pm$ helicity state. This method of extracting the asymmetry at the focal plane has the advantage of an ``exact'' cancellation of the false asymmetry, but it does not give the physical polarization transfer observables, because the precession of the proton spin in the HMS has not been accounted for.
\subsection{Spin Precession}
\label{precessionsection}
\paragraph{}
In equations \eqref{asym_real} and \eqref{phidiff}, the polarizations $P_y^{fpp}$ and $P_x^{fpp}$ represent the transverse polarization of the scattered proton after undergoing precession through the magnets of the HMS. They are related to the physical transferred polarization components by a rotation that depends on the trajectory of the scattered proton entering the HMS. In the Heisenberg picture in non-relativistic quantum mechanics, the time evolution of the spin operator $\mathbf{S}$ of a particle of mass $m$ and gyromagnetic ratio $g$ in a magnetic field $\mathbf{B}$ is given by 
\begin{eqnarray}
  \frac{d \mathbf{S} }{d \tau} &=& \frac{ge}{2m} \mathbf{S} \times \mathbf{B} \label{HeisenbergEOM}
\end{eqnarray}
This equation of motion applies in the rest frame of the particle. To obtain the corresponding equation of motion for protons moving at relativistic speeds in the lab frame, two relativistic effects must be accounted for. 

First, the Heisenberg equation of motion \eqref{HeisenbergEOM} is assumed to hold in the instantaneously comoving rest frame of the proton. The magnetic field in \eqref{HeisenbergEOM} is understood to be the magnetic field in the proton rest frame, which is modified by the boost from the lab frame, in which it is static. Secondly, since the proton is accelerating in the magnetic field, its instantaneous rest frame is also accelerating, which leads to the effect known as Thomas precession in which the rest frame spin precesses in the plane defined by the velocity and acceleration vectors of the particle. The acceleration of the proton is also determined by the magnetic field. Combining the effects of Thomas precession and the non-relativistic quantum-mechanical evolution of the spin in the rest-frame magnetic field, the fully relativistic equation of motion for the spin precession\cite{BMTequation} is obtained\footnote{Equation \eqref{ThomasBMTformula} is true for the precession of a particle with charge $e$ such as the proton. The equation of spin precession in a magnetic field for the neutron, on the other hand, is simpler. Since the neutron does not accelerate in a magnetic field, its spin precession is given by \eqref{HeisenbergEOM}, up to a factor $\gamma$ for time dilation and with $\mathbf{B}$ understood to mean the rest frame magnetic field, which is obtained from the lab magnetic field by a Lorentz boost along the neutron momentum.}.
\begin{eqnarray}
  \frac{d\mathbf{S}}{dt} &=& \frac{e}{m\gamma} \mathbf{S} \times \left[\frac{g}{2} \mathbf{B}_{\|} + \left(1+\gamma\left(\frac{g}{2}-1\right)\right)\mathbf{B}_{\bot}\right] \label{ThomasBMTformula}
\end{eqnarray}
In equation \eqref{ThomasBMTformula}, which is known as the Thomas-B.M.T. equation after Bargmann, Michel, and Telegdi, the proper time $\tau$ has been replaced by the lab time $t$, $\gamma$ is the relativistic boost factor of the particle, $\mathbf{B}_{\|}$ is the component of the magnetic field parallel to its velocity, and $\mathbf{B}_{\bot}$ is the component of the magnetic field perpendicular to its velocity. The rotation of the proton spin in the magnetic field of the HMS is calculated by integrating \eqref{ThomasBMTformula} over the full magnetic length of the spectrometer traversed by each proton. In the following discussion, two model-independent approximate methods for calculating the precession will be presented alongside the full calculation of the precession using the COSY model of the HMS.
\subsubsection{Ideal Dipole Approximation}
\paragraph{}
The proton trajectory after transport through the HMS is measured very precisely, and its trajectory at the target is reconstructed precisely using the well-known transport matrix of the spectrometer. The evolution of the proton trajectory in the HMS is governed by the Lorentz force law. Since the magnetic field does no work on the proton, its velocity vector changes but its speed does not change. The equation of motion of the unit trajectory vector $\hat{v}$ is 
\begin{eqnarray}
  \frac{d\hat{v}}{dt} &=& \frac{e}{m\gamma} \hat{v} \times \mathbf{B} \label{veleom}
\end{eqnarray}
Comparing the equation of motion for the proton's velocity to the equation of motion for its spin, there is a term $\frac{e}{m\gamma} \mathbf{S} \times \mathbf{B}_\bot$ which behaves identically to the velocity\footnote{Recall $\mathbf{B}_\bot$ was defined as the magnetic field component perpendicular to the velocity, so that $\hat{v} \times \mathbf{B} = \hat{v} \times \mathbf{B}_\bot$.}. The quantity of interest is the precession of the orientation of the proton polarization relative to its velocity, so it is natural to subtract the term which corresponds to a rotation identical to that of the velocity. Neglecting the magnetic field component parallel to the velocity, which is a reasonably good approximation for a spectrometer with small angular acceptance such as the HMS, the spin precession relative to the velocity becomes
\begin{eqnarray}
  \frac{d\mathbf{S}}{dt} &=& \gamma\left(\frac{g}{2}-1\right)\left[\frac{e}{m\gamma}\mathbf{S} \times \mathbf{B}_\bot\right] \label{relativeprecessionnoBparallel}
\end{eqnarray}
which differs from \eqref{veleom} only by a factor of $\gamma \kappa_p$, where $\kappa_p \approx 1.79$ is the proton's anomalous magnetic moment. In an idealized picture of the HMS, its horizontal angular acceptance is infinitesimally small, and the HMS dipole is a region of uniform magnetic field with no fringe field at its edges, and all protons move on planar, circular trajectories in this idealized field. The total deflection angle of the proton trajectory in this magnetic field is given by the 25-degree central bend angle of the HMS plus the difference in vertical angle between its initial (reconstructed) and final (measured) trajectories: $\theta_{bend} = \Theta_0 + \theta_{tar} - \theta_{fp}$, where $\Theta_0 = 25^\circ$ is the HMS central bend angle, $\theta_{tar} = \arctan x'_{tar}$ is the vertical angle of the proton trajectory at the target, and $\theta_{fp} = \arctan x'_{fp}$ is the vertical angle of the proton trajectory at the focal plane.

According to equation \eqref{relativeprecessionnoBparallel}, the polarization precesses relative to the velocity by an angle $\chi_\theta \equiv \gamma \kappa_p \theta_{bend}$. Elastically scattered protons have transferred polarization parallel and perpendicular to their momentum in the reaction plane, and no polarization normal to the reaction plane. In the ideal dipole approximation, the transverse component of the polarization does not rotate, and the longitudinal component of the polarization rotates through an angle $-\chi_\theta$ about the $y$ axis of the transport coordinate system:
\begin{eqnarray}
  \left(\begin{array}{c} P_t^{fpp} \equiv P_y^{fpp} \\ P_n^{fpp} \equiv -P_x^{fpp} \\ P_l^{fpp} \equiv P_z^{fpp} \end{array}\right) &=& \left(\begin{array}{ccc} 1 & 0 & 0 \\ 0 & \cos \chi_\theta & \sin \chi_\theta \\ 0 & -\sin \chi_\theta & \cos \chi_\theta \end{array}\right)\left(\begin{array}{c} P_t \\ P_n = 0 \\ P_l \end{array}\right) \label{idealdipoleprecess}
\end{eqnarray}
The identification of $P_n^{fpp}$ with $-P_x^{fpp}$ and $P_t^{fpp}$ with $P_y^{fpp}$ in equation \eqref{idealdipoleprecess} reflects the fact that in the coordinate system of the reaction plane (which approximately coincides with the horizontal plane), $P_t$ is measured along the $+x$ axis, which (approximately) coincides with the $+y$ axis of transport coordinates, and $P_n$ is measured along the $+y$ axis, which corresponds to the $-x$ axis of transport coordinates. In this idealized approximation, the small rotation of the reaction plane with respect to the horizontal plane due to the out-of-plane angle of the scattered proton is ignored.

The azimuthal scattering angle $\varphi$ is reconstructed from the measured track in the FPP drift chambers with respect to a coordinate system which is different for each focal plane trajectory, with the $z$ axis along the proton trajectory $\hat{v}$, the $y$ axis parallel to the transport $yz$ plane and perpendicular to $\hat{v}$, and the $x$ axis defined as $\hat{x} = \hat{y} \times \hat{v}$. In the ideal dipole approximation, the small rotations of the focal plane trajectory relative to the transport coordinate system are ignored, and $(P_x^{fpp}, P_y^{fpp})$ of equation \eqref{idealdipoleprecess} are assumed to be identical to those appearing in the expression \eqref{phidiff} for the focal plane asymmetry. In this approximation, the helicity difference distribution becomes 
\begin{eqnarray}
  f_{diff}(\varphi) &=& \frac{h\overline{A_y}}{\pi}\left[P_t \cos \varphi + \left(\cos \chi_\theta P_n + \sin \chi_\theta P_l\right)\sin \varphi\right] \nonumber \\
  &=& \frac{h\overline{A_y}}{\pi}\left[P_t \cos \varphi + \sin \chi_\theta P_l\sin \varphi\right]
\end{eqnarray}
where on the second line the Born approximation result of zero normal polarization in elastic scattering has been included.
\begin{table}[h]
  \begin{center}
    \begin{tabular}{|c|c|c|}
      \hline $Q^2$, GeV$^2$ & $p_0$, GeV/c & $\chi_\theta$, $^\circ$ \\ \hline
      2.5 & 2.0676 & 108.5 \\ \hline
      5.2 & 3.5887 & 177.2 \\ \hline
      6.7 & 4.4644 & 217.9 \\ \hline
      8.5 & 5.4070 & 262.2 \\ \hline
    \end{tabular}
  \end{center}
  \caption{\label{chicentraltable} Central precession angles $\chi_\theta$ for the HMS as a function of central momentum.}
\end{table}

Table \ref{chicentraltable} shows the value of $\chi_\theta$ corresponding to the central bend angle of the HMS for the kinematics of this experiment. The ideal dipole approximation shows that the optimal value of $\chi_\theta$ to measure the form factor ratio is $\left|\sin \chi_\theta \right| \approx 1$, since the sensitivity of the focal plane asymmetry to $P_l$ is maximized. This condition corresponds to maximal rotation of the longitudinal polarization at the target into transverse polarization at the focal plane, which can be measured. At $Q^2 = 8.5$ GeV$^2$, this condition is very nearly satisfied with a central $\chi_\theta$ near 270 degrees. On the other hand, at 5.2 GeV$^2$, $\chi_\theta$ is near 180 degrees, which is the worst case for measuring $P_l$. This situation is not as hopeless as it may seem, however, since there is sufficient variation of $\chi_\theta$ within the acceptance of the HMS to measure $P_l$ with reasonable precision even at the least favorable central precession angle. 
\subsubsection{Geometric Approximation}
\paragraph{}
The ideal dipole approximation is a good approximation to the full precession only to the extent that the precession angle in the non-dispersive plane either vanishes or is symmetrically distributed such that its acceptance-averaged value is near zero. In practice, neither condition is met, since the angular acceptance and the target length are both finite, and the elastic scattering cross section is not uniform, but varies strongly with the scattering angle, which is closely related to the trajectory bend angle (and thus the precession) in the non-dispersive plane. 

In order to derive a model-independent approximation to the full dispersive and non-dispersive precession, the assumption $\mathbf{B}_\| = 0$ is retained so that the equation of motion for the precession relative to the velocity is still given by \eqref{relativeprecessionnoBparallel}, and several additional assumptions are made. For an infinitesimal deflection of the trajectory by $d\theta$ in the dispersive plane or $d\phi$ in the non-dispersive plane, the spin precesses by $\gamma \kappa_p d\theta$ or $\gamma \kappa_p d\phi$ in the same plane (relative to the velocity). If the trajectory is planar, then the small rotations of both the trajectory and the spin are additive, and this microscopic version of the ideal dipole approximation also holds for the total trajectory bend and the total precession; i.e., $\chi_\theta = \int_0^{\Delta \theta} \gamma \kappa_p d\theta = \gamma \kappa_p \Delta \theta$ and $\chi_\phi = \int_0^{\Delta \phi} \gamma \kappa_p d\phi = \gamma \kappa_p \Delta \phi$. In general, the full trajectory through the QQQD magnet arrangement of the HMS is not strictly planar, but instead consists of simultaneous, coupled deflections in the dispersive and non-dispersive directions, and the full equation of motion \eqref{ThomasBMTformula} must be numerically integrated over the entire trajectory to obtain the spin rotation for a given incident trajectory.

In the geometric approximation, however, it is assumed that the trajectory experiences small, independent, additive deflections in the dispersive and non-dispersive planes, and the dipole approximation is assumed to hold separately for the precession in each plane. In this approximation, the full rotation of the spin is characterized by angles $\chi_\theta$ and $\chi_\phi$ given by\footnote{The reason for the sign inversion between $\theta_{tar} - \theta_{fp}$ in the expression for $\chi_\theta$ and $\phi_{fp} - \phi_{tar}$ in the expression for $\chi_\phi$ is that both angles are defined to be positive along the deflection direction of the trajectory. For $\phi$, the deflection direction is the same as the $+\phi$ direction, but since $\Delta \theta$ is positive for \emph{downward} deflection angles, and the central bend angle $\Theta_0$ is vertically upward, $\chi_\theta$ must be defined with the signs of $\theta_{tar}$ and $\theta_{fp}$ reversed.}
\begin{eqnarray}
  \chi_\theta &\equiv& \gamma \kappa_p \left(\Theta_0 + \theta_{tar} - \theta_{fp}\right) \\
  \chi_\phi &\equiv& \gamma \kappa_p \left(\phi_{fp} - \phi_{tar}\right)
\end{eqnarray}
For the HMS, this turns out to be a quite reasonable approximation. Most of the deflection of the trajectory in the non-dispersive plane takes place in the three quadrupoles preceding the dipole and most of the deflection of the trajectory in the dispersive plane takes place in the dipole. And although the trajectory is also deflected in the dispersive plane in the quadrupoles, these rotations are generally small compared to the main dipole precession, and their \emph{average} effect on the non-dispersive precession is very nearly zero, since the distribution of the out-of-plane angle of tracks entering the HMS is very nearly symmetric, unlike the generally asymmetric distribution of the in-plane angle. 

The total rotation of the spin corresponding to dispersive and non-dispersive precession angles $\chi_\theta$ and $\chi_\phi$ becomes
\begin{eqnarray}
  \left(\begin{array}{c} P_t^{fpp} = P_y^{fpp} \\ P_n^{fpp} = -P_x^{fpp} \\ P_l^{fpp} = P_z^{fpp} \end{array}\right) &=& \left(\begin{array}{ccc} \cos \chi_\phi & 0 & \sin \chi_\phi \\ -\sin \chi_\phi \sin \chi_\theta & \cos \chi_\theta & \cos \chi_\phi \sin \chi_\theta \\ -\sin \chi_\phi \cos \chi_\theta & - \sin \chi_\theta & \cos \chi_\phi \cos \chi_\theta \end{array}\right)\left(\begin{array}{c}P_t \\ P_n = 0 \\ P_l \end{array}\right) \nonumber \\
  & & \label{geometricapproximation} 
\end{eqnarray}
which is simply the mathematical expression of the approximation that the total spin precession in the HMS consists entirely of a rotation in the horizontal plane by angle $\chi_\phi$ in the three quadrupoles, which mixes $P_t$ and $P_l$, followed by a rotation in the vertical plane by angle $\chi_\theta$, which mixes $P'_l$ and $P_n$. Again, this turns out to be a very good approximation to the full precession because of the QQQD magnet arrangement of the HMS, for which the assumptions leading to \eqref{geometricapproximation} are more nearly satisfied than in other, more complicated magnetic systems\cite{PentchevSpinHRS}. The advantage of this approximation is its model independence. The spin transport matrix for each event is a simple function of well-known quantities--the total trajectory bend angles in the dispersive and non-dispersive planes, and the momentum (through the $\gamma$ factor), making it particularly useful for the estimation of systematic uncertainties.

Working again under the assumption of zero normal polarization in elastic scattering, the helicity-difference asymmetry at the focal plane in terms of $\chi_\theta$ and $\chi_\phi$ becomes:
\begin{eqnarray}
  f_{diff}(\varphi) &=& \frac{h\overline{A_y}}{\pi}\Big[\left(\cos \chi_\phi P_t + \sin \chi_\phi P_l\right)\cos \varphi + \nonumber \\
    & & \left(\cos \chi_\phi \sin \chi_\theta P_l - \sin \chi_\phi \sin \chi_\theta P_t  \right)\sin \varphi \Big]
\end{eqnarray}
\subsubsection{COSY Spin Transport Matrix}
\paragraph{}
To perform the integration of equation \eqref{ThomasBMTformula}, the COSY\cite{COSY} model of the HMS was used. Coefficients of the polynomial expansion of the forward spin transport matrix in powers of $x_{tar}$, $y_{tar}$, $x'_{tar}$, $y'_{tar}$, and $\delta$ were calculated up to fifth order. The expansion is given by 
\begin{eqnarray}
  M_{ij}(x_{tar},y_{tar},x'_{tar},y'_{tar},\delta) &=& \sum_{\alpha,\beta,\lambda,\mu,\nu=0}^{\alpha+\beta+\lambda+\mu+\nu \le 5} C_{ij}^{\alpha\beta\lambda\mu\nu}(x_{tar})^\alpha (y_{tar})^\beta (x'_{tar})^\lambda (y'_{tar})^\mu (\delta)^\nu \nonumber \\ 
  & & \label{COSYspinexpansion}
\end{eqnarray}
where $M_{ij}$ with $i,j=x,y,z$ is the matrix element coupling the $i^{th}$ component of the spin at the focal plane to the $j^{th}$ component of the spin at the target. 

The rotation expressed by the expansion \eqref{COSYspinexpansion} takes place in the fixed transport coordinate system rather than the comoving coordinate system of the proton trajectory. To obtain the correct expression for the total rotation of the spin from the target to the focal plane, two additional rotations must be accounted for. First, the reaction-plane polarization components must be expressed in transport coordinates before calculating the COSY sums for each matrix element. This rotation, denoted $R_1$, multiplies $M$ from the right. The combined rotation $M R_1$ transports the polarization components from the reaction plane to the focal plane in transport coordinates. Finally, since the polarization components $P_x^{fpp}$ and $P_y^{fpp}$ appearing in the expression for the asymmetry are in the comoving coordinate system of the proton trajectory coinciding with the definition of the azimuthal angle $\varphi$, one additional rotation is required, which is denoted $R_2$.

The reaction plane is defined by the beam momentum $\mathbf{k}$ and the momentum transfer $\mathbf{q}$. The definitions of $P_t$ and $P_l$ in the derivation in chapter 1 are such that the normal polarization is taken to be positive along the direction of $\mathbf{q} \times \mathbf{k}$, and the transverse polarization is taken to be positive along the direction $\left(\mathbf{q} \times \mathbf{k}\right) \times \mathbf{q}$, which points in the direction of decreasing proton scattering angle $\theta_p$. To obtain the rotation $R_1$, the $\hat{t}$, $\hat{n}$, and $\hat{l}$ axes must be expressed in transport coordinates. These unit vectors are defined in terms of $\hat{k} = \mathbf{k}/\left|\mathbf{k}\right|$ and $\hat{q} = \mathbf{q}/\left|\mathbf{q}\right|$ as follows:
\begin{eqnarray}
  \hat{l} &\equiv& \hat{q} \\
  \hat{n} &\equiv& \frac{\hat{q} \times \hat{k}}{\left|\hat{q} \times \hat{k}\right|} \\
  \hat{t} &\equiv& \hat{n} \times \hat{l}
\end{eqnarray}
In transport coordinates, $\hat{k}$ and $\hat{q}$ are given by 
\begin{eqnarray}
  \hat{k} &=& \left(0, \sin \Theta_{HMS}, \cos \Theta_{HMS} \right) \\
  \hat{q} &=& \frac{1}{\sqrt{1+x_{tar}'^2+y_{tar}'^2}} \left(x'_{tar}, y'_{tar}, 1\right)
\end{eqnarray}
The rotation $R_1$ from reaction plane coordinates to transport coordinates $(x,y,z)$ is then given by the matrix whose \emph{columns} are equal to the unit vectors $(\hat{t}, \hat{n}, \hat{l})$:
\begin{eqnarray}
  \left(\begin{array}{c} P_x \\ P_y \\ P_z \end{array}\right) &=& \left(\begin{array}{ccc} \hat{t}_x & \hat{n}_x & \hat{q}_x \\ \hat{t}_y & \hat{n}_y & \hat{q}_y \\ \hat{t}_z & \hat{n}_z & \hat{q}_z \end{array}\right)\left(\begin{array}{c} P_t \\ P_n \\ P_l \end{array}\right)
\end{eqnarray}
A similar procedure is used to calculate $R_2$. The unit vectors $\hat{x}$, $\hat{y}$, and $\hat{z}$ of the local coordinate system defined by the proton trajectory are given in transport coordinates by
\begin{eqnarray}
  \hat{z} &=& \frac{1}{\sqrt{1+x_{fp}'^2 + y_{fp}'^2}}\left(x'_{fp},y'_{fp}, 1\right) \\
  \hat{y} &=& \frac{\hat{z} \times \hat{x}_{transport}}{\left|\hat{z} \times \hat{x}_{transport}\right|} \\
  \hat{x} &=& \hat{y} \times \hat{z} 
\end{eqnarray}
where $\hat{x}_{transport} = (1,0,0)$ is the fixed $x$ axis of the transport coordinate system. The rotation $R_2$ is then simply the matrix whose \emph{rows} are equal to the unit vectors $\hat{x}$, $\hat{y}$, and $\hat{z}$:
\begin{eqnarray}
  \left(\begin{array}{c} P_x^{fpp} \\ P_y^{fpp} \\ P_z^{fpp} \end{array}\right) &=& \left(\begin{array}{ccc} \hat{x}_x & \hat{x}_y & \hat{x}_z \\ \hat{y}_x & \hat{y}_y & \hat{y}_z \\ \hat{z}_x & \hat{z}_y & \hat{z}_z \end{array}\right)\left(\begin{array}{c} P_x \\ P_y \\ P_z \end{array}\right)_{transport}
\end{eqnarray}
The product of rotations $S \equiv R_2 M R_1$ gives the total rotation of the spin from the reaction plane to the focal plane in the correct coordinate system. In terms of the total rotation $S$, the polarization components measured by the FPP are related to the physical polarization transfer observables as follows:
\begin{eqnarray}
  \left(\begin{array}{c} P_x^{fpp} \\ P_y^{fpp} \\ P_z^{fpp} \end{array}\right) &=& \left(\begin{array}{ccc} S_{xx} & S_{xy} & S_{xz} \\ S_{yx} & S_{yy} & S_{yz} \\ S_{zx} & S_{zy} & S_{zz} \end{array}\right)\left(\begin{array}{c} P_t \equiv P_x^{tar} \\ P_n \equiv P_y^{tar} \\ P_l \equiv P_z^{tar} \end{array}\right) \label{totalrotation}
\end{eqnarray}
where the identification of $(\hat{t},\hat{n},\hat{l})$ with the $(x,y,z)$ axes of the reaction-plane coordinate system described in chapter 1 has been made explicit.
\begin{figure}[h]
  \begin{center}
    \includegraphics[width=.98\textwidth]{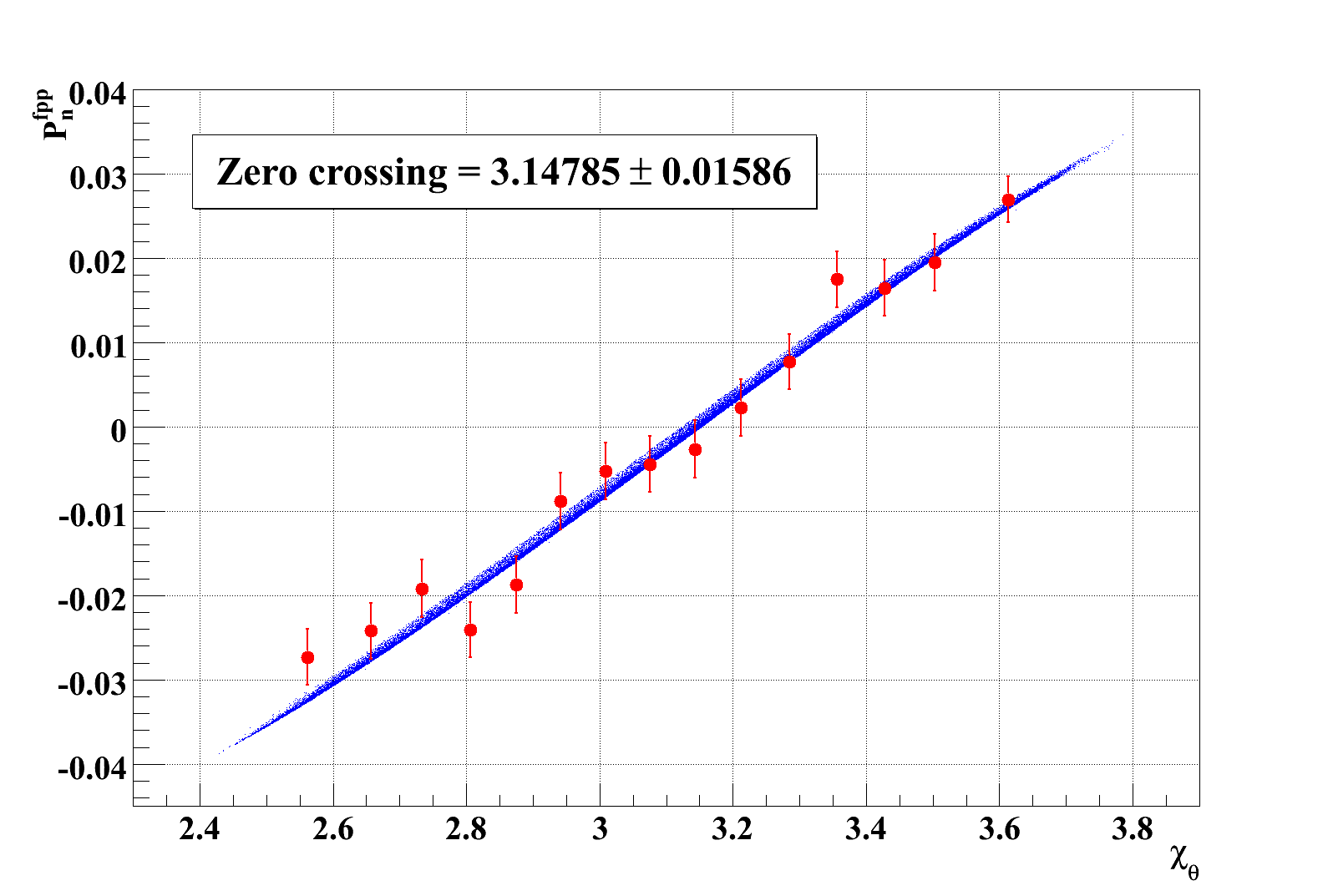}
  \end{center}
  \caption{\label{ZeroCrossing} Normal asymmetry as a function of the dispersive plane precession angle $\chi_\theta$, $Q^2 = 5.2\ GeV^2$. Red points are the data with statistical errors. Blue points are $h\overline{A_y}\left(SxzP_z + SxxP_x\right)$, where $\overline{A_y}$ is the average analyzing power of the sample, $h$ is the beam polarization, $P_z = P_l$ and $P_x = P_t$ are the longitudinal and transverse polarization components at the target, and the matrix elements $S_{xz}$ and $S_{xx}$ are those calculated by COSY.}
\end{figure}

To see that the spin really behaves according to equation \eqref{ThomasBMTformula}, it is instructive to bin the data as a function of the dispersive and non-dispersive precession angles $\chi_\theta$ and $\chi_\phi$ defined above and plot the measured asymmetries as a function of these angles. Figure \ref{ZeroCrossing} shows the measured normal asymmetry $P_x^{fpp}$ at the focal plane as a function of $\chi_\theta$ at $Q^2 = 5.2\ GeV^2$, for which the precession angle corresponding to the 25$^\circ$ central bend angle of the HMS is close to 180 degrees. In the ideal dipole approximation, the longitudinal polarization at the target precesses by $\chi_\theta$ in the $xz$ plane, the transverse polarization does not rotate, and the normal polarization is zero. Therefore, in this approximation, the normal polarization at the focal plane should cross zero at $\chi_\theta = 180^\circ$, which indeed appears to be the case. The data in figure \ref{ZeroCrossing} were fit to the form $P_n = a \sin(\chi_\theta + \delta )$, and the phase shift $\delta$ differs from $\pi$ by $6.3 \pm 15.9$ mrad, i.e., the zero crossing is consistent with $\pi$. Small corrections to the zero crossing arise due to deviations from the ideal dipole approximation. The primary mechanism for a displacement of the zero crossing is the extent to which the transverse polarization mixes with normal polarization in the quadrupoles before the main dipole precession, but this effect is quite small, because of the optical properties of the HMS and because in this case the magnitude of the transverse polarization is approximately a factor of ten smaller than the longitudinal polarization. Any possible offset of the expected or measured zero crossing from $\pi$ is certainly not distinguishable within the statistical uncertainty of the data shown in figure \ref{ZeroCrossing}. The figure also shows the expected variation of the asymmetry across the acceptance calculated using COSY, which agrees very well with the data.  
\begin{figure}[h]
  \begin{center}
    \includegraphics[width=.98\textwidth]{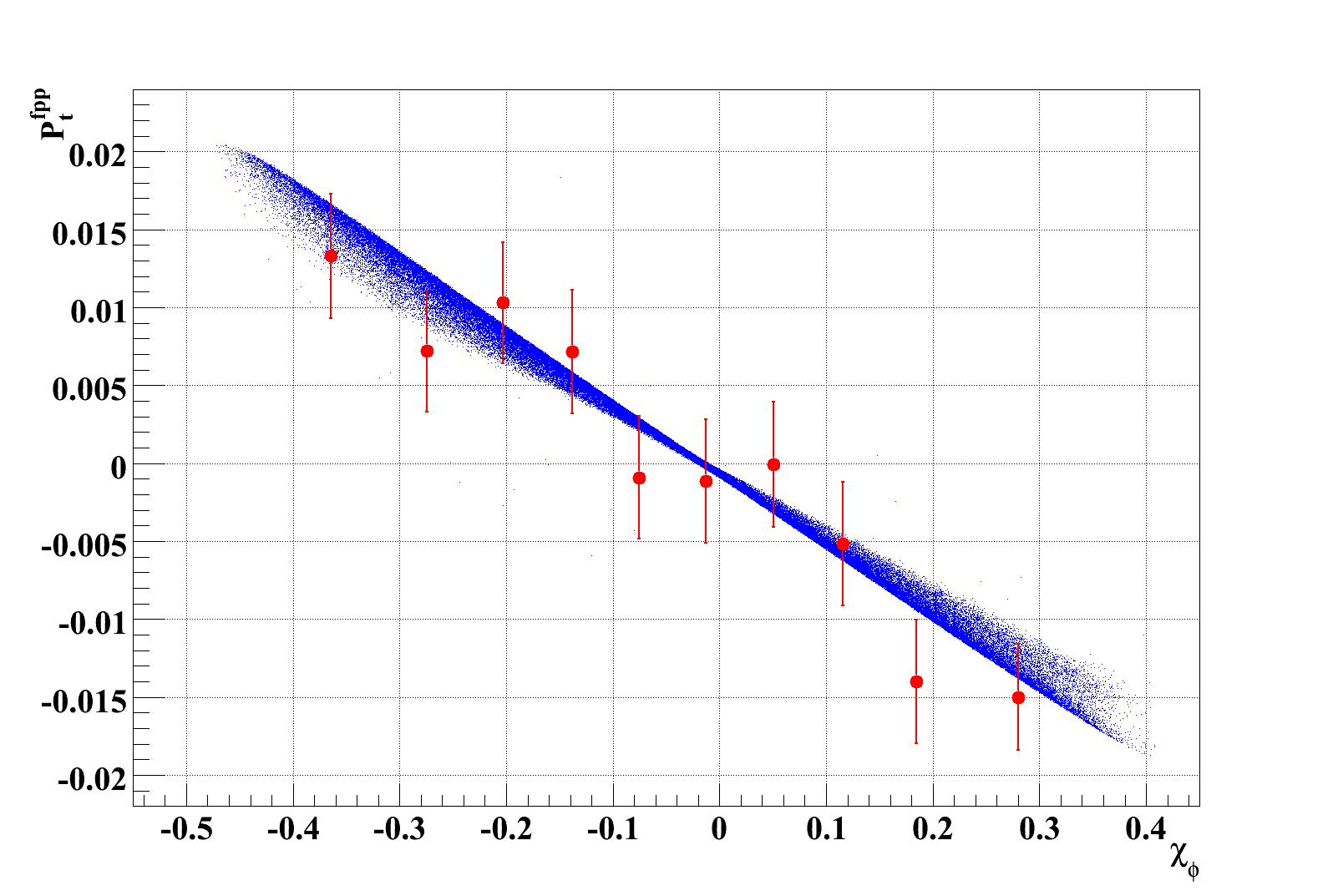}
  \end{center}
  \caption{\label{PtChiPhi} Transverse asymmetry as a function of the non-dispersive precession angle $\chi_\phi$, $Q^2 = 8.5\ GeV^2$. Red points are data with statistical errors. Blue points are $h\overline{A_y}\left(S_{yz}P_z + S_{yx}P_x\right)$, where $\overline{A_y}$ is the average analyzing power, $h$ is the beam polarization, $P_x = P_t$ and $P_z = P_l$ are the transverse and longitudinal polarization components at the target, and the matrix elements $S_{yx}$ and $S_{yz}$ are calculated by COSY.}
\end{figure}
The transverse asymmetry also varies across the acceptance in the predicted fashion. Figure \ref{PtChiPhi} shows the measured transverse asymmetry as a function of $\chi_\phi$ at $Q^2 = 8.5\ GeV^2$ (red points) together with the variation predicted by COSY.

Before discussing the extraction of the polarization observables from the measured asymmetry and the calculated spin precession, some additional remarks on the geometric approximation are needed. The precession angles $\chi_\theta$ and $\chi_\phi$ as defined above are understood to represent precession relative to the trajectory. The rotation matrix given in \eqref{geometricapproximation} does not include the rotations $R_1$ from the reaction plane to transport coordinates and $R_2$ from transport coordinates to comoving coordinates at the focal plane. As presented, the geometric approximation should not include these extra rotations, since the dispersive and non-dispersive precessions $\chi_\theta$ and $\chi_\phi$ are understood to take place in the comoving coordinates of the proton. This approach is not entirely satisfactory, since the orientations of the trajectory and the polarization components with respect to the fixed dispersive and non-dispersive planes of the HMS are constantly changing, and even the initial, reaction-plane longitudinal (transverse) polarization component does not lie strictly in the dispersive (non-dispersive) plane. Furthermore, it is not at all obvious that $P_x^{fpp}$ and $P_y^{fpp}$ correspond exactly to the components of the polarization assumed to undergo precession by $\chi_\theta$ and $\chi_\phi$. Though it is perfectly reasonable to neglect these conceptually troubling issues since \eqref{geometricapproximation} is only meant as an approximation, an alternative approach is possible that improves its accuracy and logical consistency. 

Instead of using equation \eqref{geometricapproximation} alone, the precession can be viewed as occuring in the fixed transport coordinate system, as in the COSY expansion. In this approach, the total rotation $S$ is given by $R_2 M R_1$, with $M$ given by a modified version of \eqref{geometricapproximation}, and with the precession angles redefined as
\begin{eqnarray}
  \chi_\theta &\equiv& \gamma \kappa_p \left(\Theta_0 + \theta_{tar} - \theta_{fp}\right) + \theta_{tar} - \theta_{fp} \nonumber \\
  \chi_\phi &\equiv& \gamma \kappa_p \left(\phi_{fp} - \phi_{tar}\right) + \phi_{fp} - \phi_{tar} \label{modifiedprecessionangledefinitions}
\end{eqnarray}
In this picture, the small rotations from reaction-plane to transport coordinates and from transport coordinates to comoving coordinates at the focal plane are applied, and since the precession is now assumed to take place in a fixed coordinate system, the precession angles $\chi_\theta$ and $\chi_\phi$ must now be understood as absolute precession angles (in transport coordinates). 

To obtain \emph{absolute} precession angles which correspond to the \emph{relative} precessions of the geometric approximation, the required modification is to simply add the trajectory bend angles in the dispersive\footnote{In transport coordinates, the central bend angle $\Theta_0 = 25^\circ$ corresponds to a trajectory bend of zero, so the correct modification to $\chi_\theta$ in this case is to add $\Delta \theta$ only, \emph{not} $\Delta \theta + \Theta_0$. Rotation by $\Theta_0$ is built in to the definition of the coordinate system.} and non-dispersive planes to $\chi_\theta$ and $\chi_\phi$, as in equations \eqref{modifiedprecessionangledefinitions}. With this modification, the coordinate system ambiguity is eliminated and the extra rotations $R_1$ and $R_2$ are no longer neglected. While neither approximation should be thought of as representing the true precession, and the differences between the two approximations are typically quite small, the latter approximation tends to give results for the form factor ratio that are in slightly better agreement with the full COSY calculation.

\subsection{Maximum-Likelihood Analysis of Polarization Observables}
\paragraph{}
The experimental asymmetry for the $\pm$ helicity state in terms of the reaction-plane polarization components $\mathbf{P}$ and the total spin rotation $S$ is given by
\begin{eqnarray}
  N^\pm(\mathbf{P}, S) &=& \frac{N_0^\pm \varepsilon}{2\pi}\Bigg[1+\left(a_1 \pm hA_y\sum_{j=x,z} S_{yj} P_j + A_yS_{yy}P_y\right)\cos \varphi + \nonumber \\
    & & \left(b_1 \mp hA_y\sum_{j=x,z} S_{xj}P_j - A_yS_{xy}P_y\right)\sin \varphi + \nonumber \\
    & & + a_2 \cos 2\varphi + b_2 \sin 2\varphi + \ldots \Bigg] \label{totalasymmetry}
\end{eqnarray}
The asymmetry \eqref{totalasymmetry} allows for the possibility of an induced polarization $P_y$, which is independent of the beam polarization and helicity state. Although $P_y$ is identically zero in elastic scattering in the Born approximation, a non-zero $P_y$ is still possible due to either inelastic background processes with non-zero induced polarization, effects beyond the Born approximation, or false asymmetries which mimic induced polarization. The $p$ and $\vartheta$ dependence of $\varepsilon$ and $A_y$ are implicit. The likelihood function of the polarizations $\mathbf{P}$ is obtained by forming the product of \eqref{totalasymmetry} over all contributing events:
\begin{eqnarray}
  \mathcal{L}(\mathbf{P}) &=& \frac{1}{2\pi} \prod_{i=1}^{N_{event}} \Bigg[ 1+\left(a_1 + h\epsilon_i A_y^{(i)} \sum_{j=x,z} S_{yj}^{(i)} P_j + A_y^{(i)} S_{yy}^{(i)}P_y\right)\cos \varphi_i + \nonumber \\
    & & \left(b_1 - h\epsilon_iA_y^{(i)}\sum_{j=x,z} S_{xj}^{(i)}P_j - A_y^{(i)}S_{xy}^{(i)}P_y\right)\sin \varphi_i + \nonumber \\
    & & + a_2 \cos 2\varphi_i + b_2 \sin 2\varphi_i + \ldots \Bigg] \label{likelihoodfunc}
\end{eqnarray}
In the likelihood function \eqref{likelihoodfunc}, the sign of the beam helicity for each event is absorbed into the factors $\epsilon_i$. The analyzing power $A_y$, the spin rotation $S$, and the azimuthal scattering angle $\varphi$ are unique for each event. The goal of the analysis is to find the values of $\mathbf{P}$ which maximize $\mathcal{L}$. As is common practice, the product is converted into a sum by forming the logarithm $\ln \mathcal{L}$. Since the natural logarithm is a monotonically increasing function of its argument, the maximum of $\ln \mathcal{L}$ (or, alternatively, the minimum of $-\ln \mathcal{L}$) coincides with the maximum of $\mathcal{L}$, with the advantage that sums over all events are much more convenient to work with for this particular problem than products over all events.

The maximum-likelihood estimates of the polarizations $\mathbf{P}$ are found by solving the three simultaneous equations 
\begin{eqnarray}
  \frac{\partial \ln \mathcal{L}}{\partial P_j} &=& 0, j=x,y,z
\end{eqnarray}
On the way to the solution of the problem, it is helpful to regroup the expression in square brackets in \eqref{likelihoodfunc} using a convenient shorthand: 
\begin{eqnarray}
  \Bigg[ \ldots \Bigg]_i &=& \Bigg[1 + \lambda_0^{(i)} + \sum_{j=x,y,z} \lambda_j^{(i)}P_j \Bigg] 
\end{eqnarray} 
where $\lambda_0$ represents the sum of all false asymmetry terms and $\lambda_j$ represents the coefficient of polarization component $P_j$ in the expression for the contribution of the $i^{th}$ event to the asymmetry:
\begin{eqnarray}
  \lambda_0^{(i)} &\equiv& a_1 \cos \varphi_i + b_1 \sin \varphi_i + a_2 \cos 2\varphi_i + b_2 \sin 2\varphi_i + \ldots \\
  \lambda_x^{(i)} &\equiv& h\epsilon_i A_y^{(i)}\left(S_{yx}^{(i)} \cos \varphi_i - S_{xx}^{(i)} \sin \varphi_i \right) \\
  \lambda_y^{(i)} &\equiv& A_y^{(i)}\left(S_{yy}^{(i)} \cos \varphi_i - S_{xy}^{(i)} \sin \varphi_i \right) \\
  \lambda_z^{(i)} &\equiv& h\epsilon_i A_y^{(i)}\left(S_{yz}^{(i)} \cos \varphi_i - S_{xz}^{(i)} \sin \varphi_i \right)
\end{eqnarray}
Using this notation, the logarithm of the likelihood function becomes
\begin{eqnarray}
  \ln \mathcal{L} &=& \ln \frac{1}{2\pi} + \sum_{i=1}^{N_{event}} \ln\left[1 + \lambda_0^{(i)} + \sum_{j=x,y,z} \lambda_j^{(i)} P_j \right]
\end{eqnarray}
and the partial derivatives become 
\begin{eqnarray}
  0 = \frac{\partial \ln \mathcal{L}}{\partial P_j} &=& \sum_{i=1}^{N_{event}} \frac{\partial}{\partial P_j} \left(\ln \left[1 + \lambda_0^{(i)} + \sum_{j=x,y,z} \lambda_j^{(i)} P_j\right]\right) \label{systemofequationslikelihood}
\end{eqnarray}
The coupled, nonlinear system of equations \eqref{systemofequationslikelihood} is formidable and does not lend itself to a simple solution. However, if the asymmetry is sufficiently ``small'', it can be converted to a system of linear equations by Taylor-expanding the logarithm about $x=0$:
\begin{eqnarray}
  \ln (1 + x) &=& \sum_{n=0}^{\infty} \left.\frac{d^n}{dx^n} \ln(1+x)\right|_{x=0} \frac{x^n}{n!} = x  - \frac{x^2}{2} + \mathcal{O} x^3 + \ldots
\end{eqnarray}
If only the terms up to second order are retained, then the partial derivatives are linear in the polarizations $P_j$:
\begin{eqnarray}
  \frac{\partial \ln \mathcal{L}}{\partial P_j} = 0 &=& \sum_{i=1}^{N_{event}} \left[\lambda_j^{(i)} - \left(\lambda_0^{(i)} + \sum_{k=x,y,z} \lambda_k^{(i)} P_k\right)\lambda_j^{(i)}\right] \\
  \Rightarrow \sum_{i=1}^{N_{event}} \lambda_j^{(i)} \left(1 - \lambda_0^{(i)}\right) &=& \sum_{i=1}^{N_{event}} \sum_{k=x,y,z} \lambda_j^{(i)} \lambda_k^{(i)} P_k 
\end{eqnarray}
This system of linear equations can be written as a matrix equation $\mathbf{b} = A \mathbf{P}$ thusly:
\begin{eqnarray}
  \sum_{i=1}^{N_{event}}\left(\begin{array}{c} \lambda_x^{(i)}\left(1-\lambda_0^{(i)}\right) \\ \lambda_y^{(i)}\left(1-\lambda_0^{(i)}\right) \\ \lambda_z^{(i)}\left(1-\lambda_0^{(i)}\right) \end{array}\right) &=& \sum_{i=1}^{N_{event}} \left(\begin{array}{ccc} \left(\lambda_x^{(i)}\right)^2 & \lambda_x^{(i)}\lambda_y^{(i)} & \lambda_x^{(i)} \lambda_z^{(i)} \\ \lambda_y^{(i)} \lambda_x^{(i)} & \left(\lambda_y^{(i)}\right)^2 & \lambda_y^{(i)} \lambda_z^{(i)} \\ \lambda_z^{(i)} \lambda_x^{(i)} & \lambda_z^{(i)} \lambda_y^{(i)} & \left(\lambda_z^{(i)}\right)^2 \end{array}\right)\left(\begin{array}{c} P_x \\ P_y \\ P_z \end{array}\right) \nonumber \\ 
  & & \label{loglikelihoodlinearized}
\end{eqnarray}
Each element of $A$ and $\mathbf{b}$ is a sum over all events of combinations of the various coefficients $\lambda$. The solution of the system is obtained by inverting the matrix $A$. The solution vector is $\mathbf{P} = A^{-1} \mathbf{b}$, and the covariance matrix is given by $A^{-1}$. In particular, the variances of the polarization components are given by the diagonal elements of the covariance matrix:
\begin{eqnarray}
  \Delta P_j &=& \sqrt{\left(A^{-1}\right)_{jj}}
\end{eqnarray}
Of interest for the extraction of the form factor ratio is the ratio of $P_x ( = P_t )$ and $P_z ( = P_l )$. Adopting the shorthand $K = \frac{E_e+E'_e}{2M_p}\tan \frac{\theta_e}{2}$ for the kinematic factor appearing in the expression for the form factor ratio and $\overline{K}$ for the average of this factor over all contributing events, the extracted form factor ratio is given by 
\begin{eqnarray}
  \mu_p \frac{G_E^p}{G_M^p} &=& -\mu_p\overline{K} \frac{P_t}{P_l}
\end{eqnarray}
The covariance\footnote{The matrix $A$ defined in \eqref{loglikelihoodlinearized} and its inverse are always symmetric} of $P_t$ and $P_l$ is given by $\rho = \left(A^{-1}\right)_{xz}$. The statistical error on $R = \mu_p G_E^p / G_M^p$ is given by 
\begin{eqnarray}
  \left(\frac{\Delta R}{R} \right)^2  &=& \sqrt{\left(\frac{\Delta P_t}{P_t}\right)^2 + \left(\frac{\Delta P_l}{P_l}\right)^2 - 2\frac{\rho}{P_t P_l}}
\end{eqnarray}
where the uncertainty in the kinematic factor $\overline{K}$ has been neglected. The correlation term is generally quite small compared to the quadrature sum of the variances of $P_t$ and $P_l$ and has a negligible impact on the statistical uncertainty in the form factor ratio.

It is reasonable to ask whether the linearization of the problem achieved by truncating the Taylor expansion of $\ln(1+x)$ at second order has any significant effect on the results. The variable $x$ corresponds to the asymmetry. In this experiment, the helicity-dependent asymmetry determined by the transferred polarization and the analyzing power, is never greater than about 15\% for events near the maximum of the analyzing power distribution, and never more than about 10\% when averaged over the full acceptance. At $x=-0.15$, the fractional difference $\frac{x-\frac{x^2}{2}}{\ln(1+x)}-1$ is -0.78\%, while at $x=+0.15$, it is -0.72\%. It appears, then, that the expansion of the log-likelihood function to second order is a very good approximation to the true log-likelihood function for the full range of asymmetries measured in this experiment. However, the relevant question is not the accuracy of the approximation to $\ln \mathcal{L}$ but rather the effect of this approximation on the accuracy of the maximum-likelihood values of the parameters. The linearization \eqref{loglikelihoodlinearized} of equations \eqref{systemofequationslikelihood} leads to estimators for the polarization $\hat{\mathbf{P}}$ that are mathematically equivalent to those of the ``weighted sum'' method described in section 2.2 of \cite{Besset}, in which the authors showed that these estimators are unbiased and very nearly as efficient as the true maximum-likelihood estimators over a wide range of values of the asymmetry which comfortably includes the $\pm15$\% region of this experiment. Therefore, it was concluded that this approximation engenders a negligible loss of information compared to an exact maximum-likelihood estimate while dramatically simplifying the calculations involved.

\subsubsection{Cancellations}
\paragraph{}
The matrix elements on the right hand side of equation \eqref{loglikelihoodlinearized} are proportional to $\lambda^2$, while the components of the vector on the left hand side are proportional to $\lambda(1-\lambda_0)$. The $\lambda$'s are proportional to the analyzing power $A_y$ and, in the case of the \emph{transferred} polarization components $P_x$ and $P_z$, the beam polarization. The beam polarization is known to within 1\% or less from the M\"{o}ller measurements (see table \ref{MollerTable}). The analyzing power, on the other hand, is not \emph{a priori} known, although dedicated measurements of the $\vec{p}+CH_2$ analyzing power covering the momentum range of this experiment have been previously performed\cite{Ay_pCH2}. For the measurement of the form factor ratio it is not even necessary to know the analyzing power since it enters the ratio $P_t/P_l$ in both the numerator and the denominator and cancels out. The same is true of the beam polarization. Given the independence of the extracted form factor ratio on the analyzing power used as input to the likelihood analysis, and the fact that the Born approximation values of $P_t$\eqref{Pt} and $P_l$\eqref{Pl} depend only on the ratio of form factors, the analyzing power can be rigorously extracted from the data of this experiment by comparing the extracted values of $P_t$ and $P_l$ to the values $P_t(r=G_E/G_M)$ and $P_l(r=G_E/G_M)$ expected in the Born approximation. The resulting analyzing power can then be used to improve the statistical precision of the likelihood analysis in a second pass over the data as described below.

In section \ref{fpasymsection}, the cancellation of the false asymmetries in the helicity difference distribution $f_{diff}(\varphi)$ was demonstrated. It is not obvious that the false asymmetry appearing in the vector on the left hand side of equations \eqref{loglikelihoodlinearized} strictly cancels in the determination of the maximum-likelihood estimators $\hat{\mathbf{P}}$ to all orders. Since the factors $\lambda_x^{(i)}$ and $\lambda_z^{(i)}$ multiplying $1-\lambda_0^{(i)}$ are proportional to the sign of the beam helicity $\epsilon_i$ and the numbers of incident protons in the $+$ and $-$ helicity states are equal for all practical purposes of this experiment, the contribution of $\lambda_0$ to the sums $\sum_i \lambda_x(1-\lambda_0)$ and $\sum_i \lambda_z(1-\lambda_0)$ must be quite small. But it cannot generally be assumed to vanish completely. The effect of the false asymmetry on the form factor ratio was studied in several ways.
\begin{itemize}
\item The result of analyzing the data assuming $\lambda_0=0$ was compared to the result of analyzing the data with false asymmetry coefficients $(a_1, b_1, a_2, b_2)$ obtained from Fourier analysis of the helicity sum spectrum $f_{sum}(\varphi)$ included. The difference between the uncorrected and corrected results indicates the size of the false asymmetry effect.
\item Both of the above results were compared to the value of the form factor ratio obtained from Fourier analysis of the helicity difference spectrum $f_{diff}(\varphi)$, for which the cancellation of the false asymmetry is ``exact''. The focal plane asymmetries and their uncertainties were obtained by the method described in section \ref{fpasymsection}, and the precession was calculated using the acceptance-averaged COSY matrix elements. Because the precession was not handled event-by-event as in the likelihood analysis, differences between this method and the likelihood method did not necessarily indicate the effect of a false asymmetry.
\end{itemize}
The cancellations of the beam polarization, the analyzing power, and the false asymmetry in the extraction of $G_E^p/G_M^p$ greatly reduce the systematic uncertainty in the result. The primary remaining sources of systematic uncertainty are 
\begin{itemize}
\item The uncertainty in the calculation of the spin precession in the HMS, which consists of the uncertainties arising from errors in the reconstructed quantities used as input to the calculation, and uncertainties in the model of the HMS used to derive the COSY expansion of the precession matrix elements.
\item The uncertainty in the correction to the result arising from the inelastic background and its polarization.
\item The uncertainty in the reconstructed scattering angles in the FPP, particularly the azimuthal angle $\varphi$.
\end{itemize}
These and other, less significant uncertainties will be addressed in the sections that follow.

\subsection{Analyzing Power}
\paragraph{}
The analyzing power is the momentum and angle-dependent constant of proportionality between the polarization of the incident proton and the measured asymmetry. If the proton polarization is known, then the size of the asymmetry measures the analyzing power. In this experiment, the analyzing power is initially unknown. The result of extracting the transferred polarization components assuming $A_y = 1$ measures the combinations $hA_yP_t$ and $hA_yP_l$. The beam polarization $h$ is known from the M\"{o}ller measurements. The form factor ratio is independent of the analyzing power. The polarization components can be expressed in terms of the form factor ratio as 
\begin{eqnarray}
  r &\equiv& \frac{G_E}{G_M} \\
  P_t &=& -2\sqrt{\tau(1+\tau)} \tan \frac{\theta_e}{2} \frac{r}{r^2 + \frac{\tau}{\epsilon}} \\
  P_l &=& \sqrt{\tau(1+\tau)} \tan^2 \frac{\theta_e}{2} \frac{E_e+E'_e}{M_p} \frac{1}{r^2 + \frac{\tau}{\epsilon}} \label{polff}
\end{eqnarray}
so that the analyzing power can be obtained from the ratio of the measured polarization (assuming $A_y=1$ in the analysis) to the predicted polarization for the measured form factor ratio:
\begin{eqnarray}
  A_y &=& \frac{P_t^{meas.}}{hP_t(r)} = \frac{P_l^{meas.}}{hP_l(r)} \label{Ayformula}
\end{eqnarray}
with an uncertainty obtained by propagating the uncertainties in $r$, $h$, and the measured transverse and longitudinal asymmetries through the formulae \eqref{Ayformula}. In the results of this experiment, the longitudinal component $P_l$ is typically ``large'' in magnitude while the transverse component $P_t$ is typically ``small'', while the absolute uncertainties on each component are similar, meaning $P_l$ is usually measured with a smaller relative uncertainty. This means that $P_l$ usually provides the more precise determination of the analyzing power. The measurements $A_y(P_l)$ and $A_y(P_t)$ are combined in a weighted average which is typically dominated by $A_y(P_l)$. 

Examining the formula for $A_y(P_l)$ in particular, the factor $r^2 + \frac{\tau}{\epsilon}$ appears in the numerator, while the denominator is independent of $r$. In this experiment, $\frac{\tau}{\epsilon}$ is a number significantly larger than 1, and $r^2$ is typically a number significantly smaller than 1, so $P_l(r)$ (and thus $A_y(P_l)$) depends mainly on kinematic factors and is not especially sensitive to $r$. $P_t(r)$, on the other hand, is highly sensitive to $r$. The uncertainty in $A_y(P_l)$ is a function of the uncertainties in $r$, $h$, and $P_l^{meas.}$:
\begin{eqnarray}
  \frac{\partial A_y}{\partial r} &=& \frac{2rP_l^{meas.}}{ha} = \frac{2}{r}\left(A_y - \frac{\tau P_l^{meas.}}{\epsilon ha}\right) \\
  a &\equiv& \sqrt{\tau(1+\tau)}\tan^2\frac{\theta_e}{2}\frac{E_e+E'_e}{M_p} \\
  \frac{\partial A_y}{\partial h} &=& - \frac{A_y}{h} \\
  \frac{\partial A_y}{\partial P_l^{meas.}} &=& \frac{A_y}{P_l^{meas.}} \\
  \Rightarrow \frac{\Delta A_y(P_l)}{A_y(P_l)} &=& \sqrt{\left(\left[1-\frac{\tau P_l^{meas.}}{\epsilon ha A_y}\right]\frac{2\Delta r}{r}\right)^2+\left(\frac{\Delta h}{h}\right)^2 + \left(\frac{\Delta P_l^{meas.}}{P_l^{meas.}}\right)^2} \nonumber \\
  & &
\end{eqnarray}

The analyzing power was obtained separately for single-track and multiple-track events. 
\begin{figure}[h]
  \begin{center}
    \includegraphics[angle=90,width=1.00\textwidth]{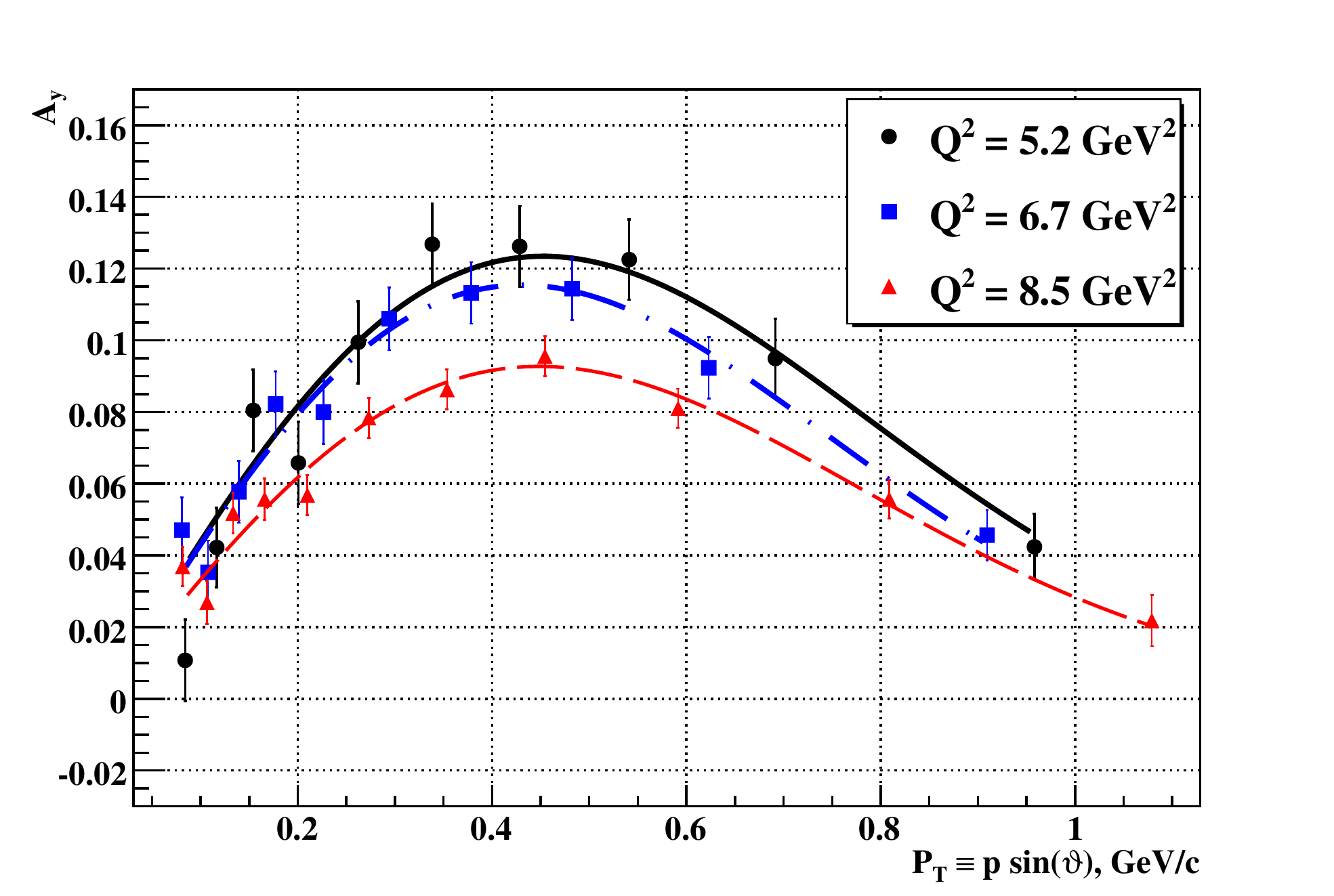}
  \end{center}
  \caption{\label{Ayfig} $A_y(p_T)$ extracted from the data of this experiment. }
\end{figure}
Figure \ref{Ayfig} shows the extracted single-track analyzing power for the kinematics of E04-108, as a function of the transverse momentum $p_T=p\sin \vartheta$. The proton momentum entering the expression for $p_T$ is corrected for energy loss. In terms of $p_T$, the angular distribution of $A_y$ has a more or less constant shape which peaks at around $p_T \approx 0.4$ GeV/c. Within uncertainties, the maximum analyzing power $A_y^{max}$ is proportional to $p^{-1}$ as expected. The errors shown are statistical only, and do not take into account the uncertainty in the beam polarization, which is assumed equal to the value given in \ref{MollerTable}. For the purpose of this analysis, the absolute normalization of the analyzing power distribution is unimportant. It is the relative variation of $A_y$ with scattering angle that is of interest. 

Using the measured $A_y(p_T)$ as an input to the likelihood analysis as an event weight improves the statistical uncertainty by giving more weight to events with large analyzing power and reducing the influence of events with lower analyzing power. This is accomplished by fitting the following parametrization to the observed $A_y(p_T)$:
\begin{eqnarray}
  A_y(p_T) &=& A_0 p_T e^{-bp_T^2}
\end{eqnarray}
with adjustable parameters $A_0$ and $b$ which determine, respectively, the height and the width of the curve. An adequate description of the measured analyzing power for the entire interesting range of $p_T$ is achieved for all three kinematics. 

\chapter{Results}
\paragraph{}
In this chapter, the results of experiment E04-108 are presented and the analysis of uncertainties in the results is discussed. The thesis concludes with a discussion of the implications of these results for the understanding of nucleon structure. For the final analysis, the maximum likelihood method was used to extract the polarization transfer observables at the target. Elastic event selection cuts were tuned to optimize the tradeoff between the number of elastic events rejected and the amount of inelastic background accepted. The spin transport model used for the final analysis is COSY. The results of analyzing the data using the geometric approximation are presented for comparison and used to estimate the model uncertainty in the precession calculation.

No radiative corrections have been applied to the results presented here. The model independent radiative corrections to polarization transfer observables in elastic $eN$ scattering were calculated to lowest order in $\alpha$ by Afanasev et al.\cite{AfanasevRadCorr}. Model-independent radiative corrections include internal and external Bremsstrahlung, vacuum polarization, and electron-photon vertex corrections, and can be calculated exactly. Reversing the beam polarization cancels the external Bremsstrahlung corrections to the polarization asymmetries, leaving only internal corrections that can affect the asymmetries. The observed cross section can be expressed as a sum of two terms
\begin{eqnarray}
  \sigma^{u,p}_{obs.} &=& (1+\delta)\sigma_0^{u,p} + \sigma_R^{u,p}
\end{eqnarray}
where $u,p$ denote unpolarized and polarized cross sections, respectively, and $\sigma_0$ is the Born approximation cross section. The radiative corrections enter through the multiplicative factor $(1+\delta)$, which comes from the vacuum polarization and vertex corrections, and the additive radiation cross section $\sigma_R$, which comes from Bremsstrahlung. The observed asymmetry is given by the ratio of the polarized and unpolarized cross sections $\sigma_{obs.}^p/\sigma_{obs.}^u$. The multiplicative factor $(1+\delta)$ can be large (10-30\%), but it cancels exactly in the numerator of the expression for the absolute correction to the asymmetry. For this reason, even in situations where the radiative correction to the cross section is large, the correction to the polarization asymmetry is still rather small. The additive portion of the radiative correction will not generally cancel and can give an important correction to the asymmetry of order several percent. However, this correction can by reduced by applying a cut on the inelasticity defined either through missing energy or missing mass.  

In \cite{AfanasevRadCorr}, the relative radiative correction to the ratio of polarization components was calculated for the kinematics of the Hall A recoil polarization experiments \cite{Punjabi05,Gayou02}, resulting in corrections no larger than 1\% for any of the kinematics. Given the high degree of similarity between the Hall A experiments and this experiment, and the tight inelasticity cuts applied in this analysis to both electron and proton kinematics (see section \ref{elasticcutssection})  to suppress the inelastic background, model-independent radiative corrections to the results presented in this thesis are expected to be at worst similar to, and probably smaller than the 1\% level predicted by \cite{AfanasevRadCorr}, although the general tendency for radiative corrections to increase with $Q^2$ may somewhat contradict this statement. Assuming relative corrections to the polarization ratio of 1\% or less, the absolute correction to the form factor ratio for the $Q^2 = 8.5\ GeV^2$ kinematics of this experiment would be $\Delta R \le 0.002$. Even assuming an order-of-magnitude larger than expected radiative correction to the polarization ratio of 10\%, the absolute correction to $R = \mu_p G_E^p/G_M^p$ would be $0.02$. Compared to the statistical and systematic uncertainties in the results of this experiment, \emph{relative} radiative corrections to the ratio $P_t/P_l$ at the level of several percent are negligible. These corrections would only become important for an experiment with much higher statistical precision. 

The argument for neglecting radiative corrections to the results of experiment E04-108 can be summarized as follows: The polarization asymmetries experience partial cancellations of radiative corrections because they are ratios of cross sections. The remaining correction comes from internal Bremsstrahlung, and is estimated to be at the level of several percent. The ratio of polarization observables $P_t/P_l$ is a ratio of asymmetries, which in turn is a ratio of ratios of cross sections, and it experiences radiative corrections which are even smaller than the corrections experienced by the asymmetries themselves due to further partial cancellations in the ratio. When inelasticity cuts are applied to the kinematics of both the measured electron and proton, these corrections can be further reduced, in principle, and the potential size of these corrections is negligible compared to the uncertainties in the results themselves. Having justified the decision not to apply radiative corrections to the data, the following sections document the analysis procedure through which the final results are obtained.
\section{Final Cuts}
\paragraph{}
The first decision taken for the final analysis is the choice of cuts applied to select events. These cuts fall into three categories.
\begin{enumerate}
\item \emph{Acceptance cuts} are applied to the reconstructed target variables $x'_{tar}$, $y'_{tar}$, $\delta$, and $y_{tar}$. The HMS angular acceptance is independent of the kinematics. The following loose cuts were applied to the reconstructed proton angles at the target:
  \begin{eqnarray}
    \left|x'_{tar}\right| &\le& 0.08 \\
    \left|y'_{tar}\right| &\le& 0.04
  \end{eqnarray}
  For all kinematics except $Q^2 = 5.2\ GeV^2$, the target was the 20 cm LH$_2$ target offset by $z_0=3.84\ cm$. For these kinematics, the following cut was applied to $z_{beam}$:
  \begin{eqnarray}
    \left|z_{beam} - z_0\right| &\le& 10\ \mbox{cm} + dz
  \end{eqnarray}
  where $dz = 0.6\ cm/\sin \Theta_{HMS}$ represents a roughly 3$\sigma$ tolerance\footnote{The background from the target endcaps after applying elastic kinematic cuts was not significant enough to necessitate applying $z$ cuts to eliminate events from the target walls, which would have resulted in a significant loss of elastic $ep$ statistics, since scattering from hydrogen overlaps with scattering from the target endcaps within the vertex resolution of the HMS.} in $z$ due to the HMS resolution in $y_{tar}$ and $y'_{tar}$. For $Q^2=5.2\ GeV^2$, the 15 cm LH$_2$ target was used. In this case, the offset was $z_0 \approx 1.5\ cm$, $dz$ was unchanged, and instead of 10 cm for the target half-length, 7.5 cm was used. Finally, a loose momentum acceptance cut was applied. This cut, however, had very little additional influence on the selection of events because it was redundant with the angular acceptance cut after elastic kinematic cuts, for which $p_p$ and $\theta_p$ are correlated. 
\item \emph{Elastic kinematic cuts} consist of cuts applied to the proton inelasticity variable $\Delta_p$ and the difference between the detected electron position at BigCal and its expected position from elastic kinematics; i.e., the two-dimensional elliptical cut defined in section \ref{elasticcutssection}. The width of these cuts was optimized to accept the largest possible fraction of elastic events for which the improvement in statistical uncertainty was not outweighed by the inelastic background contamination. 
\item \emph{FPP track cuts} consist of minimum and maximum scattering angles $\vartheta_{min}$ and $\vartheta_{max}$, a maximum distance of closest approach $s_{max}$, and $z_{close}$ limits chosen to correspond to the physical extent of the CH$_2$ with an extra tolerance of several cm to allow for the resolution of $z_{close}$, which blows up approximately as $1/\tan \vartheta$ at small angles. Only the $\vartheta$ cuts depend on kinematics. The final cuts are given as follows:
  \begin{eqnarray}
    s_{close} &\le& 3(6)\ \mbox{cm, for FPP1(2)} \\
    107 \le z_{close} &\le& 171\ \mbox{cm (FPP1)} \\
    205 \le z_{close} &\le& 270\ \mbox{cm (FPP2)} \\
    \sin \vartheta_{min} &=& \frac{0.07\ GeV/c}{p_0} \\
    \sin \vartheta_{max} &=& \frac{1.2\ GeV/c}{p_0}
  \end{eqnarray}
  where $p_0$ is the HMS central momentum in GeV/c. The minimum scattering angle cut is determined by the width of the Coulomb peak, which is inversely proportional to the momentum, while the maximum scattering angle cut is determined by the angle beyond which no significant asymmetry is observed in the data, which to a good approximation occurs at a constant transverse momentum defined by $p_T^{fpp} \equiv p \sin \vartheta_{max}$.
\end{enumerate}

To determine the optimal elastic event selection cuts, the background estimate described in section \ref{elasticcutssection} was performed for each kinematic setting for cut widths of 3, 5, and 6$\sigma$ in both $\Delta_p$ and $\Delta x$, with a fixed $3\sigma$ cut width in $\Delta y$. For a given inelastic contamination $f$, the observed polarization corresponds to a linear combination of the signal and background polarizations:
\begin{eqnarray}
  P_{i}^{obs.} &=& (1-f)P_i^{el.} + fP_i^{inel.} \\
  \Rightarrow P_i^{el.} &=& \frac{P_i^{obs.}-fP_i^{inel.}}{1-f} \label{correctedpolformula}
\end{eqnarray}
Neglecting momentarily the uncertainty in the background-corrected polarization and the uncertainty in $f$, it is clear that the statistical uncertainty in the corrected polarization is magnified by a factor $1/(1-f)$ relative to that of the total sample, which includes $N(1-f)$ elastic events and $Nf$ inelastic events, with $N$ the total number of events accepted by the cut. For large values of $f$, the uncertainty in $f$ and the uncertainty in the background polarization can no longer be neglected. Propagating the uncertainties through the formula \eqref{correctedpolformula}, the statistical error on the corrected polarization is given by
\begin{eqnarray}
  \Delta P_i^{el.} &=& \frac{1}{1-f}\sqrt{\left(\Delta P_i^{obs.}\right)^2 + f^2\left(\Delta P_i^{inel.}\right)^2}
\end{eqnarray}
The corrected polarization also has a systematic uncertainty due to the uncertainty in the estimated background $\Delta f$:
\begin{eqnarray}
   \left.\Delta P_i^{el.}\right|_{\Delta f} &=& \left|\frac{P_i^{obs.}-P_i^{inel.}}{(1-f)^2}\right|\Delta f
\end{eqnarray}

The cuts were chosen to minimize both uncertainties. For sufficiently small $f$, minimizing the uncertainty in the corrected polarization is equivalent to maximizing $(1-f)\sqrt{N}$, where $N$ is the total number of events accepted by the cut. For the final analysis, relatively tight $3\sigma$ cuts in $\Delta_p$, $\Delta x$, and $\Delta y$ were chosen. If the experimental resolution were purely Gaussian, $3\sigma$ cuts would accept 99.99\% of elastic events. However, the experimental resolution is not strictly Gaussian, and radiative effects redistribute significant numbers of events away from the main peak. Slightly lower statistical errors can be obtained with looser cuts for $Q^2=$5.2 GeV$^2$ and 6.7 GeV$^2$. Given the assumptions involved in correcting the measured form factor ratio for inelastic background contamination, the true systematic uncertainty in the size of the background correction is probably larger than what is estimated from the formulae above even for conservative estimates of $\Delta f$. For this reason, the choice of cuts was weighted more heavily toward maximal background suppression. Additionally, applying tight cuts to the variables of section \ref{elasticcutssection} has the added benefit of reducing the already quite small radiative corrections to the ratio $P_t/P_l$, which are not applied to the results presented here. Table \ref{finalelasticcuttable} shows the chosen final cuts and the estimated inelastic background contamination of these cuts. 
\begin{table}[h]
  \begin{center}
    \begin{tabular}{|c|ccc|c|}
      \hline $Q^2$, GeV$^2$ & $x_{cut}$, cm & $y_{cut}$, cm & $\Delta_p$ cut, \% & f, \% \\ \hline
      5.2 & 4.3 & 15.3 & 0.852 & 1.1 \\
      6.7 & 3.1 & 10.2 & 0.9 & 0.8 \\
      8.5 & 4.3 & 13.9 & 0.6 & 5.6 \\ \hline 
    \end{tabular}
  \end{center}
  \caption{\label{finalelasticcuttable} Final elastic event selection cuts and estimated inelastic contamination.}
\end{table}
\section{Statistical Uncertainties}
\paragraph{}
The covariance matrix of the maximum likelihood estimate of the polarizations $\mathbf{P}$ is given by the inverse of the matrix $A$ appearing on the right hand side of equation \eqref{loglikelihoodlinearized}. The elements of $A$ are proportional to the total number of events contributing to the asymmetry and the square of the product $hA_y$. Therefore, the statistical uncertainty behaves as $\Delta P \propto \frac{1}{\sqrt{Nh^2A_y^2}}$. This error also depends on the precession angle since the sensitivity of the asymmetry to $P_l$ is proportional to $\sin \chi_{\theta}$, meaning the error on $P_l$ is typically larger than the error on $P_t$, which is proportional to $\cos \chi_\phi \approx 1$. Table \ref{staterrtable} shows the absolute statistical uncertainty in $R$ for the kinematics of E04-108. Presentation of the final results is deferred until after the discussion of systematic uncertainties and background subtraction.
\begin{table}
  \begin{center}
    \begin{tabular}{|c|c|}
      \hline $Q^2$, GeV$^2$ & $\Delta R$ \\ \hline
      5.2 & 0.067 \\ \hline
      6.7 & 0.11 \\ \hline
      8.5 & 0.18 \\ \hline
    \end{tabular}
  \end{center}
  \caption{\label{staterrtable} Absolute statistical uncertainties in the results of E04-108.}
\end{table} 
\section{Systematic Uncertainties}
\subsection{Spin Precession}
\paragraph{}
The calculation of the spin precession in the HMS is one of the most important sources of systematic uncertainty in the extraction of the ratio $G_E/G_M$. The uncertainties in the precession calculation can be separated into two categories. Errors in the reconstructed quantities $x'_{tar}$, $y'_{tar}$, $\delta$, and $y_{tar}$ lead to errors in the precession matrix, since the matrix elements are calculated as polynomials in these quantities. This kind of error is the most important. The second kind of error in the precession calculation is the uncertainty in the COSY model used to calculate the precession matrix, which is generally much smaller than the effect of reconstruction errors.

To see how reconstruction errors propagate to the uncertainty in the form factor ratio, it is useful to work with the geometric approximation of section \ref{precessionsection}. Neglecting the small rotations from the reaction plane to transport coordinates and from transport coordinates to the proton trajectory coordinates at the focal plane, the following approximate expressions for the polarization components at the focal plane in terms of the dispersive precession angle $\chi_\theta$ and the non-dispersive precession angle $\chi_\phi$ are obtained:
\begin{eqnarray}
  P_y^{fp} &=& \cos \chi_\phi P_t + \sin \chi_\phi P_l \\
  P_x^{fp} &=& -\cos \chi_\phi \sin \chi_\theta P_l + \sin \chi_\phi \sin \chi_\theta P_t
\end{eqnarray}
Solving these equations for $P_t$ and $P_l$ and forming their ratio gives
\begin{eqnarray}
  \frac{P_t}{P_l} &=& -\frac{1 + \sin \chi_\theta \cot \chi_\phi \frac{P_y^{fp}}{P_x^{fp}}}{1-\sin \chi_\theta \tan \chi_\phi \frac{P_y^{fp}}{P_x^{fp}}} \tan \chi_\phi 
\end{eqnarray}
In this approximate form, it is immediately clear that the form factor ratio is highly sensitive to the non-dispersive precession angle $\chi_\phi \equiv \gamma \kappa (\phi_{fp} - \phi_{tar})$. This is because precession in the non-dispersive plane mixes $P_t$ and $P_l$. Because the non-dispersive bend angle for the central ray is zero, the average of $\chi_\phi$ over the full acceptance is usually close to zero, so that the formula above reduces to 
\begin{eqnarray}
  \frac{P_t}{P_l} &\xrightarrow[\chi_\phi \rightarrow 0]{}& -\chi_\phi - \sin \chi_\theta \frac{P_y^{fp}}{P_x^{fp}} 
\end{eqnarray}
\begin{eqnarray}
  \Delta \left(\frac{P_t}{P_l}\right)_{\chi_\phi} &\approx& \Delta \chi_\phi
\end{eqnarray}
In terms of the error in the total non-dispersive bend angle $\Delta \phi \equiv \phi_{fp} - \phi_{tar}$, this becomes 
\begin{eqnarray}
  \Delta \left(\frac{P_t}{P_l}\right) &\approx& \gamma \kappa \Delta(\Delta \phi)
\end{eqnarray}
The error in the form factor ratio due to this uncertainty is given by 
\begin{eqnarray}
  \Delta R &\approx& \gamma \kappa K \Delta (\Delta \phi) \label{nondispersivebenduncertainty}
\end{eqnarray}
where $K$ is the kinematic factor multiplying $P_t/P_l$ in the expression for $R$.

Equation \eqref{nondispersivebenduncertainty} shows that any uncertainty in the total non-dispersive bend angle in the HMS is magnified by a factor $\gamma \kappa K$ in the form factor ratio. The factor $\gamma \kappa K$ grows rather large at high $Q^2$, as shown in table \ref{gammakappaKtable}.
\begin{table}[h]
  \begin{center}
    \begin{tabular}{|c|c|}
      \hline $Q^2$, GeV$^2$ & $\gamma \kappa K$ \\ \hline
      5.2 & 32.2 \\ 
      6.7 & 40.2 \\ 
      8.5 & 72.5 \\ \hline
    \end{tabular}
  \end{center}
  \caption{\label{gammakappaKtable} $\gamma \kappa K$ for E04-108 kinematics.}
\end{table}
In order to keep this uncertainty manageable, the total non-dispersive bend angle must be determined with relatively good accuracy. The main source of uncertainty in $\Delta \phi$ is a misalignment of the quadrupoles with respect to the optical axis of the HMS. The stated accuracy in the alignment of the quadrupoles is $\pm1.0$ mm. A quadrupole offset in the non-dispersive direction manifests itself as a non-zero $\Delta \phi$ for the central ray. To estimate the shift in $\Delta \phi$, quadrupole offsets were introduced in the HMS COSY model, and the first-order couplings $(\phi_{fp}|s_i),\ (i=1,2,3)$ representing the displacement in $\phi_{fp}$ for a quadrupole shift $s_i$ for the nominal point-to-point tune of the HMS were calculated. These couplings are shown in table \ref{phiscouplingstable}.
\begin{table}
  \begin{center}
    \begin{tabular}{|c|c|}
      \hline Quadrupole magnet & $(\phi|s)$ (mrad/mm) \\ \hline
      Q1 & +0.246 \\ 
      Q2 & -0.796 \\
      Q3 & +0.273 \\ \hline
    \end{tabular}
  \end{center}
  \caption{\label{phiscouplingstable} First-order couplings $(\phi|s_i)$ calculated by the HMS COSY model for the nominal tune.}
\end{table}
Assuming an uncertainty of $\pm 1$ mm in the positioning of the quadrupoles with respect to the optical axis of the HMS, the maximum possible range of $\Delta \phi$ is $\pm 1.3$ mrad. An error of $\pm 1.3$ mrad in $\Delta \phi$ translates into an absolute error on the form factor ratio of $\Delta R = \pm(0.042, 0.052, 0.094)$ for $Q^2 = (5.2, 6.7, 8.5)\ GeV^2$. The actual error is larger or smaller depending on any possible higher-order effects which manifest themselves in the full COSY calculation but not in the geometric approximation.

To reduce the uncertainty in $\Delta \phi$, a dedicated study of the HMS optics in the non-dispersive plane was performed. Using an unrastered electron beam impinging on a thin carbon target positioned at the origin and the sieve slit collimator, with the HMS positioned at an angle of 12 degrees and set for a central momentum of 1.022 GeV with negative polarity to detect electrons, electron scattering data were obtained for seven different settings of the HMS magnets other than the nominal setting, and the movements of the image of the central sieve hole at the focal plane were observed. The central sieve hole is smaller than the other holes and it is aligned with the optical axis of the HMS. The position of the central sieve hole and the distance from the origin to the central sieve hole are known to very high accuracy. For a point target such as the thin carbon foil used for the study, the ray from the intersection of the beam with the target to the central sieve hole provides a precise measurement of $\phi_{tar}$ for electrons going through the central hole. If the position of the target with respect to the sieve slit and the position of the beam with respect to the target are also known, then the angle $\phi_{tar}$ is determined.

In the first phase of the study, all the quadrupoles were turned off and ``de-gaussed'' by first setting each magnet to a large positive (relative to the sign of the current for the nominal tune) current, then ramping to zero, changing the polarity of the magnet, ramping to a small negative current, and then back to zero, ensuring that no residual fields remained in the quadrupoles. This setting is referred to as the ``dipole only'' setting. Since the quadrupoles were all turned off, the displacement of $y_{fp}$ and $\phi_{fp}$ could not be affected by any quadrupole offsets. By scanning the beam position horizontally across the target, thus varying $\phi_{tar}$ for the ray going through the central sieve hole, the observed displacements of $y_{fp}$ and $\phi_{fp}$ could be used to constrain the offsets $\phi_0^{fp}$ and $y_0^{tar}$, the zero offsets in, respectively, the angle $\phi_{fp}$ reconstructed by the HMS drift chambers and the position of the target with respect to the HMS optical axis. 

The measurement of $y_0^{tar}$ using this technique is very precise because the dipole is defocusing in the non-dispersive plane. The first-order COSY coupling between the measured position in $y$ at the focal plane and $\phi_{tar}$ is $(y_{fp}|\phi_{tar}) = +25.6\ mm/mrad$ for the dipole-only setting. This large dispersive effect increases the sensitivity of $y_{fp}$, which is measured with a precision of $\sigma_{y_{fp}} \approx 200\ \mu m$ and an absolute accuracy\footnote{The accuracy of $y_{fp}^0$ is based on the surveyed position of the chambers and the fine-tuning of this position in software during the fitting of the reconstruction matrix elements} of (conservatively) $\pm 1$ mm, to $\phi_{tar}$, which is given by $\phi_{tar} = \frac{y_{hole} - y_{tar}}{z_{hole}}$. Since $y_{hole}$ and $z_{hole}$ are known very precisely, the displacement of the image of the central sieve hole at the focal plane serves as a precise measurement of $y_0^{tar}$. Figure \ref{dipoledispfig} shows how this image moves as the beam position changes. The size of the spot corresponding to the central hole is large because of the large defocusing effect and because of multiple scattering in S0, which is more pronounced for the relatively low-momentum (1 GeV) electrons used for the study.
\begin{figure}
  \begin{center}
    \includegraphics[width=.50\textwidth]{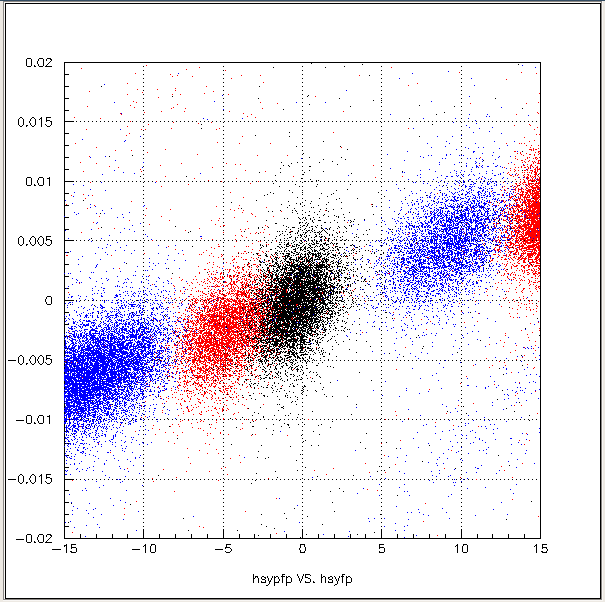}
  \end{center}
  \caption{\label{dipoledispfig} $y_{fp}$ versus $y'_{fp}$ for the dipole-only setting for $x_{beam}=$ +4.6 mm(blue), -3.2 mm(red), and -0.28 mm (black). Beam positions are according to EPICS.}
\end{figure}

In addition to the dipole-only setting and the nominal tune, six other tunings of the quadrupoles were studied. These settings are shown in table \ref{opticstable}. The dipole was at its nominal, momentum-defining field for all settings of the quadrupoles. Each quadrupole was set to its nominal current with the other two quadrupoles off, and also set to $70\%$ of its nominal current (which approximately corresponds to point-to-parallel focusing) with the other two quadrupoles at their nominal settings.
\begin{table}
  \begin{center}
    \begin{tabular}{|c|ccc|}
      \hline Setting name & $Q1/Q1_{nom.}$ & $Q2/Q2_{nom.}$ & $Q3/Q3_{nom.}$ \\ \hline
      Dipole & 0 & 0 & 0 \\ 
      Q1 & 1 & 0 & 0 \\ 
      Q2 & 0 & 1 & 0 \\
      Q3 & 0 & 0 & 1 \\ 
      Q1R & 0.7 & 1 & 1 \\
      Q2R & 1 & 0.7 & 1 \\ 
      Q3R & 1 & 1 & 0.7 \\ \hline 
    \end{tabular}
  \end{center}
  \caption{\label{opticstable} Quadrupole current settings relative to nominal tune for the various settings.}
\end{table}
For each setting, the first order optical couplings $(y|y)$, $(y|\phi)$, $(\phi|y)$, and $(\phi|\phi)$ were calculated in COSY. The isolation of the central sieve hole at the focal plane is relatively straightforward. The central row of sieve holes is selected by plotting $x'_{fp}$ as a function of $x_{fp}$. A series of stripes appears with each stripe corresponding to a different row of sieve holes. The central row of holes corresponds to the stripe whose peak is closest to $(0,0)$ in the $(x,x')$ phase space. This selection is illustrated in figure \ref{xxpopticsselection} for the Q1R setting.

After selecting the central row, the location of the central hole is usually obvious upon plotting $y_{fp}$ (or $y'_{fp}$ versus $y_{fp}$), as illustrated in figure \ref{yypopticsselection}. Fewer events go through the central hole than the other holes because it has half the diameter and $1/4^{th}$ the area of the other holes. Once the events that went through the central sieve hole are selected, $y_{fp}$ and $y'_{fp}$ are determined by Gaussian fits to these events, as shown in figure \ref{yfpypfpfit}. The reason the analysis is restricted to the central hole is to minimize deviations from the central ray, so that the modeling of the optics of the different magnet settings within COSY to linear order is a good approximation, making the analysis of the data much simpler.
\begin{figure}[h]
  \begin{center}
    \subfigure[$x'_{fp}$ vs $x_{fp}$ for the Q1R setting.]{\label{xxpopticsselection}\includegraphics[width=.49\textwidth]{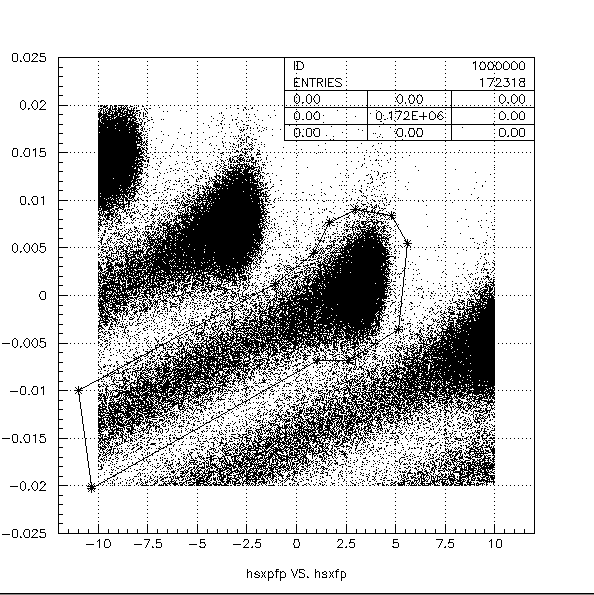}}
    \subfigure[$y'_{fp}$ vs $y_{fp}$ for the Q1R setting.]{\label{yypopticsselection}\includegraphics[width=.49\textwidth]{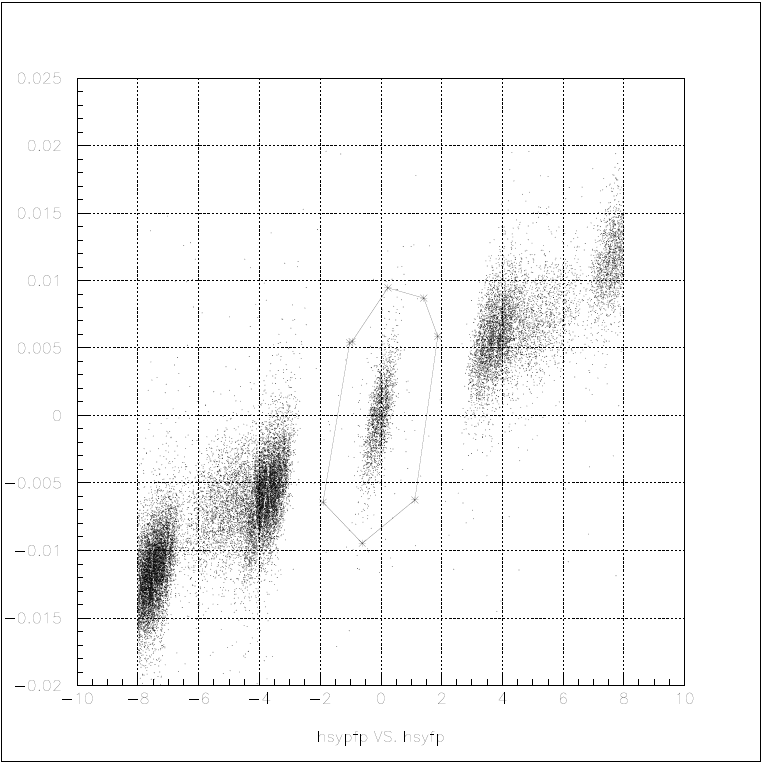}}
  \end{center}
  \caption{\label{centralholeisolation} Procedure to isolate the image of the central sieve hole at the focal plane, for the Q1R setting. In the $x'_{fp}$ versus $x_{fp}$ distribution (a), each stripe corresponds to a different row of sieve holes, with a peak and a tail corresponding to elastic and inelastic electron scattering from carbon. In the $y'_{fp}$ vs. $y_{fp}$ distribution (b) of the selection of events in (a), the location of events that went through the central hole becomes obvious.}
\end{figure}

\begin{figure}[h]
  \begin{center}
    \includegraphics[width=.49\textwidth]{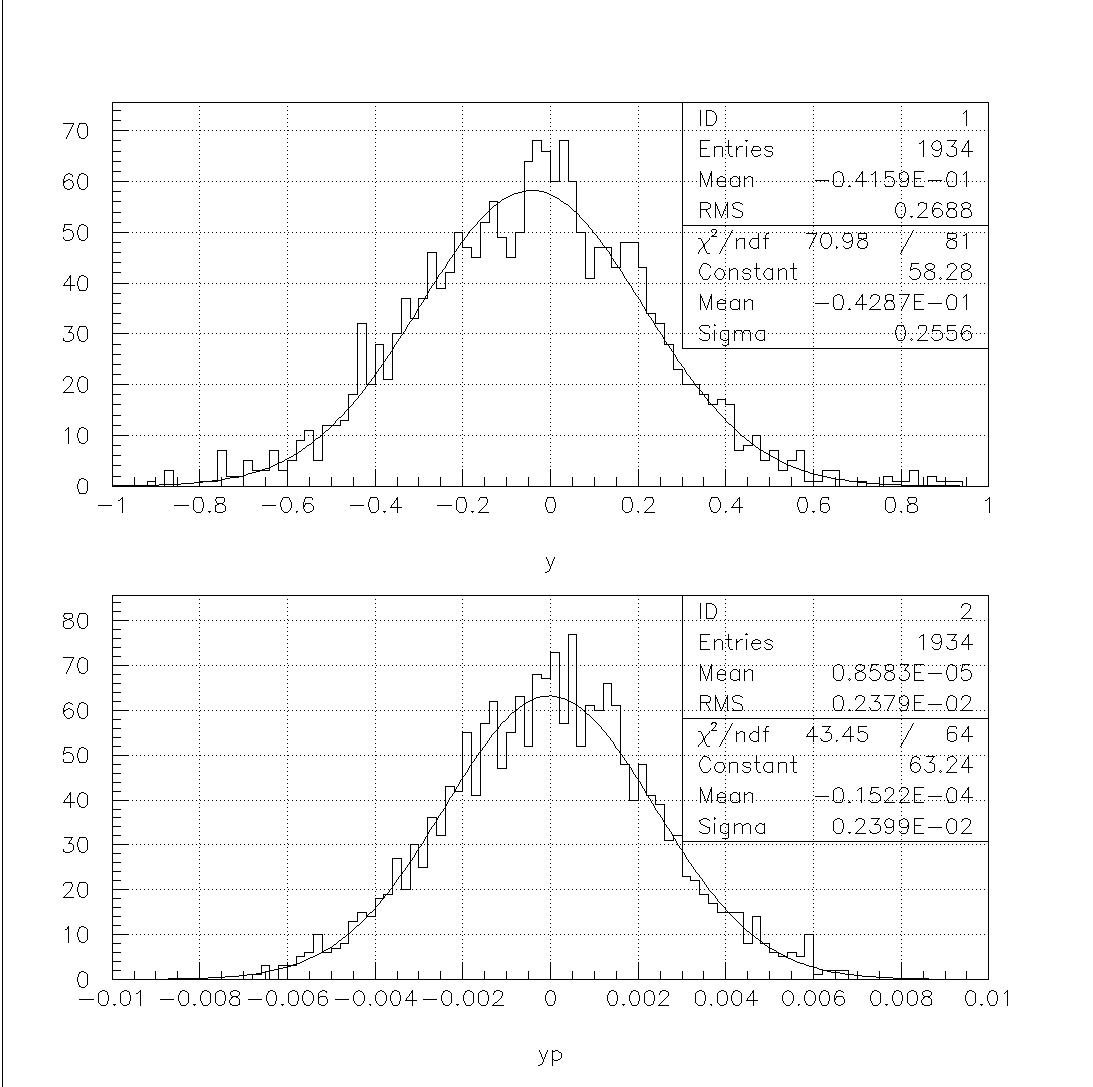}
  \end{center}
  \caption{\label{yfpypfpfit} Gaussian fits to determine $y_{fp}$ and $y'_{fp}$.}
\end{figure}
The observed spatial and angular displacements of the central ray at the focal plane are related to the quadrupole misalignments and the zero offsets $y_{tar}^0$, $y_{fp}^0$, and $\phi_{fp}^0$ through the first order couplings as follows:
\begin{eqnarray}
  y_{tar} &=& y_{tar}^0 - x_{beam} \cos \Theta_{HMS} \label{linearoptics1} \\
  \phi_{tar} &=& \frac{y_{hole} - y_{tar}}{z_{hole}} \\
  y_{fp} &=& y_{fp}^0 + (y|y) y_{tar} + (y|\phi) \phi_{tar} + \sum_{i=1}^3 (y|s_i) s_i \\
  \phi_{fp} &=& \phi_{fp}^0 + (\phi|y) y_{tar} + (\phi|\phi) \phi_{tar} + \sum_{i=1}^3 (\phi|s_i) s_i \label{linearoptics}
\end{eqnarray}
where the $+x_{beam}$ axis points toward beam right; i.e., toward the HMS or in the $-y$ direction in transport coordinates. The position $y_{hole}$ of the central hole and the distance $z_{hole}$ from the origin to the sieve slit are taken from survey results. With the measured displacements in $y$ and $\phi$ for all seven (non-standard) settings in addition to the nominal tune of the HMS, with two different beam positions for the nominal tune and five different beam positions for the dipole-only tune, there are twenty-four linear equations in five ``unknowns''\footnote{$y_{fp}^0$, the zero offset of the measured position at the focal plane, is treated as known and is fixed in the analysis.}, so the offsets are overdetermined, and the goal of the analysis is to find the set of offsets which minimizes the $\chi^2$ of all accepted data points.

In principle, it is possible to determine the three quadrupole offsets, the target position offset, and the focal plane angle and position offsets simultaneously. In practice, however, it is more informative to fix $y^0_{fp}$ and use the dipole-only data to determine $y_{tar}^0$ and $\phi_{fp}^0$, because the data do not determine $y_0^{fp}$ to any better accuracy than it is already known. The extracted value of $y_{tar}^0$, though very precise, is actually only a determination of the beam position, since the beam position is only known to within an overall accuracy of $\pm 1$ mm. For example, the value of $y_{tar}^0$ extracted using the EPICS value of $x_{beam}$ differs by approximately $1$ mm from the value obtained using the BPM ADCs in the data stream, reflecting the approximate 1 mm difference between the EPICS beam position and the ADC beam position. This uncertainty does not affect the results for the quadrupole offsets, however, since it is the determination of $y_{tar}$ and $\phi_{tar}$ that matters for this purpose, and in this respect the dipole setting still does the trick, regardless of which beam position is used in the analysis\footnote{The position of the physical target foil with respect to the ideal origin and the mispointing of the HMS do introduce uncertainty into the determination of $y_{tar}$ and $\phi_{tar}$. These quantities were also surveyed, and are assumed equal to their surveyed values. An error in the foil position is equivalent to an error in $z_{hole}$. The error in $\phi_{tar}$ due to an error in $z_{hole}-z_{foil}$ is $d\phi/\phi = dz/z$. Since $z_{hole} = 166.00$ cm is large compared to any possible uncertainty in $z_{foil}$, errors of even several mm in $z_{foil}$ introduce only small \emph{relative} errors in $\phi_{tar}$, which are negligible in comparison to the other uncertainties involved. Therefore, when combined with the determination of $y_{tar}$ using the dipole setting, the uncertainty in $\phi_{tar}$ is really quite small.}. 

The preliminary results of this study are shown in table \ref{quadoffsetstable}. Uncertainties of $1$ mm in $y_{fp}^0$ and $0.2$ mm in $x_{beam}$ (relative) are assumed, and the uncertainties in $y_{hole}$, $z_{hole}$, the HMS pointing, and the physical location of the carbon foil relative to the origin are neglected by comparison. The result of 0.1 mm uncertainty in $y_0^{tar}$ leads to a 0.06 mrad uncertainty in $\phi_{tar}$. Statistical uncertainties in the measured $y_{fp}$ and $y'_{fp}$ positions of the central ray are included.
\begin{table}
  \begin{center}
    \begin{tabular}{|c|c|c|}
      \hline $y_{tar}^0$, mm & $\phi_{fp}^0$, mrad & $\chi^2$/d.f. \\ \hline 
      $s_1$, mm & $s_2$, mm & $s_3$, mm \\ \hline 
      0.52 $\pm$ 0.10 & -0.06 $\pm$ 0.11 & 0.68 \\ \hline  
      0.51 $\pm$ 0.10 & 0.41 $\pm$ 0.46 & 0.15 $\pm$ 0.54 \\ \hline 
    \end{tabular}
  \end{center}
  \caption{\label{quadoffsetstable} Best fit quadrupole offsets $s_1$, $s_2$, $s_3$, $y_{tar}^0$, $\phi_{fp}^0$ and $\chi^2$. Preliminary.}
\end{table}
The data point from the most extreme beam position used during the dipole setting ($x_{beam} = +4.6$ mm) was omitted because the observed $y_{fp}$ of the central ray was approximately 2 cm away from its expected position (based on $x_{beam}$ and the first-order optics). This point gave by far the worst contribution to the $\chi^2$ of the any of the data points and it was deemed reasonable to assume that higher-order terms in the HMS optics matrix have a non-negligible effect on such an extreme ray. All other data points from all other settings were included. After using the dipole-only data to solve for $y_{tar}^0$ and $\phi_{fp}^0$, the quadrupole offsets were obtained by fixing these quantities and solving the simultaneous equations \eqref{linearoptics1}-\eqref{linearoptics} for $s_1$, $s_2$, and $s_3$.

The quadrupole offsets resulting from this study are compatible with the stated uncertainty in their positioning. Assuming the small deflections of the central ray induced by the offsets are additive, the results of method 1 in table \ref{quadoffsetstable} result in a non-dispersive bend angle $\Delta \phi_s \equiv \sum_{i=1}^3 (\phi|s_i) s_i = -0.17 \pm 0.20$ mrad due to the observed shifts and their uncertainties. Given the preliminary status of this analysis and the neglect of other uncertainties such as the model uncertainties in the first-order couplings, the small uncertainties in $y_{hole}$ and $z_{hole}$, and any possible uncertainty in the zero offset in the reconstructed $\phi_{tar}$ due to the fitted HMS reconstruction matrix elements, a systematic uncertainty of $\pm0.5$ mrad was assigned to $\Delta \phi$ for the results presented in this thesis. The effect of a $\pm0.5$ mrad offset in $\Delta \phi$ is approximated by an offset in $\phi_{tar}$ of $\pm$0.36 mrad, since the optics of the HMS give $\Delta \phi \approx -1.4\phi_{tar}$.

Uncertainties in the other reconstructed variables such as $x'_{tar}$, $y_{tar}$, and $\delta$ also affect the form factor ratio, but none of these uncertainties are magnified to the same extent as $\phi_{tar}$. The momentum deviation $\delta$ and the out-of-plane angle $x'_{tar}$ determine the difference of the main dispersive precession from the dominant 25$^\circ$ central bend, which enters the expression for $P_t/P_l$ through the term $-\sin \chi_\theta P_y^{fp}/P_x^{fp}$, so the form factor ratio is much less sensitive to small offsets in these variables compared to $\Delta \phi$. The ratio is somewhat more sensitive to $y_{tar}$ than to $\delta$ or $x'_{tar}$, because a systematic error in the value of $y_{tar}$ used as input to the calculation of the COSY precession matrix is more or less equivalent to an unknown quadrupole offset from the optical axis, which introduces a non-zero deflection of the central ray. The sensitivity of the matrix element $S_{yz} \approx \sin \chi_\phi$ to a $1$ mm offset in $y_{tar}$ is roughly a factor of five smaller than its sensitivity to a 1 mrad offset in $\phi_{tar}$, to first order in COSY, such that even a conservative estimate of $\pm0.5$ mm systematic uncertainty in $y_{tar}$ does not translate to an uncertainty in $R$ anywhere near as large as the dominant $\phi_{tar}$ contribution\footnote{Strictly speaking, a systematic uncertainty in $y_{tar}$ is part of the systematic uncertainty in $\Delta \phi$, since the precession only depends on $y$ indirectly through its effect on $\Delta \phi$, which in first order is given by $(\phi|y) \approx -.25$ mrad/mm.}. To determine the systematic uncertainty in $R = \mu_p G_E^p/G_M^p$, the shifts given in table \ref{systoffsetstable} were applied in both directions to the reconstructed variables used to calculate the precession matrix, and the observed shifts in $R$ were recorded.
\begin{table}[h]
  \begin{center}
    \begin{tabular}{|c|c|}
      \hline $\theta_{tar} (=\arctan x'_{tar})$ & $\pm 2$ mrad \\ \hline
      $\phi_{tar} (=\arctan y'_{tar})$ & $\pm 0.36$ mrad \\ \hline
      $\delta$ & $\pm 0.3\%$ \\ \hline
      $y_{tar}$ & $\pm 0.5$ mm \\ \hline
    \end{tabular}
  \end{center}
  \caption{\label{systoffsetstable} Offsets applied to reconstructed quantities involved in the precession calculation for systematic error analysis.}
\end{table}
\begin{table}[h]
  \begin{center}
    \begin{tabular}{|c|c|c|c|c|}
      \hline $Q^2$, GeV$^2$ & $\Delta R(\Delta \phi_{tar})$ & $\Delta R(\Delta \theta_{tar})$ & $\Delta R\left(\Delta \delta \right)$ & $\Delta R(\Delta y_{tar})$ \\ \hline 
      5.2 & $\pm 0.016$ & $\pm7.8\times 10^{-4}$ & $\pm6.0\times 10^{-5}$ & $\pm3.7 \times 10^{-3}$ \\
      6.7 & $\pm0.020$ & $\pm3.36\times 10^{-3}$ & $\pm1.67 \times 10^{-3}$ & $\pm4.9 \times 10^{-3}$ \\
      8.5 & $\pm0.035$ & $\pm1.50\times 10^{-3}$ & $\pm8.55 \times 10^{-3}$ & $\pm9.5\times 10^{-3}$ \\ \hline
    \end{tabular}
  \end{center}
  \caption{\label{systresultstable} Absolute systematic uncertainty in $R = \mu_p G_E^p/G_M^p$ due to systematic errors in reconstructed event kinematics.}
\end{table}

The sensitivity of $R$ to the systematic uncertainties in the reconstructed proton kinematics is shown in table \ref{systresultstable}. If the spin precession is handled correctly, the extracted form factor ratio does not depend on these variables. Depending on the slope of $G_E/G_M$ as a function of $Q^2$, some variation of the form factor ratio with $Q^2$ within the finite $Q^2$ acceptance of the HMS is possible, but since the ratio was not determined with sufficient statistical precision for a meaningful measurement of the $Q^2$ dependence of $R$ within the acceptance of a single data point, the results are quoted as a single value of $R$ at the acceptance-averaged value of $Q^2$. 
\begin{figure}[h]
  \begin{center}
    \includegraphics[angle=90,width=.98\textwidth]{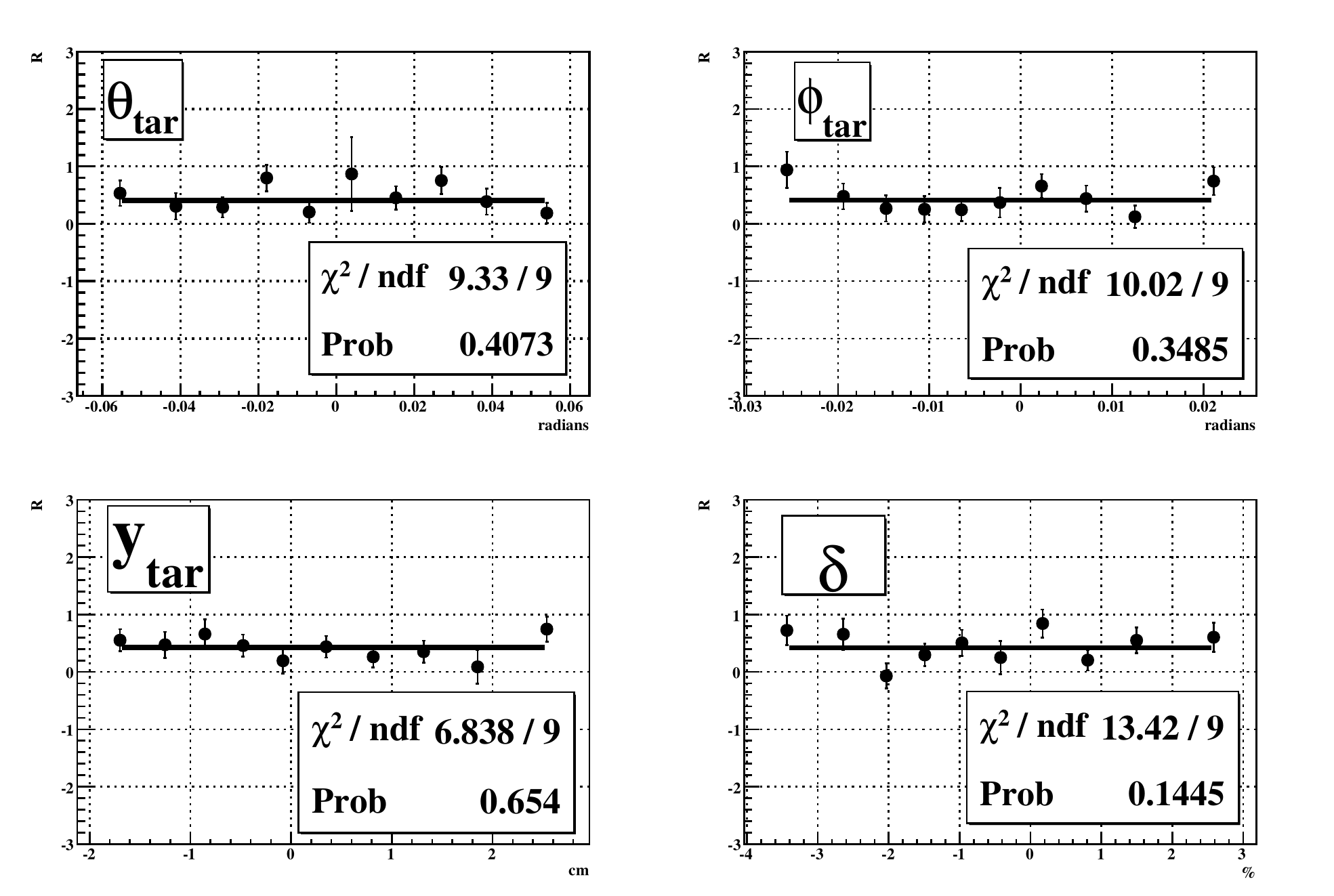}
  \end{center}
  \caption{\label{Rtargdep52} Extracted form factor ratio as a function of reconstructed proton kinematics, $Q^2=5.2$ GeV$^2$.}
\end{figure}
\begin{figure}[h]
  \begin{center}
    \includegraphics[angle=90,width=.98\textwidth]{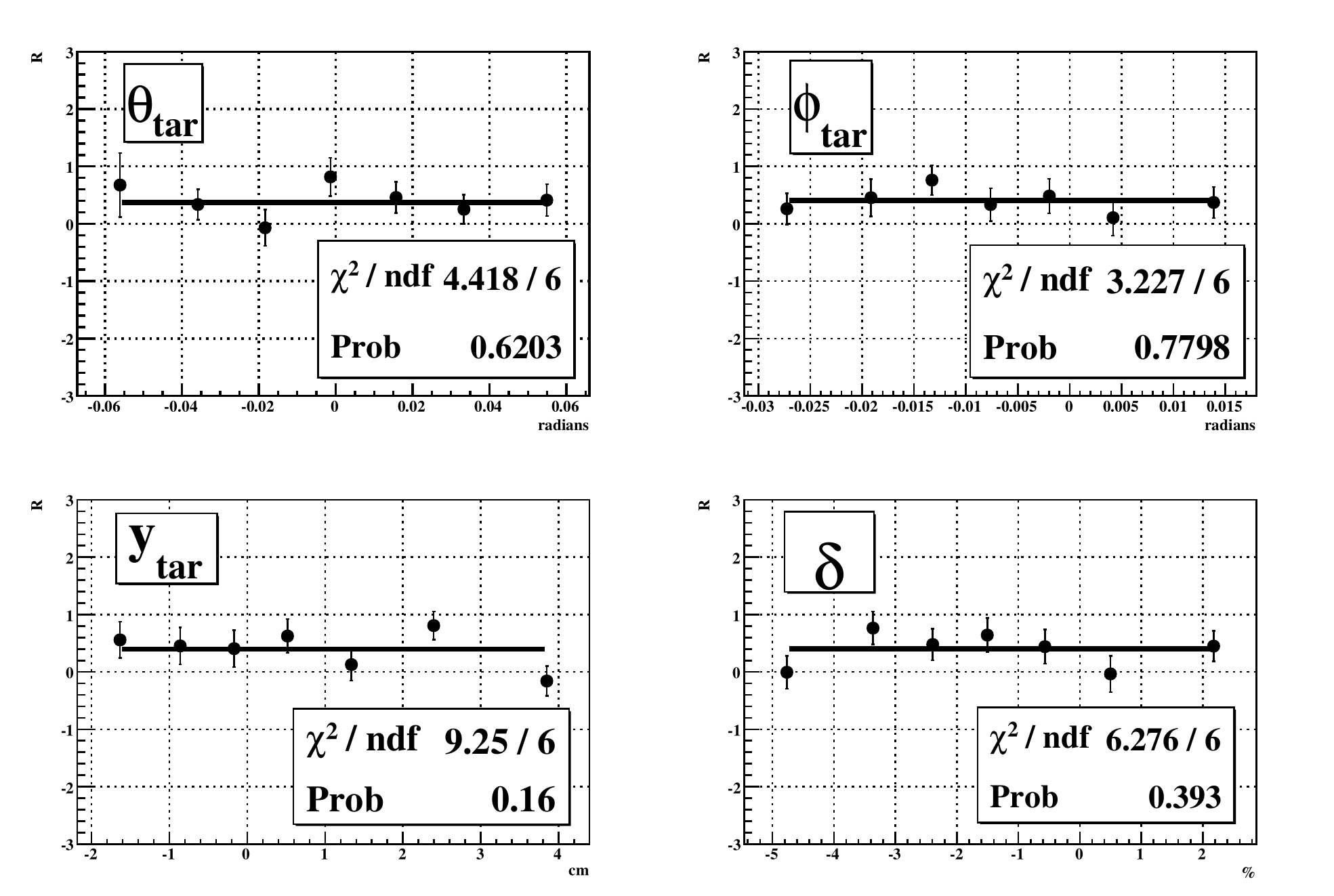}
  \end{center}
  \caption{\label{Rtargdep67} Extracted form factor ratio as a function of reconstructed proton kinematics, $Q^2=6.7$ GeV$^2$.}
\end{figure}
\begin{figure}[h]
  \begin{center}
    \includegraphics[angle=90,width=.98\textwidth]{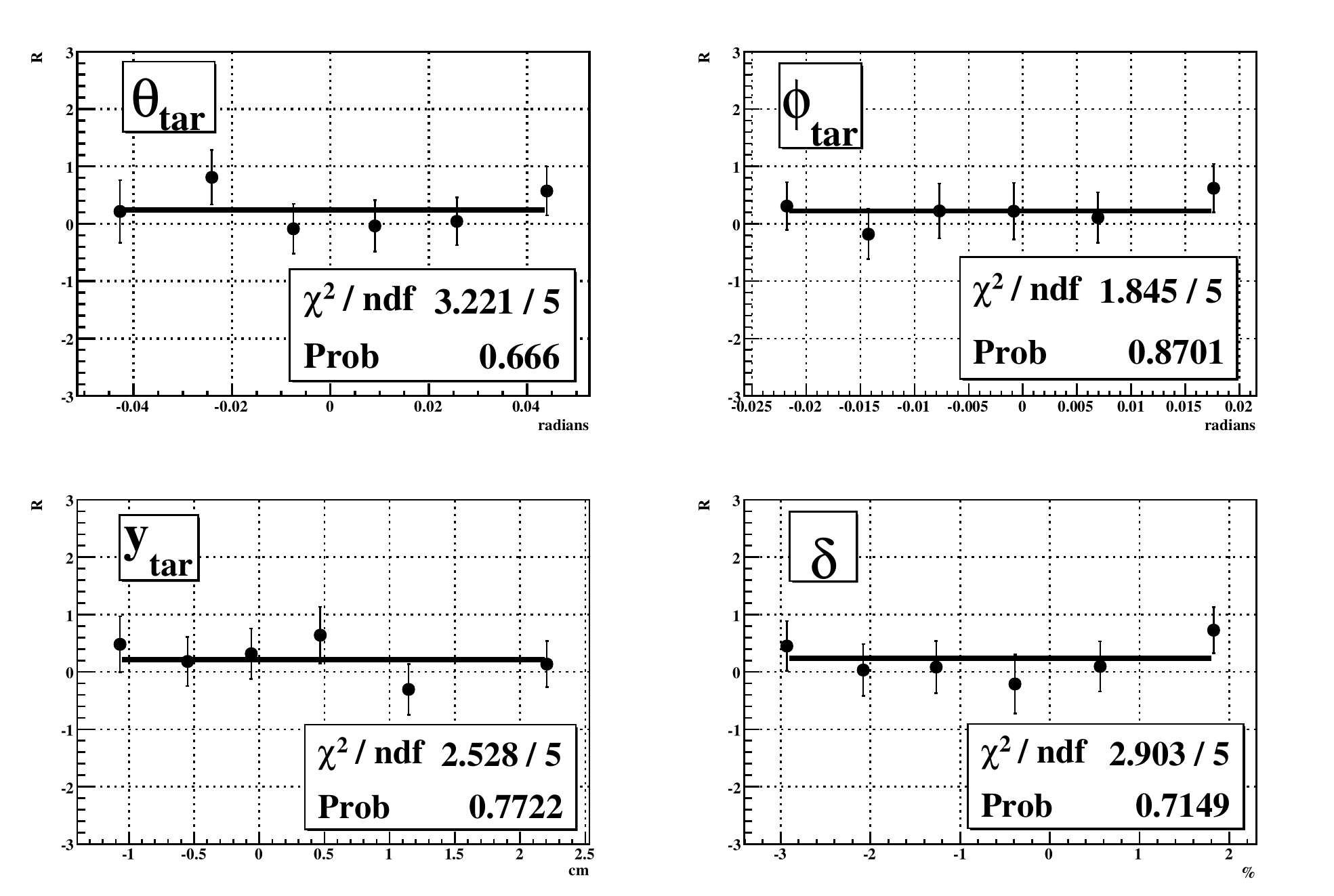}
  \end{center}
  \caption{\label{Rtargdep85} Extracted form factor ratio as a function of reconstructed proton kinematics, $Q^2=8.5$ GeV$^2$.}
\end{figure}
Figures \ref{Rtargdep52}-\ref{Rtargdep85} show the dependence of $R$ on the reconstructed proton kinematics for each of the data points of E04-108. In all three cases, the $\chi^2$ per degree of freedom of a constant fit to the data is at most 1.5, indicating that the precession calculation is well understood. 

The other source of precession-related systematic uncertainty in $R$ comes from the model of the HMS used to calculate the polynomial expansion of the integration of the BMT equation in COSY. For the HMS, this uncertainty turns out to be quite small. As a first estimate of this uncertainty, the results obtained using COSY are compared to the results obtained using the model-independent geometric approximation.
\begin{table}
  \begin{center}
    \begin{tabular}{|c|c|}
      \hline $Q^2$, GeV$^2$ & $R(geom.) - R(COSY)$ \\ \hline
      5.2 & 0.022 \\ \hline
      6.7 & 0.036 \\ \hline
      8.5 & 0.030 \\ \hline 
    \end{tabular}
  \end{center}
  \caption{\label{geomcosycomp} Difference in $R$ between the geometric approximation and the full COSY calculation.}
\end{table}
Notably, the geometric approximation agrees with the full COSY calculation for all three kinematics of E04-108 at a level well below the statistical uncertainty. This is a result of the simple QQQD structure of the HMS magnets for which the dipole approximation is satisfied separately in the dispersive and non-dispersive planes, at least when averaged over the full acceptance. This approximation depends only on the accuracy of the reconstructed precession angles $\chi_\theta$ and $\chi_\phi$, and differs from the true precession only to the extent that the assumption of small, additive trajectory rotations in the magnets is violated. Uncertainties in the full COSY calculation arise due to differences between the assumed fields in the magnets and the true magnetic fields present during the experiment, misalignment of the various elements and, to a much lesser extent, the numerical accuracy of the calculation.

These uncertainties manifest themselves as differences in the HMS transport matrix from the focal plane to the target and vice versa between COSY and the optimized reconstruction matrix elements of the HMS (for the $fp \rightarrow tgt.$ case), which can be regarded as most nearly representing its ``true'' optics. To account for all of these possible COSY model uncertainties in a rigorous and self-consistent manner requires shifting parameters of the models, re-fitting the forward transport and reconstruction matrices with COSY, and re-optimizing the reconstruction matrix using optics data, ultimately tuning the COSY model to reproduce the optics data ``exactly''. The net result of such efforts is that the COSY model of the HMS has been improved over time to the point where the differences between the COSY transport matrices and the optimized reconstruction matrix elements are quite small. For this reason, the uncertainty in the precession calculation is dominated by how well or poorly the ``true'' optics of the HMS in the non-dispersive plane is known, and the study described above has effectively reduced this uncertainty by at least a factor of two.

\subsection{Scattering Angle Reconstruction}
\paragraph{}
Systematic errors in the reconstruction of the scattering angles $\vartheta_{fpp}$ and $\varphi_{fpp}$ have been minimized by the alignment procedure of section \ref{fppalignmentsection}, in which tracks reconstructed in both sets of FPP chambers during straight-through runs (runs with both FPP doors opened) were aligned with HMS tracks. The systematic error in the FPP track slopes $x'$ and $y'$ is estimated at no more than $\Delta x' = \Delta y' = 0.1$ mrad. The most important source of uncertainty due to angle reconstruction errors in the FPP comes from the azimuthal angle $\varphi_{fpp}$. Errors in $x'_{fpp}$ and $y'_{fpp}$ are magnified in the error in $\varphi_{fpp}$ at small polar angles $\vartheta_{fpp} \rightarrow 0$, since $\Delta \varphi_{fpp} \approx \frac{1}{\sin \vartheta_{fpp}} \sqrt{(\Delta x')^2+ (\Delta y')^2}$. To estimate the error in $R$ due to an error in the reconstructed track slopes in this manner, a $\vartheta_{fpp}$-dependent shift was applied to $\varphi_{fpp}$ in the analysis, and the change in the result for $R$ was recorded. The results are shown in table \ref{systphifpptable}.
\begin{table}[h]
  \begin{center}
    \begin{tabular}{|c|c|}
      \hline $Q^2$, GeV$^2$ & $\Delta R(\varphi \rightarrow \varphi\pm .14\ \mbox{mrad}/\sin \vartheta_{fpp})$ \\ \hline
      5.2 & $\pm 2.9\times 10^{-4}$ \\ \hline
      6.7 & $\pm 5.6\times 10^{-3}$ \\ \hline
      8.5 & $\pm 0.017$ \\ \hline 
    \end{tabular}
  \end{center}
  \caption{\label{systphifpptable} Systematic uncertainty in $R$ due to azimuthal angle reconstruction.}
\end{table}
\subsection{False Asymmetries}
\paragraph{}
Helicity-independent false asymmetries are caused by misalignments of the drift chambers and $\varphi$-dependent variations in acceptance and efficiency, where inefficiencies may come from either the detection efficiency of individual wires or from $\phi$-dependent inefficiencies and/or biases in track reconstruction. Since the FPP chambers were new and very gas-tight, all wires were very close to 100\% efficient, so the latter category of false asymmetry is dominant. In elastic scattering, the induced polarization is zero. Therefore, the only asymmetries present in the helicity sum spectrum $f_{sum}(\varphi)$ are the false asymmetries. By fitting the sum spectrum, the terms $a_1$, $b_1$, $\ldots$ for $\lambda_0$ are obtained, and can be subtracted on the left hand side of equation \eqref{loglikelihoodlinearized} in the likelihood analysis. In general, this asymmetry changes as a function of the polar angle $\vartheta$, so the false asymmetry coefficients become functions of $\vartheta$, $a_i \rightarrow a_i(\vartheta)$ and $b_i \rightarrow b_i(\vartheta)$. Table \ref{falseasymtable} shows the false asymmetry coefficients averaged over the entire range of $\vartheta_{fpp}$ and the change in $R$ when $\vartheta$-dependent false asymmetry coefficients are included in the likelihood analysis.
\begin{table}[h]
  \begin{center}
    \begin{tabular}{|c|c|c|c|c|c|}
      \hline $Q^2$, GeV$^2$ & $a_1$ & $b_1$ & $a_2$ & $b_2$ & $\Delta R$ \\ \hline
      5.2 & $2.3\times 10^{-3}$ & $-7.3\times 10^{-3}$ & $-3.7\times 10^{-2}$ & $-6.7 \times 10^{-3}$ & $8.6\times 10^{-3}$ \\ \hline
      6.7 & $6.6\times 10^{-3}$ & $-1.4\times 10^{-3}$ & $-3.7\times 10^{-2}$ & $-9.2\times 10^{-3}$ & $1.15\times 10^{-2}$ \\ \hline
      8.5 & $5.9\times 10^{-3}$ & $-1.0\times 10^{-2}$ & $-4.4\times10^{-2}$ & $-6.7\times10^{-3}$ & $1.1 \times 10^{-2}$ \\ \hline
    \end{tabular}
  \end{center}
  \caption{\label{falseasymtable} False asymmetry coefficients and their effect on the form factor ratio.}
\end{table}

The false asymmetries are roughly independent of kinematics, indicating that they are primarily determined by the geometry of the chambers and the azimuthal dependence of track reconstruction inefficiencies and/or biases. Including the fitted false asymmetry terms as $\lambda_0$ in the analysis results in a positive shift in $R$ with an absolute magnitude of roughly $.01$, a result that is also roughly independent of the kinematics. The nature of the asymmetry deserves a brief comment. The $\cos \varphi$ and $\sin \varphi$ terms $a_1$ and $b_1$ are no more than one percent for any of the kinematics. These terms are mainly determined by misalignment, but can also include the influence of the helicity-independent induced polarization of the inelastic background, and for the highest $Q^2$ point in particular, the false asymmetry component is difficult to separate from the background polarization component in the region dominated by elastic scattering. The $\cos (2\varphi)$ and $\sin (2\varphi)$ terms $a_2$ and $b_2$, while relatively large (approximately four percent for the $\cos(2\varphi)$ term), come from reconstruction biases and inefficiencies. They should not contain geometric/acceptance effects because of the cone test, but some residual acceptance effects may remain since the cone test is calculated at the reconstructed value of $z_{close}$, assuming an $s_{close}$ value of zero. Since the $z_{close}$ resolution is relatively poor, particularly at small angles, the cone test cannot completely eliminate acceptance related false asymmetries. Though they can be large, these and other, higher-order terms in the false asymmetry distribution should have no effect on the helicity-difference asymmetry, since $\int_0^{2\pi} \cos(m\varphi) \cos(n\varphi) d\phi = 0$ for $m \ne n$.

\subsection{Background Corrections}
\paragraph{}
The procedure for estimating the inelastic contamination for a given set of cuts was described in section \ref{elasticcutssection}. To determine the effect of this contamination on the form factor ratio, the polarization of the background was measured. To obtain this polarization, the methods of section \ref{polextractsection} were applied to events failing the elastic event selection cuts. To be more specific, events were required to fail both the coplanarity cut and the inelasticity cut in the $\Delta x$-$\Delta y$ plot. Although such a cut certainly rejects many inelastic events as well, it has the advantage of achieving a selection of inelastic events which is free of elastic contamination, when compared to a simple anticut of the elastic cut, which is satisfied by events in the elastic radiative tail rejected by this cut, which appear along the $+\Delta x$ axis for $\Delta y \approx 0$. The abundance of inelastic events for the $Q^2=8.5\ GeV^2$ kinematics allows the measurement of the polarization of the background with much greater precision than the polarization of the elastic events. However, it cannot generally be assumed that the polarization of the background in the region where it overlaps with the elastic peak is the same as the average polarization of inelastic events over the full acceptance.

For a given elastic cut width, the contamination from inelastic background is a function mainly of the proton inelasticity $\Delta_p$. From figure \ref{SIMCdatacomp} it is clear that the ratio of the background to the total number of events is small in the super-elastic region, where only the target endcaps contribute, and rises rapidly for $\Delta_p < 0$, as the Bremsstrahlung flux increases with $E_{beam}-E_\gamma$ and the available phase space for $\pi^0$ photoproduction also increases. In the Monte Carlo calculation of the background, the $\pi^0$ contribution becomes larger than the elastic contribution at $\Delta_p \approx -0.4\%$ or approximately $-2\sigma$, a behavior which is confirmed in the data by examining the $\Delta x$ vs. $\Delta y$ plot for different bins in $\Delta_p$. 

The ratio of the background to the sum of signal and background is only large enough to significantly impact the form factor ratio in the region where $\pi^0$ production dominates the background. One might therefore reasonably assume that the polarization of the background can be assumed constant in the region of $\Delta_p$ where it overlaps with the elastic peak. However, this assumption is not even required, since the background polarization can be measured in the overlap region in $\Delta_p$, albeit with lower precision than the average over the whole inelastic region. In attempting to measure the polarization of the \emph{background} in the region close to the elastic peak, it is very important to obtain a sample of inelastic events that is free of contamination from elastic events in order to avoid misinterpreting an increasing contamination from elastic events as a variation of the polarization of inelastic events near the peak; hence, the requirement that events fail $\Delta x$ and $\Delta y$ cuts significantly wider than those used to select elastic events.
\begin{figure}[h]
  \begin{center}
    \includegraphics[angle=90,width=.98\textwidth]{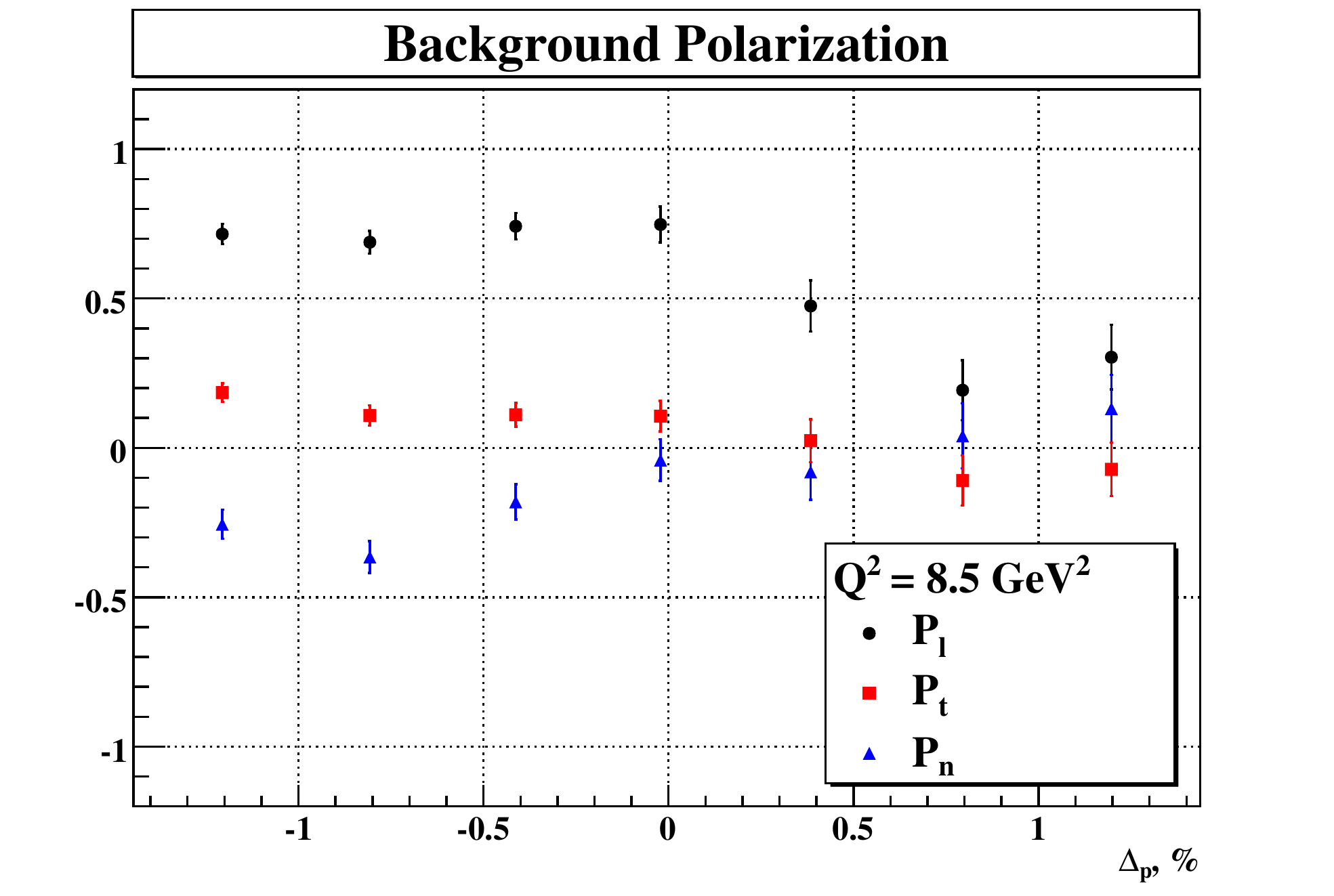}
  \end{center}
  \caption{\label{backgroundpol85fig} Measured background polarization components vs. $\Delta_p$, $Q^2 = 8.5$ GeV$^2$.}
\end{figure}

Figure \ref{backgroundpol85fig} shows the results of the polarization analysis of events failing both elastic cuts $\Delta \phi$ and $\Delta_e$ at $Q^2 = 8.5$ GeV$^2$ in the $\Delta_p$ region of the elastic peak. The assumption of constant background polarization holds up fairly well for the helicity-dependent components $P_t$ and $P_l$ for $\Delta_p \le 0$, since $\pi^0$ production is the dominant background in this region. At $\Delta_p \approx 0$, the primary reaction mechanism for the background changes from $\pi^0$ production to quasi-elastic $(e,e'p)$ from the target endcaps, so a change in the polarization is expected. The polarization components shown in figure \ref{backgroundpol85fig} were extracted using the analyzing power curve of figure \ref{Ayfig}, which was obtained from the measured polarization of elastic events using the $3\sigma$ cuts of table \ref{finalelasticcuttable}. The values of the helicity-independent induced polarization $P_n$ shown in figure \ref{backgroundpol85fig} have not been corrected for the false asymmetry, which is to be measured by the elastic events. Such corrections are unimportant for this analysis, since only the transferred polarization components $P_t$ and $P_l$ affect the determination of $G_E/G_M$.

The measured background polarization in the region of overlap with the elastic peak was combined with estimates of the inelastic contamination as a function of $\Delta_{p}$ using the same binning as the polarization measurements. This information was then used to construct a modified likelihood function using the following replacement of the polarization components $\mathbf{P}$ in the expression for $\mathcal{L}$:
\begin{eqnarray}
  P_j &\rightarrow& (1-f)P_j + fP_j^{inel.}
\end{eqnarray}
With this replacement, the coefficients $\lambda_j$ become
\begin{eqnarray}
  \lambda_j^{(i)} &\rightarrow& (1-f) \lambda_j^{(i)}
\end{eqnarray}
and a new term $\lambda_{bg}$ is added for the background asymmetry,
\begin{eqnarray}
  \lambda_{bg} &\equiv& fh\epsilon_i A_y^{(i)}\left(S_{yx}^{(i)} \cos \varphi_i - S_{xx}^{(i)} \sin \varphi_i \right) P_x^{inel} + \nonumber \\
  & & fA_y^{(i)}\left(S_{yy}^{(i)} \cos \varphi_i - S_{xy}^{(i)} \sin \varphi_i \right) P_y^{inel} + \nonumber \\
  & & fh\epsilon_i A_y^{(i)}\left(S_{yz}^{(i)} \cos \varphi_i - S_{xz}^{(i)} \sin \varphi_i\right) P_z^{inel}
\end{eqnarray}
which modifies the left-hand side of equation \eqref{loglikelihoodlinearized} to 
\begin{eqnarray}
  \lambda_j^{(i)}\left(1-\lambda_0\right) &\rightarrow& \lambda_j^{(i)}\left(1-\lambda_0-\lambda_{bg}\right) \label{LHSmodified}
\end{eqnarray}
\begin{figure}[h]
  \begin{center}
    \subfigure[$Q^2=5.2$ GeV$^2$]{\label{Rcutdep52}\includegraphics[angle=90,width=.49\textwidth]{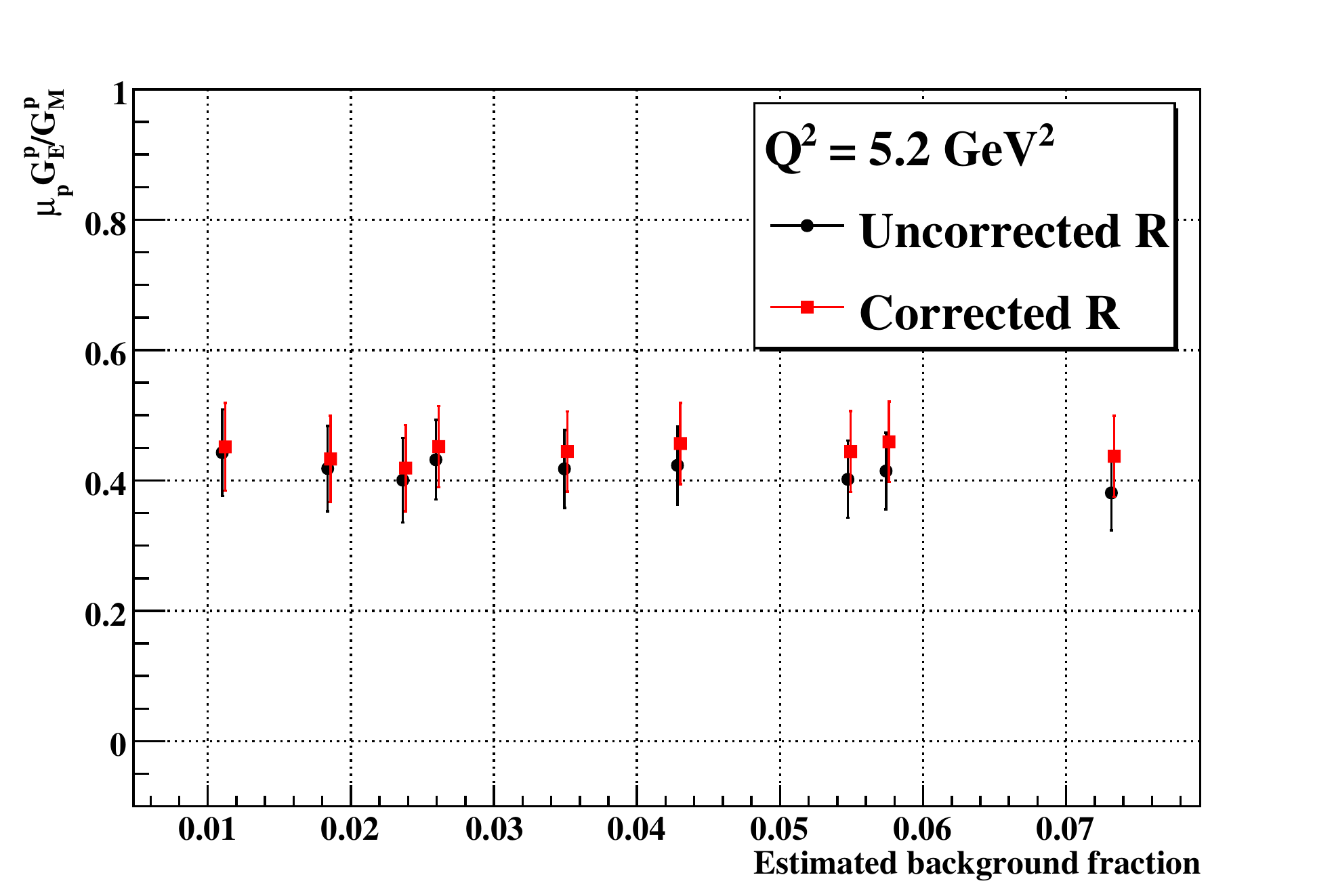}}
    \subfigure[$Q^2=6.7$ GeV$^2$]{\label{Rcutdep67}\includegraphics[angle=90,width=.49\textwidth]{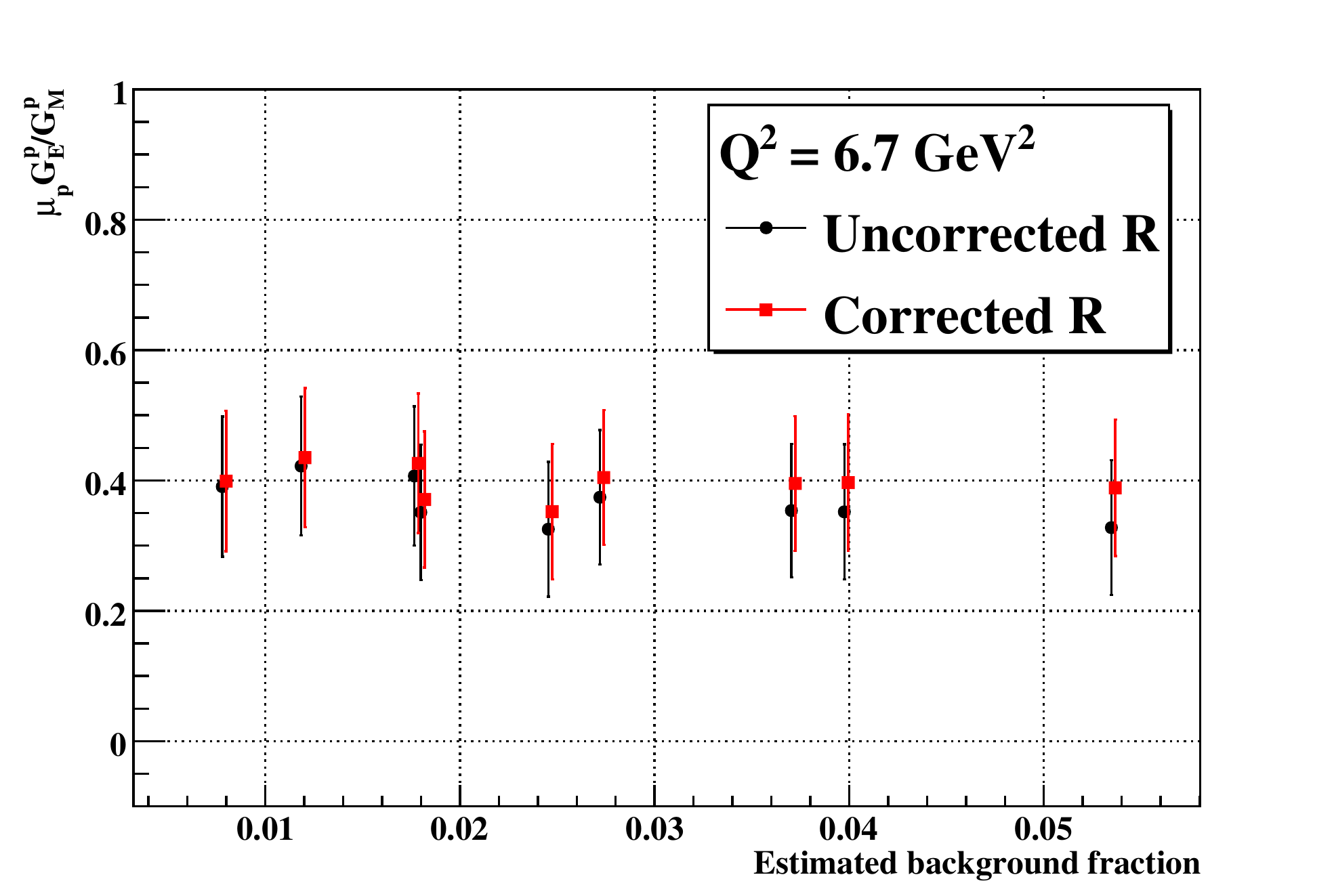}}
  \end{center}
  \caption{\label{RcutdeplowQ} Cut width dependence of the extracted form factor ratio for the $Q^2=5.2$ and 6.7 GeV$^2$ kinematics.}
\end{figure}
\begin{figure}[h]
  \begin{center}
    \includegraphics[angle=90,width=0.98\textwidth]{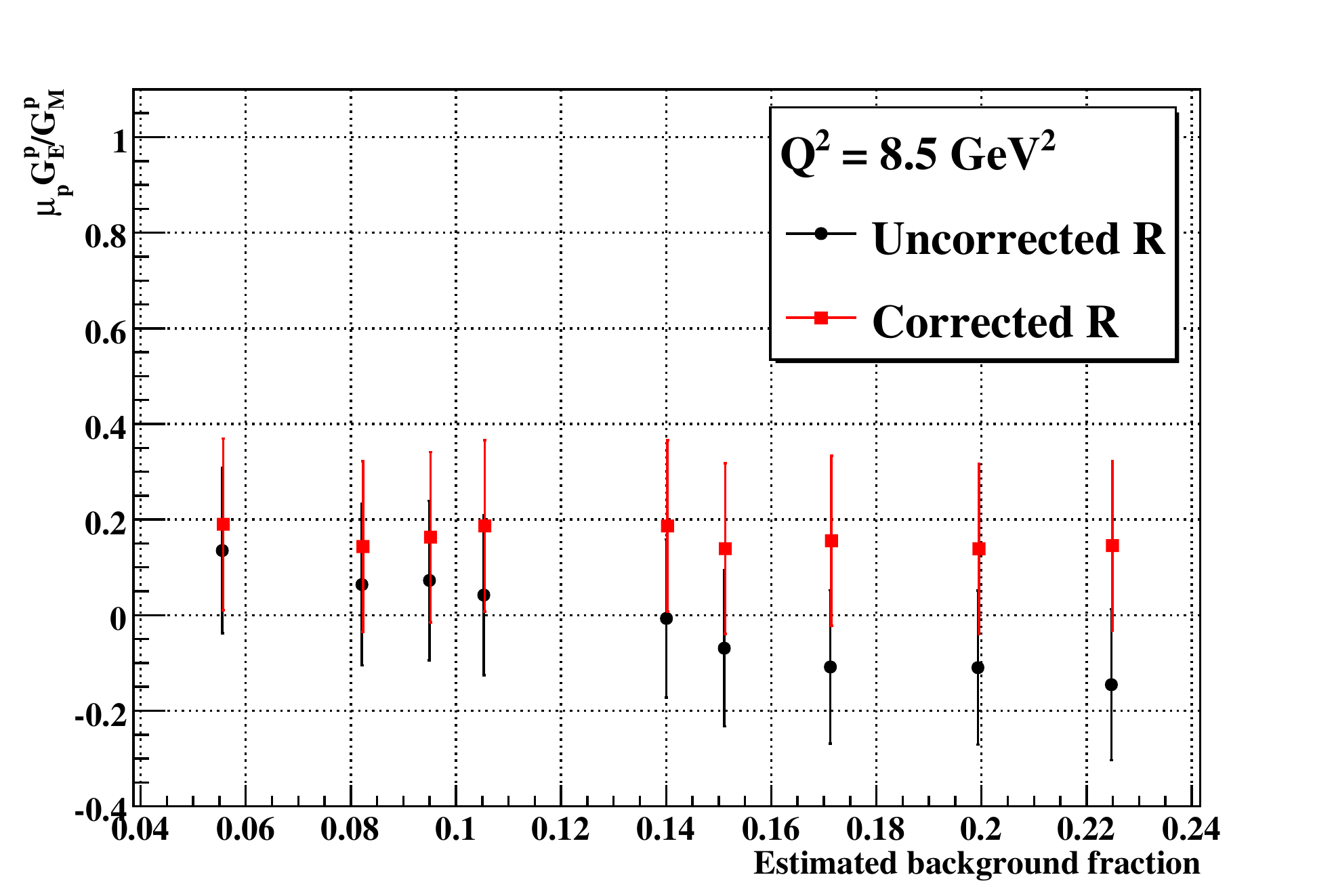}
  \end{center}
  \caption{\label{Rcutdep85} Cut width dependence of the extracted form factor ratio, $Q^2 = 8.5$ GeV$^2$.}
\end{figure}
The inclusion of helicity-dependent background polarization terms in equation \eqref{LHSmodified} which are not cancelled by the reversal of the beam polarization is responsible for the correction to $R$. The direct inclusion of the background subtraction in the likelihood analysis gives results that are essentially identical to correcting the results after the fact using equation \eqref{correctedpolformula}, but it allows for the treatment of the inelastic contamination estimate with arbitrarily fine binning in $\Delta_p$ in a straightforward manner, making it preferable to the latter method. 

Figure \ref{RcutdeplowQ} shows the variation of the extracted form factor ratio as a function of the estimated background accepted by the cuts for the two lower $Q^2$ points of E04-108. In both graphs, the black circles represent the results of the uncorrected analysis, and the red squares represent the result of the background-subtracted likelihood analysis. For both of these kinematics, the result for $G_E/G_M$ is quite stable even as the cuts are varied between $3\sigma$ and $6\sigma$, and the size of the background correction is small. The data for some of the different cut widths exhibit variations which are not explained by the background correction alone; these variations are compatible with purely statistical fluctuations within the extra events accepted by different cuts. The fact that the correction brings the points with the highest accepted background into very good agreement with the points with the lowest accepted background supports this conclusion. Figure \ref{Rcutdep85} shows the same dependence for the highest $Q^2$ point. For this setting, the background increases much more rapidly as a function of the cut width, which is apparent in the rapid decrease of the uncorrected form factor ratio as a function of $f$. The corrected form factor ratio in this case is still independent of the background correction up to statistical fluctuations in the polarization of the extra events accepted by wider cuts and the uncertainty in the background polarization used for the correction (see figure \ref{backgroundpol85fig}), demonstrating that the background subtraction is handled correctly.

In order to account for the uncertainties in the estimated background fraction $f$ and the extrapolation of the measured background polarization under the elastic peak, a relative uncertainty of $\pm15\%$ is assigned to $f$, which leads to the uncertainties in $P_t$, $P_l$, and $R$ shown in table \ref{systbackgroundtable}.
\begin{table}
  \begin{center}
    \begin{tabular}{|c|c|c|c|c|}
      \hline $Q^2$, GeV$^2$ & $f$, \% & $\Delta P_t$ & $\Delta P_l$ & $\Delta R$ \\ \hline
      5.2 & 1.1 & $3.2\times 10^{-4}$ & $5.7\times 10^{-4}$ & $1.3 \times 10^{-3}$ \\ \hline
      6.7 & 0.8 & $3.1\times 10^{-4}$ & $9.1\times 10^{-4}$ & $1.3 \times 10^{-3}$ \\ \hline
      8.5 & 5.5 & $1.8\times 10^{-3}$ & $3.6\times 10^{-3}$ & 0.012 \\ \hline
    \end{tabular}
  \end{center}
  \caption{\label{systbackgroundtable} Systematic uncertainty in $R$ due to $\pm$15\% relative uncertainty in $f$.}
\end{table}
Note that the uncertainty in the correction to $R$ is significantly smaller than what would be implied by substitution of the uncertainty in the corrected $P_t$ and $P_l$ into the formula for the error in $R$. This is because of partial cancellations occuring in the expression for the ratio:
\begin{eqnarray}
  \frac{P_t^{corr}}{P_l^{corr}} &=& \frac{P_t^{obs} - fP_t^{inel}}{P_l^{obs}-fP_l^{inel}} \\
  \Delta \left(\frac{P_t^{corr}}{P_l^{corr}}\right) &=& \frac{1}{1-f}\left(P_l^{inel}\frac{P_t^{corr}}{P_l^{corr}} - \frac{P_t^{inel}}{P_l^{corr}}\right)\Delta f 
\end{eqnarray}
The results for the uncertainty in $R$ related to the background correction show that even with the relatively pessimistic assumption $\Delta f/f = 15\%$, the uncertainty in $R$ due to the inelastic background is relatively small.
\subsection{Total Systematic Error Budget}
\paragraph{}
Table \ref{totalsysttable} summarizes the contributions to the total systematic uncertainty and reports the final systematic uncertainty as the quadrature sum of all the individual contributions.
\begin{table}[h]
  \begin{center}
    \begin{tabular}{|c|c|c|c|}
      \hline Uncertainty & $Q^2=5.2$ GeV$^2$ & $Q^2=6.7$ GeV$^2$ & $Q^2=8.5$ GeV$^2$ \\ \hline
      $\Delta R(\Delta \phi)$ & .016 & .020 & .035 \\
      $\Delta R(\Delta \theta)$ & $7.8\times 10^{-4}$ & $3.4\times 10^{-3}$ & $1.5\times 10^{-3}$ \\ 
      $\Delta R(\Delta \delta)$ & $6.0\times 10^{-5}$ & $1.67\times 10^{-3}$ & $8.55\times 10^{-3}$ \\
      $\Delta R(\Delta y_{tar})$ & $3.7\times 10^{-3}$ & $4.9\times 10^{-3}$ & $9.5\times 10^{-3}$ \\
      $\Delta R(\varphi_{fpp})$ & $2.9\times 10^{-4}$ & $5.6\times 10^{-3}$ & $0.017$ \\
      $\Delta R(\mbox{background})$ & $1.3 \times 10^{-3}$ & $1.3 \times 10^{-3}$ & $0.012$ \\ \hline 
      Total $\Delta R_{syst}$ & 0.0165 & 0.022 & 0.043 \\ \hline
    \end{tabular}
  \end{center}
  \caption{\label{totalsysttable} Total systematic uncertainty in $\mu_p G_E^p/G_M^p$ for the results of E04-108.}
\end{table}
\section{Discussion of the Results}
\paragraph{}
Table \ref{finalresultstable} shows the results of experiment E04-108, together with the acceptance averaged $Q^2$, the total accepted range of $Q^2$, and the statistical and systematic uncertainties. 
\begin{table}[h]
  \begin{center}
    \begin{tabular}{|c|c|c|c|}
      \hline $<Q^2>$ & $Q^2_{min}$ & $Q^2_{max}$ & $R \pm \Delta R_{stat} \pm \Delta R_{syst}$ \\
      \hline 5.171 & 4.904 & 5.476 & 0.436 $\pm$ 0.067 $\pm$ 0.0165 \\
      \hline 6.703 & 6.451 & 7.255 & 0.389 $\pm$ 0.114 $\pm$ 0.022 \\
      \hline 8.488 & 8.135 & 8.868 & 0.205 $\pm$ 0.181 $\pm$ 0.043 \\ \hline 
    \end{tabular}
  \end{center}
  \caption{\label{finalresultstable} Results of experiment E04-108.}
\end{table}
E04-108 represents the first high $Q^2$ recoil polarization data for $G_E^p/G_M^p$ other than the Hall A experiments\cite{Jones00,Punjabi05,Gayou02}. It is a very similar experiment, studying the coincidence $^1H(\vec{e},e'\vec{p})$ reaction using a magnetic spectrometer and a proton polarimeter to detect and reconstruct scattered protons and measure their polarization. However, it is different in important ways. First and most importantly, it used a completely different magnetic system than the Hall A High Resolution Spectrometers (HRSs). Secondly, it used a polarimeter of substantially different design. Finally, the character of the cuts used to select elastic events differs between E04-108 and the Hall A experiments because of a difference in the dominant sources of experimental resolution. 

For the three highest $Q^2$ points from Hall A, the electron was detected with a calorimeter similar to BigCal, but with a larger block size of 15$\times$15 cm$^2$, almost four times the transverse size of the lead-glass bars of BigCal. This important difference leads to significantly worse coordinate resolution for the calorimeter used in the Hall A experiments compared to BigCal. On the other hand, the momentum, angular and vertex resolution of the HRSs is significantly better than that of the HMS. The momentum resolution in particular is almost an order of magnitude better than that of the HMS. This means that the dominant (worst) resolution in the cuts used to select elastic events was different between the two experiments. In Hall A, the electron angle resolution was dominant, while the proton momentum resolution was small by comparison. In Hall C, on the other hand, the proton momentum resolution was dominant, while the electron angle resolution was small by comparison, at least when expressed in terms of the ``missing momentum'' $\Delta_e$, which is measured by BigCal with excellent resolution when the large Jacobian of the reaction is taken into account.
\begin{figure}[h]
  \begin{center}
    \includegraphics[angle=90,width=.98\textwidth]{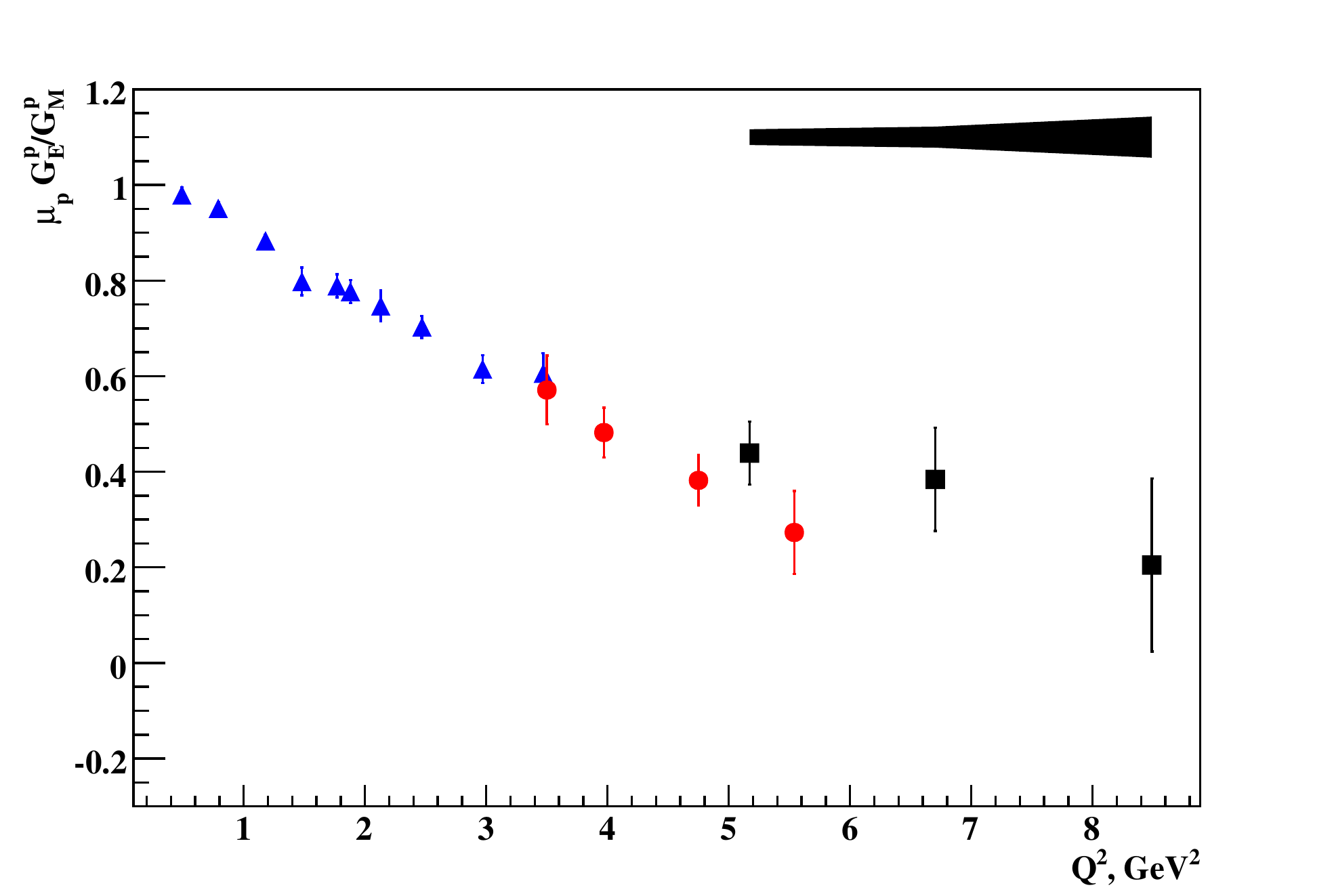}
  \end{center}
  \caption{\label{ResultsFig} Results of experiment E04-108 shown as $\mu_p G_E^p/G_M^p$. The results of experiments E93-027 and E99-007 are shown for comparison. The error bars on the data points are statistical, while the systematic errors are displayed as the band at the top of the figure.}
\end{figure}

Because it is the first recoil polarization experiment to use a completely different apparatus in a $Q^2$ range where direct comparison with the Hall A recoil polarization results is possible, it provides an important test of the reproducibility of the recoil polarization technique. Figure \ref{ResultsFig} shows the results of E04-108 (black squares) together with the results of the Hall A experiments \cite{Jones00,Punjabi05}(blue triangles) and \cite{Gayou02}(red circles). The error bars shown are statistical. The systematic uncertainty is displayed as the black band at the top of the figure. The overlap point at $5.2$ GeV$^2$ is in reasonably good agreement with the two surrounding points from \cite{Gayou02}, and confirms the continuing decrease of $G_E^p/G_M^p$ with $Q^2$. The two new data points extend the knowledge of $G_E^p/G_M^p$ to yet higher $Q^2$.

The first interesting feature of the results is that the linearly decreasing trend observed in the previous data appears to be slowing in the Hall C data. Although $R$ still decreases with $Q^2$, all of the new data points come in higher than the linear fit to the Hall A data. The statistical errors of the new data points are such that it is difficult to draw strong conclusions with respect to a change in the behavior of $R$ with $Q^2$, but the new data, when combined with planned future experiments to measure $R$ to $Q^2$ of 15 GeV$^2$ following the upcoming 12 GeV upgrade of Jefferson Lab, will answer the question definitively. 

One may reasonably ask whether the new Hall C data are in disagreement with the published data. To answer this question in terms of the probability of obtaining the value of $R$ measured at 5.2 GeV$^2$, a simple Monte Carlo calculation was performed assuming a linear $Q^2$ dependence of $R$ in the region from 3.0 to 6.0 GeV$^2$. The data points of \cite{Gayou02} were randomly sampled from a Gaussian distribution centered at the published values with a $\sigma$ equal to the published errors. Then, a straight-line fit was performed to the randomized Hall A data points. A new data point at $Q^2=5.17$ GeV$^2$ was then sampled from a Gaussian distribution centered at the expected value from the linear fit to the randomized Hall A data points, with a $\sigma$ equal to the error of the new Hall C data point. By performing a large number of such trials, it was determined that a value of $R$ greater than or equal to the new Hall C result occured in approximately 8\% of trials, indicating that while the new result is somewhat improbable, the disagreement is certainly not significant enough to imply a fundamental disagreement between the measurements, particularly given the crude assumption of linear $Q^2$ dependence of the form factor ratio used to calculate the probability. If the same analysis were performed with respect to different possible $Q^2$ dependences of the form factor ratio, such as the VMD model of Lomon\cite{Lomon2002} or the GPD model of Guidal et al.\cite{GuidalGPD2005}, or any model in which the rate of decrease of $R$ with $Q^2$ is slower than linear, the probability of the new result would be higher.
\begin{table}[h]
  \begin{center}
    \begin{tabular}{|c|c|c|}
      \hline  Experiment & $Q^2$, GeV$^2$ & $R \pm \Delta R_{stat}$ \\ \hline
      E93-027 & 2.47 & 0.703 $\pm$ 0.023 \\ \hline
      E04-019 & 2.5 & 0.694 $\pm$ 0.004 \\ \hline
    \end{tabular}
  \end{center}
  \caption{\label{prelimtwogamma} Preliminary result of E04-019 compared to the published result of E93-027 at similar $Q^2$. Errors are statistical only. The E04-019 result is the weighted average of the data taken at three different values of $\epsilon$ for a fixed $Q^2$. The variation of $R$ with $\epsilon$ was found to be very small.}
\end{table}

Additionally, the preliminary results of the high-statistics survey of the $\epsilon$-dependence of $G_E^p/G_M^p$ at $Q^2=2.5$ GeV$^2$ (experiment E04-019) are in excellent agreement with the Hall A results\cite{Punjabi05} at $Q^2=2.47$ GeV$^2$, as shown in table \ref{prelimtwogamma}. Since experiments E04-019 and E04-108 were performed using identical apparatus and the data were analyzed in exactly the same way, a neglected systematic error in the experiment is all but ruled out.
\subsection{Comparison to Theoretical Predictions}
\paragraph{}
The slowing decrease of $G_E^p/G_M^p$ with $Q^2$ apparent in the new results was correctly anticipated by the VMD-type models of Lomon\cite{Lomon2002} and Iachello and Bijker\cite{Iachello2004}, as shown in figure \ref{VMDcompfig}. In\cite{Iachello2004}, the slowing decrease of $R_p$ with $Q^2$ is a consequence of the inclusion of a direct coupling term in the isovector Pauli form factor in addition to the $\rho$ meson pole term. This direct coupling term is attributed to an intrinsic three-quark structure of the nucleon and its inclusion results in a much better prediction for the neutron spacelike form factors compared to the 1973 model\cite{Iachello1973}, which predicted a faster decrease of $R_p$ with $Q^2$. 
\begin{figure}[h]
  \begin{center}
    \includegraphics[angle=90,width=.98\textwidth]{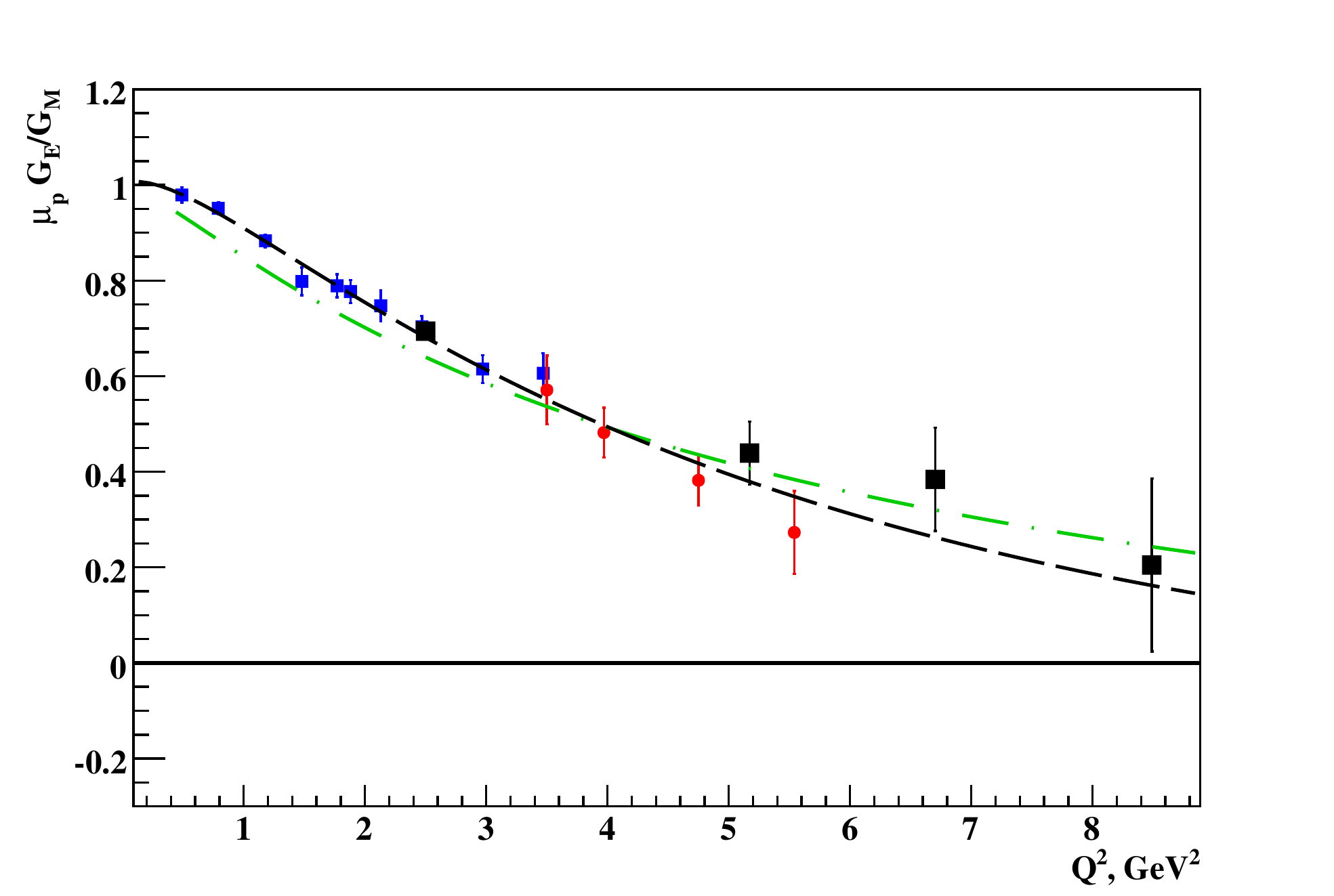}
  \end{center}
  \caption{\label{VMDcompfig} Comparison of the results of E04-108, together with E93-027 and E99-007, to the VMD models of Lomon(dashed curve) and Iachello and Bijker (dot-dashed curve). See text for references.}
\end{figure}
Both of the VMD models shown, and particularly the model of Lomon\cite{Lomon2002}, achieve quite reasonable fits to the data for all four nucleon form factors over the entire $Q^2$ range in which they are known. 

While these models are somewhat phenomenological in nature, involving a number of free parameters which can be adjusted to fit the data, the values of the best fit parameters provide important insight into the contribution of various mesons to the spectral functions presented in chapter 2 and the transition from meson dynamics known to be important at low $Q^2$ to the dimensional scaling behavior expected from perturbative QCD at high $Q^2$. The model of Lomon, for example, uses five vector mesons to fit all four nucleon form factors\footnote{In addition to the usual $\rho$, $\omega$, and $\phi$ mesons, the $\rho'(1450)$ and $\omega'(1420)$ mesons were added to the fit.}. The model achieves a sufficiently accurate description of all nucleon form factor data in the whole available $Q^2$ range that it can reasonably be applied where the form factors are required as input to the interpretation of other experiments, and the new results for $G_E^p/G_M^p$ should not significantly alter the parameters of the model\cite{Lomon2002}. In the context of this model, the data for all four nucleon form factors taken together suggest a relatively slow approach to the asymptotic behavior required by perturbative QCD, as characterized by the parameter $\Lambda_2 \approx 2.8$ GeV which controls the suppression of spin-flip; i.e., the suppression of $F_2$ relative to $F_1$.

\begin{figure}[h]
  \begin{center}
    \includegraphics[angle=90,width=.98\textwidth]{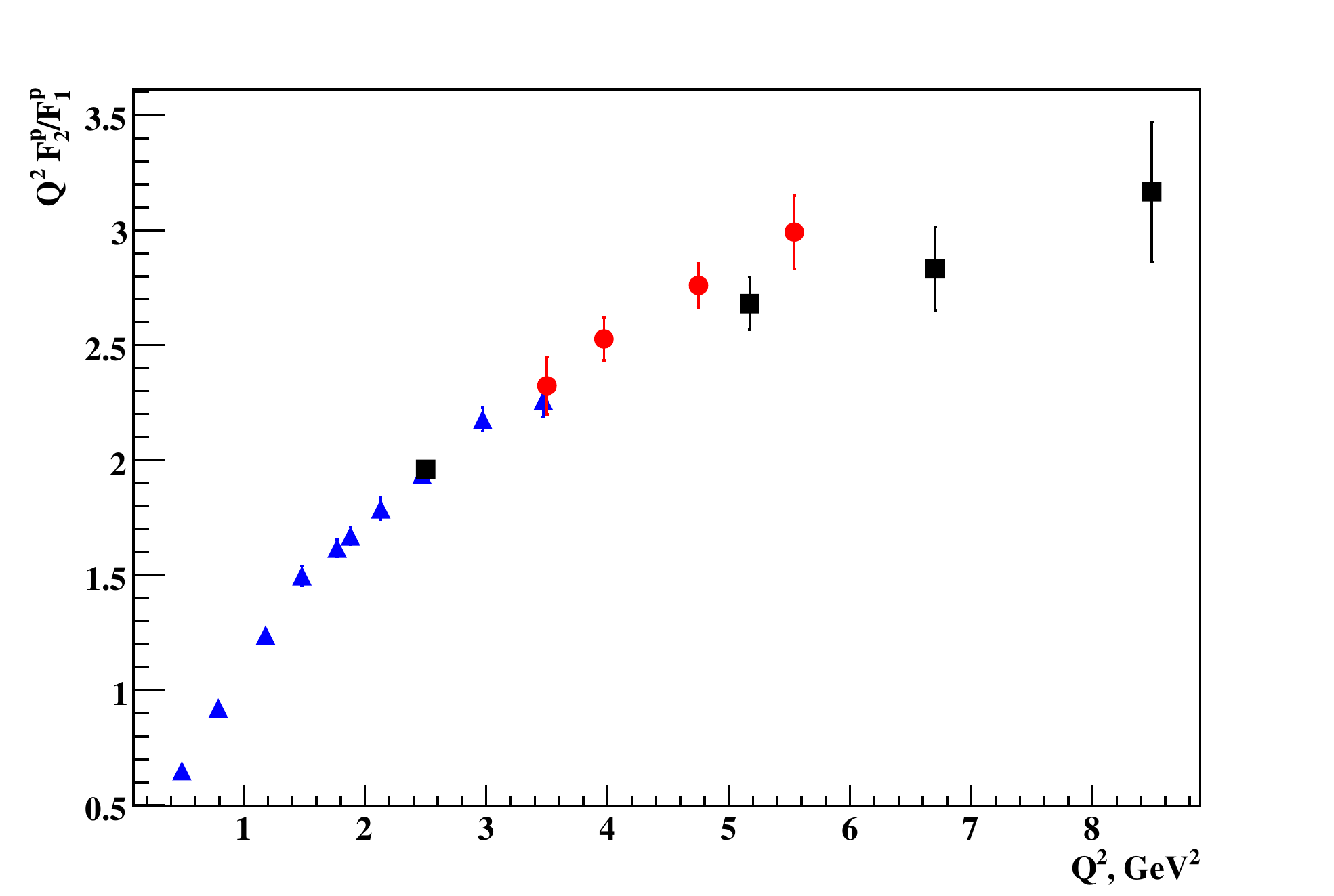}
  \end{center}
  \caption{\label{dimensionalscaling} $Q^2 F_2^p/F_1^p$ from recoil polarization data.}
\end{figure}
\begin{figure}
  \begin{center}
    \includegraphics[angle=90,width=.98\textwidth]{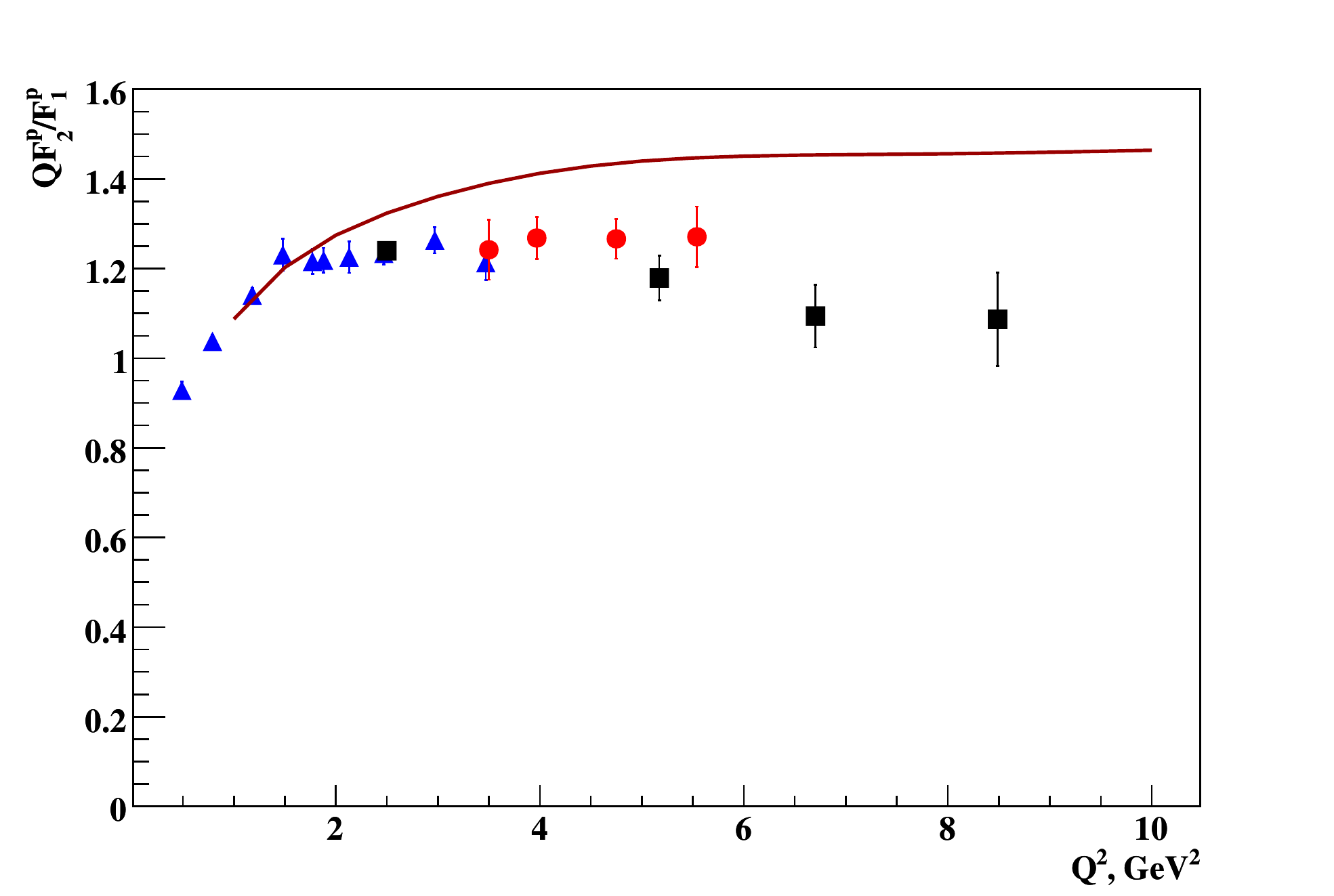}
  \end{center}
  \caption{\label{ralstonscaling} $Q F_2^p/F_1^p$ from recoil polarization data. The burgundy curve shows the LFCBM prediction for this ratio.}
\end{figure}
\begin{figure}[h]
  \begin{center}
    \includegraphics[angle=90,width=.98\textwidth]{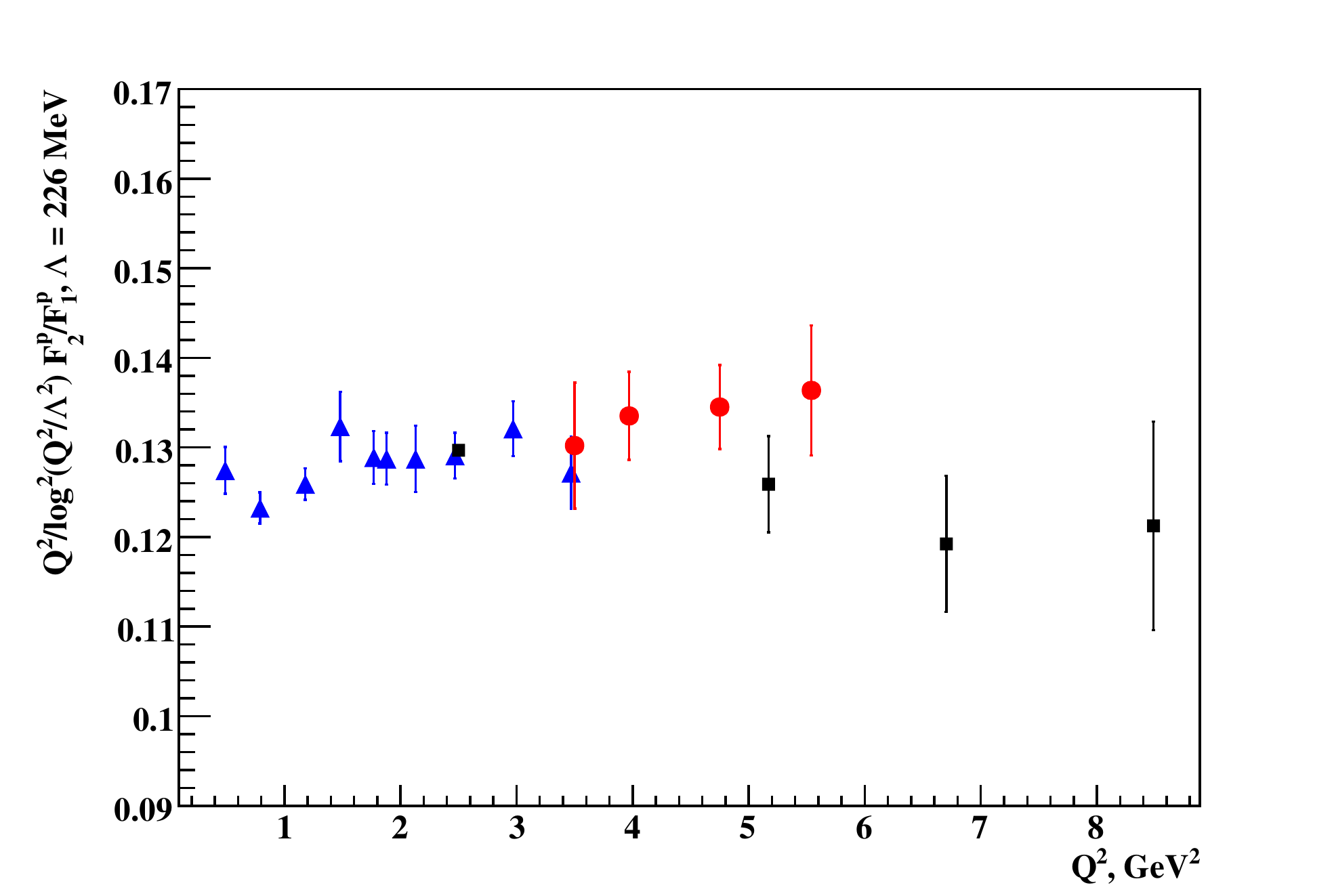}
  \end{center}
  \caption{\label{belitskyscaling} $Q^2 /\log^2\left(Q^2/\Lambda^2\right) F_2^p/F_1^p$ from recoil polarization data, with $\Lambda=226\pm25$ MeV resulting in the best fit.}
\end{figure}
The dimensional scaling law\cite{BrodskyFarrar1975} for nucleons consisting of three valence quarks predicts that $F_2$ should scale as $F_1/Q^2$ for $Q^2 \rightarrow \infty$. The ratio $Q^2 \frac{F_2^p}{F_1^p}$ is shown for the results of E04-108 in figure \ref{belitskyscaling}. Although the new data do show hints of a flattening of the ratio $Q^2 F_2/F_1$, the onset of dimensional scaling cannot be conclusively inferred from the new data. If, on the other hand, the flattening trend is confirmed by future experiments at higher $Q^2$, one may begin to discuss the onset of perturbative behavior of the proton form factors. 

Miller's light-front cloudy bag model(LFCBM) \cite{MillerLFCBM} predicts a violation of hadron helicity conservation due to the relativistic effect of the Melosh rotations required to boost the nucleon wave function to the light-front, which appears in the form factor ratio as a constant scaling of the ratio $Q\frac{F_2^p}{F_1^p}$. Such a scaling behavior was observed in the Hall A recoil polarization data (although the LFCBM significantly overpredicts the value of $QF_2^p/F_1^p$). The results of E04-108, on the other hand, appear to diverge from this behavior, as they exhibit a sagging of $QF_2^p/F_1^p$ at higher $Q^2$.

Figure \ref{belitskyscaling} shows that the new recoil polarization data are also compatible with the modified logarithmic perturbative QCD scaling of $F_2/F_1$ found by Belitsky et al.\cite{BelitskyJiYuan2003} with the value $\Lambda = 226\pm 25$ MeV giving the best fit to the data. In the context of perturbative QCD, the measurement of $G_E^p/G_M^p$ to higher $Q^2$ can be interpreted as an experimental determination of the fundamental scale parameter of QCD $\Lambda$ through asymptotic scaling relations such as \cite{BelitskyJiYuan2003}. However, the question of the minimum momentum transfer for which a pQCD analysis of the nucleon form factors is valid is far from settled, and awaits further advances in both theory and experiment. The GPD model of Guidal et al.\cite{GuidalGPD2005} also predicts a continued decrease of $G_E^p/G_M^p$ with $Q^2$ which is slower than linear, and a zero crossing at approximately 9 $GeV^2$. The results of E04-108 are certainly compatible with this model within uncertainties, and it will be interesting to determine the consequences for this and other GPD models when the results of E04-108 are included in the form factor data used to constrain the zero$th$ $x$ moments of the GPDs $H$ and $E$.
\begin{figure}[h]
  \begin{center}
    \includegraphics[angle=90,width=.98\textwidth]{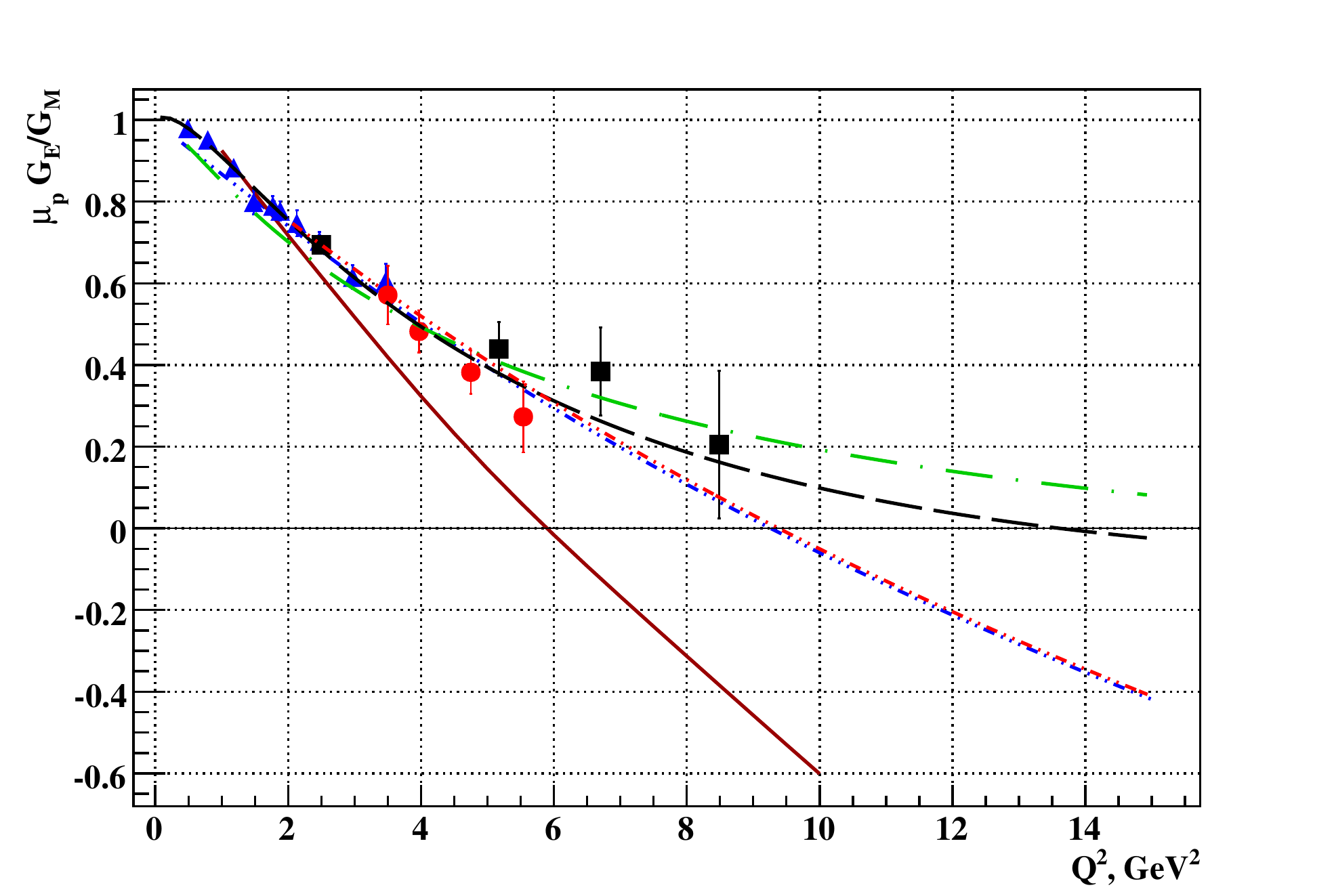}
  \end{center}
  \caption{\label{finalratiofig} Results of E04-108 with a sampling of theoretical predictions projected to the $Q^2$ region of the JLab 12 GeV upgrade. The theory curves are the VMD model of Iachello and Bijker (green dot-dashed), the Lomon VMD model (black dashed), the pQCD prediction of Belitsky et al. (red double dot-dashed), the GPD model of Guidal et al. (blue triple dot-dashed), and Miller's LFCBM (burgundy solid).}
\end{figure}
Figure \ref{finalratiofig} shows the predictions of several models for the ratio $\mu_p G_E^p/G_M^p$ in the higher $Q^2$ region that will be explored following the 12 GeV upgrade of Jefferson Lab. 
\subsection{Conclusion}
\paragraph{}
Experiments E04-108 and E04-019, respectively, extended the recoil polarization measurements of the proton electromagnetic form factor ratio $G_E^p/G_M^p$ to $Q^2=8.5$ GeV$^2$ and measured the $\epsilon$ dependence of $G_E^p/G_M^p$ at a fixed $Q^2$ of 2.5 GeV$^2$. As the first such measurements at high $Q^2$ to take place outside of Jefferson Lab Hall A, these results are an important test of the reproducibility of the recoil polarization technique, and they are in reasonable agreement with the Hall A results where they overlap, particularly at $Q^2=2.5$ GeV$^2$. The preliminary results of experiment E04-019 show no significant $\epsilon$-dependence of the proton $G_E^p/G_M^p$ ratio at fixed $Q^2$. Various models of the effect of two-photon exchange on the extraction of the proton form factors from Rosenbluth separation and double-polarization experiments predict a significant $\epsilon$-dependence of this ratio. The absence of such a dependence in the recoil polarization results is interpreted as a strong validation of the method, as the true form factors entering the Born-approximation expression for the nucleon electromagnetic current depend only on $Q^2$. The results are also expected to provide severe constraints on models attempting to explain the discrepancy between cross section and polarization experiments in terms of TPEX effects.

The new data at higher $Q^2$ show a slowing decrease of $G_E^p/G_M^p$ with $Q^2$ relative to the linear decrease observed in the Hall A data for $Q^2 \le 5.6$ GeV$^2$. Vector meson dominance (VMD) models correctly predicted this behavior. Although the statistical significance of this change in behavior is somewhat marginal, its physical implications are interesting to consider. A constant ratio $G_E^p/G_M^p$ at asymptotically large $Q^2$ is a signature of the onset of the dimensional scaling expected from perturbative QCD for a nucleon consisting of three quarks interacting weakly through gluon exchange. The planned experiments covering the $Q^2$ range from 10-15 GeV$^2$ following the 12 GeV upgrade of the CEBAF accelerator will answer the question of whether the flattening hinted at by the E04-108 results is real or whether $G_E^p/G_M^p$ will continue to decrease and eventually cross zero. The results presented in this thesis are the most precise measurements to date of the proton electric form factor $G_E^p$ in this range of $Q^2$, and as such they represent a significant advancement of the experimental knowledge of the structure of the nucleon.

\appendix
\chapter{Details of Cross Section and Polarization Transfer Derivations}
\section{Conventions for Dirac Algebra, $\gamma$-matrices and Spinors}
\paragraph{}
The convention used throughout the derivations in chapter 1 for Dirac spinors and $\gamma$-matrices is the same as that of \cite{QandL}, the so-called Dirac-Pauli representation. In this representation, the $\gamma$-matrices are given by
\begin{eqnarray}
  \gamma^0 &=& \left(\begin{array}{cc} 1 & 0 \\ 0 & -1 \end{array}\right) \\
  \vec{\gamma} &=& \left(\begin{array}{cc} 0 & \vec{\sigma} \\ -\vec{\sigma} & 0 \end{array} \right) \\
  \gamma^5 &\equiv& i\gamma^0\gamma^1\gamma^2\gamma^3 = \left(\begin{array}{cc} 0 & 1 \\ 1 & 0 \end{array}\right)
\end{eqnarray}
where each block represents a $2\times 2$ matrix. These matrices obey the anti-commutation relations 
\begin{eqnarray}
  \{ \gamma^{\mu},\gamma^{\nu} \} & = & 2 g^{\mu \nu} \\
  \{ \gamma^5,\gamma^{\mu} \} & = & 0
\end{eqnarray}
It is useful to write down some identities involving the Pauli matrices that will be helpful when deriving the polarized scattering amplitude: 
\begin{eqnarray}
  \left[\sigma_i, \sigma_j \right] &=& 2i \epsilon_{ijk} \sigma_k \\
  \left \{ \sigma_i, \sigma_j \right \} &=& 2 \delta_{ij} \\
  \sigma_i \sigma_j &=& \delta_{ij} + i\epsilon_{ijk}\sigma_k
\end{eqnarray}
From the commutation and anticommutation relations of the Pauli matrices, some useful relations for inner products of Pauli matrices and arbitrary vectors immediately follow: 
\begin{eqnarray} 
  (\vec{\sigma}\cdot \mathbf{a})(\vec{\sigma} \cdot \mathbf{b}) &=& \mathbf{a} \cdot \mathbf{b} + i \vec{\sigma} \cdot (\mathbf{a} \times \mathbf{b}) \label{paulimatr1} \\
  (\vec{\sigma}\cdot \mathbf{a})(\vec{\sigma} \cdot \mathbf{b})(\vec{\sigma} \cdot \mathbf{c}) &=& (\mathbf{b} \cdot \mathbf{c})(\vec{\sigma} \cdot \mathbf{a}) - (\mathbf{a} \cdot \mathbf{c})(\vec{\sigma} \cdot \mathbf{b}) + (\mathbf{a}\cdot \mathbf{b})(\vec{\sigma}\cdot \mathbf{c}) + \nonumber \\ & & i\mathbf{a}\cdot (\mathbf{b} \times \mathbf{c}) \nonumber \\
  \Rightarrow (\vec{\sigma}\cdot \mathbf{a})(\vec{\sigma} \cdot \mathbf{b})(\vec{\sigma} \cdot \mathbf{a}) &=& 2(\mathbf{a}\cdot \mathbf{b})(\vec{\sigma}\cdot \mathbf{a}) - a^2(\vec{\sigma}\cdot \mathbf{b}) \label{paulimatr2}
\end{eqnarray}
The following identities involving the $\gamma$-matrices are useful in deriving the polarized scattering amplitude: 
\begin{eqnarray}
  (\gamma^0)^2 &=& 1 \label{gammaidentityfirst} \\
  \gamma^5\gamma^0 & = & \left(\begin{array}{cc} 0 & -1 \\ 1 & 0 \end{array}\right) \\
  \gamma^0\vec{\gamma} & \equiv & \vec{\alpha} = \left(\begin{array}{cc} 0 & \vec{\sigma} \\ \vec{\sigma} & 0 \end{array}\right) \\
  \gamma^5\gamma^0\vec{\gamma} & = & \left(\begin{array}{cc} \vec{\sigma} & 0 \\ 0 & \vec{\sigma} \end{array}\right) \\ 
  \vec{\gamma}\gamma^5 & = & \left(\begin{array}{cc} \vec{\sigma} & 0 \\ 0 & -\vec{\sigma} \end{array}\right) \\
  \frac{1}{2}\gamma^5(1 + \gamma^0)\vec{\gamma} & = & \left(\begin{array}{cc} 0 & 0 \\ 0 & \vec{\sigma} \end{array}\right) \label{gammaidentitylast}
\end{eqnarray}
To derive the elastic electron-nucleon scattering amplitude in the Born approximation, only free-particle spinors are needed (no antiparticles), so it suffices to write down the free-particle solutions of the Dirac equation with positive energy: 
\begin{eqnarray}
  \left[\not{p} - M \right]u(p) &=& 0 \\
  u^{(s)}(p) &=& \sqrt{E+M} \left(\begin{array}{c} \chi^{(s)} \\ \frac{\vec{\sigma} \cdot \mathbf{p}}{E + M} \chi^{(s)} \end{array}\right) \\
  \bar{u} &\equiv& u^\dagger \gamma^0 
\end{eqnarray}
The normalization constant for the spinors reflects the covariant normalization convention in which
\begin{eqnarray}
  u^{\dagger(s_1)}u^{(s_2)} &=& 2E \delta_{s_1 s_2} \\
  \bar{u}^{(s_1)}u^{(s_2)} &=& 2M \delta_{s_1 s_2}
\end{eqnarray}
\section{Polarized and Unpolarized Spin Sums}
\paragraph{}
For unpolarized spins, the completeness relation for the spin states $\chi$ is $\sum_s \chi^{(s)} \chi^{\dagger (s)} = 1$, while for polarized spins the completeness relation is given by (\ref{spinensemble}), $\sum_s \chi^{(s)} \chi^{\dagger (s)} = \frac{1}{2}\left(1 + \vec{\sigma} \cdot \mathbf{h} \right)$, where $\mathbf{h}$ is the polarization vector. Therefore, the relevant completeness relation for Dirac spinors is
\begin{eqnarray}
  \sum_s u^{(s)}(p) \bar{u}^{(s)}(p) &=& (E+M) \left(\begin{array}{cc} 1 & -\frac{\vec{\sigma} \cdot \mathbf{p}}{E + M} \\ \frac{\vec{\sigma} \cdot \mathbf{p}}{E + M} & -\frac{p^2}{(E+M)^2}\end{array}\right) \nonumber \\
  &=& \left(\begin{array}{cc} E+M & -\vec{\sigma} \cdot \mathbf{p} \\ \vec{\sigma} \cdot \mathbf{p} & -E + M \end{array}\right) \nonumber \\
  &=& \gamma^\mu p_\mu + M = \not{p} + M \label{unpolarizedspinsum}
\end{eqnarray}
for unpolarized spins. For polarized spins, things get a bit more complicated. The first term in the completeness relation $\sum_s \chi^{(s)} \chi^{\dagger (s)} = \frac{1}{2}\left(1 + \vec{\sigma} \cdot \mathbf{h} \right)$ just gives the usual $\not{p}+M$ term above, with a factor of $1/2$. The $\vec{\sigma}\cdot \mathbf{h}$ term contains the polarization effects. This term evaluates to 
\begin{equation}
  \sum_s u^{(s)}(p) \bar{u}^{(s)}(p) = \frac{1}{2}\left(\begin{array}{cc} (E+M)(\vec{\sigma}\cdot \mathbf{h}) & -(\vec{\sigma}\cdot \mathbf{h})(\vec{\sigma} \cdot \mathbf{p}) \\ (\vec{\sigma} \cdot \mathbf{p})(\vec{\sigma}\cdot \mathbf{h}) & -\frac{1}{E+M}(\vec{\sigma} \cdot \mathbf{p})(\vec{\sigma}\cdot \mathbf{h})(\vec{\sigma} \cdot \mathbf{p})\end{array}\right) \nonumber 
\end{equation}
Using identities (\ref{paulimatr1}) and (\ref{paulimatr2}) allows the following rearrangements of the various terms:
\begin{eqnarray}
  (\vec{\sigma}\cdot \mathbf{h})(\vec{\sigma} \cdot \mathbf{p}) &=& \mathbf{h}\cdot \mathbf{p} + i\vec{\sigma} \cdot (\mathbf{h} \times \mathbf{p}) \nonumber \\
  (\vec{\sigma}\cdot \mathbf{p})(\vec{\sigma} \cdot \mathbf{h}) &=& \mathbf{p}\cdot \mathbf{h} + i\vec{\sigma} \cdot (\mathbf{p} \times \mathbf{h}) \nonumber \\
  -\frac{1}{E+M}(\vec{\sigma} \cdot \mathbf{p})(\vec{\sigma}\cdot \mathbf{h})(\vec{\sigma} \cdot \mathbf{p}) &=& -\frac{2}{E+M}(\mathbf{p}\cdot \mathbf{h})(\vec{\sigma} \cdot \mathbf{p}) + \nonumber \\ & & \frac{p^2}{E+M}(\vec{\sigma}\cdot \mathbf{h}) \nonumber \\
  &=& -\frac{2}{E+M}(\mathbf{p}\cdot \mathbf{h})(\vec{\sigma} \cdot \mathbf{p}) + \nonumber \\ & & (E-M)(\vec{\sigma}\cdot \mathbf{h}) \nonumber 
\end{eqnarray}
With these relations in hand, the spin sum can be grouped into several terms which can be reduced to a more convenient form: 
\begin{eqnarray}
  \sum_s u \bar{u} &=& \frac{1}{2}\left[E\mathbf{h}\cdot \left(\begin{array}{cc} \vec{\sigma} & 0 \\ 0 & \vec{\sigma} \end{array}\right) + M \mathbf{h}\cdot \left(\begin{array}{cc} \vec{\sigma} & 0 \\ 0 & -\vec{\sigma} \end{array}\right) + (\mathbf{p}\cdot\mathbf{h})\left(\begin{array}{cc} 0 & -1 \\ 1 & 0\end{array}\right) \nonumber + \right. \\ & & \left. i(\mathbf{p}\times \mathbf{h})\cdot \left(\begin{array}{cc} 0 & \vec{\sigma} \\ \vec{\sigma} & 0 \end{array}\right) - 2\frac{(\mathbf{p}\cdot \mathbf{h})}{E+M}\mathbf{p} \cdot \left(\begin{array}{cc} 0 & 0 \\ 0 & \vec{\sigma} \end{array} \right) \right] 
\end{eqnarray}
Using the identities (\ref{gammaidentityfirst})-(\ref{gammaidentitylast}), the polarized spin sum can be expressed entirely in terms of $\gamma$-matrices, which will come in handy for the calculation of the polarized electron and nucleon current tensors:
\begin{eqnarray}
  2\sum_s u \bar{u} &=& E\gamma^5 \gamma^0 (\vec{\gamma} \cdot \mathbf{h}) + M (\vec{\gamma}\cdot \mathbf{h})\gamma^5 + (\mathbf{p}\cdot \mathbf{h})\gamma^5\gamma^0 + \nonumber \\ & & i\gamma^0 (\vec{\gamma} \cdot (\mathbf{p}\times \mathbf{h})) - \frac{(\mathbf{p}\cdot \mathbf{h})}{E+M} \gamma^5(1+\gamma^0)(\vec{\gamma}\cdot \mathbf{p}) \label{polarizedspinsum}
\end{eqnarray}

\section{Trace Identities and the Evaluation of Current Tensors\label{tracetech}}
\paragraph{}
When evaluating the electron and nucleon tensors $L_e^{\mu\nu}$ and $W_N^{\mu\nu}$, one is essentially forming the outer product of a four-vector and its complex conjugate. For the nucleon tensor, this is given by:
\begin{eqnarray}
  \mathcal{J}_N^\mu &\equiv& \bar{u}(p')\Gamma^\mu u(p) \nonumber \\
  W_N^{\mu \nu} &\equiv& \mathcal{J}_N^\mu \mathcal{J}_N^{\nu *} \nonumber \\
  \mathcal{J}_N^{\nu *} &=& \left[\bar{u}(p')\Gamma^\nu u(p)\right]^* = \left[\bar{u}(p')\Gamma^\nu u(p)\right]^\dagger \nonumber \\
  \mathcal{J}_N^{\nu *} &=& u^\dagger(p) (\Gamma^\nu)^\dagger \bar{u}^\dagger(p') \nonumber \\
  &=& u^\dagger(p) \gamma^0 \gamma^0 (\Gamma^\nu)^\dagger \gamma^0 u(p') \nonumber \\
  &=& \bar{u}(p) \gamma^0 (\Gamma^\nu)^\dagger \gamma^0 u(p') \nonumber 
\end{eqnarray}
$\Gamma^\nu$ is a linear combination of $\gamma^\nu$ and $p^\nu + p'^\nu$. The latter is self-adjoint, while the former satisfies $\gamma^0(\gamma^\nu)^\dagger \gamma^0 = \gamma^\nu$, so that $\gamma^0 (\Gamma^\nu)^\dagger \gamma^0 = \Gamma^\nu$. This leaves
\begin{eqnarray}
  \mathcal{J}_N^{\nu *} &=& \bar{u}(p) \Gamma^\nu u(p') \nonumber \\
  \Rightarrow W_N^{\mu \nu} &=& \bar{u}(p')\Gamma^\mu u(p) \bar{u}(p)\Gamma^\nu u(p') \nonumber \\
  &=& \sum_{i=0}^3 \bar{u}_i(p') \sum_{j=0}^3 Q^{\mu \nu}_{ij} u_j(p') \nonumber \\
  Q^{\mu \nu} &\equiv& \Gamma^\mu u(p)\bar{u}(p) \Gamma^\nu \nonumber \\
  W_N^{\mu \nu} &=& \sum_{i=0}^3\sum_{j=0}^3 Q^{\mu \nu}_{ij} u_j(p')\bar{u}_i(p') \nonumber \\
  W_N^{\mu \nu} &=& Tr\left[\Gamma^\mu u(p)\bar{u}(p) \Gamma^\nu u(p')\bar{u}(p')\right] \label{Casimirtrick}
\end{eqnarray}
\paragraph{}
Finally, the following trace theorems will help in evaluating the lengthy expression for the polarized and unpolarized tensors encountered in the derivations of chapter 1. These relations follow from the anti-commutation relation for $\gamma$-matrices:
\begin{eqnarray}
  Tr(\gamma^\mu) &=& 0 \nonumber \\
  Tr\ 1 &=& 4 \nonumber \\
  Tr(\gamma^\mu \gamma^\nu) &=& 4 g^{\mu\nu} \nonumber \\
  Tr(\gamma^\mu \gamma^\nu \gamma^\alpha \gamma^\beta) &=& 4\left[g^{\mu \nu} g^{\alpha \beta} - g^{\mu \alpha} g^{\nu \beta} + g^{\mu \beta} g^{\nu \alpha}\right] \nonumber \\
  Tr(\gamma^5) &=& 0 \nonumber \\
  Tr(\gamma^5 \gamma^\mu \gamma^\nu) &=& 0 \nonumber \\
  Tr(\gamma^5 \gamma_\mu \gamma_\nu \gamma_\alpha \gamma_\beta) &=& 4i\epsilon_{\mu\nu\alpha\beta} \nonumber 
\end{eqnarray}
The trace of the product of an odd number of $\gamma$-matrices is zero. 
\chapter{Nucleon Current in the Breit Frame}
\label{BreitCurrentAppendix}
The general expression for the nucleon current was given in equation \ref{SachsNvertex}:
\begin{eqnarray}
  \mathcal{J}_N^\mu &=& \bar{u}(p') \left[G_M \gamma^\mu +\frac{G_E-G_M}{2M(1+\tau)}(p+p')^\mu\right]u(p)
\end{eqnarray}
The explicit Dirac free-particle spinors for the nucleon in its initial and final state are as follows:
\begin{eqnarray}
  \bar{u}(p') &=& \sqrt{E'+M} \left(\chi'^\dag, \chi'^\dag \frac{\mathbf{\sigma}\cdot \mathbf{p'}}{E'+M}\right)\gamma^0 \\
  u(p) &=& \sqrt{E+M} \left(\begin{array}{c} \chi \\ \frac{\mathbf{\sigma}\cdot \mathbf{p}}{E+M}\chi \end{array}\right)
\end{eqnarray}
The defining properties of the Breit frame simplify the spinor products:
\begin{eqnarray}
  (p+p')^\mu &=& 2M\sqrt{1+\tau} g^{\mu 0} \\  
  E' &=& E = M\sqrt{1+\tau} \\
  \mathbf{p'} &=& -\mathbf{p}
\end{eqnarray}
There are two spinor products required, $\bar{u}(p')\gamma^\mu u(p)$, and $\bar{u}(p')u(p)$. For $\mu=0$, one finds:
\begin{eqnarray}
  \bar{u}(p') \gamma^0 u(p) &=& (E+M)\left[\chi'^\dag \chi - \chi'^\dag \frac{(\mathbf{\sigma}\cdot \mathbf{p})^2}{(E+M)^2}\chi\right] \\
  &=& (E+M)\chi'^\dag \chi \left[1-\frac{p^2}{(E+M)^2}\right] \\
  &=& \chi'^\dag \chi \left[(E+M)-(E-M)\right] = 2M \chi'^\dag \chi 
\end{eqnarray}
For $\mu \equiv k \neq 0$ one finds:
\begin{eqnarray}
  \bar{u}(p') \gamma^k u(p) &=& (E+M)\left(\chi'^\dag, -\chi'^\dag \frac{\mathbf{\sigma}\cdot \mathbf{p}}{E+M} \right) \gamma^0 \gamma^k \left(\begin{array}{c} \chi \\ \frac{\mathbf{\sigma}\cdot \mathbf{p}}{E+M} \chi \end{array}\right) \\
  &=& (E+M)\left(\chi'^\dag, -\chi'^\dag \frac{\mathbf{\sigma}\cdot \mathbf{p}}{E+M} \right) \left(\begin{array}{cc} 0 & \sigma^k \\ \sigma^k & 0 \end{array}\right) \left(\begin{array}{c} \chi \\ \frac{\mathbf{\sigma}\cdot \mathbf{p}}{E+M} \chi \end{array}\right) \\
  &=& (E+M)\left(\chi'^\dag, -\chi'^\dag \frac{\mathbf{\sigma}\cdot \mathbf{p}}{E+M} \right) \left(\begin{array}{c} \sigma^k \frac{\mathbf{\sigma}\cdot \mathbf{p}}{E+M} \chi \\ \sigma^k \chi \end{array}\right) \\
  &=& \chi'^\dag\left(\sigma^k \mathbf{\sigma}\cdot \mathbf{p} - \mathbf{\sigma} \cdot \mathbf{p} \sigma^k \right)\chi \\
  &=& \chi'^\dag p_i \left[\sigma_k, \sigma_i\right] \chi \\
  &=& 2i \chi'^\dag \varepsilon_{ijk}p_i \sigma_j \chi \\
  &=& 2i \chi'^\dag \mathbf{p} \times \mathbf{\sigma} \chi
\end{eqnarray}
Finally, $\bar{u}(p')u(p)$ is given by:
\begin{eqnarray}
  \bar{u}(p')u(p) &=& (E+M)\left(\chi'^\dag, -\chi'^\dag \frac{\mathbf{\sigma}\cdot \mathbf{p}}{E+M} \right)\left(\begin{array}{c} \chi \\ -\frac{\mathbf{\sigma}\cdot \mathbf{p}}{E+M}\chi \end{array}\right) \\
  &=& \chi'^\dag \chi \left[(E+M) + (E-M)\right] = 2E \chi'^\dag \chi
\end{eqnarray}
The timelike component of the current reduces to:
\begin{eqnarray}
  \mathcal{J}_N^0 &=& \left[2M G_M + \frac{G_E-G_M}{2M(1+\tau)}4M^2(1+\tau) \right]\chi'^\dag \chi \\ 
  &=& 2M G_E \chi'^\dag \chi \label{J0Breit}
\end{eqnarray}
while the spacelike components become:
\begin{eqnarray}
  \vec{\mathcal{J}}_N &=& 2i G_M \chi'^\dag (\mathbf{p}\times\mathbf{\sigma})\chi \label{JkBreit}
\end{eqnarray}
\chapter{F1 TDC Decoding\label{F1decode}}
\paragraph{}
In order to measure the drift times for signals from the FPP drift chambers using the F1 TDCs, one must determine the amount of time elapsed between the time of a hit from the given wire and the time at which a particle passed through the drift chamber. The latter is of course determined by the scintillators. Each F1 TDC module records the time of the stop signal that triggers it to read out its data, attaching this time to its header word. However, this ``trigger'' time is recorded with 7 bits fewer resolution. Whereas the individual TDC channels have approximately 125 ps/count, the stop/trigger time has approximately 16 ns/count, which is insufficiently precise to determine the time difference needed for the drift time. The solution, of course, was to take a copy of the HMS trigger signal and record its time in the F1 TDCs as well, once per VME crate. Then the relevant time difference is the difference between the wire hit time and the measured trigger time. However, since the F1 TDCs are free-running counters, one must carefully account for the possibility of rollover when the TDC count reaches the full-scale range of the TDC. Though the trigger TDC is recorded with 7 bits lower resolution, it does roll over at the same time as the signal TDC, so it can be used to detect rollover unambiguously.

The signal delays are configured so that the hit signals always arrive earlier than the trigger/stop signals. This means that if the TDC does not roll over between the hit and the stop, the raw TDC value of the hit is smaller than that of the stop signal. If the TDC rolls over between the arrival of the hit and the stop, then the hit TDC value will be larger than that of the stop signal. Since the seven most significant bits of the trigger signal are not recorded, the trigger time must be scaled to the full count resolution. This is accomplished by multiplying the trigger TDC value by $2^7=128$ and adding $2^7-1=127$. The reason for setting all of the seven most significant (unknown) bits to 1 is to preserve the chronological ordering of the signal and trigger TDCs (early signal, late trigger) for time differences smaller than the 16 ns resolution difference. Once the trigger TDC has been scaled to the same resolution as the signal TDC in this fashion, the detection of rollover is straightforward. If the rollover occurs between the arrival of the signal and the arrival of the trigger, then the raw signal TDC value will be larger than the trigger TDC value. In the analysis, the full size of the programmed TDC window is added to the signal when the signal TDC is \emph{less} than the trigger TDC, so that in practice, the TDC value is always corrected unless the rollover occurs between the arrival of the signal and the trigger. Alternatively, one could subtract the full window size when the signal TDC exceeds the trigger TDC. The choice is irrelevant as the only thing that matters is for all TDC hits to be in correct chronological order.

\chapter{SIMC: Hall C Monte Carlo \label{SIMCappendix}}
\paragraph{}
The standard Hall C Monte Carlo package SIMC\cite{SIMC} is designed to calculate acceptance functions for the Hall C spectrometers (HMS and SOS) for cross section measurements of coincidence $(e,e'p)$ and inclusive $(e,e')$ or $(e,p)$ reactions. It contains detailed models of the magnetic optics (COSY) and physical apertures of the HMS, including those of the collimator, the magnets, the vacuum enclosure, and lastly, the detectors. SIMC is well suited to simulation of elastic scattering from hydrogen, with BigCal replacing the SOS in the role of the electron arm. 

Since measuring the scattering cross section with any significant accuracy was not the goal of this experiment, a detailed calculation of the acceptance for this purpose was not needed. On the other hand, a realistic simulation of the experiment including elastic scattering and the background processes of $\pi^0$ photoproduction, quasi-elastic $(e,e'p)$ from the target endcaps, and Compton scattering was needed to understand the detailed shape of the signal and background in the space of variables to which elastic event selection cuts were applied, including the proton inelasticity $\Delta_p$ and the position differences $(\Delta x, \Delta y)$ at BigCal. The SIMC background estimate provided an independent check on the methods used to estimate the background directly from the data and helped to determine the uncertainty in the estimated background by comparing the results obtained using different methods. 

SIMC does not simulate the response or efficiency of individual detector channels. SIMC does simulate the effects of energy loss and multiple scattering in all materials traversed by the scattered particles, as well as radiation and the decay of unstable particles. It also simulates the finite resolution of the detectors. Relevant properties of the electron beam are also simulated, including the rastered spot size, the energy spread, and energy loss and multiple scattering in the target up to the scattering vertex. In SIMC, all unique kinematic variables that define an event are generated uniformly within limits calculated to exceed the experimental acceptances. For all events, the position of the interaction vertex is generated uniformly along the target length and within the size of the (rastered) beam spot. For elastic $ep$ scattering, the angles of the electron are generated, and all other quantities are calculated from the two-body reaction kinematics. In the case of quasi-elastic $(e,e'p)$ scattering from heavy nuclei, the electron and proton angles and momenta are all generated. For pion photoproduction and Compton scattering, the proton angles and the energy of the Bremsstrahlung photon radiated from the incident beam are generated, and the scattering angles and momentum of the $\pi^0$ (for photoproduction) or the Compton-scattered $\gamma$ are calculated from the two-body reaction kinematics. 

For each event, a weight is assigned which takes into account the cross section of the reaction in question as a function of the relevant kinematic variables, the luminosity, and the phase-space volume of the generation region. In the case of elastic $ep$ scattering and quasi-elastic $(e,e'p)$ scattering from the target endcaps, internal and external radiative corrections to the cross section are calculated\cite{hornthesis,EntRadCorr} based on the formalism of Mo and Tsai\cite{MoTsai} by randomly generating the energy of the radiated photon from each of the incident and scattered particles, and then adjusting the kinematics of the particles from which radiation occured accordingly. The weight of each radiated event is then adjusted according to the cross section for radiation at the generated energy. For pion photoproduction and Compton scattering, the flux of incident Bremsstrahlung photons at the generated photon energy as a function of the position along the target and the current and energy of the incident electron beam is calculated and included in the event weight along with the reaction cross section. No radiative corrections are applied to the pion photoproduction and Compton scattering cross sections. 

The cross section model for elastic $ep$ scattering in SIMC uses the Bosted fit to the proton electromagnetic form factors\cite{BostedFit} together with the Rosenbluth formula to calculate the (unradiated) cross section. For the target endcaps, the spectral function for carbon built into SIMC is used to calculate the $^{12}C(e,e'p)$ cross section\footnote{Since the goal of the simulation was not to understand the detailed nuclear physics of the the target endcap background but merely to obtain the lineshape of this background, and since the well-known carbon-12 cross section is already built in to SIMC, it was deemed sufficient to use carbon instead of aluminum for this study.}. Although data were taken with the dummy target to measure the aluminum contribution at $Q^2 = 8.5$ GeV$^2$, the data were insufficient to determine the detailed shape of the background, as only $\approx 50$ events from the dummy target passed all the elastic cuts. For pion photoproduction, the cross section was parametrized in terms of Mandelstam $s$ and the production angle $\Theta_{CM}$ of the $\pi^0$ in the photon-proton center-of-mass frame as 
\begin{eqnarray}
  \frac{d\sigma}{dt} &=& C(1+\cos \Theta_{CM})^{-4}(1-\cos \Theta_{CM})^{-5}\left(\frac{s_0}{s}\right)^7
\end{eqnarray}
with parameters $C$ and $s_0$ determined by the data of \cite{Anderson1976}. For Compton scattering, an even simpler approximation to the cross section was used. Since there is no way to reliably separate Compton scattering from elastic $ep$ scattering in the data of this experiment, and since the Compton scattering cross section is reasonably assumed to be quite small compared to the photoproduction cross section, a very rough extrapolation from the data of \cite{Danagoulian2007} to the kinematics of this experiment was performed assuming no $t$-dependence over the range of extrapolation in $t$ and an $s^{-8}$ scaling of $d\sigma/dt$ in order to obtain an order-of-magnitude estimate of the Compton contribution to the data, which was indeed found to be quite small.

After generating and ``radiating'' an event, the scattered proton is transported forward through the HMS to the detector hut, checking all physical and detector apertures along the way. The forward and reverse transport matrices used in SIMC are those of the HMS COSY model described above for use as the starting point in fitting the true optics matrices and for calculating the proton spin precession. The role of S0 in the trigger and as a source of multiple scattering is accounted for. The trajectory of the scattered electron (elastic/quasi-elastic) or photon(s) (photoproduction/Compton scattering) is projected to BigCal. In $\pi^0$ photoproduction events, the angle of the back-to-back decay photons is generated isotropically in the $\pi^0$ rest frame and both photons are subsequently boosted to the lab frame. 

If the proton passes through the HMS and hits all the detectors required in the trigger (S1X, S1Y and S0) and the electron (or at least one photon) hits BigCal with an energy exceeding the trigger threshold, the event is considered successful. In order to save computing time, SIMC calculates the cross section weight only after determining whether an event will contribute. Events detected in both the electron arm (BigCal) and the proton arm (HMS) are then reconstructed, taking detector resolution and multiple scattering into account, and the target quantities $x'_{tar}$, $y'_{tar}$, $y_{tar}$, and $\delta$ are reconstructed using the reverse COSY matrix elements. Just as in the reconstruction of the data, the variables are reconstructed a second time to correct for the $x_{tar}$ effect discussed in section \ref{opticssection}. More detailed documentation of SIMC can be found in \cite{SIMC,hornthesis} and references therein.

\begin{singlespace}
\bibliography{main}

\newcommand{\noopsort}[1]{} \newcommand{\printfirst}[2]{#1}
  \newcommand{\singleletter}[1]{#1} \newcommand{\switchargs}[2]{#2#1}
\begin{thebibliography}{100}

\bibitem{Stern1933}
O.~Stern.
\newblock {\em Nature}, 132:169, 1933.

\bibitem{BlochAlvarez1940}
F.~Bloch and L.~W. Alvarez.
\newblock A quantitative determination of the neutron moment in absolute
  nuclear magnetons.
\newblock {\em Physical Review}, 57:111, 1940.

\bibitem{GeigerMarsden1909}
H.~Geiger and E.~Marsden.
\newblock On a diffuse reflection of the $\alpha$ particles.
\newblock {\em Proc. Royal Society of London, Series A}, 82:495, 1909.

\bibitem{Rutherford1911}
E.~Rutherford.
\newblock The scattering of $\alpha$ and $\beta$ particles by matter and the
  structure of the atom.
\newblock {\em Philosophical Magazine}, 21:669, 1911.

\bibitem{GeigerMarsden1913}
H.~Geiger and E.~Marsden.
\newblock The laws of deflexion of $\alpha$ particles through large angles.
\newblock {\em Philosophical Magazine}, 25(148), 1913.

\bibitem{Hof55}
Robert Hofstadter and Robert McAllister.
\newblock Electron {S}cattering from the {P}roton.
\newblock {\em Physical Review}, 98:217, 1955.

\bibitem{Hof56}
Robert Hofstadter and Robert McAllister.
\newblock Elastic {S}cattering of 188-{M}e{V} {E}lectrons from the {P}roton and
  the {A}lpha {P}article.
\newblock {\em Physical Review}, 102:851, 1956.

\bibitem{HofRev56}
Robert Hofstadter.
\newblock Electron {S}cattering and {N}uclear {S}tructure.
\newblock {\em Reviews of Modern Physics}, 28(3):214, 1956.

\bibitem{PDG2008}
C.~Amsler et~al.
\newblock ({P}article {D}ata {G}roup) 2008 {R}eview of {P}article {P}hysics.
\newblock {\em Physics Letters B}, 667:1, 2008.

\bibitem{PeskinSchroeder}
Michael~E. Peskin and Daniel~V. Schroeder.
\newblock {\em An Introduction to Quantum Field Theory}.
\newblock Westview Press, 1995.

\bibitem{QandL}
Francis Halzen and Alan~D. Martin.
\newblock {\em Quarks and Leptons: An Introductory Course in Modern Particle
  Physics}.
\newblock John Wiley and Sons, 1984.

\bibitem{Rosenbluth1950}
M.~N. Rosenbluth.
\newblock High energy elastic scattering of electrons on protons.
\newblock {\em Physical Review}, 79(4):615, 1950.

\bibitem{AkhiezerRekalo1}
A.~I. Akhiezer and M.~P. Rekalo.
\newblock Polarization phenomena in electron scattering by protons in the
  high-energy region.
\newblock {\em Doklady Akademii Nauk SSSR}, 180(5):1081, 1968.

\bibitem{AkhiezerRekalo2}
A.~I. Akhiezer and M.~P. Rekalo.
\newblock Polarization effects in the scattering of leptons by hadrons.
\newblock {\em Sov. J. Part. Nucl.}, 3:277, 1974.

\bibitem{ArnoldCarlsonGross}
Raymond~G. Arnold, Carl~E. Carlson, and Franz Gross.
\newblock Polarization transfer in elastic electron scattering from nucleons
  and deuterons.
\newblock {\em Physical Review C}, 23(1):363, 1981.

\bibitem{Dombey1969}
Norman Dombey.
\newblock Scattering of polarized leptons at high energy.
\newblock {\em Reviews of Modern Physics}, 41(1):236, 1969.

\bibitem{DonnellyRaskin86}
T.~W. Donnelly and A.~S. Raskin.
\newblock Considerations of polarization in inclusive electron scattering from
  nuclei.
\newblock {\em Annals of Physics}, 169:247, 1986.

\bibitem{PerdrisatPunjabiVanderhaegen2007}
C.~Perdrisat, V.~Punjabi, and M.~Vanderhaeghen.
\newblock Nucleon electromagnetic form factors.
\newblock {\em Progress in Particle and Nuclear Physics}, 59:694, 2007.

\bibitem{Andi94}
L.~Andivahis et~al.
\newblock Measurements of the electric and magnetic form factors of the proton
  from {$Q^2 = 1.75$ to $8.83\ (GeV/c)^2$}.
\newblock {\em Physical Review {D}}, 50(9):5491, 1994.

\bibitem{Berger71}
Ch. Berger et~al.
\newblock Electromagnetic form factors of the proton at squared four-momentum
  transfers between 10 and 50 fm$^{-2}$.
\newblock {\em Physics Letters B}, 35(1):87, 1971.

\bibitem{Bork75}
F.~Borkowski et~al.
\newblock Electromagnetic form factors of the proton at low four-momentum
  transfer ({II}).
\newblock {\em Nuclear Physics B}, 93:461, 1975.

\bibitem{Christy04}
M.~E. Christy et~al.
\newblock Measurements of electron-proton elastic cross sections for
  {$0.4<Q^2<5.5\ (GeV/c)^2$}.
\newblock {\em Physical Review C}, 70:015206, 2004.

\bibitem{Janssens66}
T.~Janssens et~al.
\newblock Proton form factors from elastic electron-proton scattering.
\newblock {\em Physical Review}, 142(4):922, 1966.

\bibitem{Price71}
L.~E. Price et~al.
\newblock Backward-angle electron-proton elastic scattering and proton
  electromagnetic form factors.
\newblock {\em Physical Review D}, 4(1):45, 1971.

\bibitem{Qattan05}
I.~A. Qattan et~al.
\newblock Precision {R}osenbluth measurement of the proton elastic form
  factors.
\newblock {\em Physical Review Letters}, 94:142301, 2005.

\bibitem{Simon80}
G.~G. Simon et~al.
\newblock Absolute electron-proton cross sections at low momentum transfer
  measured with a high pressure gas target system.
\newblock {\em Nuclear Physics A}, 333:381, 1980.

\bibitem{Walker94}
R.~C. Walker et~al.
\newblock Measurements of the proton elastic form factors for {$1\leq Q^2 \leq
  3\ (GeV/c)^2$} at {SLAC}.
\newblock {\em Physical Review D}, 49(11):5671, 1994.

\bibitem{Bartel73}
W.~Bartel et~al.
\newblock Measurement of proton and neutron electromagnetic form factors at
  squared four-momentum transfers up to 3 {$(GeV/c)^2$}.
\newblock {\em Nuclear Physics B}, 58:429, 1973.

\bibitem{Sill93}
A.~F. Sill et~al.
\newblock Measurements of elastic electron-proton scattering at large momentum
  transfer.
\newblock {\em Physical Review D}, 48(1):29, 1993.

\bibitem{Crawford07}
C.~Crawford et~al.
\newblock Measurement of the proton's electric to magnetic form factor ratio
  from {$^1\vec{H}(\vec{e},e'p)$}.
\newblock {\em Physical Review Letters}, 98:052301, 2007.

\bibitem{Gayou02}
O.~Gayou et~al.
\newblock Measurement of {$G_E^p/G_M^p$} in {$\vec{e}p \rightarrow e \vec{p}$}
  to {$Q^2 = 5.6\ GeV^2$}.
\newblock {\em Physical Review Letters}, 88(9):092301, 2002.

\bibitem{Jones00}
M.~K. Jones et~al.
\newblock {$G_E^p/G_M^p$} ratio by polarization transfer in {$\vec{e}p
  \rightarrow e\vec{p}$}.
\newblock {\em Physical Review Letters}, 84(7):1398, 2000.

\bibitem{Punjabi05}
V.~Punjabi et~al.
\newblock Proton elastic form factor ratios to {$Q^2 = 3.5\ GeV^2$} by
  polarization transfer.
\newblock {\em Physical Review C}, 71:055202, 2005.

\bibitem{Jones06}
M.~K. Jones et~al.
\newblock Proton {$G_E/G_M$} from beam-target asymmetry.
\newblock {\em Physical Review C}, 74:035201, 2006.

\bibitem{Maclachlan06}
G.~Maclachlan et~al.
\newblock The ratio of proton electromagnetic form factors via recoil
  polarimetry at {$Q^2=1.13\ (GeV/c)^2$}.
\newblock {\em Nuclear Physics A}, 764:261, 2006.

\bibitem{Milbrath98}
B.~D. Milbrath et~al.
\newblock Comparison of polarization observables in electron scattering from
  the proton and deuteron.
\newblock {\em Physical Review Letters}, 80(3):452, 1998.

\bibitem{Ron07}
G.~Ron et~al.
\newblock Measurements of the proton elastic form factor ratio {$\mu_p
  G_E^p/G_M^p$} at low momentum transfer.
\newblock {\em Physical Review Letters}, 99:202002, 2007.

\bibitem{GuichonVanderhaeghenTPEX}
P.~A.~M. Guichon and M.~Vanderhaeghen.
\newblock {\em Phys. Rev. Lett.}, 91:142303, 2003.

\bibitem{Anklin94}
H.~Anklin et~al.
\newblock Precision measurement of the neutron magnetic form factor.
\newblock {\em Physics Letters B}, 336:313, 1994.

\bibitem{Anklin98}
H.~Anklin et~al.
\newblock Precise measurements of the neutron magnetic form factor.
\newblock {\em Physics Letters B}, 428:248, 1998.

\bibitem{Brooks06}
W.~K. Brooks and J.~D. Lachniet.
\newblock {(for the CLAS Collaboration)} {P}recise determination of the neutron
  magnetic form factor to higher {$Q^2$}.
\newblock {\em Nuclear Physics A}, 755:261, 2005.

\bibitem{Bruins95}
E.~E.~W. Bruins et~al.
\newblock Measurement of the neutron magnetic form factor.
\newblock {\em Physical Review Letters}, 75(1):21, 1995.

\bibitem{Gao94}
H.~Gao et~al.
\newblock Measurement of the neutron magnetic form factor from inclusive
  quasielastic scattering of polarized electrons from polarized {$^3He$}.
\newblock {\em Physical Review C}, 50(2):546, 1994.

\bibitem{JLab07}
Jefferson {L}ab {E}95-001~{C}ollaboration.
\newblock Extraction of the neutron magnetic form factor from quasielastic
  {$^3\vec{He}(\vec{e},e')$} at {$Q^2=0.1$-$0.6\ (GeV/c)^2$}.
\newblock {\em Physical Review C}, 75:034003, 2007.

\bibitem{Kubon02}
G.~Kubon et~al.
\newblock Precise neutron magnetic form factors.
\newblock {\em Physics Letters B}, 524:26, 2002.

\bibitem{Lung93}
A.~Lung et~al.
\newblock Measurements of the electric and magnetic form factors of the neutron
  from {$Q^2=1.75$} to {$4.00\ (GeV/c)^2$}.
\newblock {\em Physical Review Letters}, 70(6):718, 1993.

\bibitem{Markowitz93}
P.~Markowitz et~al.
\newblock Measurement of the magnetic form factor of the neutron.
\newblock {\em Physical Review C}, 48(1):5, 1993.

\bibitem{Rock82}
S.~Rock et~al.
\newblock Measurement of elastic electron-neutron cross sections up to {$Q^2 =
  10\ (GeV/c)^2$}.
\newblock {\em Physical Review Letters}, 49(16):1139, 1982.

\bibitem{Rock92}
S.~Rock et~al.
\newblock Measurement of elastic electron-neutron scattering and inelastic
  electron-deuteron scattering cross sections at high momentum transfer.
\newblock {\em Physical Review D}, 46(1):24, 1992.

\bibitem{Xu00}
W.~Xu et~al.
\newblock Transverse asymmetry {$A_{T'}$} from the quasielastic
  {$^3\vec{He}(\vec{e},e')$} process and the neutron magnetic form factor.
\newblock {\em Physical Review Letters}, 85(14):2900, 2000.

\bibitem{Xu03}
W.~Xu et~al.
\newblock Plane-wave impulse approximation extraction of the neutron magnetic
  form factor from quasielastic {$^3\vec{He}(\vec{e},e')$} at {$Q^2=0.3$} to
  {$0.6\ (GeV/c)^2$}.
\newblock {\em Physical Review C}, 67:012201, 2003.

\bibitem{Bermuth03}
J.~Bermuth et~al.
\newblock The neutron charge form factor and target analyzing powers from
  {$^3\vec{He}(\vec{e},e'n)$} scattering.
\newblock {\em Physics Letters B}, 564:199, 2003.

\bibitem{Eden94}
T.~Eden et~al.
\newblock Electric form factor of the neutron from the
  {$^2H(\vec{e},e'\vec{n})^1H$} reaction at {$Q^2=0.255\ (GeV/c)^2$}.
\newblock {\em Physical Review C}, 50(4):1749, 1994.

\bibitem{Geis08}
E.~Geis et~al.
\newblock Charge form factor of the neutron at low momentum transfer from the
  {$^2\vec{H}(\vec{e},e'n)^1H$} reaction.
\newblock {\em Physical Review Letters}, 101:042501, 2008.

\bibitem{Glazier05}
G.~I. Glazier et~al.
\newblock Measurement of the electric form factor of the neutron at
  {$Q^2=0.3$-$0.8\ (GeV/c)^2$}.
\newblock {\em The European Physical Journal A}, 24:101, 2005.

\bibitem{Herberg99}
C.~Herberg et~al.
\newblock Determination of the neutron electric form factor in the
  {$D(e,e'n)p$} reaction and the influence of nuclear binding.
\newblock {\em The European Physical Journal A}, 5:131, 1999.

\bibitem{Meyerhoff93}
M.~Meyerhoff et~al.
\newblock First measurement of the electric form factor of the neutron in the
  exclusive quasielastic scattering of polarized electrons from polarized
  {$^3He$}.
\newblock {\em Physics Letters B}, 327:201, 1994.

\bibitem{Ostrick99}
M.~Ostrick et~al.
\newblock Measurement of the neutron electric form factor {$G_E^n$} in the
  quasifree {$^2H(\vec{e},e'\vec{n})p$} reaction.
\newblock {\em Physical Review Letters}, 83(2):276, 1999.

\bibitem{Passchier99}
I.~Passchier et~al.
\newblock Charge form factor of the neutron from the reaction
  {$^2\vec{H}(\vec{e},e'n)p$}.
\newblock {\em Physical Review Letters}, 82(25):4988, 1999.

\bibitem{Plaster06}
B.~Plaster et~al.
\newblock Measurements of the neutron electric to magnetic form factor ratio
  {$G_E^n/G_M^n$} via the {$^2H(\vec{e},e'\vec{n})^1H$} reaction to {$Q^2=1.45\
  (GeV/c)^2$}.
\newblock {\em Physical Review C}, 73:025205, 2006.

\bibitem{Rohe99}
D.~Rohe et~al.
\newblock Measurement of the neutron electric form factor {$G_E^n$} at {$0.67\
  (GeV/c)^2$} via {$^3\vec{He}(\vec{e},e'n)$}.
\newblock {\em Physical Review Letters}, 83(21):4257, 1999.

\bibitem{Warren04}
G.~Warren et~al.
\newblock Measurement of the electric form factor of the neutron at {$Q^2=0.5$}
  and {$1.0\ GeV^2/c^2$}.
\newblock {\em Physical Review Letters}, 92(4):042301, 2004.

\bibitem{Zhu01}
H.~Zhu et~al.
\newblock Measurement of the electric form factor of the neutron through
  {$\vec{d}(\vec{e},e'n)p$} at {$Q^2=0.5\ (GeV/c)^2$}.
\newblock {\em Physical Review Letters}, 87(8):081801, 2001.

\bibitem{KellyChargeDist}
J.~J. Kelly.
\newblock {\em Phys. Rev. C}, 66:065203, 2002.

\bibitem{MitraKumari}
A.~N. Mitra and I.~Kumari.
\newblock {\em Phys. Rev. D}, 15:261, 1977.

\bibitem{Iachello1973}
F.~Iachello, A.~D. Jackson, and Land\'{e}.
\newblock {\em Phys. Lett. B}, 43:191, 1973.

\bibitem{GK1985}
M.~F. Gari and W.~Kr\"{u}mpelmann.
\newblock {\em Z. Phys. A}, 322:689, 1985.

\bibitem{GK1992}
M.~F. Gari and W.~Kr\"{u}mpelmann.
\newblock {\em Phys. Lett. B}, 274:159, 1992.

\bibitem{GK1992E}
M.~F. Gari and W.~Kr\"{u}mpelmann.
\newblock {\em Phys. Lett. B}, 282:483, 1992.
\newblock Errata.

\bibitem{Lomon2001}
E.~L. Lomon.
\newblock {\em Phys. Rev. C}, 64:035204, 2001.

\bibitem{MergellMeissnerDrechsel}
P.~Mergell, Ulf-G. Meissner, and D.~Drechsel.
\newblock {\em Nucl. Phys. A}, 596:367, 1996.

\bibitem{Lomon2002}
E.~L. Lomon.
\newblock {\em Phys. Rev. C}, 66:045501, 2002.

\bibitem{Iachello2004}
R.~Bijker and F.~Iachello.
\newblock {\em Phys. Rev. C}, 69:068201, 2004.

\bibitem{Hohler}
G.~H\"{o}hler et~al.
\newblock {\em Nucl. Phys. B}, 114:505, 1976.

\bibitem{HammerMeissner2004}
H.-W. Hammer and U.-G. Meissner.
\newblock {\em Eur. Phys. J. A}, 20:469, 2004.

\bibitem{Belushkin2007}
M.~A. Belushkin, H.-W. Hammer, and Ulf-G. Meissner.
\newblock {\em Phys. Rev. C}, 75:035202, 2007.

\bibitem{IsgurKarl}
N.~Isgur and G.~Karl.
\newblock {\em Phys. Rev. D}, 18:4187, 1978.

\bibitem{DiracRelativisticDynamics}
P.~A.~M. Dirac.
\newblock {\em Rev. Mod. Phys.}, 21:392, 1949.

\bibitem{SchlumpfCQM}
F.~Schlumpf.
\newblock {\em Phys. Rev. D.}, 47:4114, 1993.

\bibitem{SchlumpfFF}
F.~Schlumpf.
\newblock {\em J. Phys. G}, 20:237, 1994.

\bibitem{FrankJenningsMiller}
M.~R. Frank, B.~K. Jennings, and G.~A. Miller.
\newblock {\em Phys. Rev. C}, 54:920, 1996.

\bibitem{MillerFrank}
G.~A. Miller and M.~R. Frank.
\newblock {\em Phys. Rev. C}, 65:065205, 2002.

\bibitem{MeloshRotation}
H.~J. Melosh.
\newblock {\em Phys. Rev. D}, 9:1095, 1974.

\bibitem{MillerLFCBM}
G.~A. Miller.
\newblock {\em Phys. Rev. C}, 66:032201, 2002.

\bibitem{Zoller1992}
V.~R. Zoller.
\newblock {\em Z. Phys. C}, 53:443, 1992.

\bibitem{Holtmann1996}
H.~Holtmann, A.~Szczurek, and J.~Speth.
\newblock {\em Nucl. Phys. A}, 596:631, 1996.

\bibitem{StructNucleon}
Anthony~W. Thomas and Wolfram Weise.
\newblock {\em The Structure of the Nucleon}.
\newblock WILEY-VCH, 2001.

\bibitem{BrodskyFarrar1975}
S.~J. Brodsky and G.~R. Farrar.
\newblock {\em Phys. Rev. D}, 11:1309, 1975.

\bibitem{BrodskyLepage1979}
S.~J. Brodsky and G.~P. Lepage.
\newblock {\em Phys. Rev. Lett.}, 43:545, 1979.

\bibitem{Kirk1973}
P.~N. Kirk et~al.
\newblock {\em Phys. Rev. D}, 8:63, 1973.

\bibitem{BelitskyJiYuan2003}
A.~V. Belitsky, X.~Ji, and F.~Yuan.
\newblock {\em Phys. Rev. Lett.}, 91:092003, 2003.

\bibitem{HardExclusiveReview2001}
K.~Goeke, M.~V. Polyakov, and M.~Vanderhaeghen.
\newblock {\em Progress in Particle and Nuclear Physics}, 47:401, 2001.

\bibitem{JiFactorization}
X.~D. Ji and J.~Osborne.
\newblock {\em Phys. Rev. D}, 58:094018, 1998.

\bibitem{CollinsFactorization}
J.~C. Collins and A.~Freund.
\newblock {\em Phys. Rev. D}, 59:074009, 1999.

\bibitem{RadyushkinFactorization}
A.~V. Radyushkin.
\newblock {\em Phys. Rev. D}, 58:114008, 1998.

\bibitem{JiGPDPRL1997}
X.~D. Ji.
\newblock {\em Phys. Rev. Lett.}, 78:610, 1997.

\bibitem{JiGPDPRD1997}
X.~D. Ji.
\newblock {\em Phys. Rev. D}, 55:7114, 1997.

\bibitem{RadyushkinGPD}
A.~V. Radyushkin.
\newblock {\em Phys. Lett. B}, 380:417, 1996.

\bibitem{GuidalGPD2005}
M.~Guidal, M.~V. Polyakov, A.~V. Radyushkin, and M.~Vanderhaeghen.
\newblock {\em Phys. Rev. D}, 72:054013, 2003.

\bibitem{MRST2002}
A.~D. Martin, R.~G. Roberts, W.~J. Stirling, and R.~S. Thorne.
\newblock {\em Phys. Lett. B}, 531:216, 2002.

\bibitem{BurkardtIJMPA}
M.~Burkardt.
\newblock {\em Int. J. Mod. Phys. A}, 18:173, 2003.

\bibitem{MillerGPDchargedensity2007}
G.~A. Miller.
\newblock {\em Phys. Rev. Lett.}, 99:112001, 2007.

\bibitem{DiehlGPDtomography2005}
Diehl. M. et~al.
\newblock {\em Eur. Phys. J. C}, 39:1, 2005.

\bibitem{StutzmanPhysToday08}
C.~Hernandez-Garcia, P.~G. O'Shea, and M.~L. Stutzman.
\newblock Electron sources for accelerators.
\newblock {\em Physics Today}, page~44, February 2008.

\bibitem{LeemannCEBAF}
C.~W. Leeman, D.~R. Douglas, and G.~A. Krafft.
\newblock The {C}ontinuous {E}lectron {B}eam {A}ccelerator {F}acility.
\newblock {\em Annual Reviews of Nuclear and Particle Science}, 51:413, 2001.

\bibitem{GueyeHallCBPM}
Paul Gueye.
\newblock Status of the actual beam position monitors in the {H}all {C}
  beamline.
\newblock Unpublished internal document, December 1995.

\bibitem{YanSuperHarp}
C.~Yan et~al.
\newblock Superharp--a wire scanner with absolute position readout for beam
  energy measurement at {CEBAF}.
\newblock {\em Nuclear Instruments and Methods {A}}, 365:261, 1995.

\bibitem{GNhallcBCM}
G.~Niculescu.
\newblock Resonant cavities used as beam current monitors.
\newblock Unpublished internal document, 1995.

\bibitem{CAhallcBCM}
C.~Armstrong.
\newblock Beam current measurement in {Hall C}.
\newblock Unpublished internal document, July 1996.

\bibitem{UnserAIP}
K.~B. Unser.
\newblock The parametric current transformer, a beam current monitor developed
  for {LEP}.
\newblock In {\em A.I.P. Conf. Proc.}, volume 252, page 266, 1992.

\bibitem{YanArcEnergy}
C.~Yan, R.~Carlini, and D.~Neuffer.
\newblock Beam energy measurement using the arc beam line as a spectrometer.
\newblock Technical report, CEBAF-PR-93-004, 1993.

\bibitem{SLIenergyspread}
P.~Chevtsov et~al.
\newblock Non-invasive energy spread monitoring for the {JLAB} experimental
  program via synchrotron light interferometers.
\newblock {\em Nuclear Instruments and Methods A}, 557:324, 2006.

\bibitem{MollerNIM}
M.~Hauger et~al.
\newblock A high-precision polarimeter.
\newblock {\em Nuclear Instruments and Methods in Physics Research {A}},
  462:382, 2001.

\bibitem{YanHallCraster}
C.~Yan et~al.
\newblock Target raster system at {CEBAF}.
\newblock {\em Nuclear Instruments and Methods A}, 365:46, 1995.

\bibitem{MeekHallCconfig}
David Meekins.
\newblock Hall {C} target configuration.
\newblock Unpublished internal report, November 2007.

\bibitem{HMSresolution}
L.~Tang, C.~Yan, and Ed.~V. Hungerford.
\newblock Systematic resolution study of the {CEBAF} {H}all {C} spectrometers -
  {HMS} and {SOS}.
\newblock {\em Nuclear Instruments and Methods A}, 366:259, 1995.

\bibitem{Leo}
W.~R. Leo.
\newblock {\em Techniques for Nuclear and Particle Physics Experiments}.
\newblock Springer-Verlag, 1987.

\bibitem{HMSdriftchamber}
O.~K. Baker et~al.
\newblock The {H}igh {M}omentum {S}pectrometer drift chambers in {H}all {C} at
  {CEBAF}.
\newblock {\em Nuclear Instruments and Methods {A}}, 367:92, 1995.

\bibitem{Ay_pp}
D.~Miller et~al.
\newblock Simultaneous measurement of the spin parameters $p$ and $c_{NN}$ in
  $pp$ elastic scattering at 2, 3, 4, and 6 {G}e{V}/c.
\newblock {\em Physical Review D}, 16:2016, 1977.

\bibitem{Ay_pCH2}
L.~S. Azghirey et~al.
\newblock Measurement of analyzing powers for the reaction $\vec{p}+${CH}$_2$
  at $p_p=$1.75-5.3 {G}e{V}/c.
\newblock {\em Nuclear Instruments and Methods A}, 538:431, 2005.

\bibitem{IHEP_leadglass}
L.~C. Bland et~al.
\newblock Electromagnetic shower profile in lead-glass calorimeter in the
  energy range of 3-23 {GeV}.
\newblock IHEP Preprint, Protvino, Russia., 2005.

\bibitem{TS_manual}
{JL}ab {CODA}~group.
\newblock Trigger {S}upervisor {M}anual.
\newblock Unpublished internal document, 1996.

\bibitem{TS_proceedings}
E.~Jastrzembski et~al.
\newblock The {J}efferson {L}ab trigger supervisor system.
\newblock In {\em Proc. 1999 IEEE Conference on Real-Time Computer Applications
  in Nuclear Particle and Plasma Physics. 11th IEEE NPSS Real Time
  Conference.}, 1999.

\bibitem{CODA_proc}
G.~Heyes et~al.
\newblock The {CEBAF} on-line data acquisition system.
\newblock In {\em Proc. of the {CHEP} conference}, April 1994.

\bibitem{ArringtonThesis}
John~R. Arrington.
\newblock {\em Inclusive electron scattering from nuclei at $x>1$ and high
  $Q^2$}.
\newblock {Ph.D.} thesis, California Institute of Technology, 1998.

\bibitem{CMOP}
K.~A. Assamagan, D.~Dutta, and P.~Welch.
\newblock Hall {C} matrix element optimization package ({CMOP}).
\newblock Unpublished internal document, May 1997.

\bibitem{COSY}
K.~Makino and M.~Berz.
\newblock {COSY} infinity version 8.
\newblock {\em Nuclear Instruments and Methods A}, 427:338, 1999.

\bibitem{JonesPrivateComm1}
Mark Jones.
\newblock Private communication.

\bibitem{Anderson1976}
R.~L. Anderson et~al.
\newblock Measurements of exclusive photoproduction processes at large values
  of $t$ and $u$ from 4 to 7.5 {GeV}.
\newblock {\em Physical Review D}, 14:679, 1976.

\bibitem{Danagoulian2007}
A.~Danagoulian et~al.
\newblock Compton-scattering cross section on the proton at high momentum
  transfer.
\newblock {\em Physical Review Letters}, 98:152001, 2007.

\bibitem{Shupe1979}
M.~A. Shupe et~al.
\newblock Neutral-pion photoproduction and proton {C}ompton scattering at large
  angles.
\newblock {\em Physical Review D}, 19:1921, 1979.

\bibitem{BMTequation}
V.~Bargmann, Louis Michel, and V.~L. Telegdi.
\newblock Precession of the polarization of particles moving in a homogeneous
  electromagnetic field.
\newblock {\em Physical Review Letters}, 2:435, 1959.

\bibitem{PentchevSpinHRS}
L.~Pentchev.
\newblock Spin transport in the {HRS}s.
\newblock Technical report, Jefferson Lab, 2003.
\newblock JLAB-TN-03-024.

\bibitem{Besset}
D.~Besset et~al.
\newblock {\em Nuclear Instruments and Methods}, 166:515, 1979.

\bibitem{AfanasevRadCorr}
A.~Afanasev et~al.
\newblock {\em Phys. Rev. D}, 64:113009, 2001.

\bibitem{SIMC}
John Arrington.
\newblock {A-B-SIMC}.
\newblock Internal {H}all {C} document., 2001.

\bibitem{hornthesis}
T.~Horn.
\newblock {\em The pion charge form factor through electroproduction}.
\newblock {Ph.D.} thesis, University of Maryland, 2006.

\bibitem{EntRadCorr}
R.~Ent et~al.
\newblock {\em Physical Review C}, 64:054610, 2001.

\bibitem{MoTsai}
L.~M. Mo and Y.~S. Tsai.
\newblock {\em Rev. Mod. Phys.}, 41:205, 1969.

\bibitem{BostedFit}
P.~E. Bosted.
\newblock {\em Phys. Rev. C}, 51:409, 1995.

\end{thebibliography}
\bibliographystyle{unsrt}
\end{singlespace}

\end{document}